\def\versionno{Principles of heliophysics, V\,2.1}
\def\version2date{March 27, 2024}
\def\dateofthisversion{\version2date}
\def\versiondate{\versionno}
\date{}

\documentclass[11pt,twoside]{book}
\usepackage[paperwidth=17.78cm,paperheight=25.4cm]{geometry}
\usepackage{latex/named,bm,amssymb,graphicx,pifont,multirow,amsmath,rotating,hhline,lscape,float,textcomp,latexsym,wasysym}
\usepackage{fancyhdr} 
\usepackage{framed,microtype,manyfoot}
\usepackage[font={small,it},indention=0.5cm,labelfont=bf]{caption}

\usepackage[compact]{titlesec}
    \titlespacing{\section}{0pt}{2ex}{1ex}
    \titlespacing{\subsection}{0pt}{1ex}{0ex}
    \titlespacing{\subsubsection}{0pt}{0.5ex}{0ex}

\usepackage{imakeidx}
\makeindex[options=-s latex/myindexstyle,title=Subject
index,columnsep=10pt,intoc,columns=3]

\usepackage[hidelinks,hyperfootnotes=false,bookmarksnumbered=true]{hyperref}

\usepackage{xcolor}
\hypersetup{
    colorlinks,
    linkcolor={red!60!black},
    citecolor={green!50!black},
    urlcolor={blue!80!black}
}
\usepackage[all]{hypcap}

\usepackage{endnotes,chngcntr}
\makeatletter
\long\def\@endnotetext#1{%
\if@enotesopen \else \@openenotes \fi
\immediate\write\@enotes{\@doanenote{\@theenmark}}
\begingroup
\def\next{#1}
\newlinechar='40
\immediate\write\@enotes{\meaning\next}%
\endgroup
\immediate\write\@enotes{/p.~\thepage/ \@endanenote}}
\makeatother

\makeatletter
\renewcommand\enoteheading{%
  \setcounter{secnumdepth}{-2}
  \mbox{}\par\vskip-\baselineskip
  \let\@afterindentfalse\@afterindenttrue
}
\makeatother

\usepackage{xparse}

\let\latexchapter\chapter

\RenewDocumentCommand{\chapter}{som}{%
  \IfBooleanTF{#1}
    {\latexchapter*{#3}}
    {\IfNoValueTF{#2}
       {\latexchapter{#3}}
       {\latexchapter[#2]{#3}}%
     \addtoendnotes{%
       \noexpand\enotedivision{\noexpand\section}%
         {\chaptername\ \thechapter. \unexpanded{#3}}}%
    }%
}
\makeatletter
\def\enotedivision#1#2{\@ifnextchar\enotedivision{}{#1{#2}}}
\makeatletter

\newcounter{notelabelcount}
\def\activity#1{
  \renewcommand{\makeenmark}{\hbox{{\tiny \,\{A\theenmark\}\,}}}
  \addtocounter{notelabelcount}{1}%
  \hyperref[notelabel\thenotelabelcount]{%
    \endnote{%
    \addtocounter{notelabelcount}{1}%
      \phantomsection\label{notelabel\thenotelabelcount}%
   {#1}}}
  \addtoendnotes{\vskip 1mm}
  \marginpar{\hyperref[notelabel\thenotelabelcount]{\rotatebox{-90}{\tiny {\{A\theenmark\}}}}}
}
\def\sactivity#1{
  \renewcommand{\makeenmark}{\hbox{{\tiny \,\{$\circledS$A\theenmark\}\,}}}
  \addtocounter{notelabelcount}{1}%
  \hyperref[notelabel\thenotelabelcount]{%
    \endnote{%
    \addtocounter{notelabelcount}{1}%
      \phantomsection\label{notelabel\thenotelabelcount}%
   {#1}}}
  \addtoendnotes{\vskip 1mm}
  \marginpar{\hyperref[notelabel\thenotelabelcount]{\rotatebox{-90}{\tiny {\{$\circledS$A\theenmark\}}}}}
}
\def\sfirstactivity#1{
  \{$\circledS$0\}
   \marginpar{\rotatebox{-90}{\tiny {\{$\circledS$0\}}}}
}
\def\enotesize{\normalsize}

\newcommand{\appropto}{\mathrel{\vcenter{
  \offinterlineskip\halign{\hfil$##$\cr
    \propto\cr\noalign{\kern2pt}\sim\cr\noalign{\kern-2pt}}}}}

\def\layout{\null}

\def\mylabel#1{\label{#1}}
\newcommand\ors[1][]{\marginpar{\rotatebox{-90}{\tiny {#1}}}}

\def\seeactivity#1{\{$\mathcal{A}$:\ref{#1}\}}

\DeclareNewFootnote{A}
\def\regfootnote#1{\footnoteA{#1}}

\newcounter{eqtag}
\def\ntag{\stepcounter{eqtag}\tag{\roman{eqtag}}}
\def\solution#1{[Solution in Sect.~\ref{#1}]}

\makeatletter
\renewcommand{\@makefnmark}{\makebox{\textsuperscript{[\@thefnmark]}}}
\renewcommand\@makefntext[1]%
     {\noindent\makebox[0pt][r]{\textsuperscript{\@thefnmark}\,}#1}
     \makeatother

\def\figindex#1{\null}

\def\citep#1{\citeauthor{#1}, \citeyear{#1}} 

\def\refcite#1{\citeauthor{#1} (\citeyear{#1})} 
\def\citet#1{\citeauthor{#1} (\citeyear{#1})}

\def\tableline{\hline}

\def\derp#1#2{{\partial #1\over\partial #2}}

\def\tderp#1#2{{\partial #1/\partial #2}}

\def\dd#1{{\rm d}#1}
\def\spose#1{\hbox to 0pt{#1\hss}}
\def\lta{\mathrel{\spose{\lower 3pt\hbox{$\mathchar"218$}}
     \raise 2.0pt\hbox{$\mathchar"13C$}}}
\def\gta{\mathrel{\spose{\lower 3pt\hbox{$\mathchar"218$}}
     \raise 2.0pt\hbox{$\mathchar"13E$}}}
\def\dt{\Delta \temp_{\rm j}}

\def\be{\begin{equation}}
\def\ee{\end{equation}}
\def\bea{\begin{eqnarray}}
\def\eea{\end{eqnarray}}

\def\la{\mathrel{\mathchoice {\vcenter{\offinterlineskip\halign{\hfil
$\displaystyle##$\hfil\cr<\cr\sim\cr}}}
{\vcenter{\offinterlineskip\halign{\hfil$\textstyle##$\hfil\cr<\cr\sim\cr}}}
{\vcenter{\offinterlineskip\halign{\hfil$\scriptstyle##$\hfil\cr<\cr\sim\cr}}}
{\vcenter{\offinterlineskip\halign{\hfil$\scriptscriptstyle##$\hfil\cr<\cr\sim\cr}}}}}
\def\ga{\mathrel{\mathchoice {\vcenter{\offinterlineskip\halign{\hfil
$\displaystyle##$\hfil\cr>\cr\sim\cr}}}
{\vcenter{\offinterlineskip\halign{\hfil$\textstyle##$\hfil\cr>\cr\sim\cr}}}
{\vcenter{\offinterlineskip\halign{\hfil$\scriptstyle##$\hfil\cr>\cr\sim\cr}}}
{\vcenter{\offinterlineskip\halign{\hfil$\scriptscriptstyle##$\hfil\cr>\cr\sim\cr}}}}}
\newcommand{\tc}[1]{{\Large \textcircled{\normalsize {#1}}}}

\renewcommand{\vec}[1]{{\bm{#1}}}
\newcommand{\di}{\vec{\nabla}\cdot}
\newcommand{\grad}{\vec{\nabla}}
\newcommand{\E}{\vec{E}}
\newcommand{\B}{\vec{B}}
\newcommand{\jj}{\vec{j}}
\newcommand{\vv}{\vec{v}}
\newcommand{\rh}{\rho}

\renewcommand{\dd}{\partial}
\renewcommand{\dt}{\partial t}

\newcommand{\avr}[1]{\overline{#1}}
\newcommand{\pr}[1]{{#1}^{\prime}}
\newcommand{\degr}{^\circ}
\newcommand{\emf}{{\vec{\mathcal{E}}}}
\newcommand{\emfa}{\avr{\vec{\mathcal{E}}}}

\newcommand{\Rm}{R_{\rm m}}

\newcommand{\curl}{\mbox{\boldmath $\nabla \times$}}

\newcommand{\lapprox} {\, \lower3pt\hbox{$\sim$}\llap{\raise2pt\hbox{$<$}}\,}
\newcommand{\gapprox} {\,
  \lower3pt\hbox{$\sim$}\llap{\raise2pt\hbox{$>$}}\,}

\newcommand\msun{M_{\odot}}

\newcommand\rsun{R_{\odot}}
\newcommand\msunyr{M_{\odot}\,yr^{-1}}

\newcommand\K{\rm K}

\newcommand\oo{\Omega_{\circ}}
\newcommand\mdot{ \, \dot{M}}
\newcommand\sd{\partial}

\newbox\grsign \setbox\grsign=\hbox{$>$} \newdimen\grdimen \grdimen=\ht\grsign
\newbox\simlessbox \newbox\simgreatbox
\setbox\simgreatbox=\hbox{\raise.5ex\hbox{$>$}\llap
     {\lower.5ex\hbox{$\sim$}}}\ht1=\grdimen\dp1=0pt
\setbox\simlessbox=\hbox{\raise.5ex\hbox{$<$}\llap
     {\lower.5ex\hbox{$\sim$}}}\ht2=\grdimen\dp2=0pt
\newcommand\simgreat{\mathrel{\copy\simgreatbox}}
\newcommand\simless{\mathrel{\copy\simlessbox}}

\def\Rm{{{\cal R}_{\rm m}}}
\def\Ro{{N_{\rm R}}}

\def\uvec{{\hat {\bf e}}}
\def\vv{{\vec{v}}}
\def\uv#1{\uvec_#1}
\def\sv#1{u_#1}

\def\Beq{B_{\rm eq}}

\def\rvec{{\bf r}}

\def\vvec{{\bf v}}
\def\bvec{{\bf B}}
\def\xvec{{\bf x}}

\def\eq{Eq.}

\def\customitemize{
  \itemsep=0pt \parsep=0pt \parskip=0pt \partopsep=0pt \topsep=0pt}

\renewcommand{\nocite}[1]{\null}

\title{PRINCIPLES OF HELIOPHYSICS: \\
  a textbook on the universal processes \\
  behind planetary habitability}
\author{by\\
  Karel Schrijver, Fran Bagenal, Tim Bastian, J{\"u}rg Beer, Mario Bisi, \\
  Tom Bogdan, Steve Bougher, David Boteler, Dave Brain, Guy Brasseur, \\
  Don Brownlee, Paul Charbonneau, Ofer Cohen, Uli Christensen, \\
  Tom Crowley, Debrah Fischer, Terry Forbes, Tim Fuller-Rowell, \\
  Marina Galand, Joe Giacalone, George Gloeckler, Jack Gosling, \\
  Janet Green, Nick Gross, Steve Guetersloh, Viggo Hansteen, Lee Hartmann, \\
  Mihaly Horanyi, Hugh Hudson, Norbert Jakowski, Randy Jokipii, \\
  Margaret Kivelson, Dietmar Krauss-Varban, Norbert Krupp, \\
  Judith Lean, Jeff Linsky, Dana Longcope, Daniel Marsh, Mark Miesch, \\
  Mark Moldwin, Luke Moore, Sten Odenwald, Merav Opher, Rachel Osten, \\
  Matthias Rempel, Hauke Schmidt, George Siscoe, Dave Siskind, \\
  Chuck Smith, Stan Solomon, Tom Stallard, Sabine Stanley, Jan Sojka, \\
  Kent Tobiska, Frank Toffoletto, Alan Tribble, Vytenis Vasyliunas, \\
  Richard Walterscheid, Ji Wang, Brian Wood, Tom Woods, and Neal Zapp \\ \\
  \versiondate \\ \\
  \\ \\
}

\def\indexit#1{\index{#1}}
\begin{document}
\renewcommand{\rh}{\rho}

\maketitle

\pagestyle{fancy}
\renewcommand{\chaptermark}[1]{ \markboth{#1}{} }
\renewcommand{\sectionmark}[1]{ \markright{#1}{} }
\renewcommand{\headrulewidth}{0.4pt}
\renewcommand{\footrulewidth}{0pt}
\fancyhf{}
\fancyhead[RE]{\leftmark}
\fancyhead[LO]{{\rightmark}}
\fancyhead[LE,RO]{\thepage}
\fancyfoot[CE,CO]{}
%-- The following block to "%--" is used to mark personal copies
\def\personalcopy#1{
  \renewcommand{\footrulewidth}{0.4pt}
  \fancyfoot[LE,RO]{{\small \em \dateofthisversion}}
  \fancyfoot[RE,LO]{{\small \em #1.}}
}
%\personalcopy{\versionno}

\def\colorfig{For a color version of this figure, see arXiv:2001.01093.}

%\clearpage
\null
\vfill
\begin{center}
\null \vspace{9cm}
  
Copyright \copyright 2024 The authors
\vspace{0.5cm}

ISBN-13: 9798847272711 (paperback) \\
ISBN-13: 9798353812982 (hardcover)

\vspace{0.5cm}
\url{https://arxiv.org/abs/1910.14022}

\vspace{0.5cm}
Cover design for the Amazon/KDP version by Karel Schrijver

\vspace{0.5cm}
Printed in the United States of America

\end{center}

\clearpage
\pagenumbering{roman}
\tableofcontents

\layout

\chapter*{\em Preface (do read it!)}
\markboth{Preface}{Preface}
\addcontentsline{toc}{chapter}{Preface (do read it!)}
\label{ch:preface}
\begin{table}[b!]
 \indexit{heliophysics|seealso{definition}}\indexit{definition!heliophysics}
  \fbox{
\vbox{
{\bf Heliophysics\/}
\begin{description}
\item[{\bf helio-, pref.,}] on the Sun and environs, from the Greek helios.
\item[{\bf physics, n.,}] the science of matter and energy and their
            interactions.
\end{description}
{\em Heliophysics is the
\begin{itemize}\customitemize
\item comprehensive new term for the science of the Sun - Solar System Connection.
\item exploration, discovery, and understanding of our space environment.
\item system science that unites all of the linked phenomena in the region
of the cosmos influenced by a star like our Sun.
\end{itemize}}
\vskip -8pt
Heliophysics concentrates on the Sun and its effects on Earth, the
other planets of the solar system, and the changing conditions in
space. Heliophysics studies the magnetosphere, ionosphere,
thermosphere, mesosphere, and upper atmosphere of the Earth and other
planets. Heliophysics combines the science of the Sun, corona,
heliosphere and geospace. Heliophysics encompasses cosmic rays and
particle acceleration, space weather and radiation, dust and magnetic
reconnection, solar activity and stellar cycles, aeronomy and space
plasmas, magnetic fields and global change, and the interactions of
the solar system with our galaxy.

{\footnotesize \em From NASA's ``Heliophysics. The New Science of the
Sun - Solar System Connection: Recommended Roadmap for Science and
Technology 2005 - 2035.''}  }}
\caption[Heliophysics: definition.]{Heliophysics: definition.}
\end{table}

Heliophysics is the system science of the physical connections between
the Sun and the solar system. As the physics of the local cosmos, it
embraces space weather and planetary habitability. The wider view of
comparative heliophysics forms a template for conditions in
exoplanetary systems and provides a view over time of the aging Sun
and its magnetic activity, of the heliosphere in different settings of
the interstellar medium and subject to stellar impacts, of the space
physics over evolving planetary dynamos, and of the long-term
influence on planetary atmospheres by stellar radiation and wind.

\begin{table}[t]
\caption[Chapters in the Heliophysics book series sorted by theme
(1).]{Chapters \indexit{Heliophysics!book chapters}and their authors in the Heliophysics book series sorted by theme (continued on the next page), not showing introductory chapters.}\label{tab:chapters}
\begin{tabular}{l}
\hline
\centerline{\bf Universal and fundamental processes, diagnostics, and methods}\\
\hline
I.2. Introduction to heliophysics \dotfill\ {\em T.\ Bogdan}\\
I.3. Creation and destruction of magnetic field \dotfill\ {\em M.\ Rempel}\\
I.4. Magnetic field topology \dotfill\ {\em D.\ Longcope}\\
I.5. Magnetic reconnection \dotfill\ {\em T.\ Forbes}\\
I.6. Structures of the magnetic field \dotfill\ {\em M.\ Moldwin {\em et al.}}\\
II.3 In-situ detection of energetic particles \dotfill\ {\em G.\ Gloeckler}\\
II.4 Radiative signatures of energetic particles \dotfill\ {\em T.\ Bastian}\\
II.7 Shocks in heliophysics \dotfill\ {\em M.\ Opher}\\
II.8 Particle acceleration in shocks \dotfill\ {\em D.\ Krauss-Varban}\\
II.9 Energetic particle transport \dotfill\ {\em J.\ Giacalone}\\
II.11 Energization of trapped particles \dotfill\ {\em J.\ Green}\\
IV.11 Dusty plasmas \dotfill\ {\em M.\ Hor{\'a}nyi}\\
IV.12 Energetic-particle environments in the solar system \dotfill\ {\em N.\ Krupp}\\
IV.13 Heliophysics with radio scintillation and occultation \dotfill\ {\em M.\ Bisi}\\
\hline
\centerline{\bf Stars, their planetary systems, planetary habitability, and climates}\\
\hline
III.3 Formation and early evol.\ of stars and proto-planetary disks \dotfill\ {\em L.\ Hartmann}\\
III.4 Planetary habitability on astronomical time scales \dotfill\ {\em D.\ Brownlee}\\
III.11 Astrophysical influences on planetary climate systems \dotfill\ {\em J.\ Beer}\\
III.12 Assessing the Sun-climate relationship in paleoclimate records \dotfill\ {\em T.\ Crowley}\\
III.14 Long-term evolution of the geospace climate \dotfill\ {\em J.\ Sojka}\\
III.15 Waves and transport processes in atmosph.\ and oceans \dotfill\ {\em R.\ Walterscheid }\\
IV.5 Characteristics of planetary systems \dotfill\ {\em D.\ Fischer \&\ J.\ Wang}\\
IV.7 Climates of terrestrial planets \dotfill\ {\em D.\ Brain}\\
\hline
\centerline{\bf The Sun, its dynamo and its magnetic  activity; past, present and future}\\
\hline
I.8. The solar atmosphere \dotfill\ {\em V.\ Hansteen}\\
II.5 Observations of solar and stellar eruptions, flares, and jets \dotfill\ {\em H.\ Hudson}\\
II.6 Models of coronal mass ejections and flares \dotfill\ {\em T.\ Forbes }\\
III.2 Long-term evolution of magnetic activity of Sun-like stars \dotfill\ {\em C.\ Schrijver}\\
III.5 Solar internal flows and dynamo action \dotfill\ {\em M.\ Miesch}\\
III.6 Modeling solar and stellar dynamos \dotfill\ {\em P.\ Charbonneau}\\
III.10 Solar irradiance: measurements and models \dotfill\ {\em J.\ Lean \&\ T.\ Woods}\\
IV.2 Solar explosive activity throughout the evol.\ of the solar system \dotfill\ {\em R.\ Osten}\\
\hline
\end{tabular}
\end{table}

\begin{table}[t]
\addtocounter{table}{-1}
\caption[Chapters in the Heliophysics book series sorted by theme
(2).]{(Continued from the previous page) Chapters and their authors in
  the Heliophysics \indexit{Heliophysics!book chapters}book series sorted by theme, not showing introductory chapters.}
\begin{tabular}{l}
\hline
\centerline{\bf Astro-/heliospheres, interstellar environment, and galactic cosmic rays}\\
\hline
I.7. Turbulence in space plasmas \dotfill\ {\em C.\ Smith}\\
I.9. Stellar winds and magnetic fields \dotfill\ {\em V.\ Hansteen}\\
III.8 The structure and evolution of the 3D solar wind \dotfill\ {\em J.\ Gosling}\\
III.9 The heliosphere and cosmic rays \dotfill\ {\em J.\ Jokipii}\\
IV.3 Astrospheres, stellar winds, and the interst.\ medium \dotfill\ {\em B.\ Wood  \&\ J.\ Linsky}\\
IV.4 Effects of stellar eruptions throughout astrospheres \dotfill\ {\em O.\ Cohen}\\
\hline 
\centerline{\bf Dynamos and environments of planets, moons, asteroids, and comets}\\
\hline
I.10. Fundamentals of planetary magnetospheres \dotfill\ {\em V.\ Vasyli{\=u}nas}\\
I.11. Solar-wind magnetosphere coupling \dotfill\ {\em F.\ Toffoletto  \&\ G.\ Siscoe}\\
I.13. Comparative planetary environments \dotfill\ {\em F.\ Bagenal}\\
II.10 Energy conversion in planetary magnetospheres \dotfill\ {\em V.\ Vasyli{\=u}nas }\\
III.7 Planetary fields and dynamos \dotfill\ {\em U.\ Christensen}\\
IV.6 Planetary dynamos: updates and new frontiers \dotfill\ {\em S.\ Stanley}\\
IV.10 Moons, asteroids, and comets interact.\ with their surround.\ \dotfill\ {\em M.\ Kivelson}\\
\hline
\centerline{\bf Planetary upper atmospheres}\\
\hline
I.12. On the ionosphere and chromosphere \dotfill\ {\em T.\ Fuller-Rowell   \&\ C.\ Schrijver}\\
II.12 Flares, CMEs, and atmospheric responses \dotfill\ {\em T.\ Fuller-Rowell  \&\ S.\ Solomon}\\
III.13 Ionospheres of the terrestrial planets \dotfill\ {\em S.\ Solomon}\\
III.16 Solar variability, climate, and atmosph.\ photochemistry \dotfill\ {\em G.\ Brasseur {\em et al.}}\\
IV.8 Upper atmospheres of the giant planets \dotfill\ {\em L.\ Moore {\em et al.}}\\
IV.9 Aeronomy of terrestrial upper atmospheres \dotfill\ {\em D.\ Siskind  \&\ S.\ Bougher }\\
\hline
\centerline{\bf Technological and societal impacts of space weather phenomena}\\
\hline
II.2 Introduction to space storms and radiation \dotfill\ {\em S.\ Odenwald}\\
II.13 Energetic particles and manned spaceflight \dotfill\ {\em S.\ Guetersloh  \&\ N.\ Zapp}\\
II.14 Energetic particles and technology \dotfill\ {\em A.\ Tribble}\\
V.2 Space weather: impacts, mitigation, forecasting \dotfill\ {\em S.\ Odenwald}\\
V.3 Commercial space weather in response to societal needs \dotfill\ {\em W.\ Tobiska}\\
V.4 The impact of space weather on the electric power grid \dotfill\ {\em D.\ Boteler}\\
V.5 Radio waves for communication and ionospheric probing \dotfill\ {\em N.\ Jakowski}\\
\hline
\end{tabular}
\end{table}

Based on a series of NASA-funded Summer Schools for early-career
researchers, this textbook is intended for students in physical
sciences in later years of their university training and for beginning
graduate students in fields of solar, stellar, (exo-)planetary, and
planetary-system sciences. The lecturers at the Summer Schools
developed a series of five volumes on \indexit{Heliophysics!book series}Heliophysics
(\href{https://cpaess.ucar.edu/heliophysics/resources/textbooks}{four
  published in printed form} by Cambridge University Press, and one
online at the Heliophysics Summer School
\href{https://cpaess.ucar.edu/heliophysics/summer-school}{website})
contain in total 1919 pages of text and figures, in 56 topical
chapters (see Table~\ref{tab:chapters}): Vol.~I: \citet{2011hppl.book.....S}; 
Vol.~II: \citet{2012hssr.book.....S}; Vol.~III: \citet{2012hesa.book.....S}; Vol.~IV: 
\citet{2016hasa.book.....S}; and Vol.~V: \citet{heliophyicsv}. The
present volume presents a selection of these texts, while adding new
text as connecting or summarizing material, with an overall text
length that is about one-fifth of the original textbooks.

The topics in this volume are organized to emphasize universal
processes from a perspective that draws attention to what provides
Earth (and similar \hbox{(exo-)}planets) with a relatively stable
setting in which life as we know it can thrive. This text aims to
serve as a textbook-style volume for which the original Heliophysics
books are the extended `readers' with much more detail, and
domain-specific topical chapters. Note that references from the
original texts were omitted here (see the original volumes for those);
references for new texts can be found in the Bibliography, where also
source references to figures are provided as needed.

This volume is intended for students in physical sciences in later
years of their university training and for beginning graduate students
in fields of solar, stellar, (exo-)planetary, and planetary-system
sciences. This contrasts with the intended audiences for the
Heliophysics volumes which included the community of mid-to-advanced
graduate students, the cohort of early postdoctoral researchers, and
those professional researchers looking for review-like introductions
into fields of heliophysics adjacent to their own. In targeting the
audience of advanced undergraduate and beginning graduate students,
many of the deeply technical details discussed in the original volumes
were omitted, introductions were broadened, and the emphasis was
placed on processes rather than on details of equations, states, or
numerical experiments.

Throughout this work the original text from the Heliophysics volumes
is directly quoted, following a volume and chapter reference, where
between double quotation marks, but with equations, units (here
cgs-Gaussian throughout with a few exceptions \regfootnote{A good
  resource for unit conversions (and many other things related to
  plasma physics) is the online
  \href{https://www.nrl.navy.mil/ppd/content/nrl-plasma-formulary}{NRL
    plasma formulary}.}), and symbols modified where needed for
homogeneity throughout this work, with edits (and some corrections)
shown between brackets, with many parenthetical notes removed, and
with citations of the professional literature left out (and those to
other sections in the books modified as appropriate).

{\em The source texts in the series of Heliophysics books are
  referenced in the margins as \ors[I:2.9]
  $\#$[roman]:$\#$[arabic].$\#$[arabic].} For example, Vol. I,
Section~9 in Chapter~2 would be referred to as ``I:2.9'' in the margin
or in captions. The original sources of all of
the figures can be found in the figure captions of the Heliophysics
books, but for many here a reference to the original publication is
included for figures not made by the Heliophysics authors but whose
original authors have given permission to have their artwork used in
this volume. A few figures were replaced by color versions or by
alternative figures.

\begin{table}%\begin{figure}
\indexit{heliophysics!and space weather}\indexit{space weather}
%\framebox[\textwidth]{
%\begin{flushleft}
\fbox{ \vbox{
%{\bf Heliophysics and space weather:}\\
%
'Space weather' is the term used to describe an ensemble of changing
conditions in the vicinity of Earth and, by extension, any other body
in a planetary system, typically occurring on time scales up to a few
days. Often, the term is implicitly taken to refer also to the
conditions from the solar dynamo outward to the furthest reaches of
the heliosphere that are involved in space weather around Earth.  Much
of what is described in this volume therefore concerns space weather:
heliophysics contains the science of space weather. However, where the
science of space weather focuses on phenomena that can impact society
through short-term variability, this text takes the long view by
putting the spotlight on evolutionary changes in the states of
star-planet systems. As such, this text does describe the foundational
processes of space weather, but is not concerned with the impacts of
space weather on technological infrastructure, does not address the
challenges of forecasting space weather, and skips coupling mechanisms
such as ground-induced currents (GICs) associated with geomagnetic
disturbances and ground-level enhancements (GLEs) of energetic
particles. This choice of focus is motivated by my desire to introduce
the reader to the science of heliophysics from the perspective of
habitability on time scales on which stellar and planetary atmospheres
change, and indeed up to time scales on which stars and their planets
evolve, and to do that in a relatively compact form. As you go through
this text, you should realize that many of the processes described
here have consequences for society, ranging from system design choices
to potentially substantial failures in one or more of the
infrastructures that we have come to rely on, including continuous and
reliable electric power, positional information, and means of
communication. Interruptions in quality or availability of any of
these can have substantial consequences that may be costly or
life-threatening on scales that may involve single individuals or
populations of millions. Descriptions of the impacts of space weather
can be found in the Heliophysics books in Chapters II:2, II:13,
II:14, H-V:2, H-V:3, H-V:4, and H-V:5; another resource is a
'\href{https://ui.adsabs.harvard.edu/abs/2015AdSpR..55.2745S/abstract}{roadmap}'
document (\citep{2015AdSpR..55.2745S}) that reviews the state of our knowledge of space weather and
its technological and societal impacts, and what is needed to advance
our abilities to forecast space weather.  }}
%\end{flushleft}
    \vspace{-0.5cm}
\caption[Heliophysics and space weather]{Heliophysics and space
  weather \label{fig:swx} }
  %}
\end{table}%\end{figure}
\section*{Activities for the reader}
New here compared to the printed volumes is the inclusion of 200
'activities' (starting in version 1.3; see the chapter on 'Version
history' for a description of changes) in the form of problems,
exercises, explorations, literature readings, and 'what if'
challenges. Many contain additional information complementing the main
text, so I suggest you read them as you go along, if not on first
reading, then at least on
review. \sfirstactivity{\null}\regfootnote{Exercises are flagged as
  \{${\rm A}\#$\} or \{$\circledS{\rm A}\#$\}, also in the margin, with
  continuous numbering throughout ({\bf this numbering is
    Version dependent!}). Activities can be found in
  Ch.~\ref{ch:activities}. For a selection of of these Activities, solutions
  and/or supplemental reading are provided in Ch.~\ref{ch:solutions}
  --~these selected Activities are marked $\circledS$ in the margin
  and in the compilation in Ch.~\ref{ch:activities}.}  Some were
developed by the teachers for the Heliophysics Summer School but most
are newly created specifically for this volume. They are meant to let
the reader look up a definition, to introduce a moment of reflection
on an equation or figure, to see connections to similarities
elsewhere, to get a feel for the magnitude of things or the relative
importance of processes, or to consider what would happen under
conditions other than those encountered in our Solar System; they are
not meant to particularly exercise mathematical skills. There are five
classes of activities as indicated at their start: ``Look up'' to
familiarize you with processes, numbers, and definitions, ``Consider''
to make you think about processes, ``Show'' for applications of
equations or numerical estimates, ``Background'' for further
information, and ``Advanced/Group'' for larger activities to explore
beyond the textbook and/or to undertake with a group of students in a
class.  At the end of the book, in
Activity~\ref{act:thefinalquestion}, the reader is asked to reflect
back on all the processes that are involved in the habitability of
Earth and, by analogy, of exoplanets elsewhere in the Universe.

\section*{Terminology}
As you go through this volume, you will encounter words that have
somewhat different meanings in different communities. For example,
'convection' is often used in the magnetospheric community to describe
movement that in astrophysics would be referred to as 'advection',
while 'convection' in that discipline is reserved for overturning
plasma motions involved in the transport of thermal energy. Another example
is that of the word 'dynamo' which in astrophysics and planetary
sciences is used to describe the ensemble of processes maintaining a
magnetic field against decay, often with an alternating temporal
character. In ionospheric physics, it is often used for processes where
differential motions of (neutral plus ionized) gas and magnetic field
exchange energy through work.

You will also note that terms may describe locations or something in a
location, or a property of what is in that location. For example,
'ionosphere' may sound like a location descriptor but actually refers
to only the ionized medium in an atmosphere (with 'thermosphere' used
for the overlapping neutral environment). The term
'chromosphere', which describes a stellar environment in some respects
not dissimilar to an ionosphere-thermosphere, encompasses both the
ionized and neutral components; it is often used as an indicator of a
volume above a stellar surface in a certain thermal range, but is
defined formally (as you will see later) by the properties of the
radiative transfer of the medium.

Finally, there are words like 'late-type star' that have nothing to do
with a temporal attribute, but which survived an older era where the
nature of stars was not yet understood and where cooler was
erroneously interpreted as older.

I hope that all terms are properly defined where first used. Here, I
want to raise your awareness that as you talk to colleagues in other
disciplines they may not only be puzzled by processes that you study,
but that communication may be hampered by misinterpretation of the
terms that you use: language can be a very precise tool, but only if
the user is aware of how the listener/reader may interpret the words
that are being used.

\section*{Online resources associated with the Summer School}
\indexit{heliophysics!online resources}
\begin{itemize}\customitemize
\item If you are looking at a paperback or hardcover version of this book from
  Amazon/KDP then look for a free e-version at
  \href{https://arxiv.org/abs/1910.14022}{https://arxiv.org/abs/1910.14022}
  which uses hyperrefs for easy navigation. It also has many figures in
  color, as does the hardcover edition, that are shown in gray-scale in the paperback.
\item The Summer School's home
    is at \href{https://heliophysics.ucar.edu}{https://heliophysics.ucar.edu}.
  \item Many of the Heliophysics lectures can be found 
    \href{https://www.youtube.com/results?search_query=heliophysics+summer+school}{on
      YouTube} by searching for 'Heliophysics Summer School'.
\end{itemize}

\section*{A few notes on other resources} 
The
\href{https://heliophysics.ucar.edu/resources/collected-figures}{figures}
  published in the Heliophysics book series are available on-line at
the \href{https://cpaess.ucar.edu/heliophysics/summer-school}{website}
of the Heliophysics Summer School (https://heliophysics.ucar.edu),
where you can also find labs (with instruction manuals) and many
recorded lectures sorted by theme (in part hosted on youtube).

There is a subject index in this volume but note that the online version of this
book can be searched with the tools of web browsers and pdf viewers
to provide an effective and entirely comprehensive alternative way to
find topics.

This volume focuses on processes, not on their measurements. For an
introduction to some of the aspects of remote and {\em in-situ} sensing
within the Heliophysics series I refer to the following chapters that
focus on that aspect in particular: Chs.~II:3, II:4,
IV:5, and IV:13.

You can find a \href{https://zenodo.org/record/3843629}{list} of
English-language textbooks, popular texts, topical monographs and book
series, related to the science of space weather and published since
1990 in \refcite{knippcade2018}.

Explore \href{https://svs.gsfc.nasa.gov}{NASA's Scientific
Visualization Studio} at https://svs.gsfc.nasa.gov for a variety of
images and movies ---~real and simulated.

\section*{Navigating the pdf version at arXiv:2001.01093}
References to sections, figures, tables, and equations in this book
are shown in red, pointers to the bibliography are shown in green,
and references to web pages are shown in blue. Clicking on any of
these jumps to that location or web page.  How you get back to reading
where you left off depends on how you are viewing this file and on
what type of device. For example, this
\href{https://helpx.adobe.com/acrobat/using/keyboard-shortcuts.html}{web
  site} shows a list of keyboard shortcuts to move around the pdf
version of this book with Acrobat Reader. Using that on a Mac, you can
return to the page you came from by pressing [command + left-arrow]
after clicking on a link to a figure, section, or activity. When using
Mac OS Preview, look under the 'Go' menu for navigation shortcuts
(where you will see that the equivalent of the above is [command +
left bracket]).

\section*{Corrections and updates}
This version of the textbook is subject to corrections and updates. I
welcome input from students, teachers, and colleagues: if you see a
typo or an explanation that you think is in error, or if you believe
a serious update is in order, please
\href{mailto:heliophysicsnutshell@gmail.com}{email me}:
heliophysicsnutshell{@}gmail.com! Be as specific
as you can about where the text is that you think should be changed,
what to change it to, and why it needs such a change. Your input will
help improve this text for all users.

\section*{Acknowledgements}
I thank all the Heliophysics authors and other teachers in the
Heliophysics Summer School for the skill with which they taught me as
one of the participants in the School as well as for their patience
with me as one of the editors of the book series. This volume was
supported by the Johannes Geiss Fellowship of the International Space
Science Institute. I am indebted to J{\"u}rg Beer, Paul Charbonneau, Terry Forbes,
Marina Galand, Dana Longcope, Sten Odenwald, and Matthias Rempel for
their insightful comments on an earlier version of the manuscript, and
to Nick Gross for working through most of the Activities, noting errors,
adding tasks, and developing new Activities.
Special thanks go to Tom Bogdan who worked through the entire draft
volume.

\vskip .5 cm
{\hfill Karel Schrijver, \dateofthisversion}

{\hfill {\em karelschrijver.com}}

\vskip .5 cm

``{\em A physical understanding is a completely unmathematical, imprecise,
and inexact thing, but absolutely necessary for a physicist.}'', R.P.\
Feynman, in {\em The Feynman Lectures on Physics, Vol.\ II.}
\clearpage

%\part{\bf Introduction}
%\addtocontents{toc}{\vspace{0.5cm}{\centerline{\bf Introduction}}\par }

\chapter{{\bf Stars, planetary systems, and the local
    cosmos}}%1
\fancyhead[RE]{\thechapter\,--\,\leftmark}

\setcounter{page}{1}\pagenumbering{arabic}
\label{ch:introduction}
{\narrower\narrower{
{\bf Chapter topics:}
\begin{itemize}
  \customitemize
\item The rationale for this book
\item Heliophysics: the system science of the physical connections
  between the Sun and the solar system
\item Magnetohydrodynamics as the basic language of heliophysics
\item Basic glossary for heliophysics
\item A timeline of exploration of (exo-)planetary systems
\end{itemize}

}}

\section{Preparing for the future}
By \indexit{solar|seealso{flare}}the
time you reach the end of this book, you will have the basic
set of tools of scientific imagination involved in understanding what
couples stars and planets. What you will learn is universal,
literally: it does not matter which stars and planets we speak of:
whether of those few nearby or of the many distant ones. Nor does it
matter whether they are those few that we are long familiar with or
the many that we know
about, so far, only in a statistical sense. It does not matter either
whether your particular interests lie within the Solar System or
beyond it: the same principles apply in our local cosmos as in the
most distant planetary system we shall ever have access to.

But looking forward to your science-based career, whether as
a researcher or as a teacher, as a journalist or as a politician, you need to
be familiar with what is known. That is particularly true in order to
discover something new. And to appreciate the value of a discovery,
you need to know how to apply what you know to what is not (yet)
known. You will need to imagine things no one has ever seen, but not
arbitrarily: science demands that you come up with what appears most
probable, not merely with things that are possible. Richard Feynman
(in {\em The meaning of it all\,}) said it this way: 'It is surprising
that people do not believe that there is imagination in science. It is
a very interesting kind of imagination, unlike that of the artist. The
great difficulty is in trying to imagine something that you have never
seen, that is consistent in every detail with what has already been
seen, and that is different from what has been thought of.'

The pace at which exoplanets are being discovered is simply
amazing. What we can learn from them, and from our Solar System,
offers so many opportunities to learn yet more. I realize that going
through the first nine chapters will be hard, because they have to
build your foundation, because they cover so many different branches
of science, and because they look at things so different from everyday
life. But these first nine chapters look for commonalities, for
'universal processes' that help create in your mind a virtual
laboratory: in the astronomy we cannot turn dials to
explore things under different conditions, but we can compare
environments and look for what they have in common and for
what sets them apart.

The final six chapters require prior digestion of the first
nine. These final six invite you to imagine, scientifically, Earth in
the distant past and future, Earth-like planets in a variety of orbits
around Sun-like stars, and the space environments and climates of
tropospheres of exo-worlds. Future discoveries have their beginnings in
lessons from the past:

\section{Considering planetary habitability}\indexit{habitability}\indexit{planet!habitability}
Planetary system are, statistically speaking, about as common as
stars. We have learned a lot about stars over the century that
followed the realization that they are huge nuclear fusion reactors
and that most, like the Sun, function also as giant dynamos. In
contrast, firm evidence that planetary systems are common companions
to stars was only obtained within the past two decades. It is
therefore no surprise that much still needs to be learned about how
planets form, how planetary systems evolve, and what the conditions
are near planetary surfaces (if indeed a solid or liquid surface
exists). The combination of exoplanetary science and the study of the
local cosmos is enlightening us as much about the history and future
of our Solar System as about the growing number of planetary
systems that have been observed in some detail. Whether life exists
anywhere beyond Earth remains to be established, but scientists are
making rapid headway in knowing about the conditions that life on
Earth has been subjected to since its genesis and also the conditions
that any life on any other planet would be subjected to depending on
the properties of their central star and companion planets.

Heliophysics deals with all of the aspects of 'living with a star' on
time scales from fractions of a second to billions of years. The
series of Heliophysics books offers an introduction to a large cross
section of that vast scientific field. In the present volume, we focus
on the universal processes that tie together the branches of
heliophysics with particular emphasis on those processes that are
relevant to what one might describe as
\indexit{habitability|seealso{definition}}'planetary habitability'. With
life \indexit{definition!habitability}having been found on only a
single astrophysical body we do not
have a particularly well-considered concept of what 'planetary
habitability' might mean, of course. But we have an intuitive feel for
it: a long-lived planet orbiting a long-lived star, with a fairly
substantial planetary atmosphere that is neither too hot nor too cool
to allow chemistry to be complex (and, in many minds, restricting that
to chemistry that involves liquid water), shielded well enough (but
not necessarily perfectly) from energetic radiation (both
electromagnetic and particulate) by that atmosphere and by a planetary
magnetic field. The star's irradiance onto a 'habitable planet' should
not vary too much, comet and asteroid impacts should be limited,
atmospheric erosion slow, \ldots\ If that sounds like we are
describing the Sun-Earth system then that is no surprise: we know it
has made the Earth habitable to a diversity of life that is on one
hand astoundingly diverse and on the other --~at the molecular
level~-- remarkably homogeneous.

As the number of known exoplanetary systems is bound to keep growing
rapidly, and as our instrumentation and methods are bringing
exoplanetary atmospheric science within our grasp, it is clear that
our understanding of the Solar System and its central star provide
crucial guidance to the study of 'planetary habitability' and --~some
day rather soon, one should anticipate~-- the study of extraterrestrial
life. That expectation has guided the selection of topics covered in
this volume.

\ors[I:2.9] \regfootnote{Throughout this work the original text from the
  Heliophysics volumes is directly quoted (with edits between
  brackets) with references shown in the margin like this:
  $\#$[roman]:$\#$[arabic].$\#$[arabic]. So, for example, Vol.\ I,
  Section~9 in Chapter~2 would be referred to as I:2.9.}``If we
gaze upon the uncountable array of stars strewn across the vault of
the heavens, one may know that the remarkable things one will come to
know about heliophysics in the pages that follow are presently
unfolding around those very stars and planetary systems that give
light to the night sky.  Heliophysics is truly a universal science.''

%\section{Preamble} 
\section{Heliophysics: unification, coupling, exploration}  \ors[I:2.1] ``Walk along an island
beach on a clear, breezy, cloudless night, or stand on the spine of a
barren mountain ridge after sunset, and behold the firmament of
stars\activity{{\em Look up} what type of astrophysical body is a true
  'star'. Contrast that to 'white dwarf star', 'brown dwarf star',
  and
  'neutron star': none of these are true stars in their present state
  and only two of which have ever been. 'Brown dwarfs' take up the
  mass interval between true stars and (exo)planets.} glittering
against the coal-black sky above. They fill the sky with their
timeless, brilliant flickering [(mostly caused by the terrestrial
atmosphere)].  With binoculars or even a small telescope one finds
that even the lacy dark matrix between the vast sea of stars is
populated with still more stars that are simply too faint to be seen
with the naked eye.  Within the Milky Way galaxy, that stretches from
horizon to horizon, the density of stars against the background sky is
even greater.

Each twinkling point of light is a star not too unlike our own
Sun. The Sun is an ordinary star that features so prominently in our
lives and on the pages of [the Heliophysics book series, as in this
volume,] because of its proximity. The next closest star,
$\alpha$\,Centauri (which is a triple system in which Proxima Centauri
is currently the closest to Earth), is almost a million times farther
away (at 4.22 light years), and the others are farther still.  We may
now say with some confidence that many of the stars are surrounded by
planets of various sizes.\activity{{\em Look up} the definition of
  'planet'. Note that, formally, the term 'planet' has only been
  defined by the International Astronomical Union for bodies within
  the Solar System; the term 'exoplanet' is reserved for bodies like
  planets in other planetary systems, although for these, and
  certainly for the joint collective, the term 'planets' is often
  used. What differentiates a planet (in the broader sense of the
  definition) from a star (hint: it is not that
  a planet orbits a star, see for example
  \href{https://en.wikipedia.org/wiki/Rogue_planet}{'rogue planets.'}
  \mylabel{act:defplanet}} Some of these orbital companions are so
immense that they are stars in their own right: double-star systems
are quite common. \activity{{\em Look up:} Many 'stars' we see in the night sky are
  binaries, including, for example, the brightest star in the night
  sky, Sirius ($\alpha$\,CMa).  More complex multiple-star systems may
  be less frequent, but are nonetheless common. (a) Look up the example of
  Castor ($\alpha$\,Gem) for an example of a sextenary, and then
  explore some more on
  \href{https://en.wikipedia.org/wiki/Star_system}{star systems} in
  general. (b) Create a table of a few of the more famous of these (such as
  Sirius, Proxima Centauri, and Polaris) and list the properties of
  the stars that they contain.  We get to the meaning of 'spectral
  type' in Ch.~\ref{ch:dynamos}. (c) You will see that stars go by many
  names: compare these and look up their roots, for example
  \href{https://www.iau.org/public/themes/naming_stars/}{at the IAU}
  or
  \href{https://en.wikipedia.org/wiki/Stellar_designations_and_names}{on
    Wikipedia}. \mylabel{act:starsystems}}

\begin{table}%\begin{figure}
\caption[Basic glossary (A--F).]{\label{fig:glossary} Basic glossary for
  domains and phenomena in heliophysics (continued on the next page).}
\indexit{glossary}\indexit{glossary|seealso{definition}}\indexit{definition|seealso{glossary}}
%\framebox[\textwidth]{
\begin{flushleft}
  \vskip -0.5cm
  \fbox{
\vbox{\small
%{\bf Essential glossary:}
\begin{itemize}
\vspace{-0.2cm}
\customitemize
\item {\em active region:} \indexit{definition!active region}a
  \indexit{active region|seealso{definition}}bipolar \indexit{active region}area of relatively strong
  magnetic flux, mostly consisting of magnetic {\em plage} (underlying
  the chromospheric {\em plage}) and, by
  definition, containing one or more {\em sunspots} at some point in
  its evolution ({\em cf.} Fig.~\ref{figure:bipoles}). Collectively
  they form \indexit{definition!active region belt}two active-region
  belts located on opposite hemisphere.
\item {\em ast(e)rosphere:} \indexit{definition!astrosphere}equivalent \indexit{astrosphere}of a heliosphere around another
  star (Sect.~\ref{sec:evolastrospheres})
\item {\em aurora:} \indexit{aurora}\indexit{definition!aurora} A
    visual phenomenon associated with geomagnetic activity visible
    mainly in the high-latitude night sky, resulting from collisions between atmospheric gases and precipitating charged particles (mostly electrons) guided by the geomagnetic field from the magnetotail.
\item {\em chromosphere:} \indexit{definition!chromosphere}domain \indexit{chromosphere}above the Sun's visible 'surface',
  with temperatures around 10,000--20,000\K  (see Table~\ref{tab:atmos-param})
\item {\em corona:} \indexit{definition!corona}the hottest \indexit{corona}domain of the Sun's atmosphere, at
  $\ge 1$\,MK (Table~\ref{tab:atmos-param})
\item {\em coronal hole:} \indexit{definition!coronal hole}formally a coronal \indexit{coronal!hole}region that is dark in
  X-rays and EUV; generally identified with a region where the Sun's
  magnetic field is 'open', {\em i.e.,}  reaches into the heliosphere 
  ({\em e.g.,} Sect.~\ref{sec:nearlystrat})
\item {\em coronal loop:} \indexit{definition!coronal loop}a high-temperature \indexit{coronal!loop}atmosphere within the
  Sun's corona, constrained to the volume of a magnetic 'flux tube'
  ({\em e.g.,} Sect.~\ref{sec:fieldlines})
\item {\em coronal mass ejection:} \indexit{definition!coronal mass
    ejection}impulsive \indexit{coronal!mass ejection}expulsion of magnetized
  material from a star
  into an astrosphere ({\em e.g.,} Fig.~\ref{fig:ophercomposite}, Sect.~\ref{sec:mhdintro})
\item {\em current sheet:}  defined in Table~\ref{fig:structures}
\item {\em exosphere:} \indexit{definition!exosphere}outermost domain \indexit{exosphere}of an atmosphere in which
  collisions are rare and ballistic trajectories dominate for
  constituent particles ({\em e.g.,} Sect.~\ref{sec:pcid})
\item {\em facula} and {\em bright point:}
  \indexit{definition!facula}a small \indexit{definition!bright point}flux \indexit{facula}tube in
  near-photospheric \indexit{bright point}layers viewed towards the solar limb or
  disk center, respectively ({\em e.g.,} Sect.~\ref{sec:energytransport})
\item {\em flare:} \indexit{definition!flare}impulsive conversion of magnetic energy in a
  stellar atmosphere into thermal and \indexit{flare}non-thermal particles and 
  bulk plasma motion, and appearing as a brightening over much of the
  stellar spectrum, although not significantly in total
  stellar brightness except for the most energetic events ({\em e.g.,} Sect.~\ref{sec:mhdintro})
\item {\em filament/prominence:} \indexit{definition!filament,
    prominence} A volume of gas at chromospheric temperatures
  suspended within the corona by magnetic forces, seen as dark ribbons
  threaded over the solar disk. A filament beyond the edge of the
  solar disk seen in emission against the dark sky is called a prominence
\item {\em flux tube/rope:}  \indexit{definition!flux tube}defined \indexit{flux!tube}in
  Table~\ref{fig:structures}
\end{itemize}
}
}
\end{flushleft}
\end{table}%\end{figure}
\begin{table}%\begin{figure}
  \caption[Basic glossary (G--Z).]{\label{fig:glossaryb} Basic glossary for
    domains and phenomena in heliophysics (continued from preceding
    page).}
\indexit{glossary}
%\framebox[\textwidth]{
\begin{flushleft}
  \vskip -0.5cm
  \fbox{
\vbox{\small
%{\bf Essential glossary:}
\begin{itemize}
\vspace{-0.2cm}
\customitemize
  \item {\em (super-)granulation:} \indexit{definition!granulation}
    granulation is \indexit{granulation}the pattern of \indexit{supergranulation}convective cells visible in the
    solar photosphere with a typical scale length of 1\,Mm
    (see Fig.~\ref{fig:bp-simple}); \indexit{definition!supergranulation}
    supergranulation is a much larger cellular pattern with a scale
    length of 20--30\,Mm that manifests primarily in velocity maps and
    in its ordering of the magnetic network (see Fig.~\ref{figure:bipoles})
\item {\em geomagnetic (sub-)storm:} \indexit{definition!geomagnetic
    (sub-)storm}Storm: A worldwide \indexit{geomagnetic!(sub-)storm}disturbance of
  the Earth's magnetic field, distinct from regular diurnal
  variations. Substorm: A geomagnetic perturbation lasting 1 to 2
  hours, which tends to occur during local post-midnight
  nighttime. The magnitude of the substorm is largest in the auroral
  zone. ({\em e.g.,} Sects.~\ref{sec:disturbances} and~\ref{sec:substorms})
\item {\em heliosphere:} \indexit{definition!heliosphere}the \indexit{heliosphere}extended region where the solar wind
  dominates over the interstellar medium ({\em e.g.,} Fig.~\ref{fig:ophercomposite})
\item {\em ionosphere:} \indexit{definition!ionosphere}the ionized \indexit{ionosphere}component of a \indexit{ionosphere|seealso{definition}}planetary
  atmosphere, largely overlapping with the thermosphere ({\em e.g.,} Sect.~\ref{sec:pcid})
\item {\em magnetosphere:} \indexit{definition!magnetosphere}a magnetic \indexit{magnetosphere}environment, generally of a
  planet, in which the intrinsic or induced magnetic field of the
  central body dominates over external fields or flows ({\em e.g.,}
  Fig.~\ref{fig:ophercomposite})
 \item {\em mesosphere:} \indexit{definition!mesosphere}layer \indexit{mesosphere}between stratosphere and
  thermosphere (Sect.~\ref{sec:nearlystrat}) 
\item {\em photosphere:} \indexit{definition!photosphere}'surface' of a star, at \indexit{photosphere}the
  rapid transition from opaque to transparent 
\item {\em stratosphere:} \indexit{definition!stratosphere}at Earth, \indexit{stratosphere}the domain between
  troposphere and mesosphere where
  temperature rises with height and convection is rare ({\em e.g.,}
  Sect.~\ref{sec:nearlystrat}) 
\item {\em solar (or sunspot or activity) cycle:}
  \indexit{definition!solar cycle}quasi-cyclic \indexit{sunspot!cycle}variation in the number of sunspots seen on the solar
  surface when averaged over time scales of months ({\em e.g.,}
  Fig.~\ref{figure:butterfly})
\item {\em (spectral, total) solar irradiance:}
  \indexit{definition!solar irradiance}solar input into a
  planetary atmospheric system in the form of photons ({\em e.g.,} Sect.~\ref{sec:nearlystrat})
\item {\em sunspot:} \indexit{definition!sunspot}a 'flux tube' in the \indexit{sunspot}near-surface layers, with
  suppressed internal convection, and large enough
  that lateral influx of radiation cannot prevent the interior from
  cooling relative to the surrounding photosphere and thereby
  appearing relatively dark ({\em e.g.,} Sect.~\ref{sec:sunstars})
\item {\em thermosphere:} \indexit{definition!thermosphere}outer layers of a planetary atmosphere \indexit{thermosphere}in
  which the temperature increases with height ({\em e.g.,}
  Sect.~\ref{sec:nearlystrat}), specifically the neutral particles
\item {\em (solar) transition region:} \indexit{definition!transition
    region}a domain \indexit{transition region}between chromosphere and
  corona with a very strong temperature gradient dominated by
  conduction (see Sect.~\ref{sec:solarouter})
\item {\em troposphere:} \indexit{definition!troposphere}the lower \indexit{troposphere}layers of a planetary atmosphere ({\em e.g.,}
  Sect.~\ref{sec:nearlystrat}) 
\end{itemize}
}
}
\end{flushleft}
\end{table}
\indexit{plasma|seealso{definition}}
% \end{figure}
With \indexit{Sun|seealso{star}}the
same measure of confidence we may assert that most of these
stars possess magnetic fields; that these magnetic fields create hot
outer atmospheres, or coronae, that drive magnetized winds from their
stars; and that these variable plasma winds blow past the orbiting
planets, distorting their individual magnetospheres, and push outward
against the surrounding interstellar medium.  Where the ram pressure
of the stellar wind becomes comparable to the surrounding pressure
(gas, magnetic, and cosmic ray) of the interstellar medium, a bow
shock forms.  This serves to mark the farthest extent of the
mechanical impact of the star on its surrounding environment: a sphere
of influence, so to speak. \regfootnote{See Tables~\ref{fig:glossary}
  and~\ref{fig:glossaryb} for definitions of the most common
  descriptors used for domains in, or phenomena related to,
  heliophysics. For a more extensive \indexit{glossary}glossary of terms used in this
  volume, see, for example
  \href{https://hesperia.gsfc.nasa.gov/sftheory/glossary.htm}{https://hesperia.gsfc.nasa.gov/sftheory/glossary.htm};
  for a glossary of terms related to space weather, see
  \href{https://www.swpc.noaa.gov/content/space-weather-glossary}{https://www.swpc.noaa.gov/content/space-weather-glossary}.}\activity{{\em
    Look up:} Review the vocabulary in Tables~\ref{fig:glossary}
  and~\ref{fig:glossaryb}. Choose a term (or, if you want, more than
  one) that you are not familiar with and look up more information on
  it. In your own words, write an expanded definition for that term in
  a few sentences. \mylabel{act:yourdefinition}}

Our Sun has and does all of these things, and we refer to the sphere of influence carved out by the solar wind as our heliosphere.  It is not really spherical and it varies in extent with solar activity.  But in broad terms we may safely think of it as extending about 100 times further from the Sun than the Earth's orbit.  We have yet to agree on the name for such spheres of influence around the other stars (for which astrospheres has been proposed), but there can be little doubt that such environments are as commonplace as the many points of light we see strewn across the sky on a dark and cloudless night.''

%%============================================================================

%\section{What is heliophysics?}
\ors[I:2.2] ``Heliophysics encompasses the
study of the various physical processes that take place within the
sphere of influence of the Sun ({\em i.e.,}  the heliosphere), and by
analogy, those environments surrounding most other typical stars.
But heliophysics also defines a specific method of study.  This
method embraces a holistic connected-system approach. It
emphasizes a comparative context in which to understand a process
by the many facets it presents in its various incarnations
throughout the heliosphere. Taken together, each diverse facet
serves to fill out a complete and physically satisfying picture of
a given process or phenomenon.

The physical processes and phenomena that we will encounter in [this
volume] are themselves especially diverse.  They include the rapid and
efficient energization of thermal particles to suprathermal energies,
the generation and annihilation of magnetic field, stellar variability
and activity cycles, space weather, turbulent transport of energy and
momentum, [the coupling between ionized and neutral atmospheres, and
atmospheric chemistry,] to name just a few. Heliophysics fills a
critical need to establish a unified science that connects these
seemingly unrelated concepts in a manner that emphasizes
complementarity over individuality, function over form, and generality
over specificity.

Along with unification, coupling provides the second
principal pillar upon which heliophysics rests.  The heliosphere is a
collection of coupled systems.  It is fortunate that many of the
linkages essentially operate only in one direction.  That is to say,
system A impacts system B, but B has little influence on A.  Under
these circumstances it is expedient to treat system A independent of
the behavior of system B.  This provides a certain economy of effort
and scale, and it often reduces the (apparent!) complexity of a
problem.  For example, complex geomagnetic activity has no impact on
solar flares, and the solar wind does not influence the Sun's cyclic
variability.

Linkages, especially when several are present and
working at cross purposes, can lead to confusion and spirited debate
over what is a root cause and what is simply a resulting effect.  The
cause and effect relationship between solar flares and coronal mass
ejections is a good case in point.  Consider, for example, what the
purported cause and effect relationship might be between a sore throat
and a fever.  Because a sore throat often starts before a fever develops
one might be tempted to assign the effect to the fever and take the
sore throat to be the cause.  Fortunately, medical research informs us
that both are effects and the root cause is the influenza virus.
Heliophysics is needed to play this very same role in sorting out the
appropriate relationships (or lack thereof) between any variety of
physical effects that often occur contemporaneously throughout the
heliosphere.

Solar variability does influence our \indexit{climate!solar variability}climate here on
Earth.  This \indexit{solar!variability and climate}fact is certainly not
negotiable in a purely scientific
context and is arguably one of the most important linkages between the
Sun and the Earth.  Satellites have confirmed that the solar
irradiance is variable on time scales from minutes to decades.  The
fluctuations are greatest on the shortest time scales. Day-to-day
irradiance changes are on the order of a percent versus tenths of a
percent over a solar cycle.  The magnitude and sense of irradiance
trends over centuries and millennia are currently difficult to
determine with any measure of certainty.  Slow but steady progress on
this question is being made through the studies of paleoclimate
records.  Over much longer time scales, stellar evolution theory
provides assurances that significant changes in solar irradiance have taken,
and will take, place with dramatic impacts on our climate and way of
life.

What is debatable, however, is precisely what the direct relationship
is between solar variability and climate change over any particular
time scale, or epoch, of interest.  For example, various opinions have
been advanced that span the entire gamut from wholly inconsequential
to complete solar responsibility for the gradual warming of the planet
that has been observed since the middle of the 20th Century. Yet, it
may not even make sense to speak of direct relationships between
drivers and the behavior of systems which are as nonlocal, nonlinear
and plagued by various hystereses as is our climate here on Earth.

The third and final pillar upon which heliophysics rests is the
exploration of Earth's neighborhood in space.  As a space-faring
civilization we have visited all the planets, [several asteroids and]
comets and numerous planetary satellites.  We have ventured to the
boundaries of the heliosphere and have flown through various parts of
our magnetosphere.  We have a spacecraft [that passed] the
Pluto/Charon system [and after that flew by Kuiper-belt object 486958
Arrokoth, provisionally known as 2014\,MU$_{69}$ and nicknamed Ultima
Thule].  Heliophysics enables our exploration to be successful and at
the same time gains in knowledge and understanding from our
exploration initiatives.

In summary, \indexit{heliophysics}heliophysics is the systems-mediated study of the
physical processes that take place within the Sun's sphere of
influence.  It is based upon the three pillars of unification of
physical processes and phenomena, coupling of distinct physical
systems, and the exploration of our neighborhood in space.  And it
is broadly applicable to the environments around most ordinary
stars.''

\section{The language of heliophysics}
\ors[I:2.3] ``The language of heliophysics is mathematics.  And the body of
literature from which heliophysics draws its substance and in turn
records its accomplishments is the physics of magnetized plasmas.
With only a rudimentary knowledge of a language, a literature is
incomprehensible, except, perhaps in translation.  And even in
translation so much of the original meaning and the nuance the author
wished to convey are inevitably lost, or worse, misinterpreted by even
the most conscientious translator.

The most precise, and intellectually demanding, literary prose of
heliophysics assigns a phase-space distribution function to each
individual species of particle.  By a
species one may simply mean free electrons, protons, or oxygen
molecules, or even photons.  In some applications it might be
necessary to distinguish between oxygen molecules in different
excited (vibrational, rotational and electronic) states, or
between iron atoms at different stages of ionization, or between
different senses of photon polarization.  In any case, the
evolution of each distribution function is obtained by setting the
total time derivative equal to the net production/loss of an
individual species by various collisional or radiative processes.
Such evolution equations are commonly referred to
as Boltzmann, or
Vlasov equations. When
there is no net gain or loss, then Liouville's Theorem asserts
that the vanishing of the total time derivative of the
distribution function conserves the phase space density for each
species [(see Sect.~\ref{sec:boltzmann})].

In specifying the total time derivative it is necessary to determine
the forces acting upon a given particle species.  For uncharged
particles, gravitational attraction is the only important
consideration.  Accordingly, to the system of equations for the
individual distribution functions one must add
Poisson's equation in order to
specify the gravitational field based on the mass distribution
provided by those particles with mass.  Charged particles are also
subject to electromagnetic interactions.  Thus we must also
include Maxwell's equations to
deduce the electric and magnetic fields based on the distribution of
charges and currents provided by the charged particle
species. \activity{{\em Look up:} Remind yourself of Maxwell's equations that are
  mathematical renditions of these properties: (1) electric monopoles
  are linked with an electric field; (2) there are no magnetic
  monopoles; (3) variations in the magnetic field are associated with
  a circulating electric field; and (4) a circulating magnetic field implies either steady
  currents or time-dependent electric fields, or both. Good news: once
  we have reached magnetohydrodynamics in Ch.~\ref{ch:universal},
  Maxwell's equations are in principle superfluous as they are
  contained within the MHD equations; if you are interested in how
  that works, see \href{https://arxiv.org/pdf/1301.5572.pdf}{Spruit (2016)}
  (Sections 1.1.1--1.1.9).  By the way, a really useful resource for
  all things related to plasma physics (and how to convert between
  different unit systems) is the online
  \href{https://www.nrl.navy.mil/ppd/content/nrl-plasma-formulary}{NRL
    Plasma Formulary}. Make a table that matches each equation (and
  add often-used names for them); use the NRL Plasma Formulary to
  compare and contrast the Gaussian and SI unit formulations. Which
  formulation do you like better? \mylabel{act:maxwellunits}}

In principle, this suffices to provide a complete description of the
grammar and syntax of heliophysics at a very elegant, learned and
precise level.  In practice the task of following through with this
program (a) is prohibitively difficult with or without the assistance
of the computer, (b) is subject to the problem that the initial
conditions are not known with any degree of certainty, (c) is
complicated by the fact that many of the collisional and radiative
transition probabilities are not even approximately known, and (d)
requires that certain conditions be fulfilled so that electromagnetic
interactions can be separated into large-scale fields and small-scale
collisions.  Finally, this comprehensive description usually provides
far more information than is usually necessary for comparing with
observations or understanding the predictions of a theory over
specified temporal and spatial scales.

At the opposite extreme from the scholarly literary prose is the
common vernacular.  For heliophysics, if high literary prose
centers on Poisson, Maxwell, Boltzmann and Vlasov, then the
vernacular is single-fluid, ideal, magnetohydrodynamics, or MHD
for short (see Ch.~\ref{ch:universal}).
MHD\indexit{MHD!description} is a continuum fluid description
that does not distinguish between particle species, averages (in
some sense) over particle collisions, ignores radiative effects
altogether, and is based on velocity moments of the underlying
distribution functions. It retains Poisson without modification,
but takes certain liberties with Maxwell.  Boltzmann and Vlasov
drop out of the picture entirely.

MHD can be rigorously derived from
Poisson/Maxwell/Boltzmann/Vlasov under various conditions that are
not altogether unreasonable for very many heliophysical
applications.  Usually this involves following the behavior of a
physical process or phenomenon over course-grained spatial and
temporal scales.  In other words, it is a useful, and indeed often
very accurate, description of the 'big picture'. Because of its
relative simplicity, ideal MHD provides a useful context in which
to interpret and understand the behavior of magnetized plasmas at
a basic and often extremely intuitive level.  On the other hand,
ideal MHD is often applied to processes or phenomena to which it
does not actually apply.  Generally speaking, if collisional and
radiative relaxation times are short compared to the
coarse-grained time scale of interest then ideal MHD is likely to
be a reasonable option.  But 'gotchas' are always present.

The successful derivation of the\indexit{MHD!closure prescription}
MHD equations requires a closure
prescription, which may be regarded as a consequence of the
familiar, 'no free lunch' maxim.  Closure entails specifying a
tractable procedure to determine the pressure tensor (second-order
velocity moment) in terms of the fluid density (zeroth-order velocity
moment), the bulk fluid velocity (first-order velocity moment), and
the magnetic field.  The so-called polytropic\indexit{polytropic
approximation} approximation --- in which the pressure is a scalar
proportional to the particle number density raised to a specified
power --- is the
simplest option.  A power law index of unity corresponds to an
isothermal process (constant temperature).  A power law index equal to
the ratio of specific heats describes an isentropic (constant specific
entropy) process that also manages to conserve energy.  More
complicated options are possible and are often tailored to accommodate
specific situations.  A successful and accurate closure scheme is
inevitably based on some additional {\em a priori} knowledge of the behavior
of the particle trajectories, or the general nature of the particle
distribution functions.

In contrast to the Poisson-Maxwell-Boltzmann-Vlasov description, ideal
MHD is a system of nine partial differential equations for nine
dependent variables [(shown in Table~\ref{fig:mhdset} and discussed in
Ch.~\ref{ch:universal})]: the gravitational potential, the fluid
density and pressure, the fluid velocity (3 components) and the
magnetic field (3 components).  These equations are (a) the Poisson
equation to describe gravity, (b) the continuity equation expressing
the conservation of mass, (c) the closure relation to specify the
pressure tensor, (d) the equation for the conservation of momentum, or
the force-balance equation (3 components), and (e) the magnetic
induction equation (3 components).

Of course, between ideal MHD and the
Poisson-Maxwell-Boltzmann-Vlasov description lies a vast real
estate filled with a plethora of compromise or hybrid
descriptions.  The number of such schemes is limited only by the
imagination and ingenuity of the investigators.  Multi-fluid
treatments allow for individual densities, velocities and
pressures associated with different particle species or groupings
of particle species, but retain a single gravitational potential
and magnetic field applicable to every fluid.  This formulation is
useful when the time scales of interest are short compared to
characteristic inter-species collisional relaxation times, but
long compared to the analogous intra-species times.

Another intermediate scheme employs high-order moment
closures.  These schemes are necessary when the species distribution
functions deviate significantly from the fully-relaxed Maxwellian.
Often this situation occurs when significant spatial gradients are
imposed on the system.  Additional partial differential equations are
then used to describe the time-evolution of the components of the
pressure tensor.  The closure is postponed to the next higher level of
the heat flux tensor (third-order velocity moment), or in extreme
circumstances to even higher-order moments.

Hybrid schemes treat some species as fluids and retain a
Boltzmann-Vlasov --~or kinetic~-- description for others.  Indeed even a
single species of particle may be partitioned in such a fashion that
some of the particles are treated kinetically (generally the high
energy suprathermal tail of the distribution function) while the
remainder are described as a fluid (the thermal core of the
distribution).
Such schemes are particularly useful in describing the energization of
charged particles; [we see this in action in Ch.~\ref{ch:conversion}].

In summary, there is a bewildering array of schemes that are presently
invoked to describe the behavior of magnetized plasmas in the
heliosphere.  They encompass an extremely wide range of complexity.
Each is specifically tailored to a given physical process and
phenomenon.  They are not simply interchangeable, but have their own
individual strengths and weaknesses.  One should always choose the
simplest description that will suffice for understanding the problem
in hand.  Use all the information and knowledge you have at your
disposal about the nature and behavior of a physical system in
selecting a scheme.  If the heliophysics concepts can
be adequately framed in the common vernacular, then eschew the
sophisticated flowery prose unless nothing less will
do.''

\section{A timeline of exploration of planetary systems}
NASA's Heliophysics Division within the Science Mission Directorate was
previously known as the Sun-Earth Connections Division. That earlier
name reflected that much of its research focused on how solar activity
impacts our home planet. As probes explored ever more of the solar
system, researchers realized that learning about the science of
terrestrial space weather and of the evolution of Earth's climate
system was boosted by the incorporation of discoveries from around the
solar system; the name change of the Division reflected the shift to a
broader perspective that was already taking place in the research
community.  As exoplanets were found to be more common than stars, the
application of the science of heliophysics to the exploration and
understanding of processes in exoplanetary systems, and in particular
to exoplanetary habitability, presents a natural development of the discipline.
The multi-disciplinary science arena that looks into star-planet couplings has
accelerated rapidly alongside astronomical exploration. 

In 1969, half a century ago, astronauts first landed on
Earth's sole moon. The first successful robotic landers touched down
on the much more distant Venus and Mars in 1970 and 1976,
respectively, and in the same decade spacecraft flybys provided the
first, fleeting close-ups of Jupiter and Saturn. It was not until two
decades later, however, that missions that explicitly targeted these
giant planets revealed how fundamentally distinct these worlds are
from our own.

The Galileo satellite started exploring the Jupiter system in late 1995, swinging by moon after moon. The Cassini-Huygens mission reached Saturn in 2004, exploring the giant planet, its rings and satellites, and even sending a lander onto Titan, the only moon in the solar system with a substantial atmosphere. These spacecraft uncovered a fascinating diversity of environments on dozens of moons: many are cold worlds enrobed in miles-thick ice; some with volcanoes spewing molten rock but others whose volcanoes somehow gush liquid water or nitrogen; and then there is Titan with its seas of liquid methane and ethane. Their pictures are as stunning and diverse as the scientific discoveries enabled by these spacecraft. The far reaches of the Solar System continue to offer surprises: dwarf planets Haumea and Makemake, objects in the distant Kuiper belt,
were not discovered until 2004 and 2005, respectively. 

As the close-up exploration of the largest planets in the solar system
got underway, a revolution was about to befall astronomers looking
much further out. It started in 1995 with the announcement of the
first exoplanet, now known as \indexit{exoplanet!51\,Pegasi\,b}51\,Pegasi\,b,
orbiting a star like our
own Sun. There are now well over 5,000 exoplanets on the books
\regfootnote{See \url{https://exoplanetarchive.ipac.caltech.edu} and
  \url{http://exoplanet.eu}.} (almost half of which were found with NASA's {\em
  Kepler} satellite), but the number expected to exist is vastly
larger: by carefully quantifying what our available methods can and
cannot observe, scientists estimate that there are over a hundred
billion planetary systems in our Milky Way galaxy alone, with perhaps
of order ten billion planets with some similarity to Earth.

Apart from its very existence, 51\,Pegasi\,b had another surprise in
store: at 150\,Earth masses and orbiting its star almost 20 times
closer than Earth does the Sun, this “hot Jupiter” should not have
existed by theories of the time. These and many subsequent
observations have changed our ideas on how planetary systems form and
evolve: we now realize that orbits can change so that planets may be
discovered well away from where they formed; planets can engage in
gravitational fights that can cause losers to be ejected as lone
'nomads' into interstellar space; planets exist that have two stars to
cast twin shadows on their surfaces; \ldots\ Many planets orbit their
stars at distances where water, if there is any, may exist in liquid
form on their surface for billions of years, as on Earth where it
enabled the development of life.

These discoveries have intensified the astronomers' hunt for
extraterrestrial life in which also solar-system scientists
participate. Organic molecules cause the haze in the icy-cold
atmosphere of Saturn's Titan and are vented in cryo-volcanic plumes
rising from the ice-locked deep ocean of nearby Enceladus. There are
many sizable moons and dwarf planets in the solar system that are rich
in water, although much of it is frozen solid. The combination of
liquids and organics in many places around our solar system fuels
theories of life and plans for space missions designed to look for
it near to home.

But exoplanet astronomers have the advantage of the vast number of
systems. Their challenge is that even the largest telescopes can image
exoplanets no better than as an unresolved blur the size of the
instrumental point spread function, if indeed they can separate the
reflected light from the exoplanets from the light of the stars that
they orbit. In fact, most of what we learn about exoplanets comes from
analyzing how their star's light is modified in brightness or color by
the exoplanets, either by adding some reflected starlight or by taking
away some light should they move in front of their star during their
orbit. Careful study of these effects as observed with the most
powerful telescopes can reveal which gases contribute to the
changes. This is receiving a big boost from NASA's James Webb Space
Telescope that started operations in 2022 and more from future
telescopes. So much was discovered in the most recent few decades;
what will the next several decades bring?

\clearpage

%\part{{\bf Fundamentals}}
%\addtocontents{toc}{\vspace{0.5cm}{\centerline{\bf Foundations}}\par }

\chapter{{\bf Neutrals, ions, and photons}}%2
\label{ch:particles}
{\narrower\narrower{
{\bf Chapter topics:}
\begin{itemize}
  \customitemize
\item Conditions in the local cosmos
\item Gravitational stratification
\item Cycle-driven variability of the solar spectral irradiance
\item Penetration depth of sunlight and its
  impact on the terrestrial atmosphere
\end{itemize}

\noindent{\bf Key concepts:}
\begin{itemize}
  \customitemize
\item Pressure scale height and differential stratification by atomic/molecular mass
\item The role of electron heat conduction in powering the solar wind
\item Collisional mean-free path
\item Optical depth
\end{itemize}

}}

\begin{figure}[p]
%\centerline{\hbox{\psfig{figure=figures/conditions,height=17.5cm,clip=}}}
\centerline{\hbox{\includegraphics[height=16.4cm,bb=56 113 481 679]{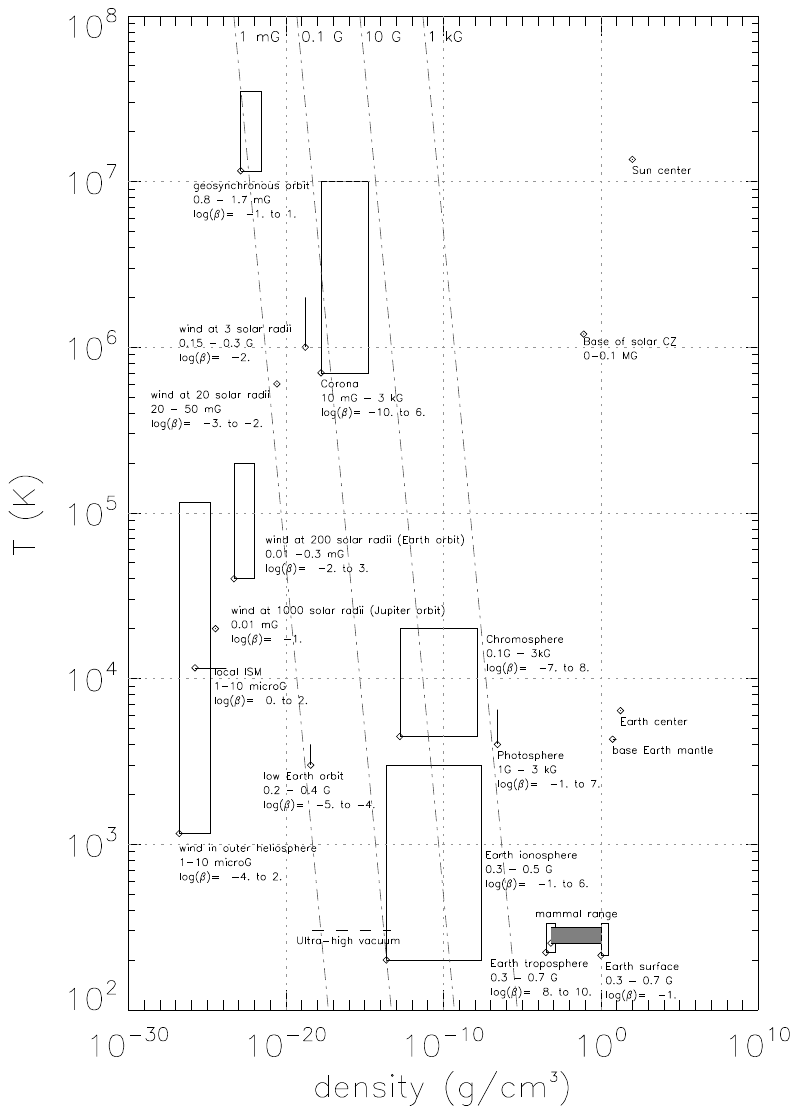}}}
\vskip -0.5cm \caption[Temperature-density diagram for
heliophysics.]{Temperature versus mass density for a variety of
  conditions within the local cosmos.
  Some typical ranges are
indicated, and labeled with \indexit{plasma!properties in different
  settings}magnetic field strengths (in Gauss)
found in that domain, followed by estimated ranges of the plasma
$\beta$, {\em i.e.,}  the ratio of energy density in plasma over that in
the magnetic field (Eq.~\ref{eq:betadef}), in this scaling for a fully ionized
hydrogen-dominated plasma. [Dash-dotted lines show where the plasma
$\beta$ equals unity for the field strengths shown near the top of the
diagram, making the same assumptions about the plasma.
 Fig.~I:1.1]  
}\label{fig:conditions}
\end{figure}
\section{Conditions in the local cosmos}
The local cosmos discussed in this book exhibits an enormous diversity
of conditions. Figure~\ref{fig:conditions} is one perspective of this
in its comparison of number densities and temperatures: densities
range over more than 28 orders of magnitude (more than the contrast
between solid rock and the 'vacuum' of low-Earth orbit) and temperatures over 5
orders of magnitude. The magnetic field, another crucial parameter
that is explored starting in Ch.~\ref{ch:universal}, provides another
dimension and adds its own physical processes. All together, these
physical parameters cover a wide range of states that include
solids, liquids, gases, and ionized and
magnetized particle ensembles called \indexit{definition!plasma}plasmas.

Matter in most of the domain of heliophysics is electrically
conducting, being generally at least partially or even fully ionized as
will be abundantly clear from the chapters in this volume. Ionization
can be a consequence of high-speed collisions between particles in a
hot medium and/or of high energies in the thermal radiation associated
with high temperatures. A hot medium can result from the transport and
conversion of different forms of energy where a balance of thermal
sources and sinks may only be reached at high temperatures. Examples
of such settings are the interior and the atmospheric domains of the
Sun. In these environments, internal collisional ionization and
recombination, as well as excitation and de-excitation processes
dominate in balancing ionization and recombination
rates. Alternatively, ionization can be the result of impacts of
externally-generated high-energy particles (such as solar energetic
particles or particles accelerated in a planetary magnetosphere) or be
caused by irradiation by solar photons of sufficiently high energy
(typically X-ray and [extreme] ultraviolet) such as occurs in
planetary ionospheres and cometary tails. \activity{{\em Consider:} Planetary lower
  atmospheres are dominated by molecular substances, transitioning to
  atomic elements with a relatively low admixture of ions and electrons
  as one moves up through the ionospheres and thermospheres, while
  magnetospheres and the solar outer atmosphere and wind are comprised
  predominantly of charged particles. (a) Compare thermal kinetic energies
  in different settings with molecular binding energies of, say, water
  and carbon dioxide. (b) Also compare the energy of X-ray and EUV photons
  with ionization energies of atomic hydrogen and oxygen. See
  Tables~\ref{tab:brain1}, \ref{tab:atmos-param}, and
  \ref{tab:wind-stats} for conditions in different settings.}

\begin{table}[t]
  \caption[Climates of terrestrial planets.]{Present
    \indexit{climate!terrestrial planets}characteristics and
    climates of the terrestrial
    planets. [Modified after Table~III:7.1, with added surface
    gravity, escape velocity, and escape energies $E_{\rm esc}$ for protons and
    atomic oxygen.] \label{tab:brain1}}
\begin{tabular}{lccc}
\hline
&	Venus	&Earth&	Mars\\
\hline
Radius	&6050\,km	&6400\,km	&3400\,km\\
Orbital radius	&0.72\,AU$^\ast$&	1\,AU	&1.52\,AU\\
Rotation period	&243\,days&	24\,hours&	24.6\,hours\\
Surface gravity   & 8.9\,m/s$^2$ & 9.8\,m/s$^2$ & 3.7\,m/s$^2$ \\
Escape velocity      & 10\,km/s & 11\,km/s & 5\,km/s \\
$E_{\rm esc}$ for H$^+$, O & 0.5, 9\,eV & 0.6, 10\,eV & 0.1, 2\,eV \\
Surface temp.	&740\,K&	288\,K&	210\,K\\
Surface pressure	&92\,bar&	1\,bar&	7\,mbar\\
Composition&	96\%\ CO$_2$&78\%\ N$_2$&95\%\ CO$_2$ \\
&	3.5\%\ N$_2$	&21\%\ O$_2$&2.7\%\ N$_2$ \\
H$_2$O content &	20\,ppm	&10,000\,ppm	&210\,ppm\\
Precipitation	&None at surface	&Rain, frost, snow	&Frost\\
Circulation	&1 cell/hemisph.;	&3 cells/hemisphere;	&1 cell/hemisphere  \\
	&quiet at surface	&local and regional 	&or patchy \\
	&but very active 	&storms	& circulation;  global\\
& aloft & &  dust storms\\
Maximum 	&	&  &	 \\
surface wind	&$\sim$3\,m/s	&$>$100\,m/s &	$\sim 30$\,m/s \\
Seasons	&None	&Comparable northern 	&Southern summer \\
	&	& and southern seasons	& more extreme\\
\hline
\end{tabular}
\newline
{\em \small $\ast$ An AU, or Astronomical Unit, is the average distance between
Sun and Earth.}
\end{table}

Much of what is described in this volume deals with the physics of
magnetized plasmas, and much of that physics is approximated by a
description known as magnetohydrodynamics, or MHD, as introduced in
Ch.~\ref{ch:universal}. In the present chapter, however, we first
look at the more familiar situation of neutral gases, also because
many of the phenomena discussed in this volume occur in the
layers of planetary atmospheres for which the concept of
hydrodynamics --~in which magnetic field is ignored~-- gives us a good
starting point. Later in this chapter, we focus on where ionization
becomes important. For now disregarding the effects of magnetic fields, the
limits of pure ('non-magneto-') hydrodynamics are reached in the high
tenuous layers of planetary atmospheres where collisions are
infrequent and other processes enter into our discussion, such as
chemical differentiation subject to gravity or even outflows from the
body in question. 
 
A great variety of phenomena in the local cosmos have their foundation
in the electrical conductivity of the media within which they
occur. This may be in the generation and maintenance of magnetic field
deep inside the Sun and in most of the planets, in the many phenomena
driven by the interaction of the magnetized flow of the solar wind with Solar-System
bodies, or even in the processes in the ionized domains of atmospheres
of many of these bodies. In most situations discussed in this volume,
that conductivity has its origin not in the metallic behavior of the
medium as it does deep inside Earth but rather in the
ionization of matter: whereas in metals ions are relatively immobile
and share some of their electrons, in a plasma both the ions and the
electrons are entirely unbound on microscopic scales. This chapter
introduces electrical conductivity in a magnetized medium, here
looking at plasma with a low degree of ionization; fully ionized
plasmas are discussed in Ch.~\ref{ch:universal}.

\section{Gravitationally stratified atmospheres and stellar
winds}\label{sec:nearlystrat}
Among the planetary atmospheres \indexit{stratification}in the Solar System, those of Venus
and Mars are most similar to those of Earth. The abundances of their
primary constituents --~mostly CO$_2$ and N$_2$, and, on Earth,
N$_2$ and O$_2$~-- are compared in Table~\ref{tab:brain1}. Note that
the order of the most abundant components as well as the absolute base
pressures differ markedly.

\begin{figure}[t]
%\centerline{\hbox{\psfig{figure=figures/Fig_1,width=7cm,clip=}}}
\centerline{\hbox{\includegraphics[width=7cm]{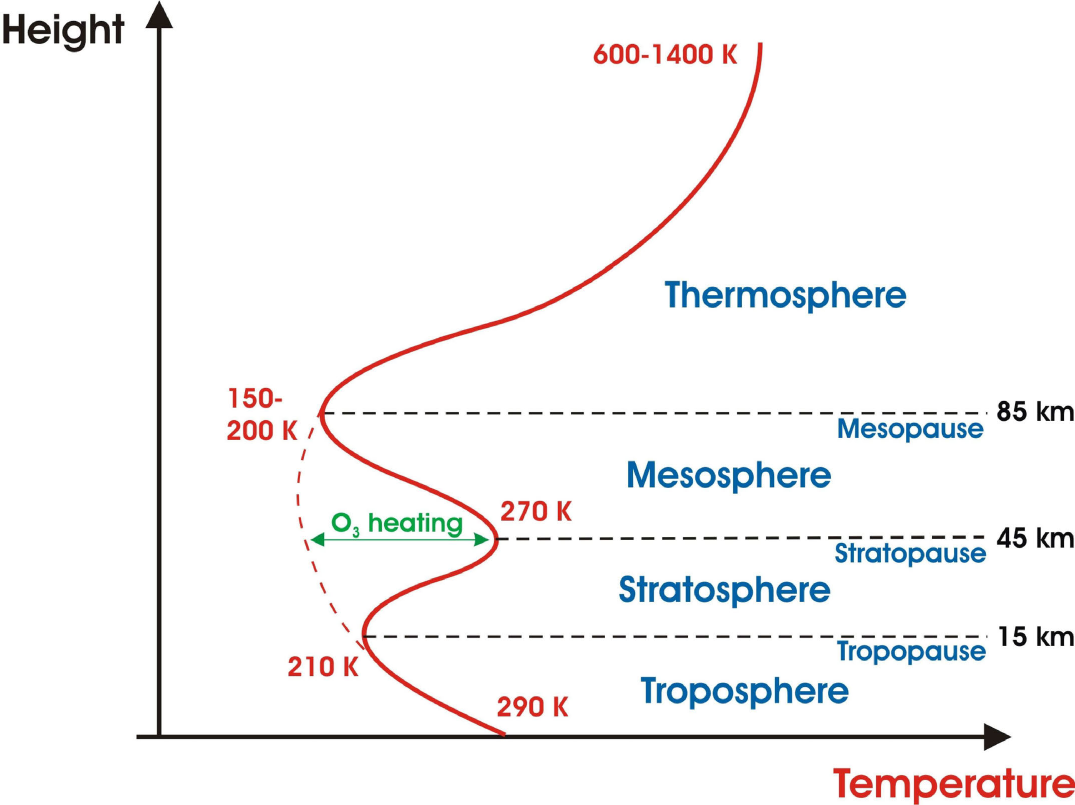}}}
\caption[Average vertical temperature profile through Earth's
atmosphere.]{Average vertical temperature profile through Earth's
  atmosphere. The general shape of the temperature profile is
  reasonably consistent to the point where it can be used to define
  the four main neutral atmosphere 'layers', from the troposphere to
  the thermosphere. The temperature of the uppermost layer, the
  thermosphere, increases steeply with altitude due to absorption of
  solar extreme ultraviolet (EUV) and far-ultraviolet (FUV)
  radiation. The thermosphere and upper mesosphere are partially
  ionized by the same EUV radiation, which varies by a factor of three
  over the solar cycle, and by auroral particle precipitation. [The
  effect of absorption by ozone is specifically
  highlighted. Fig.~I:12.1] \colorfig}\label{fig:fr1}
\end{figure}
A sketch of the Earth's atmospheric vertical thermal structure is
shown in Fig.~\ref{fig:fr1}. The temperature gradually drops from the
surface --~where the bulk of the conversion of solar irradiance into
heat occurs~-- through the troposphere due to adiabatic expansion.  At
greater altitudes the absorption of short-wavelength sunlight by
tenuous gas that is less efficient in cooling through radiation leads
to increased temperatures in the stratosphere (mainly by photons
between about 2,000\,\AA\ and 3,000\,\AA) and in the thermosphere (for
wavelengths mostly short-ward of 2,000\,\AA). Energy leaves the
Earth's atmospheric domains mainly by infrared radiation from the
lower regions, which also leads to a decrease in temperature above the
stratosphere by radiation from the mesosphere. The densities in the
thermosphere are so low, and the dominant chemical constituents such
inefficient radiators, that downward \indexit{thermal conduction}thermal conduction exceeds
radiative losses above about 100\,km (see
Ch.~IV:9). Table~\ref{tab:brain3} compares the properties of the
upper atmospheres of the three terrestrial planets (with significant
atmospheres), {\em i.e.,} the thermospheres, the ionized constituents
referred to as the ionospheres that largely overlap with the
thermospheres, and the exospheres beyond that; the reasons for the
apparent chemical mismatch between the neutral molecular and the
ionized components are discussed in
Ch.~\ref{ch:transport}. \regfootnote{This volume focuses on
  terrestrial planets; we refer to Ch.~IV:8 for an introduction to
  the upper atmospheres of the giant planets.}

\begin{table}
  \caption[Chemical species in the upper atmosph.\ of Venus, Earth,
  and Mars.]{Extent and important species for
    \indexit{atmosphere!composition!terrestrial planets}upper atmospheric regions
    of the terrestrial planets. [Table~IV:7.3; added planetary radii
    $R_{\rm p}$ (km). More detailed information is provided in the
    text and figures of Ch.~\ref{ch:transport}.] \label{tab:brain3}}
\begin{tabular}{lccc}
\hline
&	Venus	&Earth&	Mars\\
& $R_{\rm p,\venus}=6052$ & $R_{\rm p,\oplus}=6378$ & $R_{\rm p,\mars}=3396$ \\
\hline
Thermosphere	&$\sim$120-250\,km &$\sim$85-500\,km & $\sim$80-200\,km \\
                	&CO$_2$, CO, O, N$_2$	&O$_2$, He, N$_2$ & CO$_2$, N$_2$, CO \\
Ionosphere	&$\sim$150-300\,km & $\sim$75-1,000\,km & $\sim$80-450\,km \\
                	&O$_2^+$, O$^+$, H$^+$ & NO$^+$, O$^+$, H$^+$ & O$_2^+$, O$^+$, H$^+$ \\
Exosphere	&$\sim$250-8,000\,km &$\sim$500-10,000\,km & $\sim$200-30,000\,km \\
                        & H	& H, (He, CO$_2$, O)	& H, (O) \\
\hline
\end{tabular}
\end{table}

For the Sun's atmosphere, there is a comparable pattern of temperature with
height: moving upward, the temperature drops throughout the lower
atmosphere (the 'photosphere' from which the bulk of the solar
irradiance is emitted; also referred to as the 'solar surface' by
astronomers, despite the fact that the Sun is entirely gaseous
throughout), but then increases again in the chromosphere (extending a
few thousand km above the photosphere) and then shoots up to form an
extended, hot corona. Some of the physical properties of these domains
(along with a rough definition of the terms) are summarized in
Table~\ref{tab:atmos-param}. The reasons behind this similarity in
pattern are partly the same, partly completely different. A similarity
is that energy is most efficiently radiated from the low, dense
atmospheric layers, and poorly from high, tenuous layers where
conductive redistribution plays an important role. But the heat input
differentiates the two: the solar chromosphere and corona are not
heated by absorption of photons from the solar surface (which is
thermodynamically impossible because the atmospheric temperature is
higher than the surface temperature) but by dissipation of electrical
currents and a variety of waves running through the plasma (both
generated by the convective flows below the solar surface, and coupled
into the outer atmosphere via the Sun's magnetic field; see
Sect.~\ref{sec:actrad}). The amount of energy converted in the solar
outer atmosphere from chromosphere to corona and solar wind is a
function of the instantaneous magnetic activity. This activity exhibits an
11-year quasi-cyclic pattern that is often referred to as 'the sunspot
cycle' because it was discovered from multi-decade records of sunspot
counts.

\begin{table}[t]
  \caption[Domains in the solar atmosphere, with fundamental
  properties.]{Basic parameters \indexit{solar!atmosphere!domains}for,
    and definitions of, domains in
    the solar atmosphere. Note that all regions of
    the solar atmosphere are {\em very} inhomogeneous and that these
    values are only meant to give a rough idea of their
    magnitudes. [Table~I:8.1, here converted to cgs-Gaussian units,
    and with solar properties added. $n_{\rm e}$ and $n_{\rm H}$ are
    the densities of electron and neutral hydrogen; the plasma $\beta$
    is defined in Eq.~(\ref{eq:betadef})]}
\label{tab:atmos-param}
\begin{center}
\begin{tabular}{lccccc}
\hline
Region & $n$ & $n_{\rm e}/n_{\rm H}$ & $T$ & $B$ & $\beta$  \\
       & [cm$^{-3}$] & & [K] & [Gauss] &   \\
\hline
Photosphere$^1$ & $10^{17}$ & $10^{-4}$ & $6\,10^3$ & $1-1500$ & $>10$ \\
Chromosphere$^2$  & $10^{13}$ & $10^{-3}$ & $2\,10^4-10^4$ & $10-100$ & $10-0.1$ \\
Transition region$^3$  & $10^{9}$ & $1$ & $10^4-10^6$ & $1-10$ & $10^{-2}$ \\
Corona$^4$ &  $10^{8}$ & $1$ & $10^6$ & $1-10$ & $10^{-2}-1$ \\
\hline\hline
\multicolumn{6}{c}{Sun: radius $R_\odot=7\,10^{5}$\,km;
  surface gravity $g_\odot=274$\,m/s$^2$; bolometric }\\
\multicolumn{6}{c}{
  luminosity $L_{\rm bol}=4\,10^{33}$\,erg/s; effective temperature $T_{\rm eff} = 5772$\,K,} \\
\multicolumn{6}{c}{
  defined such that $L_{\rm bol}\equiv \sigma T_{\rm eff}^4 4\pi R_\odot^2$} \\
\hline
\end{tabular}
\end{center}
{\em \small  Definitions: $^1$ the {\em photosphere}\indexit{photosphere} is
the layer from which the bulk of the electromagnetic radiation leaves
the Sun (this layer has an optical thickness $\tau_\nu \la 1$ in the
near-UV, visible, and near-IR spectral continua, but it is optically
thick in all but the weakest spectral lines); $^2$ the
{\em chromosphere}\indexit{chromosphere} is optically thin in
the near-UV, visible, and near-IR continua, but optically thick in
strong spectral lines -~it is often associated with temperatures
around $10,000-20,000$\,K; $^3$ the {\em transition
region}\indexit{transition region} is a thermal domain
between chromosphere and corona in which thermal conduction leads to a
steep temperature gradient; $^4$ the {\em corona}\indexit{corona}
is optically very thin over the entire electromagnetic spectrum except at
radio wavelengths and in a few spectral lines -~the term is often used to describe the
solar outer atmosphere out to a few solar radii with temperatures
exceeding $\sim 1$\,MK.}
\end{table}

\begin{figure}
%\centerline{\psfig{figure=figures/Fig1_solar_spec_var_bb_grey,width=12cm,clip=}}
\centerline{\includegraphics[width=11.8cm]{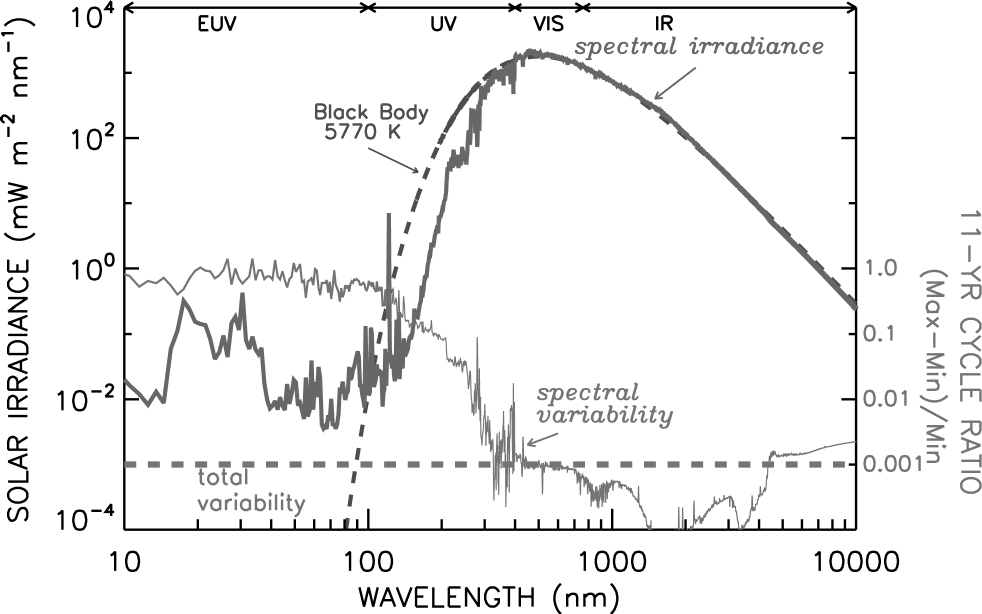}}
\caption[Solar {\em vs.}\ 5770\,K black body spectrum, and solar-cycle
variability.]{\label{lean:f1} Comparison of the solar spectrum and the
  black body spectrum for radiation at 5770\,K (the approximate
  temperature of the Sun's visible surface). Also shown is an estimate
  of the variability of the solar spectrum during the 11-y solar
  cycle, inferred from measurements (at wavelength below 4000\,\AA)
  and models (at longer wavelengths) and, for reference (dashed line),
  the solar cycle 0.1\%\ change in the total solar
  irradiance. [Fig.~III:10.1]}
%\end{figure}
%\begin{figure}[t]
%\centerline{\psfig{figure=figures/brf4,width=12cm}}
\centerline{\includegraphics[width=12.5cm]{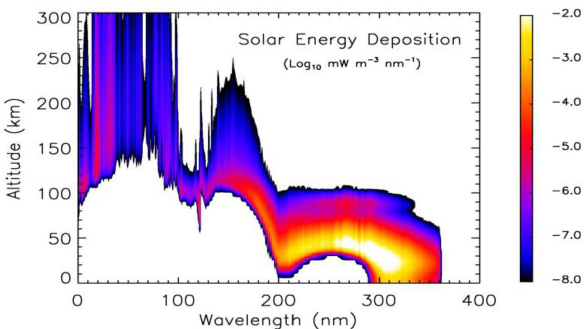}}
\caption[Altitude of penetration of the solar radiation with wavelength.]{[Altitude of penetration of the solar radiation as a
  function of wavelength [from X-rays through 3600\,\AA]. The
color range shows the amount of energy deposited in the different layers of the
atmosphere for the different parts of the solar spectrum (on
a logarithmic scale, in units of mW/m$^3$/nm [or $10^{-3}$\,erg/s/cm$^3$/\AA]). [Fig.~III:13.3]
%(From S.C.\ Solomon, private communication.)
\colorfig \label{fig:br4}}
\end{figure}
The Sun's radiative input into the Earth's atmosphere (known as the
spectral irradiance, $S(\lambda)$) \indexit{solar!irradiance!solar cycle}exhibits a significant variability
depending on solar magnetic activity (Figure~\ref{lean:f1}). The
overall emission from the solar photosphere varies little with
magnetic activity, that from the warm chromosphere mildly, and that
from the hot corona strongly.  \activity{{\em Look up} and compare
  images of the Sun's magnetic field and atmosphere in different
  phases of the solar cycle, such as
  \href{http://sdowww.lmsal.com/suntoday_v2/?suntoday_date}{those}
  obtained with the {\em HMI} and {\em AIA} instruments on {\em SDO}
  (NASA's {\em Solar Dynamics Observatory}). Note that such images are
  typically in false color, and with non-linear intensity scales to
  accommodate brightness contrasts. You could go to the
  \href{http://sdowww.lmsal.com/suntoday_v2/?suntoday_date}{AIA and
    HMI summary site} of the Solar Dynamics observatory
  (http://sdowww.lmsal.com/), use the
  \href{https://helioviewer.org}{Helioviewer} interface
  (https://helioviewer.org) to combine with other instruments and have
  a longer time base, or explore
  \href{https://iswa.gsfc.nasa.gov/IswaSystemWebApp/}{cygnet tools} at
  the \href{https://ccmc.gsfc.nasa.gov/iswa/}{iSWA} (Integrated Space
  Weather Analysis System at the CCMC: https://iswa.gsfc.nasa.gov),
  for example. For the largest contrast, compare inactive phases in
  2008-2009 and 2018-2019 to active phases in 2001-2003 and 2013-2014
  and also for the rising and falling phases in between. Make a table
  of the different properties that you notice from magnetic maps and
  from coronal images for each phase, including at least sunspot
  prevalence at different latitudes and coronal
  brightness. \mylabel{act:imagecycle}} As a result, the relative
variability in $S(\lambda)$ through the solar cycle increases markedly
short-ward of about 3,000\,\AA:
$(S_{\rm max}-S_{\rm min})/S_{\rm min}$ climbs from below one part in
1,000 long-ward of 3,000\,\AA\ to near unity short-ward of 1,000\,\AA.
The absorption of the most variable segment of the spectral irradiance
in Earth's atmosphere occurs primarily above about $50-100$\,km
(Fig.~\ref{fig:br4}), causing the high atmosphere to evolve strongly
in temperature and density in response to the solar sunspot cycle (see
Fig.~\ref{fig:fr2}), further modulated as Earth goes through its
weakly elliptical orbit around the Sun and its rotation about a tilted
axis, and with variable contributions from geomagnetic activity (see
Chs.~\ref{ch:evolvingplanetary} and~\ref{ch:transport}).

Wavelengths \indexit{solar!irradiance!photo-dissociation}short-ward of about 2400\,\AA\ and about 1250\,\AA\ can
dissociate O$_2$ and N$_2$, respectively, and short-ward of about
900\,\AA\ can ionize, {\em e.g.,} O atoms. Consequently, the atomic and
ionic components in the Earth's atmosphere do not show up
significantly below around 100\,km in altitude because all ionizing
and dissociating wavelengths have been absorbed by that depth into the
atmosphere (see Ch.~\ref{ch:transport}); \indexit{solar!irradiance!penetration depth} above that altitude, the
abundances of the ionic and
atomic components all reflect solar, orbital, and diurnal cycles.

%Dissociation energy: O2: 5.12eV~2400A; N2: 9.8eV~1265A.
%Ionization energy:    O: 13.6eV~912A; N: 14.5eV~860A

The density stratification \indexit{stratification}in much of the lower atmosphere of the
terrestrial planets (defined as below about 100\,km for Earth) can be
understood to first order by looking at the behavior of a stationary
gas subject only to gravity. \ors[I:12.2.1] ``The frequent collisions of
molecules in a gas close to thermal equilibrium enable the Maxwellian
[velocity distribution (with corresponding exponential energy
distribution)] of the individual particles to be characterized by the
basic fluid properties of pressure, $p$, temperature, $T$, number
density, $n$, and mass density, $\rho$, that are related by the
perfect\indexit{perfect gas law} gas law:
\begin{equation}
p = n k T = (\rho/m) k T  = \rho {\cal R} T / \mu,
\end{equation}
where $k$ [($1.4\times 10^{-16}$\,erg/K)] and ${\cal R}$
[($8.31\times 10^7$\,erg/K/mol)] are the Boltzmann and universal gas
constants, respectively, and $m$ is the mean molecular mass [while
$\mu$ the mean molecular mass in atomic units].\activity{{\em
    Consider} why for a fully-ionized, hydrogen-dominated plasma we
  see $p=2nkT$. For the answer, see below
  Eq.~(\ref{eq:mom}).}\activity{{\em Consider:} At the solar surface
  we see a mean 'molecular mass' of $m\approx 1.3 m_{\rm p}$ while in
  the fully-ionized corona $m\approx 0.6 m_{\rm p}$ (for proton mass
  $m_{\rm p}$). Explain why. (A hint: see
  Fig.~\ref{fig:ionprofiles}.)}  The fluid concept of pressure in the
atmosphere represents the weight of the column of gas above.

\begin{figure}[t]
\centerline{\includegraphics[width=\textwidth]{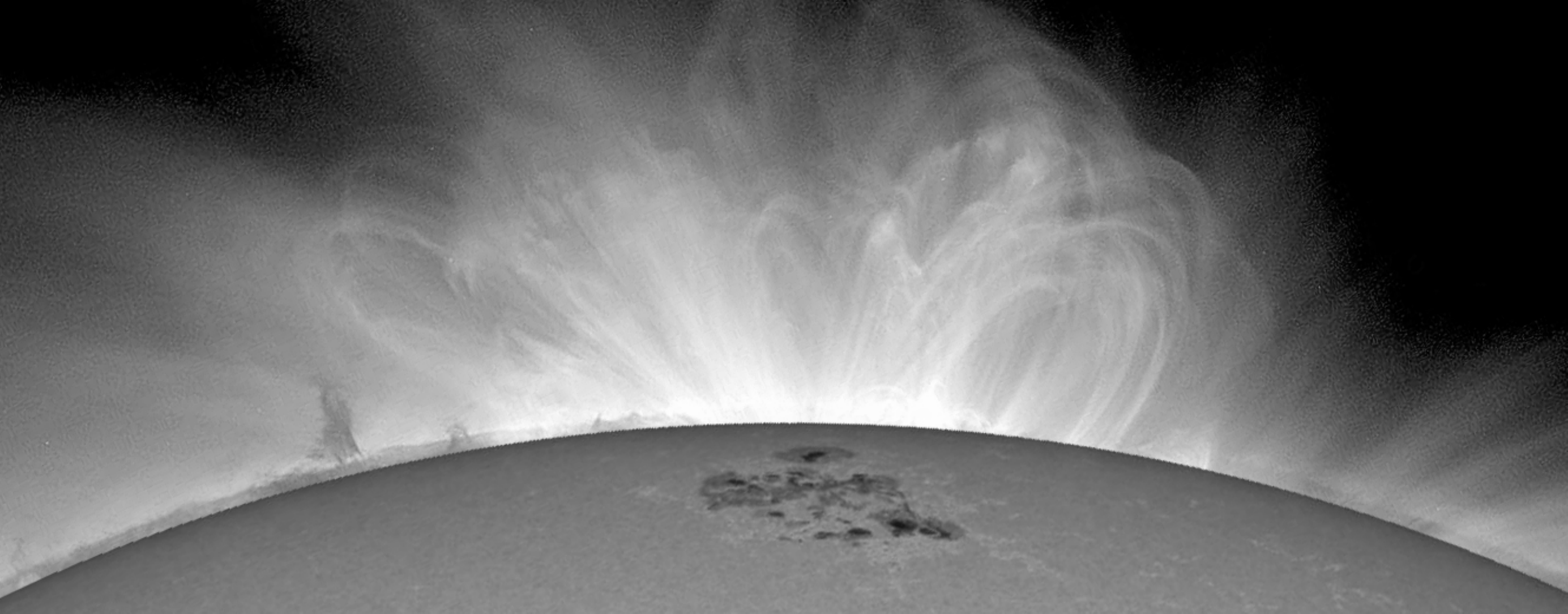}}
\caption[Solar corona over the solar limb.]{\label{fig:suncomp}
  Composite image of solar \indexit{coronal!loop}coronal loops over an active region
  (observed in the SDO/AIA 193\,\AA\ channel, most sensitive to
  coronal temperatures around 1.5\,MK). The foreground image that
  masks the on-disk corona is an SDO/HMI intensity image showing a
  large sunspot group approaching the limb from which most of the
  coronal loops in this composite image emanate. The dark band between
  the solar limb in visible light and the overlying corona is the
  chromosphere which is itself EUV-dark and opaque to coronal
  emissions, thus blocking the EUV emission from the corona behind it,
  showing up partially masked by foreground coronal emission. The
  images were taken on 2022/09/04 around 20\,UT, and were rotated by
  $115^\circ$ counterclockwise from solar north.}
\end{figure}
The neutral gas under the influence of the planet's gravitational
force gives rise to the concept of hydrostatic balance, which states
that the change in pressure with height, ${\rm d}p$, is closely
balanced by the weight of the fluid, $nmg{\rm d}h$ (where $m$ is the
mean molecular mass in [grams] and $h$ is the height), under the
action of the planet's gravitational acceleration, $g$. The concept is
expressed\indexit{hydrostatic balance} mathematically as:
\begin{equation}\label{eq:hydrobal}
\frac{{\rm d} p}{{\rm d} h} = -\rho g = -p / H_p.
\end{equation}
This basic equation describes the exponential decrease in gas density
with altitude, and results in the concept of the\indexit{pressure
scale height} pressure \indexit{definition!pressure scale height}scale height,
\begin{equation}\label{eq:hp}
H_p = k T / m g,
\end{equation}
which represents the [height difference] through which the gas pressure [in an
isothermal atmosphere] will decrease by a factor of $e=\exp{(1)}$. Earth's upper
atmosphere extends for about a dozen scale heights above 100\,km
altitude, with scale heights changing from about 5\,km to 50\,km with
increasing altitude, as the temperature increases from about 180\,K to
over 1000\,K (see Fig.~\ref{fig:fr1}).\activity{(a) {\em Show:} Compute
  scale heights $H_p$ in the Earth's atmosphere for molecular nitrogen
  (the dominant component) at a range of temperatures, and compare
  these with the value $H_{p\odot}$ for the atomic hydrogen-dominated gas in the
  solar photosphere, and for the CO$_2$-rich atmospheres of Venus and
  Mars. Use the data in Tables~\ref{tab:brain1}
  and~\ref{tab:atmos-param}. (b) Consider how the value of
  $H_{p\odot}/R_\odot$ contributes to the appearance of the Sun as having
  a well-defined surface. (c) Also, consider why neutral, atomic hydrogen dominates
  in the solar photosphere (see Fig.~\ref{fig:ionprofiles} for the answer).}

[The] quasi-equilibrium implied by hydrostatic balance does not
exclude the possibility of vertical winds. The assumption simply
demands that the rate of [flow] is such that the atmosphere adjusts at
a comparable rate. The term\indexit{quasi-hydrostatic balance}
quasi-hydrostatic balance is the more correct expression in the case
of accommodating vertical\indexit{vertical winds} winds in the
system. [\ldots] Vertical winds in Earth's upper atmosphere of the
order of 100\,m/s can be accommodated within the quasi-hydrostatic
assumption.'' \activity{{\em Show:} One way to quantify the 'strength'
  of storms in different planetary atmospheres is to compare the
  dynamic pressure $\rho v^2$ for the maximum surface winds listed in
  Table~\ref{tab:brain1} (which also lists base pressure, temperature,
  and dominant molecule, from which $\rho$ can be derived). Compare
  those values with the dynamic pressure in the solar wind using
  Table~\ref{tab:wind-stats}. Note: 1\,bar =
  10$^6$\,dyne/cm$^2$.\mylabel{act:stormstrength}} \activity{{\em
    Show:} The fastest flows (in any direction, not only in the
  vertical, gravitationally stratified direction) that can be
  accommodated in a quasi-hydrostatic situation can be estimated from
  the fact that the gas pressure $p=\zeta nkT$ (with $\zeta=1$ for a
  neutral gas and $\zeta=2$ for a fully ionized hydrogen gas) should
  well exceed the flow's dynamic pressure $\rho v^2$. Look at
  Fig.~\ref{fig:conditions} and add the horizontal lines where the two
  pressure terms are equal for a variety of flow velocities and
  corresponding temperatures; compare with the conditions discussed
  later in this chapter for the solar wind.}

The quasi-hydrostatic description applies not only to planetary
atmospheres but is also used for the interior of the gas giants, for
the interior and lower-atmosphere of the Sun, and --~as we shall see
later~-- even inside magnetic 'containers' in the solar atmosphere
that are known as flux tubes (see Table~\ref{fig:structures} for a
definition), one incarnation of which are 'coronal loops' --~which is
the general term describing the emitting structures seen in EUV and
X-ray images of the Sun's hot outer
atmosphere. Table~\ref{tab:atmos-param} summarizes characteristic
physical parameters for the domains within the solar atmosphere from
the solar 'surface' (photosphere) up into the corona (see also
Fig.~\ref{fig:suncomp} for an example image showing coronal loops
above a sunspot region).  These numbers
should be seen as characteristic values only: all these domains span a
few orders of magnitude in density and all are very dynamic at any
given location, while moreover the solar magnetic field plays a key
role in them as it structures multitudes of adjacent distinct
atmospheres along magnetic field bundles (Sect.~\ref{sec:fieldlines}).
The solar corona is visible at X-ray and EUV wavelengths up to several
hundred thousand kilometers. The coronal plasma is mostly contained in
magnetic structures relatively low down, but increasingly with height
the gas pressure forces the magnetic field to 'open' into the
heliosphere. The plasma on 'open field' streams out to form the solar
wind, resulting in a low-density and consequently dark lower coronal region
known as a 'coronal hole'.

The quasi-hydrostatic description even forms a useful, albeit very
crude, approximation for that part of extended atmosphere of the Sun
that is the inner-heliospheric domain of the solar wind: whereas there
is in fact an outflow, this 'vertical wind' leaves the \indexit{stratification}stratification
nearly hydrostatic for many solar radii above the solar surface, as we
shall see shortly.

Table~\ref{tab:wind-stats} summarizes a few characteristics of the
solar wind near Earth orbit. Outside of dynamic coronal mass ejections
(Ch.~\ref{ch:mhd}), the solar wind is predominantly in one of two
states, referred to as the 'fast wind' and the 'slow wind'. These
states originate from distinct environments on the Sun, and because
the Sun rotates underneath the radially outflowing wind, slow and fast
streams unavoidably interact --~see Sect.~\ref{sec:gosgmir}. For what
follows here, we focus on domains where only one of these types of
wind prevail for several days, which is the time it takes to flow from
Sun to Earth (the geometry of the magnetic field that it carries is
discussed in Sect.~\ref{sec:parker-spiral}).\sactivity{$\circledS$
  {\em Show:} (a) With the values in Table~\ref{tab:wind-stats}, how long
  do the slow and fast solar-wind streams take to reach Earth? (b) How
  many degrees does the Sun rotate between the moment these wind
  streams leave the Sun and the moment they arrive at Earth? (c) How long
  for Neptune? (d) Given that the wind flows out essentially radially,
  what is the apparent direction of the wind relative to the direction
  of the Sun as observed from the orbiting Earth (with an orbital
  velocity of about 30\,km/s)?  \mylabel{act:windquestions}
  \solution{windquestions}}

\begin{table}[t]
  \caption[Fast and slow solar wind; basic parameters.]{Basic
    parameters \indexit{solar!wind!near-Earth conditions}of the fast
    and slow solar wind [near Earth; modified
    after Table~I:9.1. Notes: (1) subscripts 'e', 'p', and 'i' are
    used to denote electrons, protons, and other ions, respectively;
    (2) 'FIP' stands for 'first ionization potential'; 'low-FIP' is a
    group of elements with first-ionization potentials below 10\,eV.]}
\begin{center}
\begin{tabular}{lcc}
\hline
Property (1~AU) & Slow wind & Fast wind \\
\hline
Speed & $430\pm 100$~km/s & $700-900$~km/s \\
Ion density  & $\simeq 10$~cm$^{-3}$ & $\simeq 3$~cm$^{-3}$ \\
Flux  & $(3.5\pm 2.5)\times10^8$~cm$^{-2}$s$^{-1}$ & $(2\pm 0.5)\times10^8$~cm$^{-2}$s$^{-1}$ \\
Magnetic field &$60\pm 30$~$\mu$G  & $60\pm 30$~$\mu$G\\
Temperatures & $T_{\rm p}=(4\pm 2)\times 10^4~{\rm K}$ & $T_{\rm p}=(2.4\pm 0.6)\times 10^5~{\rm K}$\\
             &  $T_{\rm e}=(1.3\pm 0.5)\times 10^5~{\rm K}>T_{\rm p}$ & $T_{\rm e}=(1\pm 0.2)\times 10^5~{\rm K}<T_{\rm p}$ \\
Anisotropies & $T_{\rm p}$ isotropic& $T_{{\rm p}\perp}>T_{{\rm p}\parallel}$ \\
Structure & filamentary, highly variable & uniform, slow changes\\
Composition & He/H$\simeq 1-30$\% & He/H$\simeq 5$\% \\
 & low-FIP enhanced & near-photospheric \\
Minor species & $n_{\rm i}/n_{\rm p}$ variable & $n_{\rm i}/n_{\rm p}$ constant\\
              & $T_{\rm i}\simeq T_{\rm p}$ & $T_{\rm i}\simeq({m_{\rm
                  i}/m_{\rm p}})T_{\rm p}$\\
              & $v_{\rm i}\simeq v_{\rm p}$ & $v_{\rm i}\simeq v_{\rm p}+v_A$\\
Associated with & streamers, transiently & coronal holes\\
                & open field & \\
\hline
\end{tabular}
\end{center}
\label{tab:wind-stats}
\end{table}
%\begin{table}%[htp]
%\caption[Solar wind parameters.]{[Statistical properties of the solar
 % wind near Earth. Modified after Table~I:11.1].}
%\begin{tabular}{|l|c|c|c|c|c|}
%\hline \hline
%Parameter                     &  mean             &  st.dev.       &  median          &  5-95\%\   & observed \\
%&  &    & &  range limit  & extremes \\
%\hline
%density $({\rm cm}^{-3})$    &   8.7             &   6.6            &   6.9            &  3-20                &  0.1-83                              \\
%velocity (km/s)               &  468              &  116             &   442            & 320-710              & 250-2000$^a$ \\
%ram press.\ $\rho v^2$ (nPa) &   7               &   5.2            &   5.5            & 0.01-14.5            & 0.05-28        \\
%magnetic field ($\mu$G)  &  62              &   29            &   56            & 22 - 99            &$0^b$ - 850 \\
%ion temp. (K)           &  1.2$\, 10^5$ & 0.9$\, 10^5$ & 1.0$\, 10^5$ & 0.1-3$\, 10^5$   & $10^4 - >3 \, 10^5$ \\
%electr.\ temp. (K)      &  1.4$\, 10^5$ & 0.4$\, 10^5$ & 1.3$\, 10^5$ & 0.9-2$\, 10^5$   & $10^5 - >2 \, 10^5$ \\
%\hline \hline
%\end{tabular}
%\begin{flushleft}
%  $^a$ Measurements of high velocities [listed here] were limited by
%  instrumental effects; $^b$ Indicates at least one interval with B
%  $<$ 1 $\mu$G.
%\label{table:1}
%\end{flushleft}
%\end{table}

The medium of the heliosphere is fundamentally distinct from that of
the lower 100\,km of the terrestrial atmosphere: the solar wind
\indexit{solar!wind} is
primarily made up of hydrogen with a lesser amount of helium, is hot
and therefore almost fully ionized, and is threaded by a magnetic
field. The dynamics of the solar wind and the ways in which it
interacts with planetary magnetospheres is modulated by that magnetic
field but, as first noted by \citet{1958ApJ...128..664P}, the basic
stratification and flow of the solar wind can be understood from its
high temperature: because it is hot and ionized, the electrons in the
solar wind are very efficient at conducting heat, and that is all it
takes to understand how it can lead unavoidably to a fast wind that
can escape solar gravity. It is not simply an 'evaporation' off the
Sun; after all, even at some millions of degrees, \ors[I:9] ``the sound
speed $c_{\rm s}$ --- essentially the mean ion speed --- is much
smaller than the [escape speed $v_{\rm esc}$ which can be derived by
equating a particle's kinetic energy with its \indexit{gravitational!potential
energy}gravitational potential energy at the surface:
\begin{equation}\label{eq:brain3}
v_{\rm esc}=\sqrt{2GM/r}.
\end{equation}
For the solar corona, the sound and escape speeds are]
\begin{equation}
c_{\rm s}\approx\sqrt{kT/m}\approx 100~{\rm {km/s}}\ll
v_{\rm esc}=\sqrt{2GM_\odot/R_\odot}=618~{\rm {km/s}},
\label{eq:cs-and-vg}
\end{equation}
where $k$ is Boltzmann's constant, $T$ the coronal temperature, $m$
the mean particle mass, $G$ the universal gravitational constant, and
$M_\odot$ and $R_\odot$ the solar mass and radius, respectively.

Mass and momentum balance radially away from the Sun [in an assumed
uniform, strictly radial flow] at heliocentric distance $r$ can be written
\begin{eqnarray}
\frac{{\rm d}}{{\rm d}r}(\rho v 4\pi r^2)&=&0 \label{eq:massv}\\
\rho v\frac{{\rm d}v}{{\rm d}r}&=&-\frac{{\rm d}p}{{\rm d}r}-
\rho \frac{GM_\odot}{r^2},
\label{eq:mom}
\end{eqnarray}
with $\rho$ the mass density, $v$ the flow speed.
%\activity{See how the
%  force balance is expressed differently in Eqs.~(\ref{eq:hydrobal})
%  and (\ref{eq:mom}) to accommodate a stationary flow, noting that
%  ${\rm d}v/{\rm d}t=({\rm d}v/{\rm d}r)({\rm d}r/{\rm d}t)$.}
Then $p=2nkT$ is the gas pressure in an electron--proton plasma with
$n$ representing the electron or proton number density, and $\rho=mn$
where $m$ is the mean particle mass which is given by
$m\approx m_{\rm p}/2$ for an electron--proton plasma.\activity{{\em
    Show:} The momentum balance in Eq.~(\ref{eq:mom}) describes a
  radially-flowing wind over a non-rotating Sun. In reality, the Sun
  is rotating, and the magnetic field reaching into the heliosphere
  enforces the wind to co-rotate with the Sun, out to a distance where
  it becomes too weak to enforce such co-rotation (somewhere between
  10 to 20 solar radii, or 0.05 to 0.1\,AU). (a) Show that for a
  sufficiently slowly rotating Sun, ignoring the centrifugal force is
  warranted. (b) At what rotation period of a star like the Sun does the
  centrifugal force at, say, $2R_\odot$ counteract gravity by more
  than 10\%? The centrifugal force in the wind would have been
  important for the very young Sun, see
  Sec.~\ref{sec:overallactivity}. Moreover, in the early phases of
  star-disk systems, centrifugal forces may be important in driving a
  cold wind; see Sect.~\ref{sec:diskwind}. \mylabel{act:centrifugal}}

[The] consequence of the \indexit{thermal conduction}thermal conduction in a million degree corona
is to extend the corona; {\em i.e.}, the temperature falls off slowly
with distance from the Sun. Thus, in a hypothetical {\em static}
atmosphere, we find a pressure at infinity given by
\begin{eqnarray}
\frac{{\rm d}p}{{\rm d}r}&=&-nm\frac{GM_\odot}{r^2},\\
p(r)&=&p_0\exp\left[-\frac{mGM_\odot}{2k}\int_{R_\odot}^r\frac{{\rm d}r}{r^2T(r)}\right].
%\Rightarrow& &\lim_{r \to \infty}n(r)T(r)> 0
\end{eqnarray}
Thus, if the temperature falls less rapidly than $1/r$, we find that
$\lim_{r \to \infty}p(r)> 0$, we expect a non-vanishing pressure at
infinity when the corona is extended. In particular, we find that for
reasonable temperatures and densities $n_0$, $T_0$ at the 'coronal
base' this pressure is much larger than any conceivable interstellar
pressure.

[The observed slow decrease of temperature with distance from the Sun,
caused by the efficient \indexit{thermal conduction}thermal conduction that is mostly carried by
electrons, implies that the solar wind must expand supersonically into
interstellar space. For a spherically symmetric, single-fluid,
isothermal outflow,] the equations of mass and momentum
conservation~(Eqs.~\ref{eq:massv}, \ref{eq:mom}) can be rewritten to
give\sactivity{$\circledS$ {\em Show:} What powers the solar wind in the basic
  model discussed here?  To see the answer, rewrite Eq.~(\ref{eq:mom}) or
  Eq.~(\ref{eq:solwind}) to an energy equation (a version of
  Bernoulli's law) with the terms for the kinetic and potential
  energy in the Sun's gravitational field, plus a term that reflects
  the work done by the expanding gas both geometrically and by
  acceleration; the energy for that expansion in the isothermal
  approximation is provided by the efficient thermal conduction by the
  electron population (see Eq.\,(\ref{eq:conduction}) and
  footnote~\ref{footspitzer}.). The real-world solar wind is not
  isothermal, certainly not far from the Sun (compare the coronal
  temperatures in Table~\ref{tab:atmos-param} with near-Earth wind
  properties in Table~\ref{tab:wind-stats}), and moreover is provided
  some additional power (in the form of heating and pressure) by waves
  and turbulence. \solution{windbasics}\mylabel{act:windenergy}}
\begin{equation}
\frac{1}{v}\frac{{\rm d}v}{{\rm d}r}\left\{ v^2-\frac{2kT}{m_{\rm p}}\right\} =
\left\{\frac{4kT}{m_{\rm p}r}-\frac{GM_\odot}{r^2}\right\}
\label{eq:solwind}
\end{equation}
[The solar wind starts slow, but is supersonic further out in the
heliosphere; such a] transonic wind\indexit{solar!wind!critical
point} passes through a critical point at 
\begin{equation}
r_{\rm c}=\frac{m_{\rm p}GM_\odot}{4kT}\quad {\rm where}\quad v_{\rm c}=\sqrt{\frac{2kT}{m_{\rm p}}}
\label{eq:criticalp}
\end{equation}
[(note the dependence on stellar
mass). \sactivity{$\circledS$ (a) {\em Show:} Estimate the mean-free path for
  collisions between electrons in the fast and slow solar wind near
  Earth based on Tables~\ref{tab:wind-stats} and
  \ref{tab:dimensionlessnumbers}.  (b) Despite these large numbers,
  the use of pressure and temperature as defined from Maxwellian
  velocity distributions is useful in the heliosphere. Discuss how
  this may come about.\mylabel{act:mfps}\solution{mfps}}\activity{{\em
    Background:} Note that there is another class of solutions allowed
  in principle by Eq.~(\ref{eq:solwind}), namely a 'solar breeze,'
  although that is incompatible with the outer boundary condition of
  negligible pressure: starting at low speed and never becoming
  transonic. Where does a 'solar breeze' reach its maximum velocity?
  By the way: in principle, Eq.~(\ref{eq:solwind}) allows for an
  inflow: where $v$ is negative, ${\rm d}v/{\rm d}r$ needs to be of
  opposite sign also. This inflow, accelerating from infinity towards
  the star, is known as Bondi accretion. However, such inflow is
  unlikely to occur as an isothermal flow from infinity because the
  interstellar medium is typically cold, with low ionization and thus
  low heat conductivity by electrons. Consequently, compression would
  raise the temperature of the inflow. Moreover, be aware that the
  quasi-hydrostatic approximation fails for the inner regions of such
  an infall, starting already well outside the critical point! So, the
  equation may allow it, but reality does
  not. \mylabel{act:parkerplot}} Formally, the equations allow such a flow] to match {\em any}
pressure as $r\to\infty$ [although in reality the reach of the flow is
limited by the existence of an interstellar medium
(Sec.~\ref{sec:impinging})].

Let us examine this transonic wind solution in somewhat greater
detail.  If we integrate the force balance, Eq.~(\ref{eq:mom}), from
the coronal base to the critical point $r_{\rm c}$ we find a density
$\rho_{\rm c}$ at the critical point given by
\begin{equation}
\rho_{\rm c}=\rho(r_{\rm c})
  =\rho_0\exp{\left\{-\frac{m_{\rm p}GM_\odot}{2kTR_\odot}+\frac{3}{2}\right\}}.
\label{eq:rho-crit}
\end{equation}
Note that this density is almost exactly the same as if there had been
{\em no} solar wind flow: {\em The subsonic corona in the solar wind
is essentially stratified \indexit{stratification}as a {\bf static} atmosphere}.

We can also find the resultant mass flux for the wind by
examining the density and the velocity at the critical point:
\begin{equation}
(nv)_r=n_{\rm c}v_{\rm c}\frac{r_{\rm c}^2}{r^2} 
            \propto \rho_0T^{-3/2}\exp\left[-\frac{C}{T}\right]
\label{eq:mflux-isotherm}
\end{equation}
where $\rho_0$ is the density at the coronal base [and $C$ a
constant]. {\em The mass flux is proportional to the density at the
  coronal base and depends exponentially on the coronal
  temperature.}'' The actual solar wind is not only driven by thermal
conduction from the coronal environment (which supplies energy for the
work of driving the wind against gravity), but also by magnetic waves,
known as Alfv{\'e}n waves, whose fluctuations act as an additional
pressure term, and whose dissipation aids in heating far above the
solar surface, all of which is particularly important for the fast
wind streams; more on that in Sect.~I:9.5. Another note on more
detail is found in Sect.~I:9.6, which begins to explain why for a
more realistic solar wind description that also allows for helium, the
exponential dependence of the solar mass loss on temperature is much
weakened into a power-law dependence of temperature.

Note that it is not only the efficient thermal conduction per se that
leads to a significant solar wind, but also the high temperature and
low particle mass, and that that is the reason for the contrast with Earth's
atmosphere.  In Eq.~(\ref{eq:hydrobal}) gravity is approximated by a
constant, leading to a formal solution for the pressure \indexit{stratification}stratification
of the terrestrial atmosphere that tends to zero exponentially even
for an isothermal atmosphere; this is not a bad approximation for an
atmosphere in which the pressure scale height (at most some 50\,km) is
well below 1\%\ of the planet's radius, so gravity changes little even
over many scale heights above the surface. But in the hot corona, the
pressure scale height for the hydrogen-dominated gas at $\approx 2$\,MK is
about $0.15R_\odot$, so gravity diminishes noticeably in
the first few pressure scale heights, hence its distance dependence
needs to be reflected in Eq.~(\ref{eq:mom}). The relatively weaker
gravity (and the correspondingly reduced escape energy) at large heights leads to a
transonic wind at coronal temperatures.

On a side note (to which we return in Ch.~\ref{ch:formation}), the
same equation Eq.~(\ref{eq:mom} also informs us about an accelerating
inflow (for which $v{\rm d}v/{\rm d}r>0$ as both $v<0$ and ${\rm
  d}v/{\rm d}r<0$) enabling the formation of stars and planetary
systems: gravity can win out over a pressure difference on very large
scales in the Galaxy on which stars form, because now gravity in fact
is built up by the infalling matter itself so that $M_\odot$ needs to
be replaced to read
\begin{equation}\label{eq:infalling}
\rho v{{\rm d}v\over {\rm d}r}=-{{\rm d}p\over {\rm d}r}-
\rho {G\over r^2} \int_0^r \rho 4\pi r^2 {\rm d}r.
\end{equation}

\ors[III:3.1] ``To make a\indexit{gravitational!collapse!condition for}
star of a given mass $M$ from a gas with temperature $T$, gravity must
overcome the pressure support. [One way to estimate the required
properties of a cloud involved in the initiation of star formation is
to look at Eq.~(\ref{eq:infalling}) and see when conditions cannot
remain in a stationary balance, {\em i.e.,}  when $v=0$ cannot be
maintained. That occurs when] the radius $R$ of the
protostellar\indexit{protostellar!cloud} cloud exceeds the critical radius]
\begin{equation}
R_{\rm c}(M,T) \ga {G M \over c_s^2} ~=~ {G M \mu m_{\rm p} \over k T} \,, \label{eqlh:rofm}
\end{equation}
where $c_{\rm s}$ is the sound speed and $m_{\rm p}$ is the mass of the hydrogen
atom.  Taking a mean molecular weight $\mu = 2.3$, appropriate for
molecular hydrogen plus helium, and a typical cold molecular cloud
temperature of $T = 10$\,K, Eq.~(\ref{eqlh:rofm}) implies that a solar
mass star must collapse from a cloud of radius $R \sim 2 \times
10^4$~astronomical units ([{\em i.e.,}  Sun-Earth distances; shorthand] AU).''

You will see the logic used in these examples applied throughout this
book, and indeed astrophysics in general: approximations in functional
forms, simplifications about geometries, and order of magnitude
estimates are used throughout to aid in the basic understanding what
is going on. With these tools, analytical and --~far more commonly~--
numerical solutions become interpretable in terms of the basic, common
processes. How much can be simplified to show the basics, however,
depends on the environment: heliophysics, as is physics in general, is
about simplifying as much as is allowed, but no more.

%\section{Separation and mixing}
\section{Photons, collisions, ionization, and differentiation}\label{sec:pcid}
In our everyday lives \indexit{atmospheric!separation and mixing}we can get away with taking it for granted that
the atmosphere around us is the same no matter where we are.
Moreover, we may take it to be true that this atmosphere is a mixture
of mostly N$_2$ and O$_2$. And that this atmosphere is a very poor
electrical conductor and that its winds are unaffected by the
planetary magnetic field. As it turns out, none of these properties
that we take for granted apply outside of the domain where we live:
the chemical mixture depends on height in planetary atmospheres and is
affected by the variable spectral irradiance from the Sun's outer
atmosphere, ions and thus electrical conductivity are important in
most of the local cosmos, and magnetic fields influence flows and vice
versa almost everywhere in space. In this section, we focus on the
processes that make the atmospheric composition dependent on location,
primarily altitude. In the next section, Section\ref{sec:pickup}, we
start looking at the role of ions in electrical conductivity and
flows, although the role of magnetic fields in that is the focus of
Ch.~\ref{ch:universal}.

The scale height for different atmospheric constituents depends on the
molecular or atomic mass, and is thus in principle different for
different chemicals. But as long as the mixing by winds and
(turbulent) convection is fast enough compared to the time scale by
which the chemical separation can occur by diffusive settling, the
atmospheric composition will remain uniform, and all major species
will share the same scale height. When \indexit{collisions!Earth atmosphere}collisions become relatively
infrequent above the homopause (at about 100\,km for Earth), and
diffusive settling exceeds mixing by flows, separation of chemicals by
molecular mass occurs; see Ch.~\ref{ch:transport}. The rate of
separation depends on the diffusion coefficients, which themselves
depend on chemical species and density, and on the chemical reactions
that couple species (and, in the ionosphere, also through ion-neutral
interactions), relative to turbulent mixing efficiency; see the
discussion in Ch.~IV:9.

Still higher in the atmosphere, where collision frequencies become so
low that the mean free path approaches or exceeds the formal pressure
scale height, the description of the medium as an ideal gas
fails. That environment, where particles essentially move
ballistically over long distances subject only to gravity (still
disregarding any effects of electric and magnetic fields), is known as
the \indexit{definition!exosphere}exosphere. The
\indexit{exosphere}exospheric
base height can be estimated by looking
at collision frequencies.

%From
%http://iopscience.iop.org/chapter/978-1-6817-4692-0/bk978-1-6817-4692-0ch6.pdf
The characteristic frequency \indexit{collision frequency} at which a
particle in a non-magnetized plasma or a non-ionized gas of identical
particles, all characterized by a temperature $T$ and at particle
density $n$, collides with other such particles is given by
\begin{equation}\label{eq:collfreq}
\nu  = \sigma_{\rm cc} v_{\rm rel} n = \sigma_{\rm cc} \left ( {kT
    \over m} \right)^{1/2} n,
\end{equation} 
where $\sigma_{\rm cc}$ is the mutual collision cross section and $v_{\rm
  rel}$ is the velocity of one particle relative to
another. In computing the mean free path, \indexit{mean free path} the velocity cancels out, leaving only
the density as a variable: 
\begin{equation}\label{eq:mfp}
\lambda_{\rm mfp} = {v_{\rm rel} \over \nu} = {1 \over  \sigma_{\rm cc} n}.
\end{equation}
By way of example, let us look at neutral atoms with a collisional
cross section of order, say, $3\times 10^{-16}$\,cm$^2$ (as for
hydrogen atoms). For these, a
density of $3\times 10^{8}$\,cm$^{-2}$ (reached at roughly 500\,km in
Earth's atmosphere, depending on solar activity) would
correspond to $\lambda_{\rm mfp} \approx 100$\,km. This
order-of-magnitude estimate shows that this density in the Earth's
atmosphere roughly forms the point at which a vertically moving atom
could jump over a scale height, or essentially through the bulk of
overlying matter, so where the assumption that we can work with the medium
as a gas of electrically neutral particles fails; this is about the
point where the Earth's atmosphere transitions into an
\indexit{exosphere}exosphere
where neutral atoms move essentially ballistically.

On the Sun, in contrast, the neutral hydrogen population could still
be described by hydrodynamics at that density because of the much
larger scales involved, if matter were largely neutral there; however,
that density is reached only in the corona where high temperatures
cause hydrogen and helium to be fully ionized (see
Table~\ref{tab:atmos-param}), and collisions occur via long-range
electromagnetic forces between charged particles (see Table~\ref{tab:5.2}
for mean-free path estimates in an ionized medium, which, with
Eq.~(\ref{eq:mfp}), shows the larger effective collision cross section
for Coulomb collisions). Lower down in the solar atmosphere where
neutrals do dominate, the mean free path lengths are significantly
smaller: the plasma throughout the Sun up to the inner corona behaves
like a gas in which (often turbulent) flows counter gravitational
separation. There are fractionation effects deep inside the Sun where
mixing by flows is negligible on solar evolutionary time
scales. Chemical differentiation is also seen in the atmosphere in the
minority species, specifically determined by the energy required for
first ionization of the atom (see Fig.~I:9.2); this differentiation,
not by diffusive settling but likely related to MHD waves and by EUV and X-ray
irradiation of the chromosphere from the higher atmosphere, is still
inadequately understood and not further discussed here.

\begin{figure}[t]
%\centerline{\hbox{\psfig{figure=figures/BougherRobleFig2.eps,width=\textwidth,clip=}}}
%\centerline{\hbox{\psfig{figure=figures/SBfig9_1,width=\textwidth,clip=}}}
\centerline{\hbox{\includegraphics[width=\textwidth]{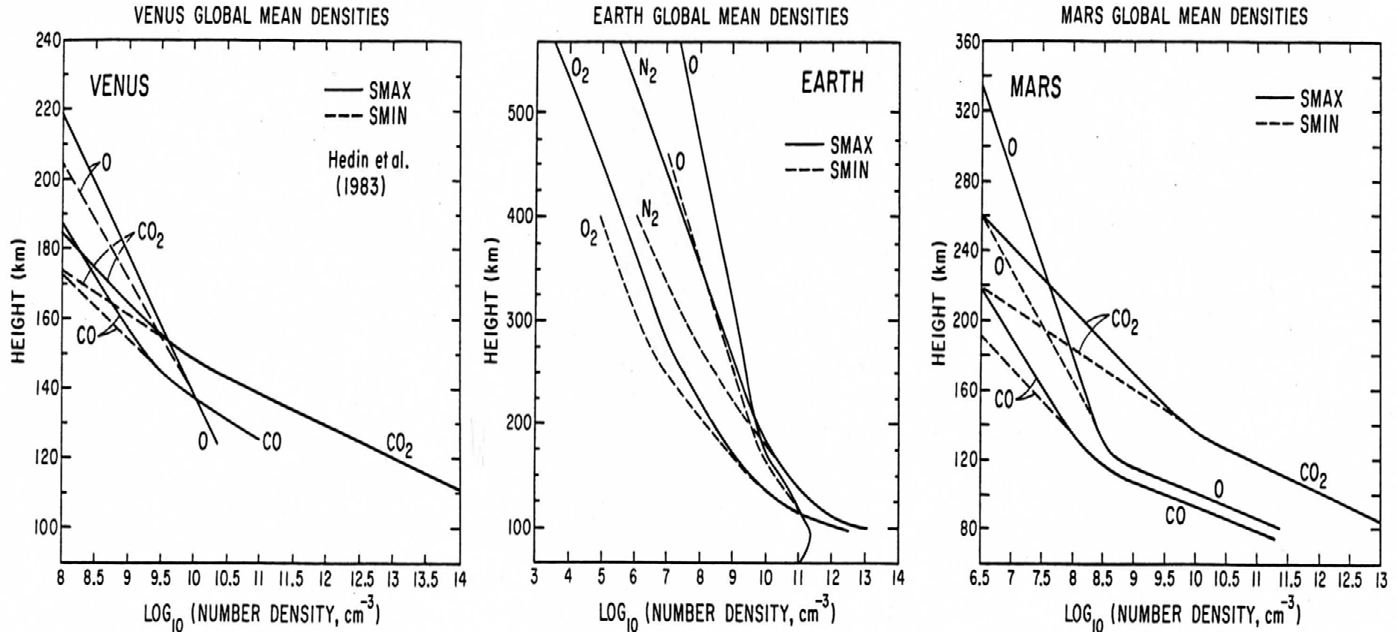}}}
%\vskip 10cm 
\caption[Profiles of major species in upper atmospheres of Venus,
  Earth, Mars.]{Comparison of the global mean vertical profiles of the
  major species in the neutral upper atmospheres of a) Venus, b)
  Earth, and c) Mars for low and high solar activity. SMIN and SMAX
  indicate solar minimum and maximum conditions. Note that the
  turbopause heights (where turbulent mixing and diffusive separation
  are comparable) are 135, 110, and 125\,km for Venus, Earth, and
  Mars, respectively.  [Note: the International Space Station orbits
  at an altitude of $\sim$400\,km. Figs.~I:12.2,
  IV:9.1;
  \href{https://ui.adsabs.harvard.edu/abs/1991JGR....9611045B/abstract}{source:
  \citet{1991JGR....9611045B}}.]}\label{fig:fr2}
%\end{figure}
%\begin{figure}
\centerline{\includegraphics[width=\textwidth]{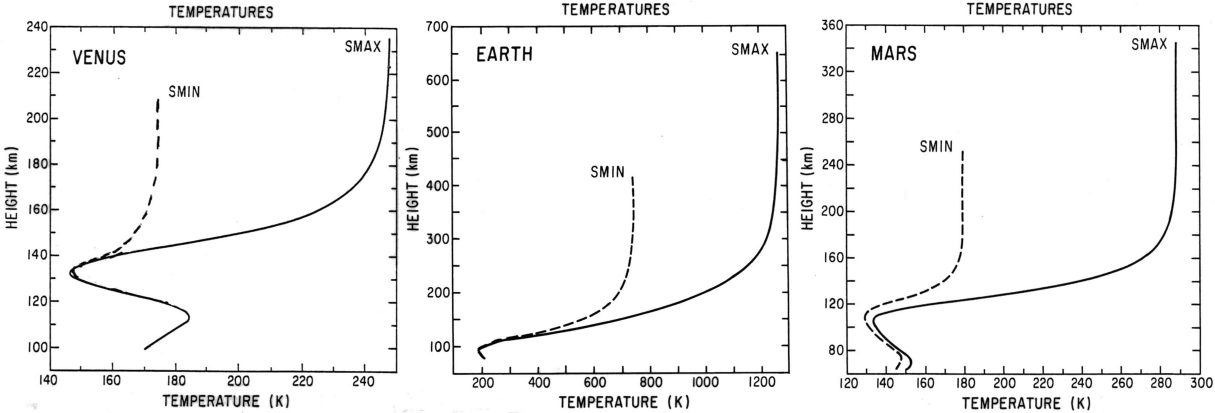}} %bougher_fig2_jgr91.eps}
\caption[Temperature profiles for solar min.\ and max.\ for Venus,
Earth, Mars.]{Three planet global mean temperature profiles for solar
  minimum (SMIN) and maximum (SMAX) conditions. [Note the differences
  in horizontal and vertical scales. Fig.~IV:9.3;
  \href{https://ui.adsabs.harvard.edu/abs/1991JGR....9611045B/exportcitation}{source:
  \citet{1991JGR....9611045B}}.]}
\label{fig:temps}
\end{figure}

Below the Earth's exosphere and above the mesosphere, in a domain
ranging from roughly 110\,km to around $500$\,km in altitude, {\em
  i.e.,} throughout much of the bulk of the thermosphere, lies a
domain where collisions are frequent enough that the gas approximation
is largely valid but not frequent enough to maintain uniform mixing of
the chemicals that make up the terrestrial atmosphere up to that
height: the atmosphere up to heights of about 110\,km \ors[I:12.3] ``is
known as the homosphere and is\indexit{homosphere} constantly being
mixed by turbulent wave eddies. It is only at altitudes above about
110\,km that turbulent\indexit{turbulent!mixing} mixing gives way to
molecular mixing processes, where each species begins to be
distributed vertically under its own pressure scale height or
hydrostatic balance, see Eq.~(\ref{eq:hydrobal}).  A heavy species,
such as carbon dioxide, will decrease in concentration with height
more rapidly than a lighter species, such as atomic oxygen (see
Fig.~\ref{fig:fr2}). Each species, $i$, will have its own
characteristic\indexit{scale height!dependence on mass} scale height
$H_{p \rm i}$, where $H_{p \rm i}=kT/{m_{\rm i}}g$, which is the
vertical distance a species will decrease in partial pressure and
number density by a [factor of $e$]. The upper atmosphere differs from
the lower atmosphere in this respect such that the mean mass of the
fluid will change with\indexit{scale height!change with altitude}
altitude, as well as other gas parameters such as the specific heat,
$c_p$. [\ldots] 
% Titan:
% https://agupubs.onlinelibrary.wiley.com/doi/epdf/10.1002/2015JA021373,
% https://eos.org/research-spotlights/how-saturn-alters-the-ionosphere-of-titan

\figindex{../siskind/art/SBfig9_3.eps}
The vertical distribution of species also has a global
seasonal/latitudinal structure from large scale\indexit{thermosphere!composition!seasons} [\ldots]
\indexit{thermosphere!interhemispheric circulation}
inter-hemispheric circulation from summer to winter. Closure of this
circulation drives an upwelling of material across surfaces of
constant pressure 
in the summer hemisphere and a downwelling in the winter
hemisphere. The upwelling causes the heavier molecular rich gas, which
had diffusively separated at lower altitudes, to be transported
upwards to increase the mean molecular mass in summer. In winter the
downwelling reduces the mean mass.''

\begin{figure}[t]
\centerline{\hbox{\includegraphics[width=10cm,trim=0 9cm 0 9cm]{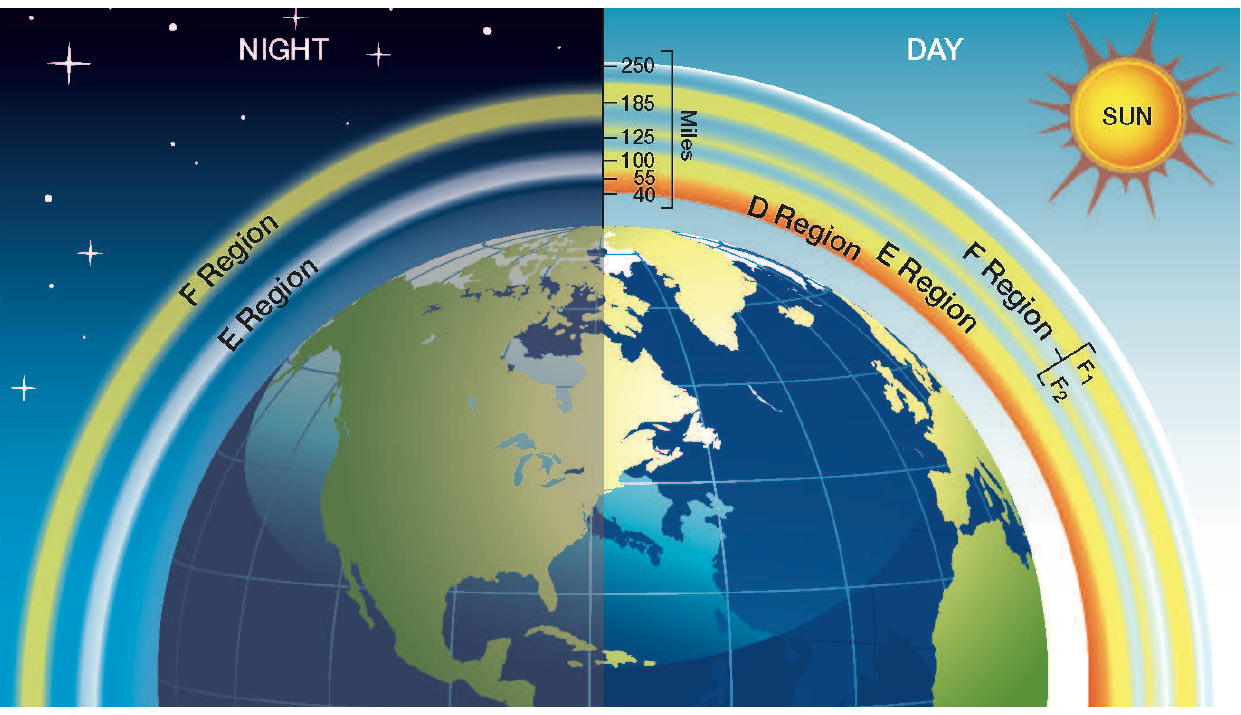}}}
\caption[Regions of the ionosphere.]{Sketch (not to scale) of the
  'regions' in the terrestrial ionosphere. Source:
  NASA/SVS. \colorfig \label{fig:ionolayers}}
\end{figure}
\begin{figure}[t]
%\centerline{\psfig{figure=figures/ssf1,width=6.5cm}\reflectbox{\rotatebox[origin=lt]{180}{\psfig{figure=figures/ssf1,height=6.1cm,width=7.7cm,angle=270}}}}
\centerline{\includegraphics[width=6.5cm]{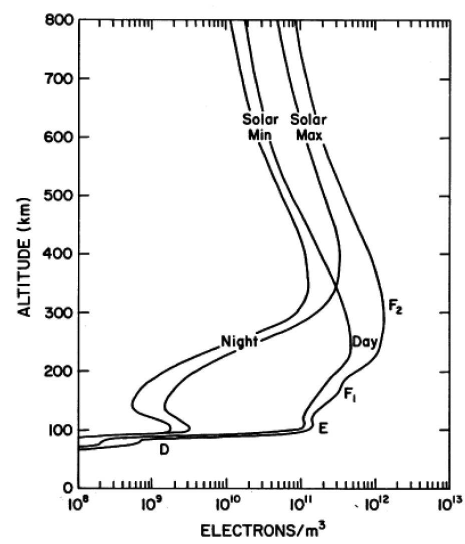}\reflectbox{\rotatebox[origin=lt]{180}{\includegraphics[height=6.1cm,width=7.7cm,angle=270]{figures/ssf1}}}}
\caption[Ionosphere for night/day, high/low solar activity.]{[{\em left:}] Overview of the altitude distribution of Earth's ionosphere
  for daytime and nighttime conditions, at high and low solar
  activity. [Fig.~III:13.1] [{\em right:} One of multiple different
  conventions between planetary scientists and astrophysicists is that the
  height coordinate is usually displayed vertically for planetary
  scientists and horizontally for stellar scientists. This flipped and
  rotated version of the figure conveys the difference in appearance.] }\label{fig:ss1}
\end{figure}

The seasonal changes in insolation and the resulting circulations
subject to Coriolis forces on the rotating planet are modulated by the
effects of space weather. These effects include the X-ray and (E)UV
part of the solar spectral irradiation, dissipation of electrical
currents, and energetic particles precipitating from the
magnetosphere. All of these (and others discussed in
Chs.~\ref{ch:flows}, \ref{ch:mhd} and~\ref{ch:evolvingexposure}) lead
to heating, ionization, and dissociation of the high atmosphere.
\ors[III:13.1] ``Early\indexit{ionosphere!domains, layers}
investigation of the terrestrial ionosphere through its effect on
radio waves resulted in description by means of layers,
\indexit{ionosphere!terminology}principally the $D$, $E$, and $F$
layers, the latter subdivided into $F_1$ and $F_2$ [(sketched in
Fig.~\ref{fig:ionolayers})].  This terminology continues to influence
our current concept of the nature of energy deposition in atmospheres,
although the misleading term 'layer' has given way to 'region'.  The
term 'layer' arose from the observation of systematic variation in the
height at which the critical frequency of reflection occurs in
ionospheric radio sounding; this method cannot detect ionization above
the peak of a region, which explains the appearance of layers.  Radar
and spacecraft measurements now give a more complete picture of peaks
and valleys and reveal the complex morphology of the
ionosphere. [\ldots] An overview of the altitude dependence and
variability of Earth's ionosphere is given in Figure~\ref{fig:ss1},
showing the diurnal and solar-cycle changes and the locations of the
named regions.''

 \ors[III:13.1] ``An
additional\indexit{ionosphere!ionization degree} historical artifact
in terminology is the word ionosphere itself.  Because the atmospheric
ionization was discovered before the neutral thermosphere in which it
is contained, anything above the stratosphere is often referred to as
the ionosphere, resulting in a common misconception that this region
of the atmosphere is mostly ionized.  \indexit{ionosphere!ionization
  fraction}In fact, it is mostly neutral,
ranging from less than a part in a million ionized during the day at
100\,km altitude to about 1\%\ ionized at the exobase ($\sim$600\,km,
depending on solar activity; compare Fig.~\ref{fig:ionprofiles}).
Even at 1000\,km, there is only of the order of 10\%\ ionization.  At
several thousand\,km, where ions (mostly protons) finally become
dominant, the region is defined as the plasmasphere. [\ldots\ In the
bulk of the terrestrial ionosphere] O$^+$ is the most important ion,
particularly in the extensive $F_2$ region above $\sim 200$\,km.  The
$F_1$ region from $\sim$150 to $\sim$200\,km appears as a mere plateau
in the profile, but is distinguished by a transition to molecular
ions, particularly NO$^+$.  The low levels of N$_2^+$, given the
dominance of N$_2$ at these altitudes, is noteworthy. \activity{{\em
    Consider:} The fact that concentrations of atomic nitrogen are not
  shown in Fig.~\ref{fig:fr2} should make you wonder given that
  molecular nitrogen is the most common species in the
  troposphere. (a) Why is atomic nitrogen rare in the upper atmosphere?
  Hint: compare the molecular binding energies of nitrogen, oxygen,
  and water.  For example, see the Wikipedia article on
  \href{https://en.wikipedia.org/wiki/Bond-dissociation_energy}{bond-dissociation
    energy.} (b) Also calculate the wavelength of a photon needed to
  dissociate oxygen and nitrogen. (c) What region of the spectrum is that
  in? \mylabel{act:bondenergy}} The $E$ region from $\sim$100 to
$\sim$150\,km exhibits a small peak, dominated by O$_2^+$ and
NO$^+$.''

\ors[I:12.4.1] ``Much of the external sources of heating, ionization,
and dissociation of a planetary atmosphere comes from the absorption
of photons or particles impinging on the neutral atmosphere. The
physics defining the altitude profile of the three processes is the
same. For example, the rate of ionization, $q$ [(cm$^{-3}$\,s$^{-1}$)],
by solar radiation intensity, $I(h)$ [(erg\,cm$^{-2}$\,s$^{-1}$)], at
some height in the atmosphere of number density, $n(h)$, can be
expressed as a product of four\indexit{ionosphere!ionization rate}
terms:
\begin{equation}\label{eq:ionrate}
%q = \eta_{\rm i} \sigma_{\rm a} n(h) I,
q = \sigma_{\rm a} I(h)  n(h) \eta_{\rm i},
\end{equation}
where $\sigma_{\rm a}$ [(cm$^{2}$) is the atomic] absorption cross section
[for a wavelength interval matching that of $I$,] and $\eta_{\rm i}$
[(erg$^{-1}$)] is the ionizing efficiency; $\eta_{\rm i}$ could
equally be the heating or dissociation efficiency. The intensity of
the radiation gradually decreases along the path through the
atmosphere starting from an initial intensity of $I(h=\infty)$. The
altitude deposition profile depends on the absorption coefficient and on
the atmospheric number density, which varies exponentially with
height. Clearly the product of the intensity of the radiation, $I$,
that decreases as the source penetrates the atmosphere, and on the
atmospheric number density, $n(h)$, that increases with increasing
depth into the atmosphere, must reach a maximum at some altitude or,
more correctly, at some pressure level [(except, of course, for visible
wavelengths for which the atmosphere is largely transparent, in which
case the surface absorption and reflection need to be taken into
account)]. The level of penetration is referred to as the
\indexit{optical depth}optical\indexit{optical
  depth|seealso{definition}}depth,
$\tau$, which is
expressed mathematically as
\begin{equation}\label{eq:optdepth}
\tau=\sigma_{\rm a} n(h) {H_{p}(h) \over \cos(\chi)},
\end{equation}
where the product of the number density $n(h)$ at height $h$ with the
scale height $H_{p}(h)$ at that level represents the integrated
content of a column of gas above that point, and $\chi$ is the angle
from the zenith at which the radiation penetrates a planar
atmosphere. [The above expression is valid as long as the curvature of
the atmosphere can be neglected, so for angles $\chi \la 75^\circ$.]
\activity{{\em Consider:} The concept of optical depth applies to any gas, be it an
  interstellar medium, a planetary atmosphere, or the solar
  atmosphere. Optical depth is an integral of absorption along a line
  of sight, and thus as useful for incoming as for outgoing
  radiation. (a) Explain why the layers contributing most to the light
  from the solar photosphere (remind yourself of its definition!) are
  geometrically higher as you look away from disk center. (b) What can you
  infer about the stratification of the solar atmosphere from the fact
  that the Sun (emitting close to black-body radiation over much of
  the optical spectrum) is brightest near disk center, darkening
  towards the limb?  (c) The solar corona seen in X-rays or extreme ultraviolet
  (EUV) has essentially double the intensity just outside the solar
  limb compared to that just inside the limb when there are no active
  regions along these lines of sight? Why does that imply that the
  corona is optically thin, {\em i.e.,} small compared to the typical
  photon mean free path through it? \mylabel{act:opticaldepth}}

\begin{figure}[t]
%\centerline{\hbox{\psfig{figure=../fullerrowell/Fig_4_bw.eps,width=7cm,clip=}}}
%\centerline{\hbox{\psfig{figure=figures/fig12_4_chapman,width=7cm,clip=}}}
\centerline{\hbox{\includegraphics[width=7cm,bb=54 360 558 720]{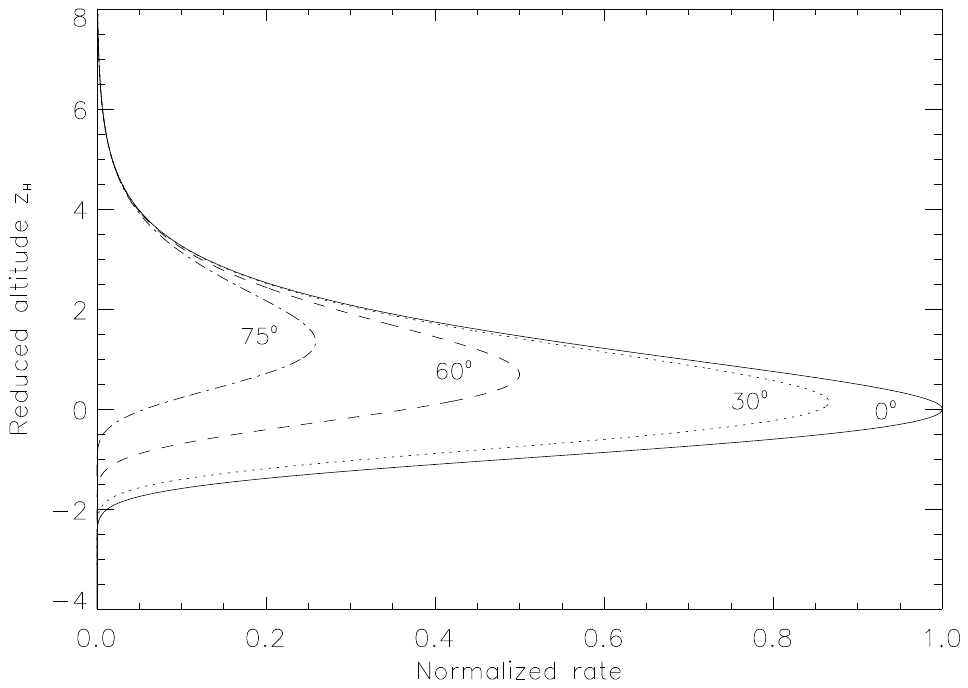}}}
\caption[Classical 
Chapman profile.]{The vertical profile of the classical 
Chapman profile appropriate for heating, ionization, or dissociation
in a stratified hydrostatic atmosphere irradiated from above, relative to a reference
height, shown for different slant angles $\chi$. [Fig.~I:12.4]}\label{fig:fr4}
\end{figure}
The profile of the rate of heating, ionization, or dissociation from
these processes takes the form of the classical\indexit{Chapman!profile} Chapman profile, as
depicted in Fig.~\ref{fig:fr4}, and is given mathematically by
\begin{equation}\label{eq:chapmanprofile}
q(h)=I_\infty \exp \left [ -\sigma_{\rm a} n(h) {H_{p}(h) \over \cos(\chi)}\right ]
\eta_{\rm i}  \sigma_{\rm a} n(h).
\end{equation}
\activity{{\em Show:} You can think of the optical depth as the mean
  number of absorbers within the cross-section along a photon's path
  from infinity to height $h$.  The probability of suffering zero
  absorptions, and thus making it to $h$, is $\exp(-\tau)$.  The
  intensity at $h$ is then an integral from infinity over the expected
  number of absorptions along the way. Combine that with
  Eq.~(\ref{eq:ionrate}) to derive Eqs.~(\ref{eq:optdepth}) and
  (\ref{eq:chapmanprofile}).}  The peak of the profile is at unit
optical depth, which depends on the mass of atmosphere above traversed
by the energetic photon or particle. This corresponds to a fixed
pressure level for a given angle of incidence. The depth of
penetration into the atmosphere of a photon or particle in pressure
coordinates therefore does not change with the gas temperature or the
degree of thermal expansion. Even with the changing heating over the
solar cycle or during a [magnetospheric] storm that might cause a
thermal expansion of the atmospheric gas, that same radiation will
still penetrate and produce heating or ionization at the same pressure
level. The altitude associated with that pressure and the local number
density would, of course, be different since they depend explicitly on
gas temperature.'' \sactivity{$\circledS$ {\em Show:} A similar
  expression to Eq.~(\ref{eq:chapmanprofile}) derived for photons
  holds for energetic particles (from, say, 1\,keV/nucleon to
  1\,GeV/nucleon) losing their energy when propagating into a
  relatively dense medium (from the Earth's magnetosphere into its
  atmosphere, from the solar corona into its chromosphere or
  photosphere, or from interplanetary space into a spacecraft
  hull). Typical solar energetic particles in the range of
  10--100\,MeV can penetrate a medium up to a column density of a few
  grams per cm$^2$ (see, {\em e.g.}, \citep{2007ipep.book.....C} on
  energetic particles passing through matter; it shows that the
  collisional energy loss rate per unit column density (expressed, for
  example, as MeV/(g/cm$^2$) --~mainly caused by transfer to bound
  electrons of the target~-- for non-relativistic energies scales as
  $\sim (Z/A)z^2/E$ for a charge $z$ of the energetic particle with kinetic
  energy $E$ and for an atomic number $Z$ of the target with atomic
  weight $A$; the book can be found online at
  \href{https://library.oapen.org/handle/20.500.12657/50879}{https://library.oapen.org/handle/20.500.12657/50879}).
  Very roughly, estimate how far down that is into Earth's atmosphere,
  into the Martian atmosphere, into the solar lower atmosphere, and
  into an aluminum shell of a spacecraft. List all of the values you
  used and all the assumptions you make (you may have to look some
  things up on outside sources). Start by making the simplest
  assumptions: an isothermal stratified atmosphere for Earth, Mars,
  and the solar photosphere given a characteristic temperature and
  gravity, and a uniform density for an aluminium shell.  Comment on
  the reasonableness of your answers.  (For context, see, {\em e.g.,}
  Sects.~II:1.6, II:13.4, and II:14.4.)
  \mylabel{act:energeticpartpen} \solution{energeticpartpen}}

\section{On collisions and currents, and on neutrals and pickup ions}\label{sec:pickup}
The terrestrial upper atmosphere is coupled to the Earth's magnetic
field through the ionized component of the atmosphere (referred to as
the ionosphere) that is in turn collisionally coupled to the
\indexit{ion-neutral coupling}neutral
molecular \indexit{ionosphere!ion-neutral coupling}and atomic medium
within which it is embedded. The dynamics
of these couplings in the overall system of solar wind, magnetosphere,
and ionosphere are discussed mostly in later Chs.~\ref{ch:flows},
\ref{ch:mhd}, and~\ref{ch:transport}. Here, we look at the consequence
of the ionized medium threaded by a dynamic magnetic field and
embedded in moving neutral gas: electrical currents. In the
terrestrial atmosphere, the effects depend sensitively on the magnetic
latitude because of the orientation of the magnetic field: at high
latitudes, where the field is predominantly vertical, the connection
with the magnetosphere dominates and the dissipated power can lead to
substantial heating. At mid and low latitudes, where the field is
mostly horizontal, internal processes dominate that provide
less dissipative power than at higher magnetic latitudes, but that do 
contribute to transport of plasma.

A moving electrical charge subject to a magnetic field experiences a
Lorentz force perpendicular to its velocity and to the magnetic field, in
a direction that depends on the sign of the charge. Also allowing for
an electrical field to be present, the total force equals:
\begin{equation}\label{eq:lorentz}
  {\bf F}_{\rm L}= m \frac{{\rm d}{\bf v}}{{\rm d}t} = q {\bf E} + \frac{q}{c} {\bf v} \times {\bf B}.
\end{equation} 
In case ${\bf E}= {\bf 0}$ and in the absence of collisions, electrons
and ions thus would spiral about the magnetic field line in opposite
directions [(much more on that in
Sect.~\ref{sec:singleparticle})]. Their gyration radii and frequencies
are very different because of their difference in mass and thermal
velocity (see Table~\ref{tab:dimensionlessnumbers}). Where the
gyration radii are well below the gradients in the magnetic field,
these opposite circular motions do not lead to a net current in the
absence of collisions. However, when \indexit{particle!gradient drift} field
gradients are substantial within the gyration radii of the particles
(most readily for the ions, in particular the more energetic ones) the
particles drift perpendicular to the field in directions opposite for
opposite charges, thus leading to a current; one important
heliophysical setting in which this occurs is in the Earth's inner
magnetosphere, where the gradient drift of primarily the energetic
ions leads to the \indexit{Earth!ring current}'ring current' (see
Sect.~\ref{sec:gcm}).

In the variety of settings in heliophysics, collisions may occur among
the electron and ion populations (see Ch.~\ref{ch:universal} for
that), or with neutral particles (the focus here). In ionospheres, the
neutral particles are atoms and molecules of a body's atmosphere.  In,
say, the environments of comets, planetary rings, or in the outer
heliospheric solar wind the neutral particles, in contrast, may be
either dust particles, escaping atmospheric gas, or inflowing neutral
interstellar atoms.

\begin{figure} 
\begin{center}
\includegraphics[width=6cm]{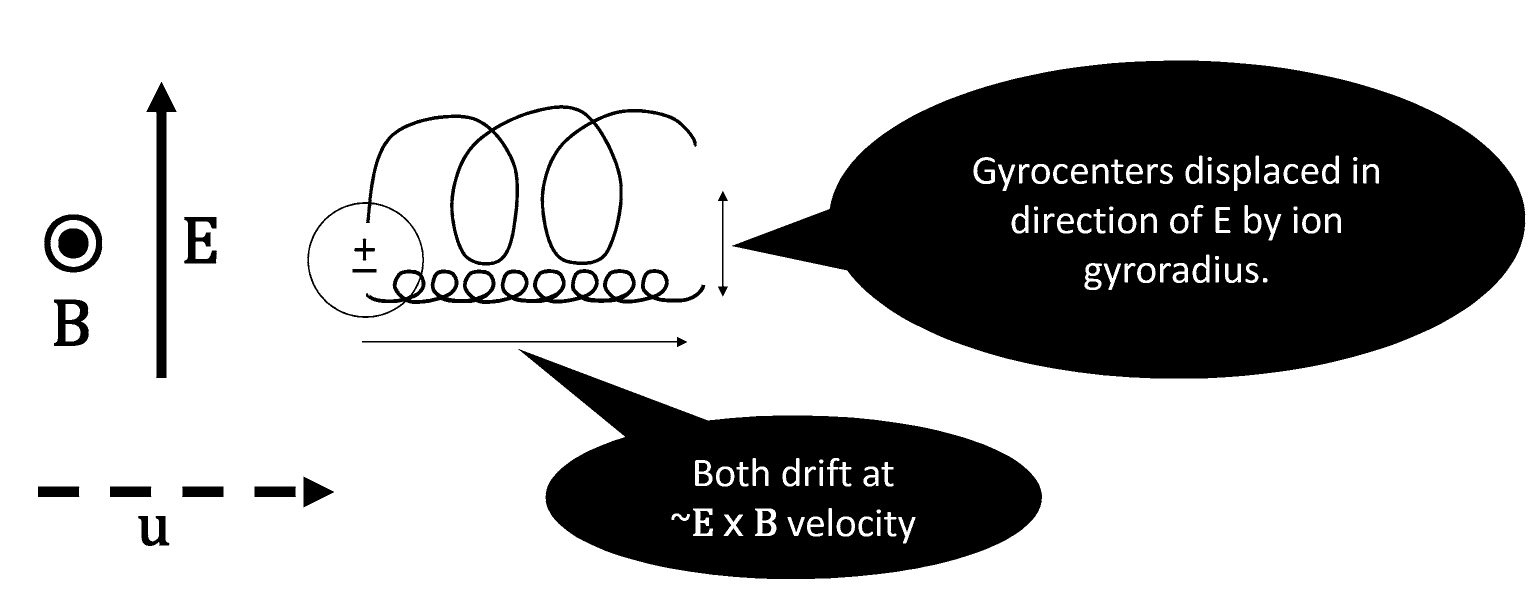}\includegraphics[width=6cm]{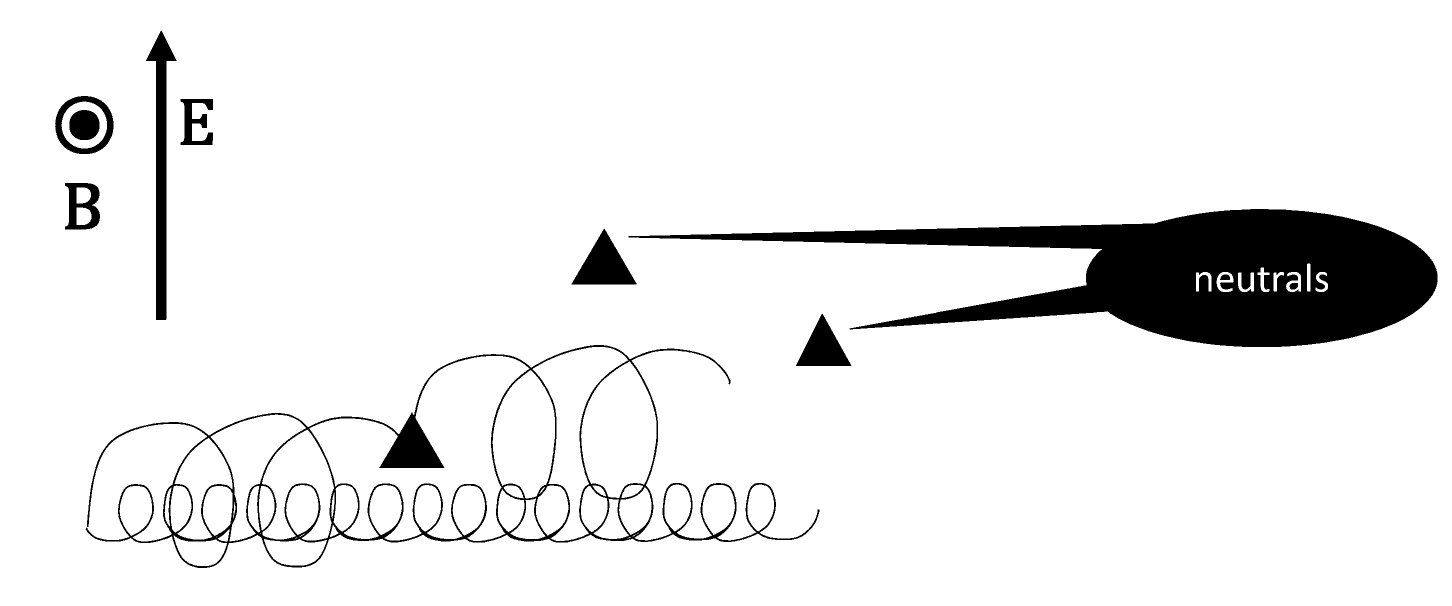}
\end{center}
\caption[Interactions of plasma with
neutrals.]{Schematic of interactions of plasma with neutrals.  {\em
    Left:} Initial motion of pickup ions and electrons.  The gray
  circle represents a neutral composed of a positively charged ion and
  a negatively charged electron.  The directions of plasma flow
  velocity, ${\bf u}$, of the magnetic field, ${\bf B}$, and of the
  electric field, ${\bf E}$, are indicated.  In the image, following
  dissociation, the ion path starts upward and the electron path
  starts downward.  Although initial motion is along ${\bf E}$ for the
  ion, the Lorentz force causes the path to twist, resulting in
  motion around ${\bf B}$ at the ion cyclotron period, leading to a
  net drift at a velocity of ${\bf E}\times {\bf B}/B^2$.  The
  electron initially moves in the $-{\bf E}$ direction.  Its motion
  also rotates around ${\bf B}$, but at the electron cyclotron frequency.  The
  net effect is a transient current in the direction of ${\bf E}$.
  {\em Right:} Schematic of the effect of collisions with neutrals for
  a case with the collision frequency of order the ion cyclotron
  frequency.  Triangles represent neutrals.  The effect of collisions
  is to slow the motion in the ${\bf E}\times {\bf B}$ direction of
  the ions but not of the electrons [(which have other collision and
  gyrofrequencies)] and to displace the ions in the direction of ${\bf
    E}$.  A net current arises, with one component along $-{\bf
    E}\times {\bf B}$ (a Hall current) and one component along ${\bf
    E}$ (a Pedersen current). [Fig.~IV:10.3]}\label{fig_10.4}
\end{figure}
\nocite{k3}\figindex{../kivelson/art/kivelson_10.4a.eps,
  ../kivelson/art/kivelson_10.4b.eps} Let us start with a collision in
which no charge-transfer occurs in a setting where the charged
particle senses both a magnetic and electric field.  In each such
collision of an electron or ion with a neutral particle, the
gyro-motion of the electron or ion involved is modified. Because of
the opposite charges of the electron and ion populations, they attempt
to gyrate about the magnetic field in opposite directions as they are accelerated by the
electric field; consequently, they exhibit
a net drift perpendicular to the magnetic field with ions and
electrons moving in the same direction and at the same rate. There
is no net current (see the left-hand side of Fig.~\ref{fig_10.4}), but if there are
collisions roughly at the same frequency as the gyrofrequency
(different for particles of different masses), the situation changes
fundamentally: collisions interrupt the gyromotion, and this results
in a net separation of the charges. A graphic example, discussed in
more detail below, is given in Fig.~\ref{fig_10.4}(right).

If collisions are very infrequent, or to be precise if the electrons
or ions can gyrate about the magnetic field many times between
collisions, the electrical conductivity \indexit{conductivity!parallel, perpendicular} across the magnetic field is very low. If
collisions are very frequent for both ions and electrons, hardly any
charge separation can occur between collisions, and the electrical
conductivity perpendicular to the magnetic field is also very
low. Peak perpendicular conductivity depends on the direction relative
to the electric field and is reached depending on the ratios of
collision and gyro-frequencies, as shown below.

%https://www.google.com/url?sa=t&rct=j&q=&esrc=s&source=web&cd=1&ved=2ahUKEwjK5NSZyM_cAhV6ITQIHYo8Dj8QFjAAegQIABAC&url=https%3A%2F%2Fwiki.oulu.fi%2Fdownload%2Fattachments%2F11767976%2Fionos_ch5.pdf&usg=AOvVaw3nrxAZpdFq3du3QNO53zPs
Collisions between the populations of charged and neutral particles in
the presence of a magnetic field while allowing for bulk flows is
described through multiple equations. One of these captures the
transfer of momentum that affects the force balance (touched upon
towards the end of this section) almost entirely by looking
at ions because they carry the bulk of the mass. Another accommodates
electrical currents that arise from the differential behavior of the
ions and electrons subject to the magnetic field. A third describes
the energy transfer through the collisional effects formulated as
Ohmic dissipation in the energy balance.

How collision frequencies influence currents in the ionosphere/thermosphere, where
the neutral component is the most common, can be approximated as follows
(collisions between charged particles are ignored here because
collisions with the abundant neutrals are far more common in the bulk of the
terrestrial ionosphere). \ors[I:12.6] ``If we take the magnetic field to be aligned
with the $z$ axis, then the generalized \indexit{Ohm's law!generalized}\indexit{ionospheric Ohm's law}
Ohm's law [(the derivation of which is shown for a fully ionized
plasma in Ch.~\ref{ch:universal})], ${\bf j} = {\bf \Sigma}_{\rm e} \cdot {\bf
  E}_0$ (where ${\bf E}_0$ is the total electric field: ${\bf
  E}_0={\bf E}+\frac{1}{c}{\bf v} \times {\bf B}$), contains the
conductivity tensor\indexit{conductivity!tensor}
\begin{equation}\label{eq:condtensor}
{\bf \Sigma}_{\rm e} = \left ( 
\begin{array}{rrr}
\sigma_{\rm P} & \sigma_{\rm H} & 0 \\
-\sigma_{\rm H} & \sigma_{\rm P} & 0 \\
0 & 0 & \sigma_{\|} \\
\end{array}
\right ),
\end{equation} 
where the \indexit{conductivity!Pedersen} Pedersen ($\perp {\bf B},
\parallel {\bf E_\perp}$), \indexit{conductivity!Hall}Hall
($\perp {\bf B}, \perp {\bf E_\perp}$), and
\indexit{conductivity}parallel ($\parallel {\bf B}$) 
conductivities 
are given [by:
\indexit{Pedersen!conductivity}\indexit{Hall!conductivity}\indexit{parallel conductivity}
\begin{eqnarray}
\sigma_{\rm P} & = & {n_{\rm e}ec \over B} \left( 
\left [ {M_{\rm e} \over 1+M_{\rm e}^2}\right ] + \left ({e \over
    q_{\rm i}} \right)  
\left [ {M_{\rm i} \over 1+M_{\rm i}^2}\right ]  \right ); \label{eq:pedersen}\\
\sigma_{\rm H} & = & {n_{\rm e}ec \over B}  \left( 
\left [ {M_{\rm e}^2 \over 1+M_{\rm e}^2}\right ] - \left ({e \over
    q_{\rm i}} \right) 
\left [ {M_{\rm i}^2 \over 1+M_{\rm i}^2}\right ] \right ); \label{eq:hall}\\
\sigma_{\rm \|} & = & {n_{\rm e}ec \over B}  \left( M_{\rm e} + \left ({e \over
    q_{\rm i}} \right)  M_{\rm i} \right ) \label{eq:parallel}
\end{eqnarray}
(where the equations from Sect.~I:12.6 were rewritten to the above 
by using the expression for $\omega_{\rm e,i}$ below). For
characteristic values of these conductivities in the terrestrial
ionosphere, see Figure~I:12.5.
%Original:
% are given by:\indexit{Pedersen!conductivity}\indexit{Hall!conductivity}\indexit{parallel conductivity}
%\begin{eqnarray}
%\sigma_{\rm P} & = & n_{\rm e}e^2 \left( {1 \over m_{\rm e}\nu_{en}}
%\left [ {1 \over 1+M_{\rm e}^2}\right ] + {1 \over m_{\rm i}\nu_{\rm in}}
%\left [ {1 \over 1+M_{\rm i}^2}\right ]  \right ); \label{eq:pedersen}\\
%\sigma_{\rm H} & = & n_{\rm e}e^2 \left( {1 \over m_{\rm e}\nu_{en}}
%\left [ {M_{\rm e} \over 1+M_{\rm e}^2}\right ] - {1 \over m_{\rm i}\nu_{\rm in}}
%\left [ {M_{\rm i} \over 1+M_{\rm i}^2}\right ] \right ); \label{eq:hall}\\
%\sigma_{\rm \|} & = & n_{\rm e}e^2 \left( {1 \over m_{\rm e}\nu_{\rm en}} + 
%{1 \over m_{\rm i}\nu_{\rm in}} \right ) \label{eq:parallel}.
%\end{eqnarray}
Here, $M_{\rm e,i}=\omega_{\rm e,i}/\nu_{\rm e,i}$ are the electron
and ion \indexit{magnetization} magnetizations, with
$\omega_{\rm e,i}= |q_{\rm e,i}|B/m_{\rm e,i}c$ the electron and ion
(with charge $q_{\rm i}$) gyro-frequencies around the field of
strength $B$, $m_{\rm e,i}$ are the electron and ion masses,
$\nu_{\rm en}$ and $\nu_{\rm in}$ the electron-neutral and ion-neutral
collision frequencies.] The effect of the collisions is to rotate the
net current from the direction of ${\bf E}$ at high altitudes towards
the negative ${\bf E} \times {\bf B}$ direction at low altitudes. In
the terrestrial ionosphere, the] current and dissipation reach a peak
at the altitude where the Pedersen and Hall conductivities are equal,
around 125\,km. For high-frequency currents, like those that may occur
in the solar chromosphere, the dissipation may increase markedly (see
Sect.~I:12.8).  Note that $\sigma_{\rm P}$ is generally dominated by
the ion term.'' \sactivity{$\circledS$ {\em Show:} (a) Work through
  Eqs.~(\ref{eq:condtensor}-\ref{eq:parallel}) to confirm that the
  effect of the collisions of charged particles in the ionosphere with
  the neutral thermospheric component is to rotate the net current
  from the direction of ${\bf E}$ at high altitudes towards the
  negative ${\bf E} \times {\bf B}$ direction at low altitudes. As the
  expressions assume ${\bf B}$ to be in the $z$ direction, you could
  chose ${\bf v}$ in the $x$ direction to describe a horizontal
  velocity near the geomagnetic pole. (b) In that same coordinate system,
  what is the direction of the current at about 125\,km in the daytime
  terrestrial ionosphere where $\sigma_{\rm P} \approx \sigma_{\rm H}$
  (see, {\em e.g.,} Fig.~I:12.5. Remember: figures from the printed
  books are accessible at
  \href{https://heliophysics.ucar.edu/resources}{https://heliophysics.ucar.edu/resources}). \mylabel{act:hallped}
  \solution{hallped}}

The collisional coupling between ions and neutrals causes momentum
exchange (through the drag force that works to reduce the velocity
difference between these two populations) and energy dissipation (in
the form of Joule heating). \ors[I:12.6] ``The electrodynamic properties
can be conveniently separated into a high [magnetic] latitude region, where the
current flow in the ionosphere is connected to the magnetospheric
current system, and a mid and low latitude region, where the majority
of the current flow and polarization electric fields are controlled
internally by the thermosphere-ionosphere conductivity and dynamics.''
\ors[I:12.6.1] ``In the ionosphere,
currents flowing perpendicular to the magnetic field are produced by
electric fields and neutral winds. Although collisions between ions
and the neutral gas are relatively infrequent [in Earth's upper
atmosphere] above $\sim$160\,km, they are sufficient to accelerate the
neutral component, {\em i.e.,}  the thermosphere, at high latitudes to many
hundreds of m/s over periods of tens of\indexit{thermosphere!winds}
minutes or more [to speeds well in excess of those associated with
solar heating]. [\ldots] At low altitude, $\sim$100\,km, the ions are
forced to move with the neutral gas, whether stationary or moving. The
large-scale wind system at this altitude is driven by the tidal and
planetary waves propagating from the lower-atmospheric terrestrial
weather system, and the mass of the atmosphere is such that ion drag
has little or no impact on the neutral dynamics. The altitude range
between 100 and 160\,km altitude is the narrow altitude range that is
responsible for most of the dissipation of electromagnetic energy from
the magnetosphere. The neutral dynamics and conductivity in this
boundary region between space and atmospheric plasma are critical.''
\ors[I:12.6.2] ``At mid and low [magnetic] latitudes the electric fields [in
Earth's ionosphere] arise largely from internal dynamo processes
driven by the conversion of neutral wind kinetic energy to
electromagnetic energy, and are typically an order of magnitude
smaller (a few mV/m) than high-latitude fields. The energy involved is
also much smaller. The importance of the small electric fields at low
latitudes is no longer the Joule heating and momentum dissipation, but
rather their role in the redistribution of plasma.'' Some of these effects are
touched upon generically in Sect.~\ref{flow}, with a more comprehensive discussion
for Earth's ionosphere in Sect.~I:12.6.

\begin{figure}[t]
%\centerline{\hbox{\psfig{figure=figures/ionprofiles,width=10cm,clip=}}}
\centerline{\hbox{\includegraphics[width=10cm,bb=54 360 558 720]{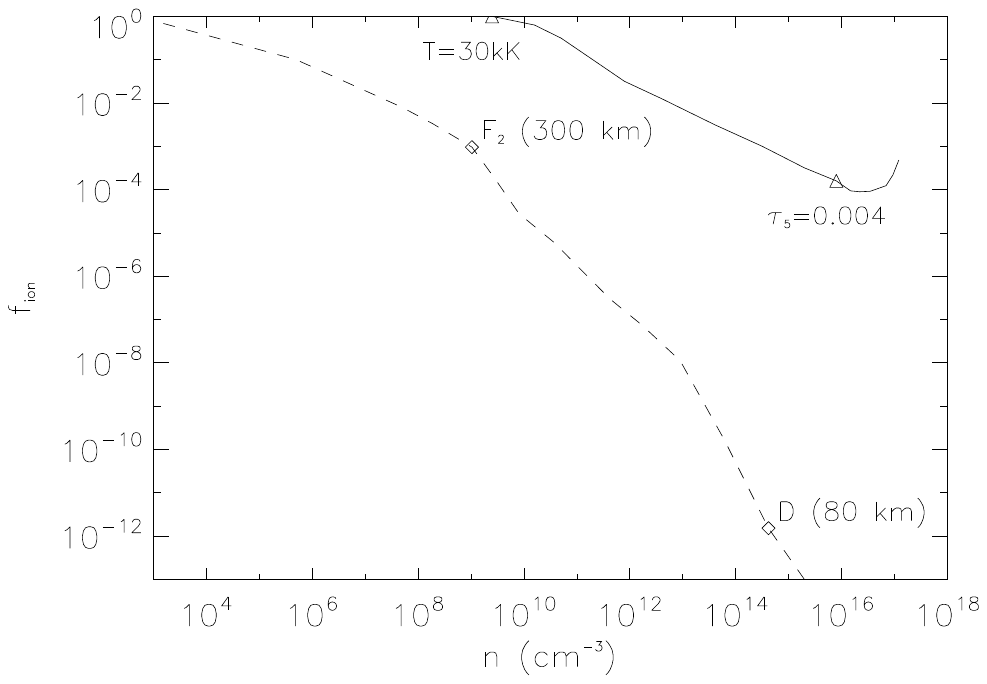}}}
\vspace*{-27ex}  % Tune this to the image height.
\begin{center}
  Terrestrial \\
  ionosphere
\end{center}
\vspace*{27ex}  % The spacing above but without the minus.
\vspace*{-50ex}  % Tune this to the image height.
\begin{center}
  \hskip 6cm Solar \\
  \hskip 6cm chromosphere
\end{center}
\vspace*{33ex}  % The spacing above but without the minus.
\caption[Typical density and ionization in
iono- and chromosphere.]{Comparison of
densities, $n$ (cm$^{-3}$), and \indexit{ionosphere!ionization
  fraction}ionization fractions,
$f_{\rm ion}$, for a \indexit{chromosphere!ionization
  fraction}characteristic dayside ionosphere (dashed) and
mean chromosphere (solid). The diamonds mark the mean values for the
ionospheric D and F$_2$ regions, centered on about 80\,km and 300\,km,
respectively. The triangles denote the base of the chromosphere (defined
here as at a continuum optical depth of $\tau_5=0.004$) and the top
of the chromosphere (where the temperature exceeds 30\,000\,K).
[Fig.~I:12.13]
}\label{fig:ionprofiles}
\end{figure}\nocite{allen72}
In much of the discussion of magnetized plasma in the
Sun's interior and atmosphere in subsequent chapters, the Hall and
Pedersen conductivities are often assumed to be negligible. A similar
approximation is often seen in the study of the heliosphere and
planetary magnetospheres. The Sun's \indexit{chromosphere! conductivity}chromosphere, however, is an
environment with a strong neutral population and with collision
frequencies not so high that Pedersen and Hall conductivities are
effectively ignorably small. The chromosphere is located immediately
above the photosphere (which itself has a thickness of roughly a
single scale height of about 100\,km), and extends over a height range
of some 2,500\,km, spanning roughly a dozen pressure scale heights in
a highly dynamic setting that is strongly patterned by the magnetic
field, before the transition region is reached in which the
temperature rapidly rises to coronal values.

\ors[I:12.8.3] ``The Earth's ionosphere has
a range of degrees of ionization, starting from the essentially
neutral troposphere below, reaching an ionization fraction of about
$10^{-4} - 10^{-3}$ around 200\,km in height, and exceeding a few
percent by 1\,000\,km. In\indexit{comparative studies!ionization
  fraction chromosphere, ionosphere} the case of the chromosphere, the
ionization fraction starts at about $10^{-4}$ around photospheric
heights, drops through $10^{-5}$ through the classical 'temperature
minimum' around 500\,km in height, and then increases through a few
percent around 1\,500\,km in height, continuing to near-complete
ionization in the solar corona. Figure~\ref{fig:ionprofiles} compares
the densities and ionization fractions for mean states characteristic
of the ionosphere and chromosphere. Note that the neutral densities in
the D--F$_2$ ionospheric region are comparable to those in the
chromosphere, but the ion densities are at least 1\,000 times lower at
any given neutral density, resulting in a much weaker ion-neutral
coupling in the ionosphere than in the chromosphere.

Let us look back at Eqs.~(\ref{eq:pedersen})-(\ref{eq:parallel}) and
assess their meaning for both chromosphere and
ionosphere.\indexit{comparative studies!conductive anisotropy} In the
limit of a weak magnetic field or a high collision frequency, the ion
and electron magnetizations $M_{\rm e,i}=\omega_{\rm e,i}/\nu_{\rm
  e,i} \rightarrow 0$, $\sigma_{\rm P} \rightarrow \sigma_{\rm
  \|},\,\sigma_{\rm H} \rightarrow 0$; hence, currents are more readily
aligned with the {\em electric} field, as expected. As the collision
frequencies with the neutral population decrease, the above
expressions would have current and {\em magnetic} field
aligned (as both $\sigma_{\rm
  P,H} \rightarrow 0$)  [\ldots]

In the chromosphere of a solar [sunspot] region,
$M_{\rm e}(500<h<2000\,{\rm km})={\cal{O}}(100)$, decreasing rapidly
towards the photosphere to $M_{\rm e}(h=0)={\cal{O}}(0.01)$ at the
solar surface. Some studies find the proton magnetization to remain
below unity throughout the chromosphere, up to the transition into the
corona (these findings depend on the atmospheric domain, of course,
and on the models used [\ldots]). Consequently, the bulk of the
active-region chromosphere has an anisotropic conductivity of at least
a factor of 10 difference between the field-aligned and transverse
components. Conduction in the corona is almost exclusively field
aligned (and thus essentially free of Lorentz forces), while
photospheric conduction is nearly isotropic. [\ldots]'' \activity{(a) {\em
    Look up} what defines a 'sunspot' and what an 'active region'. Are
  all sunspots associated with active regions? Are all active regions
  associated with sunspots? (b) A record of sunspot counts over many
  decades is shown in Fig.~\ref{figure:butterfly}: what is the typical
  latitudinal range over which sunspots and sunspot groups occur? In
  Sect.~\ref{sec:stellarwinds} you will read about high-latitude and
  even polar starspots on rapidly-rotating, active stars, as the Sun
  would have been in its first few hundred million
  years. \mylabel{act:spotvsar}}

Now, let us look at different environments, and illustrate not only
currents but also the effects of momentum transfer.
\ors[IV:10.3.2] ``At comets and in the vicinity of moons, such as Io and
Enceladus, that are significant sources of neutral gas, various
processes that convert neutral atoms or molecules into ions are
important to consider.  Neutrals can be ionized by photons
(photoionization) or by collisions with other particles, typically
electrons (impact ionization). An additional process that affects the
interaction region is \indexit{charge exchange}charge exchange in which a neutral
gives up a charge to an ion.  The original ion, now neutral, carries
off its incident momentum while the original neutral becomes an ion at
rest in the frame of the neutral gas.

The ions introduced into the plasma by ionization of neutrals modify
the bulk properties of the plasma.  Consider a situation in which the
neutrals are at rest relative to [a location] towards which the
plasma flows at (bulk) velocity ${\bf u}$.  Photoionization and impact
ionization add mass to the plasma whereas charge exchange between the
ionized or neutral form of the same element does not change the mass
density.  All three processes slow the bulk flow because the new ions
must be accelerated so that their average motion matches that of the
bulk plasma and the process extracts momentum from the incident
plasma.  These processes also change the thermal energy of the plasma
and may modify the plasma composition.  The complex effects associated
with \indexit{pickup}pickup can significantly modify the interaction region
surrounding a moon or a comet.

The relation between pickup and currents is shown schematically in
Fig.~\ref{fig_10.4}a. The newly ionized ion senses the electric field
of the flowing plasma and begins to move in the direction of this
electric field.  The electron that has separated from the ion is
initially accelerated in the opposite direction.  After one
gyroperiod, the average separation of the gyrocenters of the two
charges is close to one ion \indexit{gyroradius}gyroradius
\begin{equation}\label{eq:10.16}
r_{\rm gi} = m_{\rm ion} v_{\rm ion} c / qB
\end{equation}
%	ρ_g=m_ion v_ion/qB ) 	(10.16)
where $m_{\rm ion}$ is the ion mass, $v_{\rm ion}$ is its thermal
velocity, and $q$ is its charge (see Sect.~\ref{sec:singleparticle} for details on
single-particle motions).\activity{{\em Show:} (a) Use
  Table~\ref{tab:dimensionlessnumbers} to show that Eq.~\ref{eq:10.16}
  yields ion gyroradii $r_{\rm gi}$ for thermal motions if
  $v_{\rm ion}$ is set to the mean thermal velocity. (b) Then estimate
  energies of (1) non-thermal particles so that their $r_{\rm gi}$ are
  comparable to the length-scale of the geomagnetic field for which
  you can use the Earth's radius (this value is important for the
  terrestrial ring-current, which is a manifestation of particles
  drifting across the magnetic field because the heavy, energetic ones
  sense the gradient in the field strength; see
  Sect.~\ref{sec:fieldlines}) and (2) for perturbations in the
  heliospheric field for which you could take, say, 0.1\,AU (these
  variations are important for the propagation of solar energetic
  particles, see Ch.~\ref{ch:conversion}, and for incoming cosmic rays, see
  Ch.~\ref{ch:evolvingexposure}). (c) Compare with values in
  Table~\ref{tab:5.1}, compare these to mean-free path lengths there,
  and bear these results in mind going into
  Ch.~\ref{ch:universal}. \mylabel{act:rgi}} The result of the
separation of charges is to produce a transient current density in the
direction of the electric field. If the pickup is occurring at a rate
$\dot{n}$, where $\dot{n}$ is the number of ionizations per unit
volume and time, then the pickup current [density] is
\begin{equation}\label{eq:10.17}
j_{\rm pickup} = q \dot{n} r_{\rm gi} .
\end{equation}
%	j_pickup=qn ̇ρ_g	(10.17)
Because pickup current flows across the background field, a cloud of
pickup ions acts much like a solid conducting obstacle in the flow and
imposes the same types of perturbations, {\em i.e.}, it slows and
diverts the incident flow'' in a way outlined in Ch.~\ref{ch:flows}. 

In this volume, we do not go into the behavior of \indexit{dusty
  plasma}dusty plasmas. The
interested \indexit{plasma!dusty}reader is referred to Ch.~IV:11, which introduces the
subject as follows: \ors[IV:11.1] ``The study of dusty plasmas bridges a
number of traditionally separate subjects, for example, celestial
mechanics, mechanics of granular materials, and plasma physics. Dust
particles, typically micron and submicron sized solid objects,
immersed in plasmas and UV radiation collect electrostatic charges and
respond to electromagnetic forces in addition to all the other forces
acting on uncharged grains. Simultaneously, dust can alter its plasma
environment by acting as a possible sink and/or source of electrons
and ions. Dust particles in plasmas are unusual charge carriers. They
are many orders of magnitude heavier than any other plasma particles,
and they can have many orders of magnitude larger (negative or
positive) time-dependent charges. Dust particles can communicate
non-electromagnetic effects, including gravity, neutral gas and plasma
drag, and radiation pressure to the plasma electrons and ions. Their
presence can influence the collective plasma behavior by altering the
traditional plasma wave modes and by triggering new types of waves and
instabilities. Dusty plasmas represent the most general form of space,
laboratory, and industrial plasmas. Interplanetary space, comets,
planetary rings, asteroids, the Moon, and aerosols in the atmosphere,
are all examples where electrons, ions, and dust particles coexist.''
\activity{{\em Show and Look up:} An intriguing property of dust is
  that, if the particles are small enough, radiation pressure is
  important in their momentum equation. Assuming neutral dust
  particles, estimate at what (density-dependent) size photon pressure
  from solar illumination exceeds solar gravity (note that this is
  independent of distance to the Sun for a completely transparent
  solar wind). There is a surprise here for dust of any size: the
  orbital motion of the dust causes photon absorption (and assumed
  isotropic re-radiation of that energy) to lead to a 'brake' on the
  orbital velocity, causing larger dust particles to spiral inward;
  look up the
  \href{https://en.wikipedia.org/wiki/Poynting–Robertson_effect}{'Poynting-Robertson
    drag'} to see how that works. From this, realize that dust needs
  to be continually replenished somehow in the Solar System, generally
  by impact collisions and by disintegrating comets.}

\section{Sources of plasma}
There are many \indexit{plasma!sources}sources
of plasma around the heliosphere: all it takes
is some neutral medium subjected to sufficient energy to ionize
particles. The bulk source medium can be the largely neutral gas in
the Sun's surface layers, but dissipation of (magnetic) waves and
currents, as well as the acceleration of particles in electric fields
result in heating and ionization of the Sun's outer atmosphere. The
larger planets have neutral atmospheres of which the top layers are
ionized by solar radiation and by suprathermal particle
precipitation. Moons may be large enough to have their own atmosphere
(as is the case for Titan at Saturn), while others may still have some
matter around their surfaces because these are subjected to sputtering
by the solar wind or, for moons within planetary magnetospheres, by
magnetospheric particles, or matter may be supplied by geysers (as on
Enceladus at Saturn) or volcanoes (as on Io at Jupiter) that
contribute molecules (including SO$_2$, SO, S$_2$, H$_2$S, \ldots) as
well as atoms. Comets have a coma of gas that sublimates off the
nucleus, along with dust. And dusty material is around in the rings of
all the giant planets. Whereas the magnetized and ionized components
of the interstellar medium cannot penetrate each other (as discussed
in Chs.~\ref{ch:universal} and~\ref{ch:flows}), neutral
interstellar-medium particles can make it deep into the heliosphere,
following free-fall trajectories in the collisionless environment
until they are subjected to a charge-exchange collision with
solar-wind ions.

\clearpage

\chapter{{\bf MHD, field lines, and reconnection}}%3
\label{ch:universal}
{\narrower\narrower{
{\bf Chapter topics:}
 \begin{itemize}
  \customitemize
\item The fundamental difference between gravity and magnetism 
\item MHD as a low-order description of single-fluid plasma dynamics
\item Alfv{\'e}n, fast-mode, and slow-mode waves
\item Processes and scales of reconnection
\end{itemize}

\noindent{\bf Key concepts:}
\begin{itemize}
  \customitemize
   \vskip -\baselineskip
\item General, force-free, and potential magnetic fields
\item Magnetic pressure and tension
\item Magnetic structures: current sheets and flux tubes
\item Reconnection: failure of ideal MHD and of frozen-in field lines
\end{itemize}

}}

\section{Introduction}\label{sec:mhdintro}

\ors[II:1] ``Absent the magnetic field, neither solar activity nor
magnetic storms -- the solar and terrestrial sources of [variable
conditions referred to as space weather \regfootnote{For introductions
  to the impacts of space weather on society and its technological
  infrastructure we refer to Chs.~II:2, II:12, and~H-V:1-5.}
-- would exist.  \ldots] Although in principle fossil magnetic fields
could have remained from the creation of the Solar System, this
appears not to be the case.  Witness the 22-year magnetic cycle of the
Sun and the reversals of the Earth's magnetic field.  On shorter time
scales, the magnetic topography of the solar surface changes so
rapidly that it must be monitored constantly as input for space
weather forecasts.\activity{{\em Look up} magnetic maps of the solar surface
  (such as made with the HMI instrument on NASA's Solar Dynamics
  Observatory) and make a movie at an image cadence of a few hours and
  a duration of one week;
  you could use
  \href{https://helioviewer.org}{HelioViewer} or
  an \href{https://ccmc.gsfc.nasa.gov/iswa/}{ISWA cygnet}. Compare one for
  2013--2014 (near cycle maximum, with multiple sunspot groups
  dispersing into the surrounding network of small-scale flux
  patterns) to 2017--2018 (around cycle minimum, with only the small
  scales on the disk). Note the changes that you see in the movie and
  time time and length scales of these changes. \mylabel{act:hmimovie}}

[The contrast between magnetic variability and gravitational
persistence has its origin in the sources of the two fields: the
magnetic field, ${\bf B}$, has its origin in a variable source, namely
the relative motion of differently charged particles, while the
gravitational field, ${\bf g}$, springs forth from a conserved
(positive definite) source.] The conserved source of the gravitational
field is mass, as can be seen in the [non-relativistic] field
equations\indexit{field equations!gravity} that apply to the
gravitational field:
\begin{equation}
\nabla \cdot {\bf g} = -4\pi G\rho, \,\,\,\,\,\,\,\, 
\nabla \times {\bf g} = {\bf 0},
\end{equation}
%\begin{eqnarray}
%\nabla \cdot {\bf g}
%&=& -4\pi G\rho, \\
%\nabla \times {\bf g} &=& {\bf 0},
%\end{eqnarray}
where $G$ is the gravitational constant and $\rho$ is the mass
density.  Thus, gravity is determined by the amount of mass present
and its distribution.  Because mass is conserved and the gravitational
force causes matter to collapse into systems in which the
gravitational force is almost perfectly balanced by thermal pressure
or inertial forces, gravitationally organized matter tends to be
stable over eons [\ldots]  In contrast, the pertinent field
equations\indexit{field equations!magnetic field} for the magnetic
field are
\begin{equation}\label{eq:bfieldeq}
\nabla \cdot {\bf B} = 0, \,\,\,\,\,\,\,\,\, 
\nabla \times {\bf B} = \frac{4\pi}{c}{\bf j}
\end{equation}
%\begin{eqnarray}
%\nabla \cdot {\bf B} &=& 0, \\ \nabla \times
%{\bf B} &=& \frac{4\pi}{c}{\bf j}.
%\end{eqnarray}
[(the second expression holds if all velocities involved are well
below light speed).]
%Spruit: https://arxiv.org/pdf/1301.5572.pdf
The source term for
the magnetic field in these equations is electrical current, ${\bf j}$, which,
unlike mass, is not a conserved quantity [(although $\nabla \cdot {\bf
  j}=0$) and which can point in any direction].  Thus we see that ${\bf B}$ is a
product of dynamo or other magnetohydrodynamic (MHD) processes that 
generate current in real
time.  The crucial distinction is that unlike the gravitational field,
which is in effect a byproduct of a conserved, definite quantity of
mass and so is inherently persistent, the magnetic field is generated by a
variety of plasma motions in the Sun, in the solar wind, and in planetary
magnetospheres on time scales shorter than what would be needed to
reach an equilibrated state.  Hence, the local cosmos is constantly
adjusting and attempting to relax, but it never gets to such a
quasi-stationary state.  The consequence of this is what we call
weather, including [\ldots] space weather.''

\begin{table}[t]%\begin{figure}
\indexit{magnetic!structures}
\caption[Structures in the magnetic field.]{\label{fig:structures} Structures in the magnetic field.}
%\framebox[\textwidth]{
%\begin{flushleft}
\fbox{
  \vbox{
    {\small
{\bf Current sheet:} {\em Examples on large scales: heliospheric
  current sheet; magnetospheric current
  sheet. {\em E.g.,}~Fig.~\ref{fig:moldwin3}.}\ [I:6.2] ``Our focus here
is mainly on current sheets\indexit{current sheet} in the form of
tangential\indexit{discontinuity: contact, rotational, tangential} discontinuities or
rotational\indexit{rotational discontinuity} discontinuities that
evolve into tangential-like discontinuities.  Tangential
discontinuities are non-propagating surfaces across which no magnetic
flux passes as the magnetic field changes direction or strength or
both, while total (magnetic plus thermal) pressure is
continuous. [\ldots] current sheets (tangential discontinuities)
inevitably form in naturally occurring turbulent plasmas; [\ldots\ they]
form in the
corona through the expansion of magnetic flux tubes that poke out
of the photosphere [and] expand until at some altitude they press
against each other forming a beehive pattern of flux tubes separated
by current sheets [unless and until the currents dissipate and the
field becomes potential]. [\ldots] Interplanetary space is a honeycomb of
outwardly advecting current sheets. [\ldots] In the
magnetospheric case, the solar wind snags magnetic field lines from
the planet's two poles on the sunward side and stretches them
anti-sunward to form the characteristic two-lobe magnetospheric tail
across which the magnetosphere's analog of the heliospheric current
sheet separates the two lobes.''

{\bf Flux tube, flux rope:} {\em Examples: compact sunspots,
  pores, 'bright points'; as an entity, they are bounded by a current
  sheet. {\em E.g.,}~Fig.~\ref{figure:solarfieldexamples}(top)
  and~\ref{fig:bp-simple}.}\ [I:6.4] ``A flux tube\indexit{flux!tube|seealso{definition}} is the volume enclosed by a set of field lines
that intersect a simple closed curve.  The frozen-in flux condition of
ideal MHD describes a parcel of plasma threaded by magnetic field
lines as a conserved entity whose motion can be followed.'' In the
solar photosphere, flux tubes may emerge as preformed entities, or may
form from by 'convective collapse'. A flux rope is a flux tube twisted
about itself (and thus carrying an internal net current); many
magnetic configurations emerge into the solar photosphere as flux
ropes; many form in the corona by the dynamics of the reconnecting
field; coronal mass ejections inject ropes into the heliosphere (there
known as 'magnetic clouds') while others form by reconnection across
current sheets; at magnetopauses, flux ropes ('flux transfer events')
form by reconnection; and flux ropes ('plasmoids') form by
reconnection across the magnetospheric current sheet.

{\bf Cell:} {\em Examples: planetary magnetosphere; heliosphere. {\em
    E.g.,}~Fig.~\ref{fig:ophercomposite}.}\ [I:6.7] ``Magnetic fields
tied to gravitating bodies will expand to fill all space unless
prevented from doing so.  [\ldots] the magnetic field's expansionist
ambition is checked by some other magnetic field-bearing plasma
expanding from somewhere else.  Each magnetic field is therefore
encased within a \indexit{magnetic!cell}definable volume, which we
refer to as a cell.\indexit{magnetic!cell|seealso{definition}} In the
Sun's case, the cell is the heliosphere.  In the other cases
mentioned, the cells are planetary magnetospheres. [\ldots\ The
cellular structure] is like a Russian nesting doll in which one cell
is encased within another.  [\ldots] within the heliosphere, the scale
sizes of the objects already mentioned cover seven orders of
magnitude'' } } }
% \end{flushleft}
\end{table}%\end{figure}
\ors[II:1] ``There is an important difference regarding the types of
volumes that the gravitational and magnetic tension forces organize.
The gravitational field has no shielding\indexit{shielding current}
currents (${\bf \nabla} \times {\bf g} = {\bf 0}$) [because its
source is the positive-definite mass density ($\nabla \cdot {\bf g} =
-4\pi G\rho$); consequently, gravity] has
no discontinuities because that would require an infinite mass
density.  Hence, the gravitational field is relatively homogeneous; it
varies smoothly and continuously in space.  On the other hand, [owing
to the fact that electrical charges can be of either sign, a
magnetic field can contain] shielding currents ($\nabla \times {\bf B} =
\frac{4\pi}{c}{\bf j}$) which spontaneously form discontinuities,''
that are commonly referred to as current\indexit{current sheet} sheets
[(see Table~\ref{fig:structures} for a definition)]
despite the fact that their geometry is generally quite complex in the
local cosmos. The combination of the distinct behaviors of
gravitational and magnetic forces yields a rich diversity of phenomena
in the local cosmos and beyond that emerge from the `universal
processes' captured in the MHD description of magnetized plasma.
 
Among the universal concepts in heliophysics one pair stands out in
particular, namely that of {\em magnetic lines of force} --~or
commonly {\em magnetic field lines}~-- and of their {\em
  reconnection}. \activity{{\em Background:} Throughout this volume we use 'field line'
  only for lines of force of the magnetic field. The concept can be
  applied to any field, however, including a flow field (such as in
  Fig.~\ref{fig:ingred}(C), then often referred to as streamlines) and
  the gravitational field. Field lines of ${\bf B}$ and ${\bf g}$ are
  fundamentally different in one key respect: a magnetic field line
  never ends (because there are no magnetic monopoles) while
  gravitational field lines start from a point of mass.
  As to magnetic field lines, note that there are
  drawings in this book, as in many other resources, where field lines
  are shown to start from one polarity and end on another. As magnetic
  field has no monopoles, such drawings should not be misread to mean
  that field lines end, but only that their rendering in the diagram
  is incomplete, {\em i.e.,} merely terminated for simplicity, for
  lack of information, or to restrict the discussion to a particular
  region of interest. \mylabel{act:fieldlinenote}} \indexit{field line}Field lines are abstractions; they are
1-dimensional virtual devices that are used to outline the geometry of
magnetic structures in the local cosmos, in a way in which the tangent
of the field line anywhere along it has the same direction as the
field there while the local field line density is equivalent to the
magnetic field strength.

In a vacuum, magnetic field lines have no intrinsic temporal
continuity. For example, consider the field between and surrounding
magnets or electrical wires at time $t_0$ and again at time $t_1$
after having moved the magnets or wires into new positions. The lines
of force used to visualize the field at times $t_0$ and $t_1$ are
completely independent, the result only of the magnetic fields at the
two instances in combination with two sets of points, one for $t_0$
and one for $t_1$, selected by a researcher from which to compute the
lines of force.  In a plasma, however, field lines can be thought of
as structures whose continuity in time derives from the ionized matter
that is contained in the flux bundle or tube that is centered on the
field line.  In our thinking, we should map these `lines of force' to
their 3-dimensional equivalent, the `flux tube': as long as the ions
and electrons once contained never move out of the flux tube, the
field line has some temporal continuity. Whenever matter does migrate
out of the flux tube, the attribute of continuity for the field line
fails. However, if the locations where this occurs are compact
compared to the field line's length, one can think of field lines
--~that can never end in the divergence-free magnetic field except
close onto other field lines~-- as being cut and
connected onto another field line. Where that happens, the concept of
`magnetic reconnection' is then introduced to salvage that of the
`field line' as something that has an identity over time, at least
while matter remains constrained to within the flux tube.

Field lines and their reconnection are but two of the concepts related to a
variety of processes that occur in ionized gases (`plasmas') that
are threaded by magnetic field. We come across such processes in the
vastly different environments of the solar interior and of the far
reaches of the heliosphere, and in the depths of planets as well as in
the most tenuous parts of their outer atmospheres. Temperatures and
densities (and, as we shall discuss later in this chapter, magnetic field
strengths) differ by many orders of magnitude; a summary of some of
the conditions encountered in the local cosmos that surrounds us is
visually represented in Figure~\ref{fig:conditions}.

In everyday life we tend to ignore the Earth's magnetic field, but we
can do so only because of the low temperatures in which we live (which
renders most material, except metals, non-conducting)
combined with the high densities; together, as we shall see in more
detail later in this chapter, these conditions make the forces exerted
by the terrestrial magnetic field utterly negligible in our day-to-day
affairs, except where we take special care to uncover them, such as in
magnetic compasses. Conditions are markedly different, however, in the layers
underneath the atmosphere of the Sun, throughout the extended solar
atmosphere, and in the outermost reaches of
atmospheres of all bodies in the Solar System: there, magnetism is
effectively coupled to matter while the inertia of that matter is in
much of the domain significantly lower in comparison to magnetic
forces than in our daily settings. There, the magnetic field is an
important player that adds a significant force to compete with
pressure, gravity, and inertia. It provides a medium for a variety of waves (which
this text merely touches upon), and changes the transport of thermal
energy and energetic particles. Add to that the fact that the magnetic
field is evolving on a range of spatio-temporal scales, and you have a
source of continual change in conditions throughout the local cosmos.

\begin{table}%\begin{figure}
\indexit{MHD!philosophy}
\caption[MHD approximation and the concept of 'closure'.]{\label{fig:mhdvalidity} MHD approximation and the concept of 'closure'.}
%\framebox[\textwidth]{
\fbox{ \vbox{ {\bf Philosophy of magnetohydrodynamics:} \\ The
    fundamental assumption underlying the MHD equations as shown in
    Table~\ref{fig:mhdset}, and the principal criterion to judge the
    applicability of that MHD approximation under given circumstances,
    is that the medium can be suitably described as a continuum. This
    presents us with a statistical criterion: MHD can be applied
    beyond a fiducial length, say $L$, such that there are sufficient
    particles in a volume $L^3$ such that statistical means --~like
    density, mean velocity, pressure and so forth~-- have small
    variances or fluctuations about them. Within that volume,
    collisions (or wave-particle interactions) result in average
    properties of the medium that transform the need to describe each
    particle separately in its interaction with all others into an
    enormously truncated set of descriptions of statistical
    averages. This truncation \indexit{MHD!closure}is
    known as 'closure': the continuum
    description requires a closure relation at some level that relates
    an unknown high-order moment of the full particle distribution
    function, such as pressure, to lower-order moments (see
    Sect.~\ref{sec:boltzmann} for more on that). An equation of state,
    as in Eq.~(\ref{eos}), is predicated on there being ample
    collisions to isotropize the random motions and achieve a
    thermodynamic equilibrium, with its characteristic Maxwellian
    velocity distribution (or more than one if a multi-fluid
    description is used). The MHD equations as in
    Table.~\ref{fig:mhdset} describe a 5-moment continuum closure
    scheme using mass density, temperature, pressure, energy density,
    and velocity. As collisions become less frequent one is required
    to enforce closure at higher levels, examples of which lead to,
    {\em e.g.,} Eqs.~(\ref{ohm-general}) and~(\ref{eq:10.9}). More generally
    dielectric and magnetization properties of the material enter in
    the definitions of $D$ and $H$.  Therefore, if by some other means
    ({\em e.g.,} by observation) you know how to close the moments (like in
    Sect.~\ref{sec:nearlystrat} for the solar wind by using the
    observationally motivated approximation that the temperature is
    constant throughout the heliosphere and the pressure is an
    isotropic scalar) then you can use the continuum fluid description
    to answer some questions even about a medium where collisions are
    a rare thing. }}
\end{table}%\end{figure}

The mathematical formulation of what happens in a magnetized plasma is
often simplified through an ensemble approximation that is equivalent
to the hydrodynamics used in the description of gases, but here
including the magnetic field in what is called magnetohydrodynamics,
or MHD for short. \indexit{MHD!introduction}MHD is a description of the
multitude of constituent particles in the local cosmos that relies on
statistical averaging carried out by the medium itself, namely through
interactions that lead to essentially Maxwellian velocity
distributions, often assumed to be isotropic (but in some formulations
distinct for directions along and perpendicular to the magnetic field)
and for velocity equilibrium between electrons and ions. To this, a
few other assumptions are made about local conditions: processes
described by MHD assume that ion and electron interactions as well as
their gyrations about the field occur on scales that are small
compared to the gradients in the magnetic field while at the same time
large compared to a distance (known as the Debye length) over which
electrical charges can exist unshielded by other particles, with
velocities well below relativistic, and only allowing for wave-like
phenomena that are slow enough that electrical neutrality is achieved
well within any time scale of interest and that are slow compared to
the plasma frequency and electron/ion gyro-frequencies. However,
interactions between particles should be infrequent enough that the
medium should allow the electron and ion populations to move
differentially with relative ease, {\em i.e.,} conditions should allow
the medium to conduct electrical currents rather effectively.

% See Ch. 1.7 in "Magnetic reconnection". From Forbes comments: "Hollweg
% (1971 -­‐ collisionless solar wind, 2, variable electron
% temperature, JGR, vol.  76, p.  7491) the solar wind becomes
% collisionless at around 10 to 20 solar radii."
MHD treats the ionized medium as a \indexit{MHD!fluid
  approximation}fluid by working with ensemble
properties. In hydrodynamics this is generally allowed because of a
high frequency of molecular collisions relative to the time scales of
the processes on macroscopic scales. In many environments in
heliophysics, however, collisions can be so rare that distances
between collisions can be comparable to the scale of the system under
consideration, while the solar wind is entirely collisionless beyond a
few dozen radii from the Sun \ldots\ and yet MHD has been shown to be
a useful approximation. The key factor in making MHD useful is that
the medium should not be able to maintain a significant electric field
in its own reference frame.  Even if collisions are rare in such a
medium, long-range flights of the particles are impeded: the gyration
of particles about the magnetic field reduces the scale of flight
perpendicular to the field, while wave-particle interactions have a
similar effect along the field. Consequently, the movement of
individual charged particles in a plasma is coupled to the collective
of its environment, resulting in a fluid-like behavior even if
collisions are rare.

%See Ch. 1.7 in "Magnetic reconnection"
However, where binary interactions are important in the MHD
description, the anisotropy imposed by the magnetic field does affect
what approximations can be made. Most importantly, these effects are
seen on gas pressure and viscosity. In a collisional plasma, these
terms are generally essentially isotropic and thus described by
scalars. But in a collisionless plasma, pressure and viscosity are
anisotropic, and thus are approximated by tensors. In this volume, we
generally use a scalar for pressure, and capture anisotropy in
conductivity in Hall and Pedersen terms (see below and
Ch.~\ref{ch:particles}).

% The MHD description, as we shall discuss in the next section,
% enables heliophysicists to approximate much of the local cosmos by
% including only the magnetic field on top of the hydrodynamic quantities. They
% infer electrical currents from the geometry of the magnetic field
% rather than having to think in terms of such currents and how these
% contribute to the magnetic field --~more on that below.

\begin{table}[tp]%\begin{figure}
  \indexit{MHD!equations}
  \caption[MHD equations.]{\label{fig:mhdset} Equations of
  magnetohydrodynamics for a fully-ionized plasma, ignoring radiative
  energy transport and radiation pressure, to be complemented by
  initial and boundary conditions to specify the solution. \indexit{MHD!equations}}
%\framebox[\textwidth]{
\fbox{
\vbox{
{\bf Single-fluid non-relativistic magnetohydrodynamics (MHD):}
\begin{eqnarray}
\text{Induction}&  \frac{\dd
  \B}{\dt}=\overset{\tc{1}}{\curl(\vv\times\B)}-\overset{\tc{2}}{\curl(\eta_{\rm
    \null}\,\curl\B)} \label{induction}\\
\text{Continuity}&  \frac{\dd\rh}{\dt}+(\vv\cdot\grad)\rh =-\overset{\tc{3}}{\rh \di \vv} +\overset{\tc{a}}{(S-L)}\label{continuity}\\
\text{Momentum}&  \rh \frac{\dd\vv}{\dt}
+\rh (\vv\cdot\grad)\vv =+\overset{\tc{4}}{\rh\vec{g}}-\overset{\tc{5}}{\grad p} 
  +\overset{\tc{6}}{\frac{1}{4\pi}(\curl\B)\times\B}
  \nonumber\\
&+\overset{\tc{7}}{\di\vec{\tau}}-\overset{\tc{b}}{\vv(S-L)}
+\overset{\tc{c}}{\left ({\bf S}_{\rm p} -{\bf L}_{\rm p} \right )} \label{momentum}\\
\text{Internal energy}&  \rh\frac{\dd e}{\dt} +\rh(\vv\cdot\grad)e=
  -\overset{\tc{8}}{p\di\vv}+\overset{\tc{9}}{\di(\kappa\grad
  T)}+\overset{\tc{10}}{\left(Q_{\nu}+Q_{\eta}\right)}\label{energy} \\
\text{Gravity}&  \nabla \cdot \vec{g} = \nabla^2 \Phi = -4\pi G\rh \label{gravity}\\
\text{EOS} &  p =(\gamma-1)\,\rh\,e \label{eos} \\
&\text{Complemented by initial and boundary conditions}
\nonumber
\end{eqnarray}
{\bf Online resources:}
\begin{eqnarray}
\text{Plasma
physics:}&\href{https://www.nrl.navy.mil/ppd/content/nrl-plasma-formulary}{\text{online
           NRL Plasma Formulary}}
\nonumber\\
\text{Vector
  calculus:}&\href{https://en.wikipedia.org/wiki/Vector_calculus_identities}{\text{Wikipedia
              page}}
              \nonumber\\
  \text{Introduction to 
  MHD}&\href{https://ui.adsabs.harvard.edu/abs/2013arXiv1301.5572S/abstract}{\text{'Essential
        magnetohydrodynamics for astrophysics'}}
\nonumber \\
 &\text{(\citep{2013arXiv1301.5572S})}
\nonumber 
\end{eqnarray}
}}
{\em \small Symbols:
  $\B$ magnetic field; $\vv$ fluid velocity; $e=C_V T$ specific
  internal energy ({\em e.g.}, energy per unit mass;
   ${3\over 2}kT/\mu$ for an ideal gas with $\mu$ the average mass per particle);
  $p$ gas pressure; $\rh$ mass density; $\Phi$ the gravitational potential and
  $G$ Newton's gravitational constant; $\vec{g}$ gravity,
  $\tau_{ik}=2\rh\nu\left(\Lambda_{ik}-\frac{1}{3}\delta_{ik}\di\vv\right)$
  the viscous stress tensor with the deformation tensor
  $ \Lambda_{ik}=\frac{1}{2}\left(\frac{\dd v_i}{\dd x_k} +\frac{\dd
      v_k}{\dd x_i}\right)$; $Q_{\nu}$ viscous heating; and
  $Q_{\eta}=\eta_{\rm \null }|\vec{j}|^2=\eta_{\rm \null
  }\left(c/4\pi\right)^2 |\curl\B|^2$ the resistive (Ohmic)
  dissipation; $\nu$, $\eta_{\rm e}$ and $\kappa$ represent the
  viscosity, magnetic diffusivity, and the thermal conductivity tensor
  (which is highly anisotropic, with heat most effectively conducted
  by electrons moving along the magnetic field); $\gamma=C_p/C_V$ is
  the adiabatic index, the ratio of specific heats for constant
  pressure and constant volume. In an ideal, mono-atomic gas with 3
  degrees of freedom $\gamma = 5/3$. $S$, $L$, ${\bf S}_{\rm p}$ and
  ${\bf L}_{\rm p}$ are source and loss terms for mass and momentum by
  introduction or loss of ions from a non-ionized reservoir.}
\end{table}%\end{figure}

\section{(Magneto-)Hydrodynamics}

% [Summarize MHD; e.g. with Mestel and Weiss from Saas-Fee 1974, and
% Spruit 2015. Then discuss approximations made, and begin to
% introduce exceptions, working towards the next two sections on
% time-length scales and dimensionless numbers.]

The equations of \indexit{MHD!continuum assumption}magnetohydrodynamics, or MHD, are based on the
assumption that the plasma can be described as a continuum; see
Table~\ref{fig:mhdvalidity} for a very concise description of what
that entails. The approximations used here lead to six equations that
describe magnetized plasma subject to gravity, as shown in
Table~\ref{fig:mhdset} (note that processes involving radiative
transfer are largely omitted from this volume). \sactivity{$\circledS$
  {\em Show:} This text uses cgs-Gaussian units. In other texts
  (including many of the chapters of the Heliophysics book series) you will find SI
  units. Look into conversions from one system to another and
  transform the momentum and induction equations,
  Eqs.~(\ref{momentum}) and~~(\ref{induction}), from cgs to
  SI. \mylabel{act:unitconv} \solution{unitconv}} Five of these are essentially equations
of hydrodynamics, namely continuity, momentum, energy, gravity, and
the equation of state (EOS), with two important modifications: the
magnetic, or Lorentz, force $(1/c){\bf j}\times {\bf B}$ \tc{6} is
added in the momentum description, and there are additional terms
\tc{10} in the energy equation. We return to these terms and equations
below, and discuss the additional equation, namely the induction
equation Eq.~(\ref{induction}), which couples the magnetic field to
macroscopic flows and microscopic collisions, in some detail in
Section~\ref{sec:induction}. \activity{{\em Show:} Note that
  $\nabla \cdot {\bf B} = 0$ is not needed to complement the MHD
  equations in Table~\ref{fig:mhdset} as long as the initial condition
  satisfies that equation. Take the divergence of
  Eq.~(\ref{induction}) to prove that. Use the same operation on
  $\nabla \times {\bf B} = \frac{4\pi}{c}{\bf j}$ to show that
  currents in MHD have no sources or
  sinks. \mylabel{act:sourcessinks}}

In order to assess the validity of the assumption made to derive
the MHD \indexit{MHD!validity}equations for the vastly
different conditions with which heliophysics concerns itself we
can
look at a variety of dimensional and dimensionless numbers. 
Table~\ref{tab:dimensionlessnumbers} lists frequently used length and
time scales, as well as some commonly used ratios, some of which have
been given a name. Some of these are pertinent to microscopic,
particle-level conditions and some are pertinent to macroscopic,
system-level conditions. We introduce them
here only briefly --~most will be looked at explicitly later on~-- in order to
give you an impression of which types of processes or relative scales
are important. For example, we can look at the length scale on which ions
gyrate around the local magnetic field relative to the gradients in
the field to assess whether the ions sense the magnetic field in an
ensemble sense such as required of a fluid or whether higher-order
descriptions are needed.  Or one can ask whether length scales
involved are large enough that the plasma can be viewed as not having
significant charge separation; the length scale on which electrostatic
potential of any particle is effectively shielded by the surrounding
plasma is known as the Debye length. Or one can look at the ratio of
the average time between collisions and the time needed to complete
one gyration around the magnetic field in order to assess whether the
magnetic field can effectively be followed by the charged particles
and whether the Hall current needs to be considered.

\label{sec:mhdequations}
\subsection{MHD equations, individual terms, and special cases}
First, let us briefly \indexit{MHD!equations!terms}review what the MHD equations express, the role
of the individual terms, and some special cases:
\begin{itemize}
  {\setlength\itemindent{25pt}
\item[\textbullet\ ] $\vv$, $p$:  The velocity $\vv$ reflects the
  mass-weighted bulk velocity (the first-order moment of the velocity
  distribution) of the electron-ion plasma. For a fully-ionized
  hydrogen plasma this equals
  $(m_{\rm i}\vv_{\rm i}+m_{\rm i}\vv_{\rm e})/(m_{\rm i}+m_{\rm e})$,
  which can be taken as the velocity of the center of mass of the
  ion-electron pairs that comprise the 'particles' of a 'single-fluid
  plasma.' The pressure $p$ is the sum of the pressures of the
  electron and ion populations.
\item[\textbullet\ ] Eq.~(\ref{induction}):  The induction equation (a
  combination of Faraday's law with Ohm's law, see
  Sect.~\ref{sec:induction}) states that any local change in the
  magnetic field is associated with a `curl', or 'circulation', in the
  component of plasma flows working perpendicular to the magnetic
  field and/or to the slippage of plasma relative to the magnetic
  field through finite diffusivity. Note that this form of the
  induction equation is linear in ${\bf B}$ so that if
  ${\bf B}(t=0)={\bf 0}$ then no field can arise at a later
  time. Sect.~\ref{sec:induction} touches on the fact that some terms
  were ignored to arrive at this form, some of which can act as a
  source term for magnetic field; this is not further discussed in
  this volume as the clouds out of which stars and planetary systems
  form initially are threaded by a galactic seed field from the outset
  (interested readers could look for 'battery effects', including the
  'Biermann battery').
  \item[\textbullet\ ] Eq.~(\ref{continuity}): 
    Continuity requires that the local plasma density changes only
    because of flow through a volume and by compression or dilation in
    doing so, unless there are sources or sinks within the volume.
  \item[\textbullet\ ] Eq.~(\ref{momentum}):  The momentum
    (or force) equation (Newton's second law in volumetric form)
    summarizes how the plasma velocity is affected, as in
    hydrodynamics, by gravity, pressure gradients, and viscosity, but
    here also by the Lorentz force associated with currents flowing
    across the magnetic field.
  \item[\textbullet\ ] Eq.~(\ref{energy}):
  The local energy density (here shown in a per-mass formulation
    of the first law of thermodynamics) is affected by flows,
    including compression or dilation, thermal conduction, and by
    viscous and resistive heating.
  \item[\textbullet\ ] Eq.~(\ref{gravity}): As mass is a positive definite quantity, it can
    only strengthen gravity, which can be represented by the gradient
    of a potential.
  \item[\textbullet\ ] Eq.~(\ref{eos}):  The
    equation of state couples pressure, density, and internal
    energy.
  \item[\textbullet\ ] Eqs.~(\ref{induction}, \ref{momentum}):
    The induction and momentum equations are derived
    from the (mass-weighted) difference (see
    Sect.~\ref{sec:induction}) and the sum of the equations of motions
    for the electrons and for the ions, each of which includes a term for their
    collisional coupling.
  \item[\textbullet\ ] \tc{1}:  In case the
    term \tc{2} is negligible, Eq.~(\ref{induction}) describes what is
    known as `ideal MHD'. In this case (see
    Sect.~\ref{sec:fieldlines}) the plasma and magnetic field must
    move with each other for velocity components perpendicular to the
    magnetic field, whereas plasma movement along the field is not
    affected by that field. In this condition, the field is said to be
    `frozen in' the plasma. In such a state, the lines of force
    (`field lines') are advected with the flow while unable to break
    their connectivity between any plasma elements along their length;
    in non-ideal, or resistive, MHD such connections can be broken
    through a process known as `reconnection'. The concepts of field
    lines and reconnection are described in
    Section~\ref{sec:fieldlines}. \activity{{\em show:} Use a vector calculus
      identity and Gauss' law to show that term \tc{1} in
      Eq.~(\ref{induction})  comprises compression, shear, and
      advection. \mylabel{act:inductcomp}}
  \item[\textbullet\ ] \tc{2}:  This
    term quantifies the effects of resistivity on the magnetic field
    by the dissipation and diffusion of the electrical current
    $\vec{j}=\frac{c}{4\pi}\curl \B$. If the magnetic diffusivity
    $\eta_{\rm \null }$ is constant throughout the medium, then term \tc{2}
    can be rewritten as
    $\eta_{\rm \null } \curl (\curl \B)=-\eta_{\rm \null } \nabla^2\B$ (because
    $\nabla \cdot \B=\vec{0}$), which shows that it causes the
    magnetic field to decay diffusively; in the absence of \tc{1},
    such as in a stationary plasma, this makes Eq.~(\ref{induction}) a
    diffusion equation for decaying magnetic field.
  \item[\textbullet\ ] \tc{3}:
     For an incompressible fluid, $\rh$ is constant as
    material flows throughout the volume under study, which
    consequently means that $\nabla \cdot \vv = \vec{0}$, {\em i.e.,}  that
    the velocity field is divergence free, and --~unless there are
    terms like \tc{a} to consider~-- Eq.~(\ref{continuity}) vanishes
    from the set.  That also removes term \tc{8} from
    Eq.~(\ref{energy}), so that the energy density of the medium can
    only change by thermal conduction \tc{9} and by viscous and
    resistive dissipation \tc{10} (disregarding here, as we do
    throughout this chapter, the effects of radiation). \activity{{\em
        Background:} The
      so-called 'Boussinesq approximation' is intermediate to fully
      compressible and incompressible, and in principle internally
      inconsistent: it assumes a fluid for which (and in numerical
      codes replaces Eq.~(\ref{continuity} by)
      $\nabla \cdot \vv = \vec{0}$ but allows density variations in
      the term in the force balance that includes gravity (and thus
      allows for buoyancy). This approximation works well if the flow
      can be characterized as 'nearly incompressible'. For settings
      where the scale of the density stratification is large compared
      to processes of interest the incompressible approximation can be
      valid; in such settings, compressibility becomes only important
      in structures like shock waves, but is ignorable if the flows
      are much slower than the sound and Alfv{\'e}n speeds.  Advanced,
      for the curious: In planetary atmospheric envelopes and stellar
      interiors alike, zones of relatively low temperature under
      relatively strong gravity are highly stratified compared to the
      scales of flows within them. In such settings, numerical codes
      have been developed under the 'anelastic' approximation. This
      approximation provides a better description of the density in
      stratified settings than the pure Boussinesq one while filtering
      out sound waves that would require much higher spatio-temporal
      resolution of the code.
      \href{https://ui.adsabs.harvard.edu/abs/2007CRMec.335..655D/abstract}{This
        article by \citet{2007CRMec.335..655D}} introduces and compares
      several 'anelastic' approximations. \mylabel{act:boussinesq}}
%\item[\tc{4}: \textbullet\ ] \ldots
\item[\textbullet\ ] \tc{5}:  As formulated here, the isotropic part of
  the pressure tensor is expressed as a
  scalar, while the other terms are captured in the stress tensor. If
  only this scalar term is carried, then particle microscopic velocity distributions
  are taken to be isotropic.
\item[\textbullet\ ] \tc{6}:  Term \tc{6} measures the interaction of the Lorentz
  force and the plasma flow. The vector product $(\curl\B)\times\B$
  can be reformulated (see Eq.~\ref{eq:presscurv}) into the sum of a
  pressure-like term (that works to expand unless countered) and a
  term that is equivalent to a tension (which works to straighten
  unless countered), showing that the magnetic field in a plasma
  behaves as if it were both like a gas and like a flexing rod or taut
  string. There is a
  special class of magnetic fields in which currents run parallel to
  the magnetic field; in that case $(\curl\B)\times\B = \vec{0}$,
  {\em i.e.,} there is no Lorentz force, and these are consequently referred
  to as \indexit{force-free field}'force-free fields', of which the potential field is a
  special (and lowest-energy) state. As the field is parallel to the
  current, there is a scalar field $\alpha$ such that $\curl \B =
  \alpha \B$. If $\alpha$ is a uniform constant, this field is called
  `linear force free' (which is mathematically easier to work with,
  but does not develop in general astrophysical settings); if not,
  the corresponding field is a `non-linear force-free' field (to
  which we return in Sect.~\ref{sec:solarimpulsive}).
%\item[\tc{7}: \textbullet\ ] \ldots
%\item[\tc{8}: \textbullet\ ] \ldots
\item[\textbullet\ ] \tc{9}:  As for term \tc{2} with uniform magnetic diffusivity
  $\eta_{\rm \null }$, here a uniform thermal conductivity $\kappa$ would allow
  rewriting of term \tc{9} to be proportional to $\nabla^2 T$,
  quantifying diffusion of thermal energy.
%\item[\tc{10:}] \ldots
\item[\textbullet\ ] \tc{a},\tc{b},\tc{c}:  These terms reflect source
  and loss terms for mass and momentum density per unit volume
  through, {\em e.g.,} (de-)ionization of neutrals (including charge
  exchange) that are important, for example, where comets add gas and
  dust or around geysers on low-gravity moons (Sec.~\ref{sec:pickup}),
  or to the inflow of neutral matter into the solar wind from outside
  the heliosphere.
%***
}\end{itemize}
A few special cases:
\begin{itemize}
  {\setlength\itemindent{25pt}
  \item[\textbullet\ ] $\B=\vec{0}$:  A field-free state (or a
  non-conducting, and thus current-free, gas in
  which the field does not apply force to the gas; see also under
  'Potential' below) transforms
  Eqs.~(\ref{induction})--(\ref{eos}) into regular hydrodynamic
  equations. 
\item[\textbullet\ ] $\vv=\vec{0}$:  A static plasma is described by
  Eqs.~(\ref{induction})--(\ref{eos}) without terms \tc{1}, \tc{3},
  \tc{7}, \tc{8}, and $Q_\nu$ in \tc{10}. Moreover, with no flows, no
  change can occur that involves bulk flows, so that, for example, the
  left-hand side of the momentum Eq.~(\ref{momentum}) has to equal
  $\vec{0}$. This yields an equation for magnetohydrostatic balance in
  which gravity, pressure gradient, and Lorentz force sum to zero. 
\item[\textbullet\ ] $\frac{\dd \null}{\dt}=\vec{0}$:  Stationary
  situation in which none of the variables can change. In particular, 
  $\frac{\dd \vv}{\dt}=\vec{0}$ is a situation with
  stationary flows, which can be maintained only for limited times. 
\item[\textbullet\ ] Potential:  In the case of a \indexit{potential
    field}potential field, there are no
  currents in the system, {\em i.e.,}  $\curl\B=\vec{0}$. Consequently, term
  \tc{6} vanishes because there is no Lorentz force. Term \tc{2} also
  vanishes, leaving only term \tc{1} in the righthand side of
  induction Eq.~(\ref{induction}) (equivalent to the infinitely
  conducting case of ideal MHD with frozen-in field, or in which the
  field is maintained from outside of a current-free volume). To see to full
  consequence of this state, however, we need to realize that
  $\curl\,\B=\vec{0}$ means that there is a magnetic potential
  $\Phi_{\rm m}$ such that $\B=-\nabla \Phi_{\rm m}$ from
  $\nabla^2\Phi_{\rm m}=0$. Such a Laplace equation, once the boundary
  condition is specified, has a unique solution. And for a
  current-free system with fixed boundary conditions that, in turn,
  means that $\B$ cannot change in
  time, such that term \tc{1} then implies that there is a scalar field
    $\Psi$ such that ${\bf v} \times {\bf B} = \nabla \Psi$, of which
    one particular case has $\vv \parallel \B$.
  \item[\textbullet\ ] Force free:  See at \tc{6} above in this
    listing.  }\end{itemize}

%\activity{{\em Advanced/Group:} {\bf What if radiative transfer were
%    included?} The MHD equations as shown in Table~\ref{fig:mhdset}
%  ignore electromagnetic radiation. In a sufficiently dense medium, in
%  which the photon mean free path is small compared to plasma and
%  field gradients, energy transport by electromagnetic radiation can
%  be described by a diffusion equation. Where the mean-free path is
%  long, however, energy can 'jump' between different locations without
%  (or with weak) coupling to the intermediate medium, in a manner that
%  depends on wavelength as well as on atomic properties. With that in
%  mind, contrast the solar interior to its atmosphere; a cloud-free
%  planetary atmosphere to a (partially) clouded one; and (maybe once
%  you get to Ch.~\ref{ch:formation}) initial to later phases of star
%  formation and of protoplanetary disks. In the context of that
%  question and other assumptions going into the MHD equations in
%  Table~\ref{fig:mhdset}: Why is the solar chromosphere the hardest
%  part of the solar interior and atmosphere to describe? And what
%  makes a terrestrial ionosphere hard to capture in equations? Some of
%  the answers to these questions will come as you read along. For an
%  introduction to radiative transfer in stellar atmospheres, see
%  \href{https://ui.adsabs.harvard.edu/abs/2003rtsa.book.....R/abstract}{this}
%  freely available online text by \citet{2003rtsa.book.....R}:
%  \href{https://www.staff.science.uu.nl/~rutte101/Radiative_Transfer.html}{URL}. \mylabel{act:radtrans}}

\subsection{The induction equation}\label{sec:induction}
\indexit{MHD!induction equation}\ors[I:3.2] ``The induction equation,
Eq.~(\ref{induction}), arises from\indexit{Ohm's law} Ohm's law
combined with the non-relativistic approximation of the Maxwell
equations. In its most general form Ohm's law is a relation between
electric current, electric field, magnetic field, plasma motions and
electron pressure gradients. Ohm's law is derived from an equation of
motion for electrons in which the interaction with ions (defining the
bulk motion of the plasma with velocity $\vec{v}$ [because the ions,
here taken to be dominated by singly-ionized species, by far outweigh
the electrons]) is described through a collisional drag term related
to the differential motion:
\begin{equation}\label{eq:emomentum}
  n_{\rm e}\,m_{\rm e}\frac{{\rm d} \vec{v}_{\rm e}}{{\rm d} t}=-n_{\rm e}\,e(\E+\frac{1}{c}\vec{v}_{\rm e}\times\B)-\grad p_{\rm e}
  +n_{\rm e}\,m_{\rm e}\frac{\vec{v}-\vec{v}_{\rm e}}{\tau_{\rm ei}}\;.
\end{equation}
Here $\vec{v}_{\rm e}$ denotes the electron velocity, $\tau_{\rm ei}$
the collision time between electrons and ions, $e$ the electron
charge, $m_{\rm e}$ the electron mass, $n_{\rm e}$ the electron
density, and $p_{\rm e}$ the electron pressure'' (omitting
gravity). $\vec{E}$ and $\vec{B}$ are the electric and magnetic vector
fields. 

By noting that the differential velocity between ions and electrons
is proportional to the current,
\begin{equation}
  \jj=n_{\rm e}\,e\,(\vec{v}-\vec{v}_{\rm e})
\end{equation}
we can reformulate Eq.~(\ref{eq:emomentum}), when combined with the
analogous version for the ions, to yield a formulation of Ohm's law
(here ignoring electron inertia and assuming pressure to be a scalar,
{\em i.e.,}  isotropic; compare with Section~\ref{sec:fieldlines}):
%\marginpar{Check units, c's}
\begin{equation}
  \jj=
  \frac{\tau_{\rm ei} n_{\rm e} e^2}{m_{\rm
      e}}(\E+\frac{1}{c}\vec{v}\times\B)-\overset{{\rm Hall\,term}}{\frac{\tau_{\rm ei} e}{m_{\rm e}c}\jj\times\B}
  +\frac{\tau_{\rm ei} e}{m_{\rm e}}\grad p_{\rm e}\;.\label{ohm-general}
\end{equation}
\activity{{\em Show:} Formulate the ion equivalent of Eq.~(\ref{eq:emomentum})
  (remember Newton's third law, and with
  $m_{\rm e}/m_{\rm i} \rightarrow 0$) and derive
  Eq.~(\ref{ohm-general}) (using
  $m_{\rm e}\vec{v}_{\rm i}+m_{\rm i}\vec{v}_{\rm e}= m_{\rm
    i}\vec{v}_{\rm i}+m_{\rm e}\vec{v}_{\rm e}+m_{\rm i}[\vec{v}_{\rm
    e}-\vec{v}_{\rm i}]+m_{\rm e}[\vec{v}_{\rm i}-\vec{v}_{\rm e}]$)
  and also the corresponding momentum equation (absent gravity):
  $\rho {\rm d}\vec{v}/{\rm d}t
  =\vec{j}\times\vec{B}-\vec{\nabla}p$. Then add gravity and compare
  to Eq.~(\ref{momentum}). \mylabel{act:derivmomentum}}  Note that when the electric field
expressed through the electron pressure gradient is ignored, this
equation can be rewritten to an equivalent Ohm's law discussed for the
ionosphere in Ch.~\ref{ch:particles} that has a conductivity tensor
with components as in Eqs.~(\ref{eq:pedersen})-(\ref{eq:parallel}),
from which terms with $M_{\rm i}$ disappear for the fully-ionized
plasma because there are only ion-electron collisions, and in which
the electron magnetization subject to collisions with neutrals,
$M_{\rm e}$, is replaced by $M_{\rm ei}= eB/(mc\nu_{\rm ei})$ for
$\nu_{\rm ei} = 1/\tau_{\rm ei}$. In other words, the Hall term in
Eq.~(\ref{ohm-general}) takes care of the anisotropic part of the
conductivity in the fully ionized plasma. The Pedersen current,
directed along ${\bf E}$ is part of the first term on the right-hand
side. \activity{{\em Show}, for a fully-ionized single-species
  plasma, the equivalence of Eq.~(\ref{ohm-general}) and
  ${\bf j} = {\bf \Sigma}_{\rm e} \cdot ({\bf E}+\frac{1}{c}{\bf v} \times
  {\bf B})$ with Eqs.~(\ref{eq:condtensor})-(\ref{eq:parallel}).}

Specifically, \ors[I:3.2] ``the second term on the right-hand side
describes the\indexit{Hall!current} Hall current, which becomes
important if the collision time is longer than the electron gyration
time, {\em i.e.,}  when $\tau_{\rm ei}\,\omega_L>1$, where $\omega_L=e
B/m_{\rm e}$ denotes the\indexit{Larmor frequency}
Larmor\indexit{gyrofrequency} \hbox{(or [electron]
  gyro-)}frequency. The Hall term leads to anisotropic plasma
conductivity with respect to the magnetic field direction and is
typically important in low-density plasmas in which $\tau_{\rm ei}$
can be very large''. In many settings in heliophysics, the last two
terms in Eq.~(\ref{ohm-general}) are ignored ``(unless high-frequency
plasma oscillations are considered), leading to the simplified Ohm's
law
\begin{equation}
  \jj=\sigma_{\rm e}(\E+\frac{1}{c}\vv\times\B)\label{ohm}
\end{equation}
with the plasma conductivity\indexit{conductivity}
\begin{equation}\label{eq:conductivity}
  \sigma_{\rm e}=\frac{\tau_{\rm ei} n_{\rm e} e^2}{m_{\rm e}}\;.
\end{equation}
Using Amp{\`e}re's law,\indexit{Amp{\`e}re's law}
%\begin{equation}\label{eq:ampere}
$\curl\B=\frac{4\pi}{c}\jj,$
%\end{equation}
yields for the
electric field in the laboratory frame
\begin{equation}
  \E=-\frac{1}{c}\vv\times\B+\frac{c}{4\pi\sigma_{\rm e}}\curl\B
\end{equation}
leading to the induction equation\indexit{MHD!induction equation}
through one of the Maxwell equations:
\begin{equation}
  \frac{\dd \B}{\dt}=-c\curl\E=\curl\left(\vv\times\B
  -\eta_{\rm \null }\,\curl\B\right)\label{induction2}
\end{equation}
with the magnetic diffusivity\indexit{magnetic!diffusivity}
\begin{equation}\label{eq:diffusivity}
  \eta_{\rm \null }=\frac{c^2}{4\pi\sigma_{\rm e}}\;.
\end{equation}

In MHD, the equations are typically expressed in terms of the magnetic
field $\B$ and flows $\vec{v}$, with electric fields and currents
eliminated from the system.  This is done primarily out of
mathematical convenience, since formulating the problem in terms of
currents leads to intractable equations involving integrals of the
currents over the entire volume under study.''

Whether a formulation in terms of the magnetic field or electrical
currents is more convenient also depends on the inhomogeneity and
anisotropy of the conductivity. The most extreme example is electrical
engineering where cables give full control over the current, and thus
a current-based description is clearly the method of choice. A
formulation in terms of currents can be easier to work with also when
currents can only flow along the field or are restricted to relatively
thin layers with high conductivity, such as is the case in the
ionosphere.  In most MHD problems with highly conducting fluids,
however, there is no {\em a priori} control over where currents flow, so
that dealing with the magnetic field is typically the better choice.
Because of this, solar and heliospheric physicists generally use
arguments primarily based on the magnetic field; in space physics,
however, and in particular in ionospheric physics, currents are often
discussed.

Of interest to the induction equation Eq.~(\ref{induction}) is the
relative importance of the advection and diffusion-like terms on the
right-hand side. One way to assess that is to reformulate it into
characteristic scales and a frequently occurring dimensionless number:
\ors[I:3.2.3] ``Let $L_{\rm t}$ be a typical length-scale and $v_{\rm
  t}$ a characteristic velocity of the problem. Expressing the time in
units of $L_{\rm t}/v_{\rm t}$ and the spatial derivatives in the
induction equation Eq.~(\ref{induction}) in units of $L_{\rm t}$ leads
to the dimensionless form of the induction equation
\begin{equation}\label{dimlessinduction}
  \frac{\dd\B}{\dt}=\curl\left(\vv\times\B-\frac{1}{\Rm}\curl\B\right)
\end{equation}
with the magnetic Reynolds number \indexit{Reynolds number, magnetic}\indexit{Reynolds number, magnetic|seealso{definition}}
\begin{equation}\label{eq:reynolds}
  \Rm\equiv\frac{v_{\rm t}\,L_{\rm t}}{\eta_{\rm \null }}\;.
\end{equation}
The limit $\Rm\ll 1$ is referred to as\indexit{diffusion!dominated
regime|seealso{definition}} diffusion \indexit{diffusion!dominated
regime}dominated regime, in which the
(dimensional) induction equation reduces to a diffusion equation
of the form \indexit{magnetic!diffusion equation}
\begin{equation}
  \frac{\dd\B}{\dt}=\eta_{\rm \null }\nabla^2\B\;.
\end{equation}
Here we made the additional simplifying assumption of a constant magnetic
diffusivity $\eta_{\rm \null }$.
Assuming that the magnetic field has a typical length scale $L_{\rm t}$, we can
estimate [its  decay time scale:]
\indexit{decay time scale}\indexit{diffusion!time scale}
\begin{equation}\label{eq:diffusiontime}
  \tau_{\rm d}\sim\frac{L_{\rm t}^2}{\eta_{\rm \null }}\;.
\end{equation}
The limit $\Rm\gg 1$ is referred to as the
advection-dominated\indexit{definition!advection-dominated regime}
regime, in which the induction equation reduces to the equation of
ideal MHD (except for possible boundary layers where diffusivity could
be still important)
\indexit{MHD!ideal}
\indexit{MHD!induction equation!ideal}
\begin{equation}\label{eq:ideal_induction}
  \frac{\dd\B}{\dt}=\curl\left(\vv\times\B\right)\;.
\end{equation}
Expanding the expression of the right-hand side of this ideal induction
equation leads to
\begin{equation}%\label{eq:adampcomp}
  \frac{\dd\B}{\dt}=-(\vec{\vv}\cdot\grad)\vec{\B}
  +(\vec{\B}\cdot\grad)\vec{\vv}-\vec{\B}\,(\di\vec{\vv})
  \label{ideal-mhd-expanded}\;.
\end{equation}
While the first term on the right-hand side describes the advection of
magnetic field, the last two terms describe the amplification by shear
(second term) and compression (third term).''

Of interest to the momentum equation Eq~(\ref{momentum})  is that a vector identity allows
us to reformulate  the Lorentz force \ors[I:3.2]
``to equal:
\begin{equation}\label{eq:presscurv}
\frac{1}{4\pi}(\curl\B)\times\B =-\nabla \frac{B^2}{8\pi}
+\frac{1}{4\pi} (\B\cdot\grad)\B,
\end{equation}
which shows that the Lorentz force is a sum of an isotropic pressure-like
force and a tension force related to the curvature of the field''
(note that both of these are insensitive to a reversal of the
direction of the magnetic field). 
Because the pressure and tension terms, as does therefore the full Lorentz force,
scale as ${\cal O}(B_{\rm t}^2/L_{\rm t})$ (where the subscript 't' denotes a
typical value of the quantity) they can be compared in magnitude to the
pressure gradient force ${\cal O}(p_{\rm t}/L_{\rm t})$; the 
ratio of magnetic and gas pressure terms in Eq.~(\ref{momentum})
yields an often-used dimensionless number in heliophysics, the plasma $\beta$:
\activity{{\em Show:} (a) Look back at Fig.~\ref{fig:conditions} and
  review the ranges shown of the value of the plasma $\beta$ from
  Eq.~(\ref{eq:betadef}) to get a feel for where plasma pressure
  gradients might dominate magnetic pressure gradients or vice
  versa. (b) Add lines (on a printed version, using a pdf editor, or a
  hand-drawn version) for unit plasma $\beta$ for field strengths of
  1~$\mu$G (as found in the outer heliosphere and interstellar medium;
  see Chs.~\ref{ch:flows} and~\ref{ch:evolvingstars}) and for 0.1~MG
  (considered characteristic of the field strength of flux bundles at
  the bottom of the solar convective envelope where the principal
  processes in the solar dynamo are considered to operate; see
  Ch.~\ref{ch:dynamos}).\mylabel{act:addlinestofig}}\sactivity{$\circledS$
  {\em Show:} (a) Take the momentum equation of Eq.~\ref{momentum}
  --~for a stationary state of the solar wind, ignoring viscosity,
  sources, and sinks~-- to assess the usefulness of the plasma $\beta$
  (Eq.~\ref{eq:betadef}, and
  Table~\ref{tab:dimensionlessnumbers}). (b) Estimate $\beta$ for the
  solar wind (1) near Earth for the slow and fast solar wind as in
  Table~\ref{tab:wind-stats} (use the electron temperatures) and (2)
  around $10 R_\odot$ (roughly the closest approach of the Solar
  Orbiter; use a temperature of 1\,MK there). For (2), assume a radial
  field and a radial flow (which is not too bad within Earth's orbit
  for these rough estimates, see Section~\ref{sec:parker-spiral}). (c)
  Use the same momentum equation as above to show that the ratio of
  dynamic (or ram) pressure to magnetic pressure also arises naturally
  (to which we return for the magnetopause standoff distance). (d) The
  simplest Parker solar wind model as discussed in
  Section~\ref{sec:nearlystrat} ignores the magnetic field; where is
  that appropriate in the Parker model?
  \solution{estimatebeta}\mylabel{act:estimatebeta}}
\indexit{plasma!$\beta$|seealso{definition}}
%N.B. defined as in Eq. 1.65 in Spruit 2013.
\begin{equation}\label{eq:betadef}
\beta \equiv \frac{8\pi p}{B^2}.
\end{equation}

\section{Waves in magnetized plasmas}\label{sec:mhdwaves}
Before we proceed with a discussion of field lines and reconnection,
\indexit{plasma!waves}we \indexit{wave!Alfv{\'e}n wave}look
\indexit{wave!fast and slow mode}into an important aspect of a magnetized
plasma, namely how it carries waves. Waves are important, among other
things, in communicating information about changes in the field's
structure or in boundary conditions or the effects of obstacles
embedded in flows, while moreover they transport energy.
\ors[IV:10.2.1] ``The waves that carry information through a
magnetized plasma differ from the sound waves of a neutral gas, partly
because of the anisotropy imposed on the fluid by a magnetic field and
partly because the waves must be capable of carrying currents that
modify the properties of both matter and magnetic field. The
properties of such waves can be derived from the MHD [equations] by
analyzing the evolution of small perturbations.

Consider a uniform plasma with constant pressure and density ($p$ and
$\rho$) whose center of mass is at rest (${\bf v} = 0$).  Assume that
a constant background field (${\bf B}$) is present and that neither
sources nor losses need be considered.  Small departures from this
background state are taken to vary with space (${\bf x}$) and time
($t$) as $e^{i ( {\bf k}\cdot {\bf x}-\omega t ) }$.  Here, ${\bf k}$
is the wave vector and $\omega$ is the angular frequency of the
wave. Perturbations occur in density d$\rho$, velocity d${\bf v}$,
pressure d$p$, current ${\bf j}$, and field ${\bf b}$. Terms linear in
small quantities in Eqs.~(\ref{continuity}) and \ref{momentum})
satisfy
\begin{equation}\label{eq:10.7}
-\omega {\rm d} \rho + \rho {\bf k}\cdot {\rm d}{\bf v} =0
\end{equation}
%	-ωdρ+ρ k∙du=0 	(10.7)
\begin{equation}\label{eq:10.8}
-\omega \rho {\rm d}{\bf v} = -{\bf k}{\rm d} p + \frac{1}{4\pi} {\bf b}({\bf k}\cdot {\bf B}) -
\frac{1}{4\pi}{\bf k}({\bf b}\cdot {\bf B}).
\end{equation}
%	  -ωρdu=-kdp+b(k∙B)/μ_o  -k( b∙B)/μ_o    	(10.8)

[If we assume an isentropic ({\em i.e.,} adiabatic and reversible) process,
then Eq.~\ref{eos} becomes $p\rh^{-5/3}=\text{constant}$, so that] the
pressure perturbation in terms of the density perturbation is
\begin{equation}\label{eq:10.9}
{\rm d}p/p = \gamma {\rm d}\rho/\rho
\end{equation}
%		dp/p= γdρ/ρ	(10.9)
and [the ideal induction equation Eq.~(\ref{induction} (with
$\eta_{\rm \null }\equiv 0$)]) 
\begin{equation}\label{eq:10.10}
\omega{\bf b} = {\rm d}{\bf v}({\bf k}\cdot {\bf B}) - {\bf B}({\bf k}\cdot {\rm d}{\bf v}).
\end{equation}
%  	ωb=du( k∙B)-B(k∙du)	(10.10)

The solutions to Eqs.~(\ref{eq:10.7}) to (\ref{eq:10.10}) are the roots of the equation
\begin{equation}\label{eq:10.11}
(\omega^2-v_{\rm A}^2 k^2 \cos^2 \theta)[\omega^4 - \omega^2 k^2 (c_{\rm s}^2 +v_{\rm A}^2)+ k^4 v_{\rm A}^2 c_{\rm s}^2 \cos^2 \theta]=0,
\end{equation}
%	  (ω^2-v_{\rm A}^2 〖k^2 \cos〗^2 ϑ)[ω^4-ω^2 k^2 (c_{\rm s}^2+v_{\rm A}^2 )+〖k^4 〖v_{\rm A}^2 c〗_s^2 cos〗^2 ϑ]=0	(10.11)
where $\theta$ is the angle between ${\bf k}$ and ${\bf B,}$ and the
Alfv{\'e}n speed ($v_{\rm A}$) and the sound speed ($c_{\rm s}$) have been
introduced.  These quantities characterize the \indexit{Alfv{\'e}n!speed}speed of propagation of
waves in a magnetized plasma and are \indexit{sound speed}defined by
\begin{equation}\label{eq:10.12}
v_{\rm A}^2 = B^2/4\pi \rho,
\end{equation}
%		v_{\rm A}^2=B^2/2μ_o ρ		(10.12)
\begin{equation}\label{eq:10.13}
c_{\rm s}^2=\gamma p / \rho.
\end{equation}
%			 c_{\rm s}^2=γp/ρ		(10.13)
The sound speed has the form familiar for a neutral gas.  The
Alfv{\'e}n speed is a second natural wave speed characteristic of a
magnetized plasma. Just as [one can work with the dimensionless] sonic
Mach number as the ratio of the flow speed to the sound speed, it is
useful to define a dimensionless Mach number, the Alfv{\'e}nic Mach
number 
\begin{equation}\label{eq:alfvenmach}
M_{\rm A} = v/v_{\rm A},
\end{equation} 
related to the Alfv{\'e}n speed.

As mentioned [at Eq.~(\ref{eq:presscurv})], the quantity $B^2/8\pi$ is the pressure
exerted by the magnetic field, so both of the basic wave speeds are
proportional to the square root of a pressure divided by a
density. [\ldots\ \indexit{plasma!$\beta$}When $\beta =8\pi p/B^2 \ll 1$,] magnetic effects dominate the
effects of the thermal plasma, but in a high-$\beta$ plasma, the
plasma effects dominate.

Equation~(\ref{eq:10.11}) is of sixth order in $\omega/k$ with three pairs of roots.  One pair results from setting the first factor in Eq.~(\ref{eq:10.11}) to zero; the resulting dispersion relation is
\begin{equation}\label{eq:10.14}
(\omega^2 - v_{\rm A}^2 k^2 \cos^2 \theta) =0.
\end{equation}
%	(ω^2-v_{\rm A}^2 〖k^2 cos〗^2 ϑ)=0	(10.14)
This solution describes waves referred to as \indexit{Alfv{\'e}n!wave}Alfv{\'e}n waves.  For this dispersion relation to apply, the magnetic perturbation must be perpendicular to both ${\bf B}$ and ${\bf k}$ (see Fig.~IV:10.1a).  This orientation implies that to first order in small quantities, the Alfv{\'e}n wave does not change the field magnitude [$({\bf B} + {\bf b})^2 = B^2 + 2({\bf B}\cdot {\bf b})^2 + b^2 \approx B^2$].
The wave phase speed is $v_{\rm ph}=\omega /k$ and $v_{\rm ph}=\pm v_{\rm A}
\cos\theta$.  [Wave packets] carry information at the group velocity,
${\bf  v}_{\rm g}={\bf \nabla}_k \omega$, where the subscript on the
gradient indicates that the derivatives are taken in ${\bf k}$ space;
the solution is $v_{\rm g}=\pm \hat{\bf B}v_{\rm A}$ where $\hat{\bf
  B}$ is a unit vector along the background field.  The remarkable
property of these waves is that they carry information only along the
background field, and they bend the field without changing its
magnitude.  These properties are of considerable importance in
interpreting the interaction of a flowing plasma with the solid bodies
of the Solar System [(discussed in Ch.~\ref{ch:flows})].

Eq.~(\ref{eq:10.11}) has two more pairs of roots, the zeroes of the fourth order polynomial in square brackets in Eq.~(\ref{eq:10.11}), {\em i.e.}, the solutions
\begin{equation}\label{eq:10.15}
v_{\rm ph}^2 = \omega^2 / k^2 = {1\over 2} \left ( c_{\rm s}^2 +
  v_{\rm A}^2 \pm [(c_{\rm s}^2+v_{\rm A}^2)^2 - 4v_{\rm A}^2c_{\rm s}^2 \cos^2 \theta ]^{1/2} \right ).
\end{equation}
%  	〖v_(ph)^2=ω〗^2⁄k^2 =□(1/2){c_{\rm s}^2+v_{\rm A}^2±[〖〖(c〗_s^2+v_{\rm A}^2)〗^2-4〖〖v_{\rm A}^2 c〗_s^2 cos〗^2 ϑ]^□(1/2)}	(10.15)
The solutions (two pairs, one positive and one negative, of roots)
correspond to what are unimaginatively referred to as
\indexit{fast-mode wave}fast-mode \indexit{magnetosonic wave}(or
magnetosonic) and \indexit{slow-mode wave}slow-mode waves.  The wave
perturbations of both modes may have magnetic perturbations along and
across ${\bf B}$ (see Fig.~IV:10.1b). Perturbations along ${\bf B}$
change the field magnitude and the thermal pressure. The fast mode
changes of thermal and magnetic pressure are in phase with each other;
this implies that the total pressure fluctuates.  The slow mode
changes of thermal and magnetic pressure are in antiphase, and the
total pressure fluctuations are very small.  For waves propagating
along the background field ($\cos\theta=\pm 1$), the solutions to
Eq.~(\ref{eq:10.15}) are $c_{\rm s}^2$ and $v_{\rm A}^2$, with the
larger of the two applying to the fast mode.  For waves propagating at
right angles to the background field ($\cos\theta=0$), the [solutions]
are $c_{\rm s}^2+v_{\rm A}^2$ and $0$, indicating that only fast-mode
waves propagate across the field.'' \activity{{\em Show:} Compare
  values for $c_{\rm s}$ and $v_{\rm A}$ for the environments listed
  in Table~\ref{tab:5.1}.  You may approximate $\gamma \approx 1$ and
  so use Eq.~(\ref{eq:cs-and-vg}); also use Eq.~(\ref{eq:10.12});
  assume a hydrogen-dominated, fully-ionized plasma for all settings
  in the table. Use km/s. Note that magnetic perturbations are faster
  in all settings except in the solar interior; equivalently, magnetic
  forces are more likely to dominate in all settings except in the
  solar interior where hydrodynamic forces dominate because (show
  this:) $\beta \sim c_{\rm s}^2/v_{\rm
    A}^2$. \mylabel{act:speedcomp}}

\section{MHD, magnetic field lines and reconnection}\label{sec:fieldlines}
\ors[I:4.1] ``One of the most idiosyncratic aspects of space physics
is the central role assigned to \indexit{magnetic!field line}magnetic
\indexit{field line}field lines.  Particularly in studies of the Sun,
the heliosphere and the magnetosphere, magnetic field lines are
treated as full-fledged physical objects with their own dynamics.  The
electrical current, when needed, is derived {\em from} the magnetic
field lines.  These practices appear at odds to the basic approach,
followed in elementary electrodynamics, of deriving the magnetic field
{\em from} a current distribution and treating magnetic field lines at
best as fictitious curiosities. However, physical laws such as
Amp{\`e}re's law\indexit{Amp{\`e}re's law} (without displacement
current [because velocities are assumed to be well below
relativistic]),
\begin{equation}
  \nabla\times\bvec = \frac{4\pi}{c}{\bf j},
    \label{eq:Ampere}
\end{equation}
do not attribute a causative nature to either side of the equality;
they simply state the equality of two quantities.  So either
approach to satisfying Eq.~(\ref{eq:Ampere}), beginning with either
${\bf j}$ or $\bvec$, is a valid one.

[The central role of $\vec{B}$ in space
physics has been furthered tremendously by the introduction of the
concept of the field line.]  A magnetic field line, sometimes
called\indexit{line of force: definition} a line of force, is a
space-curve $\rvec(\ell)$ which is everywhere tangent to the local
magnetic field vector, $\bvec(\xvec)$.  This description can be cast
as the differential equation\indexit{field line!equation}
\begin{equation}
  {{\rm d}\rvec\over {\rm d}\ell} ~=~ {\bvec[\rvec(\ell)]\over
  |\bvec[\rvec(\ell)]|},
    \label{eq:fl_eq}
\end{equation}
whose solution, starting from some initial point $\rvec(0)$, is a magnetic
field line.  [\ldots] 
A field line is a curve, and therefore has zero volume.
A {\em flux tube}\indexit{flux!tube}
may be \indexit{flux!tube}constructed by bundling together a group of
field lines.  The net flux, $\Phi$, of the tube is the integral
$\int{\bvec}\cdot {\rm d}{\bf a}$ over any surface pierced by
the entire tube.  Because $\nabla\cdot\bvec=0$, the tube must have the
same flux at every cross section.

The only way, in general, to find a field line is to integrate the
differential \eq~(\ref{eq:fl_eq}).
\label{sec:phys-sig}
\indexit{field line!applicability of concept}
A solution to the field line equation, Eq.~(\ref{eq:fl_eq}), can in
principle be found for a magnetic field at any instant.  What is not
immediately evident is why such a curve should be physically
significant, even if one concedes that the magnetic field itself is
significant.  There is, in fact, no single reason that field lines
will be significant under general circumstances --- this is why
students are often warned not to attribute undue importance to them.
There are, however, numerous circumstances arising in
space physics whereby a magnetic field line can achieve a degree of
[utility].  The following is a brief list of the most common, applicable
to a wide variety of plasma regimes from general \ref{sec:SPM}, to
the fluid regime \ref{sec:Thermal}, to MHD \ref{sec:Characteristics}, to
ideal MHD \ref{sec:Frozen-field-line}.

\begin{enumerate}
\item {\em General: single particle motion.}
\label{sec:SPM}
Subject to no other forces than a relatively stationary magnetic
field, [the guiding center of a charged particle will remain tied to a
single field line while the particle gyrates about that] according to
its mass and charge [(discussed in detail in
Sect.~\ref{sec:singleparticle})].  Drifts will displace the particle's
guiding center by several gyro-radii after it has traversed a length
comparable to the field's curvature radius or gradient scale.  Global
scales of space plasmas are typically much, much greater than the
gyro-radii of their electrons, and to a lesser extent of their heavier
ions (the Earth's geomagnetic ring current is a counterexample to
this[; see Activity~\ref{act:rgi}]). Waves in the field may scatter
particles (important in, {\em e.g.,} the Earth's radiation belts, [and for
solar and galactic cosmic rays propagating through the heliosphere,
see Ch.~\ref{ch:conversion}]), but this too is generally
unimportant. Field lines therefore serve as excellent approximations
of the electron orbits. [\ldots]

\item {\em Fluid regime: thermal conductivity and solar
coronal loops:} \label{sec:Thermal}
\indexit{thermal conduction!coronal loop}
In a diffuse, high-temperature plasma, thermal energy is conducted
principally by electrons.  When electrons are strongly magnetized
({\em i.e.,}  the cyclotron frequency is much greater than the collision
frequency) their orbits will follow field lines over long distances
between collisions at which point they scatter a perpendicular
distance no greater than a single gyro-radius.  The huge disparity
between parallel and perpendicular scattering distances makes thermal
conductivity highly anisotropic.  Consequently,
heat is conducted parallel to the magnetic field far more
readily than perpendicular to the field.

Due to this anisotropic conductivity, heat deposited somewhere in a
plasma is rapidly and efficiently conducted to all points on the same
field line, at least while collision frequencies remain relatively
low.  [In the coronal setting, for example, the] \indexit{plasma!$\beta$}plasma $\beta$ is
also generally low, so plasma flows are mechanically confined by the
field.  This means that a bundle of field lines will behave as a
one-dimensional autonomous atmosphere, at least as long as
reconnection is relatively unimportant.  [\ldots]

\item {\em MHD: Alfv\'en wave propagation:}
\label{sec:Characteristics}
Low-frequency waves in a magnetized plasma [(see
Section~\ref{sec:mhdwaves})]
comprise three branches:
slow magnetosonic, fast magnetosonic and shear Alfv\'en waves.  The
group velocity of the shear Alfv\'en wave is exactly parallel to the
local magnetic field.  In the limit of very short wavelengths, any
small localized disturbance will therefore propagate along a path
following a magnetic field line.  This means that a given field line
will 'learn' of perturbations anywhere along itself at the Alfv\'en
speed.  In this sense the magnetic field line has a dynamical
integrity similar to that of a piece of string.  Indeed, it is common
to derive the Alfv{\'e}n speed intuitively using the analogy to a
string under tension. [\ldots] 

\item {\em Ideal MHD: frozen-in field lines:}
  \indexit{frozen-in field!theorem}\label{sec:Frozen-field-line} [\ldots] At its
  simplest, the frozen-in-field-line theorem states that if two fluid
  elements lie on a common field line at one time, then they lie on a
  common field line at all times past and future.  This follows
  directly from the ideal induction equation, ([Eq.~(\ref{induction})
  with $\eta_{\rm \null }\equiv 0$]), and from the fact that fluid elements move at
  the same velocity $\vec{v}$ that appears in it.''

  The mathematics of ideal
  MHD is such \ors[I:4.1] ``that differentiation along a field line is
  interchangeable with differentiation along a flow trajectory.  From
  this it follows that a field line linking two fluid elements can be
  traced either before or after following the flow of those elements.
  That is a restatement of the [frozen-in-field-line] theorem
  introduced above.  One can thereafter imagine 'labeling' all the
  fluid elements along a given field line and then following those
  fluid elements as they move at their own velocities, $\vvec$.  These
  material elements, which are manifestly real, will trace out a
  single field line at all times, so that [the field line is a useful
  concept in thinking about plasma motions. Wherever
  $\eta_{\rm \null }\ne 0$ in Eq.~(\ref{induction}) field lines lose their
  nature as coherent entities; more on this below where we discuss
  reconnection.]''
\end{enumerate}

Field lines and flux tubes have taken on a remarkable degree of
utility in the thinking of many working in the various branches of
heliophysics. \ors[I:2.5] ``The motion of plasma along the magnetic
field does not stress the field and incurs no dynamic back-reaction on
the plasma through the action of the Lorentz force. Magnetic field
lines therefore serve as conduits for moving energy, mass, momentum
and energetic particles from point to point in the
heliosphere. Heliophysics accordingly focuses on the magnetic
connectivity of the Earth to the Sun, of the magnetotail to the polar
caps, of the Io plasma torus to the Jovian magnetic field, and so
forth. Magnetic field lines are truly the interstate highway system,
the Autobahn network, the autostrada web of the
heliosphere.''\sactivity{$\circledS$ {\em Show:} In solar physics, the simple
  concept of flux tubes are commonly used as an approximation of the
  state of the magnetic field in near-photospheric layers: embedded in
  a field-free atmosphere is a bundle of field separated from its
  surroundings by a thin current envelope. (a) Assuming an ideal plasma
  without flows, show that the atmosphere within a thin flux tube
  ({\em i.e.}, one that is small compared to the scale of variations
  of condition along the tube) is in
  hydrostatic equilibrium regardless of the path of the flux tube
  through the atmosphere. (b) Show how pressure balance (incorporating
  both gas and magnetic components) determines the cross section of
  the tube. \mylabel{act:buoy}\solution{buoy}}

Field lines as true, persistent entities have their greatest utility
in ideal (non-resistive) MHD. But ideal MHD, in which field lines
always connect the same parcels of plasma, fails when magnetic
diffusivity becomes important in the MHD approximation, or when the
basic assumptions of MHD itself fail on the smallest time or length
scales. Then field is no longer 'frozen in' wherever that happens,
and the very concept of continuity of field lines in space and time
loses validity. Failure of the field-line
concept as it is discussed above is captured by the term 
\indexit{definition!reconnection}\indexit{reconnection!failure of
  frozen-in field condition}'reconnection.'
This term, widely used, turns out to be very loosely
defined. \ors[II:1] ``It can be used to refer to the changing connectivity in a
vacuum potential field as much as to the decoupling of particle
motions from the background magnetic field by any number of concepts,
ranging from inertia to wave-particle interactions, or from
resistivity to infinitesimal current sheets. It is thus as much a
culturally accepted term for something that we really do not
understand, as a descriptor of a well-understood consequence: we can
say that reconnection occurs whenever the approximation of frozen-in
flux fails.''

Non-ideal MHD sees
reconnection as a consequence of resistivity. \ors[I:5.2.2] ``To
determine a realistic resistivity for a collisionless plasma requires
consideration of the \indexit{Ohm's law!generalized}generalized Ohm's law.
For a fully ionized plasma it can be written as
\begin{equation}\label{eq:gol}
  \vec{E} =
  - \overset{\tc{i}}{\frac{1}{c}\vec{v} \times \vec{B}}
  + \overset{\tc{ii}}{{ \vec{j} \over \sigma_{\rm e}}}
  + \overset{\tc{iii}}{{m_{\rm e} \over ne^{2}} \left[ {\partial \vec{j} \over \partial t}
            + {\bf \nabla} \cdot (\vec{v}\vec{j} +\vec{j}\vec{v}) \right]}
  + \overset{\tc{iv}}{{{\vec{j} \times \vec{B}} \over nec}}
  - \overset{\tc{v}}{{{\bf \nabla} \cdot \underline{\bf p}_{\rm e} \over ne}},
\end{equation}
where $\vec{v}\vec{j}$ and $\vec{j}\vec{v}$ are dyadic tensors [(with
components $v_n j_m$ and $v_m j_n$)] and $\underline{\bf p}_{\rm e}$
is the electron stress tensor.  Term \tc{i} on the right-hand side of
this equation is the convective electric field, while the term \tc{ii}
is the field associated with Ohmic dissipation caused by electron-ion
collisions.  The conductivity, $\sigma_{\rm e}$, is the inverse of the
electrical resistivity, $\eta_{\rm \null }$.  The next group of terms
\tc{iii} describes the effects of electron inertia [(which is ignored
in Eq.~(\ref{ohm-general}) as another approximation of Ohm's law,
while the latter describes also simplifies pressure by assuming it to
be isotropic). The next term, \tc{iv},] is the\indexit{Hall!effect}
Hall effect.  Ion inertia is negligible because the large mass of the
ions means that they do not contribute significantly to a change in
the current density.  Finally, the term \tc{v} includes the electron
gyro-viscosity, which is considered by many to be important at [any
point where the magnetic field vanishes, {\em i.e.,}  at a] magnetic null.
For a partially ionized plasma, collisions between charged particles
and neutrals lead to additional terms associated with
ambipolar\indexit{ambipolar!diffusion} diffusion.

Although all of the terms on the right-hand side of the generalized
Ohm's law, other than the first, allow field lines to slip through the
plasma, \indexit{reconnection!terms in generalized Ohm's law}they
do not all produce dissipation.  For example, the
inertial terms in \tc{iii} do not cause the entropy of the plasma to increase.
Thus, even though one may speak of inertial effects as creating an
effective resistivity, this\indexit{resistivity and dissipation}
resistivity does not necessarily lead to dissipation.

Which terms are important in a particular situation depends not only
upon the plasma parameters, but also upon the length and time scales
for variations of these parameters.  For magnetic reconnection, we
normally want to know which non-ideal terms are likely to be
significant within the current sheet where the frozen-flux condition
is violated. Because each non-ideal ({\em i.e.,}  diffusion) term in the
generalized Ohm's law contains either a spatial or temporal gradient,
we can estimate the significance of any particular term by computing
the gradient scale-length, $L_{\rm t}$, required to make the term as
large as the value of the convective electric field,
$\frac{1}{c}\vec{v} \times \vec{B}$, outside the diffusion region.

Consider, for example, the three inertial components of term \tc{iii}.
If we assume that ${\bf \nabla}
\approx 1/L_{\rm t}, \, |\vec{j}| \approx (c/4\pi) B_{\rm t}/ L_{\rm t}$ and
$\partial /
\partial t \approx v_{\rm t}/L_{\rm t}$, say, where $L_{\rm t}$ is a typical length-scale and $v_{\rm t}$ a typical velocity, then these
three components of \tc{iii} will be of the same order as the convective electric field if
\begin{eqnarray}
{c m_{\rm e} \over 4\pi ne^{2}} {v_{\rm t}B_{\rm t} \over L_{\rm t}^{2}}
\approx {V_{\rm t}B_{\rm t} \over c}.
\end{eqnarray}
In other words, in order for the inertial terms to be important in a current sheet,
its thickness $\ell_{\rm inertia}$ should be
\begin{eqnarray}
\ell_{\rm inertia} \approx \left( {c^2 m_{\rm e} \over 4\pi ne^{2}} \right)^{1/2} \approx
\lambda_{\rm e} ,
\end{eqnarray}
 where
\begin{eqnarray}
\lambda_{\rm e} = {c \over \omega_{\rm pe}} = \left( {c^2 m_{\rm e} \over
    4\pi n e^{2}} \right)^{1/2} = 5.3 \times 10^{5} \, n^{-1/2}
\end{eqnarray}
is the electron-inertial length\indexit{electron-inertial length}
or\indexit{electron skin depth} skin-depth [(which characterizes the
depth in a plasma into which electromagnetic radiation can
penetrate)], $c$ is the speed of light
and 
\begin{equation}\label{eq:plasmafrequency}
\omega_{\rm pe} =(4\pi n e^2 / m_e)^{1/2} =5.6\,10^4 n_{\rm e}^{1/2}
\end{equation}
is the \indexit{electron plasma frequency}electron plasma
frequency.

Similarly, for the Hall term \tc{iv}
\begin{eqnarray}
{B_{\rm t}^{2} \over 4\pi ne L_{\rm t}} \approx {V_{\rm t}B_{\rm t} \over
  c}
\end{eqnarray}
or 
\begin{equation}
\ell_{\rm Hall} \approx {c \over M_{\rm A}} \left( {\mu m_{\rm p} \over
4\pi ne^{2}} \right)^{1/2} \approx { \lambda_{\rm i} \over M_{\rm A}} ,
\end{equation}
 where
\begin{equation}\label{eq:ioninertial}
\lambda_{\rm i} = {c \over \omega_{\rm pi}} = \left( {\mu c^2
    m_{\rm p} \over 4 \pi n e^{2} }
\right)^{1/2} = 2.3 \times 10^{7} \left ({\mu \over  n} \right)^{1/2}
\end{equation}
%https://ipfs.io/ipfs/QmXoypizjW3WknFiJnKLwHCnL72vedxjQkDDP1mXWo6uco/wiki/Plasma_parameters.html
is the ion-inertial length\indexit{ion-inertial length}
 or\indexit{ion skin depth} skin-depth [(below which ions decouple
 from electrons, and the magnetic field may no longer be frozen into
 the plasma overall but instead into the 
 electron fluid), and $\mu=m_{\rm i}/m_{\rm p}$]. The Alfv{\'e}n Mach
number equals [$ M_{\rm A} = V_{\rm t}/v_{\rm A},$] and $\omega_{\rm pi}
=(4\pi n e^2 / m_i)^{1/2}$ the ion
plasma\indexit{ion plasma frequency} frequency, and $v_{\rm A}$ the
Alfv\'{e}n speed [(see Eq.~\ref{eq:10.12} and Table~\ref{tab:dimensionlessnumbers})].

For the electron-stress term \tc{v} we can write
\begin{eqnarray}
{nkT_{\rm t} \over ne \, L_{\rm t}} \approx {V_{\rm t}B_{\rm t}
  \over c} 
\end{eqnarray}
if we assume $|{\bf p}_{\rm e}| \approx nkT_{\rm e}$ and $T_{\rm e} \approx T_{\rm i} \approx
T$.  Solving for $L_{\rm t}$ leads to
\begin{equation}
\ell_{\rm stress} \approx { \beta^{1/2} \over M_{\rm A}} r_{\rm gi} ,
\end{equation}
where [the plasma $\beta$ is given by Eq.~(\ref{eq:betadef}) and the ion-gyro radius
for the average thermal velocity ($v_{T{\rm i}}$) equals $r_{\rm gi}=
(kTm_{\rm i})^{1/2}c/eB$; 
%v_{T{\rm i}}/\omega_{c{\rm i}} = (kT/m_{\rm i})^{1/2} (m_{\rm i}c / eB)=
see also Table~\ref{tab:dimensionlessnumbers}.]

Finally, for the collision term \tc{ii}, $\vec{j} / \sigma_{\rm e}$,
\begin{eqnarray}
{ cB_{\rm t} \over 4\pi \sigma_{\rm e} L_{\rm t}} \approx {M_{\rm A} v_{\rm A}
  B_{\rm t} \over c}. 
\end{eqnarray}
[where $\sigma_{\rm e}^{-1}$ is] also the magnetic
diffusivity, $\eta_{\rm \null }$.  Using Spitzer's formula for the collisional
resistivity, $\eta_{\rm \null }$, of a plasma (see \ors[I:3]) we obtain
\begin{equation}
\eta_{\rm \null } = {(k m_{\rm e} T_{\rm e})^{1/2} \over ne^{2} \lambda_{\rm mfp}} ,
\end{equation}
where
\begin{equation}
\lambda_{\rm mfp} = \frac{3}{4\pi^{1/2}} {(k T_{\rm e})^{2} \over ne^{4} \: {\rm
ln} \Lambda} = 1.1 \times 10^{5} {T_{\rm e}^{2} \over n \: {\rm ln} \Lambda}
\end{equation}
is the mean-free path \indexit{mean free path!electron-ion}for electron-ion collisions.
Combining these expressions with those for the
electron and ion inertial lengths we obtain [an estimate for the
length scale below which effects of collisions become important to
field diffusion:] 
\begin{equation}
\ell_{\rm collisions} \approx {\beta^{1/2} \over M_{\rm A}} {\lambda_{\rm e} \lambda_{\rm i} \over
\lambda_{\rm mfp}} .
\end{equation}
Note that the length-scale, $\ell_{\rm collision}$, of the spatial
variations required to achieve significant field-line diffusion is
inversely proportional to the mean-free path, $\lambda_{\rm mfp}$.  As
$\lambda_{\rm mfp}$ increases, the diffusion caused by collisions
becomes less effective, and increasingly sharper gradients are
required to maintain the size of the dissipation term, ${\bf j} /
\sigma_{\rm e}$.

\begin{table}
\begin{center}
  \caption[Plasma parameters in different
  environments.]{\label{tab:5.1} \label{tab:5.2} Comparison of
    order-of-magnitude plasma
    \indexit{plasma!properties in different settings}parameters in different environments
    (cgs-Gaussian units --~{\em i.e.,}  length scales in cm, $n$ in cm$^{-3}$,
    $T$ in K, $B$ in $Gauss$, electric fields in
    statV\,cm$^{-1}$). [Modified after I:5, merging two tables in SI units]}
\begin{tabular}{lllll}
\hline \hline 
Parameter & Laboratory & Terrestrial & Solar & Solar \\
& experiment$^{1}$ & magneto- & corona$^{3}$ & interior$^{4}$ \\
&  & sphere$^{2}$ & & \\
\hline
region scale $L_{\rm s}$ & $10$ & $10^{9}$ & $10^{10}$& $10^{9}$ \\
density $n_{\rm t}$ & $10^{14}$ & $10^{-1}$ & $10^{9}$ & $10^{23}$ \\
temperature $T_{\rm t}$ & $10^{5}$ & $10^{7}$ & $10^{6}$ & $10^{6}$ \\
field strength $B_{\rm t}$ & $10^{3}$ & $10^{-4}$ & $10^{2}$ & $10^{5}$ \\
\hline
Debye length $\lambda_{\rm D}$ & $10^{-4}$& $10^{5}$ & $10^{-1}$ & $10^{-8}$ \\
ion gyro radius\ $r_{\rm gi}$ & $10^{-1}$ & $10^{7}$ & $10$ & $10^{-2}$ \\
ion inertial length $\lambda_{\rm i}$ & $1$ & $10^{8}$ & $10^{3}$ & $10^{-4}$ \\
Coulomb logarithm $\ln{(\Lambda)}$ & 11 & 33 &  19 & 3 \\
coll.\ mean-free path $\lambda_{\rm mfp}$ $^5$& $1$ & $10^{18}$ & $10^{6}$ & $10^{-7}$ \\
\hline
$\ell_{\rm inertia} (\lambda_{\rm e})$ & $10^{-2}$ & $10^{6}$ & $10$ & $10^{-6}$ \\
$\ell_{\rm Hall} (\lambda_{\rm i})$ & $1$ & $10^{8}$ & $10^{3}$ & $10^{-4}$ \\
$\ell_{\rm stress}$ & $10^{-1} $ & $10^{7}$ & $10^{-1}$& $1$ \\
$\ell_{\rm collision}$ & $10^{-2}$ & $10^{-5}$ & $10^{-5}$ & $10^{-1}$ \\
\hline
plasma $\beta$ & $10^{-2} $ & $10^{-1} $ & $10^{-4}$ & $10^{4}$ \\
Lundquist no. $L_{\rm u} (\approx \Rm)$ & $10^{3}$ & $10^{14}$ & $10^{14}$ & $10^{10}$ \\
Dreicer field $E_{\rm D}$ & $10^{-1}$ & $10^{-17}$ & $10^{-6}$ & $10^{7}$ \\
$E_{\rm A} (=v_{\rm A} B/c)$ & $1$ & $10^{-6}$ & $10^{1}$ & $1$ \\
$E_{\rm SP} (=E_{\rm A} / \surd \Rm)$ & $10^{-2}$ & $10^{-13}$ & $10^{-7}$ & $10^{-6}$ \\
\hline \hline
%\end{tabular}
%\vspace{1cm}
%\begin{table}
%\begin{center}
%\caption[Diffusion lengths from the generalized Ohm's
%law.]{\label{tab:5.2}Diffusion lengths (cm) from the
%generalized Ohm's law. (Modified after I:5 which is in MKS units,
%but expressed in cgs-gaussian units here)}
%\end{center}
%\begin{tabular}{lllll}
%\hline \hline
% Characteristic & Laboratory & Terrestrial & Solar & Solar \\
%length & experiments$^{1}$ & magnetosphere$^{2}$ & corona$^{3}$ & interior$^{4}$ \\
%\hline
%$L_{\rm inertia} (\lambda_{\rm e})$ & $10^{-2}$ & $10^{6}$ & $10$ & $10^{-6}$ \\
%$L_{\rm Hall} (\lambda_{\rm i})$ & $1$ & $10^{8}$ & $10^{3}$ & $10^{-4}$ \\
%$L_{\rm stress}$ & $10^{-1} $ & $10^{7}$ & $10^{-1}$& $1$ \\
%$L_{\rm collision}$ & $10^{-2}$ & $10^{-5}$ & $10^{-5}$ & $10^{-1}$ \\
%\hline \hline
\end{tabular}
\end{center}

%\vspace{1cm}

{\em \small \noindent $^{1}$ The Magnetic Reconnection eXperiment (MRX) at
Princeton Plasma Physics Laboratory; $^{2}$ plasma sheet; $^{3}$ above
a solar active region; $^{4}$ at the base of the solar convection
envelope [at a depth of about 200,000\,km around which many consider
primary dynamo mechanisms to operate; $^5$ note that this is a purely
collisional mean-free path, ignoring other couplings that may occur
through the magnetic field].}
%\end{table}
%\noindent $^{1}$ MRX at Princeton Plasma Physics Laboratory; $^{2}$ plasma sheet; $^{3}$ above a solar active region; $^{4}$ at the base of the solar convection zone
\end{table} [Table~\ref{tab:5.1} lists] various plasma parameters
along with the characteristic scale-lengths for four different regions
where \indexit{reconnection!important length scales}reconnection is
thought to occur.  The parameter $L_{\rm s}$ is the global
(system-level) scale-size of the region, and the fundamental
quantities from which all other parameters are derived are the density
$n$, temperature $T$, and magnetic field $B$.  For convenience, we
assume that the Alfv\'{e}n Mach number $M_{\rm A}$ is unity and that
the electron and ion temperatures are roughly equal.  The most extreme
plasma environments listed in Table~\ref{tab:5.1} occur in the
magnetosphere, which is completely collisionless, and in the solar
interior which is highly collisional.

In addition to the parameters discussed above, Table~\ref{tab:5.1}
also lists the\indexit{Debye length} value of the {\em Debye length}
[whose expression is shown in Table~\ref{tab:dimensionlessnumbers}.] 
The number of particles within a Debye sphere ({\em i.e.,} $4 \pi
n \lambda_{\rm D}^{3}/3)$ must be larger than unity in order for
the generalized Ohm's\indexit{Ohm's
law!generalized} law to hold.  Otherwise, the collective behavior
which characterizes a plasma breaks down.  The number of particles
in a Debye sphere for the environments shown in
Table~\ref{tab:5.1} ranges from $10^{14}$ for the magnetosphere to
only about four for the solar interior at the base of the
convection zone.  Also shown in the table is the\indexit{Lundquist
number} Lundquist number, $L_{\rm u}$, which is the same as the
magnetic Reynolds number, $\Rm$ [introduced in
Eq.~(\ref{eq:reynolds})], 
when the flow and
Alfv\'{e}n speeds are the same.  For a collisional plasma the
Lundquist number based on $L_{\rm s}$ can be expressed as
\begin{equation}
L_{\rm u} = {v_{\rm A} \over v_{d}} = {L_{\rm s} v_{\rm A} \over \eta_{\rm \null }}
= 2. \times 10^{8} {L_{\rm s} T_{\rm e}^{3/2} B_{\rm t} \over
(\mu n)^{1/2} {\rm ln} (\Lambda)} 
\end{equation}
[\ldots] In the expression on
the right, $\eta_{\rm \null }$ has been replaced by Spitzer's formula for the electrical resistivity of collisional plasma.

The \indexit{plasma!characteristic scale-lengths}characteristic
scale-lengths in Table~\ref{tab:5.1} provide an
indication of which terms in the generalized Ohm's law of Eq.~(\ref{eq:gol}) are likely to
be important for reconnecting current sheets.  As with MHD shocks and
turbulence, the large-scale dynamics of the flow cause the current
sheet to thin until it reaches a length-scale where field-line
diffusion is effective.  Thus, in principle, the term with the largest
characteristic length-scale in Table~\ref{tab:5.2} is the one that
will be most important.  Because the Hall term has the largest length in
every environment except the solar interior, one might conclude that
it is generally the most important. However, this conclusion does not
take into consideration the fact that the Hall term tends to zero in
the region of a magnetic null point or sheet. The Hall term on its own
does not contribute directly to reconnection, since it freezes the
magnetic field to the electron flow.  [\ldots] An excessively small
scale does indicate that any process associated with that term is
unlikely to be important.  Therefore, on this basis, we can conclude
that collisional diffusion is not important in the terrestrial
magnetosphere or the solar corona, and that the electron-inertial
terms and the Hall term are not important in the solar
interior. [\ldots\ On the other hand, if a term is not associated with
an obviously 'excessively small' scale, it is difficult to know whether a
particular term is really as important as suggested by its relative
length scale; evaluating such cases] requires a complete analysis of
the kinetic dynamics, which is a rather formidable task.

Although the collision length-scale, $\ell_{\rm collision}$, is equally
small in both the magnetosphere and the corona, the general
importance of collisions for these two regions is quite different.
In the magnetosphere the collision-mean-free path,
$\lambda_{\rm mfp}$, is nine orders of magnitude larger than the
global scale-size, $L_{\rm s}$, but in the corona it is four orders of
magnitude smaller than the global scale.  Thus, we can be
confident that collisional transport theory applies to large-scale
structures in the corona even though it is not applicable within
thin current sheets or dissipation layers.  By contrast, in the
magnetosphere, collisions are so few that collisional transport
theory does not apply at any scale.

%Dreicer field etc. in cgs, see, e.g.:
%http://solar.physics.montana.edu/dana/solar_numbers.pdf
Another important issue concerning the applicability of
collisional theory is the strength of the electric field in a
frame moving with the plasma.  If this field exceeds the {\em
Dreicer electric field} defined\indexit{Dreicer electric field} by
\begin{equation}\label{eq:edreicer}
E_{\rm D} = {e \, {\rm ln} (\Lambda) \over \lambda_{\rm D}^{2}} = \frac{4 \pi e^3}{k}
\, {{\rm ln} (\Lambda) \, n \over T_{\rm e}} = 
10^{-11} {n \, {\rm ln} (\Lambda) \over T_{\rm e}} ,
\end{equation}
runaway \indexit{runaway acceleration}acceleration of electrons will
occur.  The most likely location for the production of runaway
electrons in a reconnection process is in a thin current sheet that
forms at the null point.  This field could be as large as the
convective electric field based on the Alfv\'{e}n speed, that is
\begin{eqnarray}\label{eq:ealfven}
E_{\rm A} = \frac{1}{c} v_{\rm A} B_{\rm t}= 7.2  {B_{\rm t}^{2} \over (\mu n)^{1/2}} , 
\end{eqnarray}
or as low as the Sweet-Parker electric field
\begin{eqnarray}\label{eq:esweetparker}
E_{\rm SP} = {E_{\rm A} \over \Rm^{1/2}} , 
\end{eqnarray}
where $\Rm$ is the magnetic Reynolds number based on the inflow
Alfv\'{e}n speed ({\em i.e.,} the inflow Lundquist number).  As
shown in Table~\ref{tab:5.1}, the Dreicer field in the magnetosphere is
much smaller than $E_{\rm A}$ or $E_{\rm SP}$, so runaway electrons will
always be generated by reconnection there.  On the other hand, in
the solar interior the Dreicer field is so large that runaway
electrons never occur.  In the intermediate regimes of the
laboratory and the solar corona, the Dreicer field lines between
$E_{\rm A}$ and $E_{\rm SP}$, so perhaps runaway electrons are only
produced when very fast reconnection occurs.

Even in completely collisionless environments like the Earth's
magnetosphere, it is still sometimes possible to express the relation
between electric field and current density in terms of an
anomalous\indexit{anomalous resistivity} resistivity.  For example,
[\ldots] the electron inertial terms \tc{iii} in the
generalized Ohm's law of Eq.~(\ref{eq:gol}) lead to an anomalous resistivity
\begin{eqnarray}
{1 \over \sigma_{\rm e}^{*}}  = { \pi B_\perp \over 2ne},
\end{eqnarray}
where $B_\perp$ is the field normal to the current sheet. This
resistivity is derived solely from a consideration of the particle
orbits, and in the magnetotail current sheet it may be larger than
any anomalous resistivity due to wave-particle interactions.  A
typical example of the latter is the anomalous resistivity due to
ion-acoustic waves''.

For some discussion of reconnection in two and three dimensions in the
Heliophysics books, see I:5.3 and I:5.4. More on the effects of
reconnection follows in Chs.~\ref{ch:mhd} and~\ref{ch:conversion}.

\section{A few notes about conditions}
\subsection{Solar atmosphere {\em vs.}\  terrestrial magnetosphere}
The scale lengths estimated for the importance of terms in Ohm's law
in Table~\ref{tab:5.1} are very much smaller than the scale of the
corona itself and even compared to any active region, but importantly
also very much smaller than the angular resolution achievable by
imaging instruments (currently about 1\,arcsec or $\sim 700$\,km
for space-based EUV imagers). Consequently, the scale on which
reconnection occurs in the corona is not observed, while the effects
of such reconnection become apparent in the magnetic geometry and
plasma atmospheres on scales well above the reconnection itself.

In contrast, in the terrestrial magnetosphere all but the length
scale, $\ell_{\rm collision}$, on which collisional effects could
contribute significantly as a term in the generalized Ohm's law are
large enough that spacecraft can scan reconnection regions as they fly
through, while constellations of spacecraft can probe reconnection in
multiple dimensions.

Another significant distinction is that in the terrestrial
magnetosphere the ion-gyro radius (particularly for relatively heavy
and energetic particles) is not small compared to the scale on which
these particles probe the magnetic field. This is an important cause
behind what is known as the ring current (see
Sect.~\ref{sec:singleparticle}) that is largely carried precisely by
such particles. For solar conditions, in contrast, such effects of
particle gyration are not directly evident on any observable scale.

\subsection{Heliosphere}\label{sec:solwindenergy}
\indexit{heliosphere!characteristic energies}
\ors[I:11.3] ``Adopting typical solar wind values near Earth of $n_{\rm t}=5\,{\rm particles/cm}^3$ for density,
$v_{\rm t}=400$\,km/s solar wind speed and $B_{\rm t}=50$\,$\mu$G magnetic field strength (values
consistent with Table~[\ref{tab:wind-stats}]) we can evaluate the expected
energy density of the solar wind, which can be broken down into three
components: flow, magnetic and thermal.  [\ldots] The flow energy density is
estimated to be
\begin{equation}
e_{v,\odot} \equiv \left( {\frac{1}{2}\rho v^2 } \right)_{\rm sw}  \approx \frac{1}{2}m_{\rm p} nv_{\rm sw}^2  \approx 7 \times 10^{ - 9} \left( {\frac{{n({\rm cm}^{ - 3} )}}{5}} \right)\left( {\frac{{v({\rm km/s})}}{{400}}} \right)^2 {\rm erg/cm}^{3}.
\end{equation}
The energy density of the solar wind's magnetic
field is
\begin{equation}\label{eq:helwind}
e_{B,\odot} \equiv \left\langle {\frac{{B^2 }}{{8\pi}}} \right\rangle  \approx 1.
\times 10^{ - 10} \left( {\frac{{B(\mu {\rm G})}}{50}} \right)^2 
\approx 0.015 e_{v,\odot}  \,{\rm erg/cm}^{3},
\end{equation}
while the thermal energy density using values from Table~[\ref{tab:wind-stats}] is
%\begin{equation}
\begin{eqnarray}
e_{T,\odot} \equiv \left\langle {\frac{3}{2}nk(T_{\rm e}  + T_{\rm i} )} \right\rangle
& \approx & 2.5 \times 10^{ - 10} \left(\frac{{n({\rm cm}^{ - 3} )}}{6}\right)\left( {\frac{{T_{\rm e} (K)}}{{1.2 \times 10^5 }}
+ \frac{{T_{\rm i} (K)}}{{1.4 \times 10^5 }}} \right) 
\nonumber \\
& \approx & 0.03 e_{v,\odot}  \,{\rm erg/cm}^{3}
\end{eqnarray}
%\end{equation}
where $T_{\rm i,e}$ are the solar wind ion and electron
temperatures. Taking the [values from Table~\ref{tab:wind-stats}], the
above [These] estimates show that the bulk of the energy in the solar wind at
Earth is in the flow:'' $e_{v,\odot} \sim 30 e_{B,\odot} \sim 70
e_{T,\odot}$ \activity{{\em Show:} (a) Make comparisons of energy densities
  for the solar wind as in Sec.~\ref{sec:solwindenergy} at other
  bodies in the Solar System (using Table~\ref{tab:fran2}). Why
  comparisons of energy densities in planetary magnetic fields
  (Table~\ref{tab:fran2}) and in the surrounding solar wind are
  informative is discussed in Ch.~\ref{ch:flows}. (b) Why would you expect
  the flow energy density and the magnetic field energy density to be
  comparable at only a few solar radii from the Sun?}

%\indexit{planetary magnetospheres!table of properties}\indexit{IMF:
%  table of properties}\indexit{solar!wind!table of properties}

%\subsection{Planetary magnetospheres}
%Table~\ref{tab:fran3} summarizes properties of magnetic fields of
%magnetized planets and the moon Ganymede.

\clearpage

\chapter{{\bf Dynamos of Sun-like stars and Earth-like planets}}%4
\label{ch:dynamos}
{\narrower\narrower{
{\bf Chapter topics:}
\begin{itemize}
  \customitemize
\item Basic stellar structure (colors, masses, radii), and (cyclic) magnetism
\item Basic structure of the terrestrial planets and the contrast of
  the rotation/convection time scales (in the Reynolds number) with stars
\item Principle of dynamos: the conversion of mechanical energy into electromagnetic
energy through induction
\item Energy sources and fluid flows that enable stellar and planetary dynamos
\end{itemize}

\noindent{\bf Key concepts:}
\begin{itemize}
  \customitemize
\item Rotation (through Coriolis forces) introduces a source term for magnetic field in the
  induction equation 
\item The mean-field dynamo approximation relies on a separation of
  scales not strictly supported by the convective spectrum
\item Back reaction of the magnetic field on the flow limits dynamo
  strengths 
\item An 'interface dynamo' relies on storage of field below the
  convective envelope, and a 'flux-transport dynamo' involves global meridional circulation
\end{itemize}

}}

\section{Dynamo settings}
Stellar and planetary \indexit{dynamo!settings}dynamos thrive wherever
sufficiently vigorous flows of a conducting medium transport
substantial thermal energy in an adequately spinning body. The energy
transported has to come from a reservoir that may date back to the
formation of the body (in planets or very young stars) or may have its
origin in nuclear fusion (in stars) or in solidification --~the latter
often accompanied by chemical separation~-- or nuclear fission (in
terrestrial planets). The flows that transport the energy may be
dominated by \indexit{Coriolis force}Coriolis forces (in planets where flows are slow compared
to spin rates) or by \indexit{stratification}stratification (including chemical gradients in
planets, while in stars pressure gradients of the compressible medium
limit how far matter can efficiently rise before overturning). The
amount of energy transported is regulated by the source in the deep
interior as well as by the sink at the top of the dynamo region.  In
Sun-like stars that sink is the stellar surface, and the properties of
radiative transfer through these surface layers are important in
determining the internal structure of the entire star as it balances
the energy produced by nuclear fusion with its luminosity. In a planet
like Earth, the energy transport in the dynamo region of the core is
determined to a large extent by the convective motions in the
enveloping mantle that transport heat to where it is ultimately lost
through the surface.

\ors[I:3.3] ``The formal difference between the type of dynamos that we
are interested in here and the self-excited dynamos in power plants is
the homogeneous distribution of conductivity (that would lead to a
short-circuit situation) that does not put any constraints on electric
currents (electric wires could be considered as special cases of a
highly inhomogeneous conductivity distribution). For this reason these
dynamos are also called\indexit{dynamo!homogeneous fluid} homogeneous
fluid dynamos.''

\begin{figure}[t]
\begin{center}
%\epsfxsize=\hsize
%\epsfbox{figures/stellarstructure.eps}
\includegraphics[width=\textwidth]{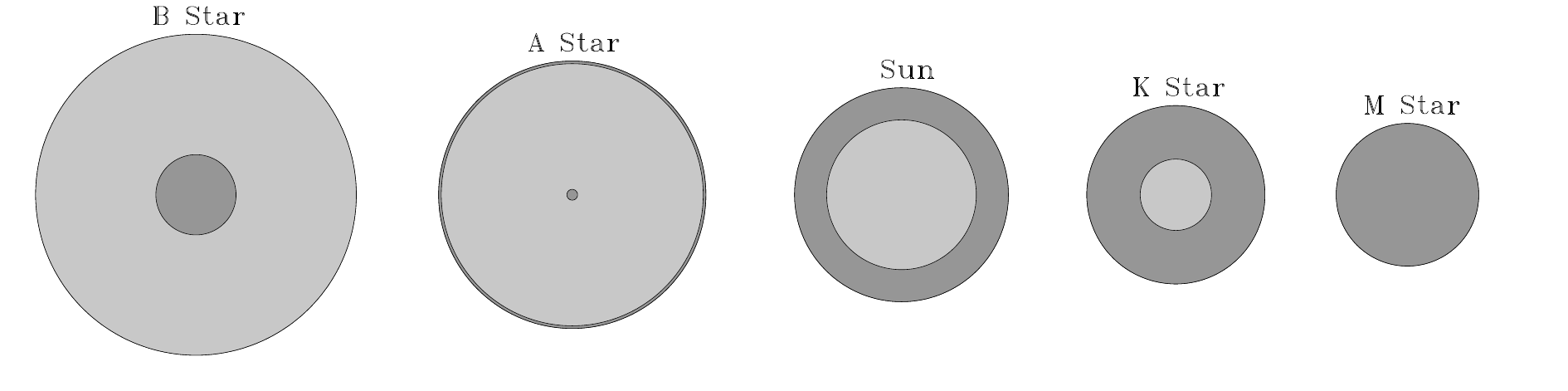}
\end{center}
\caption[Radiative
and convective internal
structure of main-sequence stars.]{
\label{fig:msstruct}
Schematic representation of \indexit{star!internal structure}the radiative (light grey)
and convective (dark grey) internal
structure of main-sequence stars. The thickness of the outer
convection zone for the A-star is here greatly exaggerated;
drawn to scale it would be thinner than the black circle delineating
the stellar surface on this drawing. Relative stellar sizes are
also not to scale: a B0~{\sc V} star has a radius of $\sim 7.5\,R_\odot$,
and and M0~{\sc V} star has a radius of $\sim 0.6\,R_\odot$, {\em i.e.,} 
12 times smaller. [Fig.~III:2.10]}
\end{figure}

\begin{figure}[h!]
  \centering
\begin{minipage}[t]{9cm}
%\centerline{\psfig{figure=figures/acthrd.eps,width=7.5cm}}
\centerline{\includegraphics[width=9.cm]{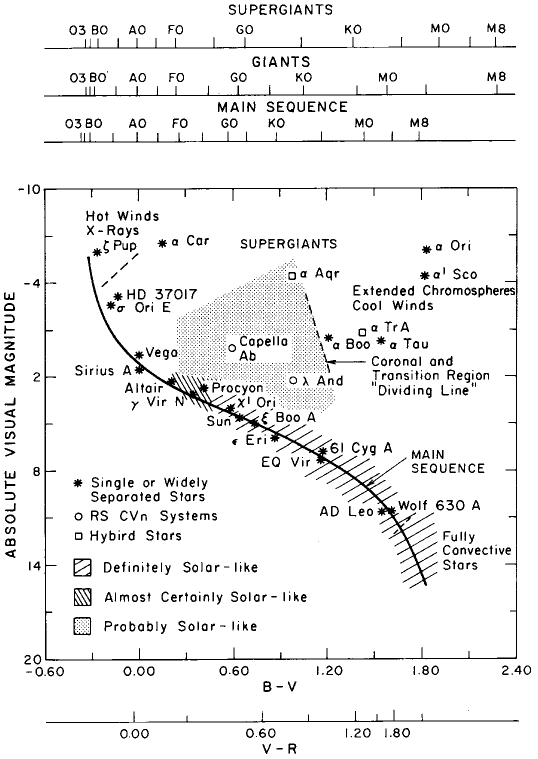}}
\end{minipage}
\raisebox{8.5cm}{
\begin{minipage}[t]{4.8cm}
\begin{tabular}{llcr}
\multicolumn{4}{c}{For main-sequence stars}\\
\hline
Sp.   &$M/M_\odot$ & $T_{\rm eff}$ & $BC$ \\
type & & & \\
\hline
$A$0&3,2 & 9600& -0.25\\
$F$0&1.7 & 7300& 0.02\\
$G$0&1.1 & 5900& -0.07\\
$K$0&0.78 & 5000& -0.19\\
$K$5&0.69 & 4400& -0.62\\
$M$0&0.60 & 3900& -1.17\\
\hline
\end{tabular}
\vskip 0.5\baselineskip
{\footnotesize \em Absolute bolometric magnitude $M_{\rm bol, \ast}$ (a logarithm of
the brightness of a star normalized to a standard distance), absolute
visual magnitude $M_V$ and stellar luminosity $L_\ast$ expressed in
solar units ($L_\odot$) are related through: $${L_\ast}/{L_\odot}
= 10^{0.4(M_{\rm bol, \odot}- M_{\rm bol, \ast})},$$ while $M_{\rm
  bol}=M_V+BC$, where $BC$ is the 'bolometric correction' that
corrects the brightness of the star in the $V$ ('visual' filter)
bandpass to the bolometric brightness.
%http://www-astro.physics.ox.ac.uk/~podsi/lec_b3_1_b.pdf
}\end{minipage}}

\caption[Activity across the Hertzsprung-Russell diagram.]{A 
Hertzsprung-Russell \indexit{Hertzsprung-Russell diagram} diagram showing
stars with \indexit{star!properties}substantial
magnetic activity in shaded or hatched domains,
which are distinguished in groups of
solar-likeness as indicated in the legend.
The main sequence\indexit{definition!star!main-sequence or dwarf}
where stars \indexit{main sequence}spend most of their lifetime fusing hydrogen into helium
in their cores is indicated by a solid curve; well above that 
lies the domain of the supergiant stars\indexit{definition!star!supergiant},
with the giant star domain in between.\indexit{definition!star!giant}
Also indicated is the region where massive winds occur
and where hot coronal plasma appears to be absent. Some frequently studied
stars (both magnetically active and nonactive) are identified by name.
The axes above the main panel show the spectral types (see footnote~\ref{note:spectraltype}) for supergiant,
giant, and main-sequence stars for the corresponding spectral
color index $B-V$ or corresponding $V-R$ index.
\label{fig:acthrd} [Fig.~III:2.8, with an added information panel on
the right;
\href{https://ui.adsabs.harvard.edu/abs/1985SoPh..100..333L/abstract}{figure
  source: \citet{1985SoPh..100..333L}}.]}
\end{figure}

In stars, \ors[III:5.1] ``[t]hermonuclear fusion in their cores
converts matter into thermal energy and electromagnetic radiation
which, in the Sun, is transported outward via the diffusion of
photons.  In the solar envelope, the \indexit{stellar!convection}plasma becomes more opaque as the temperature drops,
which inhibits radiative diffusion and steepens the temperature
gradient relative to the adiabatic temperature gradient.  The
stratification \indexit{stratification}soon becomes superadiabatic [(i.e., has a temperature
gradient steeper than for adiabatic conditions in which no energy
enters or leaves a volume of gas)] and thermal convection [gradually]
takes over as the primary mechanism for transporting energy to the
solar photosphere where it is radiated into space.\activity{{\em
    Advanced/Group:'} The transition from radiative diffusion to
  convective enthalpy transport in the deep convective envelope is
  gradual: the fraction of total energy carried as a diffusive
  radiative flux gradually drops while that of the enthalpy flux
  (total heat content: internal energy plus work term) smoothly
  increases, making convection the dominant transport about 35,000\,km
  above the bottom of the convective envelope, or roughly after a
  single pressure scale height (see Sect.~\ref{sec:fluid}). (a) Can
  you think of other terms that would be involved in the energy
  transport equation in a stellar convective envelope? A fair idea of
  the answer, along with a quantitative comparison of the relative
  importance of the processes involved in carrying energy through the
  convective envelope, can be found, for example, in
  \href{https://ui.adsabs.harvard.edu/abs/2004ApJ...614.1073B/abstract}{this
    analysis by \citet{2004ApJ...614.1073B}}, in particular their
  Fig.~3 (note that they show transport by convection that is resolved
  by their model and by (parameterized) unresolved
  --~'subgrid-scale'~-- convection). (b) Argue why the dominant energy
  transport shifts from radiative, via large-scale enthalpy, to
  'unresolved eddy flux' going
  outward. \mylabel{act:convenergytransport}} [All stars with a mass
of somewhat above that of the Sun or less than that have such a
convective \indexit{convective envelope}envelope during their
'main-sequence' (equilibrated hydrogen-fusing) phase (see
Figs.~\ref{fig:msstruct} and~\ref{fig:acthrd}); the least massive
stars are fully convective. All of these stars power a dynamo during
the longest-lived mature phase, and all stars do during their initial
birth phases and in the last phases of their lives, both of which are
short compared to the mature phase (Ch.~\ref{ch:evolvingstars}).
Stars cool enough to have a convective envelope reaching into their
surface layers are known as 'cool
stars'. \regfootnote{\mylabel{note:spectraltype} Astronomers
  characterize the properties of stars based on their spectrum. The
  overall shape gives an indication of the surface temperature, while
  details of spectral lines (generally in absorption, but some in
  emission) provide finer detail used in classification schemes. One
  such scheme frequently used is that of 'spectral type' in the
  Morgan-Keenan (MK) scheme: only after the classes were introduced
  was a monotonic mapping to temperature established, going from hot
  to cooler: $O$, $B$, $A$, $F$, $G$, $K$, $M$, $L$, and $T$ (with the
  last two fairly recent additions for very cool, very faint stars,
  with $T$ reaching the domain of 'brown dwarfs'). The letter is
  followed by a subclass from 0 to 9, and commonly an indicator of
  'luminosity class': a roman numeral indicative of the size of the
  star: I, II, III, IV, and V for supergiants, bright giants, giants,
  subgiants, and main-sequence or dwarf stars. The term 'main
  sequence' refers to a band in brightness-color diagrams, such as
  Fig.~\ref{fig:acthrd}, within which stars spend most of their lives,
  as long as they are steadily
  \indexit{HR diagram|see{Hertzsprung-Russell diagram}}fusing hydrogen into
  helium.}\activity{{\em Show:} Figure~\ref{fig:acthrd} is a
  brightness-color diagram \indexit{Hertzsprung-Russell diagram}(known as a Hertzsprung-Russell, or HR,
  diagram) using typical astronomical units: absolute visual magnitude
  $M_V$, which is a logarithmic measure of stellar brightness, and
  spectral color $B-V$, which is the logarithm of the ratio of two
  brightnesses measured in different color bands (often using
  logarithmic brightness $B$ and $V$, or less commonly $R$ for blue,
  visual, and red). The table in that figure maps spectral type (see
  footnote~\ref{note:spectraltype}), $B-V$, effective temperature
  $T_{\rm eff}$ and a correction factor $BC$ that relates visual and
  bolometric brightnesses (see equations below that table). Using this
  information, estimate stellar radii $R_\ast$ for Sirius~A,
  $\epsilon$~Eri, 61~Cyg~A, and AD~Leo, realizing that
  $L_\ast=(\sigma T_{\rm eff}^4) (4\pi R_\ast^2)$, with the
  Stefan-Boltzmann constant
  $\sigma=5.7\times 10^{-5}$\,erg/cm$^2$/sec/deg$^4$. Sketch a
  double-logarithmic $L$--$T_{\rm eff}$ version of the HR diagram and
  draw lines of constant radius in it. Then compare that to
  Fig.~\ref{figure:evolmodel}.}

The solar convection \indexit{dynamo!strength}zone occupies
approximately the outer 30\% of the Sun by radius.  It is here where
[a small fraction of the] internal energy of the plasma is converted
to kinetic energy and then [a small fraction of that] to magnetic
energy, aided by radiation and gravity.  Radiative heating [of the
bottom of the] convection zone and radiative cooling in the
photosphere maintain a superadiabatic temperature gradient that
sustains convective motions by means of buoyancy.  In a rotating star,
convection transports momentum as well as energy, establishing
shearing flows and global circulations.  These mean flows work
together with turbulent convection to amplify and organize magnetic
fields through hydromagnetic dynamo action, giving rise to the rich
display of magnetic activity so striking in modern solar
observations.''

The Sun's large scale \indexit{Sun!large-scale field}magnetic field exhibits a quasi-periodic modulation on a
roughly 11-year basis during which the level of magnetic activity
waxes and wanes as a pattern of activity migrates from mid to low
latitudes, then to pick up again at higher latitudes, with some
temporal overlap in the early and late phases of these cycles. For
stars like the Sun, the mean level of activity as expressed
by the surface-averaged absolute magnetic flux density ranges over more than
three orders of magnitude, depending on the stellar rotation rate,
age, and internal structure (more on that in Sect.~\ref{sec:actrad}; see also
III:2).  \ors[III:6.1] ``[T]he {\em existence} of solar and stellar
magnetic fields is in itself not really surprising; any large-scale
fossil field present at the time of stellar formation would still be
there today at almost its initial strength, because the Ohmic
dissipation timescale is extremely large for most astrophysical
objects [(Eq.~\ref{eq:diffusiontime})].  The challenge is instead
to reproduce the various observed spatiotemporal patterns [\ldots],
most notably the cyclic polarity reversals on decadal timescales.''

As to planetary \indexit{dynamo!planetary}dynamos, \ors[III:7.1]
``[s]pace missions revealed that most planets in the Solar System have
internal magnetic fields (see Ch.\,I:13), but there are exceptions
(Venus, Mars). Some planets seem to have had a field that is now
extinguished ({\em e.g.,} Mars). In many cases with an active dynamo
the axial dipole dominates the field at the planetary surface, but
Uranus and Neptune are exceptions. Saturn is special because its field
is extremely symmetric with respect to the planet's rotation axis.
The field strengths at the planetary surfaces differ by a factor of
1000 between Mercury and Jupiter [({\em cf.}
Table~\ref{tab:fran3})]. A full understanding of this diversity in the
morphology and strength of planetary magnetic fields is still lacking,
but a number of promising ideas have been suggested and backed up by
dynamo simulations.  Some of the differences can be explained by a
systematic dependence of the dynamo behavior on parameters such as
rotation rate or energy flux, whereas others seem to require
qualitative differences in the structure and dynamics of the planetary
dynamos.''

\ors[III:7.4.1] \indexit{Earth!interior} ``Earth serves as the prototype
for the terrestrial planets. [\ldots] There is a core
\indexit{Earth!core} with radius $R_{\rm core} \approx 0.55R_{\rm planet}$,
[the outer part of which is liquid]. The small inner core,
with\indexit{Earth!core} a radius
$0.35 R_{\rm core}$, is [solid].  The core appears to
consist predominantly of iron. [\ldots]

The total internal heat flow at the Earth's surface is $4.6\,10^{20}$\,erg/s
(although a large number, it is only 0.03\%\ of the total power
coming into the Earth's atmosphere by insolation).
\indexit{Earth!heat flow}
 Roughly one half
of it is balanced by the heat generated by the decay of
uranium, thorium and the potassium isotope $^{40}$K inside
the Earth. The remainder of the heat flow is due to the
cooling of the Earth. The loss of
gravitational potential energy associated with the contraction of the Earth
contributes a modest amount, but is much less important than it is in young stars
or in gas planets. How much of the Earth's heat flow comes from the core is rather
uncertain. Recent estimates that are based on different lines of evidence mostly fall into
the range $(0.5-1.5)\,10^{20}$\,erg/s, although values as low as $(0.3-0.4)\,10^{20}$\,erg/s
have also been discussed. Most of the radioactive elements reside in the silicate crust
and mantle. Some amount of potassium may be present in the core, but the majority of
the core heat must be due to cooling. It is important to note that
the heat loss from the core is regulated by the slow solid-state convection in
\indexit{mantle convection!Earth}
the mantle.\indexit{Earth!plate tectonics}
The core, which convects vigorously in comparison to the mantle and
which is thermally well-mixed, delivers as much heat as the mantle is able to carry away.''

\begin{figure}
\begin{center}
\includegraphics[width=9cm]{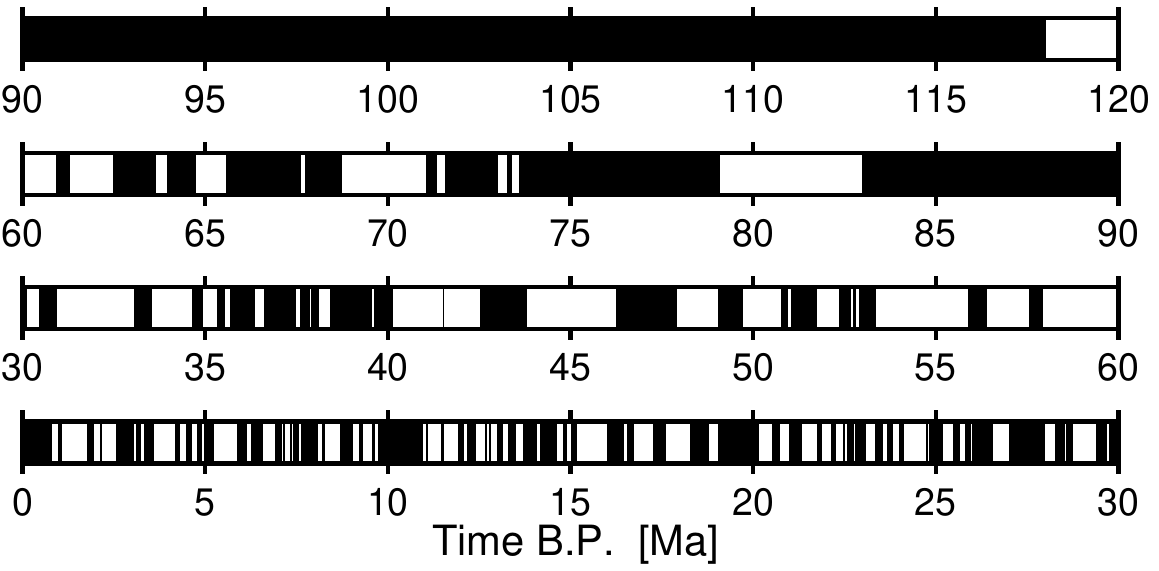}
%\vspace*{1cm}
\end{center}
\caption[Polarity of the geomagnetic field for the past 120 million years.]{Polarity of the geomagnetic field for the past 120 million years, with time running backward from left to right in each row
(before present -~B.P., {\em i.e.,}  1950~- in units of millions of years).
Dark regions indicate times when the dipole polarity was the same as
today, in white regions it has been opposite. [Fig.~III:7.4]}
\label{fig:chron120}
\end{figure}
\subsection{Earth and other terrestrial planets}
\label{intro-planets}
\indexit{magnetic!field!planets}
Solar System bodies that have a present-day active dynamo include Mercury and
Earth among the terrestrial planets, the jovian moon Ganymede, and all
the giant planets; see Table~\ref{tab:fran3} for their global properties.

\ors[I:3.1.1] ``The surface \indexit{Earth!magnetic field}magnetic
field of the Earth has a strength of about $0.5$\,G with mainly
dipolar character. [The dipole axis is tilted by a variable amount
over time with respect to the axis of rotation, such that the magnetic
north pole has wandered from as far south as about 70 degrees in
geographic latitude to within a few degrees from the geographic north
pole over the past two centuries]. From studies of rock
magnetism\indexit{magnetic!field!Earth} (when rocks cool below the
Curie point they preserve the magnetic field that was present in them
at that time) it is known that the Earth had a magnetic field over the
past $3.5\times 10^9$\,years and that the strength and orientation of
the field varied significantly on time scales of $10^3$ to
$10^4$\,years.  A given polarity typically dominates for about
$200\,000$\,years with quick reversals on a time scale of a few
thousand years in between [(see Fig.~\ref{fig:chron120})]. While the
orientation of the axis of the dipole changes significantly with time,
the dipole moment is aligned with the axis of rotation when averaged
over $\sim 10^4$\,years.''

In contrast to the case of cool stars, \ors[III:7.4.1] ``[r]adiative
heat transfer is not an issue in planetary cores, \indexit{radiative!heat flow} but liquid metal is a good thermal conductor. The heat
flux that can be transported by conduction along an adiabatic
temperature gradient, $({\rm d}T/{\rm d}r)_{\rm ad} = T/H_T$, is sometimes called
the \lq adiabatic heat flow\rq\, \indexit{adiabatic heat flow} ($T$ is
absolute temperature, $H_T=c_p/(\zeta g)$ is the temperature scale
height \indexit{temperature scale height} with $c_p$ the heat
capacity, $\zeta$ the thermal expansivity and $g$ the local gravitational
acceleration).  In terrestrial planets, the adiabatic heat flow can be
a large fraction of the actual heat flow, or it may exceed the actual
heat flow, in which case at least the top layers of the core would be
thermally stable. Near the top of Earth's core approximately
$(0.3-0.4)\,10^{20}$\,erg/s can be conducted along the adiabat,
{\em i.e.,}  close to the minimum estimates for the entire core heat flow. But
even if all the heat flux near the core-mantle boundary were carried
by conduction, a convective\indexit{Earth!core} dynamo can exist
thanks to the inner core.  \indexit{Earth!core} At the inner
core boundary, the adiabatic temperature profile of the convecting
outer core crosses the melting point of iron.  The latter increases
with pressure more steeply than the adiabatic gradient, which is the
reason why the Earth's core freezes from the center rather than from
above. As the core cools, the inner core grows with time by freezing
iron onto its outer boundary.  This has two important implications for
driving the dynamo. The latent heat that is released upon
solidification is an effective heat source, which contributes to the
heat budget approximately the same amount as the bulk cooling of the
core. [\ldots] A second, perhaps more important effect is that the
light elements in the outer core are preferentially rejected when iron
freezes onto the inner core.  Hence, they become concentrated in the
residual fluid near the inner core boundary.  This layering is
gravitationally unstable because of the reduced density, which leads
to compositional convection that homogenizes the light elements in the
bulk of the fluid core. Compositional convection
\indexit{convection!compositional} contributes as much as, or more
than, thermal convection to the driving of the geodynamo in recent
geological times.

Most predictions for the inner core growth rate imply that the inner
core did not exist for most of the history of the Earth. Rather, it
would have nucleated between 0.5 and 2 billion years ago. In the
absence of an inner core, only thermal convection by secular cooling
of the fluid core (and perhaps radioactive heating) can drive a
dynamo, which is less efficient than the present-day setting. A change
in the geomagnetic field properties might be expected upon the
nucleation of the inner core, but no clear indication for such an
event has been found in the paleomagnetic record.''

\ors[III:7.4.2] ``No direct evidence\indexit{compositional convection}
on\indexit{convection!composition} the existence or non-existence of a
solid inner core is available for any planet other than Earth. But the
possible absence of an inner core could explain why Venus and Mars do
not have an active dynamo. On Earth, \indexit{Earth!plate tectonics}
mantle convection reaches the surface in the form of plate tectonics,
\indexit{plate tectonics} which is a fairly efficient mode of removing
heat from the interior. None of the other terrestrial planets have
plate tectonics. In their cases, mantle convection is confined to the
region below the lithosphere, a rigid lid of some $100 - 300$\,km
thickness through which heat must be transported by
conduction. Without plate tectonics, the heat flow is expected to be
significantly lower not only at the surface, but also at the top of
the core, where it is very probably subadiabatic.  If no inner core
exists to provide latent heat, it is then subadiabatic throughout the
core. Furthermore, compositional convection is also unavailable to
drive a dynamo. The slower cooling of the planetary interior in the
absence of plate tectonics concurs with the idea that an inner core
has not (yet) nucleated in the cases of Mars and Venus. Early in the
planets' history the cooling rate was probably much higher and the
associated core heat flow large enough for thermal convection. The
demise of the dynamo must have occurred when the declining heat flow
dropped below the conductive threshold.''

\begin{figure}[t]
\begin{center}
%\centerline{\hbox{\psfig{figure=figures/MDI20010101_000003.eps,height=6.5cm,clip=}\psfig{figure=figures/bipolefigure.ps,height=6.5cm,clip=}}}
\centerline{\hbox{\includegraphics[height=6.5cm]{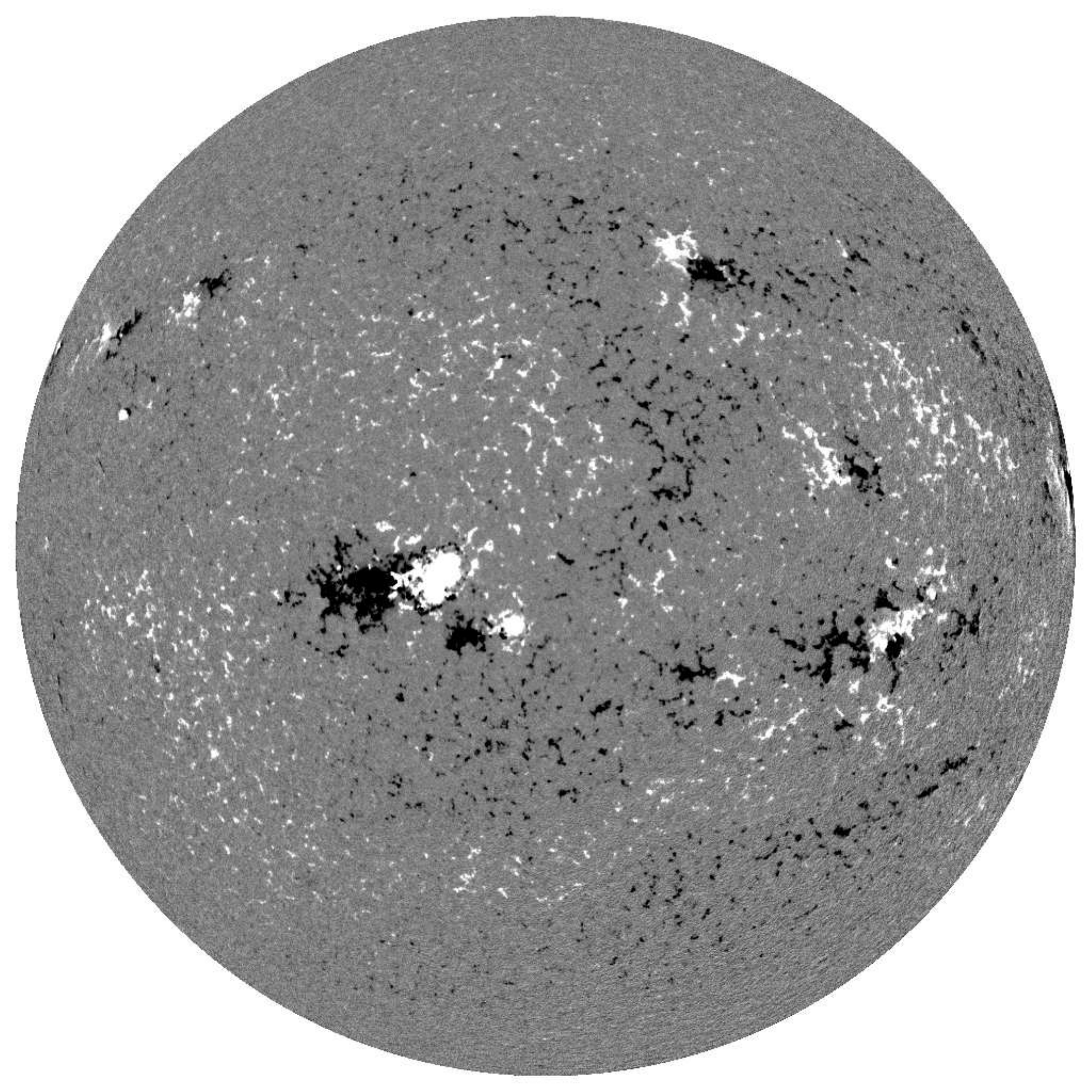}\includegraphics[width=6.5cm,bb= 54 360 450 756]{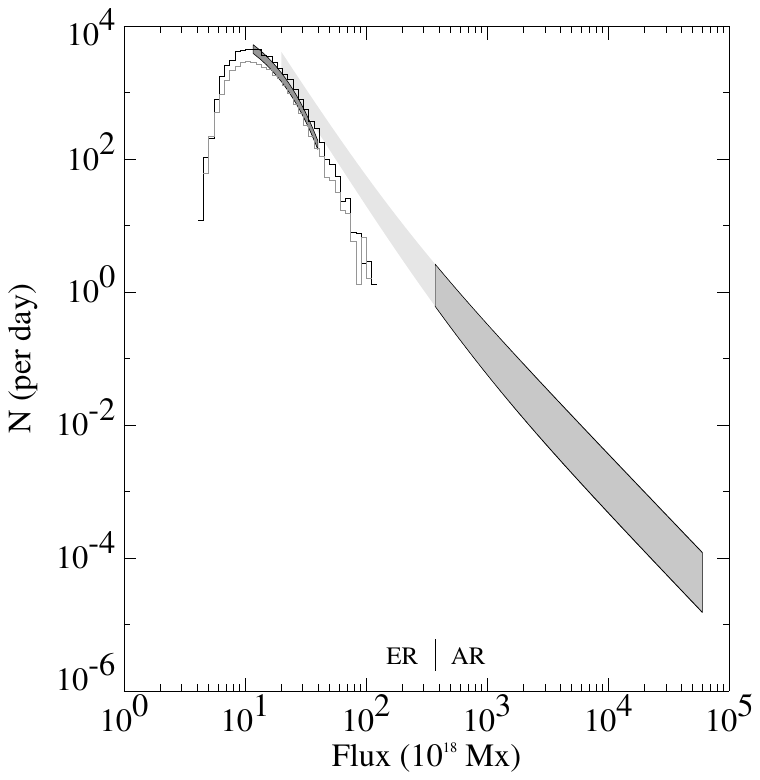}}}
\caption[Solar magnetogram, and frequency of emerging magnetic
bipoles.]{ {\em Left:\/} First solar
  \indexit{solar!magnetogram}magnetic
  \indexit{magnetogram|see{solar}}map (magnetogram) of the
  current millennium, taken by {\em SOHO}'s {\em MDI} on 2001/01/01 00:03\,UT. The
  magnetogram (with white/black for negative/positive line of sight
  polarity) shows a variety of active regions, embedded in patches of
  largely unipolar enhanced supergranular network, mixed-polarity
  quiet Sun regions, and low-flux polar caps (weak at this
  near-maximum phase of the cycle, and weakened further in the
  line-of-sight flux map because of projection effects on the
  near-vertical magnetic field).  {\em
    Right:\/} Distribution function of emerging magnetic bipolar
  regions on the Sun, showing the emergence frequency per day per flux
  interval of $10^{18}$\,Mx, estimated for the entire solar
  surface. The shaded region on the right envelopes the range of the
  active-region spectrum for solar cycle 22 (for half-year intervals
  around sunspot minimum and maximum). The histograms on the left are
  for the ephemeral regions; the shaded band shows where observations
  are least affected by spatial (lower cutoff) and temporal (upper
  cutoff) biases. The spectrum for regions below $\sim 10^{19}$\,Mx
  has yet to be determined; the cutoff here is caused by the limited
  resolution of the {\em SOHO/MDI} magnetograph.
  [Fig.~III:2.1] \label{figure:bipoles}}
\end{center}
\end{figure}

For discussion of dynamos in non-terrestrial planets, see Ch.~III:7.

\begin{figure}[t] 
\begin{center}
%\centerline{\hbox{\psfig{figure=figures/bfly.eps,width=\textwidth,clip=}}}
\centerline{\hbox{\includegraphics[width=\textwidth]{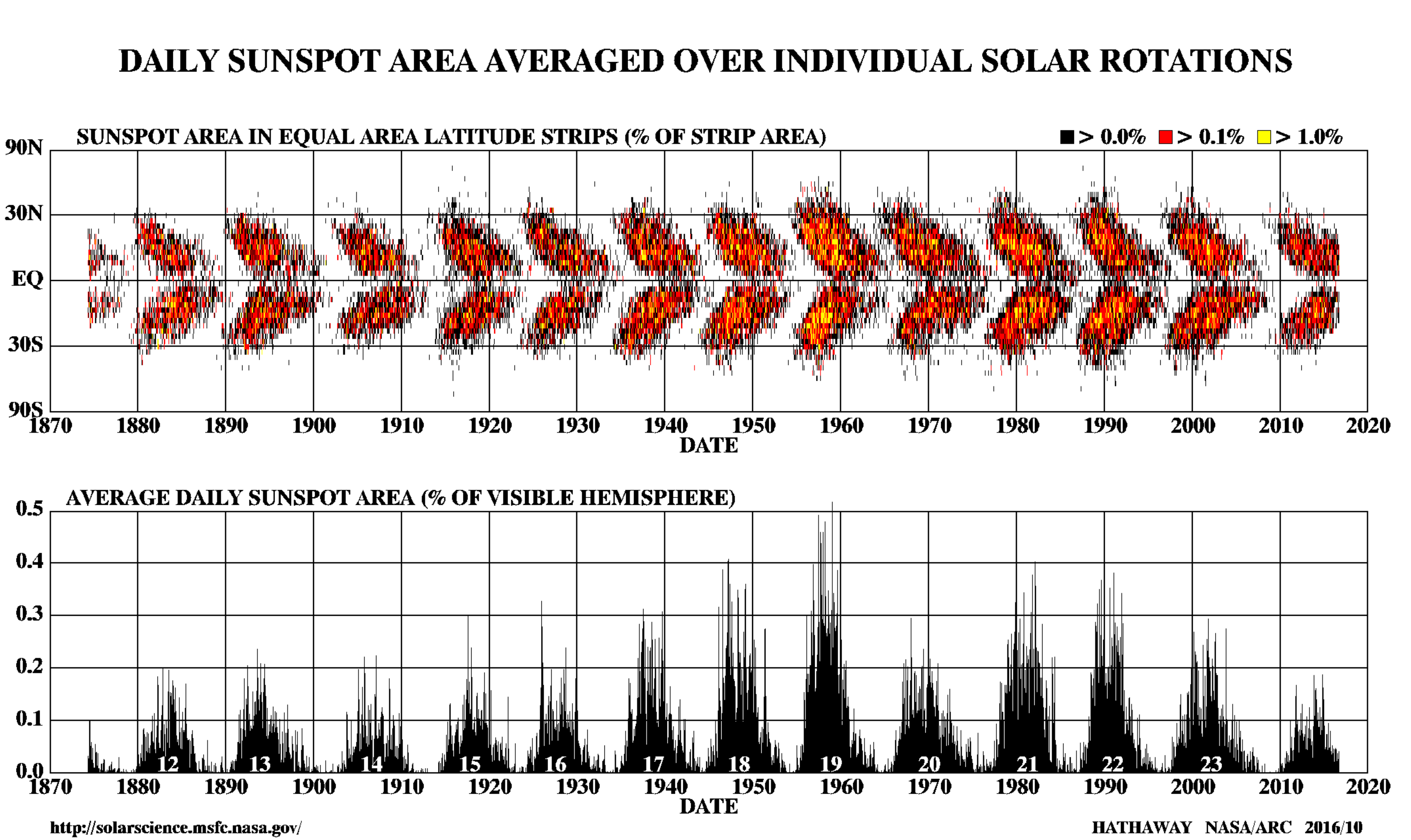}}}
\caption[Butterfly diagram: sunspot latitudes versus time.]{`Butterfly
  diagram' \indexit{butterfly diagram}showing (top) sunspot latitudes (also
  activity belts)\indexit{activity!belt} 
  and (bottom) total fractional area coverage as a function
  time. [{\em updated with data through
    2018}] \colorfig \label{figure:butterfly}}
\end{center} 
\end{figure} 
\subsection{The Sun and other stars}\label{sec:sunstars}
\indexit{magnetic!field!Sun|see{solar}} 

\ors[I:3.1.2] ``The Sun \indexit{solar!magnetic field}shows magnetic field on all observable scales
[(Fig.~\ref{figure:bipoles})] with a significant range in field
strength, from individual sunspots with magnetic field strengths of
$2\,500$ to $3\,000$\,G to the average field strength of the global
field of only a few Gauss. 

The\indexit{sunspot!cycle} most \indexit{activity!cycle}prominent
feature of solar magnetism is the $11$-year sunspot cycle (if one
considers the field reversals the full period is $22$\,years), which
is reflected in the changing number of sunspots appearing on the
surface of the Sun. In the beginning of a cycle spots appear at
latitudes of about $35\degr$, while close to the end they appear
almost at the equator. This property is commonly summarized in the
so-called\indexit{butterfly diagram|seealso{definition}} solar butterfly
diagram [(Fig.~\ref{figure:butterfly})].  During the epoch of minimum,
the large-scale field of the Sun is most dipolar; the reversal of the
poles takes place during solar maximum. On a longer time scale the
magnetic activity changes significantly in amplitude and is
interrupted by epochs of $100 - 200$ years in duration where sunspots
are [infrequent or] completely absent [(such as during the
\indexit{Maunder Minimum}Maunder Minimum, about 1645--1715). \ldots]
Observations of the stellar luminosity or of chromospheric
(UV/optical) and coronal (X-ray) emission show that a majority of
solar-like stars are magnetically active and around a third to a half
show cyclic activity with periods in the range from $3$ to
$30$\,years.'' \activity{{\em Look up:} (a) How is the Sun's
  magnetic field observed? Look up the effects on photons propagating
  through a plasma threaded by a magnetic field. This results in the
  'Zeeman effect' of line splitting and of circular and linear
  polarization. For relatively weak field or relatively low
  wavelengths, the Zeeman splitting of 'magnetically sensitive'
  spectral lines is generally less than the thermal line width (and
  less than the Doppler width for rapidly rotating stars), so that
  what is in principle line splitting for individual atoms becomes
  line broadening when averaging over populations of atoms and over
  entire stellar disks. (b) Estimate at what field strength $B_\Delta$ (G)
  the Zeeman splitting $\Delta \lambda_B$ (\AA) equals the thermal
  line broadening; use
  $\Delta \lambda_B=4.7\times 10^{-13} g \lambda^2 B$, assuming a
  Land{\'e} factor $g=2$, $\lambda=5000$\,\AA, and using a thermal
  line broadening associated with the thermal velocity for iron
  atoms. Note that $B_\Delta$ is inversely proportional to
  $\lambda$. \mylabel{act:zeeman}} \activity{{\em Look up:} Compare a
  series of solar magnetograms over the past $\sim$22\,years (using,
  {\em e.g.,} {\em SOHO/MDI} and --~since mid-2010~-- {\em SDO/HMI} observations, which can
  be accessed with, e.g., helioviewer.org, the JHelioviewer app, or at
  iswa.gsfc.nasa.gov; a sampling of once per Carrington rotation suffices). How do
  the magnetic patterns change over time in terms of overall activity,
  latitudinal distribution, polarity patterns on the northern versus
  southern hemisphere, \ldots? \mylabel{act:magpatterns}}

\section{Dynamo principles}
\ors[IV:6.1] ``Dynamo \indexit{dynamo!principles}action refers to the
conversion of mechanical energy into electromagnetic energy through
induction. In [stars and in planets alike], the mechanical energy is
supplied by fluid motions in electrically conducting regions inside
[these bodies] and the electromagnetic energy produces the observed
[\ldots] magnetic fields. A dynamo is referred to as {\em
  self-sustaining} if it does not require any external magnetic field
contributions for regeneration (except initially for a starting seed
field).

The fundamental equation governing this induction process is known as
the \indexit{MHD!induction equation!dynamo}{\em Magnetic Induction Equation} [Eq.~(\ref{induction}) in
Table~\ref{fig:mhdset}; its derivation and its limitations are
described in Sect.~\ref{sec:induction}. That equation is
complemented by the requirement that the currents and the driving
flows that are associated with the
magnetic field are entirely contained within the body, and that the
transition to outside the body for the field is smooth (compare
I:3.3). \ldots] 

By inspecting the two terms on the right-hand side of
Eq.~(\ref{induction}) we see that magnetic field can grow or decay in
time through two processes. The first term \tc{1} involves interactions of
the velocity and magnetic fields through electromagnetic induction and
acts as a source/sink term for field generation. The second term
\tc{2} 
represents diffusion due to Ohmic dissipation. To ensure magnetic
field does not decay away in time, field must be generated as fast as or
faster than its diffusion. A necessary condition for self-sustained
dynamo action is therefore that the induction term \tc{1} be larger than the
diffusion term \tc{2} in Eq.~(\ref{induction}). By using characteristic scales for
the variables in the Magnetic Induction Equation ({\em i.e.}, $B_{\rm t}$ for
the magnetic field scale, $v_{\rm t}$ for the velocity scale and $L_{\rm t}$ for a
length scale) we derive a common measure of the ratio of field
generation to field diffusion known as the \indexit{magnetic!Reynolds number}{\em magnetic Reynolds
  Number}: $\Rm\equiv{v_{\rm t}\,L_{\rm t}}/{\eta_{\rm \null }}$, see
Eq.~(\ref{eq:reynolds}).]

\begin{figure}[t]
  \center{\scalebox{0.4}{\includegraphics{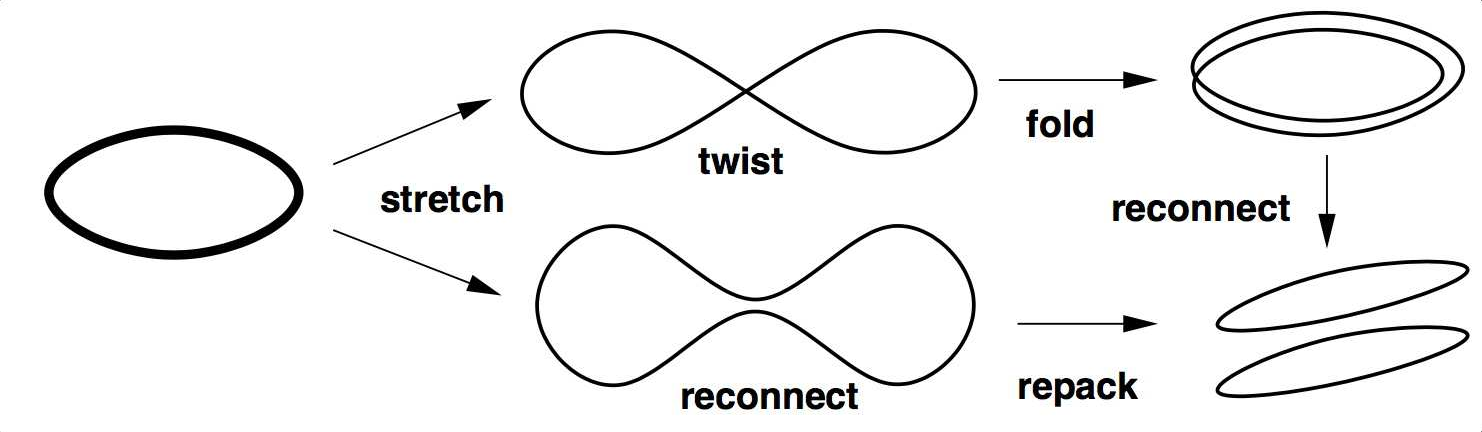}}}
  \caption[Two possible flux-rope dynamos.]{Illustration
    \indexit{dynamo!stretch-twist-fold!illustration}of two possible flux-rope dynamos. In both cases the
    field amplification takes place during the stretch operation. The
    twist-fold (top) and reconnect-repack (bottom) steps are required
    to remap the amplified flux-rope into the original volume element so
    that the process can be repeated. Magnetic diffusivity is essential to
    allow for the topology change required to close the cycle. Each cycle
    increases the field strength by a factor of 2. [Fig.~H-1:3.3]
    \label{fluxrope-dyn}
  }
\end{figure}

Upon first glance, it seems reasonable that the magnetic Reynolds
number must be larger than unity for dynamo action to be
possible. However, more rigorous theoretical analyses suggest that the
lower bound for $\Rm$ is instead closer to $\pi^2$ and planetary
numerical dynamo simulations typically find $\Rm$ must be larger than
$\sim 20-50$ for self-sustained dynamo action to occur. 
These higher values are due to the complexities in the
velocity field morphologies that cannot be captured in the simple
estimate given in Eq.~(\ref{eq:reynolds}):'' after all, it is a big
leap from small-scale field generated on the scale of the flow (such as
sketched in Fig.~\ref{fluxrope-dyn}) to a large-scale field. In cool stars, $\Rm$ 
typically far exceeds critical values for dynamo
action because of the large scales and relatively fast flows involved
(see Sect.~III:5.3.2).

A perspective of \indexit{dynamo!power source}what actually supplies
the energy to power the dynamo is provided by integrating the
induction equation Eq.~(\ref{induction}) over the object's volume to
establish the total energy in the system:
\begin{equation}\label{eq:dynamoenergy}
  {{\rm d} \over {\rm d}t} \int_V {B^2 \over 4\pi} {\rm d}V =
  - \oint_{\partial V} {\bf S}\cdot {\bf \hat{n}} {\rm d}S -
  {\eta_{\rm \null }} \int_V j^2 {\rm d}V -
   \int_V {\bf v}\cdot ({\bf j}\times {\bf B}) {\rm d}V.
\end{equation}
The first term on the right is the \indexit{Poynting flux}Poynting flux
${\bf S}=(1/4\pi) {\bf B}\times ({\bf v} \times {\bf B})$, which is
the energy via the electromagnetic field through a surface into or out
of the system across the closed boundary surface $\partial V$
(ignorable if the stellar wind does not take too much power away
compared to the total).  The second term is the dissipative loss
(assuming here that $\eta_{\rm \null }$ is uniform). The final term
shows that the magnetic energy in the system can be maintained against
the dissipative losses only if there are sufficient flows working
against --~{\em i.e.,} have an antiparallel component relative to~--
the Lorentz force ${\bf F}=(1/c) {\bf j}\times {\bf B}$.
\activity{{\em Show:} Work through how Eq.~(\ref{eq:dynamoenergy}) is
  obtained by taking the dot product of Eq.~(\ref{induction}) with
  ${\bf B}$, integrating over the total volume of the system, and
  assuming no Poynting flux or currents (or at most only a force-free
  field) leave the volume.  Use vector identities
  (${\bf a} \cdot ( {\bf \nabla} \times {\bf b}) = ({\bf \nabla}
  \times {\bf a}) \cdot {\bf b} - {\bf \nabla} \cdot ({\bf a} \times
  {\bf b})$,
  ${\bf a}\cdot ({\bf b}\times{\bf c})={\bf b}\cdot({\bf c}\times{\bf
    a})$), Eq.~(\ref{eq:bfieldeq}), and Gauss's theorem. Note that
  there is a summary of vector calculus identities in Table~\ref{tab:vecident} and
  more 
  on
  \href{https://en.wikipedia.org/wiki/Vector_calculus_identities}{vector
    calculus identities} on Wikipedia.}

\section{Essentials of fluid motions in dynamos}\label{sec:fluid}
In essence, to drive a large-scale stellar or planetary dynamo, the
magnetic field must be subjected to a combination of flow components
of a different nature that have their origin in convection and
rotation.  \ors[1:3.3.4] ``Fig. \ref{fluxrope-dyn} illustrates the basic
ingredients required to amplify a closed magnetic field loop. After a
full cycle, the magnetic field strength and the flux have doubled (two
loops, each with the original magnetic flux) and the process can be
repeated. This very simple illustration points out already a few
fundamental properties of a dynamo process. To be able to
remap the magnetic field configuration into the original volume
element, three-dimensional motions are required.  Amplification
through stretching is possible in a strictly two-dimensional domain,
but there is no way to move the resulting field to return to the
right-hand side of the image. The two examples also point out the
crucial role of diffusivity in changing the topology of the field. The
'stretch-twist-fold' \indexit{dynamo!stretch-twist-fold} mechanism
(excluding diffusive steps) leads to loops of increased complexity,
while the 'stretch-reconnect-repack' process explicitly involves
magnetic diffusivity and ends up with two flux ropes [(see
Table~\ref{fig:structures} for a definition)] of similar topology. A
reconnection step at the end of the 'stretch-twist-fold' process leads
to a similar result. In the case of the 'stretch-twist-fold' dynamo
the sign of the twist does not matter.''

\begin{figure}[t]
%\centerline{\psfig{figure=figures/convection1.eps,width=\linewidth}}
\centerline{\includegraphics[width=\linewidth]{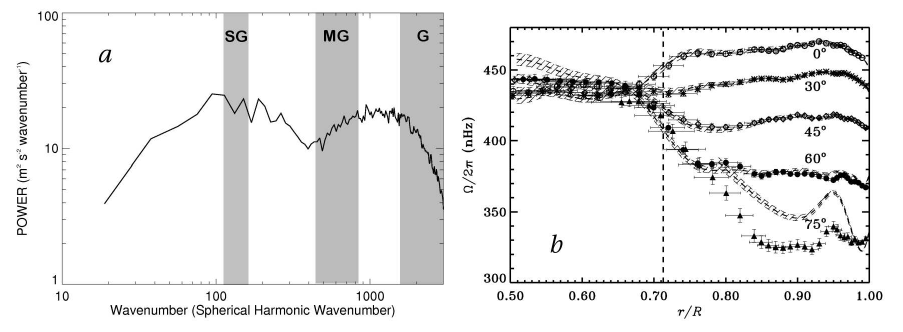}}
\caption[Photospheric convection spectrum and solar internal
rotation.]{Panel {\em a\/}: Power spectrum of the convective velocity
  field in the solar photosphere obtained from Doppler measurements,
  plotted as a function of spherical harmonic degree $\ell$.  Mean
  flows and $p$-modes are filtered out.  The falloff beyond $\ell \sim
  1500$ reflects the resolution limit of the {\em Michelson Doppler Imager}
  ({\em MDI}) instrument onboard the {\em SOHO} spacecraft from which these data
  were obtained and is therefore artificial.  Shaded areas indicate
  the approximate size ranges of supergranulation (SG),
  mesogranulation (MG) and granulation (G).  Note that the expected
  granulation spectral peak at $\ell \sim 4400$, corresponding to $L
  \sim $ 1 Mm, is not resolved.  Panel {\em b\/}: The solar internal
  rotation profile inferred from helioseismic inversions.  Angular
  velocity $\Omega/2\pi$ is shown as a function of fractional radius
  $r/R_\odot$ for several latitudes as indicated.  Symbols and dashed
  lines denote different inversion methods, known as subtractive
  optimally localized averages (SOLA) and regularized least squares
  (RLS) respectively.  Vertical 1-$\sigma$ error bars (SOLA) and bands
  (RLS) are indicated and horizontal bars reflect the resolution of
  the inversion kernels.  The vertical dashed line indicates the base
  of the convection zone. [Fig.~III:5.1; panel {\em a} is based on data
  from this
  \href{https://ui.adsabs.harvard.edu/abs/2000SoPh..193..299H/abstract}{source:
  \citet{2000SoPh..193..299H}};
%\href{https://ui.adsabs.harvard.edu/abs/2003ARA%26A..41..599T/abstract}{
  source
    panel {\em b}: \citet{2003ARA&A..41..599T}.]\label{convection1}}
\end{figure}
The driving flow of dynamos in stars and planets is
energy-transporting convection.  \ors[III:5.2]
``Thermal\indexit{convection} convection is familiar to most of us
from our daily experience; warm air rises and cooler air sinks.  When
a fluid is heated from below it overturns, provided the temperature
gradient is large enough, which here means that it must not only be
greater than the adiabatic temperature gradient (the Schwarzschild
criterion)\indexit{convection!Schwarzschild criterion} but it must
also overcome stabilizing influences such as thermal and viscous
diffusion, rotation, compositional gradients (the Ledoux
criterion),\indexit{convection!Ledoux criterion} and magnetic flux.
An intuitive way to think about convection (and to derive the
Schwarzschild and Ledoux criteria) is to consider a small isolated
volume, or {\em parcel}, of fluid that will buoyantly rise like a hot
air balloon if its density is less than that of its surroundings or
sink like a stone if its density is greater (the parcel is assumed to
be in pressure equilibrium with its surroundings so density and
temperature are anticorrelated).  [For a compressible medium, t]his is
the conceptual framework behind mixing length theory\indexit{mixing
  length theory} which goes on\indexit{convection!mixing length
  theory} to say that the parcel will lose its identity, dispersing
into the background, after traveling a vertical distance of order a
pressure scale height $H_p$.  [\ldots]

With this intuitive picture in mind, we may expect that the vertical
scale of solar convection should vary tremendously from the deep
convection zone where the \indexit{stratification}stratification is relatively gentle
($H_p \sim 35$ Mm) to the solar surface layers where the density and
pressure drop precipitously ($H_p \sim 36$ km) as [radiation escapes
freely into space]. The associated drop in temperature near the
surface triggers the recombination of hydrogen and other ions, which
modifies the opacity, decreases the particle number density, and
releases latent heat, altering the thermodynamics (in particular the
equation of state and the specific heats) and contributing to the
convective enthalpy transport.  Add in radiative energy transfer and
the result is what we call solar granulation; the continually shifting
pattern (lifetime $\sim $ 5 min) of small-scale convection cells (with
a horizontal extent $\sim$ 1 Mm) that blankets the solar surface and
accounts for the dappled appearance of the solar photosphere
(Fig.~I:8.3).'' \activity{{\em Look up} sample images of solar
  granulation, the most easily detectable pattern of convection
  reaching into the solar surface layers. What are the characteristic
  length and time scales of granulation? Also look up the larger-scale
  flow patterns of mesogranulation and supergranulation, and compare
  these in a table. \mylabel{act:convcells}}

Also the global-scale flows are important in the solar dynamo. The
solar surface exhibits a \indexit{solar!differential rotation}differential rotation: the equator rotates
faster than the poles, with a smooth latitudinal gradient between
these. \activity{{\em Show:} Estimate the time it takes for the solar
  equator to execute one more full rotation than the poles in the same
  time. Use the data shown in
  Fig.~\ref{convection1}b. \mylabel{act:diffrot}} \ors[III:5.2.3]
``Helioseismology now reveals that this monotonic decrease in angular
velocity with increasing latitude persists throughout the convection
zone, with an abrupt transition to nearly uniform rotation in the
radiative interior (Fig.\,\ref{convection1}$b$).  The transition
region near the base of the convection zone is known as
the\indexit{differential rotation!tachocline}
solar\indexit{tachocline} tachocline [\ldots].  There is also a less
dramatic but no less significant {\em near-surface shear layer}
in\indexit{differential rotation!near-surface shear layer} which
the rotation rate systematically decreases by about 10-20\,nHz from
$r = 0.96 R_\odot$ to the photosphere.  This is most apparent at low
latitudes but may also occur at higher latitudes.
[\ldots]\activity{{\em Background:} Helioseismology uses resonant
  waves that run through the solar interior. These pressure-mode (or p
  mode) waves (generated by the turbulent convective motions) probe a
  range of depths depending on the wavelength and resonance
  conditions. At depth, downward traveling waves refract upward as the
  sound speed increases with temperature. If their frequency is below
  the 'acoustic cutoff period' around the photosphere upward traveling
  waves are reflected back into the interior, even as they are
  detectable around their upper turning point both in brightness (by
  compression and dilation) and velocity (through the Doppler effect
  on spectral lines). The combination of refraction and reflection
  leads to a cavity in which resonances occur. Intuitively, the cutoff
  frequency comes about because if the wavelength of a pressure wave
  exceeds a few pressure scale heights, there is essentially no
  restoring pressure force as the bulk of the atmospheric mass is
  simply lifted and lowered in response to the wave so that the solar
  surface acts as an open boundary condition and reflects the
  wave. Based on that argument, make a rough estimate of the acoustic
  cutoff period for the solar photosphere at around 5800\,K (a later
  Activity will let you develop the relevant equations for an
  isothermal atmosphere). Waves with shorter periods continue to
  travel upward, while those with longer periods mostly reflect but
  partly tunnel through into the hotter
  chromosphere. \mylabel{act:cutoff}}\activity{{\em Background:} To
  hear how helioseismology can measure rotation rates of stars (and,
  with enough different modes, of layers within stars) you can do the
  following experiment: Hold up a bell dangling from a string, strike
  it, and listen. Then twist up the string, let the bell spin freely,
  hit it and listen once more. The modulation in intensity that you
  hear for the spinning bell results from the beat of the Doppler
  effect working differentially on waves running with and against the
  spin direction. This is the essence of how helioseismology measures
  the Sun's internal rotation.}

\begin{figure}
\begin{center}
\includegraphics[width=5.5cm]{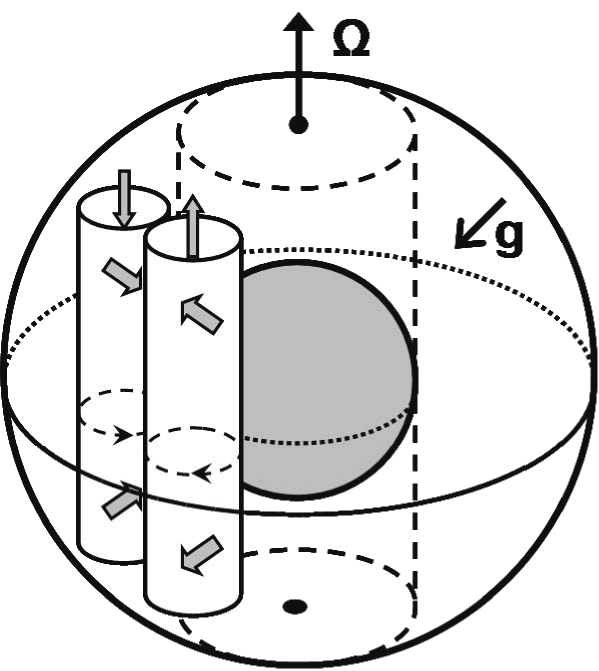}
%\vspace*{1cm}
\end{center}
\caption[Columnar convection in a rotating spherical shell.]{Columnar convection in a rotating spherical shell near onset. The
inner core tangent cylinder is shown by broken lines. Under Earth's core conditions
the columns would be much thinner and very numerous. [Fig.~III:7.6]}
\label{fig:bussesphere}
\end{figure}

The striking difference in the rotation profile of the convective
envelope and that of the radiative interior implicates convection as
the primary source of the \indexit{solar!differential rotation}differential rotation.  Furthermore, it
tells us that giant cells are large enough and slow enough to be
influenced by the rotation of the star.  The magnitude of nonlinear
advection [(${\bf v} \cdot \nabla {\bf v}$)] relative to the \indexit{Coriolis force}Coriolis force
[(${\bf \Omega} \times {\bf v}$)] is quantified by the [Rossby number:
\indexit{Rossby number}
\begin{equation}
\label{eq:ross}
\Ro \, = \, \frac{v_{\rm t}}{\Omega L_{\rm t}} \, ,
\end{equation}
where $v_{\rm t}$ and $L_{\rm t}$ are characteristic velocity and
length scale, respectively.] In the deep solar convection zone it is
of order unity or less whereas it is much greater than unity in the
solar surface layers.\activity{{\em Show:} Compare the Rossby number
  for deep solar convection to that of the Earth's core for a flow
  velocity of order 1\,mm/s and for a scale equal to the core radius
  or shell thickness. \mylabel{act:georossby}} Coriolis-induced
velocity correlations in the convection redistribute angular momentum
via the Reynolds stress, generating a substantial rotational shear:
$\Delta \Omega/\Omega \sim $ 30\%\ where $\Omega(r,\theta)$ is the
angular velocity and $\Delta \Omega$ is the angular velocity
difference between equator and pole.\activity{{\em Show:} If we take
  the Sun's polar field --~averaging at cycle minimum to be about
  5\,Gauss~-- how long would it take to wind that field into a
  strength of some $10^5$\,G --~which is the estimated minimum field
  strength for flux bundles to coherently survive their rise through
  the convective motions in the Sun's envelope to emerge as active
  regions~-- if the rotational shear were maximally used and if no
  back-reaction on that flow occurred?  Hint: remember the field line
  stretch-and-fold from Fig.~\ref{fluxrope-dyn}, look at the
  illustration in Fig.~\ref{figure:rotmagsim}, remember that field
  line density is equivalent to characteristic field strength, use the
  result from Act.~\ref{act:diffrot}, and consider 'compound
  interest.'  \mylabel{act:fieldwrap}} Furthermore, the nature of the
redistribution is such that the angular velocity increases away from
the rotation axis, $\partial \Omega / \partial d_\theta > 0$ where
$d_\theta= r \sin(\theta)$ is [the distance to the axis of rotation].
This is in stark contrast to the behavior one would expect from
isotropic turbulent diffusion (if $\Delta \Omega/\Omega \ll 1$) or
from fluid parcels that tend to locally conserve their angular
momentum as they move ($\partial \Omega / \partial d_\theta< 0$),
[which would behave as sketched in Fig.~\ref{fig:bussesphere}].  Giant
cells must be a global phenomenon distinct from supergranulation.''

These solar flow patterns are in striking contrast to what fluid
motions in the planets are thought to look like: \ors[III:5.5.4] ``the
latter often tend to be quasi-two-dimensional.  This is largely a
consequence of rapid rotation.  Planets are smaller than stars and
generally spin faster (with the exception of compact remnants such as
pulsars).  In the fluid cores and mantles of terrestrial planets and
the extended atmospheres of many gas giant planets, the convective
time scales are much longer than the rotation period, implying very
low Rossby numbers [\ldots\ T]his gives rise to elongated, quasi-2D
convective structures such that the flow is relatively invariant in
the direction parallel to the rotation axis
(Fig.~\ref{fig:bussesphere}).  In the atmospheres and oceans of
terrestrial planets, on the other hand, quasi-2D dynamics arises
simply by virtue of the geometry; global-scale horizontal motions are
confined to thin spherical shells.''

\section{Insights from approximate stellar dynamo models} 
Astrophysical \indexit{dynamo!approximating models}dynamos have been
studied for many decades, and whereas the fundamental ingredients may
be known, there is no proper theory of dynamo action in stars and
planets: there is no validated dynamo model that matches all stellar
observations or that has been demonstrated to successfully forecast
the Sun's magnetism over multiple sunspot cycles, nor do planetary
dynamo models successfully reproduce, for example, the quasi-irregular
reversals in the terrestrial magnetic field. Nonetheless, dynamo
concepts do guide our thinking as to the important ingredient
processes as well as the possible internal structure and dynamics of
both the magnetic field and the plasma/magma flows involved. The
remainder of this chapter is an exploration of some of these to create
a sense of how dynamos in stars and planets are thought to function.

\ors[III:6.1] ``All solar and stellar dynamo models to be considered
in this chapter operate within a sphere of electrically conducting
fluid embedded in vacuum.  We restrict ourselves here to
\indexit{magnetic!field!axisymmetric} axisymmetric mean-field-like
models, in the sense that we will be setting and solving evolutionary
equations for the large-scale magnetic field, and subsume the effects
of small-scale fluid motions and magnetic fields into coefficients of
these partial differential equations.  Working in spherical polar
coordinates $(r,\theta,\phi$), we begin by writing:
\def\mvarpi{d_\theta}
\begin{eqnarray}
\label{eq:axiU}
{\bf v}(r,\theta)&=&{\bf v}_{\rm
  p}(r,\theta)+\mvarpi\Omega(r,\theta)\uv{\phi}~, \\
%\end{eqnarray}
%
%\begin{eqnarray}
\label{eq:axiB}
{\bf B}(r,\theta,t)&=&\nabla\times ({\cal A}(r,\theta,t)\uv{\phi})
+{\cal B}(r,\theta,t)\uv{\phi}~,
\end{eqnarray}
where
$\mvarpi=r\sin(\theta)$,  
${\bf v}_{\rm p}$ is a notational shortcut for the component of the large
scale flow in meridional planes, and $\Omega$ is the angular velocity
of rotation, which in the solar interior varies with both depth and
latitude, and is now well-constrained by helioseismology.  Note that
in this prescription neither of these large-scale flow components is
time dependent.  \indexit{dynamo!kinematic approximation} This {\bf
  kinematic approximation} is an assumption that is tolerably
well-supported observationally.  Substituting these expressions in the
MHD induction equation in Eq.~(\ref{induction}) allows separation
into two coupled 2D partial differential equations for the scalar
functions ${\cal A}$ and ${\cal B}$ defining respectively the poloidal and
toroidal components of the magnetic field:
\begin{eqnarray}
\label{eq:cowa}
\derp{{\cal A}}{t}&=&
\eta_{\rm \null }\left(\nabla^2-{1\over\mvarpi^2}\right){\cal A}
-{{\bf v}_{\rm p}\over\mvarpi}\cdot\nabla (\mvarpi {\cal A})~, \\
%\end{eqnarray}
%\begin{eqnarray*}
\derp{{\cal B}}{t}&=&
\eta_{\rm \null }\left(\nabla^2-{1\over\mvarpi^2}\right){\cal B}
+{1\over\mvarpi}\derp{(\mvarpi {\cal B})}{r}\derp{\eta_{\rm \null }}{r}-
               \nonumber \\
%\end{eqnarray*}
%\begin{eqnarray}
\label{eq:cowb}
&& \mvarpi\nabla\cdot \left({{\cal B}\over\mvarpi}{\bf v}_{\rm p}\right)
+\mvarpi(\nabla\times ({\cal A}\uv{\phi}))\cdot\nabla\Omega~,
\end{eqnarray}
where we retain the possibility that $\eta_{\rm \null }$ varies with depth.
The shearing term ($\propto\nabla\Omega$) on the right-hand side of Eq.~(\ref{eq:cowb})
acts as a source of toroidal field.
However, no such source term appears in Eq.~(\ref{eq:cowa}). This is the essence of
\indexit{dynamo!Cowling's theorem}
Cowling's theorem 
%(see Sect.~3.3.8 in Vol.~I)
which in fact guarantees that an axisymmetric flow of the general form
given by Eq.~(\ref{eq:axiU}) {\em cannot} act as a dynamo for an
axisymmetric magnetic field
as described by Eq.~(\ref{eq:axiB}). The construction
of solar and stellar dynamo models, therefore, hinges critically
on the addition of an extraneous source term in Eq.~(\ref{eq:cowa}).
The physical origin of this source term is what fundamentally distinguishes
the various classes of solar and stellar dynamo models described [below].

\begin{figure}[t]
\begin{center}
%\centerline{\hbox{\psfig{figure=figures/rotatesim,height=4.5cm,clip=}\psfig{figure=figures/magsim,height=4.5cm,clip=}}}
\centerline{\hbox{\includegraphics[height=4.5cm]{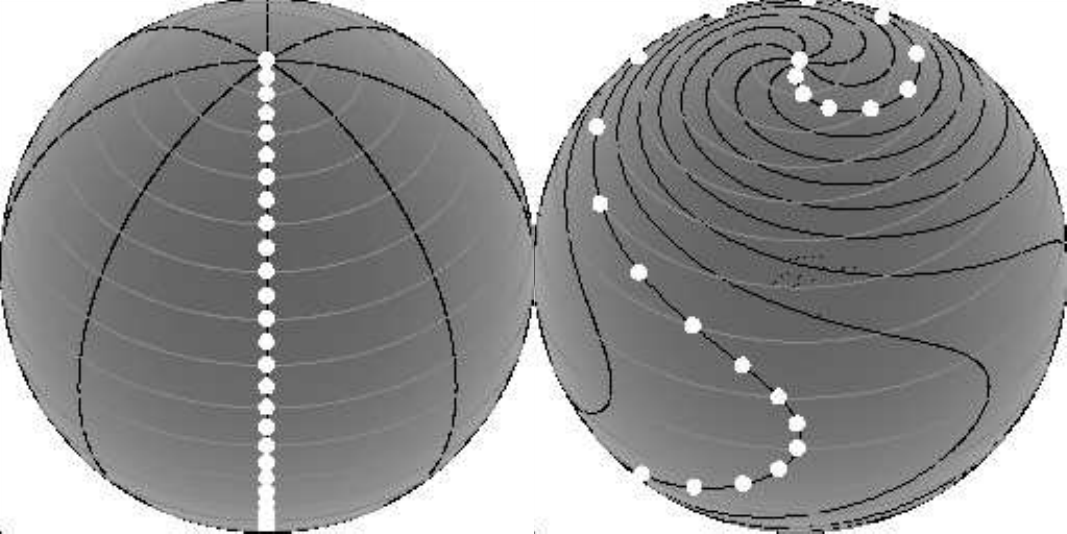}\includegraphics[height=4.5cm]{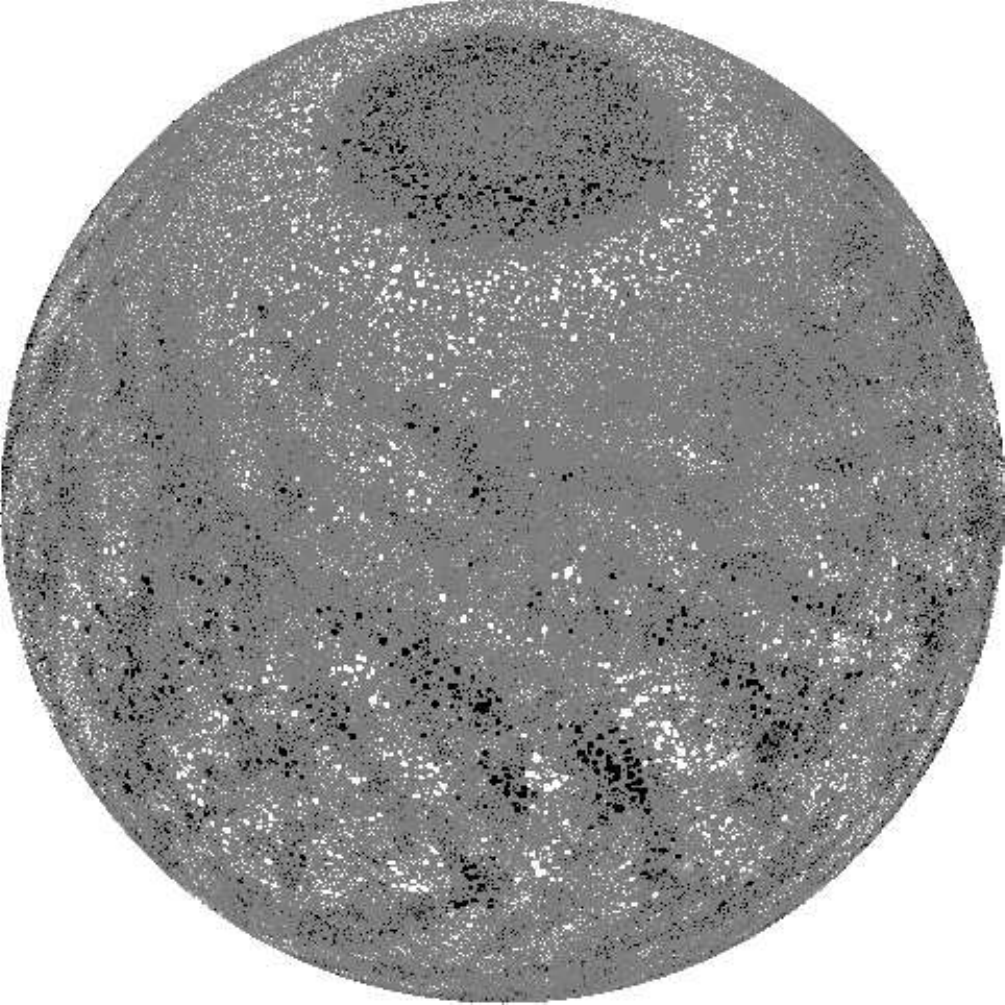}}}
\caption[Effects of large-scale flows, and a young-Sun magnetogram.]{ {\em Left and center:\/} Visualization of the effects of
  differential \indexit{solar!differential rotation!illustration}rotation and equator-to-pole meridional flow for
  Sun-like conditions: lines of equal longitude (with markers) are
  distorted into a spiral pattern. The center panel shows the
  distorted lines after 3 months.  {\em Right:\/} Simulated
  magnetogram for a star like the Sun, [simulated with a
  flux-transport model with parameters as observed for the Sun,] but
  with an active-region emergence rate 30 times larger. The simulated
  star is shown from a latitude of 40$^\circ$ to better show the
  polar-cap field structure.  [Fig.~III:2.3] \label{figure:rotmagsim}}
\end{center}
\end{figure} 
Shearing of the poloidal magnetic field into a strong toroidal
component by differential rotation [(as illustrated in
Fig.~\ref{figure:rotmagsim})] is an essential ingredient of all solar
cycle models discussed below.  The growing magnetic energy of the
toroidal field is supplied by the kinetic energy of the rotational
shearing motion, which makes for an attractive field amplification
mechanism, because in the Sun and stars the available supply of
rotational kinetic energy is immense (unless the dynamo were entirely
confined to a very thin layer, for example the tachocline, [the shear
layer just below the Sun's convective envelope into which convection
overshoots]).  Moreover, a strong, axisymmetric and temporally
quasi-steady internal \indexit{solar!differential rotation}differential rotation is likely responsible for
the observed high degree of axisymmetry observed in the Sun's magnetic
field on spatial scales comparable to its radius. This situation is
very different from that encountered in planetary core dynamos, where
differential rotation is believed to be much weaker, and energetics
pose a much stronger constraint on dynamo action. Lacking the
large-scale organization provided by differential rotation, planetary
core dynamos also \indexit{magnetic!field!non-axisymmetric}tend to
produce non-axisymmetric large-scale fields.  The one outstanding
exception appears to be Saturn, and indeed in this case the high
axisymmetry of the observed surface field may well reflect the
symmetrizing action of differential rotation in the envelope overlying
the metallic-hydrogen core. The important point remains that in the
solar dynamo context, the assumption of an axisymmetric large-scale
magnetic field is consistent with the observed and
helioseismically-inferred axisymmetry and quasi-steadiness of internal
differential rotation.''

\def\av#1{\left<#1\right>}
\def\rd{{\rm d}}
\def\ce{{\cal E}}

\def\bfomega{\setbox2=\hbox{$\omega$}
\hbox{{$\omega$}\hskip-.97\wd2
{$\omega$}\hskip-.97\wd2
{$\omega$}\hskip-.97\wd2 {$\omega$}}}

\def\bfxi{\setbox2=\hbox{$\xi$}
\hbox{{$\xi$}\hskip-.97\wd2
{$\xi$}\hskip-.97\wd2
{$\xi$}\hskip-.97\wd2 {$\xi$}}}

\def\div{\nabla \cdot}

%\section{Solar dynamo models\label{sec:sun1}}

\section{Mean-field dynamo models}\label{ssec:mfsoldyn} 

\ors[III:6.2.1] ``Turbulence at a high magnetic Reynolds number $\Rm$
[(Eq.~\ref{eq:reynolds})] is known to be quite effective at producing
a lot of \indexit{magnetic!field!small-scale} small-scale magnetic
fields, where 'small-scale' is roughly $\Rm^{-1/2}$ times the length
scale of the flow.  In addition, under certain conditions,
solar/stellar convective turbulence can also produce magnetic fields
with a mean component building up on large spatial scales. These
\indexit{dynamo!mean-field} {\bf mean-field dynamo models} remain
arguably the most 'popular' descriptive models for dynamo action in
the Sun and stars, but also in planetary metallic cores, stellar
accretion disks, and even galactic disks.

Under the assumption that a good separation of scales exists between
the large-scale 'laminar' magnetic field $\avr{{\bf B}}$ and the flow
$\avr{{\bf v}}$, and the small-scale turbulent field $\pr{\bf B}$ and
flow $\pr{\bf v}$, it becomes possible to express the inductive and
diffusive action of the turbulence on $\avr{{\bf B}}$ in terms of the
statistical properties of the small-scale flow and
field. \activity{{\em Background:} Consider that the assumption of a separation of
  scales as for the 'mean field dynamo theory' is also made in
  hydrodynamics when 'internal energy' (which includes the kinetic
  energy of the random motions of the gas particles) is 'separated
  from' the 'kinetic energy' of bulk motion. This assumption is
  commonly made with little consideration of why it works: there must
  be a scale that is small compared to flows of interest but large
  enough that low-order moments of the velocity distribution (like
  temperature and pressure) are defined by so many particles that
  there is negligible random noise when determined for a 'small'
  volume. Consider that in the context of the words in Table~\ref{fig:mhdvalidity}. }
The corresponding theory of mean-field electrodynamics is discussed in
detail in Ch.~I:3.  The turbulent flow introduces on the right-hand
side of the induction equation Eq.~(\ref{induction}) a term of the
form $\nabla\times\emfa$, where $\emfa$ is a \indexit{turbulent!electromotive force} mean-electromotive force.''

For a quick introduction to the origin of that term we can see what
happens when \ors[I:3.4] ``we decompose the magnetic field into a large-scale 'mean' field
and the small-scale components through an averaging procedure. We
assume in the following that the averaging procedure obeys
the\indexit{Reynolds rules} Reynolds rules: For any
function $f$ and $g$ decomposed as $f=\avr{f}+\pr{f}$ and
$g=\avr{g}+\pr{g}$, where the bar indicates the averaged and the
prime the fluctuating quantity, we require that \indexit{Reynolds
rules}
\begin{eqnarray}
  \avr{\avr{f}}&=&\avr{f} \longrightarrow  \avr{\pr{f}}=0\\
  \avr{f+g}&=&\avr{f}+\avr{g}\\
  \avr{f\avr{g}}&=&\avr{f}\avr{g} \longrightarrow \avr{\pr{f}\avr{g}}=0\\
  \avr{\dd f/\dd x_i}&=&\dd \avr{f}/\dd x_i\\
  \avr{\dd f/\dd t}&=&\dd \avr{f}/\dd t\;.
\end{eqnarray}
The averaging procedures that are of interest in the context of
mean-field theory are the ensemble average (meaning a chaotic system
is averaged over several representations of the chaotic system) and
the longitudinal average, in which $\avr{\B}$ reflects the
axisymmetric component of the large-scale magnetic field (multipole
series with $m=0$).'' \ors[I:3.4.1] ``In order to derive an equation for
the time evolution of the mean field we apply the averaging procedure
to the induction equation Eq.~(\ref{induction}) which leads to
\begin{equation}
  \frac{\dd\avr{\B}}{\dt}=\curl\left(\avr{\pr{\vv}\times\pr{\B}}
    +\avr{\vv}\times\avr{\B}-\eta_{\rm \null }\curl\avr{\B}\right)\label{mfi}\;.
\end{equation}
The new term which enters this equation compared to the original induction
equation is the second order correlation electromotive force (EMF)
\indexit{electromotive force}
\begin{equation}
  \emfa\equiv\avr{\pr{\vv}\times\pr{\B}}\label{emf}\;.
\end{equation}
While the fluctuating velocity component $\pr{\vv}$ is assumed to be known
(kinematic approach),
$\pr{\B}$ has to be computed from the induction equation. An equation for
$\pr{\B}$ can be derived by subtracting the mean-field induction equation
Eq.~(\ref{mfi}) from the microscopic induction equation Eq.~(\ref{induction}),
which leads to
\begin{equation}
  \frac{\dd\pr{\B}}{\dt}=\curl\left(\pr{\vv}\times\avr{\B}+\avr{\vv}
  \times\pr{\B}-\eta_{\rm \null }\curl\pr{\B}+\pr{\vv}\times\pr{\B}-\avr{\pr{\vv}
    \times\pr{\B}}\right)\label{fli}\;.
\end{equation}

It is in general only possible to solve this equation by making strong
assumptions, primarily because of the terms that are quadratic in the
fluctuating quantities (closure problem).'' For a more detailed
description, see Sect.~I:3.4.3. Here, we proceed with one particular
such assumption that leads to the conclusion that for 
\ors[III:6.2.1] ``mildly inhomogeneous and near-isotropic
turbulence, $\emfa$ can be expressed in terms of the large-scale field
$\avr{{\bf B}}$ as:
\begin{eqnarray}
\label{eq:emf2}
\emfa=\alpha{\avr{\bf B}}-\beta\nabla\times{\avr{\bf B}}~,
\end{eqnarray}
with
\begin{eqnarray}
\label{eq:SOCA}
\alpha = -{1\over 3}\,\tau_{\rm corr}
\avr{\pr{\bf v}\cdot(\nabla\times\pr{\bf v})}
~~~[{\rm cm}\,{\rm s}^{-1}]~,
\qquad
\beta = {1\over 3}\,\tau_{\rm corr}
\avr{{\pr{\bf v}}^2}~~~[{\rm cm}^2{\rm s}^{-1}]~,
\end{eqnarray}
where $\tau_{\rm corr}$ is the correlation time for the turbulent
flow. Note that the $\alpha$-term is proportional to the (negative)
\indexit{kinetic helicity} kinetic helicity
[($\pr{\bf v}\cdot(\nabla\times\pr{\bf v})$)] of the turbulence, which
requires a break of reflectional symmetry.  In stellar interiors and
planetary metallic cores alike, this anisotropy is provided by the
Coriolis \indexit{Coriolis force}force.  Small-scale turbulence thus impacts the induction
equation for the mean-field in two ways: it introduces a field-aligned
electromotive force (the $\alpha$-term), which acts as a source term
and is called the \indexit{dynamo!$\alpha$ effect}'$\alpha$-effect', and an
enhanced \indexit{turbulent!diffusion} 'turbulent diffusion' (the
$\beta$-term), associated with the folding action of the turbulent
flow.  In principle, the $\alpha$ and $\beta$ coefficients can be
calculated from the lowest-order statistics of the turbulent flow.  In
practice, more often than not they are chosen {\em a priori}, although
with care taken to embody in these choices what can be learned from
mean-field theory.\sactivity{$\circledS$ {\em Show:} (a) Take the mean-field induction
  equation Eq.~(\ref{mfi}) and the expression for Eq.~(\ref{emf}) as
  approximated in Eq.~(\ref{eq:emf2}) to find a mean-field form of the
  general induction equation Eq.~(\ref{induction}). Group the
  'diffusive' ($\eta$, $\beta$) terms together. (b) Using the values for the solar
  meridional flow and random-walk diffusive dispersal below, estimate
  the order of magnitude of the advection, $\alpha$, and diffusive
  terms. For these order of magnitude comparisons, approximate for the
  mean field $\nabla \approx 1/R_\odot$; in solar near-surface layers
  'small-scale' supergranular random walk with $\pr{\bf v}$ or order
  0.1\,km/s leads to $\beta \approx 300$\,km$^2$/s;
  the large-scale advective term of the surface meridional flow has an
  average value of order $\avr{\vv} =5$\,m/s (reaching a mid-latitude
  maximum of about 15\,m/s); estimate $\tau_{\rm corr}$ from this
  value of $\beta$ with Eq.~(\ref{eq:SOCA}), which corresponds to the
  characteristic evolutionary time scale of the dispersing
  supergranular convection; with that, estimate $\alpha$ using the
  characteristic supergranulation length scale of $30,000$\,km; then
  compare the order-of-magnitude values of the three terms expressed
  as time scales for the magnetic field. Note that the 'turbulent
  diffusivity' $\beta$ far exceeds the 'resistive diffusivity'
  $\eta_{\rm \null }$ in stellar dynamos (and see
  Activity~\ref{act:surfacediffusion} how the above helps in
  understanding how surface flux dispersal can be described quite well
  by a random-walk diffusive
  description).\mylabel{act:meanfieldequation} \solution{meanfieldequation}}

Under mean-field dynamo theory, Eqs.~(\ref{eq:cowa})---(\ref{eq:cowb})
are now taken to apply to an axisymmetric large-scale mean magnetic field.
With the inclusion of the mean-field $\alpha$-effect and turbulent diffusivity,
scaling all lengths in terms of the radius $R$ of star or planet, and time in
terms of the
\indexit{magnetic!diffusion time}
diffusion time
\begin{equation}\label{eq:difftime}
\tau_{\rm d}=R^2/\eta_{\rm \null }
\end{equation}
based on the (turbulent)
diffusivity in the convective envelope, these expressions become
\begin{eqnarray}\label{eq:mfa}
\derp{{\cal A}}{t}&=&\eta_{\rm \null
                      }\left(\nabla^2-{1\over\mvarpi^2}\right){\cal A}
-{\Rm\over\mvarpi}{\bf v}_{\rm p}\cdot\nabla (\mvarpi {\cal A})
+C_\alpha\alpha {\cal B}~, \\
%\end{eqnarray}
%
%\begin{eqnarray*}
\derp{{\cal B}}{t} &=&  \eta_{\rm \null
                       }\left(\nabla^2-{1\over\mvarpi^2}\right){\cal B}
+{1\over\mvarpi}\derp{(\mvarpi {\cal B})}{r}\derp{\eta_{\rm \null }}{r}
- \Rm\mvarpi\nabla\cdot\left({{\cal B}\over\mvarpi} {\bf v}_{\rm p}\right)+
                \nonumber \\
%\end{eqnarray*}
%
%\begin{eqnarray}
\label{eq:mfb}
&&C_\Omega\mvarpi(\nabla\times ({\cal A}\uv{\phi}))\cdot(\nabla\Omega)
+C_\alpha\uv{\phi}\cdot\nabla\times [\alpha\nabla\times({\cal A}\uv{\phi})]~.
\end{eqnarray}
We continue to use the symbol $\eta_{\rm \null }$ for the total diffusivity,
with the understanding that within the convective envelope this now includes
the (dominant) contribution from the $\beta$-term of mean-field theory.
Three non-dimensional numbers have
materialized:
\begin{eqnarray}\label{eq:dynnum}
C_\alpha={\alpha_{\rm t} R\over\eta_{\rm \null }}~,\qquad
C_\Omega={\Omega_{\rm t} R^2\over\eta_{\rm \null }}~,\qquad
\Rm={\sv{{\rm t}} R\over\eta_{\rm \null }}~,
\end{eqnarray}
with $\alpha_{\rm t}$,
$\sv{{\rm t}}$, and $\Omega_{\rm t}$
as reference values for the $\alpha$-effect, meridional
flow and envelope rotation, respectively. 
The quantities $C_\alpha$ and $C_\Omega$ are
\indexit{dynamo!number} {\em dynamo numbers}, measuring the importance
of inductive versus diffusive effects on the right-hand side
of Eqs.~(\ref{eq:mfa})--(\ref{eq:mfb}). The
\indexit{magnetic!Reynolds number}
magnetic Reynolds number $\Rm$
here measures the relative importance of advection
versus diffusion in the transport
of $A$ and $B$ in meridional planes. \activity{{\em Show:} Relate the Rossby
  number in Eq.~(\ref{eq:ross}) to the dynamo number $C_\Omega$ and
  the magnetic Reynolds number $\Rm$ in Eq.~\ref{eq:dynnum}:
  $\Ro=\Rm/C_\Omega$. \mylabel{act:rossby}} 
Structurally, Eqs.~(\ref{eq:mfa})--(\ref{eq:mfb})
only differ from Eqs.~(\ref{eq:cowa})---(\ref{eq:cowb})
by the presence
of two new source terms on the right-hand side, both associated with the
$\alpha$-effect. The appearance of this term in Eq.~(\ref{eq:mfa})
is crucial for evading Cowling's theorem.''

%\subsubsection{Linear $\alpha\Omega$ dynamo solutions\label{ssec:lmf}}
In what follows in this section, we first look at a simplified, linear
mean-field dynamo model to illustrate the geometry and temporal
evolution. Later, we look at non-linearities that lead to
amplification and saturation of the field, and to the modulation of
the magnetic cycles. First, the linear model: \ors[III:6.2.1.1] ``In
constructing mean-field dynamos for the Sun, it has been a common
procedure to neglect meridional circulation, because it is a very weak
flow.  It is also customary to drop the $\alpha$-effect term on the
right-hand side of Eq.~(\ref{eq:mfb}) on the grounds that with
$R\simeq 7\times 10^{10}\,$cm, $\Omega_{\rm t}\sim 10^{-6}\,$rad
s${}^{-1}$, and $\alpha_{\rm t}\sim 10^2\,$cm\,s$^{-1}$, one finds
$C_\alpha/C_\Omega\sim 10^{-3}$, independently of the assumed (and
poorly constrained) value for $\eta_{\rm \null }$.  Equations
(\ref{eq:mfa})---(\ref{eq:mfb}) then reduce to the so-called
\indexit{dynamo!$\alpha\Omega$ model} {\boldmath$\alpha\Omega$ \bf
  dynamo equations}. In the spirit of producing a model that is
solar-like we use a fixed value $C_\Omega=2.5\times 10^4$, obtained by
assuming [an equatorial angular velocity of]
$\Omega_{\rm Eq}\simeq 10^{-6}\,$rad s${}^{-1}$ and
$\eta_{\rm \null }=50\,$km$^2$s$^{-1}$, which leads to a diffusion
time $\tau_{\rm d}=R^2/\eta_{\rm \null }\simeq 300\,$yr.

\begin{figure}[t]
\begin{center}
%\epsfxsize=\hsize
%\epsfbox{figures/ingred.eps}
\includegraphics[width=\hsize]{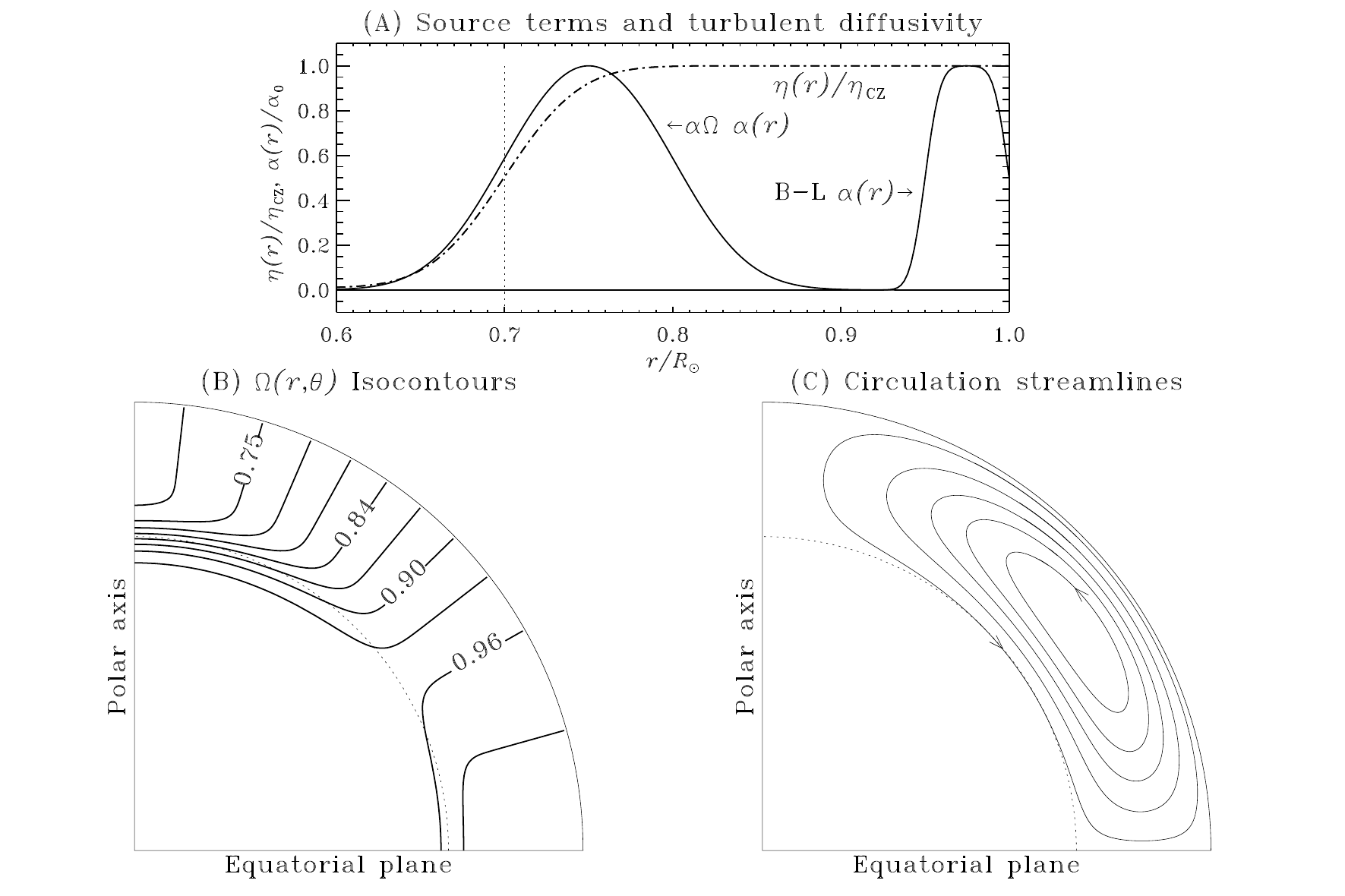}
\end{center}
\caption[Ingredients of dynamo models.]{
\label{fig:ingred}
Various 'ingredients' for the dynamo models constructed in this
chapter.  Part {\rm (A)} shows radial profiles of the total magnetic
diffusivity $\eta$ and poloidal source [terms: $\alpha(r)$ for the
$\alpha\Omega$ dynamo and for the Babcock-Leighton (B--L) dynamo].
Part {\rm (B)} shows contour levels of the rotation rate $\Omega(r,\theta)$
normalized to its surface equatorial value.  The dotted line is the
core-envelope interface at $r/R_\odot=0.7$.  Part {\rm (C)} shows streamlines of
meridional circulation, included in some of the dynamo models
discussed below. [Helioseismic studies suggest that the meridional
flow in the Sun is more complex than a single 'roll' of the flow, but
that there may be (at least) two stacked on top of each other. A key point
for a flux-transport dynamo is that the meridional flow at the base of
the convective envelope is equatorward. Fig.~III:6.1]}
\end{figure}

For the total magnetic diffusivity, we use
\indexit{magnetic!diffusivity}
a steep but smooth variation of $\eta_{\rm \null }$ from a high value ($\eta_{\rm \null CZ}$)
in the convection zone to a low value ($\eta_{\rm \null core}$) in the underlying
core [\ldots] A typical
profile is shown in Fig.~\ref{fig:ingred}A (dash-dotted line).
In practice, the core-to-envelope diffusivity ratio
$\Delta\eta_{\rm \null }\equiv \eta_{\rm \null core}/\eta_{\rm \null CZ}$ is treated as a model parameter,
with of course $\Delta\eta\ll 1$,
because we associate
$\eta_{\rm \null core}$ with the microscopic magnetic diffusivity, and
$\eta_{\rm \null CZ}$
with the presumably much larger mean-field turbulent diffusivity.
Taking at face values estimates from mean-field theory, one
should have $\Delta\eta_{\rm \null }\sim 10^{-9}$ to $10^{-6}$. The solutions
discussed below have $\Delta\eta_{\rm \null }=10^{-3}$ to $10^{-1}$,
which is still
small enough to illustrate
important effects of radial gradients in total magnetic diffusivity.

All solar dynamo models discussed in this chapter utilize the
helioseismically-calibrated solar-like parametrization
\indexit{Sun!differential rotation} of\indexit{solar!differential
  rotation} solar differential rotation [\ldots].  The
corresponding angular velocity contour levels are plotted in
Fig.~\ref{fig:ingred}B.  \indexit{Sun!differential rotation}
Such a solar-like differential
rotation profile is \indexit{differential rotation|seealso{solar}}quite complex from the point of view of dynamo
modelling, \indexit{differential rotation}in that it is characterized by multiple partially
overlapping shear regions: a rotational shear layer, straddling the
core-envelope interface, known as the {\em tachocline}, with
\indexit{Sun!tachocline} a strong positive radial shear in its
equatorial regions and an even stronger negative radial shear in its
polar regions, as well as a significant latitudinal shear throughout
the convective envelope and extending partway into the tachocline; for
a tachocline of half-thickness $w/R_\odot=0.05$, the mid-latitude
latitudinal shear at $r/R_\odot=0.7$ is comparable in magnitude to the
equatorial radial shear, and its potential contribution to toroidal
field production cannot be casually dismissed.

For the dimensionless function
\indexit{dynamo!$\alpha$ effect}$\alpha(r,\theta)$ we
use an expression [\ldots\ that] concentrates the $\alpha$-effect in the bottom half of the
envelope, and lets it vanish smoothly below, just as the net magnetic
diffusivity does (see Fig.~\ref{fig:ingred}A).  Various lines of
argument point to an $\alpha$-effect peaking in the bottom half of the
convective envelope, because there the convective turnover time is
commensurate with the solar rotation period, a most favorable setup
for the type of toroidal field twisting at the root of the
$\alpha$-effect (see Fig.~I:3.5).  [The choice made here for
$\alpha(r,\theta)$ scales with latitude as $\cos\theta$, which]
reflects the hemispheric dependence of the \indexit{Coriolis force}Coriolis force,
which also suggests that the $\alpha$-effect should be positive in the
Northern hemisphere.  The dimensionless number $C_\alpha$, which
measures the strength of the $\alpha$-effect, is treated as a free
parameter of the model. [\ldots]

In such linear $\alpha\Omega$ models the onset of dynamo activity
turns out to be controlled by the {\em product} of $C_\alpha$ and
$C_\Omega$:
\begin{eqnarray}\label{E5.13}
D\equiv C_\alpha\times C_\Omega ={\alpha_{\rm t}\Omega_{\rm t} R^3\over\eta_{\rm \null CZ}^2}~.
\end{eqnarray}
with positive growth rates materializing above a
threshold value known as the
\indexit{dynamo!number}
{\em critical dynamo number}.
[\ldots]

\begin{figure}[t]
\begin{center}
%\epsfxsize=4.0truein
%\epsfbox{figures/aoeigfunc.eps}
\includegraphics[width=4.0truein]{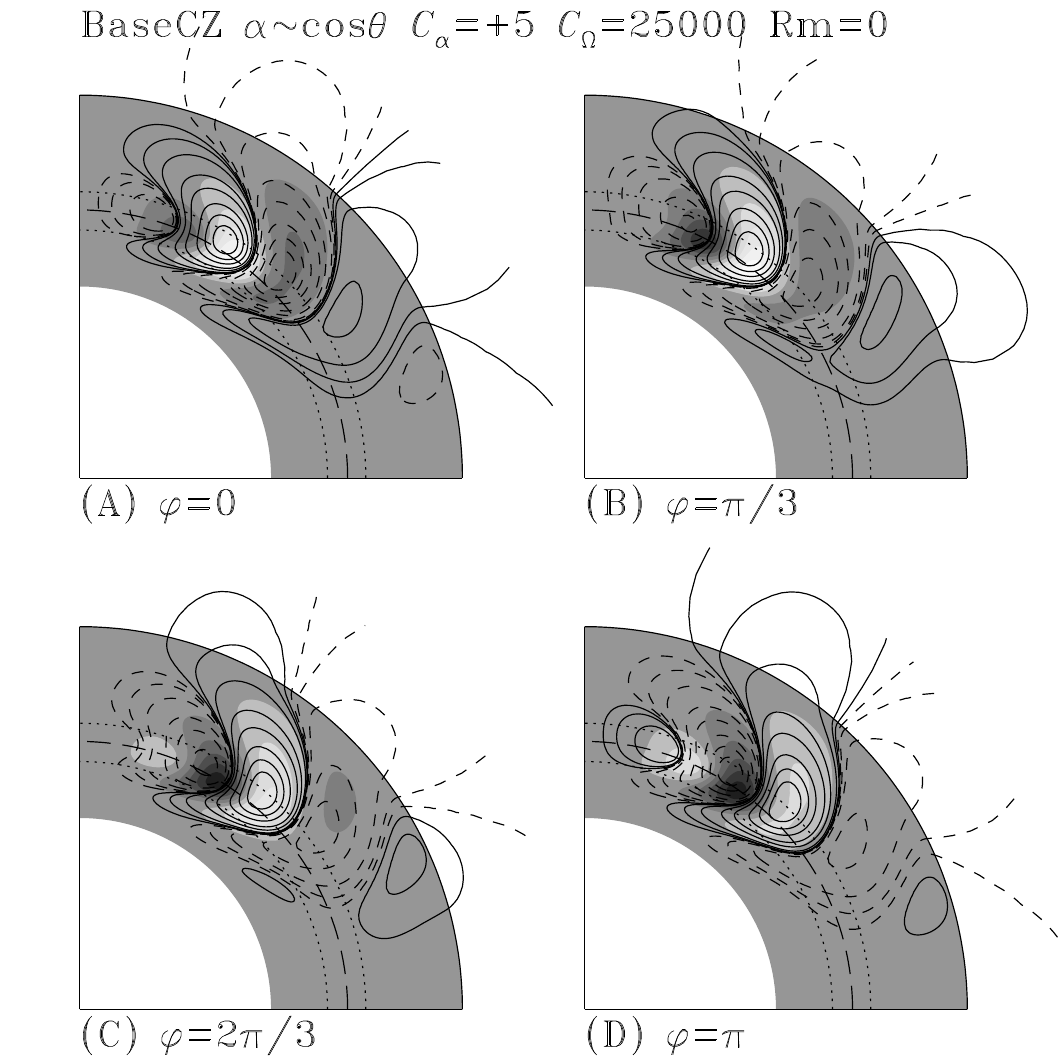}
\end{center}
\caption[A minimal linear $\alpha\Omega$
dynamo solution.]{\label{fig:aolin}
Four snapshots in meridional planes of our minimal linear $\alpha\Omega$
dynamo solution with defining parameters
$C_\Omega=25000$, 
%(see Eq.~6.11),
%(see Eq.~\ref{eq:dynnum}),
$\Delta\eta_{\rm \null }=0.1$, 
%(see Eq.~6.12),
%(see Eq.~\ref{eq:eta}),
and $\eta_{\rm \null CZ}=50\,$km$^2$/s. With
$C_\alpha=+5$, this is a mildly supercritical solution, with
oscillation frequency $\omega\simeq 300\,\tau_{\rm d}^{-1}$ (see
Eq.~\ref{eq:difftime}).  The toroidal field is plotted as filled
contours (gray to black for negative $B$, gray to white for positive
$B$, normalized to the peak strength and with increments
$\Delta B=0.2$), on which poloidal field lines are superimposed (solid
for clockwise-oriented field lines, dashed for counter-clockwise
orientation). The long-dashed line is the core-envelope interface at
$r/R_\odot=0.7$. [Fig.~III:6.2]}
\end{figure}

Figure \ref{fig:aolin} shows half a cycle of the dynamo solution, in
the form of snapshots of the toroidal (gray scale)
\indexit{dynamo!eigenfunction} and poloidal eigenfunctions (field
lines) in a meridional plane, with the symmetry axis defined by the
stellar rotation oriented vertically.  The four frames are separated
by a phase interval $\varphi=\pi/3$, so that panel (D) is identical to
panel (A) except for reversed magnetic polarities in both magnetic
components [halfway through the cycle with period
$P_{\rm cycle}=2\pi/\omega$]. Such linear eigensolutions leave the
absolute magnitude of the magnetic field undetermined, but the
relative magnitude of the poloidal to toroidal components is found to
scale approximately as $|C_\alpha/C_\Omega|$.
\indexit{dynamo!cycle period}

The [model's magnetic field] is concentrated in the vicinity of the
core-envelope interface, and has very little amplitude in the underlying,
low-diffus\-iv\-ity radiative core.
This is due to the oscillatory nature
of the solution, which restricts penetration into the core to a distance
of the order of the
\indexit{electromagnetic skin depth}
electromagnetic skin depth $\ell_{\rm skin}=\sqrt{2\eta_{\rm \null core}/\omega}$.
Having assumed $\eta_{\rm \null CZ}=50\,$km$^2$s$^{-1}$, with
$\Delta\eta_{\rm \null }=0.1$,
a dimensionless dynamo frequency
$\omega\simeq 300$ corresponds to $3\times 10^{-8}\,$s${}^{-1}$,
so that $\ell_{\rm skin}/R\simeq 0.026$, quite small indeed.

Careful examination of Fig.~\ref{fig:aolin}(A)$\to$(D) also reveals
that the toroidal-poloidal flux systems
present in the shear layer first show up at high latitudes,
and then {\em migrate equatorward} to
finally disappear at mid-latitudes in the course
of the half-cycle.
\indexit{dynamo!waves}
These {\em dynamo waves} travel in a direction given by
%%
%\begin{eqnarray}
%\label{eq:wave}
$\alpha\nabla\Omega\times\uvec{_\phi},$
%\end{eqnarray}
%%
{\em i.e.,}  along contours of equal angular velocity, a result known as
the\indexit{Parker-Yoshimura sign rule} {\em Parker-Yoshimura sign
  rule}.  Here with a negative $\tderp{\Omega}{r}$ in the
high-latitude region of the tachocline, a positive $\alpha$-effect
results in an equatorward propagation of the dynamo wave, in
qualitative agreement with the observed equatorward drift of the
latitudes of sunspot emergences as the solar cycle unfolds (see
Fig.~\ref{figure:butterfly}).''

%\subsubsection{Kinematic $\alpha\Omega$ models with $\alpha$ quenching}\label{ssec:nlmf}
\label{sec:alphaquenching}

``Obviously, \ors[III:6.2.1.2]  the exponential growth characterizing
supercritical linear solutions must stop \indexit{dynamo!nonlinearity}
once the Lorentz force associated with the growing magnetic field
becomes dynamically significant for the inductive flow.  Because the
solar surface and internal differential rotation show little variation
with the phase of the solar cycle, it is usually assumed that magnetic
back-reaction occurs at the level of the $\alpha$-effect. In the
mean-field spirit of {\em not} solving dynamical equations for the
small-scales, it has become common practice to introduce an {\it ad
  hoc} algebraic nonlinear quenching of $\alpha$ directly on the
mean-toroidal field $B$ by writing:
\begin{eqnarray}
\label{eq:alpquench}
\alpha\to\alpha(B)={\alpha_{\rm t}\over 1+({B}/\Beq)^2}~.
\end{eqnarray}
where $\Beq=(4\pi\rho u_t^2)^{1/2}$ is the equipartition field
strength, of order $10^4$\,G at the base of the solar convective
envelope.  Needless to say, this simple \indexit{dynamo!$\alpha$ quenching}
{\bf \boldmath$\alpha$-quenching} formula is an {\em extreme}
oversimplification of the complex interaction between flow and field
that is known to characterize MHD turbulence, but its wide usage in
solar dynamo modeling makes it the nonlinearity of choice for the
illustrative purpose of this [chapter: with this description, the
only MHD equation that needs solving to experiment with dynamo action
--~as we do here~-- is the induction equation Eq.~(\ref{induction})
that is now subjected to a parameterized coupling between the small-scale
flow and field that may or may not be an appropriate approximation of
reality. Note that $\alpha$ can, and in many models now is, time
dependent, leading to what is called 'dynamical $\alpha$-quenching'.]

Introducing $\alpha$-quenching in our model renders the
$\alpha\Omega$ dynamo equations nonlinear, so that solutions
are now obtained as initial-value problems starting from
an arbitrary seed field of very low amplitude, in the sense that
$B\ll B_{\rm eq}$ everywhere in the domain. [\ldots] 
At early times, $B\ll\Beq$ and the equations are effectively
linear, leading to exponential
growth [\ldots].
Eventually, however, $B$ becomes comparable to
$\Beq$ in the region where the $\alpha$-effect operates, leading to a break
\indexit{dynamo!amplitude saturation}
in exponential growth, and eventual saturation.

The saturation energy level increases with increasing $C_\alpha$,
an intuitively satisfying behavior because solutions with larger $C_\alpha$
have a more vigorous poloidal source term.
\indexit{dynamo!cycle period}
The cycle frequency for these
solutions is
very nearly independent of the dynamo number, and
is slightly {\em smaller} than the frequency of the linear critical
mode (here by some $10-15$\%), a behavior that is typical of
kinematic $\alpha$-quenched mean-field dynamo models.
Yet the overall form of the dynamo solutions very closely resembles that of
the linear eigenfunctions plotted in Fig.~\ref{fig:aolin}.''

\ors[III:6.2.1.3] ``The $\alpha$-quenching expression in Eq.~(\ref{eq:alpquench})
implies that dynamo
action saturates once the mean, dynamo-generated large-scale magnetic
field
reaches an energy density comparable to that of the driving
small-scale turbulent fluid motions.
However, various
calculations and numerical simulations have indicated that
long before the mean toroidal field $B$ reaches this strength,
the helical turbulence reaches equipartition with the
\emph{small-scale} turbulent component of the magnetic field.
Such calculations also suggest that the ratio between
the small-scale and mean magnetic components should
itself scale as $\Rm^{1/2}$, where $\Rm=v_{\rm t}L_{\rm t}/\eta_{\rm \null }$
is a magnetic Reynolds number based on the turbulent speed but
\emph{microscopic}
magnetic diffusivity. This then leads to the
alternative quenching expression
\begin{eqnarray}
\label{eq:stralpquench}
\alpha\to\alpha(B)={\alpha_{\rm t}\over 1+\Rm(B/\Beq)^2}~,
\end{eqnarray}
known in the literature as {\bf strong \boldmath$\alpha$-quenching} or
\indexit{dynamo!$\alpha$ quenching} \emph{catastrophic quenching} (see
Ch.~I:3 in Vol.~I).  Because $\Rm\sim 10^{8}$ in the solar
convection zone, this leads to quenching of the $\alpha$-effect for
very low amplitudes of the mean magnetic field, of order
$0.1\,$G. Even though significant field amplification is likely in the
formation of a toroidal flux rope from the dynamo-generated magnetic
field, we are now a very long way from the $10^4-10^5$\,G demanded by
simulations [needed for buoyantly rising flux ropes to survive
emergence and to eventually lead to] sunspot formation.

[One] way out of this difficulty exists
in the form of
\indexit{dynamo!interface} {\bf interface dynamos}.
The idea is beautifully
simple: to produce and
store the toroidal field away from where the $\alpha$-effect
is operating. [\ldots]
in a situation where a radial shear and $\alpha$-effect are
segregated on either side of a discontinuity in magnetic
diffusivity taken to coincide with the core-envelope
interface,
the constant coefficient, cartesian form of the $\alpha\Omega$
\indexit{dynamo!$\alpha\Omega$ model}
dynamo equations
support solutions
in the form of traveling
\indexit{dynamo!waves}
surface waves localized on the
discontinuity in diffusivity. For
supercritical dynamo waves,
the ratio of peak toroidal field strength
on either side of the discontinuity surface is found to
scale as $(\eta_{\rm \null CZ}/\eta_{\rm \null core})^{-1/2}$.
With
the core diffusivity $\eta_{\rm \null core}$ equal to the microscopic value, and if
the envelope diffusivity is of turbulent origin so that
$\eta_{\rm \null CZ}\sim L_{\rm t} v_{\rm t}$, then the toroidal field strength
ratio scales as $\sim (v_{\rm t}L_{\rm t}/\eta_{\rm \null core})^{1/2}\equiv \Rm^{1/2}$.
This is precisely the factor
needed to bypass strong $\alpha$-quenching, at least
as embodied in Eq.~(\ref{eq:stralpquench}).''

\indexit{meridional circulation} So far, this discussion has ignored
the large-scale flow system known as meridional circulation. Such a
flow \ors[III:6.2.1.4] ``is unavoidable in turbulent, compressible
rotating convective shells.  The $\sim 15\,$m s$^{-1}$ poleward flow
observed at the surface has been detected helioseismically, down to
$r/R_\odot\simeq 0.85$ without significant departure from the poleward
direction, except locally and very close to the surface, in the
vicinity of active region belts.  Mass conservation requires
an equatorward flow deeper down [(helioseismic measurements suggest
that there may be two meridional overturning cells stacked within the
convective envelope, but confirmation is still pending of what is a
challenging measurement close to the noise levels of
helioseismology)].

Meridional circulation can bodily transport the dynamo-generated magnetic
field (terms $\propto {\bf v}_{\rm p}\cdot\nabla$
in Eqs.~(\ref{eq:cowa})--(\ref{eq:cowb})).
At low circulation speeds,
the primary effect is a Doppler shift of the
\indexit{dynamo!waves}
dynamo wave, leading to
a small change in the
\indexit{dynamo!cycle period}
cycle period and equatorward concentration of
the activity belts.
However, for a (presumably) solar-like equatorward return flow that
is vigorous enough, it
can overpower the Parker-Yoshimura propagation rule
and produce equatorward propagation no matter
what the sign of the $\alpha$-effect is.
The behavioral turnover from dynamo wave-like solutions
sets in when
the circulation speed in the dynamo region becomes comparable to the propagation
speed of the dynamo wave.
In this
\indexit{dynamo!advection dominated}
advection-dominated regime,
the cycle period loses sensitivity to the assumed turbulent
diffusivity value, and becomes determined primarily by
the circulation's turnover time.
Solar cycle models achieving
equatorward migration of activity belts in this manner are often called
\indexit{dynamo!flux transport}
{\bf flux transport dynamos}. [\ldots]

One interesting consequence [of meridional circulation] is that
induction of the toroidal field is now effected primarily by the {\it
  latitudinal} shear within the tachocline, with the radial shear,
although larger in magnitude, playing a lesser role because
$B_r/B_\theta\ll 1$.
\begin{figure}[t]
%\epsfxsize=\hsize
%\epsfxsize=10.2cm
%\epsfbox{figures/medp6cuts-BL.eps}
\centerline{\includegraphics[width=10.2cm]{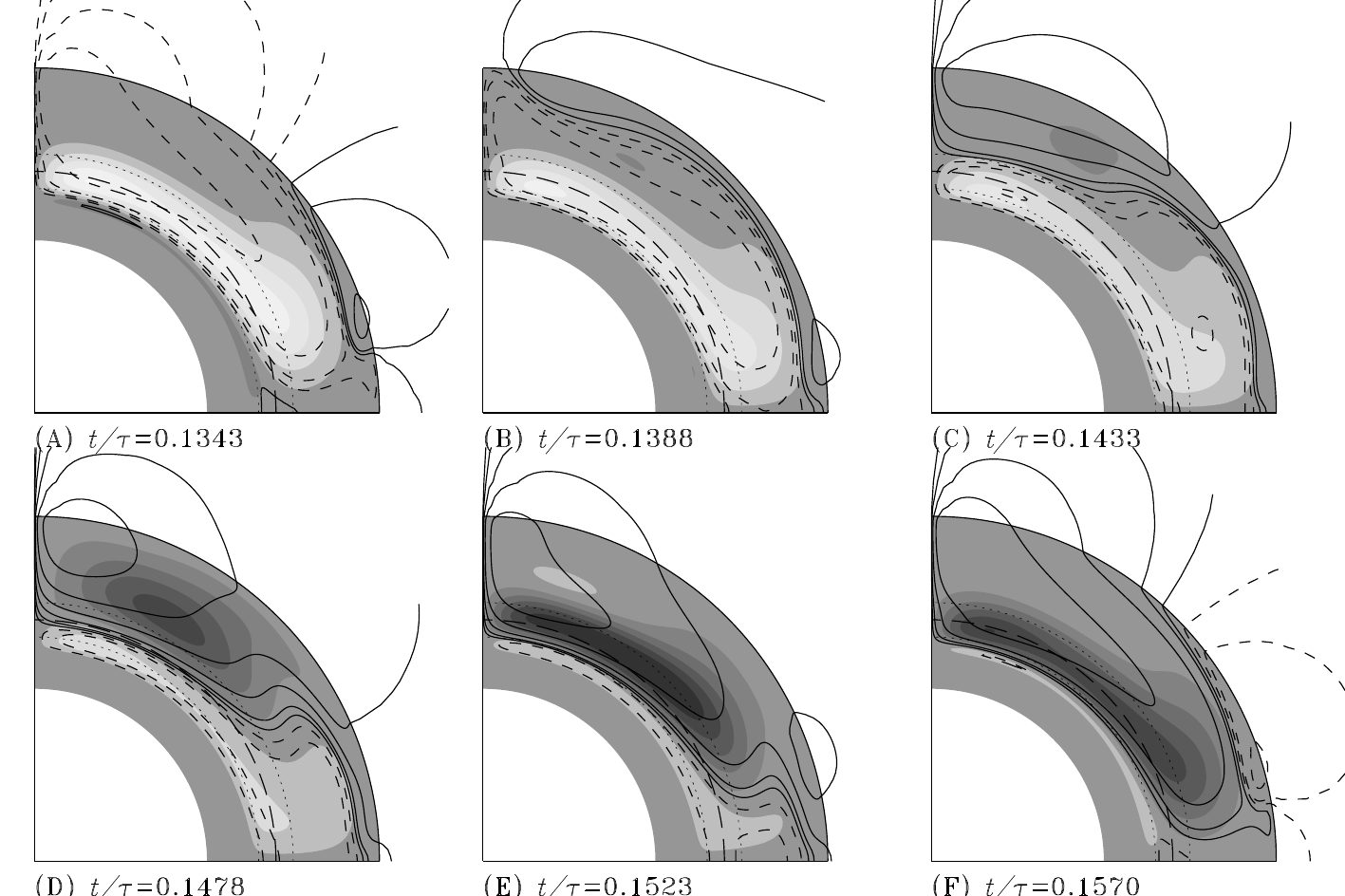}}
\caption[A Babcock-Leighton
dynamo solution.]{
[Snapshots covering half a cycle of a Babcock-Leighton
dynamo solution.
The grey-scale coding of the
toroidal field and poloidal field lines is as in Fig.~\ref{fig:aolin}.]
This solution uses the same differential
rotation, magnetic diffusivity and meridional circulation profile as
for the advection-dominated $\alpha\Omega$ solution
%of Sect.~6.2.1.4, %Sect.~\ref{ssec:aocm}, 
but now with the non-local surface source term
as shown in Fig.~\ref{fig:ingred}(A) in the curve labeled 'B--L',
% defined through Eq.~(6.21), 
%Eq.~(\ref{eq:BLsource}), 
with parameter values
$C_\alpha=5$, $C_\Omega=5\times 10^4$, $\Delta\eta_{\rm \null }=0.003$, $\Rm=840$.
Note the strong amplification of the surface polar fields, and
the latitudinal stretching of poloidal field lines by the meridional
flow at the core-envelope interface. [Fig.~III:6.6]
%Snapshots covering half a cycle of a Babcock-Leighton dynamo solution.
%Grey-scale coding of the toroidal field and poloidal field lines as in
%Fig.~\ref{fig:aolin}.  This solution uses the same differential
%rotation, magnetic diffusivity and meridional circulation profile as
%for the advection-dominated $\alpha\Omega$ solution of
%Sect.~\ref{ssec:aocm}, but now with the non-local surface source term
%defined through Eq.~(\ref{eq:BLsource}), with parameter values
%$C_\alpha=5$, $C_\Omega=5\times 10^4$, $\Delta\eta=0.003$, $\Rm=840$.
%Note again the strong amplification of the surface polar fields, and
%the latitudinal stretching of poloidal field lines by the meridional
%flow at the core-envelope interface.
\label{fig:BLsoln}}
\end{figure}
The meridional flow also has a profound impact on the magnetic field evolution
at $r=R$, as it concentrates the poloidal field in the polar regions. This leads to
a large amplification factor through magnetic flux conservation, so
[these dynamo models] are typically characterized
by very large
\indexit{Sun!polar magnetic field}
polar field strengths, here some 20\% of the toroidal field
magnitude in the tachocline, even though
we have here $C_\alpha/C_\Omega =10^{-6}$.
This concentrated poloidal field,
when advected downwards to the polar regions of the tachocline, is
responsible for the strong polar branch often seen in the time-latitude
\indexit{butterfly diagram} diagram of dynamo solutions including
a rapid meridional flow.
This difficulty can be alleviated, at least in part, by a number
of relatively minor modifications to the model, such as the addition
of a high-$\eta_{\rm \null }$ subsurface layer, or displacement of the meridional
flow cell towards lower latitudes, thus reducing the degree of polar
convergence. [\ldots]''

%\subsection{Solar cycle models based on active region decay\label{sec:BLsoldyn}}

Yet another incarnation of solar cycle models is based on active
region decay and dispersal. These go back to 1961 when
\indexit{dynamo!Babcock-Leighton}Babcock
\ors[III:6.2.2] ``suggested that the polarity reversals
\indexit{Sun!polar magnetic field} of the high-latitude surface
magnetic field are driven by the accumulation of magnetic fields
released at low latitudes by the decay of bipolar magnetic regions.
\indexit{Sun!active regions} Figure~\ref{figure:rotmagsim} shows a
numerical simulation illustrating this process, which leads to the
buildup of a net poloidal hemispheric flux because the trailing member
of the pair tends to be located at higher latitudes than the leading
component, a pattern known as \indexit{Joy's rule} Joy's rule, and
therefore are subjected to less transequatorial dissipative flux
cancellation than the leading members of the bipolar pair. Babcock
went on to argue that in conjunction with shearing by
\indexit{solar!differential rotation}differential
rotation, this could explain the observed patterns of solar cycle
polarity reversals.  In subsequent years [he] turned this idea into a
{\it bona fide} solar cycle model, known since as
\indexit{dynamo!Babcock-Leighton} the {\bf Babcock-Leighton
  model}.  [\ldots] The key point, from the dynamo perspective, is
that the Babcock-Leighton mechanism taps into the (formerly) toroidal
flux in the bipolar magnetic region to produce a poloidal magnetic
component, and so can act as a source term on the right-hand side of
Eq.~(\ref{eq:cowa}).  [\ldots] \sactivity{$\circledS$ {\em Show:} One of the basic
  concepts behind Babcock's idea is that magnetic field at the solar
  surface is largely advected like a scalar quantity. Consequently,
  the field disperses in the random motions of the surface convection
  (with an equivalent diffusion coefficient of
  $D\approx 250$\,km$^2$/s) subject to the large-scale advection of
  the differential rotation and meridional flow. (a) To see how this can
  be, use the 'ideal' version of Eq.~(\ref{induction}) (which ignores
  resistivity) and assume that the field is always vertical to the
  surface (a good approximation to the observed photospheric field,
  except during emergence and cancellation; a result of the buoyancy
  of flux bundles --~see Activity~(\ref{act:buoy})~-- and show that it
  has the same properties as Eq.~(\ref{continuity}) for the advection
  of a scalar (without the source and loss terms in that
  version). Note that this formulation is linear, so that you can
  think about N and S polarities as diffusing separately, and can sum
  these results to obtain the net result; this helps visualize why the
  active-region tilt angle is important in reversing the polar fields
  from cycle to cycle. (b) Question: with this value of $D$, what is the
  characteristic time scale for flux to disperse over the solar
  surface (hint: Eq.~\ref{eq:diffusiontime})?  (c) With that in mind, how
  important is the meridional advection from equator to pole (with a
  characteristic velocity of 10\,m/s) in transporting the field within
  the duration of a solar cycle?  \mylabel{act:surfacediffusion}
  Remember that Activity~\ref{act:meanfieldequation} shows how the
  diffusion coefficient $\beta$ associated with (super-)granular
  random walk adds to the molecular/resistive diffusion coefficient
  $\eta_{\rm \null }$.\mylabel{act:bl}\solution{bl}}\activity{{\em Show:} Joy's
  rule, that the leading polarities (in the direction of rotation) of
  active regions emerge statistically closer to the equator than the
  trailing polarities (see Fig.~\ref{figure:bipoles}, for example), is
  the reason why eventually flux of the trailing polarity builds up a
  polar cap that reaches its maximum strength at cycle minimum. For
  some interval around that time, the bulk of the heliospheric field
  originates from the polar caps. Estimate the total flux in the solar
  wind (assuming an isotropic flux density at Earth orbit; use
  Table~\ref{tab:wind-stats}); what is the corresponding magnetic flux
  density averaged over the solar surface? This total flux is the equivalent of only a few
  large active regions (Fig.~\ref{figure:bipoles}) although it is in
  fact composed of a fraction of the flux from the ensemble of all
  bipolar regions emerging over a cycle.\mylabel{act:joy}}

To the degree that a positive dipole moment is being produced from a
toroidal field that is positive in the N-hemisphere, this is
operationally equivalent to \indexit{dynamo!$\alpha$ effect}a positive
$\alpha$-effect in mean-field theory. In both cases the Coriolis force
is the agent imparting a twist on a magnetic field; with the
$\alpha$-effect this process occurs on the small spatial scales and
operates on individual magnetic field lines. In contrast, the
Babcock-Leighton mechanism operates on the large scales, the twist
being imparted via the \indexit{Coriolis force}Coriolis force acting on the flow generated
along the axis of a buoyantly rising \indexit{magnetic!flux rope}
magnetic flux tube that, upon emergence, gives rise to sunspot
pairs. \activity{{\em Consider:} (a) Why are solar photospheric flux tubes
  buoyant (hint: look back at Activity~(\ref{act:buoy})? What is the
  maximum density contrast between interior and exterior for a 1\,kG
  flux tube? (b) For an essentially zero-density flux tube at the solar
  surface, show that the buoyancy force per unit length (causing the
  tube to buoy towards vertical) dominates the dynamic pressure force
%  $2a \rho v^2$ 
  exerted by a convective flow (which could bend the tube away from
  vertical) of $v=1$\,km/s for any tube with diameter $2a$ exceeding
  just a few km. Indeed, observations show flux tubes to be
  essentially vertical to the photosphere (except around emergence and
  collisional cancellation when magnetic curvature forces of the field
  arching from one polarity to the opposite one are strong). \mylabel{act:buoy2} } 

Numerous dynamo models based on this mechanism of poloidal field
regeneration have been constructed, based on the axisymmetric
mean-field dynamo equations but with the $\alpha$-effect replaced by a
suitably designed source term on the right-hand side of
Eq.~(\ref{eq:mfa}).  One important difference with the mean-field
$\alpha\Omega$ models considered earlier is that the two source
regions are now spatially segregated: production of the toroidal field
takes place in or near the tachocline, as before, but now production
of the poloidal field is restricted to the surface layers. A transport
mechanism is then required to link the two source regions for a dynamo
loop to operate.  \indexit{dynamo!Babcock-Leighton}
\indexit{dynamo!flux transport} [\ldots\ Most 
Babcock-Leighton models use the 
\indexit{meridional circulation} meridional circulation for this, which acts]
as a form of conveyor belt, concentrating to high latitudes the
surface magnetic fields released by the decay of active regions, and
dragging it down to the tachocline where shearing by differential
rotation leads to the buildup of a new toroidal flux system, and thus
to the onset of a new sunspot cycle. [\ldots] 

Figure~\ref{fig:BLsoln} shows a series of meridional-plane snapshots
of one such Babcock-Leighton dynamo solution, covering one sunspot
cycle and starting approximately at sunspot maximum (based on magnetic
energy as a proxy for sunspot number). Surface poloidal flux from the
current cycle has begun to build up at low latitudes, and is rapidly
swept to the pole, with polarity reversal of the polar field taking
place shortly thereafter (panel B).  As with the advection-dominated
$\alpha\Omega$ solution discussed above, this solution is
characterized by strong surface polar fields resulting from the
poleward transport by the meridional flow of the poloidal component
produced at lower latitudes, and the equatorward propagation of the
toroidal field in the tachocline is also driven by the meridional
flow.  \indexit{dynamo!cycle period} The turnover time of the
meridional flow is here again the primary determinant of the cycle
period. With $\eta_{\rm \null }=30\,$km$^2\,$s$^{-1}$, this solution
has a nicely solar-like half-period of 12.4 yr.  All in all, this is
once again a reasonable representation of the cyclic spatiotemporal
evolution of the solar large-scale magnetic field.'' \activity{{\em
    Advanced/Group:} Further study:
  \href{https://ui.adsabs.harvard.edu/abs/2017ApJ...834..133L/abstract}{The
    work} by \citet{2017ApJ...834..133L} describes an interesting
  dynamo experiment in which the Babcock-Leighton concept is combined
  with surface flux transport modeling (see Activity~\ref{act:bl}) to
  create a quasi-regular dynamo in which convection-induced
  fluctuations on the tilt angle of emerging active regions
  (perturbations on Joy's rule, see Activity~\ref{act:joy}) provide
  the stochastic noise that can lead to cycle-to-cycle differences and
  even extended periods of weak cycling (as in the Maunder Minimum
  period for the Sun), something also reported on in
  \href{https://ui.adsabs.harvard.edu/abs/2017ApJ...847...69K/abstract}{a
    paper} by \citet{2017ApJ...847...69K}.  }

There are yet other non-linearities that can be considered in
solar/stellar dynamo models. For example, in \ors[III:6.2.3] ``the
presence of \indexit{stratification}stratification and rotation, a number of hydrodynamical
(HD) and magnetohydrodynamical (MHD) instabilities associated with the
presence of a strong toroidal field in the stably stratified,
radiative portion of the tachocline can lead to the
\indexit{dynamo!instability-driven} growth of disturbances with a net
helicity, which under suitable circumstances can produce a toroidal
electromotive force, and therefore act as a source of poloidal
field. Different types of solar cycle models have been constructed in
this manner.  In nearly all cases the resulting dynamo models end up
being described by something closely resembling the axisymmetric
mean-field dynamo equations, the novel poloidal field regeneration
mechanisms being once again subsumed in an $\alpha$-effect-like source
term appearing of the right-hand side of Eq.~(\ref{eq:mfa}).'' More on
this in Sect.~III:6.2.3. \activity{{\em Look up:} Make a summary of the essential
  distinctions between the dynamo concepts discussed up to this point:
$\alpha\Omega$ (with or without $\alpha$ quenching, which itself can
be strong/catastrophic or not); interface; flux-transport; and Babcock-Leighton.}

The models discussed thus far all lead to a steadily repeating
magnetic cycle, where appropriate after an initial growth phase. The
Sun, however, displays a rather erratic modulation of its activity
from one cycle to the next, \ors[III:6.3] ``and certain aspects of the
observed fluctuations may actually hold important clues as to the
physical nature of the dynamo process.''  Section~III:6.3 discusses
some of these processes, including those that could be responsible for
long-term modulations of the solar cycle pattern (such as the Maunder
Minimum): stochastic effects (with strong evidence from both models
and observations for the importance of the scatter on tilt angles of
active regions that reflect the influence of random convective flows
during the rise of the flux to the surface), back reaction of the
field on the flow patterns, and time delays through transport
processes. Section~III:6.3.5 discusses issues related to the
forecasting of the solar cycle based on precursor signatures.

%\section{Stellar dynamos}
\section{Dynamos in other stars}\label{chap:dynstar}

\ors[III:6.4] ``Figure \ref{fig:msstruct} illustrates, in schematic
form, the internal structure of main-sequence stars, more
\indexit{dynamo!stars}specifically the presence or absence of
\indexit{star!dynamo}convection zones.  \indexit{star!convection} A
$G$-star like the Sun has a thick outer convection zone, spanning the
outer 30\% in radius. As one moves to lower masses, the relative
thickness of the convective envelope increases until, somewhere around
spectral type $M$5, stars become fully convective [(see
Fig.~\ref{fig:acthrd} for an HR diagram and indications of spectral
types)].  Moving from the Sun to higher masses, the convective
envelope becomes ever thinner, until somewhere around spectral type
$A$0 it essentially vanishes. However, at around the same spectral
type hydrogen [fusion] switches from the proton-proton (or $p$-$p$)
chain to the CNO cycle, for which nuclear reaction rates are much more
sensitively dependent on temperature. Core energy release becomes
strongly depth-dependent, leading to convectively unstable temperature
gradients.  The resulting small convective core grows in size as one
moves up to larger masses. In an early $B$-star of solar metallicity,
the convective core spans the inner 25\% or so in radius of the star.

%\subsection{Early-type stars}\label{ssec:Etypedyn}

Main-sequence stars of the $O$ and $B$ [spectral type] combine
vigorous \indexit{star!early-type, core convection} core-convection
and high rotation rates, which makes dynamo action more than
likely. This expectation has been amply confirmed by 3D MHD numerical
simulations of dynamo action in the convective cores of massive
stars. [\ldots] All these core dynamo models have one thing in common:
the large [diffusivity contrast
$\eta_{\rm \null core}/\eta_{\rm \null envelope}$ between the
convective core and the stably \indexit{stratification}stratified envelope] leads to a 'trapping' of
the magnetic field in the lower part of the radiative envelope, a
direct consequence of the difficulty experienced by an
externally-imposed magnetic field to diffusively penetrate a good
electrical conductor [(analogous to, but here the inverse of, the
'skin depth' issue that was discussed in Section~\ref{ssec:mfsoldyn}  for a
cooler star) \ldots] This long-recognized property of stellar core
dynamos represents a rather formidable obstacle to be bypassed if the
magnetic fields generated by dynamo action in convective cores are to
become observable at the stellar surface [\ldots\ In fact, in] a
time-dependent situation where the core dynamo 'turns on' at or
shortly before [a young star settles into a stable equilibrium
represented by] the arrival on the \indexit{main sequence}zero-age main sequence, the time
needed for the magnetic field to resistively diffuse to the surface
can become larger than the star's main-sequence lifetime, for masses
in excess of about $5\,M_\odot$.'' (More on formation and evolution of
stars in Ch.~\ref{ch:evolvingstars}.)

%\subsection{A-type stars\label{ssec:Adyn}}

\ors[III:6.4.2] ``Stars with spectral types ranging from late-$B$ to
early-$F$ \regfootnote{In stellar parlance, 'late' means 'cooler' and
  'early' hotter. 'late-$B$' thus refers to $Bn$-type stars on the
  cooler side of the HR diagram, with digits $n$ closer to 9 than to
  0. 'Late type stars' is often used synonymously with 'cool stars',
  which refers to stars with convective envelopes immediately below
  their surface; see Fig.~\ref{fig:acthrd}. \label{note:earlylate}} stand out
\indexit{star!intermediate mass} as the least likely to support
dynamo action, because they lack a convective region of substantial
size. This squares well with various lines of observations; in
particular, main-sequence $A$-stars are among the most 'magnetically
quiet' stars in the HR diagram. A subset of late-$B$ and $A$ stars,
namely the slowly-rotating, chemically peculiar $A$p/$B$p stars, do
show strong magnetic fields, but even those show no sign of anything
even mildly analogous to solar activity. The single pattern of
temporal evolution noted is a decrease, by factors of 2 to 3, in the
overall strength of the surface field, most prominent in the early
stages of main-sequence evolution.  This seems compatible with the
idea of diffusive decay of residual higher-degree eigenmodes, and slow
decreases associated with flux conservation as the stars slowly expand
in the course of their main-sequence evolution.  For these reasons,
the fossil field hypothesis remains the favored explanatory
\indexit{fossil magnetic fields} model for the magnetic field of $A$p
stars.  It is also quite striking that the high field strength
observed in $A$p stars (a few times $10^4$\,G), in magnetized white
dwarfs ($\sim 10^9\,$G), and in the most intensely magnetized neutron
stars ($\sim 10^{15}\,$G) all amount to [a] total surface magnetic
flux $\sim 10^{27}\,$Mx, lending support to the idea that these high
fields can be understood from simple flux-freezing arguments [along an
evolutionary timeline for these objects] (see Ch.~I:3). [\ldots]''

%\subsection{Solar-type stars\label{ssec:solarstars}}

\ors[III:6.4.3] ``Until strong evidence to the contrary is brought to
the fore, we are allowed to assume that late-type stars $^{\ref{note:earlylate}}$ with a thick
convective envelope overlying a radiative core host a solar-type
dynamo.  \indexit{star!solar-type} Observationally, a lot of what we
know regarding dynamo activity in solar-type stars comes from the Mt.\
Wilson Ca\,H$+$K survey[, a survey that focuses on a pair of strong
resonance lines, which are known as the H and K lines, so named by
Fraunhofer during early spectroscopic studies, and which were later
found to be associated with singly ionized calcium; their signal
reflects the chromospheric activity of a star.]  Two important pieces
of information can be extracted from these data, as constraints on
dynamo models. The first is [that the overall level of Ca\,H$+$K emission,
which is taken as a measure of overall magnetic flux in the
photosphere,] is found to increase with rotation up to
$5-10$ times the solar rotation rate, after which saturation sets in
(see Ch.~\ref{ch:evolvingstars}).  The second is of course the cycle
period, for [the minority of stars that exhibit a regular cycle.]

The preponderance of strong magnetic field concentrated at high
latitude in rapidly rotating solar-type stars (see
Ch.~\ref{ch:evolvingstars}) is also a potentially interesting
discriminant. This can arise through channeling of buoyantly rising
toroidal flux ropes along the polar axis [prior to surfacing], or
efficient poleward transport of surface magnetic flux [after
surfacing. \ldots]'' \regfootnote{As computers continue to grow more
  powerful, 3D MHD dynamo simulations are advancing towards generating
  cycling large-scale fields in modeled stellar convection zones. An
  entry point for that literature is provided, for example, in the
  \href{https://ui.adsabs.harvard.edu/abs/2014ARA%26A..52..251C/abstract}{review
    of ``Solar Dynamo Theory''} by \citet{2014ARA&A..52..251C}. His contribution to
  \href{https://ui.adsabs.harvard.edu/abs/2010LRSP....7....3C/abstract}{Living Reviews in
    Solar Physics (\citep{2010LRSP....7....3C})} provides a description of advanced Babcock-Leighton
  type models that can now take observed magnetograms to provide
  forecasts of long-term trends of solar activity.}

%\subsection{Fully convective stars\label{ssec:fullconv}}

\ors[III:6.4.4] ``With fully convective stars we encounter potential
deviations from a solar-type dynamo mechanism; without a stably
stratified \indexit{stratification}tachocline and radiative core to store and amplify toroidal
flux ropes, the Babcock-Leighton mechanism, the tachocline
$\alpha$-effect and the flux-tube $\alpha$-effect all become
problematic.  Mean-field models based on the turbulent $\alpha$-effect
remain viable, but the dynamo behavior becomes dependent on the
presence and strength of internal \indexit{differential rotation!stars}differential rotation, about which
we really don't know very much in stars other than the Sun. The
full-sphere MHD simulations of an 'M-star in a box' are particularly
interesting in this respect, as they indicate that fully convective
stars do produce significant internal differential rotation and
well-defined patterns of hemispheric kinetic helicity, both supporting
the growth of a spatially well-organized large-scale magnetic
component.

Moving to even cooler stars, as the luminosity drops and surface
temperature falls below a few thousand K, the
\indexit{magnetic!Reynolds number}magnetic Reynolds number
in the surface layers is expected to eventually fall back towards
values approaching unity [because of the low degree of ionization at
such temperatures]. Small-scale turbulent dynamo action may shut
down, with magnetic activity then reflecting only the operation of a
deep-seated, large-scale dynamo. Whether this transition is sharp or
gradual, and whether it leads to well-defined observational
signatures, remain open questions.  There is certainly no {\it a
  priori} reason to presume that dynamo action should cease.  Indeed,
in some ways rapidly rotating very low-mass stars are getting closer
to the physical parameter regime characterizing the geodynamo.''

\section{Dynamos in terrestrial planets} 

\ors[III:7.5] ``Planetary \indexit{dynamo!terrestrial planets}dynamos share
with stellar dynamos that the basic physical concept for their
description is that of convection-driven magnetohydrodynamic flow in a
rotating spherical shell combined with the associated magnetic
induction effects. [\ldots]

Inside a shell of depth $d$ with an electrical conductivity
$\sigma_{\rm e}$ the fluid must move
with a sufficiently large characteristic velocity $v_{\rm t}$, so that the magnetic Reynolds number
\indexit{magnetic!Reynolds number} [in Eq.~(\ref{eq:reynolds})]
exceeds a critical value ${\cal R}_{\rm m,crit}$ in order to have a self-sustained dynamo.
The flow pattern must also be favorable for dynamo action, which requires a certain
complexity. In particular helical
\indexit{helicity!flow}
(corkscrew-type) motion with a large-scale order
in the distribution of right-handed and left-handed helices is suitable. The Coriolis force
\indexit{Coriolis force}
plays a significant part in the force balance of the fluid motion and influences
the pattern of convection. With this the requirement for \lq flow complexity\rq\,
seems to be satisfied and self-sustained dynamo action is possible above
${\cal R}_{\rm m,crit} \approx 40 - 50$.

At greater depth in the solar convection zone, the magnetic Reynolds
number reaches values of order $10^{9}$ for molecular values of the
magnetic diffusivity.  In the geodynamo $\Rm$ is approximately
1000. This fairly moderate value allows for the direct numerical
simulation of the magnetic field evolution without the need to use an
\lq effective diffusivity\rq\, or a parameterization of the induction
process through a turbulent $\alpha$-effect. [\ldots]
\indexit{dynamo!$\alpha$ effect}

The density in the Sun varies by many orders of magnitude and the
convection region spans many density scale heights. The density
changes associated with radial motion are thought to be
important. Flow helicity [(${\bf v}\cdot(\nabla\times{\bf v})$)]
arises in the Sun because of the action of the \indexit{Coriolis force}Coriolis force on
rising expanding and sinking contracting parcels of plasma.  Strong
magnetic flux tubes\indexit{flux!tube} have their own dynamics,
because the reduction of fluid pressure that compensates magnetic
pressure reduced their density and makes them buoyant. In contrast,
the dynamo region in Jupiter covers approximately one density scale
height \indexit{scale height!density} and much less in terrestrial
planets.  The two compressibility effects mentioned before probably do
not play a significant role in planetary dynamos.  Present geodynamo
models usually neglect the small density variation and assume
incompressible flow in the Boussinesq approximation
\indexit{Boussinesq approximation} (where density differences are only
taken into account for the calculation of buoyancy forces; see
Activity~\ref{act:boussinesq}).

Many models of the solar dynamo assume that most of magnetic field
generation occurs at the tachocline, \indexit{tachocline} the shear
layer between the radiative deep interior and the convection zone of
the Sun. For planetary dynamos the process of magnetic field
generation is thought to occur in the bulk of the convecting
layer. [\ldots] The relevant equation of motion for an incompressible
fluid [in a corotating frame of reference] is
\begin{equation}
\label{eq:motion}
\rho\frac{\partial\vec{v}}{\partial t} + \rho (\vec{v}\cdot{\bf \nabla})\vec{v}
+ 2\rho \Omega \, \uv{z} \times \vec{v} 
\, = \, \rho\zeta T g \uv{r} - {\bf \nabla} p^\prime + 
\frac{1}{4\pi} (\nabla \times {\bf B}) \times {\bf B} + \rho\nu {\bf \nabla}^2 \vec{v} \; ,
\end{equation}
where $\vec{v}$ is velocity, $\Omega$ rotation rate, $\rho$ density, $p^\prime$ non-hydrostatic
pressure, $\nu$ kinematic viscosity, $\zeta$ thermal expansivity, $g$ gravity, $T$ temperature,
${\bf B}$ magnetic field,
%${\bf j} = \frac{c}{4\pi} {\bf\nabla}\times{\bf B}$ current density,
$r$ radius and $z$ the direction parallel to the rotation axis. The terms in
Eq.~(\ref{eq:motion}) describe, in order, the linear and non-linear parts of inertial
forces, Coriolis force, buoyancy force, pressure gradient force,
Lorentz force, and viscous force [(compare with Eq.~(\ref{momentum}))].
\indexit{Lorentz force}
\indexit{viscous force}

In the non-magnetic and rapidly rotating case, the primary force
balance is between the pressure gradient force and the \indexit{Coriolis force}Coriolis force
(geostrophic balance), \indexit{geostrophic!balance} similar as for
large-scale weather systems in the Earth's atmosphere.  [Assuming a
stationary flow, and ignoring all other terms on the right of]
Eq.~(\ref{eq:motion}) and taking the curl, we arrive at the
Taylor-Proudman theorem, \indexit{Taylor-Proudman theorem} which
predicts the flow to be two-dimensional with
$\partial\vec{v}/\partial z = 0$. The only type of perfectly
geostrophic flow in a sphere, {\em i.e.,} a flow that satisfies this
condition, is the differential rotation of cylinders that are
co-aligned with the rotation axis \indexit{differential
  rotation!geostrophic cylinders}(geostrophic cylinders).
\indexit{geostrophic!cylinders} Such a flow can neither transport heat
in the radial direction, nor can it act as a dynamo.  Convection
requires motion away from and towards the rotation axis. This must
violate the Taylor-Proudman theorem, because a column of fluid that is
aligned with the $z$-direction will then stretch or shrink because it
is bounded by the outer surface of the sphere. Hence the velocity
cannot be independent from $z$. The necessity to violate the
Taylor-Proudman theorem inhibits convection and requires that some
other force, such as viscous friction, must enter the force
balance. In order for viscosity to do so, the length scale of the flow
must become small, at least in one direction. But the flow maintains a
nearly geostrophic structure as far as possible. At the onset of
convection it takes the form of columns \indexit{convection!(Busse)
  columns} aligned with the rotation axis
(Fig.~\ref{fig:bussesphere}). They surround the inner core tangent
cylinder like pins in a roller bearing. The tangent cylinder is
parallel to the $z$-axis and touches the inner core at the equator. It
separates the fluid core into dynamically distinct regions.

The primary\indexit{Earth!core!tangent cylinder} circulation is around the axes of these columns. However, in addition there
is a net flow along the column axes which diverges from the equatorial plane in anticyclonic
vortices and converges towards the equatorial plane in columns with a cyclonic sense of
rotation. The combination implies a coherently negative flow helicity in the northern
hemisphere and positive helicity in the southern \indexit{dynamo!$\alpha^2$ model}hemisphere, [which] can serve as an
efficient dynamo of the {\bf \boldmath$\alpha^2$-type}.

When the motion becomes more vigorous at highly supercritical convection and when a
strong magnetic field is generated, other forces such as inertia (advection of momentum)
and the Lorentz force can affect the flow. However, one difference between the solar
dynamo and planetary dynamos is the different role of inertial forces versus the Coriolis
force. Their ratio is measured by the Rossby number (Eq.~\ref{eq:ross}).
Deep in the solar convection zone $\Ro \approx 1$ when the pressure
scale height is taken for $L_{\rm t}$.
With typical estimates for the flow velocity in the Earth's core (1~mm$\,$s$^{-1}$),
the Rossby number is of order $10^{-6}$ when a global scale such as the core radius or shell
thickness is used for $L_{\rm t}$. Therefore, fluid motion in the geodynamo is often considered to
be largely unaffected by inertial forces.
The general force balance is believed to be that between \indexit{Coriolis force}Coriolis force, pressure
gradient force, Lorentz forces and buoyancy forces.
However, at small scales inertial
forces may become important also in planetary dynamos and can potentially feed back
on the large scale flow.

Like rotation, the presence of an imposed uniform magnetic field
inhibits convection in an electrically conducting fluid. However, the
combination of a magnetic field and rotation reduces the impeding
influence that either effect has separately. This constructive
interference is most efficient when the \indexit{Coriolis force}Coriolis force and the Lorentz
force are in balance. [\ldots\ Applied to dynamos, it is argued that
when the Coriolis force exceeds the Lorentz force the field will
strengthen, and when the Lorentz force exceeds the Coriolis force the
convection will weaken. Hence, it is assumed that the field
equilibrates when the forces match (referred to as a magnetostrophic
balance). The field strength inside the geodynamo or in Jupiter's
dynamo seems to agree with that argument.]  However, numerical dynamo
simulations put some doubt on its validity.'' More on that in
Ch.~III:6.

\begin{figure}
\begin{center}
\includegraphics[width=6cm]{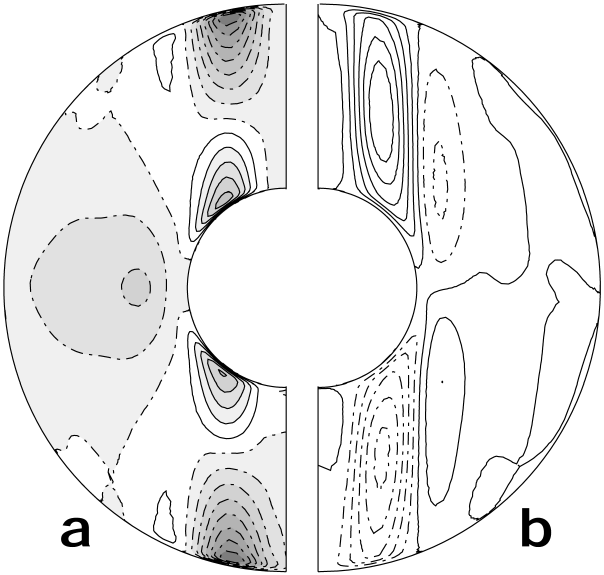}\includegraphics[width=6cm]{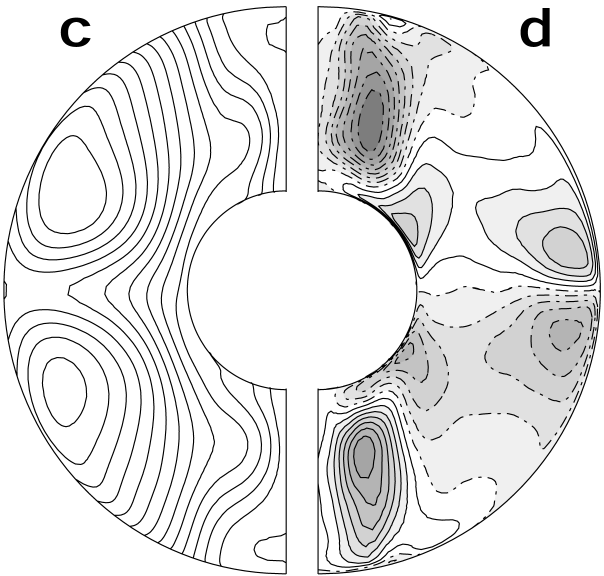}
\end{center}
\caption[Velocity and
magnetic field for a planetary dynamo model.]{Time-averaged
  axisymmetric components of velocity and magnetic-field components
  for a planetary dynamo model with [Rayleigh number] $R_a^*$=0.225,
  [Ekman number] $E=3\times10^{-4}$, $P_r=1$, [magnetic Prandtl
  number] $P_m=3$, [Reynolds number] $\Rm \approx$ 250 and [Rossby
  number] $\Ro \approx$ 0.1. The grey-scale indicates absolute
  intensity.  (a) Azimuthal velocity, broken lines are for retrograde
  flow, (b) streamlines of meridional velocity, full lines for
  clockwise circulation, (c) poloidal magnetic field lines, (d)
  azimuthal (toroidal) magnetic field, broken lines westward directed
  field. [Fig.~III:7.8]}
\label{fig:azmean}
\end{figure} \ors[III:7.6.3] ``The stretching of magnetic field lines by
differential rotation in the case of the solar dynamo, particularly
at\indexit{Earth!core!tangent cylinder} the tachocline, is thought to
be of major importance for the generation of a toroidal magnetic field
that is much stronger than the poloidal field.  In most geodynamo
models, in contrast, differential rotation does not contribute much to
the total kinetic energy and the toroidal and poloidal magnetic field
components have similar strength.  As mentioned before, the flow is
strongly organized by rotational forces and the vortices are elongated
in the $z$-direction. Even at a highly supercritical Rayleigh number
[(which measures the time scale of conductive relative to convective
transport)] and in the presence of a strong magnetic field, the flow
outside the inner core tangent cylinder is reminiscent of the helical
convection columns found at onset. Inside the tangent cylinder, the
flow pattern is different and often exhibits a rising plume near the
polar axis (Fig.~\ref{fig:azmean}b). \activity{{\em Background:} The dynamo model in
  Fig.~\ref{fig:azmean} is characterized by five dimensionless
  numbers. Two, the magnetic Reynolds number and the Rossby number,
  are defined in Eqs.~(\ref{eq:reynolds}) and~(\ref{eq:ross}),
  respectively. Look up the meaning of the other three: Ekman,
  magnetic Prandtl, and Rayleigh. These three are important numbers
  when computing the flows and their coupling, but not encountered
  until this Figure in this text because the solar dynamo models that
  were discussed and shown in the figures are kinematic, relying on a
  given, not consistently computed, flow pattern. The dynamo model
  behind Fig.~\ref{fig:azmean}, in contrast, computes flow, field, and
  their interaction, and thus also needs these remaining three
  dimensionless numbers specified.} The plume is accompanied with a
strong vortex motion (called a \lq thermal wind\rq\,) \indexit{thermal
  wind} with a retrograde sense of rotation near the outer surface
changing to prograde rotation at depth (Fig.  \ref{fig:azmean}a),
because the \indexit{Coriolis force}Coriolis force acts on the associated converging flow near
the inner core boundary and diverging flow near the outer boundary.

[\ldots] There is general agreement that the axial dipole
field is generated from the axisymmetric toroidal field by an
$\alpha$-effect \indexit{dynamo!$\alpha$ effect} associated with the helical
flow in the convection columns outside the tangent cylinder. In
mean-field theory as it is used in astrophysics, the $\alpha$-effect
is associated with unresolved turbulent eddies.  In
the geodynamo models a \lq macroscopic\rq\, $\alpha$-effect is
observed.

The mechanism for generating the axisymmetric toroidal field is less
clear and both an $\alpha$-effect and differential rotation seem to
play a role. Often two flux bundles in the azimuthal direction are
found outside the tangent cylinder, with opposite polarity north and
south of the equatorial plane (Fig.~\ref{fig:azmean}d).  [T]hey are
generated from the axisymmetric poloidal field by a similar
macroscopic $\alpha$-affect associated with the helical convection
columns ($\alpha^2$-dynamo).  Other authors show that the
$\Omega$-effect \indexit{dynamo!$\Omega$ effect} (the shearing of poloidal
field by differential rotation) contributes strongly to the generation
of axisymmetric toroidal field, even though the kinetic energy in the
differential rotation is rather limited. While in weakly driven
numerical dynamo models the regions inside the tangent cylinder, north
and south of the inner core, are nearly quiescent, vigorous flow is
found here in more strongly driven models.  In these cases a strong
axisymmetric toroidal field is found inside the tangent cylinder
region, produced by the shearing of poloidal field lines in the polar
vortex (Fig.~\ref{fig:azmean}a,c,d).''\activity{{\em Look up:} Summarize the contrast
  between the dynamo of a terrestrial planet with that of stars as
  discussed in this Ch.~\ref{ch:dynamos}: consider, among others,
  flow speed, rotation period, stratification, differential rotation,
  meridional advection, and Rossby and magnetic Reynolds numbers.}

\clearpage
	    
\chapter{{\bf Flows, shocks, obstacles, and currents}}%5
\label{ch:flows}
{\narrower\narrower{
{\bf Chapter topics:}
\begin{itemize}
  \customitemize
\item The behavior of sub- or super-Alfv{\'e}nic plasma flows around
  (non-)conducting bodies and around bodies with an intrinsic magnetic field
\item The formation, importance, and occurrence of shock waves 
\item The magnetized solar wind and its Parker spiral
\item Interaction of the magnetized solar wind with a planetary
  magnetic field 
\end{itemize}

\noindent {\bf Key concepts:}
\begin{itemize}
  \customitemize
\item Rankine-Hugoniot jump conditions for shock waves
\item Stellar magnetic braking 
\item Open solar magnetic field into the heliosphere
\item Planetary magnetospheres
\end{itemize}

}}

\section{Introductory overview}
Much of what happens in the heliosphere originates in the interaction
of an object or a plasma-filled volume with flows of magnetized plasma
directed at it. The scales of the phenomena discussed here range from
comets and asteroids up to the entire heliosphere where the solar wind
couples to the interstellar medium.  The interactions often involve
shocks, such as in cases when a fast solar wind stream catches up with
a significantly slower one, or where the solar wind envelops a
magnetosphere.  In other settings, they may involve smooth
sub-Alfv{\'e}nic adjustments in the flow, such as happens around many
of the moons orbiting within the plasma-filled magnetospheres of the
giant planets.  The flow of magnetized plasma around a body may be
affected by a magnetic field that it induces in that body's conducting
deep interior or near-surface shell, or that induced field may add to
an already present dynamo-generated field.  The flow may pick up
matter from an object's outer atmosphere through reconnection
processes, or through ionization of neutral matter that enters it from
outside, such as happens when the solar wind engulfs a comet or
because of interstellar-medium neutrals entering the heliosphere.

Despite this great variety of conditions, \indexit{flow!around
  planetary bodies}common patterns emerge. These are the focus of this
chapter which reviews the effects of a plasma flow around objects from
two different perspectives. One is to look into what happens to the
external flow, the other is concerned with what happens to the
atmosphere or magnetic field of the body that is cocooned by that
flow.  The first can be summarized by looking at what response is
induced in the body by the magnetized external flow. Electrodynamics
teaches us that the moving external magnetized plasma induces a
current system in a conductor. In the extreme of a perfect conductor,
that induced \indexit{magnetic!field!induced}current corresponds to a
magnetic field that counters the continuation of the external field
inside the conductor so that there is no net field there. In the
opposite extreme of a perfect insulator no field is induced and the
external field permeates the body as in its surroundings.
Intermediate conductivity induces an intermediate response.

Should the conductivity be limited to part of the body, the resulting
field external to that body provides clues as to the location of the
conducting medium and the magnitude of the conductivity. For example,
a planetary system body with a relatively small conducting core that
is enveloped by a non-conducting shell will have an induced field near
that core that is comparable, but opposite, to the external field. 
The strength of that field (generally largely bipolar when a
large-scale field flows by the object) decreases through the envelope
towards the body's surface so that the external signature may be
weak. If there is a conducting layer in a body that lies at or just
above or below the surface (such as an electrolyte-laden ocean or a substantial
ionosphere) the induced field can be strong if the conductance is
high, leading to a net field outside the body that is markedly
distinct from that of the incoming plasma. If the body has a
substantial, sustained intrinsic magnetic field ({\em i.e.,}  a dynamo in
its interior, and thus a conducting volume somewhere in its interior)
an induced field distorts the intrinsic field. For the incoming flow,
the 'object' that it encounters is bounded by the permanent and/or
induced field, and thus lies anywhere between the body's surface (or
high in the atmosphere, if it has one) or where the magnetic field of
that body is strong enough to withstand the momentum of the incoming
flow of ionized, magnetized matter.

The consequence of the dynamo and induced field for the incoming flow
is communicated by waves. \indexit{flow!modified by waves}If the
incoming flow is relatively slow, specifically if it is
sub-magnetosonic (see Sect.~\ref{sec:mhdwaves}), the flow can be
deflected well before approaching the 'object' and largely flow around
it; how much flows around it depends on the magnetic field. If, in
contrast, the flow comes in super-magnetosonically, the incoming flow
is unaware of the object until forced to realize its presence, either
(essentially) at the surface of the body or where its magnetic field
is strong enough; there will be a shock that decelerates and deflects
the flow. The interplay between the incoming field and the body's
magnetic field through compression and reconnection drives
magnetospheric and ionospheric processes, resulting in much of what we
know as space weather, further modified by planetary rotation.

But 'obstacles' to flows in the local cosmos are not limited to
planets, moons, and other bodies. One example of another type of
obstacle is the \indexit{flow!around astrosphere}outflowing wind from
a Sun-like star as it is encountered by the interstellar medium.
Another is that of (relatively) fast wind streams and coronal mass
ejections that catch up with slower wind plasma ahead of them; such
pileups often include shocks, here with the potential of plasma
becoming compressed in the collision zone rather than flowing around
the 'obstacle' because the scales are such that there generally is no
way 'around' the 'obstacles' within the characteristic time scale of
the passage of such flows through a large part of the heliosphere.

Generally, when a flow interacts with another volume of magnetized
plasma, the magnetic field is distorted in both volumes (if a field
exists, which in heliophysics is commonly the case) and magnetic (Lorentz)
forces play into the balance of forces, as expressed in the momentum
equation Eq.~(\ref{momentum}). One can view this as a result of the
pressure and tension forces ascribed to the magnetic field or,
equivalently, to induced currents -~the equations do not care about
our perspective in this matter (see Ch.~\ref{ch:universal}). 

One key differentiating factor in how the flow and the enveloped
volume interact is whether the magnetic fields in these two domains
can connect or not. In the \indexit{flow!in ideal MHD}ideal-MHD
approximation, in which effects of resistivity are ignored, the
induction equation Eq.~(\ref{induction}), through the frozen-in flux
paradigm, leads to the conclusion that the two plasmas involved cannot
interpenetrate: the flow moves around the impacted plasma as a wind
that flows past a solid object. This can still lead to very complex
dynamics, as diverse as for a wind flowing past a flag or around a
supersonic jet-plane. If the magnetic field can \indexit{flow!with
  reconnection}reconnect, however, the plasmas can interact in
entirely different ways, that include, for example, a variety of
magnetospheric phenomena. The differentiator here is not solely the
plasma resistivity but the effect of such resistivity within the
interaction time scale of the flow passing by the enveloped volume as
expressed by the magnetic Reynolds number (Eq.~\ref{eq:reynolds}). The
geometry of the interaction is also set by the Alfv{\'e}n Mach number
(Eq.~\ref{eq:alfvenmach}).

In~Sect.~\ref{lh:4} we encounter another type of flow into a
magnetosphere. In very young stars that are still surrounded by a
gaseous disk, matter spirals gradually towards a still growing
star. \indexit{flow!accretion disk}Close to the star, this accreting
matter will diffuse into, and then be locked onto the magnetic field
of the star. This allows the material to 'fall' through the stellar
magnetosphere, while being channeled by the magnetic field, to end up
near the surface of the star in what are known as 'accretion columns'
(sketched in Fig.~\ref{figlh:magnetoscheme}). But that is for later.

This chapter takes you through the following situations throughout the
heliosphere: \activity{{\em Advanced/Group:} To build a comparison of the different
  conditions, keep pen(cil) and paper at hand to sketch the various
  configurations as you read about how flows interact with bodies in
  the planetary system. Working in a reference frame in which the
  body is at rest, assume a spherical object, and let a flow move past
  it from left to right. Then prepare to make drawings in two
  orthogonal planes: the first plane is defined by the flow vector and
  the magnetic field carried in the flow (you may assume the field to
  be normal to the flow), while the second plane is normal to the
  first. Draw streamlines of the flow and subsequently add magnetic
  field lines. If you are good at 3-D renderings, also try a
  visualization such as in Fig.~\ref{fig:14Venus.eps}. }

\begin{itemize}\customitemize
\item Sects.~\ref{sec:flowshock}, \ref{sec:shocks} and
~\ref{sec:parker-spiral} are {\em introductions}: they discuss,
respectively, {\bf low-velocity interactions versus shocks}, the {\bf
elementals of shocks and discontinuities}, and {\bf the magnetized
solar wind and the Parker spiral} that forms as the wind flows out
from the rotating Sun. Staying with the {\em solar wind}, Sect.~\ref{sec:gos9.5}
reviews {\bf solar-wind stream interactions}.

\item Next come discussions of {\em flows around bodies in the
heliosphere}, beginning in Sect.~\ref{sec:flownonconducting} with {\bf
a non-conducting body without atmosphere}, then in
Sect.~\ref{sec:flowconducting} a {\bf flow around a conducting body}
without an intrinsic magnetic field.

\item By Sect.~\ref{sec:flowmagnetized} we reach {\em bodies with
dynamos} and look at {\bf plasma flow around a permanently magnetized
body}, after which we can discuss {\em magnetospheres}: first, in
Sect.~\ref{mp} {\bf a closed magnetosphere} which exists only in the
world of ideal MHD, but then in Sect.~\ref{open} we introduce {\bf the
open magnetosphere} such as happens in the real world.

\item Finally, we move on to what happens {\em within the magnetosphere of
a planet or moon} as a consequence of the variable solar wind
coupling to the body's magnetic field: in Sect.~\ref{flow} we talk
about the overall system of {\bf solar wind-magnetosphere-ionosphere
interaction}, including the effects of rotation and advection.

\item Finally, we return to the solar wind, looking at the outermost
  regions of the heliosphere, where the outflow meets the interstellar
  medium: Sect.~\ref{sec:impinging} explores what happens when we have
  {\bf a flow impinging on a fast outflow} but now on scales such that
  the flow can find a way around the outflowing plasma, in contrast to
  what happens in the case of wind streams interacting with comparable
  scales discussed in Sect.~\ref{sec:gos9.5}.
  
\end{itemize}

\section{Low-velocity interactions versus shocks}\label{sec:flowshock}
In view of the great diversity in conditions encountered throughout
the local cosmos \ors[IV:10.2] ``it may seem unlikely that general rules
can describe the interaction regions.  We are rescued from the need to
treat each case as totally distinct by recognizing that physical
theories often incorporate a small set of dimensionless parameters
that control important aspects of a system, even if such properties as
spatial scale, temperature, and flow velocity vary by many orders of
magnitude.  For a flowing plasma incident on an obstacle, the form of
the interaction depends critically on how the flow speed is related to
the speed of waves that transmit information about changes of plasma
properties from one part of the system to another.  An analogy to
waves in neutral gases helps to clarify the concept.  In the frame of
an airplane in flight, the atmosphere flows onto the plane at some
velocity, call it $v$.  As the gas encounters the plane, pressure
perturbations develop.  Pressure perturbations launch sound waves that
travel at the sound speed, $c_{\rm s}$.  If such waves can move away [in the
forward direction] from the plane, they can divert the atmosphere
upstream of the plane.  But the waves are swept back toward the plane
at the flow speed of the plasma. Only if $v < c_{\rm s}$ is it possible for
the waves to begin to divert the atmosphere well upstream of the
plane.  If $v > c_{\rm s}$, as for a supersonic jet, the waves pile up in
front of the plane, causing a shock to develop upstream. Only
downstream of the shock is the flow diverted.  Assuming that the plane
is large compared with distances characteristic of atmospheric
properties, the parameter that determines whether or not a shock will
form is the (dimensionless) sonic Mach number of the surrounding
atmosphere, $v/c_{\rm s}$. [Shocks are described in Sect.~\ref{sec:shocks}.]

%Table 10.1.  
\begin{table}
  \caption[Plasma properties upstream of solar-system
  bodies.]{Properties \indexit{solar!wind!properties!around small solar-system bodies}of the plasmas upstream of selected small bodies of the Solar System. [Listed are the Alfv{\'e}n and
    magnetosonic Mach numbers and the plasma $\beta$
    (Eq.~\ref{eq:betadef}).
    Table~IV:10.1].} 
\begin{center}\begin{tabular}{lllll}
\hline
Obstacle& Ambient plasma	& $M_{\rm A}$	& $M_{\rm ms}$	& $\beta$ \\
\hline
%CME & slow solar wind & & \\
%Heliosphere & Interstellar medium & & \\
%Earth & solar wind & & \\
%\ldots & \ldots & & \\
%\\
Io, Europa, Ganymede & jovian magnetosph. & $<1$ & $<1$ & $>1$ \\
Asteroids& solar wind & $>1$ & $>1$ & $\sim 1$ \\
Comets & solar wind & $>1$ & $>1$ & $\sim 1$ \\
Moon & Earth's magnetosph. & $>1$ & $>1$ & $\sim 1$ \\
         & or solar wind & \,or $<1$ & \,or $<1$ & \,or $<1$ \\
\hline
\end{tabular}\end{center}\label{tab:10.1}
\end{table}

In a plasma, much as in a neutral gas, compressional perturbations
develop when there is an obstacle in the flow. [\ldots] Having
identified [in Section~\ref{sec:mhdwaves}] some of the waves that
carry information through a magnetized plasma, we are now able to
introduce the dimensionless parameters that help us understand aspects
of flow and field perturbations.  The magnetosonic Mach number
($M_{\rm ms}$) is the ratio of the flow speed to the fast mode speed,
taken as $(c_{\rm s}^2+ v_{\rm A}^2)^{1/2}$. $M_{\rm ms}$ reveals whether or not a
shock is likely to form upstream in the flow.  When $M_{\rm ms} < 1$,
compressional waves can travel upstream from the obstacle faster than
the flowing plasma can sweep them back.  These waves, moving upstream,
can divert the incident flow around the obstacle, much as the bow wave
of a ship diverts water to the sides, and no shock develops.  However,
as in the situation discussed in the context of supersonic flight, if
$M_{\rm ms} > 1$, compressional waves are unable to propagate upstream
faster than they are swept back by the flow. They pile up to form a
shock. Most bodies in the super-magnetosonic solar wind [\ldots]
create shocks standing somewhat upstream on their sunward sides.
Downstream of the shock, plasma is heated, compressed, and diverted
around the obstacle.

The Alfv{\'e}n Mach number ($M_{\rm A}$, [Eq.~\ref{eq:alfvenmach}]) is
the ratio of the speed with which the ambient plasma flows towards an
obstacle divided by the Alfv{\'e}n speed. We will see that this
quantity controls the shape of the interaction region in planes
containing the unperturbed plasma flow and the background magnetic
field. The plasma beta ($\beta$, [see Eq.~\ref{eq:betadef}]) is the
ratio of the thermal pressure to the magnetic pressure. This quantity
enables us to understand how significantly the magnetic field
structure can be modified by changes of the plasma pressure.

The plasma environment differs greatly among the small bodies of the
Solar System. Some of the bodies are embedded in the solar wind,
others in the plasma of a planetary magnetosphere, and some [\ldots]
move from one environment to another [(such as Earth's Moon, which
spends part of each lunar orbit in Earth's magnetotail and the rest of
the month in the solar wind)].  Table~\ref{tab:10.1} lists some plasma
properties relevant to the environment of selected bodies.''

\begin{figure}[th]
\centering
\includegraphics[width=12.5cm]{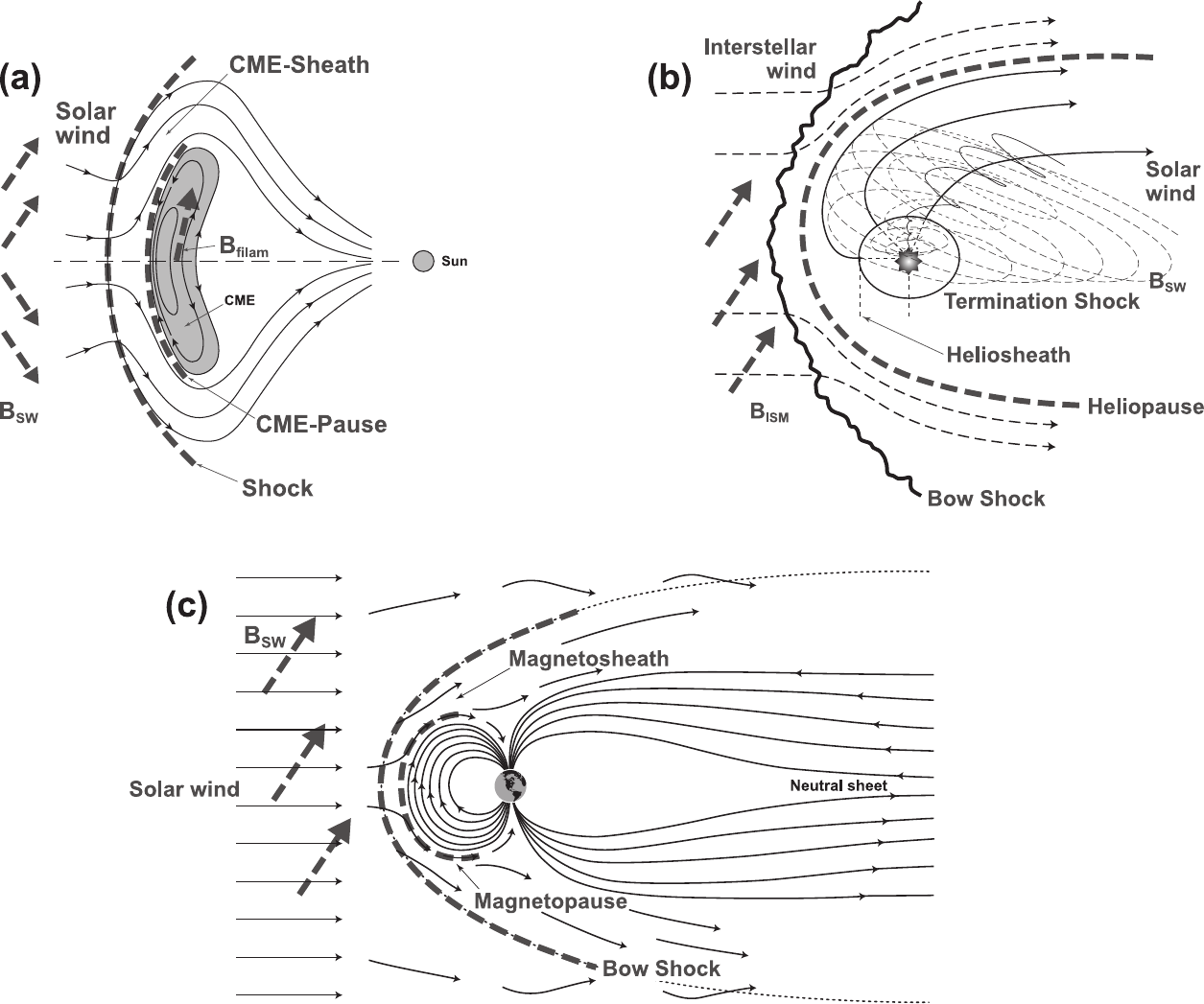}
\caption[Shocks around CMEs, the heliosphere, and Earth's
magnetosphere.]{Schematic \indexit{shock!illustration in different settings}comparison of shocks around [coronal mass
  ejections (CMEs, magnetically-driven explosions from the solar
  atmosphere into the heliosphere)],
the heliosphere, and the terrestrial magnetosphere [(all seen from
around the equatorial plane)]. The figure shows some of the
types of shocks and sheaths that exist in the heliosphere and
their universal basic structures: {\rm (a)} a CME [pushing its way
super-Alfv{\'e}nically into the solar wind]; {\rm (b)} the outer heliosphere,
and {\rm (c)} Earth's magnetosphere. The
same basic structures appear: shocks where the solar wind becomes
subsonic; the sheaths that separate the subsonic solar wind from the
obstacle ahead; and the 'pause' where there is a pressure equilibrium
between the subsonic solar wind and the obstacle's environment.
In the case of a CME these three structures are the shock, CME-sheath,
CME-pause and the obstacle is the magnetic filament that drives
the CME. In the case of the outer heliosphere the
structures are the termination shock, heliosheath, and heliopause. The
obstacle is the interstellar wind and the magnetic field it is
carrying. If the interstellar wind is supersonic there is an
additional shock, the bow shock. In the case of the Earth's magnetosphere
the structures are the shock, the magnetosheath, the
magnetopause and the obstacle is the Earth's dipolar magnetic field. [Fig.~II:7.1]}
\label{fig:ophercomposite}
\end{figure}
\section{Elementals of shocks and other discontinuities}\label{sec:shocks}
Shocks \indexit{flow!shock}that \indexit{shock!wave}we discussed up to here
develop when a flow speed exceeds the speed of waves that can serve as
a warning to the flow that an obstacle lies ahead. Shocks can also
form if non-linear effects in the propagation of a wave become
important, such as when a wave runs into a medium in which strong
gradients in density or temperature cause the wave amplitude to grow
more rapidly than dissipation can limit that growth; examples of such
shock waves are found in upward traveling pressure waves in
atmospheres, including the Earth's and the Sun's, and also in very
large and long-lived wind streams in the heliosphere.

\ors[II:7.2] ``In the small-amplitude limit, the profile of a
magnetohydrodynamic (MHD) wave does not change as it propagates, but
even a small-amplitude wave will eventually distort due to {\em wave
  steepening}.\indexit{wave!steepening} The wave steepening happens
when gradients of pressure, density and temperature become so large
that dissipative processes ({\em e.g.,} viscosity, thermal conduction) are
no longer negligible. In the steady state, a steady wave-shape --~a
{\em shock wave}~-- is\indexit{shock!wave} formed in which the
steepening effect of nonlinear advective terms balance the broadening
effects of dissipation. The shock waves move at speeds larger than the
ambient intrinsic speed, which for magnetized ionized matter in the
heliosphere, is the magnetosonic speed. If the shock moves much faster
than the magnetosonic wave, it is called a strong shock; if it moves
just slightly faster, it is called a weak shock. The dissipation
inside the shock front leads to a gradual conversion of the energy
being carried by the wave into heat.  In the heliospheric plasma, we
have collisionless shocks in which the thermalization happens through
wave-particle interactions. [\ldots]

\begin{figure}[t]
\centering
\includegraphics[width=5.cm]{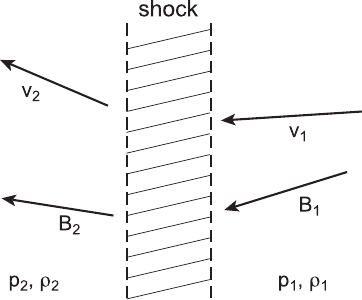}
\caption[Diagram: upstream and downstream of a
shock.]{Diagram showing the region upstream (left) and downstream of a
shock. [Fig.~II:7.3]} \label{fig:opher2}
\end{figure}
A propagating wave described by the ideal fluid equations leads to
infinite gradients in a finite time. There is no solution for the
ideal MHD equations. This is not surprising: ideal equations are valid
when scales of variations are larger than the mean free path. The
breakdown in ideal equations occurs in a very thin region, while the
fluid equations are valid everywhere else. In this very thin region,
it is difficult to describe the plasma in detail. The simple picture
is a discontinuity dividing two roughly uniform fluids. An important
aspect is that the simple picture of a discontinuity dividing two
roughly uniform fluids is not usually applicable in a plasma. Shocks
can involve turbulence for example. For this initial discussion, we
make the simplifying assumption that there is a planar discontinuity
of zero thickness that separates two uniform fluids, as depicted in
Figure~\ref{fig:opher2}. We also assume that the shock is stationary
[or, in other words, that we are in the co-moving frame of reference
\ldots] The transition must be such as to conserve mass, magnetic flux,
and energy. The MHD jump conditions are independent of the physics of
the shock itself and are known as the {\em Rankine-Hugoniot jump
  conditions}.''\indexit{Rankine-Hugoniot jump conditions}

\label{sec:jump} \ors[II:7.3] ``It is straightforward to obtain the
Rankine-Hugoniot jump conditions from [Maxwell's equations and] the
MHD equations. Assuming steady state in the frame of reference of the
shock, the equation for the conservation of mass [in
Eq.~(\ref{continuity}) in the absence of sources and sinks,]
%\begin{equation}
%\frac{\partial {\rho}}{\partial t} + \bf{\nabla}\cdot (\rho \bf{v})=0~,
%\end{equation}
gives
\begin{equation}
\rho_{1} {\bf v_{1}}\cdot {\bf \uv{\perp}} =\rho_{2} {\bf v_{2}}\cdot {\bf \uv{\perp}},
\end{equation}
[(where ${\bf \uv{\perp}}$ is a unit-length vector pointing in the direction normal to the shock)] or in a different notation
\begin{equation}\label{eq0}
\{ \rho {\bf v}\cdot {\bf \uv{\perp}} \} =0~,
\end{equation}
where the symbol $\{\ldots\}$ represents differences between the two
sides of the discontinuity.

Conservation of momentum, [with Eq.~(\ref{momentum}) without sources,
sinks, or viscosity,] 
%\begin{equation}
%\frac{\partial (\rho {\bf v})}{\partial t} + {\bf \nabla}\cdot [ \rho {\bf v}{\bf v}+\left( p + \frac{B^{2}}{2\mu_{0}}\right) {\bf I} - \frac{\bf{B}\bf{B}}{\mu_{0}} ] =0~,
%\end{equation}
yields
\begin{equation}
\left \{ \rho {\bf v} ({\bf v}\cdot {\bf \uv{\perp}})+ \left( p {\bf \uv{\perp}}
    +\frac{B^{2}}{8\pi} {\bf \uv{\perp}}-\frac{({\bf B}\cdot {\bf \uv{\perp}})}{4\pi}{\bf B} \right) \right \} =0~.
\end{equation}

Conservation of energy, [\ldots]
%\begin{equation}
%\frac{\partial}{\partial t} \left( \frac{1}{2} \rho v^{2} \frac{P}{\gamma - 1} + \frac{B^{2}}{2\mu_{0}}+\bf{\nabla}\cdot \left( \frac{1}{2} \rho v^{2}\bf{v} \right) +\frac{\gamma P}{\gamma - 1} \bf{v} +\frac{1}{\mu_{0}} \bf{E} \times \bf{B} \right) =0~,
%\end{equation}
results in
\begin{equation}
\left \{ \left( \frac{1}{2} \rho v^{2} + \frac{\gamma p}{\gamma - 1}
  \right) ({\bf v}\cdot {\bf \uv{\perp}}) + \frac{c}{4\pi}({\bf E} \times
  {\bf B})\cdot {\bf \uv{\perp}} \right \} =0.
\end{equation}
[Note that ${\bf S} = (c/4\pi) {\bf E} \times {\bf B}$ is the Poynting
  flux, which measures the directional energy transfer in an
  electromagnetic field; compare with Eq.~(\ref{eq:dynamoenergy})
  where that is expressed for a plasma with infinite conductivity, as
  it is below in Eq.~(\ref{eq3}).]

Conservation of magnetic flux, [\ldots]
%\begin{equation}
%\bf{\nabla} \cdot \bf{B} =0~,
%\end{equation}
gives
\begin{equation}
\{ {\bf B} \cdot {\bf \uv{\perp}} \} =0~.
\end{equation}

The equation
\begin{equation}
{\bf \nabla} \times {\bf E} = -\frac{1}{c}\frac{\partial {\bf B}}{\partial t}
\end{equation}
[for a steady state] can be written as
\begin{equation}
\{ \bf{E}\times {\bf \uv{\perp}} \} = 0.
\end{equation}

Let us consider, now, the normal {$\perp$} and the tangential $\parallel$ components relative to the shock's surface so that the\indexit{shock!jump conditions} jump conditions can be written as:
\begin{equation}
\left \{ \rho v_{\perp}^{2} + p + \frac{B_{\parallel}^{2}}{8\pi} \right \} =0 \label{eq1}
\end{equation}
\begin{equation}
\left \{ \rho {\bf v}_{\parallel} v_{\perp} - \frac{{\bf B}_{\parallel}B_{\perp}}{4\pi} \right\} =0 \label{eq2}
\end{equation}
\begin{equation}
\left \{ \left( \frac{1}{2}\rho v^{2} +  \frac{\gamma p}{\gamma - 1} +\frac{B^{2}}{4\pi}  \right) v_{\perp} - ({\bf v}\cdot {\bf B} ) \frac{B_{\perp}}{4\pi} \right\} =0 \label{eq3}
\end{equation}
\begin{equation}
\{ B_{\perp} \} =0 \label{eq4}
\end{equation}
\begin{equation}
\{ {\bf v}_\perp \times {\bf B}_{\parallel} + {\bf v}_{\parallel} \times {\bf B}_{\perp} \}
={\bf 0}~. \label{eq5}
\end{equation}

\def\thetabn{\theta_{\rm B_\perp}}
\begin{figure}[t]
%\centerline{\hbox{\psfig{figure=figures/DKV_fig_1.ps,width=11cm,clip=}}}
\centerline{\hbox{\includegraphics[width=11cm,bb= 0 0 536 225]{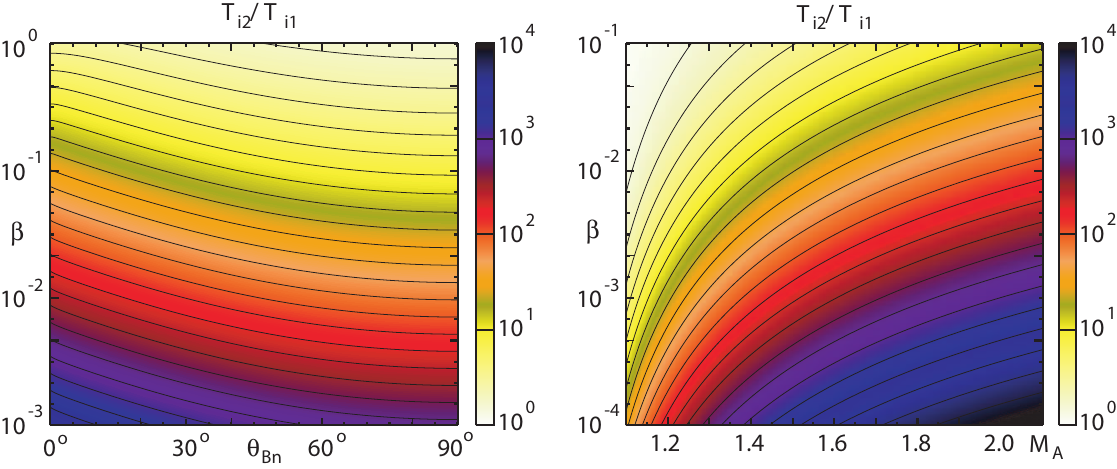}}}
\caption[Iso-contours of shock heating as a
function of $\thetabn$ and $M_{\rm A}$.]{Iso-contours of shock 
heating, expressed as the ratio between
downstream to upstream ion temperature $T_{i2} / T_{i1}$, as a
function of shock-normal angle $\thetabn$  [(the angle between the
shock normal and the upstream magnetic field)]  (fixed $M_{\rm A} = 2$) and
Alfv{\'e}n Mach number $M_{\rm A}$ (fixed $\thetabn = 45^\circ$) for
low $\beta$ plasmas.  Derived from standard Rankine-Hugoniot conditions for
fast shocks, assuming a specific heat ratio $\gamma = 5/3$.  The graphs show
that for a wide range of angles, there can be very substantial
downstream heating at sufficiently low plasma $\beta$, as present in
much of the solar corona.  Such extreme heating may help form a seed
population for further acceleration [into energetic particle
populations that will be discussed in
Ch.~\ref{ch:conversion}]. [Fig.~II:8.1] \colorfig }\label{fig:kv0}
\end{figure}
Equations~[(\ref{eq0}) and] (\ref{eq1})--(\ref{eq5}) are the
Rankine-Hugoniot jump conditions that describe all types of shocks''
and also allow for three types of discontinuities that are not
shocks. An example of the heating associated with shocks is given in
Fig.~\ref{fig:kv0}. \regfootnote{A note on terminology: a ``parallel
  shock'' propagates along the magnetic field, {\em i.e.,}  has the vector
  $\uv{\perp}$ normal to the shock front aligned along the magnetic
  field, or $\uv{\perp} \parallel {\bf B}$. A ``perpendicular
  shock'' has $\uv{\perp} \perp {\bf B}$. \mylabel{note:shockterm}}
\activity{{\em Show:} Write the Eqs.~(\ref{eq0}) and (\ref{eq1})--(\ref{eq5}) for
  the hydrodynamic limit, and derive the temperature ratio between the
  post- and pre-shock media. You should find that the density contrast
  $r_\rho=((\gamma +1)M_{\rm s}^2)/(2+(\gamma -1)M_{\rm s}^2)$ and the
  pressure ratio $r_p=(2\gamma M_{\rm s}^2-(\gamma-1))/(\gamma+1)$
  where $M_{\rm s}=v_1/c_{{\rm s}1}$ for sound speed $c_{\rm s}$. Note
  there is a maximum value for $r_\rho$ but not for $r_p$ as function
  of $M_{\rm s}$. What are the values for $r_\rho$ and $r_p$ for
  $\gamma=5/3$ for $M_{\rm s}$ near unity and for $M_{\rm s}\gg 1$?}
%https://www.cfa.harvard.edu/~namurphy/Lectures/Ay253_2016_08_Shocks.pdf

\begin{figure}[ph!]
  \centering
  \includegraphics[width=14.2cm]{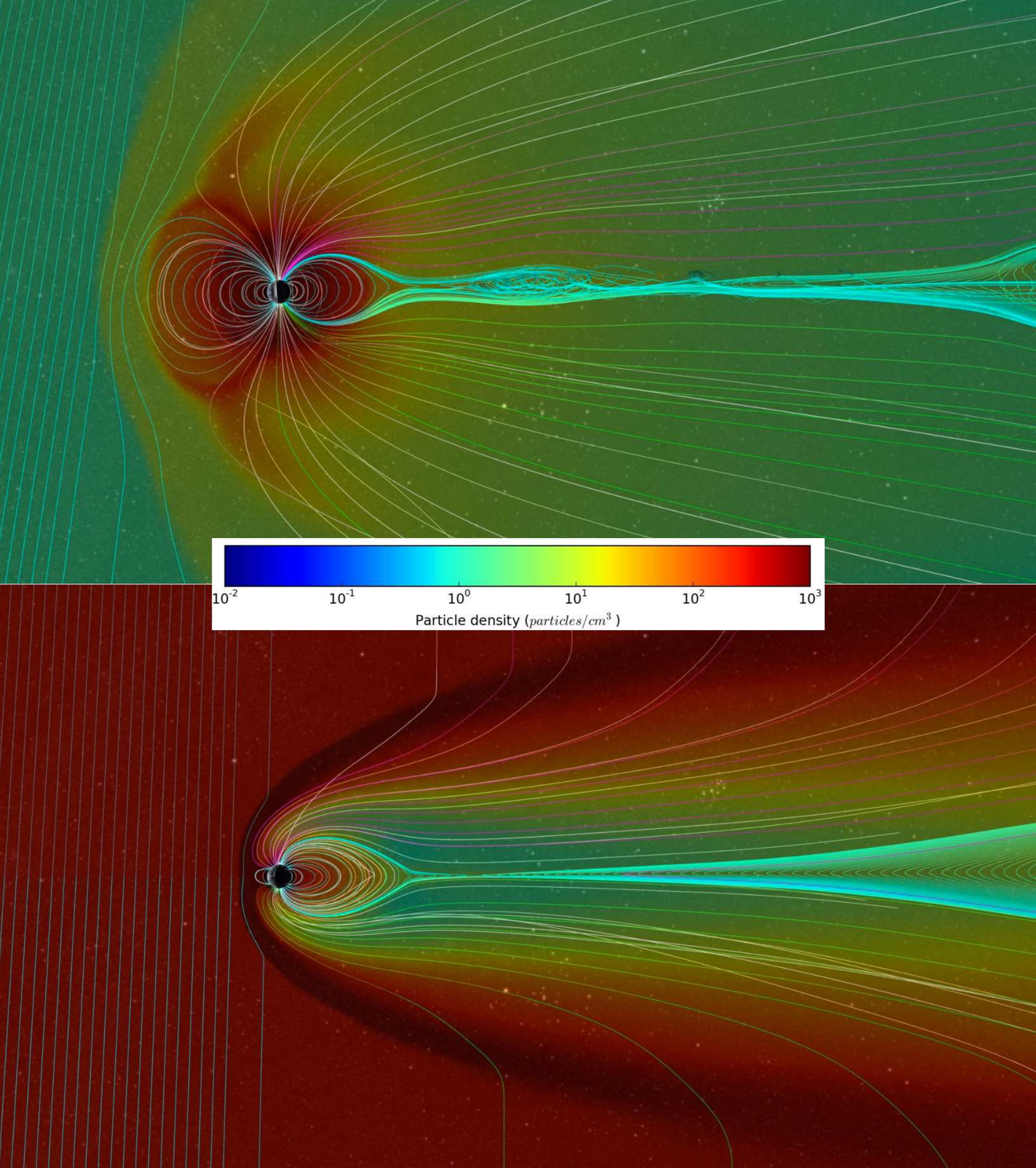}
\caption[Magnetospheric field subject to a Carrington-level storm.]{An
  MHD simulation of the Earth's magnetic field subject to ({\em top})
  a typical background solar wind and ({\em bottom}) about 3.5\,h
  after subjecting that field to a coronal mass ejection similar to
  that thought to have hit the Earth after the Carrington-Hodgson
  solar flare of 1859. The current sheet, extending from the cusp in
  the field in the downwind direction, clearest in the bottom panel,
  is the interface between oppositely-directed magnetic fields. The
  color bar in the center shows how the particle density is encoded in
  color. Source: first
  and last frames of a \href{https://svs.gsfc.nasa.gov/4189}{NASA
    animation (https://svs.gsfc.nasa.gov/4189)}, based on
  \href{https://ui.adsabs.harvard.edu/abs/2014JGRA..119.4456N/abstract}{the}
  study by \cite{2014JGRA..119.4456N}. See Fig.~I:6.3 for a comparison of 
  heliospheric and magnetospheric field configurations. \colorfig \label{fig:moldwin3}}
\end{figure}
%\begin{figure}[t]
%\includegraphics[width=8.5cm]{figures/figure6.3.eps}
%\caption[Heliospheric and
%magnetospheric current sheets.]{A comparison of the heliospheric current sheet
%(labeled 'field reversal' in this figure) and its plasma sheet (top) and
%the magnetospheric current sheet (here 'neutral sheet')
%and its plasma sheet (bottom) [Fig.~I:6.3].\label{fig:moldwin3}}
% \end{figure}

\nocite{1981JGR....86.5438G}\nocite{1969RvGSP...7...97N} \ors[II:7.4]
``Discontinuities\indexit{discontinuity: contact, rotational, tangential} can be classified as
either contact or rotational discontinuities. Contact discontinuities
happen when there is no flow across the discontinuity, {\em i.e.,}
$v_{\perp}=0$, but $\{ \rho \} \neq 0$. A classic example is the
contact discontinuity of a mix of vinegar and olive oil. If
$\{B_\perp\}\neq 0$ at a contact discontinuity then only the density
changes across the discontinuity, which is rarely observed in plasmas.
A tangential discontinuity occurs
when $\{B_{\perp}\}=0$, then $\{v_\parallel\} \neq 0$ and
$\{B_{\parallel} \} \neq 0$ and $\{p +B^{2}/8\pi\}=0$.  This means
that the fluid velocity and magnetic field in this case are parallel
to the surface of the discontinuity but change in magnitude and
direction, and that the sum of thermal and magnetic pressures is
constant. [\ldots\ Large heliophysical examples of tangential
discontinuities with $\{B_\perp\}= 0$ are the heliospheric current
sheet and the magnetospheric current sheet (illustrated in
Fig.~\ref{fig:moldwin3}).]

A rotational discontinuity occurs when $\{v_{\perp}\} \neq 0$ and
$\{\rho\}=0$. From the jump conditions this implies that $\{v_{\perp}\}=0$
and $\{p+B^{2}_\parallel/8\pi\}=0$ so ${\bf v}_{1}\cdot {\bf \uv{\perp}}={\bf v}_{2}\cdot
{\bf \uv{\perp}}=v_{\perp}$ and $\rho_{1}=\rho_{2}$. After some math, we find that
$v_{\perp}^{2}=B_{\perp}^{2}/4\pi \rho$, and that $B_{\parallel}$ remains constant in
magnitude but rotates in the plane of the discontinuity. [\ldots]

The Rankine-Hugoniot jump conditions have 12 variables. Four upstream
parameters are specified ($\rho$, $v$, $B_{\parallel}$, and
$B_{\perp}$), so we have 7 equations for 8 unknowns. Therefore we need
to specify one more quantity, namely the strength of the shock
$\rho_{2}/\rho_{1}$.''

Examples of tangential discontinuities are the heliopause and
planetary magnetospheres when there is little reconnection (see
Sect.~\ref{mp}). A rotational discontinuity occurs for example when
reconnection at the magnetopause is relatively efficient
(Sect.~\ref{open}). Shocks occur upstream of where the solar wind
meets planetary system objects
(Sects.~\ref{sec:flownonconducting}--\ref{open}) as well as where it
encounters the interstellar medium (Sect.~\ref{sec:impinging}), and
also where fast wind streams plow into slower ones ahead as well as at
the leading edge of relatively fast explosions called coronal mass
ejections (Sect.~\ref{sec:gos9.5}).

\section{The magnetized solar wind and the Parker spiral}
\label{sec:parker-spiral}
Before we can discuss the interplay of the \indexit{solar!wind!Parker
  spiral}solar wind with objects throughout \indexit{Parker!spiral}the
heliosphere, we need to introduce two properties of the solar wind
itself: the geometry of the magnetic field that it carries, and the
consequence of wind gusts running at different velocities. First, the
magnetic field:

\ors[1:9.2] ''Let us briefly consider [a steady-state] outflow of ionized,
magnetized gas from a {\em rotating} star with [a magnetic field that
scales with distance from the star like a monopole, which is to say
with a radially-flowing wind that stretches the field out from its effective base
that lies, say, a few stellar radii above its actual surface \regfootnote{This
base is what is meant by the term 'source surface' introduced in
Activity~\ref{act:sourcesurface}; within that surface, the field is
approximated as corotating rigidly with the star.})] The salient aspects of such a flow
are found even when only considering the equatorial plane and
restricting attention to solutions where all variables are functions
of $r$ only. We make use of spherical coordinates $r$, $\phi$, $\theta$.

We find that the equation of mass conservation (for plasma mass
density $\rho$ moving at velocity ${\bf v}$ and carrying a field
${\bf B}$) then can be written
\begin{equation}
{1\over r^2}{\partial\over\partial r}\left(\rho v_rr^2\right)=0,
\end{equation}
while the $\phi$ component of the momentum equation is given by
\begin{equation}
\rho\left(v_r{\partial v_\phi\over\partial r}+v_\phi{v_r\over r}\right)=
  {1\over4\pi}\left(B_r{\partial B_\phi\over\partial r}+B_\phi{B_r\over r}\right)
\end{equation}
or
\begin{equation}
\rho v_r{1\over r}{d\over dr}(r v_\phi)={1\over4\pi}B_r{1\over r}{d\over dr}(rB_\phi)
\label{eq:phi-momentum}
\end{equation}
Mass conservation implies that $\rho v_rr^2$ is constant, while the
divergence-free magnetic field requires that $B_rr^2$ is constant. Multiplying
Eq.~(\ref{eq:phi-momentum}) with $r^3$ we see that
\begin{equation}
rv_\phi-{B_rr^2\over\rho v_rr^2}{1\over 4\pi}rB_\phi={\rm constant}=L.
\label{eq:wind-spin}
\end{equation}

Under the assumptions above, the [ideal] induction equation is
\begin{equation}
{1\over r}{d\over dr}\left(r[v_rB_\phi-v_\phi B_r]\right)=0.
\label{eq:phi-induction}
\end{equation}
For a star rotating with angular velocity $\Omega$, radius $R_{\rm s}$
and with a [simplified monopolar] field so that $B_{\phi {\rm s}}\approx 0$ [(with the
index 's' meaning at the solar wind base or source surface)] we find
that the induction equation implies
\begin{equation}
r(v_rB_\phi-v_\phi B_r)={\rm constant}\approx-R_{\rm s}(R_{\rm
  s}\Omega)B_{r {\rm s}}=-\Omega r^2B_r .
\label{eq:wind-induction}
\end{equation}

\begin{figure}[t]
%\centerline{\psfig{figure=figures/IMF_Spiral_BW_invert,width=\textwidth}} 
%\centerline{\psfig{figure=figures/Cohenfiles/f3l.eps,width=4.3cm}\psfig{figure=figures/Cohenfiles/f3e.eps,width=4.3cm}\psfig{figure=figures/Cohenfiles/f3f.eps,width=4.3cm}} 
\centerline{\includegraphics[width=4.3cm,bb=0 0 592 527]{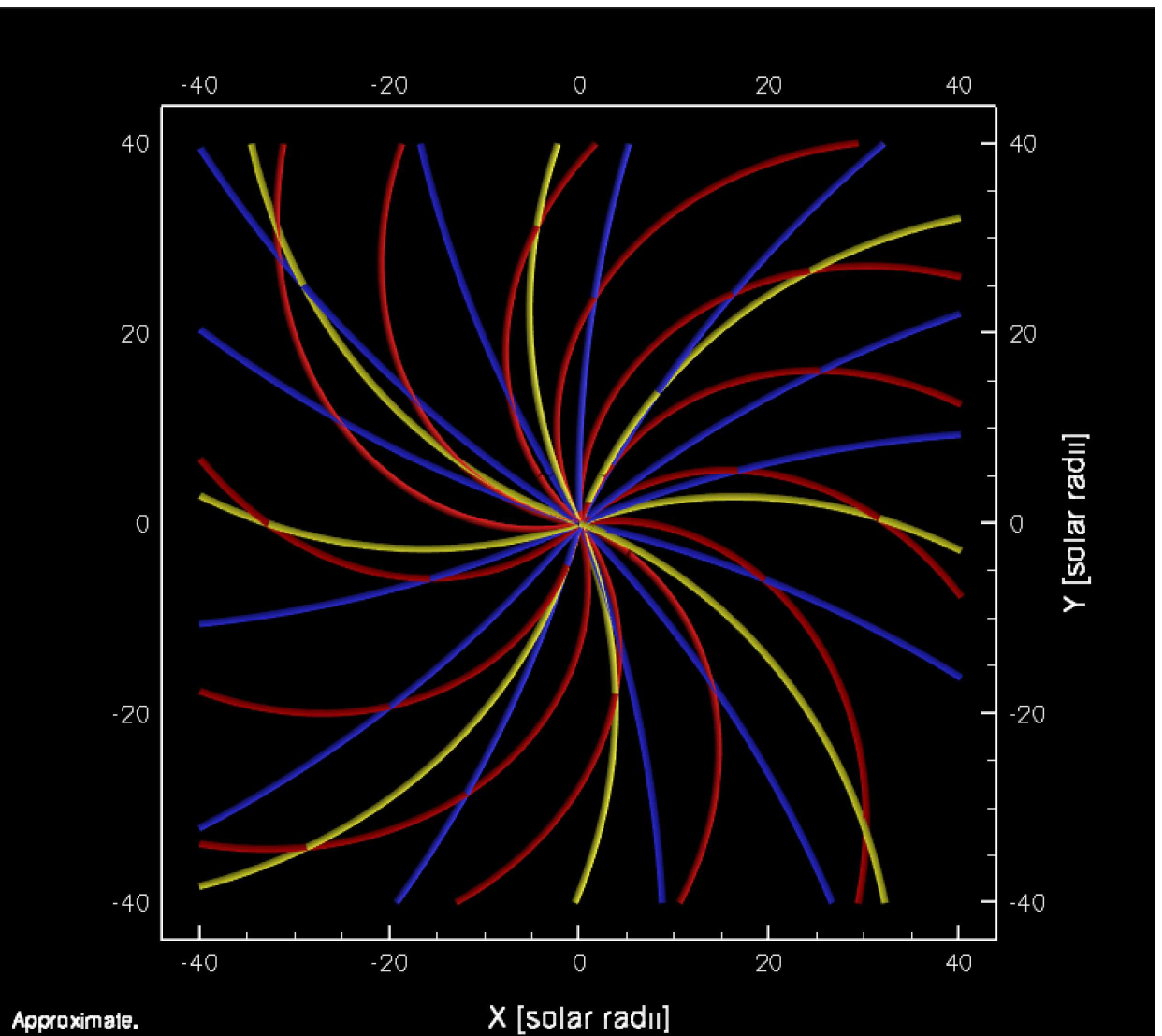}\includegraphics[width=4.3cm,bb=0 0 188 168]{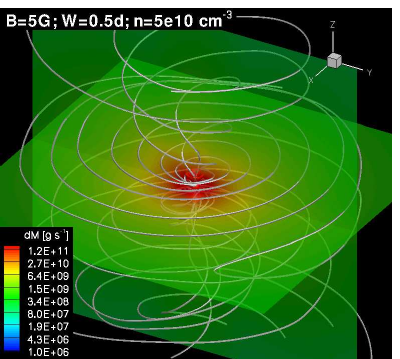}\includegraphics[width=4.3cm,bb=0 0 189 168]{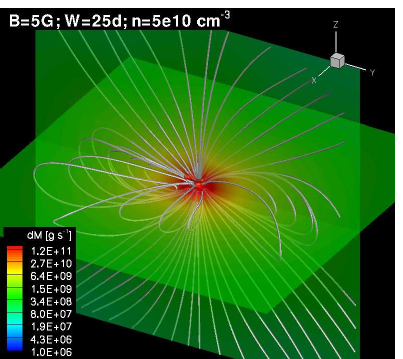}} 
%\centerline{{\bf [Missing figure file]}} 
\caption[Spiraling asterospheric magnetic field for different rotation
periods.]{{\em Left:} Conceptual display of different stellar-wind
  magnetic field spirals for a Sun with a 4.6\,day rotation period
  (red), a 10\,day period (yellow), and a 26\,day period (blue), as a
  function of distance in solar radii. {\em Center/right:} Results
  from numerical simulations for the stellar coronae of solar analogs
  with rotation period of 0.5\,day (middle) and 25\,days (right), [but
  assuming the same base field strengths and densities.
  Fig.~IV:4.1; sources:
  \href{https://ui.adsabs.harvard.edu/abs/2012ApJ...760...85C/abstract}{left
  from \citet{2012ApJ...760...85C}}
  and
  \href{https://ui.adsabs.harvard.edu/abs/2014ApJ...783...55C/abstract}{center
    and right from \citet{2014ApJ...783...55C}}.] \colorfig}
\label{fig:AMFSpiral}
\end{figure}
We can now solve Eq.~(\ref{eq:wind-spin}) and
Eq.~(\ref{eq:wind-induction}) for $v_\phi$ and $B_\phi$ and using
$M_{\rm A}^2=(v_r^2/v_{\rm A}^2)$ with $v_{\rm A}^2=(B_r^2/4\pi\rho)$ we find
\begin{equation}
v_\phi=\Omega r{M_{\rm A}^2({L/r^2\Omega})-1\over M_{\rm A}^2-1}\,;\,
%\end{equation}
%and
%\begin{equation}
B_\phi=-{B_r\Omega r\over v_r}\left({1-({L/r^2\Omega})\over M_{\rm A}^2-1}\right)M_{\rm A}^2.
\end{equation}
Both expressions show that we must have $1-({L/r^2\Omega})=0$ when
$M_{\rm A}^2-1=0$ [(formally, going to zero in such a way that their
ratio remains finite)]. We define $r\equiv r_{\rm A}$ where
$M_{\rm A}^2=1$. Thus, we must have $L=r_{\rm A}^2\Omega$. Notice
[that $v_r$ tends to a constant for large $r$, and thus
$M_{\rm A}^2\propto \Omega r^2$ and as a result
\begin{equation}\label{eq:spiralforlarger}
{\rm for\, large\,} r:\,\,
v_\phi\approx{\Omega r_{\rm A}^2\over r}\rightarrow 0\,;\,
%\end{equation}
%and
%\begin{equation}
B_\phi\approx -{B_r\Omega r\over v_r},
\end{equation}
while 
\begin{equation}
{\rm close\, to\, the\, star:}\, v_\phi\approx \Omega r\,;\,
B_\phi\approx {-{B_r \Omega r \over v_{\rm A}}}.
\end{equation}
In] other words, the magnetic field and stellar wind rotate like a
solid body out to the critical point $r_{\rm A}$ where the radial flow
speed is equal to the 'radial' Alfv{\'e}n speed. Beyond this point the
field is pulled along the wind into a spiral, the {\em Parker spiral},
as the flow becomes nearly radial far from the
star''. Figure~\ref{fig:AMFSpiral} shows this spiral and also two MHD
simulations of stellar winds (discussed in
Sect.~\ref{section:4_1}). Note that whereas $B_r$ decreases as
$\sim 1/r^2$, $B_\phi$ decreases as $\sim 1/r$. \activity{{\em Show:}
  At what distance from the Sun does the solar-wind model discussed in
  Sect.~\ref{sec:parker-spiral} have $|B_r| = |B_\phi|$ for typical
  values of the slow and fast solar wind? What are typical values for
  $B_\phi/B_r$ at 1\,AU, 5\,AU ({\em cf.} Fig.~\ref{fig:gf9.5}), and
  at the ice giants?  }\activity{{\em Show:} Compare the relative
  radial dependence of the magnetic fields in the solar wind model of
  Sect.~\ref{sec:parker-spiral} with the values listed in
  Table~\ref{tab:fran2}. Also: use these scalings to demonstrate that
  the plasma $\beta$ tends to a constant value far from the Sun. You
  may assume the temperature of the solar wind to be essentially
  constant compared to the changes in distance and density.}

\begin{figure}[t]
%\centerline{\psfig{figure=figures/alfven_currentsheetsketch.eps,width=8cm}}
%\centerline{\psfig{figure=figures/Heliospheric-current-sheet.eps,width=10cm}}
\centerline{\includegraphics[width=10cm]{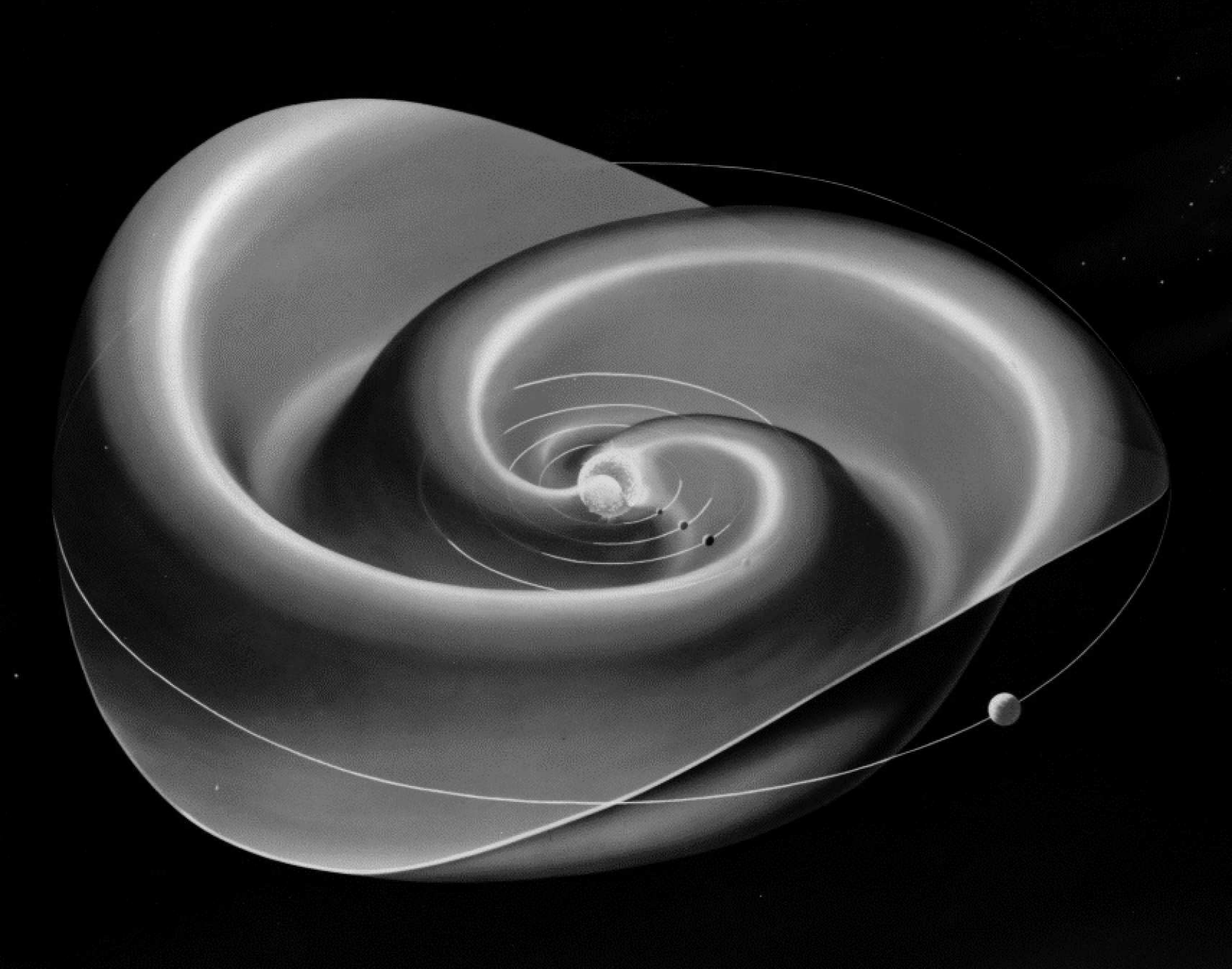}}
\caption[Sketch of the heliospheric current sheet.]{The heliospheric current sheet forms a warped, undulating structure
extending from the top ridge of the helmet \indexit{streamer belt}streamer belt [\ldots] 
%and~\ref{fig:modelcs}) 
that sweeps by the
planets as the Sun rotates once per 27 days (synodic period). The
magnetic field changes direction across the current sheet.
[\href{https://commons.wikimedia.org/wiki/File:Heliospheric-current-sheet.gif}{image
  source: Wikipedia}; see also Fig.~I:9.3] \label{fig:currentsheetsketch} }
%from http://www.mps.mpg.de/solar-system-school/lectures/marsch/4.ppt
\end{figure}

\begin{figure}[ph!]
%\centerline{\hbox{\psfig{figure=figures/tunnelcomposite.eps,width=11cm,clip=}}}
\centerline{\hbox{\includegraphics[width=13.4cm]{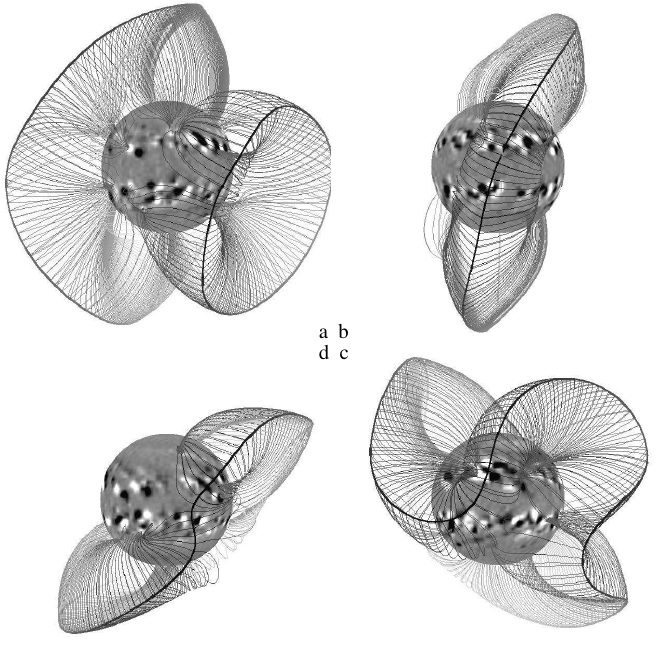}}}
\caption[The Sun's magnetic field and its extension into the 
heliosphere.]{\label{fig:modelcs} The Sun's surface magnetic field is
comprised of a multitude of dipolar regions of widely different fluxes,
whose numbers wax and wane with the solar cycle. The large-scale
coronal magnetic field, the foundation of the heliospheric field, expands
from regions of partly open magnetic field that enclose the closed-field
corona. This diagram shows the global topology of the Sun's field in
a so-called potential-field source-surface (PFSS) approximation. In particular,
it shows four realizations of \indexit{streamer belt}the 'streamer
belt' for a solar magnetic model; each belt separates opposite
polarities of the field reaching into the heliosphere. Shown are four phases of the
simulated magnetic cycle: clockwise from the top left, $t=3.1,
\,3.6, \, 4.5, \,6.0$\,years into a sunspot cycle of 11.\,years. Each
panel shows a magnetogram of the solar surface, the neutral
line(s) at the source surface, and the highest closed field lines
that reach up to the neutral line(s); the lines are colored so
that the darkest colors are nearest to the 'observer.' The
panels show, clockwise, an example of a near-quadrupolar
situation; a strongly tilted dipolar case; a strongly warped
current sheet; and another nearly dipolar case with less tilt
relative to the solar equator. [Fig.~I:8.1] For a short discussion
how the 'streamer belts' in this figure map to `helmet streamers' seen
in coronagraph images and during eclipses, see Sect.~\ref{pfssplots}.}
\end{figure}\nocite{cjssolwind11}\indexit{PFSS model!approximation}

\begin{figure}[t]
\begin{center}
%\centerline{\hbox{\psfig{figure=figures/gf9.4.eps,width=6cm,clip=}}}
\centerline{\hbox{\includegraphics[width=6cm]{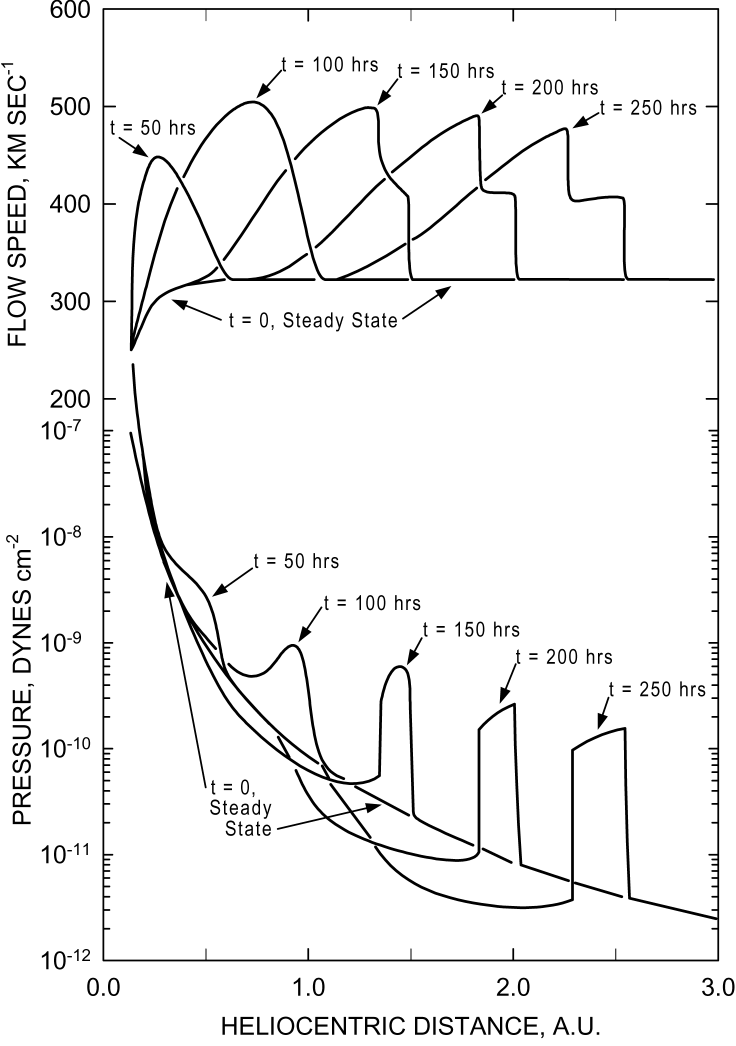}}}
\caption[Solar wind $v$ and $p$ (1D model) for a high-speed stream.]{Snapshots
of solar wind flow speed and pressure as
functions of heliocentric distance at different times during the
outward evolution of a high-speed stream as calculated using a
simple 1D gas-dynamic code. After obtaining a steady-state solar
wind expansion that produced a flow speed of 325\,km/s far from
the Sun, a high-speed stream was introduced into the calculation
by linearly increasing and then decreasing the temperature (and
thus also the pressure) by a factor of four at the inner boundary
at 0.14\,AU over an interval of 100\,hrs. [Fig.~III:8.4]
\label{fig:gf9.4}}
\end{center}
\end{figure}
\section{Flow-based interactions in heliophysics}\label{sec:environments}
\subsection{Solar-wind stream interactions}\label{ch:windwind}\label{sec:gos9.5}

\subsubsection{A 1D model of high-speed stream evolution}\label{sec:gosgmir}
The \indexit{solar!wind!stream interactions}solar wind stretches the
high-coronal field nearly radially into the heliosphere, there to
deform subject to solar rotation into the Parker spiral. The magnetic
field in the solar wind has its roots in the two magnetic polarities
on the solar surface. Although the solar surface field has myriad
adjacent regions of either polarity, further from the Sun the low
orders dominate, so that the heliospheric field often resembles that
of a distorted bipolar pattern stretched out far beyond the
planets. In this field the opposite polarities straddle a transition
in field direction in a warped skirt around the Sun, known as the
heliospheric current sheet (sketched in its fundamental properties in
Fig.~\ref{fig:currentsheetsketch}; see Fig.~\ref{fig:modelcs} for a
representation of an approximating (potential) field model at the
foundation of that current sheet based on solar
observations). \sactivity{$\circledS$ {\em Show:} The solar wind
  stretches the high-coronal magnetic field into the heliosphere into
  a roughly radial field below the Alfv{\'e}n radius. This enables an
  analogy with electrostatics: the field of electric charges placed
  above a flat perfect conductor can be computed by placing mirror
  charges opposite to the conducting surface, which then naturally has
  the electric field perfectly normal to the conducting
  surface. Analogously, in a magneto-static consideration above the
  spherical Sun (of unit radius) called the potential-field
  source-surface (or PFSS) model, the magnetic field can be
  approximated by placing mirror 'charges' on a sphere at distance
  $d_{\rm SS}^2$ which then has the field perfectly radial at
  $d_{\rm SS}$. Empirically, $d_{\rm SS}\approx 2.5$ (where that
  'source surface' is taken as the foundation of the heliospheric
  field; the virtual surface with mirror charges used to compute the
  potential field below $d_{\rm SS}$ is then at $d_{\rm SS}^2$). This
  model (introduced by
  \href{https://ui.adsabs.harvard.edu/abs/1969SoPh....6..442S}{\citet{1969SoPh....6..442S}})
  works remarkably well below $d_{\rm SS}$ on large scales. The
  heliospheric field is approximated by a continuation from that
  source surface, subject to
  the Parker spiral. \\
  (a) For illustration, simplify the source-surface model by a 2-d
  sketch of a potential field involving charges (of alternating
  polarity: N-S-N-S-N-S) placed on a straight line and another line
  parallel to it involving mirror charges of opposite polarity. Sketch
  the equivalent of the foundation of the heliospheric current sheet
  and examples of 'closed' field lines (the equivalent of coronal
  loops closing back onto the solar surface) and 'open' field lines
  (the equivalent of field stretched out into the heliosphere), at the
  base of which we find dark 'coronal holes' in X-ray images of the
  Sun. Consider how the X-points that form on the midpoints between
  these lines are like the 'helmet streamers' seen
  in coronagraphs and during eclipses.\\
  (b) For a sphere, show that the mirror charges on the surface at
  $d_{\rm SS}^2$ are $-d_{\rm SS}$ times the strength of those on the
  solar surface. \solution{pfssplots}\mylabel{act:sourcesurface}}
Before going into this geometry, let us look into the simplified case
of a 1-D radial outflow.

Different magnetic regions on the Sun lead to different speeds in the
wind. [Outside of eruptive phases,] this typically manifests itself in so-called fast
streams and slow streams (see Table~\ref{tab:wind-stats} for their
properties). Because the Sun rotates, the radially flowing fast and
slow winds cannot avoid but to run into each other.  \ors[III:8.5]
``Because radially aligned parcels of plasma within a stream originate
from different locations on the Sun, they are threaded by different
magnetic field lines and thus cannot interpenetrate one another
[without reconnection, and such reconnection proceeds relatively
slowly in the solar wind compared to the characteristic time the wind
takes to traverse much of the heliosphere].  Figure~\ref{fig:gf9.4},
which shows the result of a simple 1D gas-dynamic simulation,
illustrates the basic reasons why high-speed streams evolve with
increasing heliocentric distance. The rising portion of the high-speed
stream steepens kinematically with increasing heliocentric distance
because gas (plasma) at the peak of the stream is traveling faster
than the slower plasma ahead. As the speed profile steepens, material
within the stream is rearranged; parcels of plasma on the rising-speed
portion of the stream are compressed, causing an increase in pressure
there, while parcels of plasma on the falling-speed portion of the
stream are increasingly separated, producing a rarefaction.

It is common to refer to the compression on the leading edge of a
high-speed stream as an\indexit{solar!wind!interaction region}
interaction region. Being a region of high
pressure, the interaction region expands into the plasma both
ahead and behind at the fast mode speed (actually at the sound
speed in the calculation shown in Figure~\ref{fig:gf9.4}). The leading edge of
the interaction region is called a forward wave because it
propagates in the direction of the solar wind flow; the trailing
edge is called a reverse wave because it propagates sunward in the
solar wind rest frame but is carried away from the Sun by the
highly supersonic flow of the wind. Pressure gradients associated
with these waves produce an acceleration of the slow wind ahead
and a deceleration of the high-speed wind within the stream. The
net result of the interaction is to limit the steepening of the
stream and to transfer momentum and energy from the fast wind to
the slow wind. [\ldots]

As long as the amplitude of a high-speed stream is sufficiently small,
it gradually dampens with increasing heliocentric distance in the
manner just described. However, when the difference in speed between
the slow wind ahead and the peak of the stream is more than about
twice the fast mode (sound) speed the stream initially steepens faster
than the forward and reverse pressure waves can expand into the
surrounding plasma; thus in such cases the interaction region at first
narrows with increasing heliocentric distance. The nonlinear rise in
pressure associated with this squeezing eventually causes the forward
and reverse waves\indexit{solar!wind!shock waves} bounding the
interaction region to steepen into shocks.  Because shocks
(Sect.~\ref{sec:shocks}) propagate faster than the fast mode (sound)
speed, the interaction region can expand once shock formation occurs.
Observations reveal that relatively few stream interaction regions are
bounded by shocks at 1\,AU, but that most are near the equatorial
plane at heliocentric distances beyond about 3\,AU because the fast
mode (sound) speed generally decreases with increasing distance from
the Sun. At heliocentric distances beyond about $5-10$\,AU a large
fraction of the mass and magnetic field flux in the solar wind at low
heliographic latitudes is found within expanding compression regions
bounded by shock waves on the rising portions of strongly damped
high-speed streams. The basic structure of the solar
wind\indexit{solar!wind!stream structure in inner {\em vs.}\ outer
  heliosphere} near the solar equatorial plane in the distant
heliosphere thus differs considerably from that observed near
Earth. Stream amplitudes are severely reduced, and short-wavelength
structure is damped out. The dominant structures at low latitudes
({\em i.e.,}  within the band of variable wind \regfootnote{The solar wind
  originating from high latitudes is typically fast as long as there
  are polar cap fields, {\em i.e.,}  in phases around solar minimum. The solar
  wind from mid-to-low latitudes is a mixture of fast and slow
  streams, particularly around solar maximum. }) in the outer
heliosphere are expanding compression regions that interact and merge
with one another to form what are commonly called global
merged\indexit{solar!wind!global merged interaction regions (GMIRs)}
interaction\indexit{global merged interaction region}
regions, GMIRs.''

\begin{figure}[t]
\begin{center}
%\centerline{\hbox{\psfig{figure=figures/gf9.5.eps,width=8.5cm,clip=}}}
\centerline{\hbox{\includegraphics[width=8.5cm]{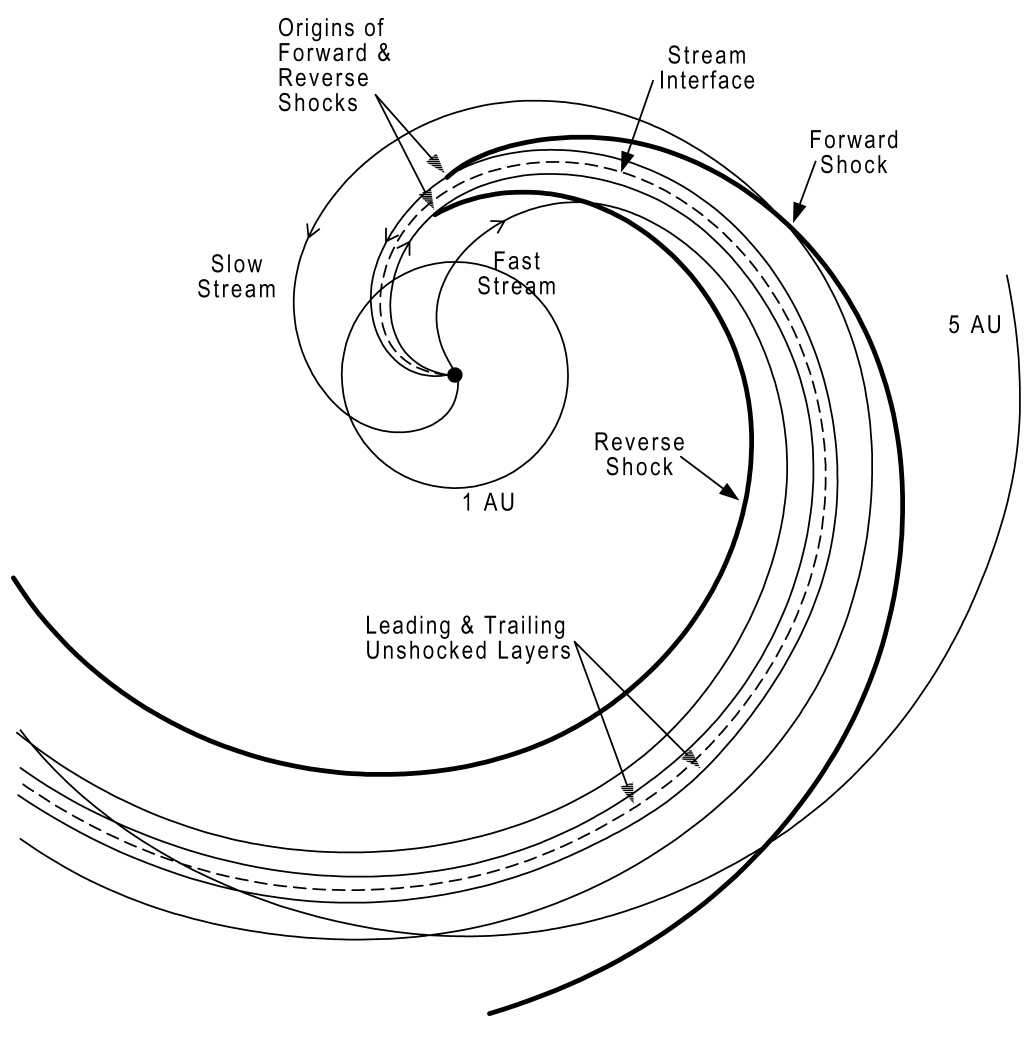}}}
\caption[A corotating
interaction region in the solar equatorial plane.]{Idealized schematic
  illustrating the basic structure of a corotating interaction region
  in the solar equatorial plane. The dashed line threading the middle
  of the corotating interacting region (CIR) denotes the stream
  interface and the solid heavy lines indicate the forward and reverse
  shocks. Plasma immediately surrounding the stream interface is
  compressed, but not shocked.  [Note that in this stationary model
  moving a radial cut through the Sun in the clock-wise direction is
  equivalent to going forward in time at any given
  angle. Fig.~III:8.5;
  \href{https://ui.adsabs.harvard.edu/abs/1999SSRv...89..179C/abstract}{source:
  \citet{1999SSRv...89..179C}}.]) \label{fig:gf9.5}}
\end{center}
\end{figure}
\subsubsection{Stream evolution in two and three dimensions} \ors[III:8.5]
``Should the coronal expansion be time-independent but inhomogeneous
in heliocentric latitude and longitude, stream evolution proceeds
similarly at all longitudes, but the state of a stream's evolution
varies with longitude. Because of solar rotation, the interaction
region on the leading edge of a high-speed stream is wound into a
spiral that at any particular heliocentric distance is inclined to the
radial direction at an angle intermediate to that of the magnetic
field threading the slow and fast wind flows respectively, as
illustrated in Figure~\ref{fig:gf9.5}. The entire pattern of
interaction co-rotates with the Sun and the compression region is
known as a corotating\indexit{corotating interaction region (CIR)}
interaction region, CIR. It is important to note, however, that it is
only the pattern that co-rotates with the Sun because each parcel of
solar wind plasma moves radially outward in this simple picture,
except within the interaction region itself where both radial and
transverse deflections of the flow occur. Because a CIR is inclined
relative to the radial direction the pressure gradients associated
with the interaction region have both radial and azimuthal
components. With increasing heliocentric distance the forward wave
propagates both anti-sunward and westward (in the direction of
planetary motion about the Sun), whereas the reverse wave propagates
both sunward (in the rest frame of the average solar wind) and
eastward. As a result, the slow wind is accelerated outward and
deflected westward within the interaction region and the fast wind is
decelerated and deflected eastward there, thus accounting for the
characteristic westward and then eastward flow deflections commonly
associated with interaction regions on the leading edges of high-speed
streams (see Figure~III:8.3 and related discussion). One consequence
of the transverse deflections is that they partially relieve the
pressure build-up induced by stream steepening by allowing the plasma
to slip aside. Thus solar wind streams steepen less rapidly than is
predicted by the simple 1D simulation shown in Figure~\ref{fig:gf9.4}.''

\ors[III:8.5] ``There is, of course, a three-dimensional aspect to stream evolution
that becomes most apparent at heliocentric distances beyond about
$3-4$\,AU and at latitudes away from the solar equatorial
plane. [O]bservations have revealed (1) that the reverse shocks
on the trailing edges of CIRs are observed both within the
low-latitude band of solar wind variability and at latitudes
$10^\circ-20^\circ$ above that band, whereas the forward shocks on the
leading edges of corotating interaction regions are generally confined
to the low-latitude band itself; and (2) that in addition to the flow
deflections already discussed, the slow wind is usually deflected in
both solar hemispheres toward the opposite hemisphere at the forward
shocks, whereas the fast wind is usually deflected poleward at the
reverse shocks.'' For more details, see Sect.~III:8.5.  
% For field around a moving sphere, see for example http://citeseerx.ist.psu.edu/viewdoc/download?doi=10.1.1.205.5366&rep=rep1&type=pdf

%[Buildup of cases:]
\subsection{A non-conducting body without
  atmosphere}\label{sec:flownonconducting}
\begin{figure} 
\begin{center}
\includegraphics[trim=0pt 221pt 0pt 0pt, clip=true, width=6.4cm]{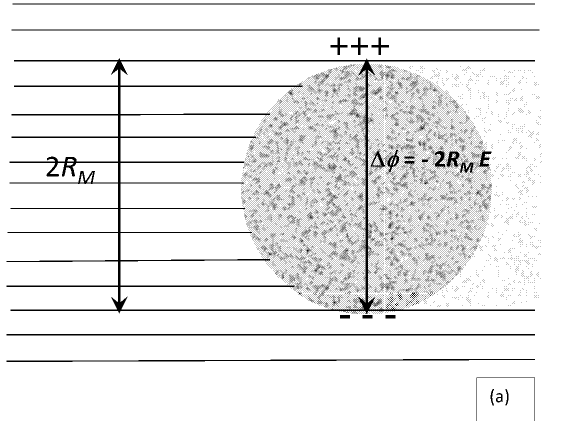}\includegraphics[trim=0pt 0pt 0pt 211pt, clip=true, width=6.4cm]{figures/kivelson_10-2}
% bbox 0 0 275.76 422.16; trim=left bottom right top, clip
\end{center}
\caption[Plasma flow (non-)conducting body.]{Schematics of plasma flow (shown by lines of flow)
  at velocity ${\bf v}$ from the left onto (a) a non-conducting body
  and (b) a conducting body [with radius $R_{\rm M}$].  In the plasma,
  ${\bf B}$ is into the paper, ${\bf E}$ is $-{\bf v}/c \times {\bf B}$
  in both cases. Diagram {\rm (a)} shows that a non-conducting body builds
  up surface charge that imposes a potential drop
  $\Delta \phi = -2 R_{\rm M}E$ across the diameter, producing an
  electric field that opposes the solar wind electric field.  Diagram
  {\rm (b)} shows the response of a conducting body that does not build up
  surface charge.  Conducting paths allow current (shown schematically
  as a dashed line) to flow through the body and close in the incident
  flow.  Heavy banded arrows identify the orientation of the resultant
  ${\bf j}\times {\bf B}$ force that diverts part of the incident
  flow. Because much of the incident flow has been diverted, the
  potential drop across the body is only $\Delta \phi = - d \, E$,
  where $d < R_{\rm M}$ is the distance in the incident flow between
  the flow lines that just graze the body. The electric field that
  penetrates the body is a fraction of the upstream field determined
  by the fraction of the upstream flow that impacts the surface. In
  the wake region, gray in both diagrams, the plasma pressure is
  reduced and the magnetic pressure is increased relative to the
  upstream values. [Fig.~IV:10.2]}\label{fig_10.2}
\end{figure}

Next, \indexit{flow!non-conducting body}we look at one of the simplest
setups for a flow encountering a body: a non-conducting sphere moving
relative to a low-density magnetized plasma. The non-conducting body
has no intrinsic magnetic field, and no currents can be induced in it
by the magnetized plasma through which it moves. For a low-density
plasma, this means that no signal is sent upstream from that body that
could modify the flow heading towards the body: the gas pressure is
insignificant and the magnetic field is not affected by the
non-conducting body, moving through it without generating reflected
waves that might move upstream from the body. Consequently, the
upstream plasma that is on a collision course with the body will, in
essence, simply crash onto the body, while the plasma to the sides of
that body continues to flow without noticing the object at all
[(Fig.~\ref{fig_10.2}a)]. \activity{{\em Show:} In case you wonder how
  much solar-wind material actually hits, say, the lunar surface,
  estimate the mass flux density per unit area over a time interval
  $\Delta t$, with ion density $n_{\rm sw}\approx 10$\,cm$^{-3}$ and
  $v_{\rm sw}\approx 500$\,km/s at Earth (and Moon), ignoring
  long-term change. Express that in units of mg/cm$^2$/Myr. } This is
true regardless of whether the body is moving sub-Alfv{\'e}nically or
super-Alfv{\'e}nically relative to the incoming plasma. Examples of
rather non-conducting bodies in the Solar System are Earth's Moon,
Jupiter's moon Callisto and Saturn's moon Rhea (which are subjected to
the sub-Alfv{\'e}nical flow of the giant planets' magnetospheric
plasma throughout their orbits), and also many asteroids, particularly
the S-type, or silicate-rich 'rocky' ones (note that MHD does not
apply for asteroids small compared to gyro-radii of solar wind ions;
among other things, this means there is no upstream shock, as that is
a collective phenomenon).

\figindex{../kivelson/art/kivelson_10.2.eps}
In such situations, there will be a \indexit{flow!wake}wake behind the
body that is void of plasma immediately downstream of the body. This
void has two primary effects. One is that the plasma pressure has
dropped away, and consequently plasma will propagate into the void to
refill it (at about the slow-mode speed), taking matter from an
outward propagating domain behind a rarefaction front that moves out
at essentially the fast-mode wave speed (somewhat anisotropically);
this leads to a wake in density behind the body, forming a somewhat
asymmetric conical V-shape, albeit with the wings ending on a
terminator-like ring defined by where the incoming plasma is just
tangent to the body's surface. The other effect is that because the
contribution of the plasma to the total pressure falls away
immediately behind the object, the field is somewhat strengthened
(mainly by a motion perpendicular to the plane spanned by field and
flow vectors) to regain total pressure balance, to adjust again
further downstream as plasma refills the void. In the case of Earth's
Moon in the solar wind, the void persists up to about a dozen lunar
radii downstream.\sactivity{$\circledS$ {\em Show: } For the solar
  wind flowing onto a non-conducting, non-magnetized sphere, use
  estimates of wave speeds to sketch the density wake, the slow-mode
  refilling, and the fast-mode rarefaction front in a plane defined by
  the flow vector and the field vector, and in a plane defined by the
  flow vector and perpendicular to the field; assume the field vector
  to be perpendicular to the flow vector. You may compare the result
  with measurements for the case of the Moon (in Fig.~IV:10.7).
  \mylabel{act:noncondsphere} \solution{noncondsphere}}

Note that the Moon is not a perfect insulator, and in fact has
something akin to a weak ionosphere because the incoming solar wind
ionizes some of the surface dust, and the ongoing process of
ionization of such 'pick-up particles' is associated with a current
that can send a magnetic signal upstream; see Chs.~IV:10{\&}11. 

\subsection{Flow around a conducting body}\label{sec:flowconducting}
When a \indexit{flow!conducting body}conducting body moves through a
magnetized plasma, the $-\vv/c \times \vec{B}$ electric field
associated with the relative motion induces a potential drop across
the body. Because that body is conducting, a current flows to attempt
to neutralize the charge buildup that would occur in the absence of
such a current.  That current closes through the incoming plasma in
such a way that the associated Lorentz forces act to bend the plasma
around the object (Fig.~\ref{fig_10.2}b), which, equivalently, sets up
an induced magnetic field that, at infinite conductivity, would keep
the external field entirely outside the conducting body. The
conducting medium can be a metallic core (which in the case of the
moons of the giant planets is generally too small to detect with
significance) or a mantle ocean of water with dissolved electrolytes,
{\em e.g.,} salts (which is seen on multiple moons of the giant
planets, such as the Galilean moon \indexit{Europa}Europa at Jupiter which is
discussed in IV:10.5.2), a magma layer (as is inferred for another
of the Galilean moons, \indexit{Io}Io) or an ionosphere in the upper layers of the
body's atmosphere (such as in the case of Venus discussed in
Sect.~\ref{sec:ss4}).

Europa moves sub-Alfv{\'e}nically within the jovian
magnetosphere. Because its orbit is inclined by about $10^\circ$
relative to Jupiter's magnetic dipole moment, Europa senses a changing
magnetic field throughout its orbit, so that not only a current system
is induced by its motion, but that current system (and thus its
associated perturbation magnetic field) evolves through the
orbit. These changes (slightly modified by pickup ions playing their
part) have revealed where the current flows, and thereby the existence
of a conducting liquid underneath the non-conducting ice mantle.

The induced current system, or equivalently the induced
magnetic field, sends out information about the obstacle into the
plasma ahead. These waves, led by the magnetosonic fast-mode type,
modify the upstream flow so that it can begin to deflect well ahead
of the body. Part of the incoming flow may impact upon the surface if
the conduction is not infinite. The rest of the flow is diverted
around the body, leading to a narrower wake behind the body than in
the case of an  insulating body. 

If the flow is coming in super-magnetosonically, however, no
significant 'warning signal' can move upstream so that much of the
flow will impact the body or flow very close to it, as it would in the
case of large iron-rich asteroids (as has been argued, for example, for
asteroid Ida, despite it being characterized as an S-type). But for
many asteroids the scale is too small for MHD to apply, so the analogy
with larger bodies fails in multiple respects.

\begin{figure}
\centering
\includegraphics[height=6.5cm]{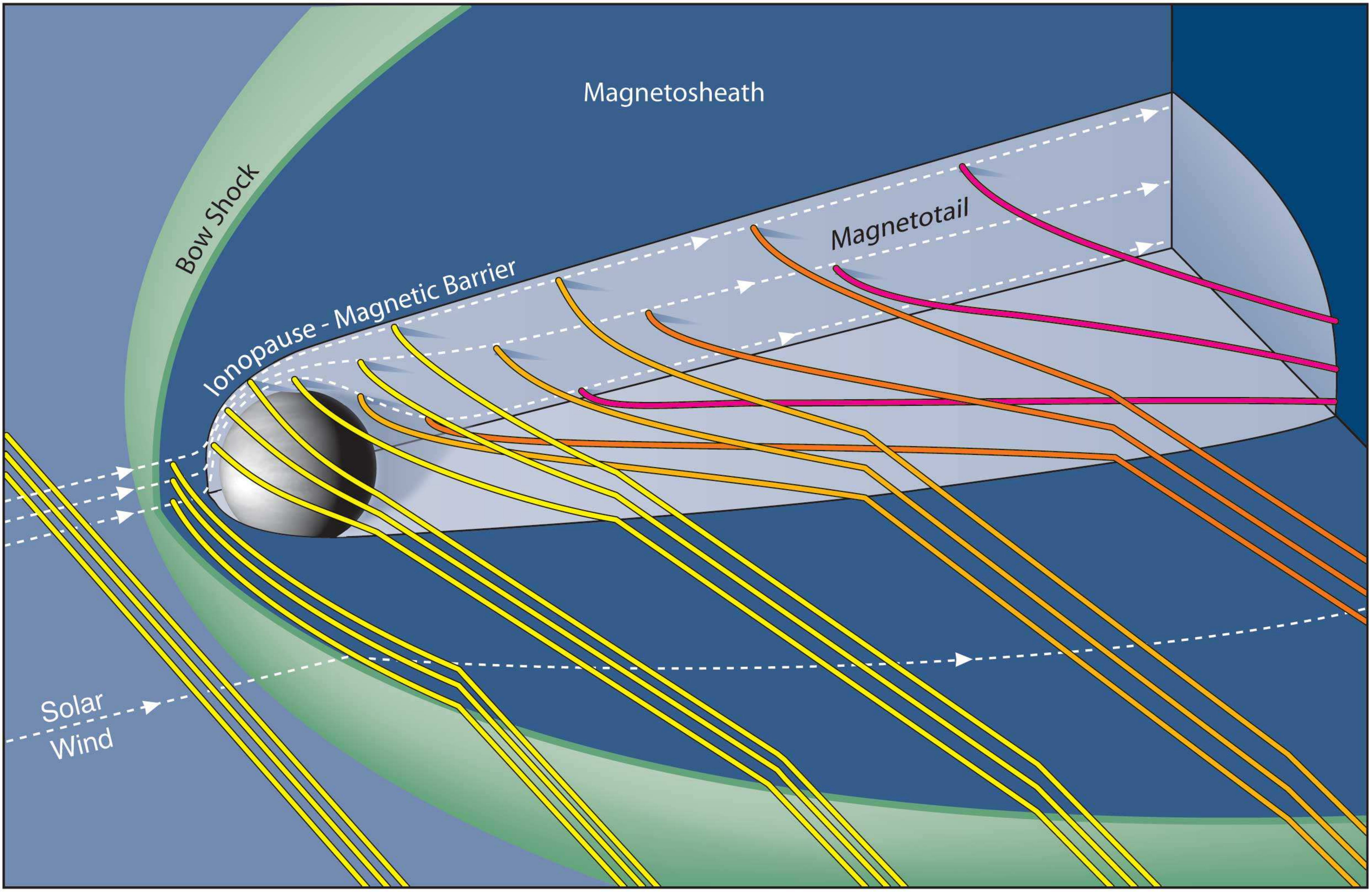}
\caption[Draping solar magnetic flux tubes around a conducting
ionosphere.]{Sketch of the draping of tubes of solar magnetic flux
  around a conducting ionosphere such as that of Venus. The flux tubes
  are slowed down and sink into the \indexit{flow!Venus}wake to form a
  tail. \indexit{Venus}[\href{http://lasp.colorado.edu/home/mop/files/2012/04/13Venus.jpg}{color
    version} of Fig.~I:13.12] \colorfig}
\label{fig:14Venus.eps}
\end{figure}
%\subsection{``Fast'' flow around a body without dynamo}
\label{sec:venus}
Venus and Mars do not have active dynamos, but they do have conducting
ionospheres. \ors[I:13.6] ``The magnetic structure surrounding Mars and
Venus is similar to that around magnetized objects because the
interaction causes the magnetic field of the solar wind to drape
around the planet. The draped field stretches out downstream (away
from the Sun), forming a magnetotail\indexit{Venus!magnetic field
  draping}\indexit{Mars!magnetic field draping}\indexit{magnetotail!Venus, Mars} (Fig.~\ref{fig:14Venus.eps}). The symmetry of the
magnetic configuration within such a tail is governed by the
orientation of the magnetic field in the incident solar wind, and that
orientation changes with time. For example, if the interplanetary
magnetic field (IMF) is oriented northward, the symmetry plane of the
tail is in the east-west direction and the northern lobe
field [has a component] away from the Sun while the southern lobe
field [has a component] 
towards the Sun. A southward oriented IMF would reverse these
polarities, and other orientations [(such as the east-west orientation
 in Fig. ~\ref{fig:14Venus.eps}) would produce corresponding] rotations of the
 tail's plane of symmetry.
 \activity{{\em Background:} On the largest scales, there may be
  a long magnetotail to the entire heliosphere, that may even be
  oblate because of the tension force of the interstellar magnetic
  field. Although alternative views propose a much shorter tail,
  making the heliosphere more like a bubble, it is illustrative to see
  how such a moderate flattening by the interstellar magnetic field
  might work. Have a look at, {\em e.g.,}
  \href{https://ui.adsabs.harvard.edu/abs/2013ApJ...771...77M/abstract}{\citet{2013ApJ...771...77M}}, in particular their Figure~9.}

The solar wind brings in magnetic flux tubes that pile up at high
altitudes at the dayside ionopause where, depending on the solar wind
dynamic pressure [($\rho v^2$)], they may either remain for extended
times, thus producing a magnetic barrier that diverts the incident
solar wind, or may penetrate to low altitudes in
localized\indexit{Mars!remanent magnetism - solar wind} bundles. Such
localized bundles of magnetic flux are often highly twisted structures
stretched out along the direction of the magnetic field. [These
bundles] may be dragged deep into the atmosphere, possibly carrying
away significant amounts of atmosphere.

While \indexit{flow!Mars}Mars' remarkably strong \indexit{remanent magnetism}remanent
magnetism \regfootnote{'Remanent magnetism' is defined as the magnetic
  field that remains after the magnetizing field is removed.} extends its influence
$>1000$\,km from the surface, the overall interaction of the solar wind
with Mars is more atmospheric than
magnetospheric. Mars interacts with the solar wind principally through
currents that link to the ionosphere, but there are portions of the
surface over which local magnetic fields block the access of the solar
wind to low altitudes. It has been suggested that
'mini-magnetospheres' extending up to 1000\,km form above the regions
of intense crustal magnetization in the southern hemisphere; these
mini-magnetospheres protect portions of the atmosphere from direct
interaction with the solar wind. [\ldots]''

\subsection{Plasma flow around a permanently magnetized body}\label{sec:flowmagnetized}
Ganymede, \indexit{flow!Ganymede}orbiting \indexit{flow!around a magnetized
  body}sub-magnetosonically in \indexit{Ganymede}Jupiter's
magnetosphere, is the only moon with a substantial, large-scale
internally maintained magnetic field. Of the planets, Earth and the
giant planets all have magnetic fields sustained by dynamos, but in
contrast to Ganymede and its surrounding plasma, they all move
super-magnetosonically relative to the solar wind. In all of these
cases, the bodies' magnetic fields are the primary 'obstacle' to the
plasma flowing around it. All deflect the plasma stream around
them. In the ideal-MHD approximation, the field-carrying plasma should
flow around the magnetic obstacle, with the distance out to which the
body's field can withstand the inflowing field dependent on the
relative strength of the forces exerted (balancing magnetic fields and
plasma inertial forces). In a realistic, non-ideal case, reconnection
between the fields is important, which depends on the plasma
parameters and on the relative directions of the two fields
involved. In case the relative motion corresponds to a
super-magnetosonic flow, a shock front develops; upstream of that, the
inflowing plasma (generally the incoming solar wind) is, so to speak,
unaware of the existence of the obstacle ahead, while the flow is
deflected only after going through the shock, then moving around the
obstacle at a reduced speed. This can still be faster than the
Alfv{\'e}n speed; see IV:10.4, which leads to a strong bending back
of the wind flow around Earth into a bullet shape, in contrast to a
V-shaped pattern for a sub-magnetosonic flow ({\em cf.}
Fig.~IV:10.4).

For those planets with a magnetic field of their own, {\em i.e.,}  those with
a dynamo, the solar wind leads to a shock-enveloped, asymmetrically-stretched
magnetosphere.  \indexit{magnetosphere|seealso{definition}} \ors[I:10.2] ``In
the most general context, we consider a {\em central object}: a
distinct well-defined body held together (in most cases) by its
gravity. It is immersed in a tenuous {\em external medium}, assumed to
be sufficiently ionized so it behaves like a plasma. The {\em
  magnetosphere} is then the region of space around the central object
within which the object's magnetic field has a dominant influence on
the dynamics of the local medium. An alternative and in some ways more
precise view is to regard the magnetosphere as the region enclosed by
its bounding surface, the \indexit{magnetopause|seealso{definition}}{\em
  magnetopause}, the latter being defined as the discontinuity of the
magnetic field where its direction changes: inside it is controlled by
the magnetic field of the central object, while outside it is
determined primarily by the magnetic field of the distant external
medium. This definition is particularly useful for the magnetospheres
of planets in the solar wind: the continual variability of the
\indexit{interplanetary!magnetic field} interplanetary magnetic field
direction in contrast to the relative constancy of the
\indexit{magnetic!dipole!planetary}planetary magnetic dipole allows
in most cases an easy observational identification of the
magnetopause.''

%\subsection{Interaction of solar wind with a planetary magnetic field}
\label{int}
\subsection{A closed magnetosphere}\label{mp}

\begin{figure}[t]
%\includegraphics[width=\linewidth,clip,viewport=30mm 160mm 240mm 300mm]
%%viewport=15mm 80mm 120mm 170mm]
%%,viewport=0.2\linewidth 0.75\linewidth 1.0\linewidth 1.4\linewidth]
%{../vasyliunas/ssvmvF1.pdf}
%\centerline{\psfig{figure=figures/ssvmvF1.eps,width=9cm}}
\centerline{\includegraphics[width=9cm]{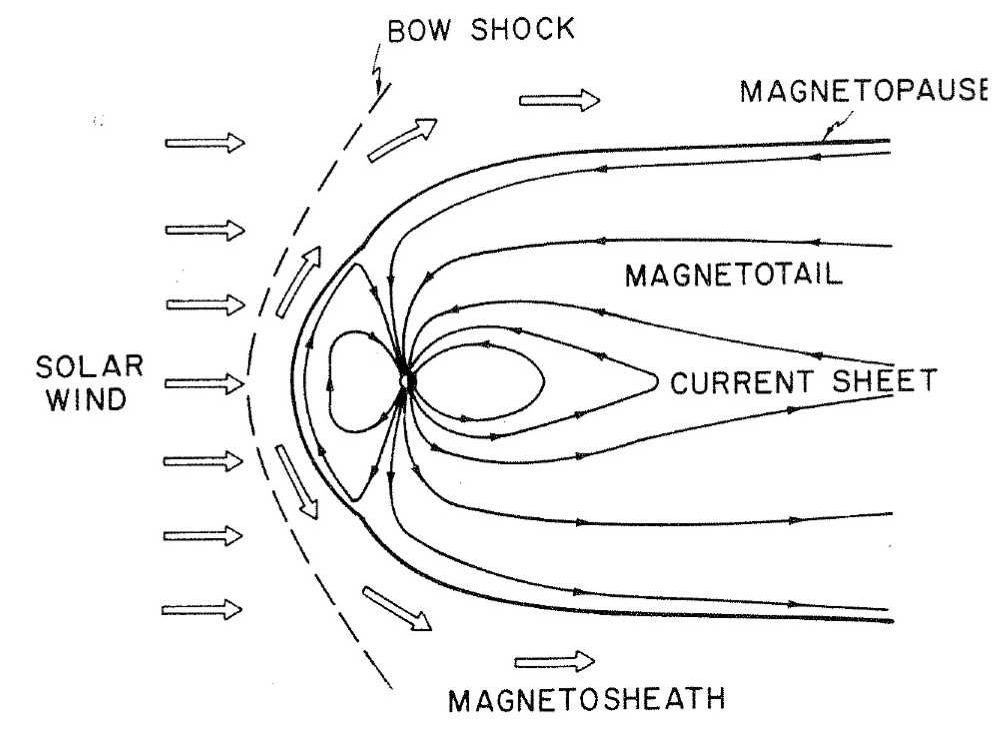}}
%{vmv_pict_organ1.jpg}
%\figurewidth{20pc}
%\figbox*{}{}{\epsfig{file=curr_clos.eps,width=\hsize}}
\caption[A magnetically closed
magnetosphere.]{\label{fig:MS}Schematic view of a magnetically closed
  magnetosphere, cut in the noon-midnight meridian plane. Open arrows:
  solar wind bulk flow. Solid lines within magnetosphere: magnetic
  field lines (direction appropriate for Earth). [Fig.~I:10.1]}
\end{figure}

\indexit{solar!wind!interaction with planetary magnetic field}\indexit{flow!closed
    magnetosphere} \indexit{magnetosphere!closed}
\ors[I:10.3] ``The basic configuration of a prototypical planetary
magnetosphere is sketched in Figure \ref{fig:MS}.
%
%\begin{figure}[tb]
%\setlength{\unitlength}{0.8cm}
%\begin{picture}(16,12)
%\thicklines
 %\put(1,1){\framebox(14.5,11){\ }}
%\end{picture}
%%%%%%%%%%%%%%%%%%%%%%%%%%%%%%%%%%%%%%%%%%%%%%%%%%%%%%%%%%%% FIGURE 1
Many of its characteristic structures can be understood on the basis
of a simple model that takes into account only the two ingredients
indispensable for the formation of a magnetosphere: the solar wind
(mass density $\rho_{\rm sw}$, bulk velocity ${\mathbf v}_{\rm sw}$)
and the planetary \indexit{dipole moment}magnetic
\indexit{dipole moment|seealso{definition}}field
(dipole moment \mbox{$\mu_{\rm p}=B_{\rm p} {R_{\rm p}}^3$}, with $B_{\rm p}$ the surface
magnetic field strength at the equator and $R_{\rm p}$ the radius of the
planet). As a consequence of constraints imposed by the
magnetohydrodynamic (MHD) approximation [\ldots], the boundary surface
between\indexit{magnetopause} the solar wind and the planetary
magnetic field --- the magnetopause --- is nearly impermeable both to
plasma and to magnetic field, resulting in a clear separation between
the two distinct regions of space: the magnetosphere itself, within
which the magnetic field lines from the planet are confined and from
which the solar wind plasma is excluded, and the exterior region
beyond the magnetopause, to which the plasma that comes from the solar
wind is confined. This simple {\em closed magnetosphere} is only a
first-order approximation (in reality the magnetopause is not
completely impermeable but allows, under certain conditions, some
penetration\indexit{magnetosphere!open}\indexit{magnetosphere!closed} of plasma and of magnetic field to produce the {\em open
  magnetosphere} described in Sect.~\ref{open}); it does, however,
describe fairly accurately the size and shape of the main structures.

The solar wind flow, initially directed away from the Sun, must be
diverted around the magnetosphere, as indicated in Figure
\ref{fig:MS}. Because the initial flow speed is supersonic and
super-Alfv\'enic (faster than both the speed of sound and the Alfv\'en
speed $v_{\rm A}$), the solar wind is first slowed down, deflected,
and heated at a detached {\em bow shock} \indexit{bow shock}standing
upstream of the magnetopause (analogous to the sonic boom in
supersonic aerodynamic flow past an obstacle). The region between the
bow shock and the magnetopause, within which the plasma from the solar
wind is flowing around the magnetosphere, gradually speeding up and
cooling, is called \indexit{magnetosheath}the \indexit{magnetosheath|seealso{definition}}{\em
  magnetosheath} [\ldots]. \activity{{\em Show:} Use Eqs.~(\ref{eq0})
  and~(\ref{eq1}-\ref{eq5}) to calculate that in the case of a strong shock
  (in which the thermal energy of the solar wind upstream of the bow
  shock can be ignored) the temperature just downwind of
  the bow shock is given by $(3m_{\rm p}/32k)v_{\rm sw}^2$ for a wind
  speed of $v_{\rm sw}$, and that the density contrast across the
  shock is a factor of 4 (show that is true anywhere along the
  shock). Use this to estimate the angle from the upwind direction
  out to which the flow remains supersonic just inside the shock front
  (remembering that the transverse component of the velocity is
  unaffected by the shock).}

The location of the magnetopause is determined primarily by the
requirement \indexit{pressure balance} of pressure balance: the total
pressure (plasma plus magnetic) must have the same value on both sides
of the discontinuity. In the simple closed magnetosphere considered
here, the plasma pressure inside the magnetopause and the magnetic
pressure outside are both neglected. The exterior pressure then scales
as the linear momentum flux density in the undisturbed solar wind,
\mbox{$\rho_{\rm sw}{v_{\rm sw}}^2$} (often called the \indexit{solar!wind!dynamic pressure} {\em dynamic pressure} [or {\em ram
  pressure}] of the solar wind), and is maximum in the sub-solar
region, where the plasma near the magnetopause is almost stagnant. The
interior pressure scales as the magnetic pressure of the dipole field,
\mbox{$(1/8\pi)(\mu_{\rm p}/{\mathrm r}^3)^2$} with $\mu_{\rm p}$ the
magnetic dipole moment of the planet, and thus varies strongly with
distance from the planet. Equating the two gives an estimate for the
distance $R_{\rm mp}$ of the sub-solar magnetopause:
\begin{equation}
R_{\rm mp} = \frac{\left( \xi \mu_{\rm p} \right)^{1/3}}{\left( 8 \pi \rho_{\rm sw} {v_{\rm sw}}^2 \right)^{1/6}}
\label{eq:PB}
\end{equation}
where $\xi$ is a numerical factor to correct for the added field from
magnetopause currents ($\xi \simeq 2 $ to first
approximation). \sactivity{$\circledS$ {\em Show:} What is the expression
  for the temperature of the gas at the stagnation point on the
  magnetopause assuming that the flow continues adiabatically after
  the shock ({\em i.e.,}  that it conserves the sum of bulk kinetic and
  thermal energies)? Calculate that temperature for $v_{\rm
    sw}=800$\,km/s. \mylabel{act:stagnation} \solution{stagnation}}

The distance given in Eq.~(\ref{eq:PB}) (with various choices of
$\xi$) is often called the \indexit{magnetopause!Chapman-Ferraro
  distance}Chapman-Ferraro \indexit{Chapman-Ferraro
  distance}distance. [Here, we] consistently use the symbol
$R_{\rm CF}$ for the distance {\em defined} by Eq.~(\ref{eq:PB}) with
$\xi = 2 $, {\em i.e.,} for the nominal distance of the sub-solar
magnetopause predicted by \indexit{solar!wind!pressure
  balance}pressure balance; [the symbol $R_{\rm mp}$ is reserved] for
the {\em actual} distance of the sub-solar magnetopause in any
particular\indexit{magnetopause!sub-solar distance} context. Thus,
\mbox{$R_{\rm mp}\simeq R_{\rm CF}$} in the present case of a simple
closed magnetosphere but not necessarily in the case of more general
models. \sactivity{$\circledS$ {\em Show:} (a) Use Eq.~(\ref{eq:PB}) to calculate the
  scaling of $R_{\rm CF}$ with orbital radius (in AU), planetary
  magnetic field at the surface, and planetary radius. For Earth and
  the giant planets, estimate for each of the planets the model-based
  magnetopause distance $R_{\rm CF}$ setting $\xi=2$. Assume the
  following: that the solar wind speed averages to roughly the same
  value at all the planets (say $v_{\rm sw} = 400$\,km/s). Use info in
  Tables~\ref{tab:fran2} and \ref{tab:fran3}. (b) For Earth and
  Saturn, compare $R_{\rm CF}$ to the orbital radii of Moon, Enceladus
  and Titan. \solution{cfradius}
  \mylabel{act:cfradius}} \activity{{\em Show: } With the fastest
recorded solar-wind gusts at $v_{\rm sw} \approx 2500$\,km/s,
calculate the required plasma density to push the magnetopause to
within geosynchronous orbit according to Eq.~(\ref{eq:PB})?}

The pressure balance condition, combined with assumptions about the
sources of the magnetic field within the magnetosphere, may be used to
calculate not only the distance to the sub-solar point, but also the
complete \indexit{magnetopause!shape of}shape of the magnetopause
surface (for discussion of such models at Earth, see
Ch.~I:11). Typically the magnetopause is roughly spherical on the
dayside of the planet, facing into the solar wind flow (the effective
center of the sphere being located behind the planet, very roughly at
a distance $\sim 0.5\ R_{\rm mp}$), and is elongated in the
anti-sunward direction.

\begin{figure}[t]
\setlength{\unitlength}{0.45cm}
%\includegraphics[width=\linewidth,clip,viewport=30mm 85mm 240mm 240mm]{ssvmvF2.pdf}
%\figurewidth{20pc}
%\figbox*{}{}{\epsfig{file=curr_clos.eps,width=\hsize}}
\begin{center}\begin{picture}(11,14)(0,1)
\put(1,1){\vector(1,0){8.5}} \put(9.7,1){\parbox{2cm}{\Large $\log
r$}} \put(1,1){\vector(0,1){13.5}} \put(0.8,14.7)
{\parbox{2cm}{\Large $\log p$}} \thicklines
\put(4,2){\line(-1,6){2}} \qbezier[75](1,2)(5.5,2)(8,2)
\put(8.1,2){\parbox{5cm}{\Large $p_{\rm external}$}}
\qbezier(3.5,5)(3.75,3.5)(4.5,2) \qbezier(3,8)(3.5,5)(5.5,2)
\thinlines \put(5,5){\vector(-1,-1){1.1}}
\put(5.1,5.1){\parbox{3cm}{\Large $\frac{\mathrm{B}^2}{8\pi}$}}
\put(6.5,4){\vector(-1,-1){1.3}} \put(6.6,4.1){\parbox{5cm}{\Large
$\frac{\mathrm{B}^2}{8\pi}+p$}} \put(1.1,2.5){\parbox{3cm}{\Large
$\frac{\mu_{\rm p}^2}{8\pi r^6}$}} \put(2.5,2.5){\vector(2,1){1}}
\put(4,1.5){\parbox{2cm}{\Large ${\mathrm{R_{\rm CF}}}$}}
\put(5.5,1.5){\parbox{2cm}{\Large ${\mathrm{R_{\rm mp}}}$}}
\end{picture}\end{center}
\caption[Total pressure (magnetic plus plasma) with distance from a planet.]{\label{fig:MP}Variation of total pressure (magnetic plus plasma) with distance from the planet
and its relation to the radial distance of the sub-solar magnetopause. Compared
are the relationship in 
%Eq.~(10.1)
Eq.~(\ref{eq:PB})
to a schematic representation of
a  more realistic plasma-filled, non-dipolar planetary magnetic
field. [Fig.~I:10.2]}
\end{figure}
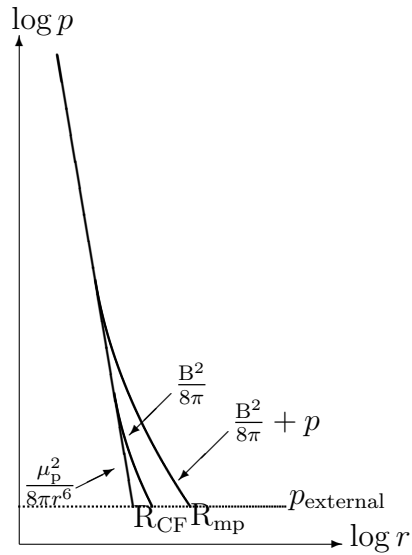

The magnetopause distance $R_{\rm mp}$ may be regarded as the
characteristic scale\indexit{magnetosphere!characteristic scale size}
for the size of a magnetosphere. Equal to $R_{\rm CF}$ in the case of
negligible plasma pressure and no magnetic field sources other than
the planetary dipole inside the magnetosphere, $R_{\rm mp}$ can be
readily calculated from Eq.~(\ref{eq:PB}) given only a few basic
parameters of the system. In the case that the plasma pressure or a
non-dipolar field in the outer regions of the magnetosphere are not
negligible, the qualitative effects on $R_{\rm mp}$ can still be
estimated from pressure balance, as illustrated in Figure
\ref{fig:MP}:
%
%\begin{figure}[tb]
%\setlength{\unitlength}{0.8cm}
%\begin{picture}(16,12)
%\thicklines
% \put(1,1){\framebox(14.5,11){\ }}
%\end{picture}
%%%%%%%%%%%%%%%%%%%%%%%%%%%%%%%%%%%%%%%%%%%%%%%%%%%%%%%%%%%% QQQ
(a) the actual distance $R_{\rm mp}$ is larger than the nominal distance
$R_{\rm CF}$ (the value $\xi=2$ instead of  $\xi=1$ is in fact a
consequence of the non-dipolar field from the magnetopause currents),
(b) a change of solar wind dynamic pressure produces a larger
change of magnetopause distance --- the magnetosphere is
less 'stiff' if plasma pressure in the interior is significant.''

\subsection{The open magnetosphere}\label{open}
\indexit{magnetosphere!open} \ors[I:10.3.3] ``At the location of
the planets, \indexit{interplanetary!magnetic field}the interplanetary
magnetic field is weak in the sense that the energy density of the
magnetic field is very small in comparison to the kinetic energy
density of solar wind bulk flow, or equivalently ${v_{\rm A}}^2 \ll
{v_{\rm sw}}^2$ [(see Sect.~\ref{sec:solwindenergy})]. The flow of
solar wind plasma past the magnetospheric obstacle deforms the
magnetic field lines within the magnetosheath and drapes them around
the magnetopause, a process well modeled at Earth ({\em cf.}\
Sect.~I:11.4). The magnetic field is amplified and may become
dynamically no longer negligible as the magnetopause is approached,
but the total pressure is in general not greatly modified, an increase
of magnetic pressure often being offset by a decrease of plasma
pressure. One might therefore anticipate that the effect of the
interplanetary magnetic field on planetary magnetospheres should be
minimal.

\begin{figure}[t]
%\includegraphics[width=\linewidth,clip,viewport=0mm 5mm 200mm 230mm]
%{../vasyliunas/ssvmv_F3.pdf}
%\centerline{\psfig{figure=figures/ssvmv_F3.ps,width=9cm}}
\centerline{\includegraphics[width=9cm,bb= 0 0 515 578]{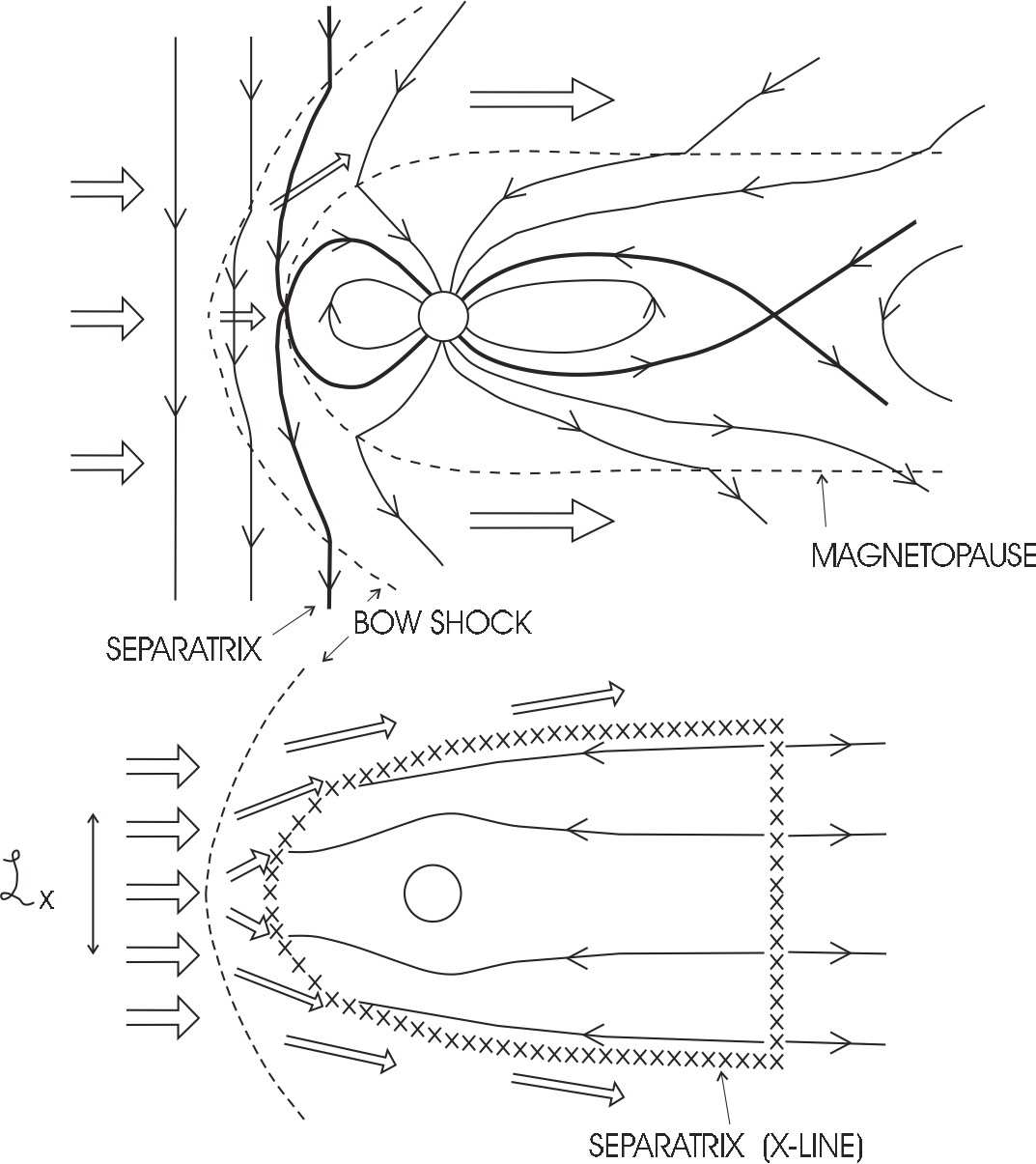}}
%\figurewidth{20pc}
%\figbox*{}{}{\epsfig{file=curr_clos.eps,width=\hsize}}
\caption[A magnetically open
magnetosphere.]{\label{fig:OMS}Schematic representation of a
magnetically open magnetosphere. {\em Top}: cut in noon-midnight meridian
plane; thick lines are magnetic field lines within the 'separatrix 
surfaces' that separate open from closed or open from interplanetary field
lines; other conventions same as in Fig.~\ref{fig:MS}. {\em Bottom}: cut in
equatorial plane; a line of $\times$ symbols represents intersection
with the two branches of the separatrix; solid lines are streamlines
of magnetospheric plasma flow, and $\mathcal{L}_\times$ represents
the projection of the dayside magnetic reconnection region along
streamlines into the solar wind. [Fig.~I:10.3]}
%(see Sect.~\ref{convec}).}
\end{figure}

What is overlooked in the above discussion is the possibility that,
through the process of \indexit{magnetic!reconnection}{\em magnetic
  reconnection} [\ldots], the magnetic field lines from the planet may
become connected with those of the interplanetary magnetic field, to
produce the magnetically \indexit{magnetosphere!open}{\em open
  magnetosphere}, sketched in Fig.~\ref{fig:OMS} for the simplest case
of the interplanetary magnetic field parallel\indexit{magnetopause} to
the planetary dipole moment [(and thus with antiparallel fields at
their interface)].
%
%\begin{figure}[tb]
%\setlength{\unitlength}{0.8cm}
%\begin{picture}(16,12)
%\thicklines
% \put(1,1){\framebox(14.5,11){\ }}
%\end{picture}
%%%%%%%%%%%%%%%%%%%%%%%%%%%%%%%%%%%%%%%%%%%%%%%%%%%%%%%%%%%%
The magnetopause is now no longer impermeable to the magnetic field,
and as a consequence it no longer need be impermeable to plasma,
either. \activity{{\em Background:} Illustrative diagrams like
  Fig.~\ref{fig:OMS} typically show the interplanetary magnetic field
  (IMF) as lying within the $x-z$ plane of such diagrams. In reality,
  the three IMF components $B_{x,y,z}$ are typically of comparable
  magnitude. Because of the relative tilt of the rotation and magnetic
  dipole axes relative to the Earth's orbital plane, the orientation
  of the Earth's magnetic axis relative to the incoming wind changes
  in the course of the year. Consider how the diagrams should look
  when drawn in three dimensions for a few different combinations of
  $B_{x,y,z}$. Look up the 'Russell-McPherron effect' which attributes
  the semi-annual variations in geomagnetic activity largely to the
  relative orientation of the Earth's dipole axis: maximum geomagnetic
  activity around the equinoxes, minimum around solstices. {\em
    Advanced/Group:} Try sketching a version of Fig.~\ref{fig:OMS}
  under these other circumstances. Consider where the reconnection
  points are and the shape of the separatrix. Keep in mind that the
  $x-z$ cut may not be the best plane to
  use. \mylabel{act:russellmcpheron}}\activity{{\em Show: } Use the
  equation for a dipole field to show that the ratio of maximum to
  minimum field strength on a sphere equals 2.  Look up the oblateness
  of Jupiter and Saturn and see if it accounts for the ratio of
  minimum to maximum surface field in Table~\ref{tab:fran3}. Interpret
  your finding. \mylabel{act:oblatedipole}}

\begin{table}[t]
{\small
  \caption[Properties of the solar wind near the planets.]{Properties
    of the solar \indexit{solar!wind!properties!near the planets}wind near the planets [after
    Table I:13.2]. Plasma $\beta$ values assume a solar-wind
    temperature of 1.5\,MK. }
\label{tab:fran2}
\begin{center}\begin{tabular}{lrrrrr}
   \hline \hline
Planet & Distance & Solar wind ion & $B_{\rm IMF}$ & $\approx \beta$ & $\approx v_{\rm A}$ \\
           & $d_{\rm p}$ (AU) $^a$ & density (cm$^{-3}$) & ($\mu$G)
           $^c$& & (km/s) \\
   \hline
Mercury & 0.39 & 53 & 410 & 2 &120\\
Venus & 0.72 & 14 & 140 & 4 & 80\\
Earth & 1 & 7$^b$ & 80 & 6 & 70\\
Mars & 1.52 & 3 & 50 & 6 & 60\\
Jupiter & 5.2 & 0.2 & 10 & 10 & 50\\
Saturn & 9.5 & 0.07 & 6 & 10 & 50\\
Uranus & 19 & 0.02 & 3 & 10 & 50\\
Neptune & 30 & 0.006 & 2 & 10 & 50\\
   \hline \hline
  \end{tabular}\end{center}

{\em \small $^a$ 1 AU = $1.5 \, 10^8$\,km; $^b$ The density of the solar wind
fluctuates by about a factor of 5 about typical values of $n_{\rm
  sw} \sim (7\, {\rm cm}^{-3})/d_{\rm p}^2$; $^c$  mean values.}
}
\end{table} 
\begin{table}%[t]
\caption[Intrinsic magnetic fields of Solar System
  bodies.]{\label{tab:fran3} [Intrinsic \indexit{magnetic!field!of
      Solar-System bodies}magnetic fields of Solar System
  bodies. After Table~I:13.3, with planetary rotation periods
  $P_{\rm p}$ and planetary radii $R_{\rm p}$].}
%\begin{center}
{\small
\begin{center}\begin{tabular}{llllllll}
\hline \hline
&Gany-&Mer-&Earth&Jupiter&Saturn&Uranus&Nep-\\
&mede &cury  &       &           &          &          &tune\\
\hline
$B_{dip,eq}$ $^a$&7.2\,mG&3\,mG&0.31\,G&4.3\,G&0.21\,G&0.23\,G&0.14\,G\\
$B_{max}/B_{min}$ $^b$	&2&2&2.8&4.5&4.6&12&9\\
dipole tilt $^c$&$-4^\circ$&$\sim 10^\circ$&11.2$^\circ$&-9.4$^\circ$	&-0.0$^\circ$&-59$^\circ$&-47$^\circ$\\
dipole offset $^d$ &-&-&0.076&0.119&0.038&0.352&0.485\\
obliquity $^e$&0$^\circ$&0$^\circ$&23.5$^\circ$&3.1$^\circ$&26.7$^\circ$&97.9$^\circ$	&29.6$^\circ$\\
$\delta \phi_{\rm sw}$ $^f$&90$^\circ$&90$^\circ$&67-114$^\circ$&87-93$^\circ$&64-117$^\circ$&8-172$^\circ$	&60-120$^\circ$\\
\hline
$P_{\rm p}$ (h) & 171 & 4223. & 24 & 9.9 & 10.7 & 17.2 & 16.1 \\
$R_{\rm p}/R_{\rm p,\oplus}$ & 0.41 & 0.38 & 1 & 11.2 & 9.4 & 4.0 & 3.9 \\
\hline \hline
\end{tabular}\end{center}
}
% \end{center}

{\em \small $^a$ Surface field at dipole equator.  Values derived from modeling
the magnetic field as an offset dipole; $^b$ ratio of maximum
surface field to minimum, which equals to 2 for a centered dipole
field (this ratio tends to increase with the planet's oblateness); $^c$
angle between the magnetic and rotation axes (positive values
correspond to magnetic field directed north at the equator;  the
magnetic dip poles of the Earth's field are currently located at
86$^\circ$N and 65$^\circ$S latitudes and moving about 10$^\circ$ per
century); $^d$ values (in planetary radii, $R_{\rm p}$); $^e$ the
inclination of a planet's spin equator to the ecliptic plane; $^f$
range of the angle between the
radial direction from the Sun and the planet's rotation axis over an
orbital period (in Ganymede's case, the angle is between the
corotational flow and the moon's spin axis).}
\end{table} 

\begin{figure}
\centering
\includegraphics[width=\textwidth]{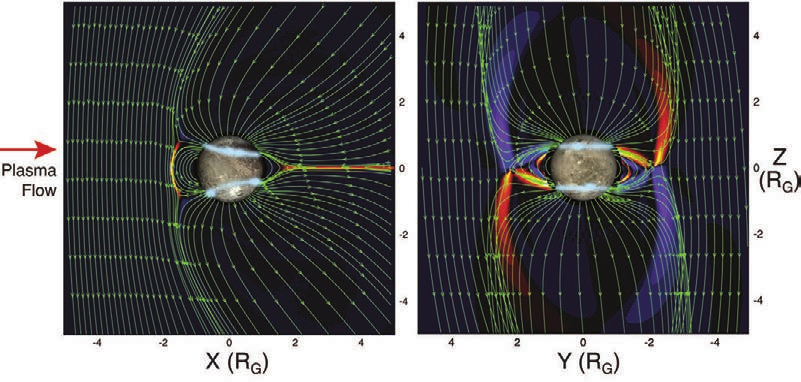}
\caption[Numerical model of the magnetosphere of Ganymede.]{Numerical
  model of the \indexit{Ganymede}magnetosphere of Ganymede with the satellite and the
  location of \indexit{aurora}auroral emissions superimposed. [The shaded areas show
  the current density perpendicular to the plane; yellow-red out of
  the plane, purple-blue into the plane.] {\em Left:} view looking at
  the anti-Jupiter side of Ganymede. {\em Right:} View looking in the
  direction of the plasma flow at the upstream side (orbital trailing
  side) of Ganymede with Jupiter to the left. The shaded areas show
  the regions of currents parallel to the magnetic
  field[; yellow-red anti-parallel, purple-blue
  parallel]. [Fig.~I:13.11;
  \href{https://ui.adsabs.harvard.edu/abs/2008JGRA..113.6212J/abstract}{source:
  \citet{2008JGRA..113.6212J}}.] \colorfig }
\label{fig:13Ganymede.eps}
\end{figure}
\nocite{Jia2008} The modifications of the magnetospheric system
implied by the open character of the magnetosphere are in some ways
minor, in other ways very far-reaching.  The location and shape of the
dayside magnetopause is for the most part not greatly modified (in
agreement with the expectations above). The component $B_\perp$ of the
magnetic field normal to the magnetopause is in general small compared
to the magnitude of the field, \mbox{$|B_\perp| \ll |B|$} (so much so
that it is often difficult to establish by direct observation that
\mbox{$B_\perp \neq 0$}, and much of the evidence for an open
magnetosphere has been indirect). On the other hand, the total amount
of \indexit{open magnetic flux}open magnetic flux $\Phi_M$ of one
polarity can (at least at Earth) become comparable to the maximum
amount that could reasonably be expected to be open (estimated as
\mbox{$\sim \mu_{\rm p}/R_{\rm mp}$}, the dipole flux beyond the
distance of the sub-solar magnetopause); despite
\mbox{$|B_\perp| \ll |B|$}, this is possible if the effective length
of the magnetotail is much larger than $R_{\rm mp}$. [\ldots That
length is in large part determined by the\indexit{magnetopause!reconnection} efficiency of the reconnection process. This] depends greatly
on the relative orientation of magnetic fields on the two sides of the
magnetopause, one result of which is that the open character of the
magnetosphere is most pronounced when the interplanetary magnetic
field is parallel to the \indexit{dipole moment!planetary}planetary
dipole moment ({\em i.e.,}  anti-parallel to the dipole magnetic field in the
equatorial plane),
\mbox{${\mathbf B}_{\rm sw}\cdot {{\bm \mu_{\rm p}}}>0$}.  Because the
direction of the interplanetary magnetic field is highly variable on
all time scales, this can lead to pronounced time-varying changes of
magnetospheric configuration as well as energy input and dissipation
[\ldots].''

Although diagrams of the terrestrial environment such as
Fig.~\ref{fig:OMS} generally include the bow shock, the processes
discussed here are generic and apply equally when a magnetized plasma
flows sub-Alfv{\'e}nically around a magnetized body. One example is a
numerical simulation of Ganymede orbiting within Jupiter's
magnetosphere as shown in Fig.~\ref{fig:13Ganymede.eps}.

%\activity{{\em Show: } Use Fig.~\ref{fig:13Ganymede.eps} to estimate
%  the Alfv{\'e}n velocity in the jovian magnetosphere near the orbit
%  of Ganymede. First, estimate the flow speed of the incoming plasma
%  relative to the moon realizing that the plasma is sub-corotating by
%  about 80\%\ of the speed of corotation with Jupiter at Ganymede's
%  orbit. Then use the geometry of the field shown in the figure to
%  estimate the Alfv{\'e}n velocity. The coordinate system for the
%  simulation has the $y$-axis pointing towards Jupiter and the
%  $z$-axis aligned with the jovian spin axis, and the units are
%  expressed in Ganymede radii. \mylabel{act:ganymede}}

\subsection{Solar wind-magnetosphere-ionosphere interaction}\label{flow}
\indexit{magnetosphere!interaction with ionosphere}
\subsubsection{Fundamental principles}\label{fun}
\ors[I:10.4.1] ``A well-known consequence of the \indexit{MHD!approximation}MHD
approximation is a constraining relation between the plasma bulk flow
and the magnetic field: plasma elements that are initially on a common
field line remain on a common field line as they are carried by the
bulk flow. Because magnetic field lines in the magnetosphere of a planet
connect to the ionosphere of the planet, any discussion of plasma flow
in the magnetosphere immediately involves questions of
magnetosphere-ionosphere
interaction.\indexit{magnetosphere-ionosphere!interaction} The field
lines extend in fact into the interior of the planet, which in many
cases is highly conducting electrically; hence it might seem that the
magnetospheric flow should be constrained by the planet itself. This
does not happen, however, because [\ldots] most planets possess an
electrically neutral (and effectively non-conducting)
\indexit{atmosphere!insulating}atmosphere, sandwiched between the
ionosphere and the planetary interior. Although very thin in
comparison to the radius of the planet, this layer suffices to break
the MHD constraints and thus allows the plasma in the ionosphere and
the magnetosphere to move without being necessarily attached to the
planet; without such an insulating layer, much of the magnetospheric
dynamics as we know it would not be possible.

While the plasma in the \indexit{ionosphere}ionosphere can thus move
relative to the planet, it remains constrained to move more or less
together with the plasma in the magnetosphere.  The conventional
formulation, however, describes the plasma flow rather differently in
the two regions. The magnetosphere is treated, to first approximation
at least, as an MHD medium, with the electric field ${\mathbf E}$
related to the plasma bulk flow ${\mathbf v}$ by the MHD approximation
and with the electric current ${\mathbf j}$ related to plasma pressure
by stress balance. The \indexit{ionosphere}ionosphere is treated, on
the other hand, as a moving conductor (the conductivity results
primarily from collisions between the ions and the neutral particles,
planetary ionospheres being\indexit{ionosphere!planetary} for the
most part weakly ionized), with ${\mathbf j}$ related by a
conductivity tensor to \mbox{${\mathbf E}+{\mathbf v}_\perp \times
  {\mathbf B}/c$}, where ${\mathbf v}_\perp$ is the bulk velocity of the
neutral medium. [\ldots\ A few comments on this coupling:]

(1) As long as ${v_{\rm A}}^2/c^2\ll1$ ({\em i.e.,}  the inertia of
the plasma is dominated by the rest mass of the plasma particles, not
by the relativistic energy-equivalent mass of the magnetic field),
${\mathbf v}$ produces ${\mathbf E}$ but ${\mathbf E}$ does not produce
${\mathbf v}$. The primary quantity physically is thus the plasma bulk flow,
established by appropriate stresses. The electric field is the result of the
flow, not the cause; its widespread use in calculations is primarily for
mathematical convenience [\ldots].

(2) The electric current in the ionosphere is not an Ohmic current in
the physical sense, and its conventional expression by
the\indexit{Ohmic current} 'ionospheric\indexit{ionospheric Ohm's
law} Ohm's law' [as discussed around Eq.~(\ref{eq:condtensor})] 
has only a mathematical significance [\ldots] 
Physically, the current is determined by the requirement that the
Lorentz force balance the \indexit{collisions!plasma/neutral
atmosphere}collisional drag\indexit{drag!collisional} between the
plasma and the neutral atmosphere when their bulk flow velocities
differ.  [\ldots] 
The current in the ionosphere is thus
governed by stress balance in the same way as the current in the
magnetosphere [\ldots\ while neglecting the time derivative term in the
momentum equation.]

(3) Underlying the neglect of the time derivative (acceleration) terms
in the momentum equations is the implicit assumption that any
imbalance between the mechanical and the magnetic stresses (which,
fundamentally, is what determines the acceleration of the plasma)
produces a bulk flow that acts to reduce the imbalance, which then
becomes negligible over a characteristic time scale (easily shown to
be of the order of the Alfv{\'e}n \indexit{wave!travel time}wave
travel time across a typical spatial scale $\mathcal{L}$, {\em e.g.,} along a field
line). The theory is thus applicable only to systems that are stable
and evolve on time scales much longer than $\mathcal{L}/v_{\rm A}$.

(4) For [phenomena well above the scale of plasma oscillations, so slow
compared to \mbox{$1/\omega_{\rm p}$} and large compared to 
\mbox{$\lambda_{\rm e}
  \equiv c/\omega_{\rm p}$},] (where $\omega_{\rm p}$ is the electron plasma frequency and
$\lambda_{\rm e} $ the electron\indexit{plasma!oscillations!time and length
  scales}\indexit{electron plasma
  frequency}\indexit{electron-inertial length}\indexit{collisionless
  skin depth} inertial length, also known as collisionless skin
depth)], ${\mathbf j}$ adjusts itself to become equal to
\mbox{$(c/4\pi) \nabla \times {\mathbf B}$} and not the other way
around; although ${\mathbf B}$ is in principle determined from a given
${\mathbf j}$ by Maxwell's equations (on a time scale of light travel
time, $\sim \mathcal{L}/c$), in a large-scale plasma any
\mbox{${\mathbf j}\neq(c/4\pi) \nabla \times {\mathbf B}$} is
immediately (on a time scale $\sim 1/\omega_{\rm p}$) changed by the action
of the displacement-current electric field on the free electrons in
the plasma.  The current continuity condition $\nabla \cdot {\mathbf
  j}=0$ is thus satisfied automatically; there is no physics in
current closure --- what is often discussed under that rubric is in
reality the coupling of the Maxwell stresses along different portions
of a field line.''

\subsubsection{Corotation}\label{corot}
\indexit{magnetosphere!corotation} \ors[I:10.4.2] ``Corotation with the
planet is the simplest pattern of plasma flow in a planetary
magnetosphere and one that plays a major role particularly in the
magnetospheres of the giant planets. [\ldots] If the planet possesses
an \indexit{atmosphere!insulating}insulating atmosphere, the rotation
of the planet itself has no {\em direct} effect on plasma flow in the
magnetosphere, as discussed in Sect.~\ref{fun}. What does affect
plasma flow is the motion of the neutral upper atmosphere
(thermosphere) at altitudes of the ionosphere (where the neutral and
the ionized components coexist and interact). 'Corotation with the
planet' is therefore not quite an accurate description. What really is
meant is co-motion with the upper atmosphere, which in turn is then
assumed to corotate with the planet, for reasons unrelated to the
magnetic field: vertical transport of horizontal linear momentum from
the planet to the neutral atmosphere ({\em e.g.,} by collisional or eddy
viscosity and similar processes), together with an assumed small
relative amplitude of neutral winds.

Any difference between the bulk flow of the neutral medium and the
ionized component of the plasma in the\indexit{drag!collisional}
ionosphere results in a \indexit{collisions!plasma/neutral
atmosphere}collisional drag that must be balanced by the Lorentz
force; without it, the drag force would soon bring the plasma to flow
with the (much more massive) neutral medium. The Lorentz force in the
ionosphere is coupled to a corresponding Lorentz force in the
magnetosphere, which in turn must be balanced locally by an
appropriate mechanical stress.  The net result is that departure from
corotation requires a mechanical stress in the magnetosphere to
balance the plasma-neutral drag\indexit{drag!plasma-neutral} in the
ionosphere; conversely, plasma will corotate if the stress in question
is negligibly small. (It is fairly obvious that the direction of the
stress must be more or less azimuthal, opposed to the direction of
rotation.) Quantitatively, the requirements for corotation of
magnetospheric plasma may be expressed by four conditions:

(1) Planet-atmosphere coupling: This is simply the assumption,
discussed above, that the upper\indexit{planet-atmosphere coupling}
atmosphere effectively corotates with the planet.

(2) Plasma-neutral coupling in the ionosphere: the collisional drag of
the neutral medium on the plasma must be sufficiently strong
to\indexit{thermosphere-ionosphere!coupling} ensure
\mbox{${\mathbf v}\simeq{\mathbf v}_{\rm n}$}. The quantitative condition is derived
in principle [from the 
momentum equation\indexit{momentum equation!ionospheric} of the 
ionospheric plasma
(horizontal components only; compare with Eq.~(\ref{fig:mhdset}) to
see what is hiding in the ellipsis)
\begin{equation}
\frac{\partial\rho {\mathbf v}}{\partial t}  + \ldots
= {\mathbf j} \times {\mathbf B}/c
- \nu_{\rm in} \rho ({\mathbf v}-{\mathbf v}_{\rm n})
\label{eq:ionmom}
\end{equation}
%Eq.~(\ref{eq:ionmom}) 
with the left-hand side set
to zero (where $\nu_{\rm in}$ is the ion-neutral collision
frequency),]
but with one complication: what is relevant for the
interaction with the magnetosphere is not the local current density
${\mathbf j}$ but the current per unit length integrated over the
extent of the ionosphere in altitude $z$, {\em i.e.,}  the
height-integrated\indexit{height-integrated current} current
\mbox{${\mathbf I}\equiv \int\, {\mathbf j}\,dz$}.  A direct
integration of Eq.~(\ref{eq:ionmom}) over height, however, is not
simple because ${\mathbf v}$ varies strongly with $z$ (even when
${\mathbf v}_{\rm n}$ is independent of $z$, as usually assumed). The
horizontal electric field, on the other hand, is essentially constant
over the entire (relatively thin) height range of the ionosphere, from
continuity of tangential components implied by Faraday's law. It is
thus convenient to first express ${\mathbf j}$ by [the ionospheric
Ohm's law as discussed around Eq.~(\ref{eq:condtensor})]
%Eq.~(\ref{eq:ionohm}) 
and then integrate over height to obtain
\begin{equation}
{\mathbf I}_{\perp}=\left(B/c\right)
\left[\Sigma_P \hat{{\mathbf B}} \times
\left({\mathbf v}_0-{\mathbf v}_{\rm n} \right) - \Sigma_{\rm H}
\left({\mathbf v}_0-{\mathbf v}_{\rm n} \right)_{\perp}\right]
\label{eq:ionhtohm}
\end{equation}
where ${\mathbf v}_0$ is the plasma flow at
the top side of the ionosphere, related to the [rest-frame electric
field ${\mathbf E}$ or the electric field ${\mathbf E}^*$ in the frame moving with the
neutral atmosphere] by
\begin{equation}
{\mathbf E}^* = - \left({\mathbf v}_0 - {\mathbf v}_{\rm n} \right) \times {\mathbf
B}/c \qquad \mbox{or equivalently}\qquad
{\mathbf E}= - {\mathbf v}_0   \times {\mathbf B}/c \ .
\label{eq:Vtop}
\end{equation}
$\Sigma_{\rm P}$ and
$\Sigma_{\rm H}$ are
the \indexit{Pedersen!conductivity} height-integrated Pedersen and the \indexit{Hall!conductivity}Hall conductances, $\Sigma_{\rm P}$ being the more important one
for magnetosphere-ionosphere interactions (Hall currents close within the
ionosphere, to first approximation).

Obviously, to ensure ${\mathbf v}_0\simeq{\mathbf v}_{\rm n}$, the ionospheric
conductance $\Sigma_{\rm P}$ must be sufficiently large in relation to the
height-integrated current ${\mathbf I}$, which scales as the current
per unit length in the magnetosphere and hence ultimately as the mechanical
stresses in the magnetosphere. For a more precise criterion, one must
consider a specific process. [\ldots]

(3) \indexit{MHD!coupling}MHD coupling from ionosphere to
magnetosphere along magnetic field lines:
conditions\indexit{ionosphere-magnetosphere coupling} (1) and (2)
ensure merely that the plasma corotates at the top side of the
ionosphere, at the foot of a magnetic flux tube within the
magnetosphere.  For corotation to extend into the magnetosphere
itself, the MHD constraining relation between the flow and the
magnetic field must hold [(Eq.~(\ref{momentum}).\ldots]

(4) Stress balance to maintain \indexit{centripetal
  acceleration}centripetal acceleration in the magnetosphere: If
conditions (1), (2), and (3) are satisfied, the plasma will be
corotating at least as far as the components of ${\mathbf v}$
perpendicular to ${\mathbf B}$ are concerned, but the flow parallel to
${\mathbf B}$ remains unconstrained. For the entire flow to be
corotational, one further condition must be satisfied: there must
exist a radial stress to balance the centripetal acceleration of the
corotating plasma. In most cases, this stress is produced by the
corotation itself, as the magnetic field lines are pulled out until
their tension force becomes sufficiently strong to balance the
centripetal acceleration.''

%%%%%%%%%%%%%%%%%%%%%%%%%%%%%%%%%%%%%%%%%%%%%%%%%%%%%%%%%%%%
\begin{figure}[t]
%\includegraphics[width=\linewidth,clip,viewport=-30mm 80mm 250mm 200mm]
%{../vasyliunas/ssvmvMC.pdf}
%\centerline{\psfig{figure=figures/ssvmv_F5.eps,width=7cm}}
%\centerline{\psfig{figure=figures/HallPedersenCurrentSystems.eps,width=\textwidth}}
\centerline{\includegraphics[width=\textwidth]{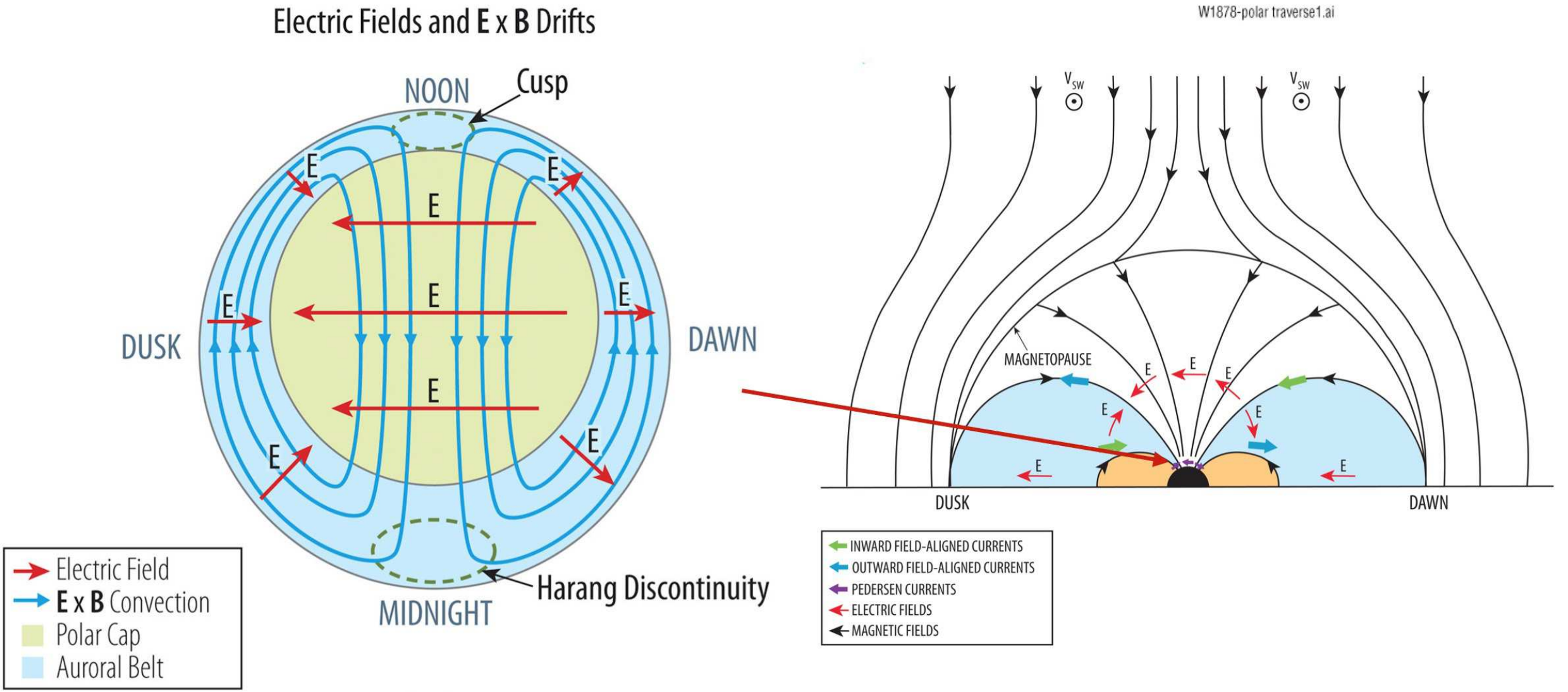}}
% \figurewidth{20pc}
%\figbox*{}{}{\epsfig{file=curr_clos.eps,width=\hsize}}
\caption[Magnetospheric convection on the Earth's
surface.]{\label{fig:MC}Schematic diagram of
  \indexit{magnetosphere!convection}magnetospheric convection
  over the Earth's north polar region (note the diagram extends in
  latitude only to the lower edge of the auroral belt).  {\em Left:}
  streamlines of the plasma bulk flow and associated electric field
  (the Sun is towards the top). {\em Right:} Magnetic field and
  current systems for the northern hemisphere, for a southward
  interplanetary magnetic field, viewed from behind the Earth looking
  towards the Sun. [The red arrow indicates the small segment of this
  image shown on the right.]
%{\em Left:} streamlines of the plasma bulk flow
%(the Sun is towards the left). {\em Right:} Electric field lines and
%associated Pedersen currents, and sketch of the magnetic-field aligned
%(a.k.a.\ Birkeland) current (large
%arrows).
See 
Section~I:10.4.3
%Section~\ref{convec} 
for a detailed description and
Sect.~I:11.6
%Sect.~\ref{sec:magcontoff} 
for corresponding MHD model results for the
electric potential. For a schematic representation of magnetospheric
convection throughout the magnetosphere, see Fig.~I:13.4. [Source:
\href{https://swrc.gsfc.nasa.gov/main/sites/default/files/10thSpaceWeather_Presentations/PfaffGDCOverviewGSFCSpaceWeatherWorkshop2018.pdf}{NASA}
;compare: Fig.~I:10.5] \colorfig}
\end{figure}
\ors[I:10.4.4] ``The four conditions for corotation \activity{{\em
    Advanced/group:} Make a list of the similarities and differences
  between 'corotation' in a planetary magnetosphere and in the solar
  wind, and consider the four conditions for corotation and how they
  compare to conditions in the solar wind. Include (a) the absence of
  a sufficiently neutral atmosphere in the Sun to decouple the motions
  between internal and heliospheric fields (associated with a concept
  called 'line tying', which we touch upon in
  Sec.~\ref{sec:storagemodels}), and (b) the very term 'corotation'
  which to a heliospheric physicist does {\em not} include the
  component of ${\bf v} \parallel {\bf B}$ but is limited to the
  pattern of the field, not the plasma
  itself. \mylabel{act:corotation}}
are all, in essence, local conditions at a given magnetic flux tube.
Deviations from corotational flow when one or another of these
conditions is no longer satisfied need not, therefore, be global but
can be confined to limited regions. Typically, plasma flow in any
particular magnetosphere may follow corotation in the inner regions,
out to a critical radial distance in the equatorial plane, and then
deviate significantly from corotation at larger distances.  The
critical distance depends on which of the four conditions is violated
and by which process.''  Section~I:10.4.4 provides more discussion.
%. Co rotation of tilted dipole.

\subsubsection{Magnetospheric convection}\label{convec}

\indexit{magnetosphere!convection}\indexit{magnetosphere!convection}
\indexit{Dungey cycle}
\ors[I:10.4.3] ``Magnetospheric convection may be considered the other canonical
pattern (besides corotation) of plasma flow in a planetary
magnetosphere, one that plays an overwhelmingly important role in the
magnetosphere of Earth.  The basic concept is that the flow of solar wind plasma
past the magnetosphere imparts some of its motion to plasma in the
outermost regions of the magnetosphere, either directly by
\indexit{MHD!coupling} MHD coupling along open field lines
or through\indexit{drag!tangential} an unspecified
\indexit{tangential drag}tangential drag near the magnetopause.

By continuity of mass and magnetic flux transport, the flow then
extends into the region of closed field lines or the interior of
the magnetosphere, setting up a large-scale circulation pattern
(which has some superficial resemblance to, but no real physical
commonality with, what is called convection in ordinary fluid
dynamics). Figure~\ref{fig:MC} illustrates the pattern, projected
on the top-side ionosphere of the planet: shown on the left-hand
side are the streamlines of the plasma bulk flow ${\mathbf v}$,
which are also the equipotentials of the electric field according
to Eq.~(\ref{eq:Vtop}). The lines of the electric field and the
associated \indexit{Pedersen!current}Pedersen currents are shown
on the right-hand side [in red], along with a sketch of the implied
\indexit{Birkeland current}Birkeland ({\em i.e.,}  magnetic
field-aligned) currents [in blue and green]. The yellow region is the polar cap, identified with the region of open field lines in
the open magnetosphere; otherwise it represents the mapping (along
field lines) of the boundary region where the solar wind motion is
being imparted to magnetospheric plasma. The equatorial-plane
counterpart of the flow outside the polar cap has been sketched in
Fig.~\ref{fig:OMS} (bottom).

A quantitative global measure of the strength of magnetospheric
convection is the EMF (maximum line integral of the electric field)
across the polar \indexit{polar cap!potential}cap, $\emf_{\rm
  PC}$. Its physical meaning is that of the rate of magnetic flux
transport (advection) through \indexit{magnetosphere!reconnection
  rate} the polar cap. In an open magnetosphere, $c\emf_{\rm PC}$
equals the rate of reconnection of magnetic flux between the
interplanetary and the planetary magnetic fields. Numerous empirical
studies at Earth have shown that for a southward interplanetary
magnetic field ({\em i.e.,}  [(at least partly) aligned with the
Earth's magnetic dipole, so that] $({\mathbf B}_{\rm sw}\cdot{{\bm \mu_{\rm p}}})>0$),
$\emf_{\rm PC}$ can be related to solar wind parameters roughly
as
\begin{equation}
c\emf_{\rm PC} \simeq v_{\rm sw}\left({\mathbf B}_{\rm sw}\cdot\hat{{\bm \mu_{\rm p}}}\right)
\mathcal{L}_\times
\label{eq:PC}
\end{equation}
where $\mathcal{L}_\times$ is a length that typically is a fraction
($\sim$0.2 to $\sim$0.5) of the magnetopause radius $R_{\rm mp}$. When
comparing different magnetospheres, one often supposes that the ratio
$\mathcal{L}_\times/R_{\rm mp}$ is a more or less universal
constant. Physically, $\mathcal{L}_\times$ may be looked at as the
length of the reconnection\indexit{reconnection!X-line} X-line on the
magnetopause at which\indexit{magnetopause!reconnection line} the
magnetic field lines from the solar wind and from the planet first
become interconnected, projected along the streamlines of the
magnetosheath flow back into the undisturbed solar wind, as
illustrated in Fig.~\ref{fig:OMS}.''
\activity{{\em Advanced/Group:} Consider the
  possible equivalent of a Dungey cycle for the heliospheric field
  subject to reconnection with an interstellar magnetic field. Think
  about the effect of the super-Alfv{\'e}nic stellar wind. What would
  things look like in case there were no coronal heating? Consider the
  Sun's global dipole and how that would then interact with the
  incoming interstellar medium. \mylabel{act:stardungey}}
\activity{{\em Advanced/Group:} {\bf What if? A discussion activity:}
  Many so-called 'hot Jupiters' have been found among the exoplanet
  population: giant planets that orbit very close to their parent
  stars. What would the estimated magnetopause distance
  $R_{\rm CF,hJ}$ be if Jupiter were orbiting the present-day Sun at
  0.05\,AU? For a younger Sun (see Ch.~\ref{ch:evolvingplanetary}) the
  solar wind would have been stronger, pushing $R_{\rm CF,hJ}$ to
  below the orbital radius of Ganymede; describe what that would mean
  for this 'hot-Ganymede' moon?}
\activity{{\em Background:} Advanced, for the curious: Things get more
  complicated when objects are smaller than the gyration radii of
  particles in the flow, or when ionization processes occur when
  neutral particles from an 'atmosphere' move into an approaching
  flow, or both. If you are interested in seeing how these
  complications play out, have a look at this study of comet 67P by
  \href{https://ui.adsabs.harvard.edu/abs/2017MNRAS.469S.396B/abstract}{\citet{2017MNRAS.469S.396B}}
  using observations by the Rosetta spacecraft.}

\subsection{A large-scale flow impinging on a fast
  outflow} \label{sec:impinging} An example of \indexit{flow!wind onto
  interstellar medium}colliding plasmas on the largest scale in
heliophysics involves the \indexit{interstellar medium}interstellar
\indexit{ISM|see{interstellar medium}}medium into which the
heliosphere is moving.  Figure~\ref{fig:ophercomposite} \ors[IV:3.3]
``shows the prevailing picture of the global heliosphere. The
structure is characterized by three flow discontinuities: the
\indexit{termination shock}termination shock (TS), the
\indexit{heliopause}heliopause (HP), and the \indexit{bow shock}bow
shock (BS).  The solar wind density and therefore its ram pressure
falls off as $r^{-2}$, where $r$ is the distance from the Sun.  The
wind speed is supersonic and super-Alfv\'{e}nic, so when the ram
pressure falls to the pressure of the ambient [interstellar medium
(ISM)], the result is a shock, specifically the ellipsoidal TS in
Fig.~\ref{fig:ophercomposite}, where the flow is decelerated.

If the ISM flow is super-Alfv\'{e}nic, it also encounters a shock as
it approaches the Sun, specifically the roughly hyperboloid shaped bow
shock in Fig.~\ref{fig:ophercomposite}, where the ISM flow is
decelerated to subsonic speeds.  However, the $v_{\rm ISM}=23-27$
km~s$^{-1}$ interstellar flow happens to yield an Alfv\'{e}nic Mach
number of $M_{\rm A}\approx 1$, making the existence or nonexistence
of a bow shock very much an open question.  Much depends on the
strength and orientation of the ISM magnetic field, $B_{\rm ISM}$.
The higher $B_{\rm ISM}$ is (and the more perpendicular to the ISM
flow), the lower $M_{\rm A}$ should be, and less likely that there is
a bow shock.

Even the seemingly small uncertainty in $v_{\rm ISM}$ is enough to
make a difference.  For many years the best assessments were believed
to be the $v_{\rm ISM}=26.3\pm 0.4$ km~s$^{-1}$ measurement of the ISM
neutral He flowing through the Solar System by {\em Ulysses}
and the $v_{\rm ISM}=25.7\pm 0.5$
km~s$^{-1}$ measurement from ISM
absorption lines.  With these relatively high values, heliospheric
modelers favored $M_{\rm A} > 1$, implying the existence of a bow shock.  However,
later He flow measurements and a new analysis of ISM absorption line
data have yielded lower velocities.  Specifically, measurements of
neutral He flow from {\em IBEX} suggest $v_{\rm ISM}=23.2\pm 0.3$
km~s$^{-1}$, and
$v_{\rm ISM}=23.84\pm 0.90$
km~s$^{-1}$ from ISM absorption lines.

This has been enough for many to argue that $M_{\rm A} < 1$ should be
preferred, though [it has been argued] that including He$^+$ density
in the calculation of sound and Alfv\'{e}n speeds instead of just
assuming a pure proton plasma would still suggest $M_{\rm A} > 1$ even if
$v_{\rm ISM}\approx 23$ km~s$^{-1}$.  With $M_{\rm A}$ so close to 1, it is
possible that the issue will not be fully resolved until an
interstellar probe mission of some sort is sent out to this region.
However, with $M_{\rm A}$ so close to 1 it is also possible that secondary
physical processes ({\em e.g.}, charge exchange interactions with
neutral particles) make it fundamentally ambiguous whether any
boundary that may exist out there should be called a true bow shock, or
whether we should instead refer to it as a 'bow wave'.

Regardless of whether or not a bow shock exists, strong plasma
interactions prevent the ISM plasma from mixing with the solar wind
plasma.  The roughly paraboloid heliopause in Fig.~\ref{fig:ophercomposite},
lying between the termination shock and the bow shock (or wave) is the
contact discontinuity separating the two plasma flows.  Representing
the boundary between solar wind and ISM plasma, the heliopause is
generally considered the true boundary of the heliosphere.''

\ors[IV:3.4] ``The basic structure in Fig.~\ref{fig:ophercomposite} is
mostly defined by plasma interactions.  The local ISM is partly
neutral, but collisional mean free paths for neutrals are large
compared to the size of the heliosphere, so their effects on
heliospheric structure were long ignored.  In essence, the assumption
was that neutrals pass through the heliosphere unimpeded, feeling only
the Sun's gravity and photo-ionizing flux.  However, in reality
neutrals do participate in heliospheric interactions through
\indexit{charge exchange}charge exchange (CX).  The CX interactions
end up providing ways to remotely explore the heliosphere that would
be impossible if the local ISM were fully ionized.

A CX interaction is a rather simple process by which an electron
hops from a neutral atom to a neighboring ion ({\em e.g.},
${\rm H^{0}} + {\rm H}^{+}\rightarrow {\rm H}^{+} + {\rm H^{0}}$).
Mean free paths for CX for most neutral ISM atoms are short enough
that they do experience significant CX losses on their way through the
heliosphere.  The exceptions are the noble gases, which have low CX
cross sections, explaining why neutral He flowing through the solar
system is considered the best local probe of the undisturbed ISM flow.

Modeling neutrals in the heliosphere is very difficult because CX
sends the neutrals wildly out of thermal and ionization equilibrium
with the ambient plasma.  Including neutrals in hydrodynamic models of
the global heliosphere therefore requires either a fully kinetic
treatment of the neutrals, or at least a sophisticated multi-fluid
approach.  The earliest models that could treat neutrals properly were
from the 1990s.  These models demonstrated that through CX, neutrals
could have significant effects on heliospheric structure.  The ISM
protons are heated, compressed, deflected, and decelerated as they
approach the heliopause, and thanks to CX [with the neutral component
of that incoming ISM] the proton properties are at least
partially imprinted on the neutral hydrogen as well, creating what has
been called a 'hydrogen wall' of higher density [neutral hydrogen]
around the heliosphere, in between the heliopause and the bow shock.''
Section~\ref{sec:evolastrospheres} (and Ch.~IV:3) discusses stellar
observations and inferences about astrospheres.

For us living deep inside the heliosphere, the consequences of the
solar wind sculpting out a cavity in the interstellar medium
are limited as no perturbations in the solar wind or its magnetic
field can propagate against the super-Alfv{\'e}nic wind. Nonetheless,
the heliosphere that is shaped by this interplay does affect our
exposure to cosmic rays that traverse it, see Ch.~\ref{ch:evolvingexposure}.

\clearpage

\chapter{{\bf Magnetic \hbox{(in-)stability} and energy pathways}}%6
\label{ch:mhd}
{\narrower\narrower{
{\bf Chapter topics:}
\begin{itemize}
  \customitemize
\item Magnetic instabilities in flares, CMEs, and
  magnetospheric (sub-)storms
\item The appearance of solar flares across the electromagnetic
  spectrum
\item Coronal and geomagnetic instabilities, chromospheric and
  ionospheric heating, charged particle precipitation, 
  EM radiation, energetic neutrals
\item Instability mechanisms (tearing mode, current-driven,
  interchange, and ballooning) that convert energy stored in a
  non-potential magnetic field
\end{itemize}

\noindent{\bf Key concepts:}
\begin{itemize}
  \customitemize
\item Solar flare ribbons and terrestrial aurorae 
\item Neupert effect
\item Force-free field
\item Reconnection 
\end{itemize}

}}

\section{Introduction}\label{sec:explintro}
Instabilities \indexit{magnetic!instabilities}occur when mild perturbations to some energy reservoir provide access to an energy conversion pathway into a significantly reduced state of that reservoir. This can happen because the pathway itself develops (such as when a condition for fast reconnection is met), because the energy reservoir changes in content (as external sources insert energy), because the surrounding conditions change (for example, the direction of the solar wind magnetic field or the makeup of the solar magnetic landscape), or because a sufficiently large perturbation occurs (such as by variations in solar wind speed or the passage of a (shock) wave associated with another impulsive event). One analogy of a purely mechanical nature is the fall of a ball that is somehow nudged over the edge of a bowl; in that process, gravitational energy contained in the reservoir (the elevated ball in the bowl) is converted into the kinetic energy of the ball's fall, and ultimately into heat and waves (that themselves eventually dissipate into the microscopic kinetic energy of heat) as the ball hits the floor. The magnetic field of volumes within the Sun's atmosphere and of planetary magnetospheres can similarly destabilize: when deformed from a potential state, the added energy may be gradually dissipated thereby avoiding a (large-scale) instability or it may be stored for some time in a growing reservoir, then to be converted impulsively through a variety of pathways, eventually ending in kinetic or electromagnetic energy that is extracted from the magnetic field and the plasma that it holds. Tracking energy reservoirs and flows is often helpful in understanding processes.
%: where in criminology and politics one often hears the adage 'follow the money', here we 'follow the energy'.

The primary \indexit{magnetic!energy storage}storage reservoir for what eventually develops into a solar impulsive event or a magnetospheric (sub-)storm is the distortion of the magnetic field away from a potential state. This elevated energy is often attributed in our thinking to electrical currents, but is stored throughout the distorted magnetic field. In quantitative terms, the maximum energy available for an impulsive event is the volume-integrated field energy in excess of the minimum level. The latter is often taken to be the potential field $B_{\rm pot}$ matching the observed surface field on the Sun or a reference dipole field $B_{\rm dip}$ for a planet, {\em i.e.,} 
\begin{equation}
\label{eq:fieldenergy}
E_{B,{\rm res}}=\frac{1}{8\pi}\int\,\left[B^2-\left(B_{\rm pot,dip}\right)^2 \right] {\rm d}V\,,
\end{equation}
although that potential-field energy level may not practically be achievable (see discussions of helicity in solar conditions, or consider continued stressing of the geomagnetic field by a sustained solar wind). Note that this energy is an integral quantity: the local quantity $B^2-\left(B_{\rm pot,dip}\right)^2$ measures the change in local energy density, but as the energy is contained in the field rather than in embedded electrical currents, regions with high values of $B^2-\left(B_{\rm pot,dip}\right)^2$ do not necessarily correspond to locations of electrical current systems. \activity{{\em 
% Slight edit pointing to induction and momentum equations made for V1.3
Consider} how a non-potential state can arise or be strengthened in the solar atmosphere and in a magnetosphere, including the roles of plasma motions and induction. Eq.~(\ref{induction}) and~\ref{momentum} provide the mathematical basis; Eq.~(\ref{eq:dynamoenergy}) is illustrative for the overall energy budget. \mylabel{eq:nonpotform}} 
 
The observable signatures of magnetically-dominated instabilities in the solar atmosphere and in planetary magnetospheres have led to the development of a colorful and often unclear and ambiguous array of terms, generally introduced well before the processes themselves were understood and incorporated into an overall view. For example, present-day understanding ascribes the impulsive and decay phases of flares, coronal mass ejections (CMEs), and terrestrial magnetospheric \hbox{(sub-)}storms and their counterparts in other planetary magnetospheres to a loss of a quasi-equilibrium in, or a departure from, a quasi-steady evolution of the magnetic field that leads to a rapid increase in the rate of reconnection. The latter is associated, among other things, with the acceleration of populations of ions and electrons that lead to observable emissions in a wide range of wavelengths in the electromagnetic spectrum, among them the terrestrial auroral emissions and their solar counterpart, the \indexit{flare!ribbon}flare ribbons. \activity{{\em Look up:} The processes of electromagnetic radiation from a plasma involve three fundamentally distinct
  processes: bound-bound (i.e., electron transitions from an excited state to a lower-energy state), free-bound (radiative recombination), and
  free-free (Bremsstrahlung) emission. Aurorae and flare ribbons are caused by collisions of downward-propagating, energetic charged particles with the atmosphere below. Aurorae \indexit{aurora}observed from the ground include both free-bound and bound-bound emission from ions and molecules, respectively (there is X-ray emission, too, but that does not penetrate to ground level). Look up which ions and molecules dominate (and at which color(s)) in the terrestrial aurora, and which emission processes are involved with these. See also Activity~\ref{act:coronallines} for the contrast with solar coronal emission. \mylabel{act:auroracolors}}

\begin{table}
\begin{center}
\caption[Solar flare classifications.]{\label{tab:class}Solar flare classifications. [Listed are the GOES flare class, the \indexit{flare!classification}corresponding flare-peak irradiance at the top of Earth's \indexit{flare!frequency}atmosphere, the class and surface footprint based on chromospheric emission patterns, the fraction of such events associated with coronal mass ejections (CMEs), and the characteristic  frequency of such events during the maximum and minimum of a typical solar cycle. Table~II:5.1]}
\begin{center}\begin{tabular}{clcrrr}
\tableline
GOES & 1-8\AA~peak & H$\alpha$ & H$\alpha$ Area  & CME &Events/year\\
class & \textit{(W/m$^2$)}           &  class                & \textit{(millionths} &fraction$^a$ & \textit{(cycle}\\
&    \textit{(kerg/cm$^2$/s)}                &   \textit{(percent)}               & \textit{of hemisphere)}&&\textit{max./min.)}\\
\tableline
A  & $>$10$^{-8}$& - & - & - & -\\
B  & $>$10$^{-7}$& S & $<$200 & - & -\\
C  & $>$10$^{-6}$& 1 & $>$200 &0.2 & $>$2000/300\\
M  & $>$10$^{-5}$& 2 & $>$500 &0.5 & 300/20\\
X  & $>$10$^{-4}$& 3 & $>$1200&0.9 &  10/one?\\
-  & $>$10$^{-3}$& 4 & $>$1200&1.0 & few?/none?\\
\tableline
\end{tabular}\end{center}
\end{center}
{\em \small $^a$ (approximate values)}
\end{table}\indexit{coronal!mass ejection!flare association}
Not
\activity{{\em Show:} The average speed of a CME between Sun and Earth is close to 500\,km/s while the fastest have speeds exceeding 3000\,km/s. (a) How long are the transit times from Sun to Earth? Compare the average and peak CME speeds to typical wind speeds (Table~\ref{tab:wind-stats}). (b) Describe qualitatively what happens in the interaction with slow and fast wind streams for average CMEs and for the fastest CMEs. (c) Draw a sketch or cartoon to accompany your description. \mylabel{act:cmecme}}\activity{{\em Background:} The phenomena discussed in Ch.~\ref{ch:mhd} are all part of what is referred to as space weather. To explore how aspects of space weather are quantified review the \href{https://www.swpc.noaa.gov/noaa-scales-explanation}{NOAA webpage} https://www.swpc.noaa.gov/noaa-scales-explanation that lists the types of 'storms,' their potential effects, and their approximate frequency within a solar cycle. Make a table of the types of measurements you would need to have awareness of these different storms; include in that table the different spacecraft and instruments that could be used. N.B. For current space weather conditions, forecasts, and more see \href{https://www.swpc.noaa.gov}{the website} of the Space Weather Prediction Center: https://www.swpc.noaa.gov. \mylabel{act:swphenomena}} only do these impulsive phenomena share many physical processes, they are also links in the chain of Sun-Earth connections: many of the more energetic solar field destabilizations are associated with both flares and CMEs (see Table~\ref{tab:class}), while CMEs that envelop a planet that has an intrinsic magnetic field often trigger (immediate or delayed) magnetospheric activity.

\subsection{Introducing solar flares and coronal mass ejections}
\begin{figure} 
\centering
\includegraphics[width=1.0\textwidth,bb=0 0 710 432]{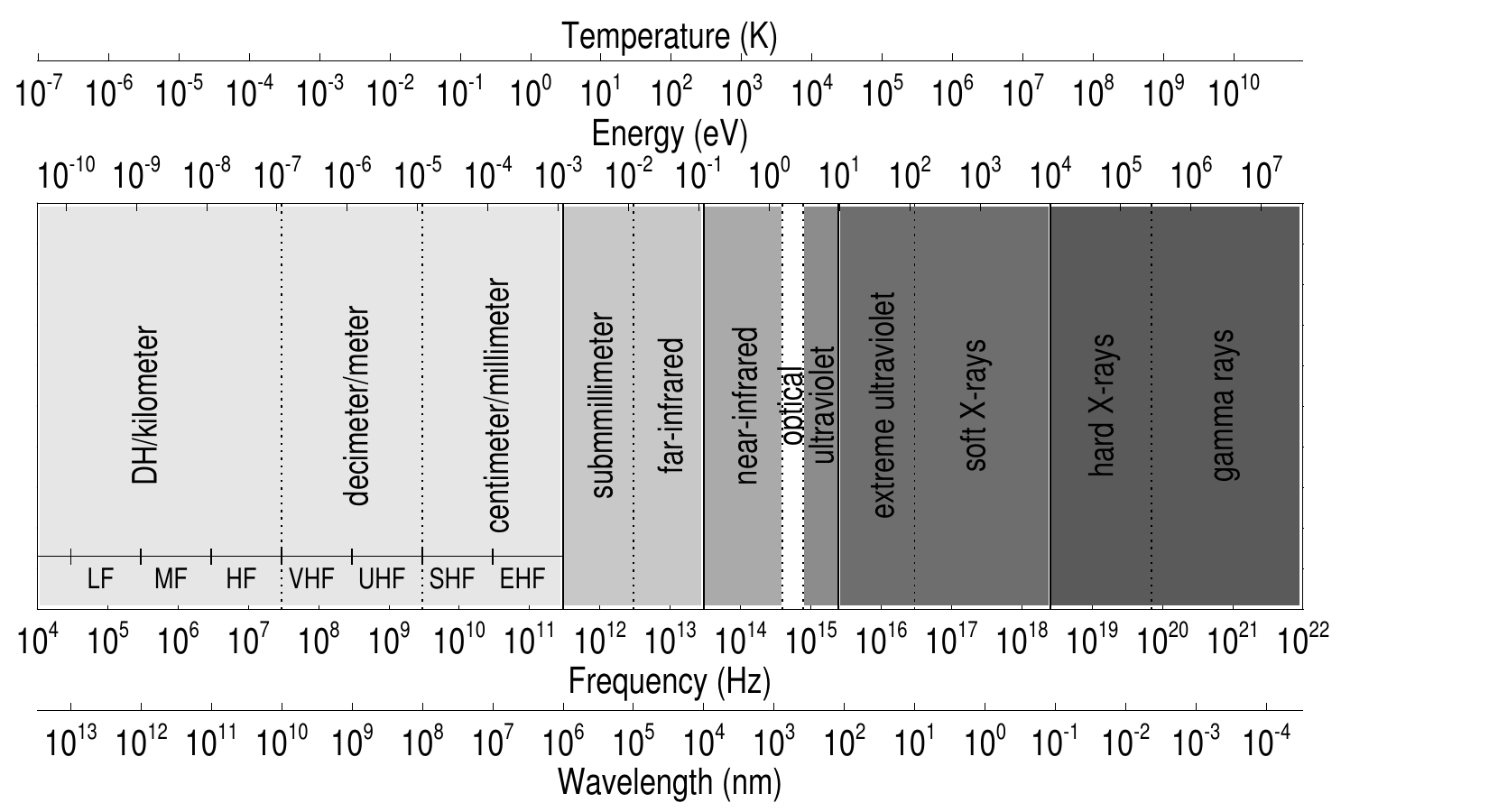}
\caption[Overview of the electromagnetic spectrum.]{Overview of the electromagnetic spectrum with \indexit{electromagnetic spectrum}energy in
electron volts and the equivalent temperature in Kelvin
(top axes), frequency in Hertz, and wavelength in
nanometers (bottom axes). Note that the AM band lies in the low- and
medium-frequency
(LF-MF) range and the FM band in the very-high frequency (VHF)
range. [Fig.~II:4.1]}  \label{fig:tsb1} 
\end{figure}

\ors[II:5.1] ``A\indexit{definition!flare} solar flare\indexit{flare|seealso{definition}}
is narrowly defined as a sudden atmospheric brightening,
traditionally in chromospheric H$\alpha$ emission [(at 656\,nm, associated with a $3\rightarrow 2$ level transition of hydrogen atoms, and thus the lowest-energy transition in the Balmer series)]
but more practically now as a coronal soft X-ray source [(Fig.~\ref{fig:tsb1} summarizes the common names used for wavelength bands from radio to gamma rays)].  The physical processes resulting in a flare include restructurings of the magnetic field, non-thermal particle acceleration, and plasma flows.  Flares have intimate relationships with other observable phenomena such as filament eruptions, jets, and coronal mass ejections [\ldots]

The energy released in a solar flare is dominated by particle acceleration, both of electrons and ions.  This means that the most direct observations are in the X-ray and $\gamma$-ray domains; note that non-thermal processes also usually dominate the emission signatures in the radio range (10$^7$-10$^{12}$~Hz; meter--submillimeter wavelengths).  Please refer to Ch.~II:4 for a fuller discussion of the remote-sensing signatures.  We will simply comment here that in general the hard X-ray spectrum (h$\nu \gapprox$10\,keV [or wavelengths shortward of about 1\,\AA]) is dominated by electrons of this energy or greater, while the soft X-ray spectrum (h$\nu \lapprox$10\,keV) is dominated by the free-bound and bound-bound transitions of a thermal plasma with assumed Maxwellian distribution functions, and also usually assuming the electron and ion temperatures to be equal, {\em i.e.,} T$_{\rm e}$~=~T$_{\rm i}$.  The free-bound process (radiative recombination) may also contribute to the hard X-ray spectrum under certain conditions.''

\begin{figure}
\centering
\includegraphics[width=7cm]{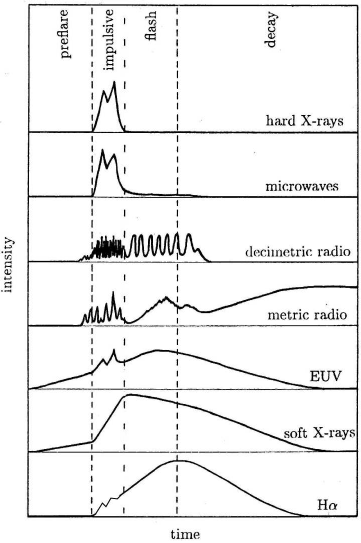}
\caption[Evolution of flare emissions, and concepts of particle acceleration.]{Schematic view of the evolution \indexit{flare!spectral evolution}of
  solar flare emissions in
different wavelengths, showing the intermingling of impulsive-phase
and gradual-phase signatures across the spectrum.
Note the wide variety of radio signatures. [In wide wavelength bands in the visible, the emission peaks in the impulsive phase of the flare. Fig.~II:5.1; \href{https://ui.adsabs.harvard.edu/abs/2002ASSL..279.....B/abstract}{source: \citet{2002ASSL..279.....B}}.] 
\label{fig:hudsonfig1}
}
\end{figure}
\nocite{2002ASSL..279.....B}

\begin{figure}[t]
\centering
\includegraphics[width=6.5cm]{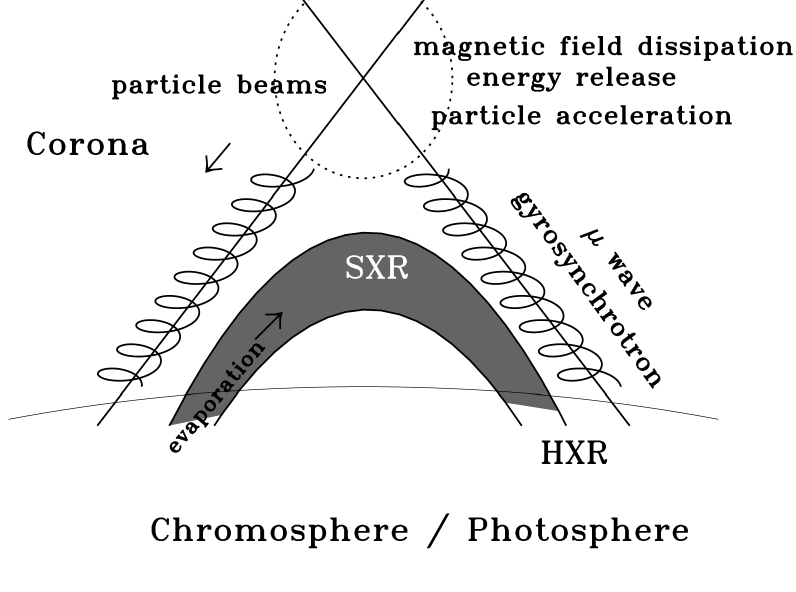}\includegraphics[clip=true,trim=0cm 0cm 0cm 0cm,width=6.5cm]{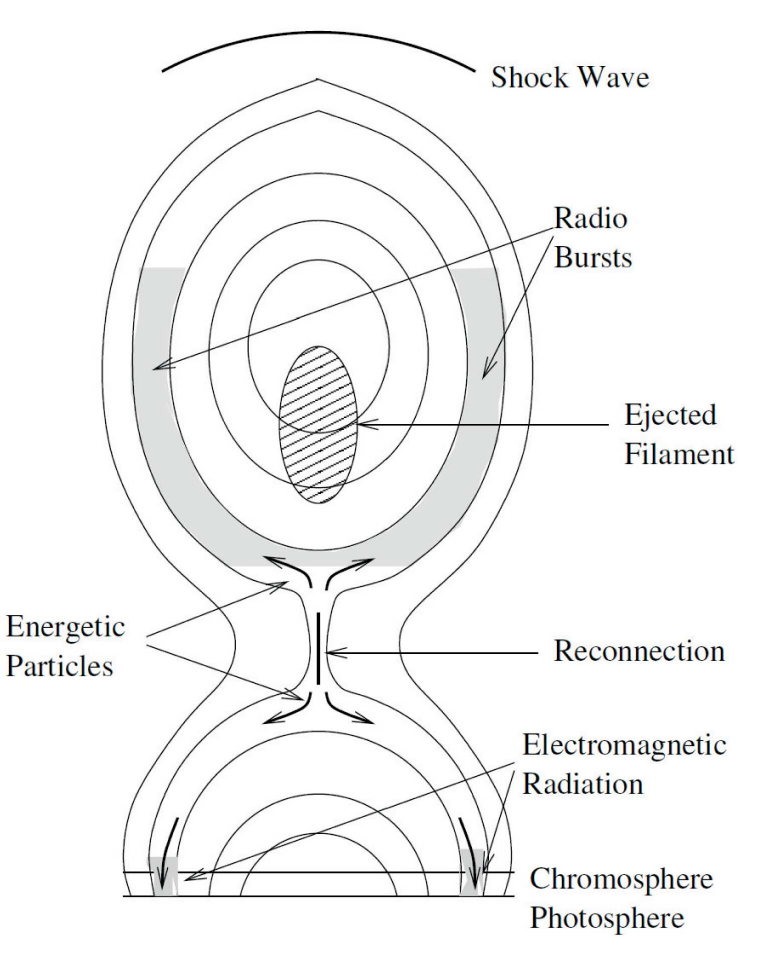}
\caption[Sketch of the flow of energy during a flare.]{{\rm (left)} Schematic
  arrangement in the outer atmosphere of the Sun or a comparable cool
  star indicating the flow of energy during a flare: a flare involves magnetic reconnection high in the atmosphere which accelerates
  particles, leading to motion along field lines upward away from or
  downward towards the visible surface.  Resulting emissions include
  hard X-rays (HXR), soft X-rays (SXR), and microwave emission. \label{fig:ostenloops} [Fig.~IV:2.1] {\em (right)} Concepts of particle acceleration and emissions in a solar [eruptive] event. [Fig.~IV:12.6; \href{https://ui.adsabs.harvard.edu/abs/2003JPhG...29..965K/abstract}{source: \cite{2003JPhG...29..965K}}.]
  \label{fig:Krupp_HelioIV_Kallenrode2003-2r}}
\end{figure}

A common observed pattern, most frequently in eruptive flares associated with coronal mass ejections into the heliosphere, is that a volume of the corona over a magnetic polarity inversion line expands explosively (often involving a large-scale shock front) as the hard X-ray and $\gamma$-ray emission brightens impulsively, with two (or more) ribbon-like brightenings at chromospheric and photospheric levels propagating away from the polarity inversion line, with a coronal mass ejection moving away while behind it the corona fills with heated plasma from the lower atmosphere, which then cools by radiation in the soft X-ray and EUV bands and by conduction into the lower, cooler atmosphere. See Fig.~\ref{fig:hudsonfig1} for the characteristic evolution of a flare in wavelength space, and Fig.~\ref{fig:ostenloops} for a sketch of the various emissions throughout the EM \indexit{flare!in the EM spectrum}spectrum.

\figindex{../osten/art/loops.eps}
\ors[II:5.2.1] ``The modern view of [solar flares] is via the soft X-ray monitoring by the GOES and other 'operational' spacecraft.  We now routinely classify solar flares by their GOES \indexit{flare!classification}classes: A, B, C, M, and~X in decades, with the X class signifying 1-8\AA~energy fluxes greater than 10$^{-4}$~W/m$^2$, on the order of 0.01\%~of the solar luminosity.  Table~\ref{tab:class} summarizes these and other properties, with very approximate correspondences between the [chromospheric] H$\alpha$ and GOES X-ray systems, and very approximate ranges for the number of flares that occur per year at maximum and minimum of the solar cycle.''

\begin{figure}[t]
\centering
  \includegraphics[width=8cm]{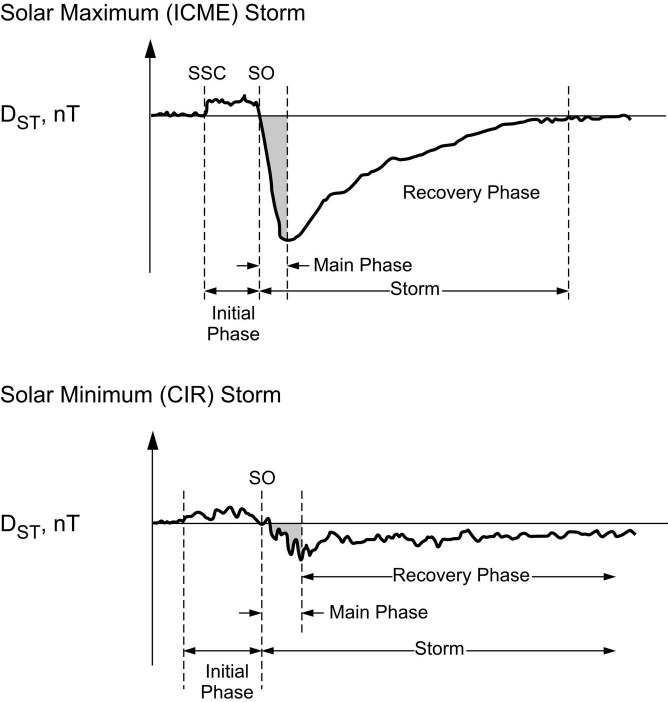}
  \caption[Geomagnetic field variation for two characteristic magnetic storms.]{\label{fig:Dst} Schematic time history of geomagnetic field variation for two characteristic magnetic storms. Time range: several days. Vertical variation range: $\sim 100-200$\,nT ($\sim 1-2$\,mG). SSC: storm sudden commencement. SO: storm onset. The top panel shows the storm development in response to a characteristic interplanetary coronal mass ejection (ICME), and the bottom panel that for the passage of a corotating interaction region (CIR).  [Fig.~II:10.1; \href{https://ui.adsabs.harvard.edu/abs/2006JGRA..111.7S01T/abstract}{source: \citet{2006JGRA..111.7S01T}}.]}\nocite{tsur}
\end{figure}
\subsection{Introducing geospace (sub-)storms}\label{sec:substorms}
In contrast to a flare or CME observed by imaging the electromagnetic radiation, a terrestrial magnetic storm is typically observed by sampling magnetic field changes at the Earth's surface or tracked by monitoring energetic particles and their effects (such as in aurorae), with their root causes in elements of the Dungey cycle in the geomagnetic tail. 
A terrestrial \ors[II:10.2] ``magnetic storm is\indexit{definition!geomagnetic storm} defined\indexit{magnetic!storm|seealso{definition}} nowadays \indexit{magnetic!storm}by the time variation of the geomagnetic Dst index, illustrated schematically in Fig.\,\ref{fig:Dst}. The Dst\indexit{definition!Dst index} index \indexit{Dst index|seealso{definition}} is a measure of a quasi-uniform magnetic disturbance field near the Earth, aligned with the dipole axis (northward [field] for Dst$>0$), such as would be produced by a ring of electric current (westward [current] if Dst$<0$) near the equatorial plane.  A prolonged (hours to days) interval of negative Dst values constitutes a magnetic storm. The peak negative excursion is often taken as a measure of storm intensity: Dst -30\,nT to -50\,nT are weak storms, -50\,nT to -100\,nT moderate, and over -100\,nT intense; storms over -300\,nT occur at most a few times during a solar cycle (Earth's dipole field at the equator is $\sim 31,000$\,nT, [or 0.31\,Gauss,] for comparison) [\ldots]

[T]he
field depression quantified by Dst is the result of plasma pressure
that inflates the dipole field. The essential phenomenon of the
magnetic storm is thus the addition of a large amount of plasma energy
to the dipolar field region of the magnetosphere. Furthermore, it is
now well established that this energy addition results from a
particular condition in the solar wind: 'a sufficiently intense and
long-lasting interplanetary convection electric field', meaning \mbox{$-{\mathrm {\bf v}}\times{\mathrm {\bf B}}/c$}, for
the [interplanetary magnetic field's (IMF's)] southward component.

In contrast to the magnetic storm, there is much less unanimity on what defines a magnetospheric substorm. Probably the most spectacular phenomenon, and the one most widely used as a unifying concept, is the auroral \indexit{aurora}substorm,\indexit{auroral substorm} [\ldots, which has a characteristic temporal development. Early to show up are] the auroral forms (light-emitting regions) during what is called the {\em expansion phase} of\indexit{substorm!expansion phase} the substorm: beginning with an initial brightening at the lowest latitudes near midnight ({\em onset}), the aurora intensifies greatly, becomes very complex in spatial structure ({\em auroral breakup}) and\indexit{substorm!auroral breakup} expands, predominantly westward and poleward but also eastward, eventually subsiding in\indexit{substorm!recovery phase} a {\em recovery phase}. This auroral development is accompanied by strong geomagnetic disturbances (commonly reaching $\sim 1500$\,nT [or 0.015\,G] and more), with a spatial distribution almost as complex as that of the aurora but describable roughly as equivalent to a current above the\indexit{electrojet} Earth ({\em auroral electrojet}) that is westward near and before midnight and eastward after midnight.  [Essentially the same sequence occurs simultaneously in the two hemispheres,] at the (more or less) magnetically conjugate locations [\ldots]''

%\regfootnote{The strength of geomagnetic disturbances is often expressed in terms of the Dst index, the 'disturbance storm time' index, which quantifies (in nanoTesla) the variation in the horizontal component of the Earth's magnetic field as derived from hourly measurements at a network of near-equatorial geomagnetic observatories.}

\section{Terrestrial magnetospheric disturbances}\label{sec:disturbances}
\subsection{Energy pathways and reservoirs}
\ors[II:10.3.2] ``For the magnetosphere and the upper regions of the ionosphere [\ldots] the solar wind is the only significant external source of energy available [because these regions are completely transparent to solar radiation \ldots] An interior source of\indexit{magnetosphere!energy from planetary rotation} energy available for a planetary magnetosphere is planetary rotation [\ldots]

When considering the solar wind as the energy source, only the kinetic energy of plasma bulk flow is of importance; the thermal and magnetic energies of the solar wind can be neglected [\ldots: they are relatively small to begin with (Sect.~\ref{sec:solwindenergy}) while moreover] at the bow shock they are overwhelmed by additional thermal and magnetic energies extracted from the flow. Furthermore, to transfer magnetic energy across the magnetopause requires [\ldots] a tangential component of the electric field which interacts with the magnetopause current to extract more mechanical energy from the plasma [\ldots] The interplanetary magnetic field does exert a dominant influence on energy conversion processes in a planetary magnetosphere, but primarily by control of magnetic reconnection processes and open field lines [\ldots]''

\ors[II:10.3.3] ``The following are among the principal loss and dissipation processes in planetary magnetospheres, energy being lost primarily to the atmosphere in (1) and (2) and being removed outside the system (to 'infinity') in (3) and (4):

(1) {\bf Collisional and Joule heating in the ionosphere}.\indexit{ionosphere!collisional heating} If\indexit{ionosphere!Joule heating} the bulk
flow of plasma differs from the bulk flow of the neutral
atmosphere (usually as a consequence of magnetospheric dynamics), there
is energy dissipation given by \mbox{${\mathbf E}^*\cdot{\mathbf J}$},
where ${\mathbf E}^{*}$ is the electric field in the frame of
reference of the neutral atmosphere. This is commonly referred to as
'ionospheric Joule heating'; [\ldots] it is primarily frictional heating by collisions between plasma
and neutral particles, Joule heating in the true physical sense
(\mbox{${\mathbf E}^{\prime}\cdot{\mathbf J}$}, where ${\mathbf
E}^{\prime}$ is the electric field in the frame of reference of the
plasma) contributing only a small fraction of the total. The energy is
removed from the magnetic field and converted (via kinetic energy of
relative bulk flow as an intermediary) to heat (thermal energy), with
the heating rate per unit volume partitioned approximately equally
between plasma and neutrals.

(2) {\bf Charged-particle precipitation}.\indexit{ionosphere! particle precipitation} Energetic\indexit{particle!precipitation} charged particles
that enter the atmosphere from above are usually said to be {\em
precipitating}. They penetrate the atmosphere to a depth that
increases with increasing energy, until their energy is lost, going
partly into heating the atmosphere and partly into ionization or other
interactions.

One source of precipitating particles is simple loss from the radiation belts or from the ring current and plasma sheet regions; \activity{(a) {\em Look up} locations and properties of the Earth's (1) electron and proton radiation belts, (2) ring current, and (3) plasma sheet. Create a table containing their locations, characteristic particle energies, and their densities. (b) What is the typical energy density for each population? \mylabel{act:bcs}} the energy deposited in the atmosphere is taken from the mechanical (thermal) energy of the respective magnetospheric particle populations. In addition to these particles that precipitate merely because their velocity vectors are oriented in the appropriate direction, there are other sources of precipitating charged particles, in which the energy and the intensity of the particles have been enhanced by an acceleration process. In particular, the auroral phenomena that occur in nearly all of the planetary magnetospheres observed to date are generally interpreted as resulting from some special acceleration process that supplies the required intensities of precipitating charged particles. A widely accepted model, developed from extensive studies at Earth and applied to \indexit{aurora}aurora at Jupiter and at Saturn, ascribes auroral acceleration to\indexit{Birkeland current} Birkeland (magnetic-field-aligned) electric currents accompanied by electric fields parallel to the magnetic field; the rate of energy supply to the precipitating particles is \mbox{$E_{\parallel}J_{\parallel}$}, hence the added energy is taken out of the magnetic field (in this model, an aurora occurs only when the Birkeland current is directed upward, corresponding to electron motion downward). Auroral acceleration has also been associated with intense Alfv\'enic turbulence (which contains fluctuating Birkeland currents) [\ldots]

(3) {\bf Emission of electromagnetic radiation}. A variety of
processes\indexit{magnetosphere!EM radiation} in planetary magnetospheres produce electromagnetic
radiations of various types: atomic and molecular line emissions (from
the aurora and from magnetospheric interactions with plasma and
neutral tori), radio waves (wideband and narrowband), a veritable zoo
of plasma waves, and even X-rays (bremsstrahlung from precipitating
electrons and, possibly, nuclear line emissions excited by very
energetic precipitating particles). [\ldots] As far as the
energetics of planetary magnetospheres are concerned, however, the
amount of energy involved is negligibly small for most emissions, with
only a few exceptions (UV radiation from the Io torus at Jupiter).

(4) {\bf Energetic neutral particle escape}. Neutral\indexit{magnetosphere!neutral particle escape} particles that remain within a magnetosphere must be gravitationally bound to the planet; plasma particles within the magnetosphere, on the other hand, typically have speeds that exceed (often by a large factor) the gravitational escape speed --- plasma is held within the magnetosphere by the magnetic field, not by gravity [\ldots] Charge-exchange collisions between ions and neutrals, in which the outgoing neutral has the velocity of the incoming ion and vice versa, thus produce fast neutrals that escape from the system immediately, with their kinetic energy. This process represents a loss (generally by quite significant amounts) both of neutral particles and of energy from the magnetosphere.

(5) {\bf Dissipation processes in the magnetosphere}. In regions of
the magnetosphere with major departures from the MHD approximation
(particularly where magnetic reconnection is occurring) dissipative
processes such as Joule heating associated with effective resistivity
may be significant.  The primary effect is not energy loss but
enhancement of conversion from magnetic to thermal energy.''

\ors[II:10.3.4] ``The field approach to energy implies that energy may be regarded as {\em stored} in space [\ldots] The primary reservoir of stored mechanical energy in a planetary magnetosphere is the thermal energy of its various plasma structures, especially the {\em plasma sheet} of the magnetotail or magnetodisk, the {\em ring current}, and the plasma and neutral {\em tori} associated with the planet's moons; the kinetic energy of bulk flow of magnetospheric plasma also plays a role, particularly for plasma tori and in the case of rapid changes [\ldots]

The primary reservoir of stored electromagnetic energy of importance
for a planetary magnetosphere is the energy of the magnetic field [\ldots] Because the energy of the
planetary dipole field itself does not change (except on time scales
of the secular variation, $\sim 10^2-10^3$ years for Earth) and thus
has no effect on the energetics of the magnetosphere, a convenient
measure of stored electromagnetic energy is the energy of the total
magnetic field minus the (unchanging) energy of the dipole field, [reflected in Eq.~\ref{eq:fieldenergy}].

The stored gravitational energy can be changed only by a net radial
displacement of matter; any such effects in the magnetosphere are for
the most part negligible in comparison to changes of mechanical or
magnetic energy.''

\subsection{What leads to explosive energy releases?}\label{explosive}
\ors[II:10.5] ``The\indexit{magnetosphere!explosive energy release} discussion so far has ignored time variations and has proceeded on
the tacit assumptions that all the energy supply, conversion, and
dissipations processes are more or less in balance. There is no
general requirement for this to be the case, and in fact often it is
not the case [\ldots] 
The prototypical example is kinetic energy from the solar wind being
converted into magnetic energy of the magnetotail at an increased rate
due to enhanced dayside reconnection (in response to changed
solar-wind conditions), but the rate of removal by conversion of
magnetic energy into mechanical energy of magnetospheric plasma plus
escape down the magnetotail not being equally enhanced (for reasons
that need to be identified); in this case, the magnetic energy
reservoir increases with time and reaches a point at which (again, for
reasons that need to be identified) the magnetic energy content can no
longer be maintained but must be converted to other forms.''

%\subsection{Magnetic topological changes}\label{deltatop}

First, let us look at topological changes involved in magnetospheric processes. \ors[II:10.5.1] ``[M]agnetic\indexit{magnetosphere!magnetic topology change} flux transport and the increase of magnetic energy by stretching the field play an important role in supplying energy to the magnetosphere. Non-equilibrium configurations of the magnetotail that change the magnetic topology and allow different paths of flux transport are therefore of particular interest.

%%%%%%%%%%%%%%%%%%%%%%%%%%%%%%%%%%%%%%%%%%%%%%%%%%%%%%%%%%%%
\begin{figure}[ph]
  \centering
  \includegraphics[width=8.2cm,angle=90]{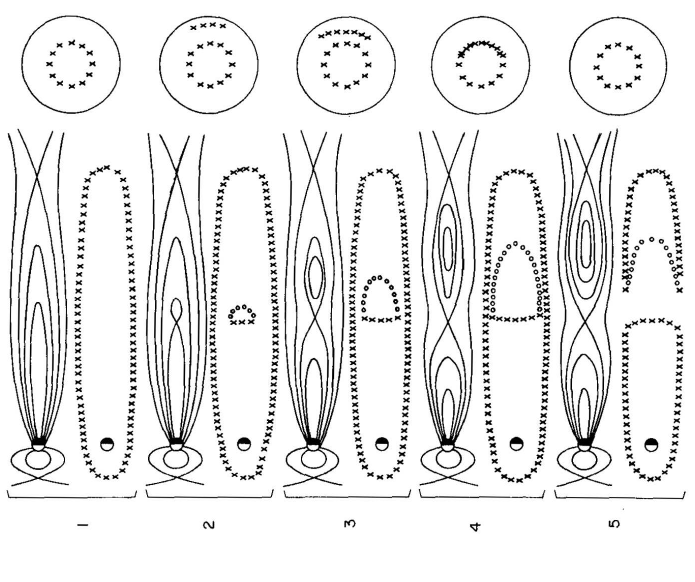}
  % could use movie:
  % e.g. frame numbers 44, 48, 50, 55, or 69, 70, 71, 72, ?
\caption[Topology of the magnetotail in a wind-dominated magnetosphere.]{\label{fig:plsmdE} Possible changes of the magnetic field
topology in the magnetotail of a solar-wind-dominated magnetosphere.
The diagram is shown [with the Sun's direction at the bottom] to facilitate
comparisons with diagrams of filament eruptions [\ldots]
Each panel in the sequence shows a side view of the magnetic field ({\em left}),
the outline of the X lines [where field of opposite directions meet] seen from above the north pole ({\em right}), and
a top-down view of the mapping of the reconnection region onto the Earth ({\em top}). [Compare this to the sketch of a solar eruption in Fig.~\ref{fig:hudsonasai}; Fig.~II:10.5; \href{https://ui.adsabs.harvard.edu/abs/1976mgpa.proc...99V/abstract}{source: \citet{1976mgpa.proc...99V}}.]}
% \cite{vmv76}.}
%\end{figure}
\nocite{vmv76}
%%%%%%%%%%%%%%%%%%%%%%%%%%%%%%%%%%%%%%%%%%%%%%%%%%%%%%%%%%%
%\begin{figure}
\centering
\includegraphics[width=20pc]{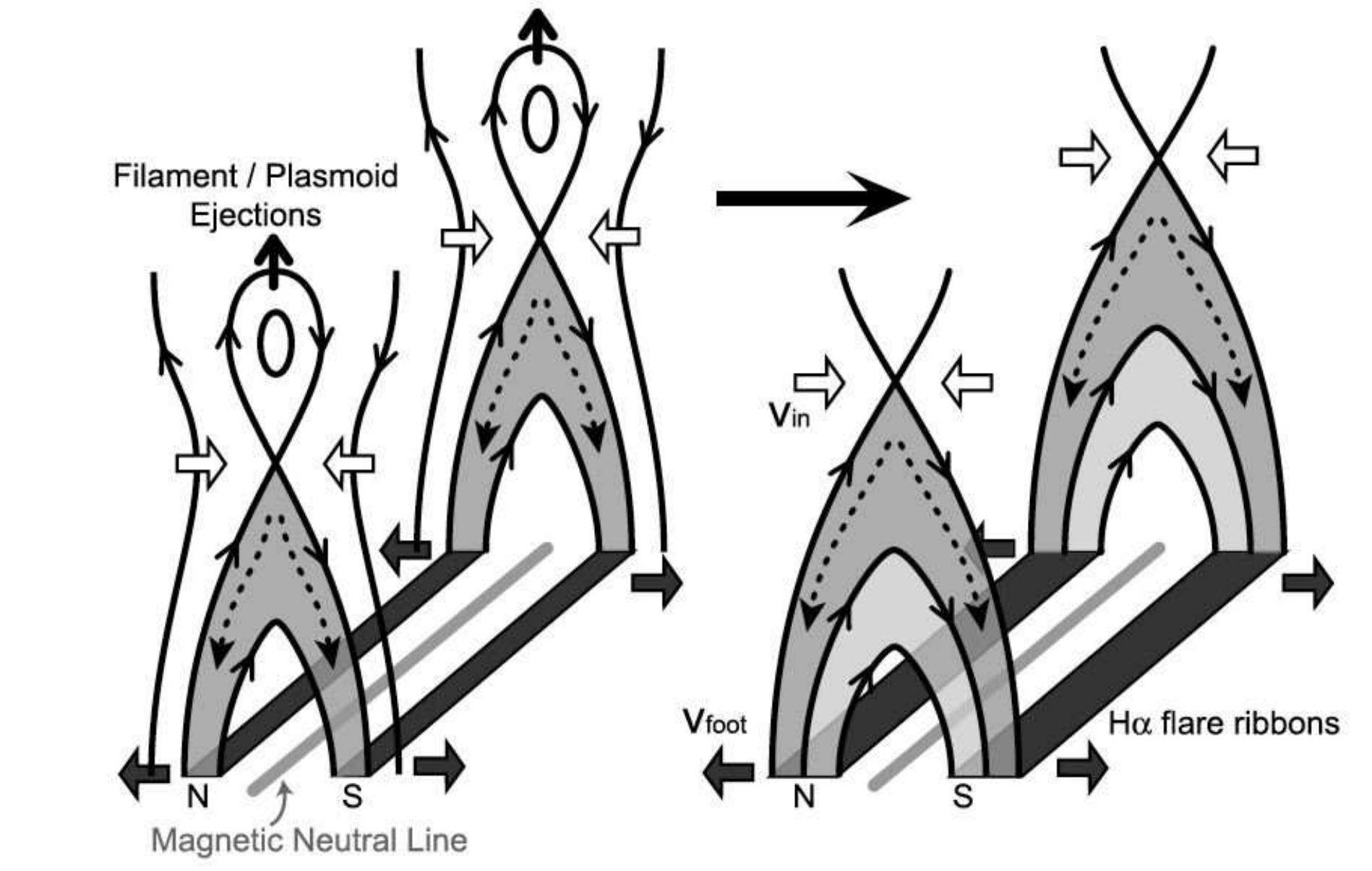}
\caption[Ribbon motion in solar flare reconnection.]{How the ribbon motion associated with solar flares sweeps out magnetic field during the
reconnection process in the standard model. [Compare this solar eruption to the sketch of the magnetospheric substorm (righthand side of each panel) in Fig.~\ref{fig:plsmdE}; Fig.~II:5.16; \href{https://ui.adsabs.harvard.edu/abs/2004ApJ...611..557A/abstract}{source: \citet{2004ApJ...611..557A}}.]
}
\label{fig:hudsonasai}
\end{figure}

A simple sketch of a model widely invoked to interpret magnetospheric substorms at Earth is shown in Fig.~\ref{fig:plsmdE},\indexit{reconnection!topology in magnetosphere} which displays a time sequence of magnetospheric configurations.  Each panel shows the magnetic field line configuration in the noon-midnight meridian plane (left) as well as the configuration of magnetic singular X [lines (where field vectors of opposite directions cross; see, for example, Fig.~\ref{fig:OMS})] and O lines [(around which field lines loop)] in the equatorial plane (right) and projected to the ionosphere (top); the equatorial projection, [\ldots] is essential for describing the three-dimensional structure of the magnetic field.  Panel~1 is the simplest topology of the open magnetosphere [(compare with Fig.~\ref{fig:OMS})]. In panel~2, a small volume usually called a {\em plasmoid} appears deep within the closed-field-line region, bounded on the earthward side by a newly formed {\em near-Earth X-line} (NEXL) and threaded by magnetic field lines that encircle the attached O-line; ideally, the field lines are confined within the plasmoid and connect neither to the Earth nor to the solar wind (what the real topology is, however, is still uncertain). For the ideal topology, the plasmoid can be visualized in three dimensions as shaped roughly like a banana, oriented approximately dawn to dusk and tapering to zero thickness at both ends, with the X-line on its surface and the O-line running through the middle of its volume.  The plasmoid grows (panel~3) by magnetic reconnection until it touches the separatrix of the open field lines (panel~4, {\em onset of lobe reconnection}); afterward (panel~5), the plasmoid is on interplanetary field lines and is carried away (presumably) by the solar wind.

A model of topological changes [internal to] a rotation-dominated magnetosphere
[(such as Jupiter's) differs] only in three respects: (1) the time
sequence has been translated into an azimuthal-angle sequence, (2)
field lines are stretched by the outflow of plasma from an internal
magnetospheric source (planetary/magnetospheric wind) [\ldots], (3) there are no
counterparts to panels 4 and 5, since field lines connected to the
solar wind are not considered. [\ldots]''

%\subsection{Role of instabilities}\label{unstable}

Next, let us look at the role of instabilities in causing rapid changes in topology. \ors[II:10..5.2] ``Instabilities have attracted much attention as a possible way of
inducing rapid change from equilibrium to non-equilibrium
configurations -- an alternative to straightforward evolution to
non-equilibrium as the result of changing boundary
conditions. [\ldots]

(1) {\bf Tearing-mode instabilities}. 'Tearing mode' is\indexit{instability!tearing mode} a\indexit{tearing mode instability} generic
term for instabilities that result in the reconnection of initially
oppositely directed magnetic fields. They are obvious candidates for
initiating topological changes of the magnetotail (in particular,
those envisaged in Fig.\,\ref{fig:plsmdE}).

(2) {\bf Current-driven instabilities}. The concept that a\indexit{current-driven instability}
sufficiently\indexit{instability!current-driven} intense electric current may bring about its own
breakdown, by creating conditions that impede current flow, was first
suggested [\ldots] as a model for solar flares. Under the
name 'current disruption' it has been widely discussed as a model
for substorm onset and expansion. Various instabilities that develop
when the current density exceeds some threshold value have been
proposed.

(3) {\bf Interchange and ballooning instabilities}. Interchange\indexit{instability!interchange}
instabilities\indexit{instability!ballooning} which do not appreciably change the magnetic field are
thought to be essential for plasma transport in rotation-dominated
magnetospheres.  Ballooning instabilities can be viewed roughly
as interchange that does change the magnetic field. As a model for
substorms, they have been invoked particularly at the transition
between the dipole field and the magnetotail, in several variants.''

\subsection{Terrestrial magnetospheric substorms}\label{substorms}

A\indexit{substorm! Earth} substorm can be summarized\indexit{substorm!growth phase} as a two-stage process. \ors[II:10.6.1] ``Stage~1 (growth phase): as a consequence of a southward interplanetary
magnetic field, the configuration of the magnetosphere changes, its
magnetic field becoming highly stretched (increased magnetic flux in
the magnetotail, reduced flux in the nightside equatorial\indexit{substorm!expansion phase}
region). Stage~2 (expansion phase, initiated by the onset): the
magnetic field changes to more nearly dipolar (increased flux on the
nightside), and there is enhanced energy input and dissipation to the
inner magnetosphere and the ionosphere/atmosphere; the process occurs
on dynamical time scales (comparable to or shorter than wave travel
times) and is accompanied (most probably) by changes of magnetic
topology.

%In terms of energy flow paths of Fig.\,\ref{fig:sumE}: during stage 1, $\mathcal{P}_{\rm I}$ (power in path I) is enhanced and is appreciably larger than the sum \mbox{$\mathcal{P}_{\rm II}+\mathcal{P}_{{\rm II}^{\prime}}+\mathcal{P}_{\rm III_{(sw)}}$}. During stage 2, $\mathcal{P}_{\rm II}$ and particularly $\mathcal{P}_{{\rm II}^{\prime}}$ are enhanced; $\mathcal{P}_{\rm III_{(sw)}}$ and $\mathcal{P}_{{\rm III}^{\prime}}$ presumably are enhanced in connection with topological changes exemplified by Fig.\,\ref{fig:plsmdE}.

[In terms of energy flow paths: during stage 1, power input from the bulk flow kinetic energy of the solar wind is enhanced and is appreciably larger than the sum of power outputs due to heating, bulk motion, and plasma escape from the geomagnetic system.  During stage 2, energy flow into mechanical energy of plasma and particularly plasma heating are enhanced; power flow through plasma and field escape and field reconfiguration presumably are enhanced in connection with topological changes]

The\indexit{substorm!growth phase} substorm growth phase is in essence the increase of open magnetic
flux in the magnetosphere, which occurs for a two-fold reason. First,
the flux addition rate at the dayside reconnection region increases as
the solar wind transports more magnetic flux, of the sense opposite to
the terrestrial dipole flux, toward the magnetosphere; the reasons for
this are assumed to lie in the physics of magnetic reconnection. Second, the flux return rate at the nightside
reconnection region does {\em not} increase to match the addition
rate; the reasons for this are not at all well understood. [\ldots] Within the
magnetosphere, the net effect of the substorm growth phase is to
remove magnetic flux from the nightside magnetosphere by flow toward
the dayside reconnection region and to add magnetic flux to the
magnetotail (enhanced stretching of magnetotail field lines).

The\indexit{substorm!expansion phase} substorm expansion phase does return the magnetic flux, rapidly
and spectacularly, from the magnetotail to the nightside magnetosphere
(dipolarization of a\indexit{dipolarization} previously stretched tail-like field); given that
plasma in the magnetotail beyond a distance typically $\sim 15-20$
Earth radii is observed to flow away from Earth, the process must
almost unavoidably proceed by topological changes of the type sketched
in Fig.\,\ref{fig:plsmdE}. The energy input into plasma, energetic
charged particles, and the \indexit{aurora}aurora can be largely accounted for by
adiabatic compression and Birkeland current effects. What remains
highly controversial is how the process starts and why it is so
sudden and catastrophic [\ldots]

A further complication is the question of external versus internal
influences.  That the growth phase is initiated by changing solar wind
conditions is the consensus view. The onset and expansion phase, on
the other hand, are regarded by the majority as basically the result
of internal dynamical processes, although subject to solar wind
influences ({\em e.g.,} if the system is evolving toward instability, it may
be pushed over the threshold by a change in the solar wind). A
substantial minority, however, considers the substorm onset
intrinsically as triggered by a solar wind change (typically toward a
more northward interplanetary magnetic field).''

\subsection{Terrestrial magnetic storms}\label{storms}

\ors[II:10.6.2] ``Our\indexit{magnetic!storm!Earth} understanding of magnetic storms has been decisively influenced by a remarkable theoretical result, the \indexit{Dessler-Parker-Sckopke theorem}Dessler-Parker-Sckopke
theorem, which relates the external magnetic field at the location of a dipole to properties of the plasma trapped in the field of the dipole. [T]he theorem states that ${\mathbf b}(0)$, the magnetic disturbance field of external origin at the location of a dipole of moment $\bm{\mu}_B$ [in an undisturbed state], satisfies
\begin{equation}
\bm{\mu}_B\cdot {\mathbf b}(0) = 2U_{\rm K}
\label{eq:DPS0}
\end{equation}
where $U_{\rm K}$ is the total kinetic energy content of plasma in the
magnetosphere. What is remarkable is that the right-hand side does not
depend on the spatial distribution, the partition between bulk-flow
and thermal energy, or any properties of the energy spectrum. [\ldots]

Although ${\mathbf b}(0)$ nominally is evaluated at the center of the
Earth, it is also equal to the (vector) average of ${\mathbf
b}({\mathbf r})$ over the surface of the globe (by a theorem for
solutions of Laplace's equation, satisfied within the globe by each
Cartesian component). The Dst\indexit{Dst index} index is the average, over a
low-latitude strip of the globe, of the disturbance field component
aligned with the dipole; after some corrections (chiefly removing the
contribution from induced earth currents), \hbox{-Dst} may be considered a
reasonable proxy for the left-hand side of Eq.~(\ref{eq:DPS0}), as
long as Dst$<0$.  The Dessler-Parker-Sckopke theorem then provides a
method of inferring the plasma energy content
%--- the energy contained
%in the box 'plasma mechanical energy' in Fig.\, \ref{fig:sumE},---
simply\indexit{Dst index!relation to magnetosphere energy} from the value of the Dst index. [\ldots] Direct
{\em in situ} observations have established that the greater part of the
energy resides in what is called the ring current region.

Geomagnetic storms, particularly the intense ones, are characterized by unusually large amounts of energy stored as mechanical energy of plasma in the ring current region, in comparison to other storage regions. This implies that during the development of an intense storm the power [going from the magnetic reservoir to the ring-current plasma kinetic reservoir] is unusually large, on the average. Whether this enhanced conversion rate from magnetic energy into mechanical energy of ring current plasma [\ldots] results from a different interaction process or simply from a different time sequence of solar wind parameters is an unresolved question. More specifically, can the energy for storms be supplied by a sequence of substorms (perhaps unusually frequent and/or unusually intense), or is some other process required?  A related question is that of {\em geoeffectiveness}: when interplanetary structures such as CME's impinge on the Earth, under what conditions do they produce intense magnetic storms?  (prolonged southward $B_{\rm sw}$ is one that is well established).''
%https://agupubs.onlinelibrary.wiley.com/doi/epdf/10.1029/2004JA010418
%https://agupubs.onlinelibrary.wiley.com/doi/epdf/10.1029/2012JA017584
\activity{{\em Show:} The energy processed by the magnetosphere during a magnetic storm is of order $E_{\rm storm}=5\times 10^{23}$--$5\times 10^{24}$\,erg from moderate storm to superstorm. Compare that to an order of magnitude estimate of the energy $E_{\rm mag,\oplus}$ contained in the geomagnetic field (by, say, using a scale of $3R_\oplus$ and a characteristic field strength of $0.1$\,G) and with the incoming total energy $E_{\rm sw}$ of the solar wind during the storm period (with typical conditions for the fast solar wind and an active cross section of $\pi R^2_{\rm CF}$, and a storm duration of 1-10\,h). What are the values of  $E_{\rm storm}/E_{\rm mag,\oplus}$ and $E_{\rm storm}/E_{\rm sw}$? Compare these values to solar equivalents when you reach Activity~\ref{act:flares}. \mylabel{act:storms}}

\section{Models of solar impulsive events}\label{sec:solarimpulsive}
\subsection{The magnetic reservoir}
\subsubsection{Storage models}\label{sec:storagemodels}
\ors[II:6.2.1] ``Although\indexit{flare!storage model} it is generally agreed that flares and CMEs derive their energy from the Sun's magnetic field, exactly how the magnetic energy is extracted remains uncertain.  One possibility is that a flare or CME occurs when a slowly evolving coronal magnetic field reaches a point where a stable equilibrium is no longer possible.  The slow evolution of the corona is driven by the changes continually occurring in the photospheric field as a result of solar convection; [in phases before solar impulsive events, these processes build up the stored magnetic energy].  The equilibrium may disappear altogether or, alternatively, a stable equilibrium may simply become unstable.  The continual emergence of new flux from the convection zone and the shuffling of the footpoints of closed coronal field lines increase the free magnetic energy in the corona.  Eventually, these stresses may exceed a threshold beyond which a stable equilibrium cannot be maintained, and the field erupts. Models based on this principle are often referred to as {\em storage models}.

\begin{figure}
\centering
\includegraphics[width=30pc]{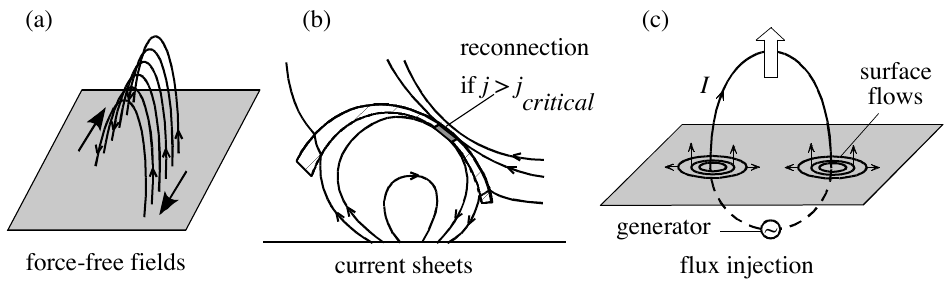}
\caption[Three different ways to use magnetic energy to power a flare or CME.]{Schematic illustration of three different types of models that use magnetic energy to power a flare or CME.  Panel {\rm (a)}:  Magnetic energy is stored in the corona in the form of field-aligned currents that eventually become unstable.  Panel {\rm (b)}:  Magnetic energy is stored in the corona in the form of a thin current sheet that is suddenly dissipated when a micro-instability is triggered within the sheet.  Panel {\rm (c)}:  An example of a directly driven flare model.  Here magnetic flux is suddenly injected from the convection zone into the corona at the onset of the flare or CME.  Such a model produces a well-organized flow pattern during the impulsive phase (small arrows at surface in panel {\em c}). [Fig.~II:6.8]}
\label{forbesCMEfig4}
\end{figure}

[\ldots] Because the plasma in the photosphere is almost $10^9$ times denser than the plasma in the corona, it is difficult for disturbances in the tenuous corona to have much effect on the photosphere and the deeper layers below it.  Field lines mapping from the corona to the photosphere are thus said to be 'inertially line-tied' which\indexit{line tying} means that the footpoints of coronal field lines are essentially stationary over the time scale of the eruption [\ldots]

Unlike models of confined flares, models of CMEs must be able to explain not only the release of magnetic energy, but also how mass is ejected into interplanetary space.  During a CME, magnetic field lines mapping from the ejected plasma to the photosphere are stretched outwards to form an extended, open field structure [\ldots]'' that resembles the sketches on the left of each of the panels in Fig.~\ref{fig:plsmdE}: plasmoids leaving the magnetosphere have been compared to filaments erupting as part of a CME.

\subsubsection{Directly driven models}

\ors[II:6.2.2] Some researchers ``have proposed models that produce a sudden energy release in the corona by means of a surface or sub-surface current generator.  In contrast to storage models, there is no build-up of magnetic energy in the corona prior to onset.  Instead, there is a sudden\indexit{flare!  injection model} injection of current or magnetic flux into the corona from below.  As a rule, the models do not address the mechanism that leads to the sudden injection of current or flux.  They simply posit that such an injection occurs, and then model the consequences of such an injection for the corona.'' Section~II:6.2.2 describes some of these models and their problems when compared to observations and physical conditions in the Sun; these are not further discussed here.

\subsubsection{Pre-eruption current sheet models}

\ors[II:6.2.3] ``Because the\indexit{flare!current sheet model}  magnetic energy in the corona is much larger than the thermal and gravitational energies, the magnetic force (${\bf j \times B}$) cannot, in general, be balanced by gravity or by a gas pressure gradient.  Thus, as a rule, the coronal field will tend to be force-free, meaning that the current will flow along the direction of the magnetic field ({\em cf.} Fig.~\ref{forbesCMEfig4}a).  An exception to this rule occurs when a current sheet is present.  In this case gas pressure within the sheet balances the strong magnetic field outside.  If the current sheet is sufficiently thin, then the high temperature or density within the sheet may not be detectable.  Thus the corona could still have the appearance of a plasma with a low gas to magnetic pressure ratio  ({\em i.e.}, plasma $\beta \ll 1$).  Figure~\ref{forbesCMEfig4}b shows a flare model with such a current sheet, where a micro-instability within the sheet triggers an eruption.

Prior to onset, the current sheet grows as a consequence of the emergence of new magnetic flux into a pre-existing magnetic loop as shown in Fig.~\ref{forbesCMEfig4}b.  As the current sheet grows, it eventually reaches a point where a micro-instability is triggered because the current density exceeds some critical value.  Once the micro-instability occurs, the electrical resistivity of the plasma in the sheet dramatically increases, and rapid reconnection ensues.''

\subsection{Two-dimensional force-free models}

\ors[II:6.2.4] ``[M]any storage\indexit{force-free field!models}\indexit{flare!force free field model} models use configurations that have currents flowing parallel to the magnetic field in the pre-eruption state.  Thus, there is no [net] magnetic force anywhere in the configuration prior to eruption.  To explain an eruption, such models need to show how a strong magnetic force can rapidly appear as a result of the slow evolution of the photospheric boundary conditions.

\begin{figure}
\centering
\includegraphics[width=0.8\textwidth]{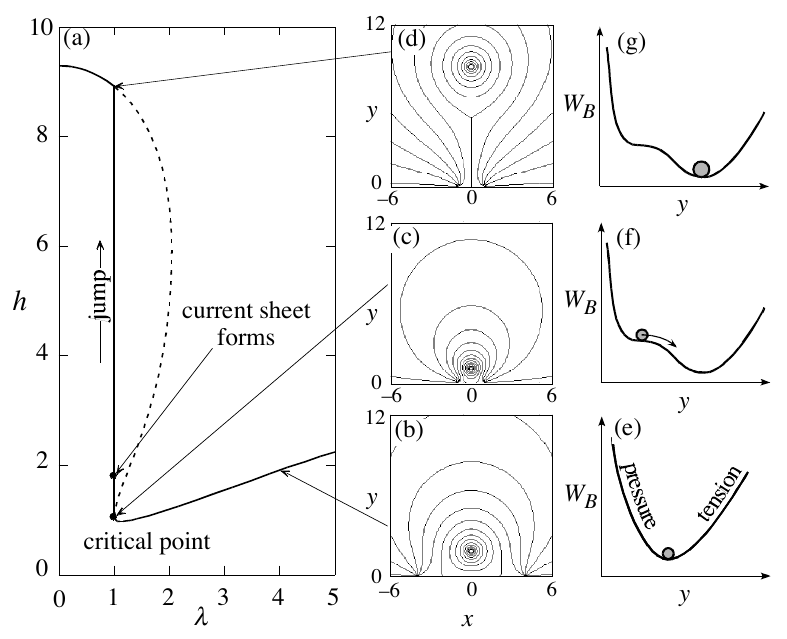}
\caption[Ideal-MHD evolution of a 2D arcade
with a flux rope.]{Ideal-MHD evolution of a two-dimensional arcade
containing a magnetic flux rope [described by Eq.~(\ref{mag_field.eq}), with values $p$ and $q$ the height of the lower and upper tips of the of the current sheet below the flux rope.] Panel {\em (a)} shows the equilibrium curve for the flux rope height, $h$, in normalized units, as function of the source separation half-distance $\lambda$.  Panels {\em (b, c)}, and {\em (d)} show the magnetic field configuration at three different locations on the equilibrium curve, and panels {\em (e, f)} and {\em (g)} show the corresponding energy schematic for each configuration.  The case shown is for a flux rope radius of 0.1 in normalized units. [Fig.~II:6.9; \href{https://ui.adsabs.harvard.edu/abs/1995ApJ...446..377F/abstract}{source: \citet{1995ApJ...446..377F}}.]}
\label{forbesCMEfig5}
\end{figure}

To illustrate the basic principles, we first consider a relatively simple flux-rope model [\ldots\ for which the external field is] prescribed by
\begin{equation}
\label{mag_field.eq}
B_y + iB_x = \frac{2iA_0\lambda(h^2 + \lambda^2)\sqrt{(\zeta^2 + p^2)(\zeta^2 + q^2)}}{\pi(\zeta^2 - \lambda^2)(\zeta^2 + h^2)\sqrt{(\lambda^2 + p^2)(\lambda^2 + q ^2)}},
\end{equation}
where $\zeta = x + iy$ and $A_0$ is the photospheric magnetic flux, or, equivalently, the magnetic vector potential at the origin.  In this expression $h$ is the height of the flux rope\indexit{flux!rope} above the surface and $p$ and $q$ are the lower and upper tips of a vertical current sheet below the flux rope as shown in Fig.~\ref{forbesCMEfig5}.  The parameter $\lambda$ is the half-distance between two photospheric field sources located at $\zeta = \pm \lambda$ on the surface [\ldots]

Application of the frozen-flux condition at the surface of the flux rope determines the current in the rope.  This condition keeps the magnetic flux between the flux rope and the surface constant in time.  It also ensures that during an eruption there is no flow of energy into the corona if the normal component of the field at the base remains invariant.  Consequently, the current in the flux rope is prescribed by 
\begin{equation}
\label{current.eq}
I = \frac{c\lambda A_0}{2\pi h}\frac{\sqrt{(h^2 - p^2)(h^2 - q^2)}}{\sqrt{(\lambda^2 + p ^2)(\lambda^2 + q^2)}}.
\end{equation}
This current decreases with time during an eruption as magnetic energy is converted into kinetic energy.  This decrease becomes apparent only when the formula giving the dependence of $q$ upon $h$ and $p$ is incorporated into the above expression [(for references, see Sect.~II:6.2.4)].  

The magnetic field configuration is shown in Fig.~\ref{forbesCMEfig5} for three different sets of parameters.  The surface at $y = 0$ corresponds to the
photosphere, and the boundary condition at this surface is
\begin{equation}
A(x,0)=A_{0}\, {\mathcal{H}} (\lambda - | x| ),
\end{equation}
where $\mathcal{H}$ is the Heavyside step-function and $A_{0}$ is the value of $A$
at the origin.  This boundary condition corresponds to two sources of
opposite polarity located at $x = \pm \lambda$.

[\ldots] Depending on the choice of model parameters, there may be three equilibria, one equilibrium, or no equilibrium for a given set of parameters.  In situations with three equilibria the magnetic energy of each equilibrium is different.  For the isolated equilibrium shown in Fig.~\ref{forbesCMEfig5}b the flux rope sits in an energy well as shown in Fig.~\ref{forbesCMEfig5}e.  If the flux rope is pushed downward toward the surface, compression of the magnetic field between the flux rope and the surface creates an upward force.  If the flux rope is pulled upward away from the surface, magnetic tension from the overlying arcade creates a downward force.  \indexit{line tying}Line-tying\regfootnote{Line-tying assumes that the surface footpoints of all magnetic field lines remain fixed at their initial positions during the eruption.}
plays a key role in creating the equilibrium because it prevents field lines from being pushed into, or pulled out of, the surface when the flux rope is perturbed.

An evolutionary
sequence is created by assuming that the distance between the two sources at $\pm \lambda$ decreases at a rate that is much slower than the Alfv\'{e}n
time-scale in the corona.  A flux rope located on the lower portion of the equilibrium curves shown in Fig.~\ref{forbesCMEfig5}a will erupt when the distance between the line sources becomes less than the height of the flux rope.  When this location is reached, the unstable and stable equilibria coincide as shown in Fig.~\ref{forbesCMEfig5}g.
Once equilibrium is lost, the flux rope rapidly moves upwards.  In the absence of reconnection ($p = 0$) the flux rope does not escape, but, instead, reaches a new equilibrium position with a vertical current sheet, as shown in Fig.~\ref{forbesCMEfig5}d.

In the absence of any reconnection the amount of energy released by the loss of equilibrium is quite small, less than 5\%\ [\ldots]  Thus, while the loss of equilibrium can account for the rapid onset of an eruption, it cannot, by itself, account for the large amount of energy released.  For this, magnetic reconnection is needed. [\ldots\ F]or typical coronal conditions a very modest rate of reconnection is sufficient to allow escape.  For reconnection rates corresponding to an inflow Alfv\'{e}n Mach number, $M_{A}$, $> 0.05$ (at the midpoint of the current sheet sides) the flux rope can escape without any deceleration [\ldots]''

MHD simulations are needed to analyze such an eruption with more realism, and also to understand the role of waves, including shocks that develop when the eruption speed exceeds the propagation speeds of any of the possible MHD waves. More on this in the Heliophysics books.

\subsection{Three-dimensional force-free models}

\ors[II:6.2.6] ``It\indexit{force-free field!models!3d} will probably come as no surprise that three-dimensional models are considerably more complex than two-dimensional ones.  Three-dimen\-sional field configurations are subject to a much greater number of instabilities.  The helical ideal-MHD kink mode is an example of an inherently three-dimensional instability that does not exist in two dimensions.  The dynamical evolution that occurs in three-dimensions is also more complicated.  Fully nonlinear three-dimensional MHD turbulence can occur and magnetic reconnection exhibits new features that have no counterpart in two dimensions.  Nevertheless, despite these additional complications, the underlying principles of the three-dimensional storage models remain the same.

\begin{figure}
\centering
\includegraphics[width=12pc]{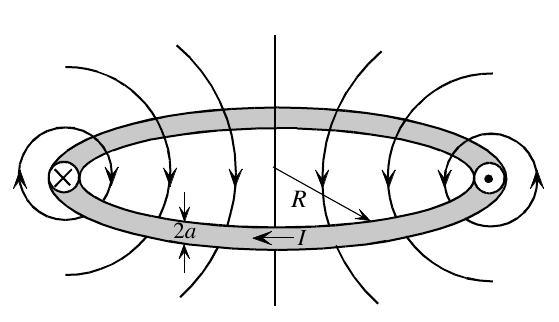}
\caption[An isolated toroidal flux rope.]{An isolated toroidal flux rope.  The flux rope has a major radius, $R$, a minor radius, $a$, and carries a net toroidal current $I$.  The antiparallel orientation of the current flowing on the opposite sides of the torus creates an outward force in the radial direction. [Fig.~II:6.16]}
\label{forbesCMEfig12}
\end{figure}

[\ldots] In order to show the relation of the relatively simple two-dimensional model of the previous section with these three-dimensional models, we take a reductionist approach.
That is, we start with a very simple three-dimensional configuration and then sequentially add new features that increase its complexity.  We start with the simple toroidal flux rope shown in Fig.~\ref{forbesCMEfig12}.  The antiparallel orientation of the current flowing on the opposite sides of the ring produces a repulsive force similar to the force between two parallel wires with antiparallel currents.  For a small minor radius, $a$, this force, sometimes referred to as the hoop force, is approximately
\begin{equation}
F  \propto  {I^{2} \over R} {\rm ln}(R/a),
\end{equation}
where $I$ is the flux-rope current, $R$ is the major radius, and $a$ is the minor radius of torus.  The right-hand side of the above expression is the lowest order term of an expansion in the parameter $a/R$, so the expression is only valid for $a \ll R$ [\ldots]

Just as for two-dimensional storage models, the three-dimensional models assume that the time scale of the eruption is so fast that any additional input of magnetic energy after the eruption starts, is completely negligible.  Therefore, the flux associated with the flux rope current is conserved.  In the limit that $a/R$ tends to zero, the flux-rope current is roughly
\begin{equation}
I \approx {I_{0} R_{0} \over R  {\rm ln}(R/a)},
\end{equation}
where $I_{0}$ and $R_{0}$ are initial values.  If one considers the torus configuration as an initial state that subsequently evolves in response to the force, then $R$ will increase to infinity, but as it does, so $I$ will decrease to zero.  In the process the magnetic energy associated with the flux rope's initial current is converted into the kinetic energy of the expanding plasma ring.

\begin{figure}
\centering
\includegraphics[width=30pc]{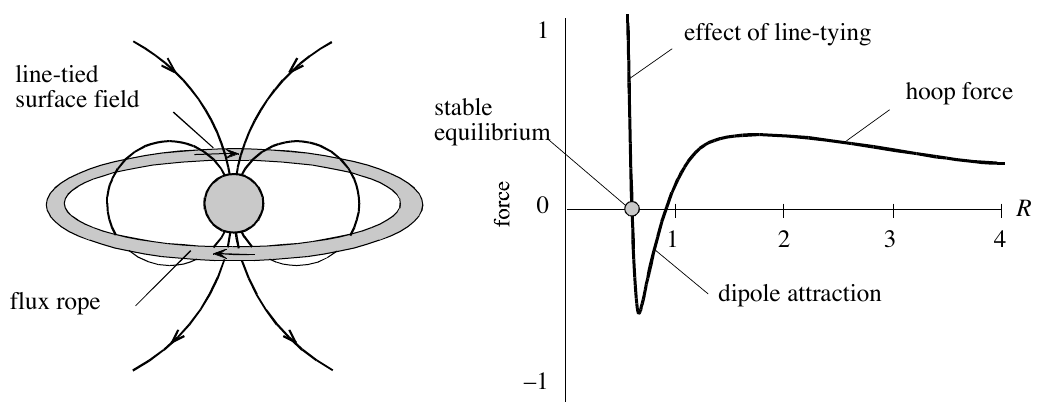}
\caption[A stable toroidal equilibrium.]{A stable toroidal equilibrium.  {\em (left)} The addition of a line-tying surface representing the surface of the Sun creates the possibility of a stable equilibrium.  Surface currents (which can be modeled using an image current) create an additional magnetic field component that gives rise to a second equilibrium position as shown on the {\em right}.  The new equilibrium is stable because displacements away from it produce a restoring force. [Fig.~II:6.18]}
\label{forbesCMEfig14}
\end{figure}
To create an equilibrium one must add an additional magnetic field of the proper orientation and strength.  In tokamak terminology such a field is called a strapping field. [\ldots\ Whereas it is possible to create a stable equilibrium by an appropriately shaped] strapping field, an alternative possibility that is more appropriate for a storage model is to introduce a line-tying surface as shown in Fig.~\ref{forbesCMEfig14}.  The effect of line-tying can be modeled by introducing a fictitious image current below the surface.  With the introduction of this additional current, a new equilibrium appears which, unlike the previous one, is stable.  Stabilization is achieved because line-tying prevents field lines from being pushed into, or pulled out of, the surface [\ldots]

\begin{figure}
\centering
\includegraphics[width=20pc]{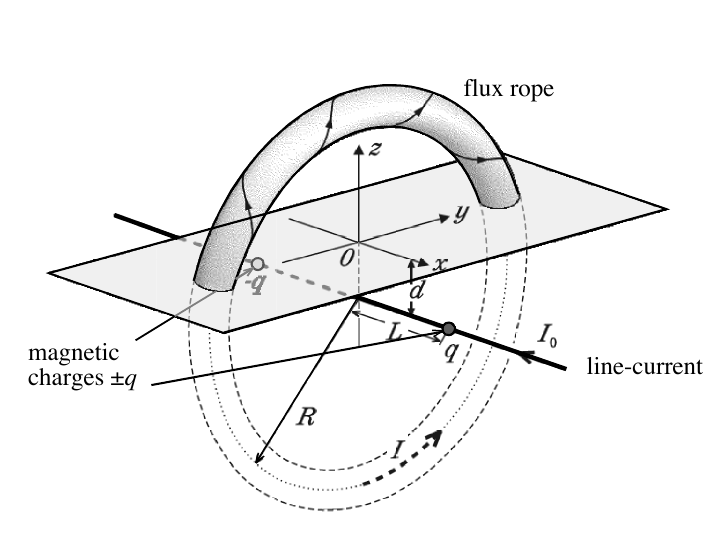}
\caption[The three-dimensional flux-rope model).]{The three-dimensional flux-rope model of Titov and D\'{e}moulin.  The coronal magnetic field is produced by three different sources consisting of a flux rope current, a pair of magnetic charges, and a line current.  The source regions located below the surface are fictitious constructs used to create the coronal field.  The model does not prescribe the form of the subsurface field. [Fig.~II:6.19; \href{https://ui.adsabs.harvard.edu/abs/1999A&A...351..707T/abstract}{source: \citet{1999A&A...351..707T}}.]}
\label{forbesCMEfig15}
\end{figure}
Although we now have an eruptive model with some degree of three-dimensionality, it still has the drawback that the flux rope is not itself anchored to the solar surface.  An analytical configuration that does have this property is shown in Fig.~\ref{forbesCMEfig15}; [\ldots] it consists of a toroidal flux rope that intersects the photospheric surface.  The flux rope, with current $I$, is held in equilibrium by an overlying arcade (not shown in the figure) which is produced by subsurface magnetic charges $\pm q$ located along the centerline at a depth $d$ below the photospheric surface at $z = 0$.   Finally, there is a subsurface line current lying along the centerline.  The strength of the current, $I_{0}$, flowing in this subsurface line controls the pitch of the coronal magnetic field.  When $I_{0}$ is varied from small to large values, the configuration changes gradually from a highly twisted flux rope resembling a slinky to one that resembles a sheared arcade without a flux rope.

\begin{figure} \centering \includegraphics[width=18pc]{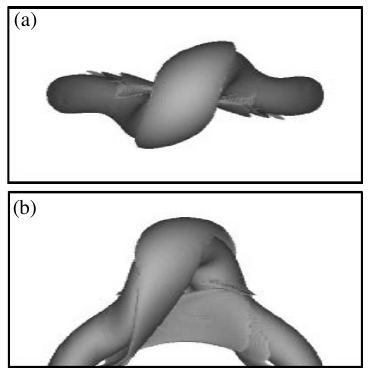} \caption[Current density surfaces for an unstable Titov-D\'{e}moulin equilibrium.]{Top view {\rm (a)} and side view {\rm (b)} of constant current density surfaces from a simulation for an unstable Titov and D\'{e}moulin equilibrium. [Fig.~II:6.20; \href{https://ui.adsabs.harvard.edu/abs/2004A&A...413L..27T/abstract}{source: \citet{2004A&A...413L..27T}}.]}
\label{forbesCMEfig16}
\end{figure}

Although the magnetic field of [what is known as a Titov-D{\'e}moulin] configuration is still azimuthally symmetric about the centerline of the torus, the solar surface no longer shares this symmetry.  Instead the surface is a flat plane that intersects the flux rope torus at some arbitrary position without influencing the field structure.  Thus, any line-tied evolution of this configuration away from the initial state necessarily creates a highly asymmetrical configuration.  An example of what such a configuration looks like is shown in Fig.~\ref{forbesCMEfig16}.  This figure shows two different views of an iso-current surface of the current density obtained from a simulation.   This simulation starts with an unstable Titov and D\'{e}moulin configuration that is given a small perturbation.  Within a few Alfv\'{e}n scale times the configuration evolves into the kinked, omega-shaped flux rope shown in the figure.  For this particular case, the initial instability is actually a helical kink instability rather than the torus instability discussed previously.  However, it is possible to construct unstable Titov and D\'{e}moulin configurations that are unstable to the torus instability rather than the helical kink [\ldots]''

\subsection{Formation of the pre-eruption field}

\ors[II:6.2.7] ``An \indexit{flare!pre-eruption field}important issue that the above flux rope models do not address is the creation and growth of the magnetic stress that causes the field to erupt.  It could be that most of the stress build-up occurs in the convection zone before the field emerges into the corona.  Alternatively, it may be that the field emerges in a nearly unstressed, current-free state, and that the stress subsequently develops in response to the observed surface flows.  In practice both possibilities are likely to occur at least at some level.

\begin{figure}
\centering
\includegraphics[width=0.8\textwidth]{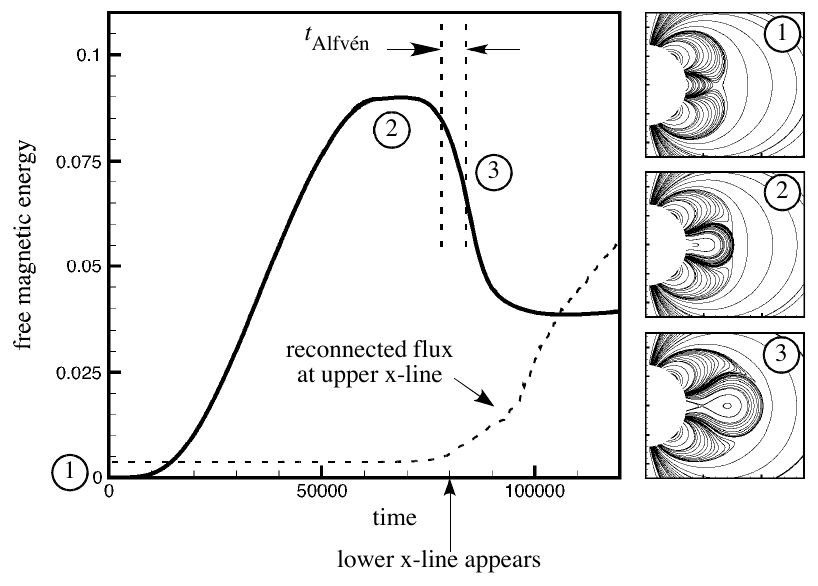}
\caption[Numerical simulation of a storage model.]{Numerical simulation of a storage model.  The panel at left shows the free magnetic energy as a function time, while the three panels at right show contours of the magnetic flux surfaces at three different times. [Fig.~II:6.21; \href{https://ui.adsabs.harvard.edu/abs/2004ApJ...614.1028M/abstract}{source for inset panels: \citet{2004ApJ...614.1028M}}.]}
\label{forbesCMEfig17}
\end{figure}

[Among the three-dimensional simulations that address this issue is one] called the {\em breakout model}\indexit{flare!breakout model}\indexit{breakout model}.  The evolution of this model is shown in Fig.~\ref{forbesCMEfig17}.  The initial state consists of a quadrupolar magnetic field that carries no current, so it contains no free-magnetic energy.  Slowly shearing the central arcade around the equator gives rise to a set of stressed loops that push outward against the overlying arcade.  As this happens, a curved, horizontal current sheet forms at high altitude at the pre-existing x-line.  Eventually, the stresses build up to a level that causes an eruption.  The nature of the mechanism that triggers the eruption has not yet been fully resolved, but it is likely that it consists of some kind of combination of both ideal and non-ideal processes.

[An alternative to this model is where a flux rope emerges into a pre-stressed field from below the photosphere.] Generally, the flux rope will tend to erupt once there are one or two turns in the portion of it that has emerged into the corona.  However, if the flux rope emerges into a pre-existing arcade, the strength and orientation of this arcade also has a strong effect on whether an eruption occurs or not.

[\ldots] One of the important issues that [various] studies address is the effect of mass loading on the emergence of a flux-rope into the low-density corona.  Most of the CME models discussed in the previous section are based on the supposition that a flux rope exists in the corona prior to onset, but it is not obvious how such a structure could be formed.  Formation of the flux rope within the convection zone followed by its buoyant rise into the corona immediately encounters the problem that mass cannot easily drain out of concave-upward portions of the magnetic field.  Unless there is a way for the mass to drain out of the flux rope, the rope will remain half buried in the solar surface.  One way around this difficulty is to suppose that the flux rope does not exist prior to the emergence of magnetic flux, but instead forms in the corona by a combination of converging flows and slow reconnection.  Most dynamo models, however, predict that large-scale flux ropes will form near the base of the convection zone and then rise buoyantly to the solar surface to form an active region.  Thus, this solution to the mass-loading problem involves both the destruction and reformation of the flux rope below and above the surface.''

\subsection{Observed signatures of flares and CMEs}
\indexit{flare!phases} \ors[II:5.2.2] ``The release of energy can either be 'impulsive', with time scales sometimes faster than 1\,s, or 'gradual.'  The impulsive and gradual signatures of a flare extend across the entire electromagnetic spectrum in a complicated way, as illustrated in Fig.~\ref{fig:hudsonfig1}.  The terminology may not seem appropriate when one considers a slowly developing flare-like event, such as a quiet-Sun \indexit{filament!eruption}filament \indexit{filament|seealso{definition}}eruption; in such a case the 'impulsive phase' may take tens of minutes to evolve, and the hard X-ray emission may be below the detection level.  Thus we don't know how 'impulsive' the energy release really is in such an event, but in other respects it has the morphology of an ordinary active-region flare.''

\ors[IV:2.1.2] ``An individual flare can be divided into two main phases: impulsive and gradual.  This generally refers to the timing of emissions relative to the processes thought to be occurring in the flare. In the standard picture, the initial energy conversion caused by magnetic reconnection powers particle acceleration and possibly --~depending on the energetics and on the magnetic configurations~-- a mass ejection.  The downward-directed particles become trapped in loops and emit non-thermal incoherent radio emission ([compare Fig.~\ref{fig:ostenloops}]).  Coherently emitting particles can be traveling either upwards out of the atmosphere or downwards into the atmosphere.  Once the trapped particles precipitate from the magnetic trap, they deposit their energy in or just above the photosphere, producing thick-target non-thermal bremsstrahlung emission.  This energy deposition results in the heating of the photospheric material to temperatures near 10$^{4}$\,K, and emissions from FUV lines.  All of this is associated with the impulsive phase of the flare.  The flow of energy at this point proceeds back into the upper atmosphere, with line emission from the lower chromosphere. Thermal X-ray emission occurs as well. As the energy input into the system decreases, emissions of all flare components return to the pre-flare level.''

\ors[II:5.2.2]\indexit{Neupert effect|see{flare}} \indexit{chromospheric evaporation} ``We understand the impulsive and gradual phases to show the main energy [conversion out of the magnetic reservoir] and its aftermath (secondary effects), with the proviso that it is really not just that simple.  \indexit{flare!Neupert effect}The most prominent 'aftermath' is the action of coronal magnetic loops as an energy reservoir, with cooling time scales that can approach hours.  This reservoir function is often described as the 'Neupert effect': the coronal manifestations of a flare tend to lag behind its chromospheric ones.  This results from the finite time scale associated with the coronal density increase during the impulsive phase, via the process of 'chromospheric evaporation.'  \activity{{\em Show:} 'Chromospheric evaporation' is a misnomer because there is no phase transition involved: the heating of chromospheric material from $\approx 10^4$\,K to of order $\approx 5\times 10^6$\,K causes the pressure and the associated pressure scale height to increase. What are the pre-heating and post-expansion scale heights for the above temperatures? (Look back at Sect.~\ref{sec:nearlystrat}.) What assumptions did you make when doing the calculation? How do these compare to the solar radius? \mylabel{act:evaporation}} The decay time scale reflects its slower cooling and return to the lower atmosphere.  The new material in the corona could be seen in the coronal emission lines, via free-free emission at radio wavelengths, or via free-free emission at soft X-ray wavelengths [\ldots]'' \sactivity{$\circledS$ {\em Show:} (a) For a given temperature, why does the coronal soft X-ray brightness scale essentially with the square of the particle density? (b) Let a given coronal loop have an initial loop-top density $n_0$ at temperature $T_0$ and let an impulsive heating event change these to $n_1$ and $T_1$. With $T_{0,1}$ within the range of about 0.4--30\,MK the radiative losses scale as $P(T)\propto T^{-2/3}$. If the temperature changes from 1\,MK to 5\,MK and the density increases by a factor of 25, show that the ratio of radiative cooling time scales is not too different from unity. Time scales for conductive losses exhibit a somewhat larger ratio; why? \mylabel{act:loopemission}\solution{loopemission}}

\ors[IV:2.1.2.1] ``The Neupert Effect relationship [\ldots] was formulated originally to
describe the integral relationship between markers in a solar flare
corresponding to the action of non-thermal particles, and the response
from the atmosphere to the deposition of
energy from these particles as it appears in coronal radiation.  Written more generally, \\
\begin{equation}
L_{\rm gradual}(t) \sim {\cal{C}}_{\lambda\lambda^\prime} \int_{t_{0}}^{t} L_{\rm impulsive}(t') \;\; dt',
\end{equation} 
where $L_{\rm impulsive}(t)$ is [evolving emission during the impulsive phase] which diagnoses the presence and action of particles accelerated in the explosive event (for stellar studies usually radio gyrosynchrotron, transition region FUV emission lines, or photospheric UV-optical continuum emissions), and $L_{\rm gradual}(t)$ is the intensity corresponding to the gradual phase (usually coronal emission, but some chromospheric emission lines display the Neupert effect as well). The interpretation is that the gradual phase emission is responding to the buildup of energy that occurs as a result of the energy deposition being diagnosed by the impulsive phase emission. [\ldots\ Note that] not all solar flares follow the standard flare scenario'': it appears to hold for some 80\%\ of large flares, but overall for about half of all flares.
% Veronig et al. 2002
The value of ${\cal{C}}_{\lambda\lambda^\prime}$ depends on the wavelength bands [$\lambda$ and $\lambda^\prime$] used for both the impulsive and gradual phases.''

\ors[II:5.2.2] ``The different atmospheric layers have a high degree of interconnectedness.  Because a flare marks a transition between one quasi-stable configuration and another, the ordinary law of hydrostatic equilibrium dictates the run of pressure up through the atmosphere.  A flare increases the gas pressure in the corona, at the expense of magnetic energy, and this can readily be detected at all levels).  The hydrostatic scale height for pressure is given by $2 k_{\rm B} T_{\rm e} / m g_\odot$, where $k_{\rm B}$ is the Boltzmann constant, $T_{\rm e}$ the temperature, $m$ the mean molecular weight, and $g_\odot$ the surface gravitational acceleration.  For a flare temperature of 10$^7$~K, this scale height is a large fraction of the solar radius, much larger than the flare loop structures.  Thus the vertical structure is isobaric in the upper chromospheric and coronal regions, and the chromosphere acts as a reservoir of mass to maintain this isobaric state as the flare loops cool, [lose pressure, and drain into the chromosphere] quasi-statically.''

\ors[II:5.3] \indexit{flare!white light} \indexit{flare!UV continuum} ``In the photospheric spectrum we see solar flares as brief flashes of white light and UV~continuum.  At present these sources are often not resolved either in space (Mm scales) or time (few sec scales).  The bright emission regions are embedded in the 'ribbon' regions that become more prominent in the chromospheric and EUV coronal lines.  In the coronal emissions one sees bright coronal loops developing slowly, with those from the highest temperatures appearing first and then cooling down through generally longer wavelengths, while at the same time shrinking in length. [\ldots]

\indexit{flare!bolometric detection}
Solar flares are not luminous on the scale of the total solar irradiance
('solar constant'), although they may produce
a localized brightening seen against the bright photosphere.
The powerful flare of November~4, 2003 was the first that could actually
be detected in the total solar irradiance, by the radiometer on
board the SORCE spacecraft.
The signal, at roughly 5$\sigma$ significance, amounted to
about 300\,ppm of the total signal, or 0.3\,millimagnitudes in astronomical terms.
There is a solar background noise level for such a measurement due
to convection and oscillations; this amounts to some 50-100~ppm spread out
over a bandwidth of a few mHz. 

\indexit{star!Vega ($\alpha$ Lyrae)} The localized brightening of a flare is much easier to see, of course, via an image even in white light.  Carrington [was the first to see a solar flare. He described] his 1859 discovery as resembling the brilliance of Vega ($\alpha$ Lyrae), for example. [The photospheric brightening is a major fraction of a flare's energy budget.]  Soft X-ray emission, for example, contains only 5-10\% as much luminosity.  This gradual component [\ldots] results from a thermal distribution (hot gas) for which the X-ray emission itself is a dominant cooling term.  The non-thermal tail of the X-ray spectrum (h$\nu >$~10\,keV), on the other hand, is due to bremsstrahlung from stopping particles.  The bremsstrahlung mechanism is very inefficient, providing a fraction of order 10$^{-5}$ of the energy losses.  The rest of the energy winds up in longer-wavelength radiation, notably the visible/UV continuum.

\indexit{flare!bulk energy} \indexit{coronal!mass ejection!without major flare} We must also consider the bulk kinetic energy [involved in major solar impulsive events: \indexit{CME|see{flare}}CME kinetic energies can rival [total photon losses] in such cases.  In rare cases a CME can occur in the absence of a major perturbation of the lower atmosphere. [\ldots] The partition of energy in a flare/CME event remains unclear physically and hard to determine observationally.

\label{sec:hudsonimpulsive}
\indexit{flare!phases!impulsive}

The impulsive phase of a flare marks the period of intense energy release
and strong non-thermal effects, including the launching of the CME.
The traditional observational tools for the impulsive phase are
hard X-ray emission and gyrosynchrotron emission at cm~to mm~radio
wavelengths.
The hard X-rays normally show two dominant footpoints embedded in
ribbon regions of opposite magnetic polarity, but we do not presently
understand why there are normally just two.
The sources are compact and rapidly variable, and we associate
them with the UV and white-light continuum emissions that also come
from the footpoint regions.
Other wavelengths show impulsive
emission components as well as gradual ones.
A clear impulsive-phase signature also appears even in the total irradiance,
but rarely exceeds the background [variability: whereas the flare brightness commonly stands out against the local quiescent surface brightness, it stands out against among the overall variability of the entire disk only in the most energetic events) \ldots]

The hard X-ray spectrum above about 10\,keV plays a central role in our
understanding of the impulsive phase because the collisional energy
losses of the bremsstrahlung-emitting electrons rival the total
flare energy itself.
This relationship can be established directly by inverting the hard
X-ray spectrum, under model assumptions.
The 'collisional thick target model'
envisions a black-box accelerator of
10-100\,keV electrons in the corona, with a directed beam penetrating
to the chromosphere or even photosphere to excite UV~and visible-light emission.
This simple model has become less tenable as spatial resolution improves, since
the WL/UV brightenings [observed with newer instruments] imply beams with extreme intensity [\ldots]

\label{sec:hudsongradual}
\indexit{flare!phases!gradual}

'Gradual phase' refers to the thermal emission from the hot coronal material evaporated during the impulsive phase, plus the strong transition-region and chromospheric emissions driven by the cooling of these coronal loops.  The loops connecting the roughly parallel ribbons form a semi-cylindrical \textit{arcade} structure, divided into many unresolved loops.  [\ldots] The hot regions eventually cool to form the H$\alpha$ loop prominence system, whence thermal instability leads to the phenomenon of 'coronal rain'.  The cooling also corresponds to shrinkage, as the gas pressure diminishes; shrinkage may also relate to the gradual release of energy as the coronal equilibrium returns to a stable configuration.  This is the process termed\indexit{dipolarization} 'dipolarization' in the geomagnetic community [\ldots]''

\begin{figure}
\centering
\includegraphics[width=14pc]{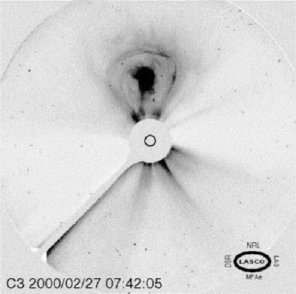}
\includegraphics[width=15pc]{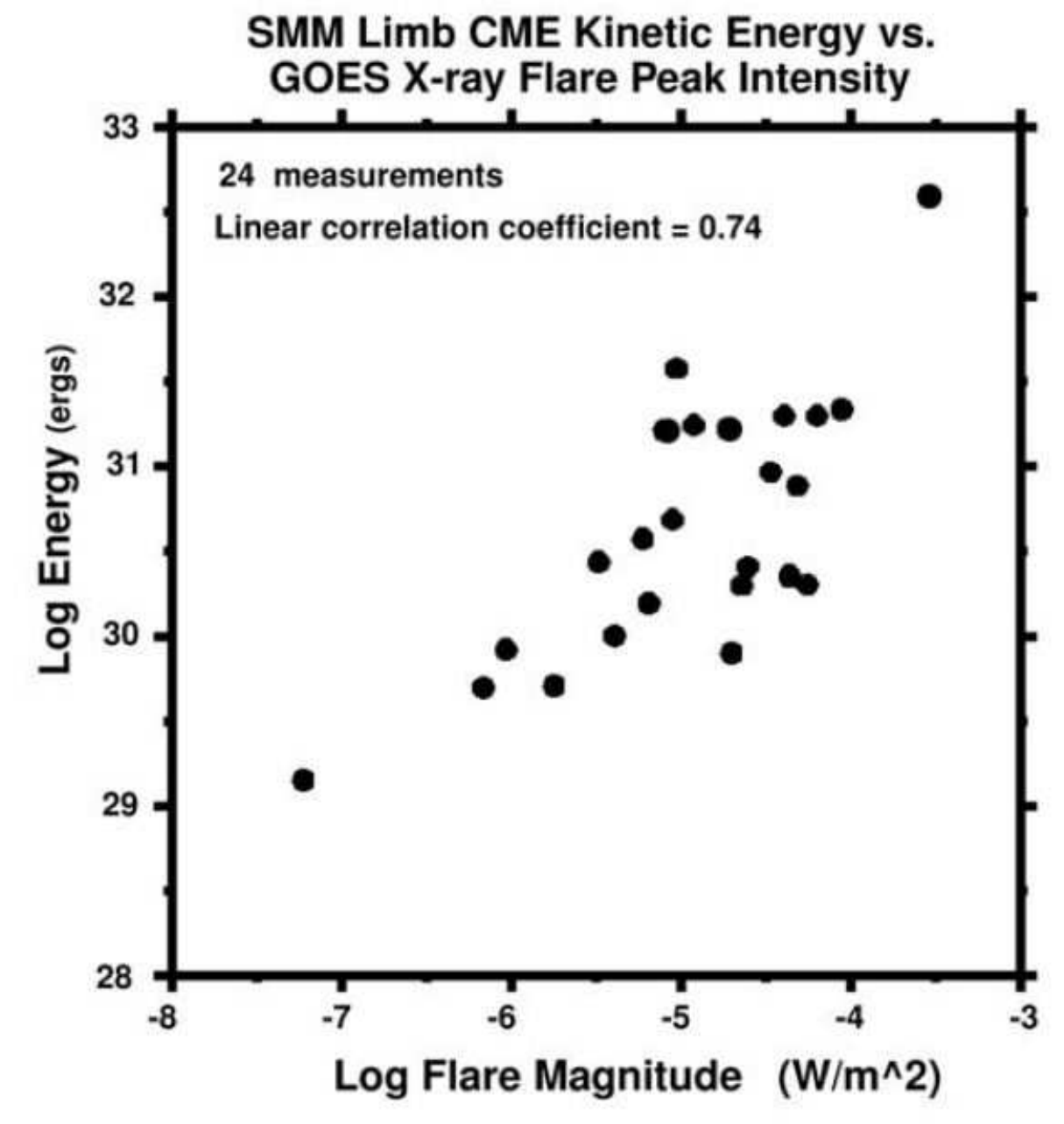}
\caption[Observation of a CME, and CME kinetic energy {\em vs.}\ flare class.]{
{\rm Left:} Coronagraph observation of a CME that nicely shows the three-part
structure: front, cavity, and (the bright core) filament (this is a file
image taken from the LASCO database, presented in a reverse greyscale).
{\rm Right:} Correlation between inferred CME kinetic energy and
peak GOES soft X-ray flux. [Fig.~II:5.5; \href{https://ui.adsabs.harvard.edu/abs/2004JGRA..109.3103B/abstract}{source: \citet{2004JGRA..109.3103B}}.]
}
\label{fig:hudson3part}
\end{figure}
\nocite{2004JGRA..10903103B}
\ors[II:5.6.4] ``The \indexit{flare!ribbon!motion} expanding motions of flare ribbons provided one of the first clues to what we think of as the standard reconnection model of a flare [T]hese motions can be interpreted as an electric field.  This is a motional or 'convective' electric field given by ${\bf E}=-{\bf v} \times {\bf B}/c$, and it is often taken as a measure of the reconnection rate.  [T]he rate the ribbons sweep out the field should correspond in some sense to the rate at which energy is released during reconnection, and that at the same time the field guides the particle or heat flux responsible for the ribbon excitation.''

\ors[II:10.2] ``[T]he resemblance\indexit{flare!substorm resemblance} of terrestrial substorms to\indexit{substorm!resemblance to flare} a two-ribbon solar flare, with ribbons of opposite magnetic polarity, has been repeatedly remarked upon'' (compare Figs.~\ref{fig:plsmdE} and~\ref{fig:hudsonasai}). Note, however, that the process leading up to the event is entirely different: the unstable field configuration is built up by stressing and/or flux injection from below in the case of solar eruptions but driven via the wind-magnetosphere interaction from the outside in the case of terrestrial substorms.

\label{sec:hudsoncmes}
\indexit{coronal!mass ejection}

\indexit{filament!cavity} \ors[II:5.3] ``Major flare events almost invariably involve the 'opening' of the magnetic field as a CME; see Table~\ref{tab:class} for the statistics [\ldots] Observationally, [\ldots] we often see a characteristic three-part structure: front, cavity, and filament (Fig.~\ref{fig:hudson3part}).  \indexit{filament}This pattern makes it clear that the CME originated in a filament cavity near the surface of the Sun.  A filament cavity consists of long, basically horizontal field, presumably more intense than its overlying 'tie-down' field that is more potential [\ldots] \nocite{chapman-bartels}

Modern images in coronal emissions such as soft X-rays allow a comparison of the coronal state before and after a CME event.  Such comparisons revealed \indexit{dimming}'dimmings,' readily interpreted as the evacuation of the mass of the corona by the CME eruption.  The soft X-ray dimmings presumably correspond to the coronal depletions found via similar before/after comparisons of the visible corona.'' \activity{(a) {\em Consider:} Describe what is seen in Fig.~\ref{fig:hudson3part}: how can a CME be imaged, and why is that best done from space, or from a very high mountain top? (b) Argue why the CME in this image is not likely to envelop Earth. (c) What would an Earth-bound CME look like? (d) Can you differentiate that from one moving in the opposite direction?}\activity{{\em Show:} The energy processed during a strong to intense solar flare and CME is of order $E_{\rm flare}= 10^{30}$--$10^{33}$\,erg. (a) Compare that to an order of magnitude estimate of the energy $E_{\rm AR,\odot}$ contained in the field of an active-region core (by, say, using a scale of $30,000$\,km and a characteristic average magnetic flux density of $300$\,G). )b) What is the value of $E_{\rm flare}/E_{\rm AR,\odot}$?  (c) How does this compare to $E_{\rm storm}/E_{\rm mag,\oplus}$ and $E_{\rm storm}/E_{\rm sw}$ in Activity~\ref{act:storms}? \mylabel{act:flares}} \activity{{\em Show:} One phenomenon associated with many CMEs is a so-called \indexit{dimming}'coronal dimming', in which a large fraction of the quiet-Sun solar corona fades for some time. Think about the potential causes: temperature change (so the signal moves from one bandpass to another), quasi-adiabatic expansion of the coronal field, and entrainment of coronal plasma in the erupting CME.  Estimate the fraction of the volume of entire quiet-Sun corona (at a density of some $10^7$\,cm$^{-3}$) that would need to be involved if it were to move out with an erupting field configuration if that made up, say, 25\%\ of the erupting mass of, for example, $10^{15}$\,g. \mylabel{act:cmevolume}}\activity{{\em Look up:} For a sense of scale: how many nuclear bombs are needed to match the energy released in a large solar flare of $10^{32}$\,erg?}  \indexit{coronal!dimming} \activity{{\em Advanced/Group:} Advances in numerical capabilities are making a big difference in understanding magnetic instabilities, how and where associated plasma heating occurs, and how combinations of plasma flows and a variety of temperatures in plasmas along a line of sight through the optically thin corona lead to observables. Such work shows how apparently non-thermal signatures in spectra can emerge from line-of-sight integration through thermal plasmas. If you would like to learn more about how observables based on numerical work help guide the interpretation of real-world observables, a paper (with illuminating graphics)  \href{https://ui.adsabs.harvard.edu/abs/2019NatAs...3..160C}{by \citet{2019NatAs...3..160C}} provides a good example.}

\section{Magnetic instabilities and reconnection}\label{sec:reconshock}
One of the mechanisms thought to be involved in the destabilization of magnetic configurations is \indexit{reconnection}reconnection.  Fast reconnection is often accompanied by shocks, and both the motions in the reconnecting field and the shocks themselves contribute to energy conversion into a mixture of thermal and non-thermal populations. There is a vast literature on the topic, including how such processes contribute to instabilities in the solar corona and the magnetosphere. Here we touch only on the fundamentals, specifically the concepts involved in steady 2-dimensional reconnection; more comprehensive material (and references to further reading) moving towards the time-dependent and 3-dimensional real world is provided in I:5 and II:6.

\begin{figure}[t]
\includegraphics[width=7cm]{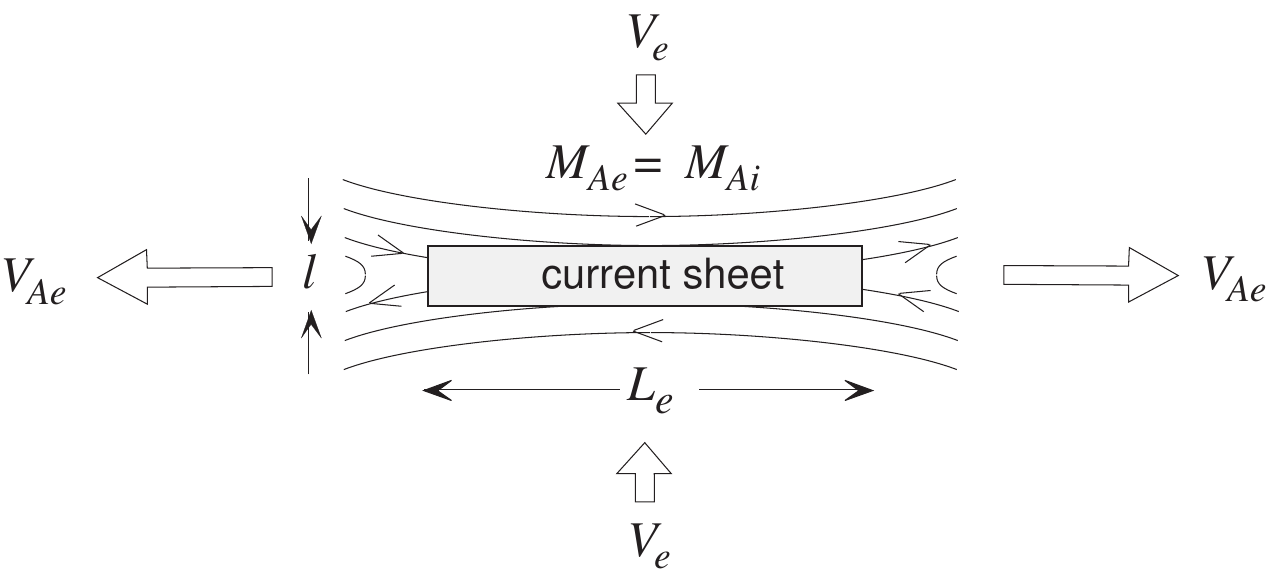}\includegraphics[width=5.5cm]{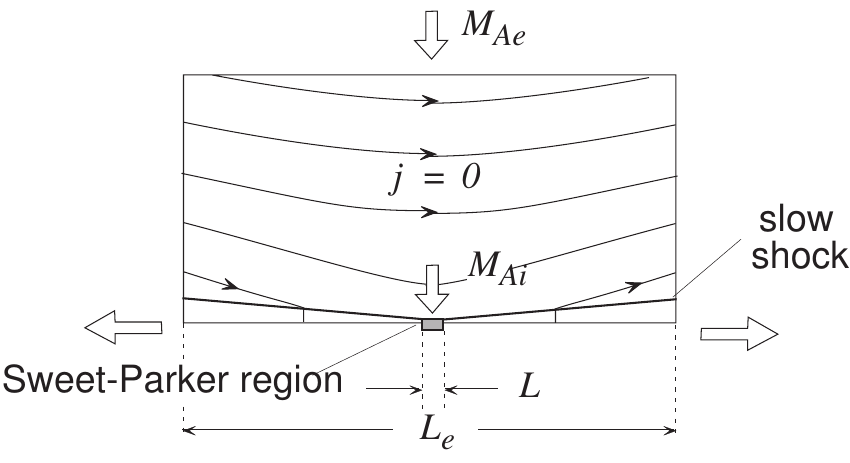}
\caption[The Sweet-Parker and Petschek field
configurations.]{[{\em (left)}] The Sweet-Parker field configuration.
Plasma flows into the upper and lower sides of a current sheet of
length $L_{\rm e}$, but must exit through the narrow tips of the sheet
of width $l$. Because the field is assumed to be uniform in the inflow
region, the external {\em Alfv\'{e}n Mach number}, $M_{\rm Ae}=v_{\rm
  e}/v_{\rm Ae}$, at large distance is the same as the internal
Alfv\'{e}n Mach number, $M_{Ai}$, at the midpoint edge of the current
sheet. [Fig.~I:5.2] [{\em (right)}] Petschek's field configuration.
  Here the length, $L$, of the Sweet-Parker current sheet is much
  shorter than the global scale length, $L_{\rm e}$, and the magnetic
  field in the inflow is nonuniform. Two pairs of standing slow-mode
  shocks extend outwards from the central current sheet.  Petschek's
  model assumes that the current density in the inflow region is zero
  and that there are no external sources of field at large
  distance. [Fig.~I:5.3]\label{forbesfig2}\label{forbesfig3}}
\end{figure}

Let us start with a highly simplified configuration, generally
referred to as \indexit{reconnection!Sweet-Parker}'Sweet-Parker reconnection'  of steady
2-dimensional reconnection in an incompressible plasma in a current
sheet with [system-level] length scale $L_{\rm e}$ \ors[I:5.3.1] ``as shown in
[the left panel of] 
Fig.~\ref{forbesfig2}.  Under these conditions [\ldots] 
the speed of the plasma flowing into the current sheet
is [approximately
\begin{equation}
v_{\rm e}  = \left( {v_{\rm Ae} \eta \over L_{\rm e} }\right )^{1/2}
\end{equation}
where $v_{\rm Ae}=B_{\rm e}/\sqrt{4\pi \rho_{\rm e}}$ is the
Alfv\'{e}n speed in the inflow region. The outflow speed of the plasma
from the current sheet is the local Alfv{\'e}n speed $V_{\rm Ae}$.]
The reconnection rate in two dimensions is measured by the electric
field at the reconnection site.  This electric field is perpendicular
to the plane of Fig.~\ref{forbesfig2}, and it prescribes the rate at
which magnetic flux is transported from one topological domain to
another. In two-dimensional steady-state models this electric field is
uniform in space. Therefore, the Alfv\'{e}n Mach number, $M_{\rm Ae} =
v_{\rm e}/v_{\rm Ae}$, provides a quantitative measure of the
reconnection rate, normalized by the characteristic electric field
$v_{\rm Ae} B_{\rm e}$. [\ldots]

In astrophysical and space plasmas [\ldots] Sweet-Parker reconnection
is usually too slow to account for phenomena such as geomagnetic
substorms or solar flares.  [A later model, known as
\indexit{reconnection!Petschek}'Petschek
reconnection', was developed to ensure much faster reconnection by
encasing the] current sheet in an exterior field with global scale
length $L_{\rm e}$, [and by introducing] two pairs of standing
slow-mode shocks radiating outwards from the tip of the current sheet
as shown in [the righthand panel of] Fig.~\ref{forbesfig3}.  In
Petschek's solution most of the energy conversion comes from these
shocks which accelerate and heat the plasma to form two hot outflow
jets.

Petschek also assumed that the magnetic field in the inflow region was
current free and that there were no sources of field at large
distances. These assumptions, together with the trapezoidal shape of
the inflow region created by the slow shocks, lead to a logarithmic
decrease of the magnetic field as the inflowing plasma approaches the
Sweet-Parker current sheet.  This variation of the field leads in turn
to Petschek's formula for the maximum reconnection rate, namely
\begin{equation}
M_{\rm Ae[Max]}  = \pi / (8\,\ln(L_{\rm e}v_{\rm Ae} / \eta))
\label{forbes_c}
\end{equation}
where [\ldots] $M_{\rm Ae}$ is the Alfv\'{e}n Mach number in the
region far upstream of the current sheet as shown in
Fig.~\ref{forbesfig3}.  [\ldots\ The] Petschek reconnection rate is many orders of
magnitude greater than the Sweet-Parker rate, and for most space and
laboratory applications Petschek's formula predicts that $M_{\rm Ae}
\approx 10^{-1}$ to $10^{-2}$. [\ldots]

It is not always appreciated that Petschek's reconnection model is a
particular solution of the MHD equations which applies only when
[\ldots] the flows into the reconnection region be set up
spontaneously without external forcing [and] that there be no external
source of field in the inflow region.  In other words, the field must
be just the field produced by the currents in the diffusion region and
the slow shocks.  In many applications of interest neither of these
conditions is met.'' \ors[I:5.3.1] ``Even in circumstances where Petschek's model would be expected to
apply it apparently does not. [Numerical simulations suggest that it
only does in case of a nonuniform, localized resistivity. This] does not
contradict Petschek's model because the model makes no explicit
assumption about whether the resistivity is uniform or not.  It is
equally valid for both cases because it assumes only that the region
where resistivity is important is localized.  The numerical
experiments carried out to date imply that the diffusion region can
only be localized by enhancing the resistivity near the $\times$-line.
Whether there might be other ways to localize the diffusion region
({\em e.g.,} a non-uniform viscosity) remains unknown.''

\begin{figure}[t]
\centering
\includegraphics[width=24pc]{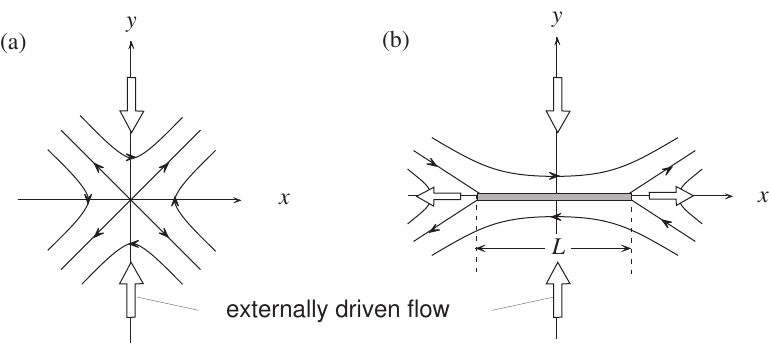}
\caption[Syrovatskii's field configuration.]{Syrovatskii's field
configuration.  Unlike Petschek's configuration, this one has
external sources which produce an $\times$-type configuration even when
local sources of current are absent.  The application of external
driving creates a current sheet whose length, $L$, depends on the
temporal history of the driving and the rate at which reconnection
operates.  The fastest reconnection rate occurs when $L$ is equal
to the external scale length, $L_{\rm e}$. [Fig.~I:5.4]} \label{forbesfig4}
\end{figure}

 \ors[I:5.3.1] ``An alternative\indexit{reconnection!Syrovatskii-Green} approach to
reconnection in current sheets was [developed by considering] what
happens when a weak flow impinges on an $\times$-line in a strongly
magnetized plasma as indicated in Fig.~\ref{forbesfig4}.  The imposed
flow creates a current sheet which achieves a steady-state when the
rate of field line diffusion through the sheet matches the speed of
the flow.  [\ldots]

For a steady-state MHD model the spatial variation of the field in the
inflow region is the key quantity which [sets the]
reconnection rate. [\ldots] For
any such model, the electric field is uniform and perpendicular to the
plane of the field.  Thus, outside the diffusion region $E_{\rm o} = -
v_{y} B_{x}/c$ where $E_{\rm o}$ is a constant, $v_{y}$ is the inflow along
the axis of symmetry ($y$ axis in Fig.~\ref{forbesfig4}), and $B_{x}$
is the corresponding field.  Thus the inflow Alfv\'{e}n Mach number,
$M_{\rm Ae}$, at large distances can be expressed as
\begin{equation}
M_{\rm Ae}  = M_{\rm Ai} \left({B_{\rm i}\over B_{\rm e}}\right)^{2}
\label{forbes_a}
\end{equation}
where $M_{\rm Ai}$ is the Alfv\'{e}n Mach number at the current sheet,
$B_{\rm i}$ is the
magnetic field at the edge of the current sheet, and  $B_{\rm e}$ is the
magnetic
field at large distance.

In Syrovatskii's model the field along the inflow axis of symmetry is
\begin{equation}
B_{x}  = B_{\rm i} (1 + y^{2}/L^{2})^{1/2}
\label{forbes_b}
\end{equation}
where $B_{\rm i}$ is the field at the current sheet, $y$ is the
coordinate along the inflow axis, and $L$ is the length of the
current sheet.  Combining (\ref{forbes_b}) with (\ref{forbes_a}) yields
\begin{equation}
M_{\rm Ae}  = M_{\rm Ai} / (1 + L_{\rm e}^{2}/L^{2})
\end{equation}
which has its maximum value when $L = L_{\rm e}$.  Thus the maximum
reconnection rate in Syrovatskii's model scales [the same as] the Sweet-Parker model.

By comparison, the field in Petschek's model along this axis varies approximately
as
\begin{equation}
B_{x}  = B_{\rm i} {1 - (4/\pi) M_{\rm Ae} \mbox{ } {\ln} (L_{\rm e}/y) \over
1 - (4/\pi) M_{\rm Ae} \, {\ln} (L_{\rm e}/l)}
\end{equation}
where $l$ is the current sheet thickness.  (This expression for
the field is only a rough estimate since the actual variation in
the region $y < L$ is more
complex.) Evaluating this at $y = L_{\rm e}$ and substituting the
result into Eq.~(\ref{forbes_a}) gives
\begin{equation}
M_{\rm Ai}  = M_{\rm Ae} /{[1 - (4/\pi) M_{\rm Ae}\,  {\ln} (L_{\rm e}/l)]}^{2}
\end{equation}
The Sweet-Parker theory can be used to eliminate $L_{\rm e}/l$, so as
to obtain an expression for $M_{\rm Ae}$ [which] has a maximum value
as given by equation (\ref{forbes_c}). [\ldots]''

Even a much longer summary than the above could not be conclusive:
\ors[I:5.5] ``There are many aspects of magnetic reconnection that have
yet to be explored.  Even well long studied topics such as
steady-state two-dimensional reconnection are not fully understood.
Many questions remain about how time-dependent reconnection works in
impulsively driven phenomena such as solar flares and geomagnetic
substorms.  For example, during the impulsive phase of eruptive solar
flares the current sheet where reconnection occurs can grow at a rate
that exceeds the Alfv\'{e}n time scale of the system.  This rapid
growth means that no steady-state reconnection theory applies during
the impulsive phase, and there are virtually no theories that predict
how the reconnection rate scales with plasma resistivity in such a
situation. [Another large challenge to our understanding of
reconnection is that, despite] growing evidence that the reconnection
process in both solar flares and the terrestrial magnetosphere may be
turbulent, there are only few studies that address the issue of
turbulent reconnection.  The occurrence of plasma turbulence in a
highly-structured environment poses a severe challenge to large scale
numerical simulations, so progress in this area may be slow for
sometime to come.''

\clearpage

\chapter{{\bf Torques and tides}}%7
\label{ch:torques}
{\narrower\narrower{
{\bf Chapter topics:}
\begin{itemize}
  \customitemize
\item Angular-momentum loss through magnetized stellar winds 
\item Planetary magnetic fields as a cause of angular-momentum loss
\item Angular-momentum transfer in planet-forming disks
\item Tidal forces, covered-ocean formation on satellites,
  powering of dynamos
\end{itemize}

\noindent{\bf Key concepts:}
\begin{itemize}
  \customitemize
\item Gravitational tides and spin-orbit synchronization
\item (Exo-)planetary atmospheric tides by solar/stellar irradiation
\item Magnetic torques in stellar winds and
  accretion disks
\item Alfv{\'e}n surface
\end{itemize}

}}

\section{Introduction}
Rotation and revolution are key properties for stars, planets, and
indeed the entirety of planetary systems. For example, rotation is one
of the essential ingredients of dynamo action in stars and planets,
while climates are determined to a large extent by the Coriolis forces
associated with planetary spin in combination with the overall
duration of insolation on a planet's dayside, by the orbital
eccentricity, and by the planetary
\indexit{definition!obliquity}obliquity,
{\em i.e.,}  the tilt of the
planetary spin axis relative to the orbital plane. But spin rates
evolve: stars slow their rotation because of their magnetic activity,
while planetary rotation can change subject to tidal coupling with
their moons, with their own atmospheres, and with their central
stars. The latter, tidal synchronization of planetary spin and orbital
motion, is likely common among exoplanets in the habitable zones of
intrinsically faint M-type dwarf star because these exoplanets would
have to have very tight orbits.

Also revolution is subject to change: planetary orbits need to be
stable over long times to offer long-term habitability, but --~as we
discuss in Sects.~\ref{sec:orbitinteraction} and~\ref{sec:exo}~-- orbits
need to evolve for juvenile planets to grow efficiently and also, for
instance, to transport water through a planetary system across the ice
line from the cold outer reaches to the habitable inner
domain. \regfootnote{The temperature at which H$_2$O freezes into a
  solid in a proto-planetary disk is dependent on the partial pressure
  of the water vapor in the overall gas mixture, and is typically
  expected to lie in the range of 145\,K to 170\,K. \label{note:iceline}} And if orbital
dynamics (quantified in angular momentum) could not be efficiently
transported through gas and dust, planetary systems and their central
stars could not form as they are observed to do (see
Sects~\ref{sec:magrot}, \ref{sec:diskwind}, and~\ref{sec:star}).
%Ice line: https://arxiv.org/pdf/1207.4284.pdf

\ors[III:1.3] ``Transport of angular momentum \indexit{angular
  momentum!transport}through the coupling of distant concentrations of
mass occurs either through gravitational tides, by \indexit{star-disk
  coupling}magnetic stresses, or by flows. Gravitational coupling has
obviously played an important part in the spin-orbit synchronization
of the Earth's single moon. This coupling continues to be important as
a stabilizer for the direction of the Earth's spin axis, even as it
causes the precession of that axis with associated climatic effects
(Chs.~III:11 and ~III:12).  [This chapter reviews these processes,
and more:] Tidal forces also act significantly on Jupiter's moon
Europa and Saturn's moon Enceladus, in which it appears to result in
liquid water in their interiors, which makes these moons interesting
objects to study from an exo-biological perspective. Tidal spin-orbit
coupling also leads to the formation of short-period, highly active
binary stars (like the so-called RS\,CVn type systems).  [\ldots]

[This chapter also highlights a]ngular momentum transport via the
magnetic field [which] is important in the coupling of proto-stars and
young \indexit{T\,Tauri star}T\,Tauri stars to their surrounding disks and magnetized stellar
winds.  \activity{{\em Look up} what T\,Tauri stars are, and what
  differentiates the 'classical' T\,Tauri star from the 'weak-line'
  variant. List a few primary properties of T\,Tauri
  stars. \mylabel{act:ttauri}} After the early formation phases of a
planetary system, the loss of stellar angular momentum continues
through a stellar wind, leading to magnetic braking of the stellar
rotation and the concomitant gradual decrease in stellar activity with
age. In tidally interacting binaries with one or more magnetically
active components, the loss of spin angular momentum by a stellar wind
drains the orbital angular momentum reservoir, eventually leading to
the merger of the component stars, leaving an old but rapidly-spinning
single star (like FK\,Comae). [The consequences of these couplings are
discussed in Chs.~\ref{ch:evolvingstars} and~\ref{ch:formation}.]

Angular momentum transport by flows inside astrophysical bodies is the
cause of the near-rigid rotation with latitude and depth of the
solar interior. But the models of the full convective envelopes of
stars and giant planets need to advance significantly before we can
use their results in, {\em e.g.,} magnetohydrodynamic dynamo models in which
the non-rigid rotation and other large-scale circulations appear to be
crucial.''

Even photons are involved in a form of tidal action: \ors[III:15.2]
``Atmospheric tides\indexit{atmosphere!tide} are the response to
periodic astronomical forcing. Atmospheric tides [on Earth] are forced primarily
by the thermal heating due to the absorption of solar radiation by
ozone and water vapor. These tides have periods which are the length
of a mean solar day and its harmonics. [\ldots]''

This chapter briefly introduces each of these processes and the
settings in which they are important, but the consequences of
evolving orbital motions and spin rates are left for later chapters.

\section{Magnetic torques}
\subsection{Stellar winds and magnetic braking}\label{sec:braking}

The \indexit{magnetic!torque}solar \indexit{magnetic!braking}wind
discussed in Ch.~\ref{ch:flows} not \indexit{torque|see{gravitational,
    magnetic}}only
  carries \indexit{solar!wind!magnetic braking}mass away from the
Sun, but also angular momentum. \ors[IV:4.2] ``The conventional
mechanism for stellar spin down is that stars lose angular momentum to
the magnetized stellar wind in the concept called 'magnetic
braking'. In this process, the mass flux carried by the accelerating
stellar wind conserves angular momentum as long as the wind speed is
below the Alfv\'en speed, $v_{\rm A}=B/\sqrt{4\pi\rho}$ (in cgs units
of cm s$^{-1}$), where $B$ is the local magnetic field strength, and
$\rho$ is the local mass density. Once the wind speed equals the
Alfv\'en speed [(at the 'Alfv\'en radius'; see
Sect.~\ref{sec:parker-spiral}) the wind is effectively decoupled] from
the star. Another way to look at this process is to think of the
magnetic field lines as rods [up to the Alfv{\'e}n radius, beyond
which the wind flows out essentially radially as if flung free from
the star only at that radius (which in reality is a gradual process),
dragging the magnetic field into a Parker spiral.]  As a result, each
field line applies a torque on the star and spins it down. This torque
is proportional to the momentum of the wind at the Alfv\'en point, to
the stellar rotation rate, and to the distance of the Alfv\'en point
(the lever arm that applies the torque). The imaginary surface that
contains all the Alfv\'en points is called the 'Alfv\'en surface' and
the integral of the mass flux through this surface is the mass loss
rate, $\dot{M}$, of the star to the stellar wind. For a spherically
symmetric wind, and [a magnetic field that is close to uniformly
distributed across the Alfv{\'e}n surface,] we can calculate the total
torque on the star and the total angular momentum loss rate,
$\dot{J}$:
\begin{equation}
\label{eq:4_02}
\dot{J}=-\Omega_\ast \dot{I}_{\rm shell} = -\frac{2}{3}\Omega_\ast\dot{M}r^2_{\rm A},
\end{equation}
where $\Omega_\ast$ is the stellar rotation rate,
[$\dot{I}_{\rm shell}$ is the moment of inertia \indexit{moment of
  inertia}of a uniform shell of mass $\dot{M}$ and radius $r_{\rm A}$,
and] $r_{\rm A}$ is the average distance to the Alfv\'en surface, and
we assume constant moment of inertia [for the star ({\em i.e.,} we
assume the time scale for angular momentum loss in this expression is
short relative to the evolution of the internal structure of the star,
which is appropriate for the long-lived 'mature' phase of the star,
see Ch.~\ref{ch:evolvingstars}). Note that Eq. ~(\ref{eq:4_02}) shows
that the near co-rotation out to $r^2_{\rm A}$ causes the solar wind
to carry a factor of $r^2_{\rm A}/r^2_\odot $ more angular momentum
away from the star than is contained in the mass that is actually
leaving the stellar surface.]

From Eq.~(\ref{eq:4_02}) we see that the mass-loss rate is necessary
to estimate the spin-down rate of a star. However, stellar winds of
cool, Sun-like stars are very weak and cannot be directly observed
(see Ch.~\ref{ch:evolvingstars}), which makes it challenging to
estimate $\dot{J}$ as a necessary input for stellar evolution models
[\ldots] Based on [measurements supported by modeling (described in
Sect.~\ref{sec:stellarwinds}) mass-loss rates in Sun-like stars] fall in the
range between $10^{-15}-10^{-11}\;M_\odot\;{\rm yr}^{-1}$ (the present-day
solar mass-loss rate is $(2-3)\times
10^{-14}\;M_\odot\;{\rm yr}^{-1}$). However, stars can also lose mass via
CMEs. In the case of the Sun, each CME carries some
$10^{13}-10^{17}$\,g into space, with an annual integrated mass-loss
via CMEs of several percents of the ambient mass-loss. Therefore, CMEs
on the Sun play very little role in the solar mass-loss. This role can
become significant if the CME rate is higher by a factor of 10 or
more. In this case, CMEs can even dominate the stellar mass-loss.''
But we know very little of CMEs of stars other than the Sun, or even
of the Sun in its distant past, so this area is left for future
exploration.

Let us make a few comparisons of energy budgets and time scales, using
rough approximations only: The above-mentioned solar mass-loss rate
can be combined with numbers in Table~\ref{tab:wind-stats} to estimate
the power needed to drive the flow of the solar wind (bulk kinetic
energy), the Alfv{\'e}n radius, and with that the rate at which
rotational energy is drained from the Sun. Assuming for this estimate
a constant mean wind velocity, a temperature of 1.5\,MK for the high
corona, an isotropic heliospheric magnetic field strength
(approximating the field as radial), a total characteristic power
associated with all forms of coronal radiative losses driven by solar
magnetic activity of order $\approx 10^5$\,erg/cm$^2$/s (averaged over
a solar cycle), and a moment of inertia of
$I_\odot \approx 7\times 10^{53}$\,g\,cm$^2$ (see
Ch.~\ref{ch:evolvingstars}), one can conclude that (1) the solar wind
kinetic energy flux is several times smaller than the coronal
radiative losses, (2) the characteristic Alfv{\'e}n radius is roughly
15 solar radii, and (3) the time scale for magnetic braking, {\em
  i.e.,} the ratio of angular momentum to loss rate of angular
momentum for the present-day Sun is of order
10\,Gyr.\sactivity{$\circledS$ {\em Show:} Verify the numbers in the
  conclusions about stellar magnetic braking for the present-day Sun
  at the end of Sect.~\ref{sec:braking}. \mylabel{act:verbrake}
  \solution{verbrake}}

In very rapidly spinning stars, the centrifugal forces that we have
ignored for the solar wind, also need to be taken into account. An
example where these dominate the process in the case of a cold wind is
discussed in Sect.~\ref{sec:diskwind}.

\subsection{Planetary magnetospheric torque}

We can make a similar \indexit{torque}comparison of \indexit{magnetosphere!torque}energy budgets and time scales for
the solar wind that delivers power to Earth's magnetosphere and
induces a torque on Earth's rotation. First, let us look at the
energy, then at the angular momentum. \ors[II:10.4.1] ``The net
rate of energy extraction (power) $\mathcal{P}_{\rm sw}$ from the solar wind
flow is equal to the difference of the solar wind kinetic energy flux
across two surfaces $A$ perpendicular to the Sun-planet line, surface 1
ahead of the bow shock and surface 2 far downstream of the entire
interaction,
\begin{eqnarray}
\mathcal{P}_{\rm sw}
 & = & \frac{1}{2} \int_1  \rho v^3 {\rm d}A\,- \frac{1}{2} \int_2  \rho v^3 {\rm d}A\,\nonumber\\
 & = & \frac{1}{2} \int  \rho v ( \overline{v}_1^2 -\overline{v}_2^2 ) {\rm d}A\,\nonumber\\
 & \approx & \dot{M}_{\rm ft}\; \overline{v} \Delta v
\label{eq:PdV}
\end{eqnarray}
(subscripts 'sw' on $\rho$ and $v$ have been omitted, for simplicity),
and the total force $F$ is similarly equal to the difference of the linear
momentum flux,
\begin{equation}
F_{\rm sw} = \int_1  \rho v^2 {\rm d}A\,- \int_2 \rho v^2 {\rm d}A\,
\approx \dot{M}_{\rm ft} \Delta v,
\label{eq:FdV}
\end{equation}
where $\Delta v \equiv \overline{v}_1 - \overline{v}_2$ and
$\overline{v} \equiv (\overline{v}_1 + \overline{v}_2)/2 $ (bars
indicate suitable averages) and
\begin{equation}
\dot{M}_{\rm ft} = \int_1 \rho v {\rm d}A\, \simeq \int_2  \rho v {\rm d}A\,
\label{eq:defS}
\end{equation}
is the amount of mass per unit time flowing through the region of
interaction between the solar wind and the magnetosphere, to be
distinguished from $\dot{M}_{\rm sw}$, the mass input rate from the
solar wind into the magnetosphere.\regfootnote{Note that the final
  expressions in Eqs.~(\ref{eq:PdV}) and~(\ref{eq:FdV} are
  approximations that assume that $v_1$ and $v_2$ are rather uniform
  across $A$. Whereas the upstream solar wind moving through 'surface
  1' is generally quite uniform, this approximation is less tenable
  for the downstream flow through 'surface 2' which is placed in the
  geotail far downstream of Earth. This assumption thus limits the
  practical use of these expressions, although upper limits for
  $\mathcal{P}_{\rm sw}$ and $F_{\rm sw}$ are readily obtained by
  ignoring the downstream terms. Note also that magnetic and thermal
  contributions to solar wind energy and momentum flux in
  Eqs.~(\ref{eq:PdV}) and~(\ref{eq:FdV}) have been neglected as small
  in comparison to those of the bulk flow; see
  Sect.~\ref{sec:solwindenergy}.}  Combining Eqs. (\ref{eq:PdV}) and
(\ref{eq:FdV}) yields a relation between the power and the force (in
the direction of solar wind flow), \activity{{\em Show:} For
  comparison: what is the approximate ratio of forces exerted on the
  Earth of the total solar irradiance onto the Earth's surface to the
  solar-wind pressure on the magnetopause? In your estimate, ignore
  reflection by and radiation from Earth's atmosphere, use an
  approximate Chapman-Ferraro radius for the cross section of the
  magnetosphere interacting with the solar wind, and maximize the
  solar wind pressure by ignoring the term involving $v_2$ in
  Eq.~(\ref{eq:FdV}). This exercise shows why solar sails are designed
  for photon pressure rather than solar-wind dynamic pressure (note
  that some are designed to couple to induced electromagnetic effects,
  not dynamic pressure). \mylabel{act:radwindpressure}}
\begin{equation}
\mathcal{P}_{\rm sw} = F_{\rm sw} \overline{v},
\label{eq:PF}
\end{equation}
which [has been used] to estimate the energy input into the
terrestrial magnetosphere, under the assumption that the relevant
force $F_{\rm sw}$ is the tangential (magnetotail) force acting
primarily on the nightside, $F_{\rm MT}$ (see
Sect.~I:10.3.2).''

\ors[I:10.3.2] ``While pressure from the external medium thus accounts
for the formation and shape of the magnetosphere on the dayside of the
planet, it cannot by itself explain the formation of the
\indexit{magnetotail}{\em magnetotail} on the night side. This
structure, shown also in Figure \ref{fig:MS}, is a region of magnetic
field lines pulled out into an elongated tail in the
anti-sunward\indexit{magnetotail!current sheet} direction, with the
magnetic field reversing direction between the two sides of
\indexit{magnetotail!plasma sheet} a {\em current sheet} or {\em 
  plasma sheet} in the equatorial region. To form this structure one
needs an appropriate stress: a tension force pulling away from the
planet. If we choose a closed volume bounded by a surface just outside
the magnetopause plus a cross-section of the magnetotail (vertical cut
at the right edge of Figure \ref{fig:MS}) and evaluate the force
[\ldots], the total tension force ${F}_{\rm MT}$ is given by
the integral over the cross-section and the total pressure
\indexit{magnetosphere!magnetotail force}force ${F}_{\rm MP}$
by the integral over the \indexit{magnetosphere!magnetopause
  force}magnetopause:
\begin{equation}
{F}_{\rm MT} \simeq \left({B_{\rm t}}^2/8\pi\right) A_{\rm t} \qquad \quad
{F}_{\rm MP} \simeq \rho_{\rm sw} {v_{\rm sw}}^2 A_{\rm t}
\label{eq:Ftot}
\end{equation}
where $B_{\rm t}$ is the mean magnetic field strength and $A_{\rm t}$ the
cross-sectional area of the magnetotail (typically, $A_{\rm t}$ exceeds $\pi
{R_{\rm MP}}^2$ by a factor 3 to 4). Both ${F}_{\rm MP}$ and
${F}_{\rm MT}$ are directed away from the Sun and are exerted
ultimately on the planet.'' \ors[II:10.4.1] ``Note: if ${F}$ [in
Eq.~(\ref{eq:PF})] is equated to the pressure force ${F}_{\rm
  MP}$ on the entire magnetopause, it can be shown that the associated
$\mathcal{P}$ does not go into the magnetosphere but represents the
power expended in irreversible heating at the bow shock.

Calculating\indexit{energy!from planetary rotation} the power extracted from planetary rotation is somewhat
simpler.  The angular momentum of the rotating planet is
\mbox{$I_{\rm p}{\Omega_{\rm p}}$} and the kinetic energy of rotation is
\mbox{$\scriptstyle{\frac{1}{2}}\displaystyle
I_{\rm p}{\Omega_{\rm p}}^2$}, where $I_{\rm p}$ is the moment of
inertia and ${\Omega_{\rm p}}$ the angular frequency of rotation [of
the planet].  With
$\mathcal{T}$ the torque on the planet (component along the rotation
axis),
\begin{equation}
\mathcal{P}_{\rm rot} = \frac{{\rm d}}{{\rm d}t}\left(\scriptstyle{\frac{1}{2}}\displaystyle
I_{\rm p}{\Omega_{\rm p}}^2\right) =
{\Omega_{\rm p}}\,\frac{{\rm d}}{{\rm d}t}\left(I_{\rm p}{\Omega_{\rm p}}\right)=\mathcal{T}{\Omega_{\rm p}}\
,
\label{eq:Ptr}
\end{equation}
a relation between the power and the torque, completely analogous to
Eq.~(\ref{eq:PF}). [\ldots]

What happens to the linear momentum extracted from the solar wind flow
is well understood: it is transferred to and exerts an added force on
the massive planet. The angular momentum extracted from the rotation
of the planet, on the other hand, can only be removed to 'infinity,'
and identifying the mechanism by which it is transported away is
indispensable for understanding the interaction. There are several
possibilities:

(a) In magnetospheres with a significant interior source $\dot{M}$ of plasma
(from moons or planetary rings), angular momentum can be advected by
the outward transport of mass [as long as the planet's rotation
period is below the orbital period of the plasma source]. For the
simple example of plasma corotating rigidly out to a distance $R_{\rm
  c}$ and coasting freely beyond $R_{\rm c}$, angular momentum is
transported outward at the rate \mbox{$\dot{M}{R_{\rm c}}^2{\Omega_{\rm p}}$},
hence from Eq.~(\ref{eq:Ptr}) the extracted power is
\begin{equation}
\mathcal{P}_{\rm rot}\simeq \dot{M} \,{\Omega_{\rm p}}^2 \,{R_{\rm c}}^2 \ ,
\label{eq:PJ}
\end{equation}
one half of which goes into the kinetic energy of bulk flow of the
outflowing plasma (in this model), and the remainder is available for
powering other magnetospheric processes (proposed for the
magnetosphere of Jupiter). 

(b) If the solar wind exerts a tangential force on the magnetosphere,
it will also exert a torque whenever the distribution of the force is
not symmetric about the plane containing the solar wind velocity and
the planetary rotation axis. The torque may be estimated as
\mbox{$\mathcal{T}\sim R_{\rm MP}\Delta{F}$}, where $R_{\rm MP}$ is
the distance to the dayside magnetopause and $\Delta{ F}$ is the
difference between the force on the dawn and on the dusk side; this
gives the ratio of power from rotation to power from solar wind flow
as
\begin{equation}
\mathcal{P}_{\rm rot}/\mathcal{P}_{\rm sw} \sim \left(\Delta{F}/{
F}\right)\left(\Omega_{\rm p} R_{\rm MP}/v_{\rm sw}\right) \ .
\label{eq:ratpow}
\end{equation}
In a slowly rotating magnetosphere such as Earth, \mbox{$\Omega_{\rm p}
R_{\rm MP}/v_{\rm sw}\equiv\epsilon<<1$} and one also expects
\mbox{$\Delta{F}/{F}$} to scale as $\sim\epsilon$;
hence the power extracted from rotation by the solar-wind torque is
negligible.'' \ors[II:10.4.1] ``In principle, ${\Omega_{\rm p}}$ decreases with time as
the result of the torque, but in practice the rate of decrease is
completely negligible.  The time for the present magnetospheric torque
to reduce appreciably the planet's rate of rotation is several orders
of magnitude longer than the [age of the Universe], both at Jupiter and at
Earth; for the latter, this implies that the magnetospheric torque is
much smaller than the lunar tidal torque.''

\ors[II:10.4.1] ``(c) In a rapidly rotating open magnetosphere, on the
other hand, magnetic field lines that extend from the planet into the
solar wind may become twisted (by a process analogous to the formation
of the Parker spiral in the solar wind), creating a Maxwell stress
that transports angular momentum outward into the solar wind.  This
mechanism of extracting energy from planetary rotation was proposed
for Jupiter (where it is now considered not important in comparison to
mass outflow) and for Uranus.

(d) If the magnetic moment of the planet is tilted relative to the
rotation axis, electromagnetic waves that carry away angular momentum
may be generated by the rotation. This is generally believed to be the
primary mechanism for energy loss from pulsars but is negligible for
systems that are very small in comparison to $c/{\Omega}$, the
radius of the speed-of-light cylinder (which is the case for all
planets in our Solar System and their magnetospheres).''

\subsection{Magneto-rotational coupling}\label{sec:magrot}

As we saw in the case of the stellar wind, magnetic fields can support
tension (Sect.~\ref{sec:induction}) and thereby can essentially
enforce co-rotation of gases at different distances from a star, at
least out to where the field is strong \indexit{magneto-rotational!coupling}compared to the
inertial forces associated with the plasma. This not only holds for
outflows such as stellar winds, but also in systems where matter is
'descending' onto the star, such as in very young proto-planetary
systems where material has shaped itself into a disk spinning around
an accreting star. More on that process in Ch.~\ref{ch:formation}, but
let us look at what a magnetic field that threads such a disk can do:
a field can be an effective agent in transporting angular momentum outwards,
thereby enabling the gas in the disk to spiral inwards and thus help
form the star.

\begin{figure}[t]
\centering
\includegraphics[width=0.8\textwidth,bb=0 490 615 720]{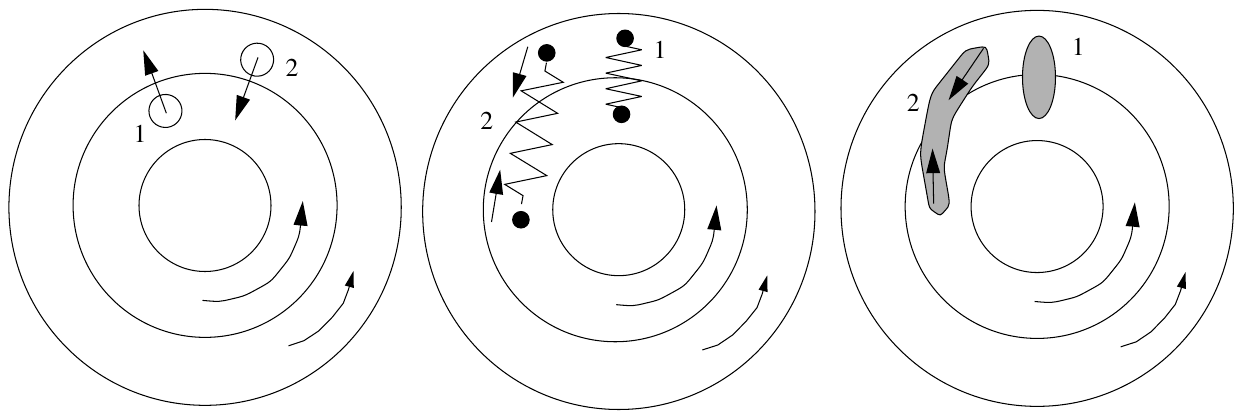}
\caption[Angular momentum transfer in a shearing disk.]{Schematic
  treatment of angular momentum transfer in a shearing disk with
  angular velocity decreasing outwards [as the orbiting material
  approximates Keplerian orbits].  An initially radial field is
  perturbed azimuthally ({\rm left panel}); these azimuthal perturbations
  grow due to the shear in the disk ({\rm middle panel} [going from
  time label '1' to time label '2']).  In the case of gravitational
  instability ({\rm right panel}), an excess of material gets sheared
  out by the differential rotation; the gravitational attraction on
  the sheared excess (spiral arm pattern) exerts a restoring force in
  the same sense as the magnetic case, again transferring angular
  momentum outward. [Fig.~III:3.4;
  \href{https://ui.adsabs.harvard.edu/abs/2009apsf.book.....H/abstract}{source:
  \citet{2009apsf.book.....H}}.]}
\label{figlh:disktorques}
\end{figure}

\ors[III:3.2] ``As shown in the middle panel of
Figure~\ref{figlh:disktorques}, if the magnetic field lines are
thought of schematically as springs tying adjacent disk annuli
together, then as differential rotation continually separates the
regions 'tied' to the field ({\em e.g.,} evolution from ['1' to '2']), the
'springs' or field lines become stretched [and bent], and the
resultant [tension] forces will work in the direction of spinning up
the outer annulus while spinning down the inner annulus.

\nocite{Hartmann2009}

The magnetic fields shown in the top-down view of the middle panel of
Figure~\ref{figlh:disktorques} cannot be stretched  indefinitely; at
some point there will be reconnection and diffusion as the flow
becomes turbulent.  [In that case, an] initially vertical field is
perturbed radially; these radial perturbations grow due to the shear
in the disk; and eventually the field lines become so stretched that
they pinch off and develop into full turbulence.

Although\indexit{magneto-rotational!instability|seealso{definition}}
there\indexit{definition!magneto-rotational!instability} is currently
some controversy over the efficiency of this 'magneto-rotational
instability', or MRI, it seems very likely that it provides a
sufficiently effective means of promoting accretion in astrophysical
disks -- provided, of course, that the magnetic field can couple
effectively to the gas; there must be a sufficient population of ions
and electrons to collide rapidly enough with neutral gas to make the
MRI work.  Protostellar\indexit{protostar!disk} disks are problematic
in this regard: with much or most of their mass heavily shielded from
ionizing radiation, and possessing temperatures far too low to
effectively ionize even low-ionization potential metals like Na and K,
it seems highly unlikely that the MRI can account for (at least
low-mass) star formation on its own.'' More on this in Ch.~\ref{ch:formation}.

\begin{figure}[t]
\centering
\includegraphics[width=0.6\textwidth]{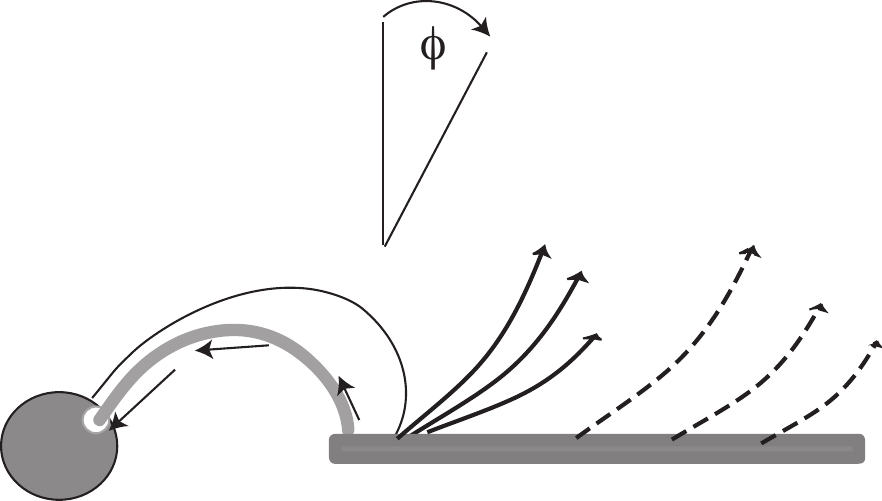}
\caption[Connected accretion disk,
stellar wind, and stellar magnetosphere.]{Schematic structure for a
  connected system of accretion disk, stellar wind, and stellar
  magnetosphere.  Magnetic fields which[, owing to a finite magnetic
  diffusivity,] penetrate the disk inside the co-rotation radius
  (where the angular velocity of the rotating disk matches the angular
  velocity of the star) allow material to accrete (gray curve);
  fields penetrating the disk outside of corotation help provide a
  spindown torque (solid dark curve).  In the $\times$-wind model, the wind
  arises from the disk just at corotation (curved solid arrows), while disk wind
  models involve mass loss from a wider range of disk radii (dashed
  arrows).  Magnetic field lines pitched at angles $\Phi_{\rm e} >
  30^{\circ}$ allow for rapid, cold mass loss (see
  text). [Fig.~III:3.7]}
\label{figlh:magnetopluswind}
\end{figure}
\subsection{Disk winds}\label{sec:diskwind}

An \indexit{disk wind}alternative to \indexit{wind|see{disk, solar, stellar}} transporting angular momentum
outward through \indexit{accretion disk!angular momentum transfer}an
accretion disk going against, and thereby enabling, matter to spiral
inward is to remove angular momentum by a variant of a stellar wind,
namely one that is cold and propelled by centrifugal
forces. \ors[III:3.3] ``The basic version of the cold,
magnetically-driven wind takes advantage of the rapid disk rotation to
fling material outward (and later collimate it). Near the disk it is
assumed that the magnetic pressure is much larger than the gas
pressure.  In this limit, the magnetic fields are stiff at the
launching region, {\em i.e.,} corotation of the inner wind is assured.
In this case the energy (Bernoulli) constant of the motion [for a unit
of mass] becomes
\begin{equation}
E~=~ {v_\phi^2 \over 2} ~+~ c_{\rm s}^2\, \ln \, \rho
~-~ {\frac{1}{2} \oo^2 r^2 } ~-~ {GM_* \over (r^2 ~+~ z^2)^{1/2}}
~=~ {v_\phi^2 \over 2} ~+~ c_{\rm s}^2 \, \ln \, \rho ~-~ \Phi_{\rm e}\,, \label{eqlh:bernst}
\end{equation}
where $v_\phi$ is the poloidal velocity, $\oo$ is the (Keplerian) angular
velocity of the disk in which the magnetic field is rooted, $c_{\rm s}$ is
the (assumed isothermal) sound speed, and $\Phi_{\rm e}$ is an effective
potential term including the effects of rotation and magnetic fields;
[the terms in the central expression measure kinetic energy (first and
third), change in internal energy in an isothermal process (second),
and gravitational potential energy (fourth) at a distance $r$ from
the rotation axis of the disk, 
and height $z$ above that disk].
The behavior of the flow depends upon the form of $\Phi_{\rm e}$,
which in turn depends upon the geometry of the flow.

In the case of a perfectly vertical field, perpendicular to the disk, any material
which flows outward must be propelled initially by gas pressure; the Keplerian rotation
is of course insufficient by itself to drive outflow.  The atmospheric structure is nearly
hydrostatic until one reaches a radial distance such that
\begin{equation}
c_{\rm s}^2 \sim {GM_* \over (r^2 ~+~ z^2)^{1/2}}\,,
\label{eqlh:critpoint}
\end{equation}
in analogy with a Parker thermal wind [(see
Sect.~\ref{sec:nearlystrat} around Eq.~\ref{eq:criticalp})].  When
the gas is cold, the flow 'starts' only at large radii; the flow
interior to this must pass through many scale heights of density,
resulting in negligible outflow.

In contrast, a field line tipped away from the rotation
axis can effectively drive a cold flow, taking advantage of the
$\frac{1}{2} \oo^2 r^2$ term in Eq.~(\ref{eqlh:bernst}).
Neglecting thermal pressure,
\begin{equation}
E ~=~ {1 \over 2} v_\phi^2 ~-~ \Phi_{\rm e}\,, \label{eqlh:bernsimp}
\end{equation}
where the 'effective' potential is
\begin{equation}
\Phi_{\rm e} ~=~ - {GM_* \over r_{\rm o}}
\left [ {1 \over 2}\, {r^2 \over r_{\rm o}^2} ~+~ {r_{\rm o} \over (r^2 ~+~ z^2)^{1/2}} \right ]\,.
\end{equation}
Consider now a small displacement along the field line, with a coordinate
given by $s$, and
\begin{equation}
{\rm d}s^2 ~=~ {\rm d}r^2~+~{\rm d}z^2\,.
\end{equation}
At the base of the flow, the disk material
is rotating at the local Keplerian velocity.  This is an equilibrium state,
because ${\rm d}\Phi_{\rm e} /{\rm d}s =0$ at $z = 0$.  However, if $\sd^2 \Phi_{\rm e}/\sd s^2 < 0$,
this equilibrium is {\em unstable}; any small perturbation along
the field line will result in an increased (outward) poloidal velocity
from Eq.~(\ref{eqlh:bernsimp}).
If $\theta$ is the angle between the field line
and the disk plane, the critical stability criterion
\begin{equation}
{\sd^2 \Phi_{\rm e} \over \sd s^2} ~=~ 0 \,\,\,\,\,\, (r~=~r_{\rm o}\,, z~=~0)
\end{equation}
%\begin{equation}
%{\sd^2 \Phi_{\rm e} \over \sd s^2} ~=~ 0 \,\,\,\,\,\, (r~=~r_{\rm o}\,, z~=~0)
%\end{equation}
requires $\tan^2 \theta_c = 3$, or $\theta_c = 60^{\circ}$.
Disk
magnetic field lines which are tipped away from the rotation axis
by an angle greater than $30^{\circ}$ result in an unstable equilibrium,
and rapid outflow will commence at the disk.'' This flow carries
angular momentum away from the disk.

\begin{figure}[t]
\centering
%\hbox{\begin{minipage}[t]{5.5cm}\vspace{0.4cm}\psfig{figure=figures/jbeer_2_1-3.eps,width=5.5cm,clip=}\end{minipage}\begin{minipage}[t]{7.5cm}\vspace{0pt}\psfig{figure=figures/jbeer_2_1-4.eps,width=7.5cm,clip=}\end{minipage}}
\hbox{\begin{minipage}[t]{5.5cm}\vspace{0.4cm}\centering
\includegraphics[width=5.5cm]{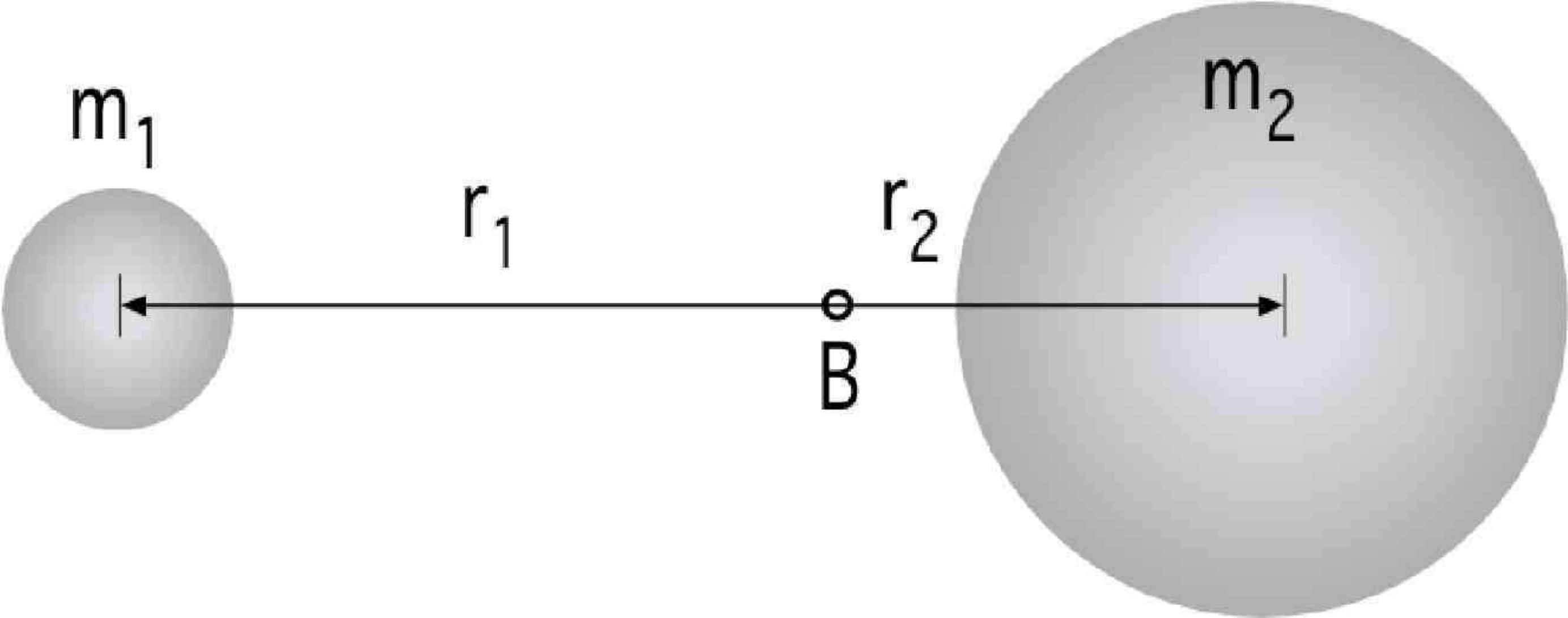}\end{minipage}\begin{minipage}[t]{7.5cm}\vspace{0pt}\includegraphics[width=7.5cm]{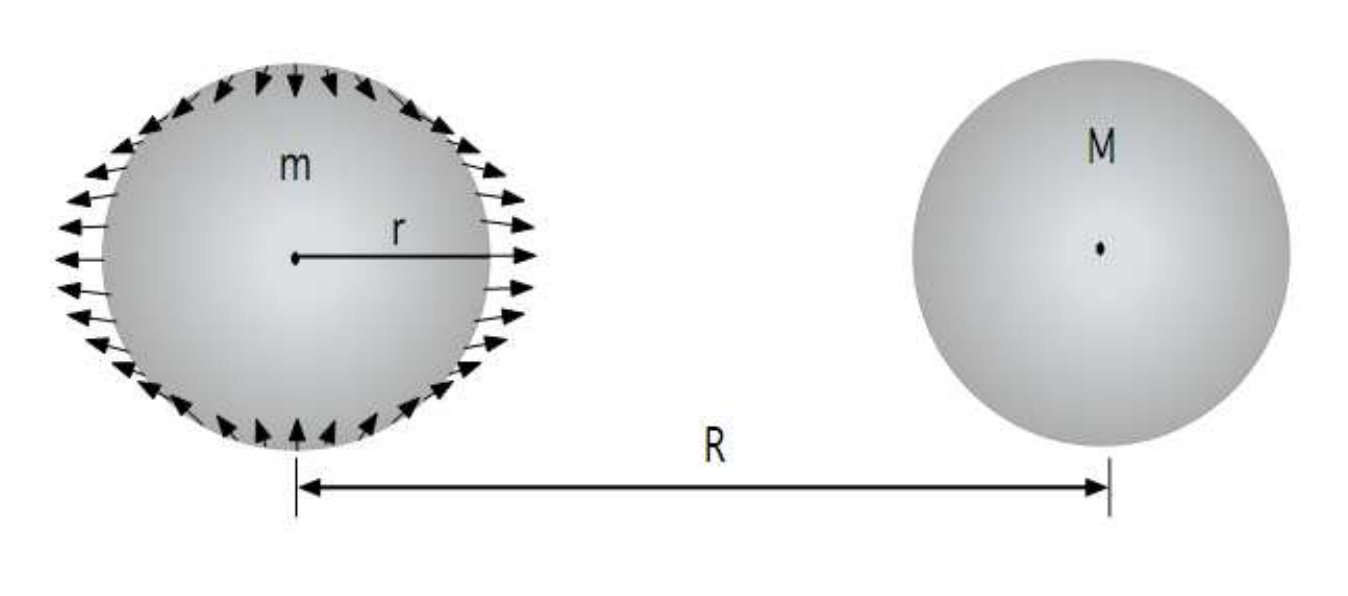}\end{minipage}}
%\hbox{\psfig{figure=figures/jbeer_2_1-3.eps,width=5.5cm,clip=}\psfig{figure=figures/jbeer_2_1-4.eps,width=7.5cm,clip=}}
%\hbox{\includegraphics[width=5.5cm]{figures/jbeer_2_1-3.eps}\includegraphics[width=7.5cm]{figures/jbeer_2_1-4.eps}}
\caption[Two bodies orbiting their barycenter,
and their tidal acceleration.]{{\em Left:} Two bodies orbiting around
  the barycenter {\em B}.
{\em Right:} Tidal acceleration induced by the body with mass $M$ on
the body with
mass $m$ with the distance $R$ between their centers. [Fig.~III:11.3]
\label{fig:2.1-3}\label{fig:2.1-4}}
\end{figure}
\section{Gravitational tides}
\subsection{Spin-orbit interactions}

\ors[III:11.2.1] ``A well \indexit{gravitational!tide|see{tide}}known
\indexit{tide!gravity}gravitational \indexit{spin-orbit coupling}influence is the tidal force of
Moon and Sun on Earth. [Similar tides occur in other planet-moon
systems throughout the Solar System --~and has led to spin-orbit
synchronization for most of the major moons~-- and also in binary
stars and in star-planet pairs with relatively tight orbits --~more on
that below.] To calculate the tidal acceleration, let us consider two
masses $M$ and $m$ with the distance $R$ between their centers as
shown in Figure~\ref{fig:2.1-4}.  According to Newton's law of
gravitation, the mass $m$ feels the gravitational acceleration $a$:
\begin{equation}
g=-G\,\frac{M}{R^2}
\label{eq:2.1-15}
\end{equation}
However, each point of a body with mass $m$ and radius $r$ feels a
different gravitational acceleration depending on the effective
distance to mass $M$ which ranges from $R-r$ to $R+r$. For the two
extreme cases we find:
\begin{equation}
g =-G\,\frac{M}{(R\pm r)^2}=-G\,\frac{M}{R^2(1\pm r/R)^2}.
\label{eq:2.1-17}
\end{equation}
[In cases for which]
$r$ is much smaller than $R$ this equation can be expanded into a Taylor
series:
\begin{eqnarray}
\frac{1}{(1+x)^2}&=&1-2x+3x^2-\dots ,
\label{eq:2.1-18}\\
\label{eq:2.1-19}
g&=&-G\,\frac{M}{R^2}\pm G\,\frac{2\,M}{R^2}\frac{r}{R}\mp \ldots.
\end{eqnarray}
The tidal\indexit{tidal!acceleration} acceleration $a_{\rm t}$ is the
difference between the effective and the gravitational acceleration
\sactivity{$\circledS$ {\em Show:} (a) Looking only at gravitational forces, how close
  to a solar-mass object would the Earth need to be to be pulled apart
  by tides? Whereas this is impossible with the Sun, an Earth-sized
  planet could be pulled apart if it approached a white dwarf or
  neutron star (and something like that is involved in 'contaminating'
  some white-dwarf atmospheres with heavy elements). (b) An object of
  lesser density can be pulled apart, however, during a sufficiently
  close approach to the Sun: estimate at what distance comet 67P (with
  a mass of about $10^{16}$\,g and characteristic dimension of 3\,km)
  would have to come to the Sun to be broken up (ignore tensile
  strengths, spin, and orbital forces, thus accounting only for
  gravitational forces to hold the object together). Some Sun-grazing
  comets (such as the Kreutz family) have been observed to go through
  this breakup process. \mylabel{act:breakup} \solution{breakup}}:
\begin{equation}
a_{\rm t}\approx\pm G\,\frac{2\,M}{R^2}\frac{r}{R}.
\label{eq:2.1-20}
\end{equation}
Note that $a_{\rm t}$ decreases with the third power of $R$. As a
result of this the tidal accelerations are relatively small. On Earth
the tidal acceleration is about $1.1\times 10^{-6}$\,m\,s$^{-2}$ due
to the Moon and $0.5\times 10^{-6}$\,m\,s$^{-2}$ due to the Sun
compared to the gravitational acceleration of about
10\,m\,s$^{-2}$. This corresponds to an expected lunar tidal effect of
about 70\,cm. In reality, the average tide is about 30\,cm because of
a slight deformation of the Earth. In the case of the Sun, the tidal
effects caused by the planets are\indexit{tidal!bulge} very small;
[t]he largest effects are due to Venus and Jupiter with a theoretical
tide in the order of 1\,mm.

%The last total solar eclipse: https://www.forbes.com/sites/startswithabang/2017/08/18/earths-final-total-solar-eclipse-will-happen-in-less-than-a-billion-years/#4619f98d635a
As a result of the friction between the tide and the planet, the
rotation [and revolution tend towards synchronizing]. In the case of
Earth this [results in a slowing down of the spin rate by] about one
second per year. Some $2.5$ billion years ago the length of a day was
only about 6\,hours. Because the angular\indexit{tidal!friction and
  rotational retardation} momentum must be conserved this leads to a
corresponding increase in the distance between Moon and Earth (4\,cm
per year) as measured by laser technique.  \activity{{\em Consider} what it
  means for solar eclipses that the Moon is moving away from the
  Earth: at some future time, the Moon will be so far away
  that no more total solar eclipses can occur anywhere on
  Earth. (a) Assuming the Moon continues to move away at 4\,cm/yr, roughly
  when will the last total solar eclipse occur? Confirm that the
  answer is somewhat more than 600 million years. (b) Describe the effect
  on lunar eclipses. \mylabel{act:eclipses}}
The tidal friction generates a power of $3\times 10^{19}$\,erg/s which
is mostly dissipated in the ocean. There are indications that this
tidal power affects the global ocean circulation which plays a crucial
role in the climate system by transporting energy from low to high
latitudes.  The tides act also in the atmosphere causing changes in
pressure, temperature, and wave propagation.

There are climatic effects on Earth related to the lunar tides. The
plane in which the moon moves is inclined to the ecliptic by about
5$^\circ$. The points where the lunar orbit crosses the ecliptic are
called nodes. As a result of the gravitational force of the Sun on the
Moon the orbital spin axis of the Moon precesses, which leads to a
continuous slight shift of the nodes. After\indexit{lunar 18.6-year
  cycle} 18.6\,years the nodes are back to their original
position.
%\activity{{em Background:} If you are interested in solar eclipses,
%  and wonder why the saros cycle has a slightly different length from
% the lunar nodal period, have a look \href{https://en.wikipedia.org/wiki/Saros_(astronomy)}{here}.}
The inclination of the Moon's rotation axis has an effect on the amplitude
of the tides. The amplitude of the lunar nodal tide is only about
5\,\% of the daily diurnal tide but integrated in space and time it
becomes significant. The 18.6\,yr cycle and sometimes also its second
subharmonic of 74\,yr have been found in the arctic ocean temperature
and sea ice extent and in drought records.

The dynamics of a multibody system such as the Solar System is largely
determined by gravitation. The bodies orbit around the barycenter. In
the case of a two-body system with a large body (Sun) and a small body
(planet) the orbit is an ellipse with the large body in one of the
focal points. \activity{{\em Show:} One of the ways in which exoplanets are
  detected is to look spectroscopically at the displacements of the
  star about the barycenter of the exoplanetary system. How large is
  the velocity amplitude, and how large the associated Doppler shift
  at visible wavelengths, for the Sun-Jupiter
  system? \mylabel{act:doppler}} In a multibody system (Solar System)
the gravitational interaction between the bodies disturbs slightly
their orbital parameters. For example the planets (mainly Jupiter and
Saturn) change the eccentricity of the Earth's orbit with
periodicities of about 100,000 and 400,000 years which has an effect
on the amount of solar radiation received from the Sun'' determined by
the orbital eccentricity; more on that in Sect.~\ref{chapter:3.3}
around Eq.~(\ref{eq:3.3-1}).

\begin{figure}[t]
 \centering
 \includegraphics[width=9.5cm]{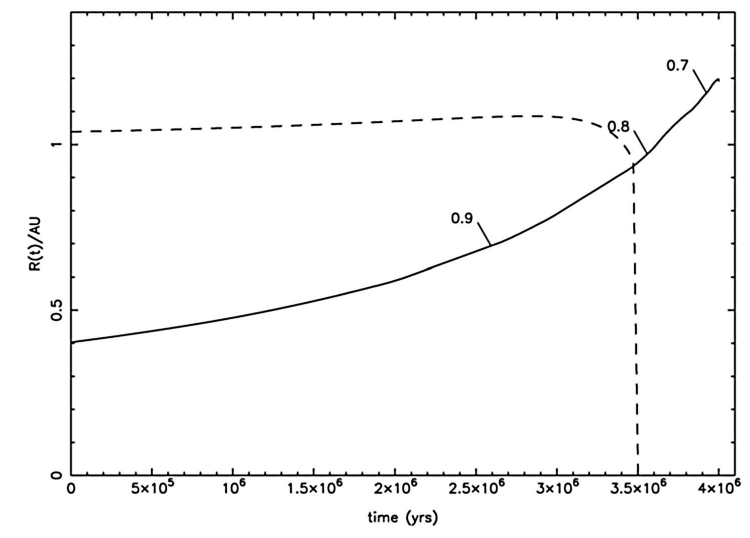}
  \caption[Distant-future
  diameter of the Sun and the size of Earth's
  orbit.]{\label{fig:brownlee6} The diameter
    \indexit{solar!evolution}of
    the future red-giant Sun
    (solid; in AU; [the labels along the curve show the Sun's mass at
    the time expressed in present-day solar masses]) and the size of
    Earth's orbit (dashed) during the 4 million years leading up to
    the phase when the Sun reaches maximum brightness.  Earth's orbit
    expands slightly as the Sun loses mass but the Sun expands to the
    point where tidal drag causes Earth's orbit to decay and intersect
    the Sun's upper layers.  These calculations predict that Earth
    will be destroyed in the Sun's atmosphere 7.59 billion years from
    present. [Fig.~III:4.6;
    \href{https://ui.adsabs.harvard.edu/abs/2008MNRAS.386..155S/abstract}{source:
    \citet{2008MNRAS.386..155S}}.]}
\end{figure} 
The effects of the solar tides on the Earth's orbit are negligible,
but that will not stay that way. Late in the life of the Sun, as it
runs out of fuel (see Ch.~\ref{ch:evolvingstars}), the Sun will swell
up into what is known as a red giant. In fact, its \ors[III:4.11]
\indexit{habitable zone!ultimate fate of} ``diameter increases by
approximately two orders of magnitude.  The physical expansion of
giant stars results in the assimilation of many of the planets that
may have formed in their formerly habitable zones [(defined as the
distance from the star where liquid water can be present on a planet's
surface)].  In the case of the Sun, current predictions indicate that
the Sun will expand (Fig.~\ref{fig:brownlee6}) to nearly 1 AU,
engulfing both Venus and Mercury.  Because the Sun loses over $40$\%\
of its mass during [its phases as a red giant], Earth's orbit will
actually expand to conserve angular momentum.  This seemingly places
Earth just beyond the presently modeled maximum diameter of the Sun
but detailed modeling indicates that Earth will be assimilated into
the Sun because of tidal effects. Tidal forces raise a bulge in the
Sun's upper layers that follows Earth and provide a retarding force
that causes Earth's orbit to decay.  \activity{{\em Show:} What is the upper limit
  to the Sun's rotation rate in its red-giant phase? Formulate your
  arguments. You may ignore solar mass loss in this estimate. Use
  Fig.~\ref{fig:Ievolve}. This upper limit shows that the Sun's outer
  layers are rotating (much?) more slowly than the Earth is orbiting
  it, so that the tidal bulge on the Sun will be traveling through,
  and dissipating energy within, the solar outer envelope.} Earth is
totally vaporized by this process due to the power generated by its
$\sim$25\,km/s entry into the Sun's upper atmospheric layers. If Earth
had formed 15\%\ further from the Sun it would have escaped
assimilation. Mars and all other planets are well beyond the effects
of gas drag and tidal effects and are safe from total destruction
although they are severely heated and rendered lifeless during the
Sun's red giant phase.''

This effect of orbital \indexit{orbital
  synchronization|see{tide}}synchronization by
\indexit{tide!orbital synchronization}gravitational tides occurs
for all close-in planets, and has a particular consequence if we look
at the evolution of the billions of planetary systems throughout the
Galaxy. \ors[III:4.11] ``In the far future, the Universe will look quite
different than it does at present. All massive bright stars will have
evolved and become invisible.  Only the slowly evolving and faint M
stars will persevere. After several tens of billions of years of
Galactic evolution, questions about habitability will only concern the
bodies that remain, these faint low mass red stars and planets that
orbit them in thin habitable zones close to their surfaces[; \ldots]
they are the most numerous stars now in the Universe and [\ldots] in
the long term they will be the only stars in the Universe.  Compared
to the Sun these low mass stars offer new challenges to understanding
habitability.  Although faint, they have pronounced flare activity
which generates both UV and energetic particle fluxes capable of
harassing life. Due to their faintness, their habitable zones are so
close to the stars that planets can be tidally locked with one side
always facing out to space.  This can cause thin atmospheres to freeze
out on the dark side of planets although sufficiently thick
atmospheres may be able to adequately distribute heat and prevent this
calamity.''

Tides also have their consequences on the multitude of double stars:
roughly one in two of the stars seen in the sky are pairs of stars
orbiting their joint center of gravity. \ors[III:2.5] ``The
gravitational tides in binaries \indexit{tide!binary stars}with periods of order a week or less
(depending on stellar masses and radii) are so strong that the orbital
and rotational periods of these stars are synchronized on time scales
much less than the main-sequence life time. Because any cool-star
components of such binaries lose angular momentum through their wind,
they will tend to spin down, but the tidal coupling replenishes the
lost rotational angular momentum from the reservoir of orbital angular
momentum. This causes the orbital separation to shrink, the locked
orbital and rotational periods to decrease, and
--~counterintuitively~-- the activity to increase with age until
eventually the stars merge into a single, rapidly-rotating but old
star (forming the class of \indexit{FK Comae stars}FK\,Comae
stars).''\activity{{\em Look up:} Between the phases of tidally-locked
  binaries and merged binaries are (semi-)contact binaries in which
  mass transfer can occur as one of the binary components becomes
  larger than its 'Roche lobe', either because the star swells up in
  late evolutionary phases (see Ch.~\ref{ch:evolvingstars}) or because
  the orbit shrinks by 'magnetic braking'. (a) Now or after reading
  Ch.~\ref{ch:evolvingstars}, look up the definition of 'Roche lobe'
  and the properties of RS~CVn, Algol, W~UMa, and FK~Com objects as
  characteristic phases in the evolution of different types of close
  binary stars toward single stars. (b) Make a list of these classes
  to summarize the masses of the two components involved in each
  class, their orbital separation, and their magnetic
  activity. \mylabel{act:binarieclasses} }

Another consequence of gravitational tides in the case of a tilted
rotation axis relative to the orbital \indexit{tide!precession}plane
\indexit{precession|see{tide}}is
precession. \ors[III:11.3] ``The \indexit{precession}precession
is a wobbling of the Earth's
axis of rotation which is caused by the tidal forces associated with
the Moon and the Sun.  Because the Earth is spinning, its shape
deviates slightly from a sphere leading to an equatorial bulge.  Tidal
forces act on the bulge and force the axis to precess. The periods of
[Earth's] precession range from 19,000 to 24,000 years.''
\activity{{\em Look up:} The Earth's equatorial bulge is nowadays used to keep
  satellites in a Sun-synchronous orbit, which is useful for
  satellites that need to scan the entire surface of the Earth, and also to
  enable Earth-orbiting satellites to have an uninterrupted view of
  the Sun throughout the year. (a) Look up and summarize how this
  works. (b) Why is this
  important for solar and heliospheric observations? \mylabel{act:bulge}}

\subsection{Orbital interaction}\label{sec:orbitinteraction}

Differential gravitational \indexit{orbital!interaction}forces are also thought to be of major
importance in the formative phases of planetary systems, specifically
acting between clumps of matter once these have condensed within the
spinning accretion disk, and in even earlier phases when gravity may
have led to unstable situations in which relatively dense areas may
form by contraction and compression.  \ors[III:3.2] ``As shown in the
right-most part of Figure~\ref{figlh:disktorques}, [such early gaseous
concentrations will be] sheared due to the differential rotation.
The gravitational attraction of one 'end' of the spiral arm pulls on
the other; this has the effect of accelerating the outer material at
the expense of decelerating the inner material -- {\em i.e.,}  transferring
angular momentum outward.  [One among several distinct mechanisms (see
Ch.~\ref{ch:formation})], it appears that this mechanism [of
gravitational instability (GI)] will prevent most of the mass from
remaining in the disk, but instead will allow accretion toward the
central object.''

These same gravitational forces are likely to play a major role in
enabling forming planets to grow into giants: growing planets set up
wave-like density disturbances in large spirals, and the interaction
of the growing planet with the matter in these spirals can cause all
of the constituent parts to change their orbits, as long as the mass
in the disk is not too small compared to that in the growing
planet. \ors[IV:1.2] ``For example, the combination of observations
and numerical experiments \indexit{planet!formation}suggests that gas giants accumulate up to a
few hundred Earth masses of material --~first the solids and then
increasingly rapidly gases~-- within a matter of a few million
years. This process is aided in its efficiency by the
\indexit{planet!migration}migration of growing planets within the
young planetary system: planets are not bound to their initial orbits,
but can migrate either inward or outward, subject to gravitational
interactions, thus having access to a large volume of the primordial
disk from which to collect material. Interestingly, it appears that it
is the very collection process of matter onto the growing planet that
causes mass redistributions within the disk so that their tidal
effects can make planets migrate, particularly if other planets are
forming elsewhere in the system, while the gravitational coupling
between multiple young planets in eccentric orbits can scatter bodies
around (both in distance from their central star and in orbital
inclination).''

\ors[IV:5.7.2] ``The realization that exoplanets are mobile during the
early stages of formation has led to many studies of dynamical
interactions.  The details of migration and the parking mechanisms
that [can lead to]  gas giant planets just a few stellar radii away
from their host stars are an active area of research.  In the younger
primordial disk with significant gas and dust density, the planet
embryos will clear gaps in the disk.  In this case, material can pile
up at both the inner and outer edges of the gap.  When the disk mass
at the edges of one of these gaps is comparable to the mass of the
planet embryo the disk will exert a torque that causes the planet [to
either tighten or widen its orbit around the parent star, {\em i.e.,} causes
the planet] to migrate. The outer edge of the disk causes inward
migration while the inner edge of the disk can produce outward
migration.  When multiple planet embryos exist in the disk it is
possible for the outer embryo to become locked into a resonant orbit
with the inner planet, a process called convergent migration.  As the
disk clears, convergent migration can leave planets in resonant orbits
that persist stably over the lifetime of the star.  This effect is
especially powerful for resonances where the ratio of the orbital
periods ($P_{\rm outer} / P_{\rm inner}$) is close to an integer number,
$N$. Planets with small $N$ are said to be in mean-motion resonance
(MMR) and the exchange of angular momentum between MMR planets is
flagged by oscillations in eccentricity and orbital periods.''

\section{Planetary atmospheric tides}

Apart from \indexit{tide!irradiation}magnetic torques and
gravitational tides, there is also a class of tides associated with
irradiation.  \ors[III:15.7.1.2] ``In general\indexit{tide} terms,
tides\indexit{atmosphere!tide} are the periodic response to
periodic astronomical forcing. In the [Earth's] atmosphere, by far the
dominant forcing agent is thermal excitation by solar radiation,
although forcing by latent heat release [({\em e.g.,} cloud
formation)] can also be important.  The dominant atmospheric tides are
the diurnal tide and the semi-diurnal tide at double the frequency.
In the lower and middle atmosphere, tides are excited primarily by the
absorption of solar UV radiation by stratospheric ozone and solar
near-IR radiation by tropospheric water vapor. The diurnal tide is
forced about one-third by water vapor absorption and about two-thirds
by ozone absorption. The semidiurnal tide is predominately forced by
ozone absorption. Although\indexit{tide!ozone, water-vapor
  absorption} the diurnal component of the diurnal variation of solar
heating is stronger than the semidiurnal component, there is a rough
parity between the two because the semidiurnal tide responds more
efficiently to ozone forcing than does the diurnal tide. This is
because the region of ozone forcing is fairly deep and main
semidiurnal modes with their comparatively long vertical wavelengths
respond in phase over the forcing regions, while the diurnal tide with
its fairly short wavelengths experience a degree of phase
cancellation.''

In \ors[III:15.12.2] ``[p]lanets with thick atmosphere [\ldots],
atmospheric tides can affect rotation. It is speculated that all
planets [in the Solar System] formed with similar rotation rates and
spun in the prograde sense (aligned with the total angular momentum of
the Solar System). \indexit{gravitational!torque}Gravitational torques can de-spin rotation toward
synchronous rotation, but cannot produce retrograde rotation. The
torques acting on the solar tidal bulge and coupling with the solid
planet, however, can cause retrograde rotation and this is what may have
produced the retrograde rotation of Venus. The present state of Venus
is thought to be an equilibrium between gravitational and thermal
atmospheric tidal torques. Clearly the resonances supported by
planetary atmospheres can affect where equilibrium states might be
found and thus the speed of retrograde rotation.''

More on tides and other large-scale wave phenomena in both oceans and
atmospheres can be found in Ch.~III:15.

\clearpage

\chapter{{\bf Particle orbits, transport, and acceleration}}%8
\label{ch:conversion}
{\narrower\narrower{
{\bf Chapter topics:}
\begin{itemize}
  \customitemize
\item Energy conversion from
large-scale dynamics of magnetized plasma into the thermal reservoir, the energetic-particle reservoir, or both
\item Single-particle motion in a magnetic field
\item Particle scattering and transport: solar energetic particles,
  galactic cosmic rays, radiation belts
\item Particle acceleration in shocks
\end{itemize}

\noindent{\bf Key concepts:}
\begin{itemize}
  \customitemize
\item Magnetic invariants
\item Phase-space density and the collisionless Boltzmann equation
\item Gradient and curvature drifts
\item Diffusive scattering
\end{itemize}

}}

\section{Introduction}
Deep \indexit{energetic!particles}inside stars and planets energy is
exchanged between particles (including photons) so frequently that the
distribution of velocities of the ions and electrons in stars, and of
the atoms and molecules in planets, are essentially pure Gaussians
(and thus the distributions of the magnitudes of the velocities pure
Maxwellians) around the mean bulk velocity.  With sufficient
collisional interactions in a neutral medium, or in an ionized medium
in the absence of magnetism, the mean bulk flows of different species
in a mixture tend to be equal. The presence of a magnetic field, in
contrast, is associated with a difference in bulk motions between
negatively-charged electrons and positively-charged ions. This, in
turn, leads to collisional interactions that convert the kinetic
energy of bulk population motions into random kinetic energy, {\em
  i.e.,} the dissipation of electrical current equates to heating.

Where collisional time scales grow to time scales approaching those of
physical processes, or even exceed these, velocity distributions can
deviate from Maxwellians. The populations of non-thermal particles of
most interest in the context of heliophysics are those of the highest
energies. Among these are radiation-belt particles, but also those
that originate from outside the Earth's environment, and referred to
as 'cosmic rays', which encompass \indexit{SEP|see{solar}}solar energetic particles (SEPs),
galactic \indexit{GCR|see{cosmic}}cosmic rays (GCRs) and 'anomalous cosmic rays'
\indexit{ACR|see{cosmic}}(ACRs). \regfootnote{Anomalous cosmic rays have a complex history:
  originally neutral particles in the interstellar medium, ionized by
  charge-exchange or photo-ionization in the solar wind, advected to
  the boundary regions of the heliosphere there to be accelerated. See
  Fig.~\ref{fig:jok2} for where they appear in the energy spectrum.}
\regfootnote{For an introduction to how energetic particles are
  detected and their properties determined, see Ch.~II:3.}

\ors[II:8.1] ``To understand the ubiquitous presence of energetic
particles it is important to realize that except for planetary
ionospheres and the lowest layers of the Sun's corona and below, most
plasmas in the heliosphere are basically collisionless.  That is, the
mean free path of charged particles is larger than most scales of
interest.  For example, in the undisturbed solar wind, the mean free
path for ions is of the order of 1\,AU [(see also
Table~\ref{tab:5.1})].  The lack of such collisions means that there
exists no primary mechanism that forces the particles to assume
thermalized Maxwellian distributions.  In fact, observed
distributions, often on top of thermal (colder) approximate 'core'
Maxwellians, almost universally contain energetic tails, which usually
can be described by power laws.  In real-world plasmas, there is a
multitude of processes responsible for generating such supra-thermal
and high-energy tails; usually, so-called wave-particle
interactions are involved.''

This chapter touches on various aspects of how energy can be converted
from large-scale dynamics of magnetized plasma into an increased
energy content in the thermal reservoir, the energetic-particle
reservoir, or both, as well as on the transport and loss of such
energy once in these reservoirs. This chapter covers topics as diverse
as GCR transport inward through the heliosphere to SEP transport
outward from the corona; all of these topics have to do with
conversion or transport of energy. The chapter starts with motions of
individual particles and their transport within magnetic environments,
then moves to mechanisms by which their energies can change to become
so-called 'energetic particles'. A description of how energy from
non-thermal particles is deposited into the thermal energy reservoir
with particular focus on the solar corona is partitioned off into Ch.~\ref{ch:heating}.

Flares, CMEs as well as magnetospheric (sub-)storms extract their
energy from what has been somehow stored in the magnetic field. This
extraction is typically enabled by the phenomenon of reconnection, and
both the total flux involved and the rate at which reconnection
proceeds help set the magnitude of energetic-particle events.  The
chapter touches on reconnection and shocks, which are essential
ingredients in both heating and impulsive phenomena, but only
introduces the basics of these complicated processes, which remain far
from understood.

\section{Single particle motion}\label{sec:singleparticle}
\ors[II:11.2.1] ``The \indexit{particle!motion!single}motion of every
individual charged particle in the heliosphere can be described by the
Lorentz force equation, Eq.~(\ref{eq:lorentz}). [\ldots]''
\ors[II:9.2.2] ``For\indexit{gyromotion} the simplest case of no
electric field and a constant magnetic field in the $z$ direction, the
solution to Eq.~(\ref{eq:lorentz}) is straightforward.  It is given by
\regfootnote{Note that the gyrating charged particle emits
  gyro-synchrotron radiation, thereby losing energy, so that this
  orbital motion approximated by Eq.~(\ref{eq:gyromotion}) --~and thus
  also in Eqs.~(\ref{eq:gcmotion}) and~(\ref{eq:derivvel})~-- is not
  sustained indefinitely.}:
\begin{equation}\label{eq:gyromotion}
v_x = +v\sin\alpha\cos(\omega_{\rm g} t - \phi) \,\,;\,\,
v_y = -v\sin\alpha\sin(\omega_{\rm g} t - \phi)\,\,;\,\,
v_z = +v\cos\alpha, 
\end{equation}
where $\omega_{\rm g}=qB/(mc)$ is the cyclotron\indexit{gyrofrequency} (gyro-)frequency, $\alpha$ is called the
pitch\indexit{pitch angle} angle (note that our definition is such that $\alpha=0$ implies
the particle is moving directly along the magnetic field), $\phi$ is
the phase angle, and  $v$ is the magnitude of the particle velocity.''

\ors[II:11.2.1] ``A very important aspect of the Lorentz equation when discussing
particle acceleration is that the electric field may change the energy
of the particle but the magnetic field does not. This relation is
shown by taking the dot product of the Lorentz equation with ${\bf v}$ giving:
\begin{equation}
{\bf F} \cdot {\bf v} = q({\bf v} \cdot {\bf E}) + {\bf v} \cdot ({\bf v} \times {\bf B}),
%\end{equation}
\,\,\,{\rm or}\,\,\,\,
%\begin{equation}
{{\rm d}W \over {\rm d}t} = q ({\bf v} \cdot {\bf E}),
\end{equation}
where\indexit{particle!energization!equation} $W$ is the kinetic
energy.

\begin{figure}[t]
\centering
\includegraphics[width=4.5cm,bb=0 0 362 358]{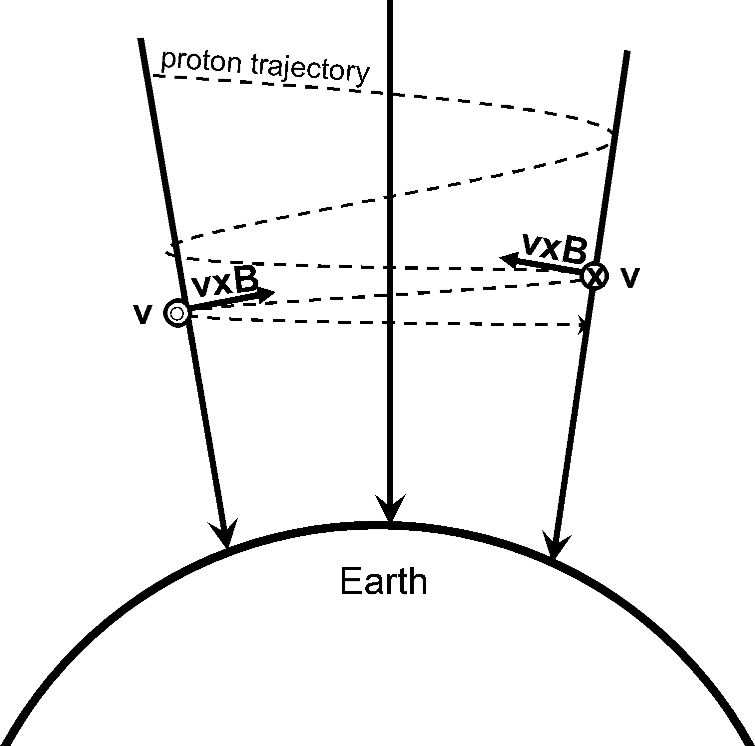}\hspace{1cm}\includegraphics[width=7cm,bb= 0 0 234 165]{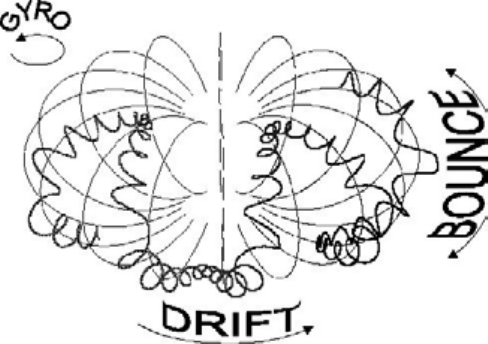}

\caption[Particle motions in magnetic field.]{\label{fig:green2.1} {\em (left)}
  Schematic diagram showing the Lorentz force as a particle moves into
  the magnetic field gradient at Earth's poles. [Fig.~II:11.2] {\em
    (right)} \label{fig:green2.3}Schematic diagram of particle motion
  in a dipole magnetic field. [Fig.~II:11.4]}
%\end{figure}
%\begin{figure}[t]
\centering
\includegraphics[width=8cm,bb= 0 0 489 146]{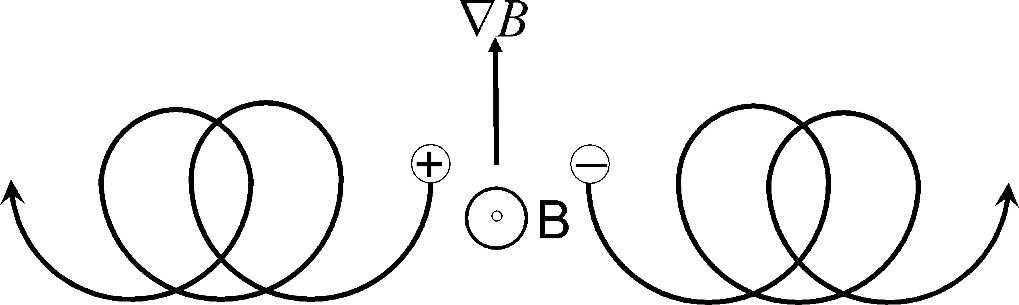}
\caption[Schematic diagram for the gradient-B
drift.]{\label{fig:green2.2}Schematic diagram for the gradient-B drift
  [(compare with a NASA/SVS animation at
  \href{https://svs.gsfc.nasa.gov/4263}{https://svs.gsfc.nasa.gov/4263})].
  Fig.~II:11.3]}
\end{figure}

[In realistic situations magnetic and electric fields rarely occur in
separate and uniform configurations. Even in the simple case of a
dipole potential field,] the motion separates into three oscillatory
types occurring at increasingly slower timescales, [visualized
together in Fig.~\ref{fig:green2.3}(right)]. On the fastest timescale,
a particle gyrates around the field line as described
above.\indexit{particle!motion! gyration}

The second oscillatory type motion in the dipole relates to the
particle's velocity parallel to the magnetic field. As the particle
follows the field line towards the poles, it moves through a gradient
because the magnetic dipole field increases [when the particle
approaches the planetary or solar] surface. The
effect of this gradient is to convert the parallel motion of the
particle into perpendicular motion as shown schematically in
Fig.~\ref{fig:green2.1}(left).  As the particle moves toward
the pole, the gradient effectively creates a Lorentz force opposite to
the parallel motion. Eventually, the parallel velocity will go to zero
and then reverse direction [ultimately] causing the
particle to bounce between the southern and northern poles. The point
at which the parallel velocity goes to zero is called the
mirror\indexit{particle!motion!mirror point} point and the
oscillation between the two poles is referred to as the bounce
motion.\indexit{particle!motion!bounce motion}

[In the case of a planetary magnetosphere dominated by a dipole, the particle will
circle the planet] in an oscillatory manner known as drift
motion. The azimuthal drift is caused by the radial gradient of the
dipole field. Intuitively, this drift can be attributed to the
changing gyroradius in different magnetic field strengths. In the
stronger magnetic field the gyroradius will decrease and in the weaker
field the gyroradius will increase creating the orbit shown in the
schematic of Fig.~\ref{fig:green2.2}. As protons and electrons gyrate
in opposite directions, they also drift in opposite directions.''

More generally speaking, for particles moving through a magnetic field
with a mixture of waves and turbulence, the latter two processes
transition from drifts to scattering --~to which we turn later in this chapter.

\subsection*{Guiding center motion} \label{sec:gcm}
\ors[II:11.2.2] ``Often, particle
motion can be described\indexit{particle!motion!guiding center}
by\indexit{guiding center motion|see{particle}} separating it into a drift velocity
with gyromotion superimposed as in the examples provided here.

{\bf ${\bf E} \times {\bf B}$ drift:} The
\indexit{particle!E$\times$B drift}${\bf E} \times {\bf B}$ drift
can be defined by including a uniform electric field in the Lorentz
equation and separating the equation into components parallel and
perpendicular to the magnetic field. [Let
${\bf B} = B {\bf \hat{z}}$.]  In the parallel direction the Lorentz
equation becomes
\begin{equation}
m\dot{v}_z = q E_z ,
\end{equation}
where $E_z$ is the component of the electric field parallel to the magnetic
field. This equation simply describes a particle accelerating along
the magnetic field. In the perpendicular direction, assuming
${\bf E} = E_x {\bf \hat{x}}+E_z {\bf \hat{z}}$ [(so that $E_y=0$)], the
Lorentz equation becomes
\begin{equation}\label{eq:gcmotion}
\dot{v}_x = +\omega_{\rm g} v_y + {q\over m}E_x \,\,;\,\,
\dot{v}_y = -\omega_{\rm g} v_x.
\end{equation}
Taking the second derivative of the velocity gives
\begin{equation}\label{eq:derivvel}
\ddot{v}_x = -\omega_{\rm g}^2 v_x\,\,;\,\,
\ddot{v}_y = -\omega^2_{\rm g} (v_y+{E_x \over B}).
\end{equation}
These equations describe gyration superimposed on [a] drift in the
${\bf E} \times {\bf B}$ direction.\sactivity{$\circledS$ {\em Show:} Use
  Eq.~(\ref{eq:derivvel}) to formulate (in a general vector
  expression) the magnitude of the ${\bf E} \times {\bf B}$ drift (in case
  you need a hint: assume the velocity can be described by an
  oscillatory component plus a constant drift). \mylabel{act:exbdrift}}

{\bf General force drift:} [If \indexit{particle!drift!general force}for the electrical force
$q{\bf E}$ we substituted another force ${\bf F}$ (such as
gravity)] into the ${\bf E} \times {\bf B}$ drift equation creates a
general force equation,\indexit{particle!general force drift}
\begin{equation}
{\bf v}_F = {1 \over \omega_{\rm g}} \left ( {{\bf F} \over m} \times {{\bf B} \over B}\right ).
\end{equation}
This equation can be used to define the drift velocity caused by any
general force. Other types of drift include curvature drift caused by
a centrifugal force related to the curvature of the dipole field
lines, polarization drift that results from a slowly varying electric
field, and a gravitational drift. [\ldots]''

Let us quantify the gradient \indexit{particle!gradient drift}and
\indexit{curvature drift|see{particle}}curvature \indexit{particle!curvature drift}drifts: [$v_{\rm G}$ and
$v_{\rm C}$] \ors{II:9.2.3} ``For
the special case in which $\nabla\times {\bf B}=0$, these [are] given
by:
\begin{eqnarray}\label{eq:drifts}
{\bf v}_{\rm G} &=& {cW_\perp\over qB^3}{\bf B}\times\nabla |{\bf
                    B}| \label{eq:Giacalone_Eq1.6} \\
%\end{equation}
%\begin{equation}
{\bf v}_{\rm C} &=& {2cW_\parallel\over qB^3}{\bf B}\times\nabla |{\bf B}|=
                         {2cW_\parallel \over qR_{\rm c}^2 B^2} {\bf R}_{\rm c} \times {\bf B}
\label{eq:Giacalone_Eq1.7}
\end{eqnarray}
where $W_\perp = (1/2)mv_\perp^2$ and
$W_\parallel = (1/2)mv_\parallel^2$. Note that these expressions are
for the case of non-relativistic particles. [The final expression in
Eq.~(\ref{eq:Giacalone_Eq1.7}) is added to explicitly show the
dependence on the curvature radius ${\bf R}_{\rm c}$ of the field;
this latter expression holds also in a non-potential
field.]\sactivity{$\circledS$ {\em Show:} Rewrite Eqs.~(\ref{eq:Giacalone_Eq1.6})
  and~(\ref{eq:Giacalone_Eq1.7}) to show that the drift velocity
  scales as the product of the particle's velocity and the gyroradius
  relative to the typical length scale in the gradient of the field,
  {\em i.e.,} as $v(r_{\rm g}/\ell_{\rm
    t})$. \mylabel{act:driftvelocity}}

However, in most applications of interest $\nabla\times {\bf B}\ne
0$. A more general expression for the particle drift can be derived by
expanding the magnetic field about the smallness parameter $r_{\rm g}/\ell_{\rm t}$
where $r_{\rm g}$ is the particle gyroradius and $\ell_{\rm t}$ is the characteristic
scale of the variation of the magnetic field. The resulting guiding
center drift velocity, in the non-relativistic limit, is given by:
\activity{{\em Consider:}
  Why do you think that bounce and drift motions are commonly ignored
  for the solar corona but are of dominant importance in the
  terrestrial magnetosphere? Hint: look at Table~\ref{tab:5.1},
  specifically $\lambda_{\rm mfp}$ and $r_{\rm gi}$ compared to
  $L_{\rm s}$.}
\begin{equation}
  {\bf v}_{\rm gc} =
       \bigg[v_\parallel + {cW_\perp\over qB} \uv{B}\cdot(\nabla\times\uv{B})\bigg]\uv{B}
    + {cW_\perp\over qB^2} \uv{B}\times\nabla |{\bf B}|
    + {2cW_\parallel\over qB} \uv{B} \times(\uv{B}\cdot\nabla)\uv{B}
\label{eq:Giacalone_Eq1.8}
\end{equation}
where $\uv{B}={\bf B}/B$. \regfootnote{Note that
  Eq.~(\ref{eq:Giacalone_Eq1.8}) is a corrected version of
  Eq.~(II:9.8).} The gradient and curvature drifts are associated
with the last two terms in this equation, which are in the direction
normal to the magnetic field; however, it is important to note that
there exists a component of the drift {\em along} the magnetic field
in addition to these. \activity{{\em Show:} Use a vector identity to show that the
  final term in Eq.~(\ref{eq:Giacalone_Eq1.8}) transforms into the
  central expression in Eq.~(\ref{eq:Giacalone_Eq1.7}) for a potential
  field.}
%http://shadow.eas.gatech.edu/~cpaty/courses/SpacePhysics2013/SpacePhysics2013/Lectures_files/Lecture13_14_15_2013.pdf
%Curvature drift: http://www.ss.ncu.edu.tw/~lyu/lecture_files_en/lyu_SPP_Book_A4format_pdf_html/pdf_2_App/lyu_SPP_Appendix_D.pdf
%GradB drift:
%http://www.ss.ncu.edu.tw/~lyu/lecture_files_en/lyu_SPP_Book_A4format_pdf_html/pdf_2_App/lyu_SPP_Appendix_E.pdf
%https://www.tcd.ie/Physics/people/Peter.Gallagher/lectures/PlasmaPhysics/Lecture4_single_particle.pdf
%http://how.gi.alaska.edu/ao/msp/chapters/chapter4.pdf
%http://mafija.fmf.uni-lj.si/seminar/files/2011_2012/motion_in_EM_fields.pdf

When Equation~(\ref{eq:Giacalone_Eq1.8}) is averaged over an isotropic
distribution of particles, one obtains the drift velocity ${\bf
  v}_{\rm d} =
(cmv^2/q)\nabla\times({\bf B}/B^2)$, which is commonly used in models
of cosmic-ray transport.''

The gradient-curvature drift in the terrestrial magnetosphere causes
one of the primary mechanisms often discussed as an agent in space
weather: the ring current. Seen from above the geographic north pole,
positive particles drift clockwise and negative particles drift
counter-clockwise. This differential motion leads to a westward ring
current between about 2 and 9 Earth radii. This ring current is
associated with a largely dipolar magnetic field with direction
opposite to the Earth's field. The variability of this current is
caused by the injection of particles into, and leakage from, the
magnetosphere associated with solar-wind variability. The Dst
(disturbance storm time) index \activity{{\em Look up} the
  magnetometer observatories that contribute to the Dst Index.  What
  do you notice about the latitude of their location?
  \mylabel{act:dstindex}} used in space weather characterizations
quantifies the strength of the ring current. The variation in the
surface magnetic field at Earth owing to the ring current is of order
$0.1-0.23$\,mG (see Table~I:13.5). The phenomenon of a ring current
is captured in an MHD description in principle, but because of the
interest in how particles of different energies and anisotropic
pitch-angle distributions behave, the inner-magnetospheric ring
current is generally studied with a custom ring-current model that
then is coupled to MHD magnetospheric and solar-wind
models. \sactivity{$\circledS$ {\em Show:} (a) Estimate the orbital
  period associated with the drift velocity as in
  Eq.~(\ref{eq:Giacalone_Eq1.6} for a purely equatorial motion for a
  proton with kinetic energy of 50\,keV orbiting, respectively, at 2
  and 10 planetary radii $r_{\rm p}$ for, for example, Mercury, Earth
  (where the ring current is contained roughly within these
  distances), and Jupiter. Use the equatorial field strengths
  $B_{\rm e}$ as in Table~\ref{tab:fran3} and
  $B(r)=B_{\rm e}(r_{\rm p}/r)^3$ for the equatorial dipole field. (b)
  Is the non-relativistic approximation warranted for this proton? (c)
  And for an electron of the same energy?
  \mylabel{act:driftvelocities} \solution{driftvelocities}}

\ors[II:11.2.3] ``The\indexit{particle!motion!invariants}
Lorentz\indexit{particle!invariants} equation and drift velocity
derivations provide a feel for how single particles [behave \ldots\ but
the analysis of] satellite measurements requires a more generalized view of
particle motion because detectors do not measure the position and
velocity of every particle in space to be propagated forward in time
using the Lorentz equation. To this end, it is instructive to describe
particle motion using aspects of the motion that are conserved when
time variations of the magnetic field are slow. For charged particles
in the magnetosphere, there are three such invariants associated with the
gyro, bounce, and drift motion.  Assuming that the invariants are
conserved confines the particle location to within a shell [in a
dipolar field such as that in the inner magnetosphere] about
Earth.

{\bf First invariant:}
The 
first\indexit{particle!motion!invariant} invariant is associated with the gyromotion
of the particle about the field line and is given by:
\begin{equation}\label{eq:firstinvariant}
\mu_{\rm m} = {p_\perp^2 \over 2mB}.
\end{equation}
Here $p_\perp$ is the relativistic momentum in the direction perpendicular to
the magnetic field, $m$ is the rest mass [\ldots], and $B$ is the
field strength.

{\bf Second invariant:}
The second invariant corresponds to the\indexit{particle!motion!invariant} bouncemotion of a particle along a field line and is given by:
\begin{equation}\label{eq:secondinvariant}
J = \oint p_\parallel {\rm d}s,
\end{equation}
where $p_\parallel$ is the particle momentum parallel to the magnetic
field and ${\rm d}s$ is the distance a particle travels along the field
line. It is convenient to rewrite the second invariant in terms of
only the magnetic field geometry by the following manipulation. If no
parallel forces act on a particle then momentum is conserved along a
bounce path and $J=2pI$ where $p$ is momentum and
\begin{equation}
I = \int_{s_{\rm m}}^{s^\prime_{\rm m}} \left ( 1- {B(s) \over B_{\rm m}} \right )^{1/2} {\rm d}s.
\end{equation}
Here $s_{\rm m}$ is the distance of the particle mirror point, $B(s)$ is the
field strength at point $s$, and $B_{\rm m}$ is the mirror point magnetic field
strength. If the first invariant is conserved then $K$, as defined
below, is also conserved.
\begin{equation}\label{eq:kdef}
K = {J \over 2 \sqrt{2m\mu_{\rm m}}} = I \sqrt{B_{\rm m}} =
\int_{s_{\rm m}}^{s^\prime_{\rm m}} \left ( B_{\rm m} - B(s) \right )^{1/2} {\rm d}s
\end{equation}
[\ldots]

{\bf Third invariant:}
The third\indexit{particle!motion!invariant} and final invariant corresponds to the drift motion of a
particle [and is given by:
\begin{equation}\label{eq:thirdinvariant}
\Phi = \oint A_\Phi {\rm d}l =\int {\bf B} {\rm d}S.
\end{equation}
In this equation $A_\Phi$ is the magnetic vector potential, ${\rm d}l$
is the curve along which lies the guiding center drift shell of the
electron, ${\bf B}$ is the magnetic field and ${\rm d}S$ is area.]
Therefore, conservation of this invariant requires that an electron
gyration always encloses the same amount of magnetic flux as it drifts
[\ldots] In a dipole field this is equivalent to saying that the
electron remains at fixed radial distance. The Roederer $L$ parameter,
commonly\indexit{particle!motion!L parameter} written as $L^\ast$,
is another useful form of the third invariant [often used for the
terrestrial magnetosphere]:
\begin{equation}\label{eq:lstar}
L^\ast = {2\pi {\bf \mu}_{\rm p} \over \Phi R_{\rm E}},
\end{equation}
where ${\bf \mu}_{\rm p}$ is the magnetic moment of the Earth's dipole field. The
$L^\ast$
parameter is the radial distance to the equatorial location where an
electron would be found if all external magnetic fields were slowly
turned off leaving only the internal dipole field.''

\section{Phase space density and Liouville's theorem}
\ors[II:11.2.4] ``Two\indexit{phase space density}
more\indexit{Liouville's theorem} concepts are needed to finally
interpret particle measurements from satellites: phase space density
and Liouville's Theorem.''
\ors[II:9.3.1] \label{sec:Giacalone_Sec3.1} 
``The number of particles per phase-space\indexit{phase space density}
volume is known as the phase-space distribution function, $f$, which
is a function of the 6-dimensions of phase space and time $({\bf
  p},{\bf r},t)$, where $\bf p$ is the particle momentum vector
(${\bf p} = m{\bf v}$).  The number density of particles at a given
location at a given time, $n({\bf r},t)$ is related to the phase
space distribution function by:
\begin{equation}
n({\bf r},t) = \int f({\bf p},{\bf r},t) {\rm d}^3{\bf p} ,
\label{eq:Giacalone_Eq1.9}
\end{equation}
where ${\rm d}^3{\bf p}$ is the volume element of phase space.  For
example, for a Cartesian geometry ${\rm d}^3{\bf p}={\rm d}p_x{\rm
  d}p_y{\rm d}p_z$ and for a spherical geometry it is ${\rm d}^3{\bf
  p}={\rm d}\phi\sin\alpha {\rm d}\alpha p^2{\rm d}p$ [\ldots]
($\alpha$ is the pitch angle).

The differential\indexit{differential intensity} intensity [\ldots] is related
to the phase-space distribution function by
\begin{equation}
J = p^2f.
\label{eq:Giacalone_Eq1.10}
\end{equation}
\noindent
Sometimes this is written as ${\rm d}J/{\rm d}E$.  This has units of
particles per area, per time, per energy, per solid angle.  If one
integrates $J$ over energy and solid angle ({\em i.e.,}  a spacecraft
detector with a given acceptance cone that sums over all energy
channels), the result is the {\em flux density} of particles, or the number of particles crossing per area per
time.''

\ors[II:11.2.4] ``Our interest in working with phase space density is
that it can be used to understand how collections of particles move
rather than individual particles. More specifically, Liouville's
Theorem states that as the system evolves or moves along a trajectory
in phase space the density must remain constant. The proof of this
theorem is illustrated intuitively by considering a volume of phase
space. As the particles in the volume are subjected to forces their
position and momentum will change but the trajectories of particles in
phase space can never cross. Trajectories crossing would imply the
physical impossibility that two particles with the same position and
momentum subjected to the same forces go in different
directions. Thus, the particles act as an incompressible fluid [in
phase space]. As
they move, the volume can change shape but the density remains the
same.

At first glance, Liouville's Theorem seems to be an esoteric statement
but in fact its application is quite powerful. The particle flux
(number of particles per ${\rm cm}^2\,{\rm s}\,{\rm str}\,{\rm
  keV}$) measured by a particle detector on a satellite,
$J(E,\alpha,\varphi,{\bf x})$ where $E$ is the energy, $\alpha$ is the
pitch angle, $\varphi$ is the gyro-phase, and ${\bf x}$ is the
position, can be directly related to the phase space density through
the relation $J(E,\alpha,\varphi,{\bf x}) = f({\bf x},{\bf
  p})/p^2$. Liouville's theorem states that the phase space density
does not change as the particles move along a trajectory. We also know
that if time variations of the magnetic field are slow, a particle's
trajectory must move along a contour of constant adiabatic
invariants. Putting these two concepts together means that
$f(\mu_{\rm m},J,L^\ast,\varphi_1,\varphi_2,\varphi_3)$ wherever it is
measured must remain constant. (Here $\varphi_{1,2,3}$ are phase
angles associated with each invariant. For simplicity, it is generally
assumed that the phase space density does not vary with the phase
angles.) Any change of phase space density implies that one of the
invariants is broken. In fact, acceleration mechanisms always violate
an invariant. Thus, an increase in phase space density expressed as a
function of the adiabatic invariants is a sign that acceleration has
occurred. Flux measurements, in contrast, can change simply because
the magnetic field topology has changed making these data very
difficult to interpret.''

% see emails from Tom Bogdan, and also:
% http://www.phy.pku.edu.cn/~fusion/down/Dia/QuasilinearA.pdf
% https://www3.nd.edu/~gtryggva/CFD-Course/2013-Lecture-2.pdf
\section{The collisionless Boltzmann equation}\label{sec:boltzmann}

Let \indexit{Boltzmann equation}us start with a general view of what
we can do with the phase space density, looking specifically at what
it takes to change it (or at what it takes to maintain it so that the
adiabatic invariants can be applied to particle trajectories). You can
review this section quickly on first pass, then revisit this when you
reach the end of Sect.~\ref{sec:gcrtransport}.

% http://www.astro.yale.edu/vdbosch/lecture7.pdf
Throughout the heliosphere, we can generally ignore collisions between
charged particles, particularly for the particles residing in
supra-thermal tails of velocity distributions. Consequently, the
distribution function for heliospheric charged particles generally
satisfies the collisionless Boltzmann equation, which is a continuity
equation in the 6D space of momentum and location coordinates
${\bf w}=[{\bf r}, {\bf p}]$ and time:
${\partial f / \partial t} + {\bf \nabla} \cdot (f \dot{{\bf w}}) =0$
(a 6D mathematical equivalent of Eq.~\ref{continuity}, absent sources
and sinks), or in another formulation ([using the index notation so
that, for example, the vector for space coordinates is written
$x_i = {\bf x} = x_x\hat x + x_y\hat y + x_z\hat z$, as for] momentum
$p_i$ (or velocity $v_i$); with acceleration $a_i$, and with implied
summation over repeated indices):
\begin{equation}\label{eq:generalboltzmann}
  {\partial f \over \partial t} =
  - {p_i \over m}{\partial f \over \partial x_i}\ 
  - {F_i} {\partial f \over \partial p_i}\ 
  + {S} - {L},
\end{equation}
where the components of the force $F_i$ are given by
\begin{equation}\label{eq:generalforce}
F_i \equiv m a_i = q\left( E_i + {1 \over mc} \epsilon_{ijk}p_jB_k
\right) + {\cal S}_i.
\end{equation}
Here, ${\cal S}_i$ is a placeholder for any other force or sum of forces that
may apply, including gravity;
even radiative energy losses or gains could be incorporated
(although we ignore these here). Sources $S$ and
losses  $L$ could represent couplings to other reservoirs, such as
neutral atoms or dust, which could happen through charge exchange or
photoionization. We ignore these terms further in this
chapter. 

%http://www.astro.uu.se/~hoefner/astro/teach/adp08_L3_notes.pdf
Note that low-order velocity moments of the Boltzmann equation for
combinations of interacting particle populations yield the equations
of fluid dynamics. Take, for example, the case of a fully ionized
hydrogen plasma with phase-space densities $f_{\rm e}$ and $f_{\rm i}$
for electrons and ions. The suitable combinations of the Boltzmann
equations Eq.~(\ref{eq:generalboltzmann}) for these phase space
densities after multiplication by $mv^\alpha$ and integration over
velocity space for $\alpha=0,1,2$ yield, respectively, the continuity
equation Eq.~(\ref{continuity}), the momentum equation
Eq.~(\ref{momentum}), and the energy equation Eq.~(\ref{energy}). A
complete set of fluid dynamics equations would continue with ever
higher moments until the entire phase space density has been
described, but that is not practical. Instead, the series is commonly
truncated by some approximation, known as 'closure'; see also
Table~\ref{fig:mhdvalidity}.

A persistent electric field (such as in reconnection processes, see
Sect.~\ref{sec:reconshock}), for example, can change a particle's
energy when that is accelerated along the field. Forces that can
change the energy of particle populations need to be retained
explicitly in whatever we do with Boltzmann's equation. Fluctuations
in the magnetic fields in space and time (such as in Alfv{\'e}n
waves), in contrast, do not change a particle's energy (more on that
in Sect.~\ref{sec:particletransport}): they do scatter a particle in
pitch angle. Repeated scattering in a perturbation field that is
symmetric in the probability of scattering a particle in either
direction can be described as diffusion. With that realization,
Eq.~(\ref{eq:generalboltzmann}) can be reformulated in a quasi-linear
approximation by separating large-scale trends from small-scale
fluctuations, denoting the large-scale average flow ${\bf u}$, and
capturing the net effects of the small scale fluctuations in diffusion
terms \activity{{\em Advanced/group:} If you are interested in the origin of
  the terms in Eq.~(\ref{eq:boltzmanndiffusion}) you could review
  classic papers with fairly 'intuitive' introductions to the equation
  by, {\em e.g.,}
  \href{https://ui.adsabs.harvard.edu/abs/1976SSRv...19..845M/abstract}{Harm
    \citet{1976SSRv...19..845M}} or
  \href{https://ui.adsabs.harvard.edu/abs/1983RPPh...46..973D/abstract}{Luke
    \citet{1983RPPh...46..973D}}, the latter also including how term \tc{d} arises. Or you
  could look at the paper by
  \href{https://ui.adsabs.harvard.edu/abs/1984SSRv...37..201Q/abstract}{John
    \citet{1984SSRv...37..201Q}} which also describes the so-called 'force-field solution'
  that you will find in Sec.~\ref{sec:beerradio} on cosmogenic
  radionuclides. For each term, identify conditions or a circumstance
  (solar corona, solar wind, inner magnetosphere, etc.)  where it
  provides an important contribution. \mylabel{act:eqorigin}}:
\begin{equation}\label{eq:boltzmanndiffusion}
  \overset{\tc{a}}{{\partial f \over \partial t}}\  =
  {\partial \over \partial x_i} \left[\overset{\tc{b}}{\kappa_{ij} {\partial f \over\partial x_j}}
    - \overset{\tc{c}}{u_i f} \right]
    + {\partial \over \partial p_i} \left[ \overset{\tc{d}}{ D_{ij} {\partial f \over
          \partial p_j}}
        - \overset{\tc{e}}{ F_i f}\right ]
  + \overset{\tc{f}}{[{\rm S} - {\rm L}]},
\end{equation}
\begin{equation}\label{eq:boltzmanndiffusionforce}
{\rm with}\,\,\,\  F_i = - {1 \over 3} p_i {\partial u_j \over \partial x_j} + \ldots
\end{equation}
Here, $u$ is the mean velocity of the scatterers, which equals the
flow speed of the bulk thermal plasma provided that comparable power
resides in waves traveling in opposite directions. The explicitly listed
term in $F_i$ above represents the adiabatic momentum change. 

The expressions \tc{b}{\&}\tc{c} and \tc{d}{\&}\tc{e} in
Eq.~(\ref{eq:boltzmanndiffusion}) reflect fluxes in physical space and
in momentum space, respectively. Terms \tc{b} and \tc{d}  
reflect  diffusive processes; in geometric space with diffusion parameter
$\kappa_{ij}$ (with diagonal elements describing diffusion parallel
and perpendicular to the magnetic field, and off-diagonal elements
quantifying particle drifts) and in momentum or velocity space with
diffusion parameter $D_{ij}$ (which includes, among other things,
pitch-angle scattering that does not affect the particles' energy);
whereas we can more readily appreciate the symmetry between these two
spaces, the physics of the scattering processes now lies hidden in the
two diffusion tensors (see, for example,
Sect.~\ref{sec:particletransport}). Terms \tc{c} and \tc{e} reflect
advection, in geometric space in \tc{c} and in momentum space (by
the forces acting on the medium) in \tc{e} (but note the mixed partial
derivatives in \tc{e} which means different groupings are possible).

Eq.~(\ref{eq:boltzmanndiffusion}) informs us on how particles move in
the coupled 6-dimensional realm of geometric space and velocity space. When
we talk about the transport of either solar energetic particles or
galactic cosmic ray particles through the heliosphere, we look primarily at transport in
geometric space which involves terms \tc{b} and \tc{c}: transport is affected by
scattering and advection
(in addition to, {\em e.g.,} geometric expansion in a spherical
geometry, which involves term \tc{e}).

When we look into acceleration (and thus also heating) mechanisms,
such as for shocks in Sect.~\ref{sec:shockacceleration}, we need to
figure out how particles move about in momentum space, {\em i.e.,}  using
expressions \tc{d}{\&}\tc{e}, while crossing the shock in geometric
space with expressions \tc{b}{\&}\tc{c}. It often helps to focus on
parts of the overall function $f$. For example, for the bulk of the
plasma well within the thermal range of Maxwellian distributions, with
relatively short mean-free paths compared to system scales and with
frequent interactions with the collective, we have the MHD description
(Ch.~\ref{ch:universal}) and, for shocks, the Rankine-Hugoniot jump
conditions (Sect.~\ref{sec:shocks}). In contrast, for a 'contaminant'
population of solar energetic particles and galactic cosmic rays
moving through, but to first order not interacting with, the
background plasma flow of the solar wind, but being scattered by
perturbations in its magnetic field, Eq.~(\ref{eq:boltzmanndiffusion})
provides a powerful tool, as we shall see next. Other descriptions
below focus on narrow parts of the overall phase-space distribution,
such as the supra-thermal particles that interact at shallow angles
with a shock and scatter in the collective of particles around it, and
for a sub-population of quite energetic particles with long mean-free
paths that bounce back and forth across a shock in a ping-pong fashion
as they are scattered by waves. With this perspective, let us look at
how this all describes the propagation of energetic particles through the
heliosphere and the creation of energetic particles at shocks.

\section{Particle scattering and transport}\label{sec:particletransport}
\ors[II:9.2.4] ``To\indexit{particle!scattering} this point we have considered only smoothly varying
electric and magnetic fields as compared to the radius
of\indexit{gyroradius} gyration of the particles,
$r_{\rm g} = v/\omega_{\rm g}$.  For such cases, the particle speed
and pitch angle change very slowly compared to the cyclotron period.
However, when the typical scale of the variation in the fields,
$L_{\rm t}$, is of the order of $r_{\rm g}$, the speed, phase, and
pitch angle can undergo more rapid changes.  This leads to a form of
scattering that is loosely analogous to classical scattering, although
it differs in important ways.  For instance, the particles do not
collide off of one another, as in the lower portions of
Earth's atmosphere, nor do they collide off of large targets, like
photons moving through a dense gas, but rather, they scatter off of
irregularities in the magnetic field.  Formally one can solve the
equations of motion under the approximation that the amplitude of the
magnetic fluctuations are small and show that there exists a
resonance\indexit{particle!scattering!resonance condition} condition,
%\begin{equation}
$v_\parallel \sim L_{\rm t}\omega_{\rm g}$,
%\label{eq:Giacalone_Eq1.8a}
%\end{equation}
for which the equations become undetermined.  At such instances, the
particle is said to 'scatter' and it reverses its pitch angle and
its phase angle becomes randomized. [\ldots]

\label{sec:Giacalone_Sec3.2} \label{sec:Giacalone_Sec3.3} \label{sec:Giacalone_Sec3.4}

Because particle scattering is a stochastic process, it is most useful
to perform a statistical analysis on a large number, or {\it
  ensemble}, of charged particles.  The relationship between the
average particle motion and the magnetic field can be determined from
the quasi-linear theory.  It is found that the dynamical behavior of
the distribution function obeys the standard diffusion equation in
classical statistical physics.  [\ldots]'' \ors[II:9.3.2]
\indexit{diffusion!equation!phase space} ``It is important to keep in
mind that this equation is strictly valid only for time scales that
are long compared to the time in between scatterings (the scattering
time) and spatial scales that are large to the distance traveled
between scatterings (the mean-free path).'' \ors[II:9.3.3] Because ``the
magnetic field in space exists in a highly electrically conductive
plasma, the field moves with the flow of the plasma (it is said to be
'frozen in'.  In the limit of ideal magnetohydrodynamics (MHD), which
is the limit we are concerned with for energetic-particle transport,
there is no electric field in the frame moving with the plasma. Thus,
as a charged particle scatters off of a magnetic irregularity, its
energy in the frame of reference moving with the plasma remains
unchanged. [Strictly speaking, this assumes that the magnetic field is stationary in this
frame of reference which is factually incorrect because of the
presence of waves with a variety of phase and group velocities, but a
good approximation in the case of the transport of energetic particles that
move much faster than the waves ({\em i.e.,}  $v\gg v_{\rm
  A}$, where $v_{\rm A}$ is the Alfv{\'e}n speed).] From the
perspective of such fast particles,
the magnetic fluctuations, which provide the scattering centers,
move with the bulk plasma.  [\ldots] In an inertial frame relative to
which the plasma moves with a velocity $u$, the evolution of $f$
satisfies the advection-diffusion] equation, which in one spatial
dimension is given by''
\begin{equation}
{\partial f\over\partial t} = 
{\partial\over\partial x}\bigg(\kappa{\partial f\over \partial
  x}\bigg) - u{\partial f\over\partial x},
\label{eq:Giacalone_Eq1.12}
\end{equation}
\ors[II:9.3.2] ``where $\kappa$ is the diffusion coefficient.  For the case
of charged particles moving in an irregular magnetic field, $\kappa$ is
related to the statistical properties of the magnetic field, in particular,
its power spectrum.'' Here, we interpret $f$ as
integrated over all velocity space, so looking only at total numbers
as a function of space. Thus, Eq.~(\ref{eq:Giacalone_Eq1.12}) is the
1D version of Eq.~(\ref{eq:boltzmanndiffusion}) for a case with
constant $u$, in which the velocity integral for expression
\tc{d} disappears because scattering under these conditions does not
change the overall energy
of the population and \tc{e} because there are no other forces
assumed to act on the plasma. 

\ors[II:9.3.2]  ``We note that for Eq.~[\ref{eq:Giacalone_Eq1.12}] we have assumed that
the distribution function varies only in one spatial direction.  This should
not be confused with [\ldots] 
the restriction on particle motion arising from fields that vary with only
one spatial coordinate.  By using Eq.~(\ref{eq:Giacalone_Eq1.12}), we
have already assumed that the process is diffusive.  If, for example, $x$
is taken to be the direction normal to a mean magnetic field, then the use
of this equation implies that the field must be fully three dimensional
for cross-field diffusion to take place.  The key is that the field
is fully three dimensional but it is also statistically homogeneous in space.''
\ors[II:9.3.4] ``In
two dimensions, there are two diffusion coefficients, one for
each direction (plus cross terms which we can ignore for now).  Consider the
motion of particles in a turbulent magnetic field \regfootnote{This
  volume does not go into the generation and properties of
  turbulence; for an introduction within the context of heliophysics,
  see I:7.} whose average points
along the $z$ direction.  Then, for example, in the $x$-$z$ plane, the
diffusion equation (neglecting the advection term discussed above
and cross terms) is given by:
\begin{equation}
{\partial f\over\partial t} = {\partial\over\partial x}\bigg(\kappa_\perp{\partial f\over \partial x}\bigg) + {\partial\over\partial z}\bigg(\kappa_\parallel{\partial f\over \partial z}\bigg),
\label{eq:Giacalone_Eq1.13}
\end{equation}
\noindent
where $\kappa_\perp$ and $\kappa_\parallel$ are the diffusion coefficients
across the magnetic field and along it, respectively.

Because the time $\tau_s$ it takes for a charged particle in the
heliosphere [or magnetosphere] to scatter is generally much longer than the time it takes
to gyrate about a magnetic field ({\em i.e.,}  $\omega_{\rm g}\tau_{\rm s}\gg 1$),
particles tend to move much more closely along the magnetic field than
across it.  As such, $\kappa_\perp$ is usually assumed to be much
smaller than $\kappa_\parallel$.  For this reason, many analyses
simply neglect perpendicular transport.  However, it is important to
note that in many astrophysical plasmas of interest, perpendicular
transport is the most important [\ldots]

The motion of a particle across a magnetic field occurs in two ways: (1)
the actual transfer of particles from one magnetic field
line\indexit{diffusion!cross field} to the next resulting from
scattering, or across the field arising from drifts, and (2) the
motion of particles along magnetic lines of force that themselves
meander in space in the direction(s) normal to the mean magnetic
field.  [\ldots]''

\label{sec:Giacalone_Sec3.6} \ors[II:9.3.6] ``In\indexit{cosmic ray!energy change} addition to scattering and advection with the flow,
the particle speed itself can change.  Principally, this can happen in
two ways: (1) by scattering within a spatially varying flow [({\em i.e.,} 
by term \tc{c} of
Eq.~(\ref{eq:boltzmanndiffusion}))], or (2) by diffusing in energy
space because of collisions with randomly moving scattering centers.
The latter of these two
[(related to $D_{ij}$ in term \tc{d} of
Eq.~\ref{eq:boltzmanndiffusion}, and related to dispersion in
scatterer speeds)] is called second-order Fermi
acceleration, or stochastic acceleration.  This is an interesting
topic, but is not considered in our discussion here.  We examine
further the first case.

Consider a \indexit{particle!acceleration}particle moving in a given direction in an inertial frame
which then scatters.  Energy is conserved in the local plasma frame,
but in the inertial frame the particle either gains or loses energy
depending on whether it is moving initially against or with the flow
${\bf u}$.
Suppose that at one scattering, it initially moves against the flow,
and gains energy in the inertial frame (this is a head-on collision).
When it next scatters, it will be moving initially with the flow and
will lose energy.  If the flow is everywhere uniform, then the
particle loses the energy it gained in the
previous scattering and there is no net energy gain.  But, if the
second scatter occurs at a different flow speed, there is a net
change in the particle's energy.  The term that accounts
for this behavior is given by
\begin{equation}
{p\over 3}\nabla\cdot{\bf u}{\partial f\over \partial p}
\label{eq:Giacalone_Eq1.16}
\end{equation}
[(as in term \tc{e} in Eq.~\ref{eq:boltzmanndiffusion} and the
expression for
$F$ following it)].
Particles gain energy if this term is negative and lose energy if it is
positive.

A particularly good example of this is particle acceleration at a
shock.  Consider the energy of a particle in a frame of reference
moving with the shock.  As a particle scatters in the flow behind the
shock, it loses energy because the particle was initially moving with
the flow.  The particle then returns upstream where it scatters off of
the incoming upstream flow leading to a gain in energy.  The energy
lost by the downstream scattering event is smaller than the energy
gained by the upstream scattering event because the upstream flow
speed is larger than that downstream.  Thus, there is a net energy
gain, which leads to an acceleration of particles [(more on that in
Sect.~\ref{sec:shockacceleration})].  Note that at a shock, the flow
goes from large to small (in the shock frame) so that the divergence
is negative and Eq.~(\ref{eq:Giacalone_Eq1.16}) is negative, giving
rise to acceleration.

It is also noteworthy that the energy change term is positive for the
case of a constant radial solar wind speed.  So, all charged particles
{\em lose energy} in the adiabatically expanding solar wind!''

\label{sec:Giacalone_Sec3.7}

\ors[II:9.3.7] ``The\indexit{transport equation!Parker's}\indexit{Parker!transport equation}
resulting\indexit{cosmic ray!transport equation} superposition of
the terms that we have discussed above, lead to the 
cosmic-ray transport equation.
It is given by
\begin{equation}
   {\partial f \over \partial t}\  =   {\partial \over \partial x_i}
\left[\kappa_{ij} {\partial f \over \partial x_j}\right]  - u_i
     {\partial f \over \partial x_i}  + {p \over 3}
{ \partial u_i \over \partial x_i} \left[ {\partial f \over \partial p}
\right] + {S} - {L}
\label{eq:Giacalone_Eq1.17}
\end{equation}
[(which is very nearly the diffusion version of the collisionless Boltzmann equation
of Eq.~\ref{eq:boltzmanndiffusion}, but with $D_{ij}=0$).]
Note that we have written the diffusion coefficient $\kappa_{ij}$ in its full tensor
form [\ldots]

The cosmic-ray equation is remarkably general.  It has been used
widely in most discussions of cosmic-ray transport and acceleration
over more than three decades.  It is a good approximation provided there
is sufficient scattering to keep the pitch-angle distribution nearly
isotropic \regfootnote{This should not to be confused with anisotropic
diffusion resulting when $\kappa_\perp \ne \kappa_\parallel$.}, and if
the particles move substantially faster than the speed of both the
background fluid and the characteristic speed of the MHD waves
contained in the plasma.''

\ors[II:9.4] ``All\indexit{diffusion!tensor} of the quantities in the transport equation, except for the
diffusion tensor, are directly observed by spacecraft or can be
accurately determined by using the hydromagnetic approximation.
Consequently, determining transport coefficients poses a
fundamental challenge in the modeling of cosmic rays.

In general, the diffusion tensor $\kappa_{ij}$ is related to the
magnetic field vector $B_i$, the diffusion coefficients parallel
and perpendicular to the mean field, $\kappa_\perp$ and $\kappa_\parallel$,
and the antisymmetric diffusion coefficient, $\kappa_A$, as
\begin{equation}
\kappa_{ij}=\kappa_\perp\delta_{ij}-{(\kappa_\perp-\kappa_\parallel)B_iB_j\over
B^2}+\epsilon_{ijk}\kappa_A{B_k\over B},
\label{eq:Giacalone_Eq1.20}
\end{equation}
\noindent
where $\delta_{ij}$ is the Kronecker delta function ($\delta_{ij}=1$
if $i=j$ and $\delta_{ij}=-$ if $i\ne j$), and $\epsilon_{ijk}$ is the
Levi-Civita symbol: $\epsilon_{ijk}=1, \,{\rm or}\, -1$ if $(i,j,k)$
is an even or odd permutation of $(1,2,3)$, respectively, and
$\epsilon_{ijk}=0$ if any index is repeated.  We have also introduced
the antisymmetric diffusion coefficient $\kappa_A$.  Note that the
symmetric terms reflect the diffusion due to small-scale turbulent
fluctuations; in contrast, the antisymmetric term contains the
particle drifts caused by the spatial variations of the large-scale
magnetic field.''

\subsection{Solar energetic particles}\label{sec:septransport}
\ors[II:9.5.1]
``A\indexit{solar!energetic particles!point-source
  evolution}\indexit{solar!energetic particles!impulsive-event
  problem} particularly simple, yet illustrative example of the use of
the cosmic-ray transport equation is the evolution of impulsively
released particles from a point source.  This is presumably a
reasonable representation of the physics of solar-energetic particle
transport after their release onto open magnetic field
lines following their rapid acceleration in the vicinity of a solar
flare.  Of course, we must recognize that the earliest arriving
particles suffer very little pitch-angle scattering, and therefore,
the transport equation is not useful for describing these particles,
but is adequate to describe the long-time behavior.

A proper treatment of the impulsive SEP problem should necessarily
include, as a minimum, the effects of diffusion, advection with the
solar wind, and adiabatic cooling.  Spherical coordinates with the
origin at the Sun would be a good choice.  The resulting equation,
even when simplified by making various assumptions about the choice of
parameters can be impossible to solve
analytically.  For our purposes here, which is simply for illustration
and by no means is meant to be directly comparable to SEP
observations, it suffices to consider a Cartesian geometry, a constant
diffusion coefficient, and to neglect both advection with the flow and
energy change. The result is simply [Eq.~(\ref{eq:Giacalone_Eq1.12})
with $v=0$], which is the 1D diffusion equation.  The
solution for an impulsive injection of particles at $x=0$ at time,
$t=0$ is given by
\begin{equation}
f(x,t) = {N_0\over \sqrt{4\pi\kappa t}}\exp\bigg(-{x^2\over 4\kappa t}\bigg),
\label{eq:Giacalone_Eq1.25}
\end{equation}
where $N_0$ is the number of particles released.

\begin{figure}
\centering
\includegraphics[width=8.cm,bb=100 340 510 600]{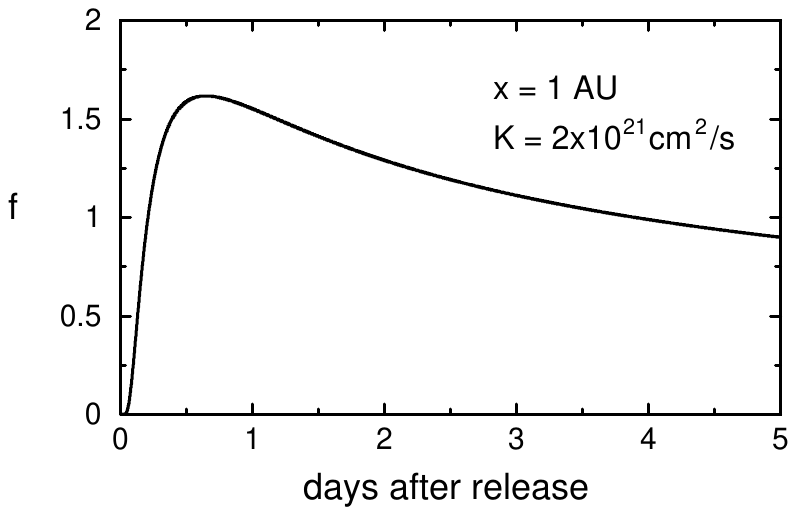}
\caption[1-dimensional particle diffusion from a point-source.]{Solution to the one-dimensional diffusion equation for a point-source
release at a position 1\,AU away from an observer: $f(1,t)$ from
Eq.~(\ref{eq:Giacalone_Eq1.25}). [Fig.~II:9.10]}
\label{Giacalone_Fig10}
\end{figure}

\begin{figure}
\centering
\includegraphics[width=11.cm,bb=0 210 612 580]{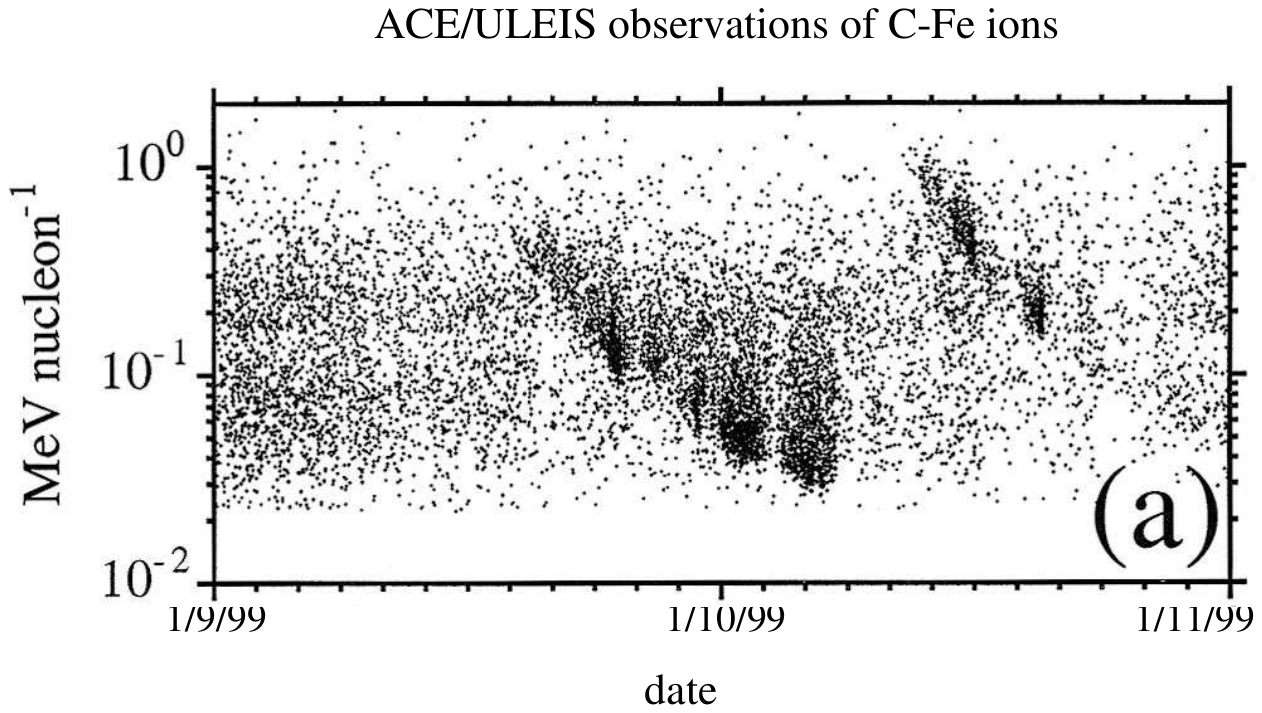}
\caption[SEP event associated with an impulsive solar flare.]{A solar energetic particle (SEP) event, associated
with an impulsive solar flare, seen by ACE/ULEIS.  Each dot represents
the detection of a particle by the detector.  Two distinct events are
shown.  [Fig.~II:9.11;
\href{https://ui.adsabs.harvard.edu/abs/2000ApJ...532L..79M/abstract}{source:
\citet{2000ApJ...532L..79M}}.]}
\label{Giacalone_Fig11}
\end{figure}

Figure \ref{Giacalone_Fig10} shows a plot of the distribution of particles,
given by Eq.~(\ref{eq:Giacalone_Eq1.25}), at the
location $x=1$\,AU, as a function of time (in days).  The diffusion coefficient
was taken to be $\kappa = 2\times 10^{21}\, {\rm cm^2/s}$, and
$N_0 = 10^{14}$.  If, for example, these are 10-MeV protons, then the
corresponding mean-free path would be about 0.1\,AU.  This profile has
similarities to those
seen at 1\,AU following a flare or CME on the Sun [\ldots]
An example of an impulsive-like solar-energetic particle event observed
at 1\,AU by the ACE spacecraft (ULEIS instrument) is shown in
Fig.~\ref{Giacalone_Fig11}.  Each dot represents a detection by
the instrument of an individual particle.  Plotted is the particle
kinetic energy versus time.  The earliest arriving particles are
the ones with the highest energy since they move with the highest speed.
The slower ones arrive later.  This velocity dispersion leads to the
characteristic profile shown in the figure.

It is clear from Fig.~\ref{Giacalone_Fig11} that particles released at
the Sun and observed near Earth undergo pitch-angle scattering in the
inner heliosphere, because at any given time there is a range of
particle energies detected.  That is, high energy particles can arrive
later because they have scattered in the medium between the source and
the observer.  Thus, the 'thickness' of the comma-shaped particle
event seen in the middle of this figure is related to the scattering
frequency of the particles.\sactivity{$\circledS$ {\em Show:} (1) Some
SEP events suggest that the particles arrive to the observer with
little or no scattering in the inner heliosphere. This most often
holds for the earliest arriving particles. Assuming this to be the
case, answer the following: (1a) For the moment ignoring diffusive
dispersal, use a ballistic model to estimate the time at which the
particles in the first event shown in Fig.~\ref{Giacalone_Fig11} were
released at the Sun. Consider whether you need to account for a
relativistic correction. Note that the display in
Fig.~\ref{Giacalone_Fig11} implicitly uses the approximation that the
energy of SEPs scales (roughly) with their mass.  (1b) With that
timing information, estimate the path length for the mean of the
particle population from the Sun to the Earth; why is that larger (by
some 10{\%}-20\%) than the Sun-Earth distance? Remember
Sect.\,\ref{sec:parker-spiral}.  (2) Now assume that the bulk of the
particles following the early arrivers propagates subject to
scattering. Use Eq.~(\ref{eq:Giacalone_Eq1.25}) together with the time
of the peak in the particle flux, with the initial event time and path
length derived in (1), to show that the diffusion coefficient $\kappa$
at $\approx$0.04\,MeV/nucleon is very roughly $10^{21}$\,cm$^2$/s.
\mylabel{act:sepprop} Note the dropouts in Fig.~\ref{Giacalone_Fig11};
these are discussed in the next-to-last paragraph of
Sect.~\ref{sec:septransport}. \mylabel{act:sepsun} \solution{sepsun}}
Aside from this, however, there are many features in this event that
are difficult to explain with a diffusive-advection-energy change
approach [\ldots]

It is noteworthy to point out another feature of the event shown in
Fig.~\ref{Giacalone_Fig11}.  There are intermittent dropouts in
intensity during each of the two distinct events shown.  These
dropouts have been interpreted as resulting from the passage of
alternatively filled and empty 'tubes' of particle flux by the
spacecraft.  The connection to the source, {\em i.e.,} the flare site,
determines which field lines are populated with particles and which
are not. [\ldots\ The dropouts suggest some braiding in the
heliospheric magnetic field such that adjacent field lines can connect
to different source regions on the Sun; see Fig.~HII:9.12.]

These observations indicate that solar-energetic particles associated
with impulsive solar flares undergo little cross-field transport,
otherwise, these intermittent dropouts would not exist.  This, of
course, leads to the interesting puzzle of why galactic cosmic rays,
or other types of energetic particles, do not exhibit such behavior.
The answer is simply that the energetic particles in impulsive SEP
events were relatively recently injected into the system and therefore
have not had time to scatter sufficiently to become more spatially
uniform.  GCRs, however, have spent much more time in the Solar System
(see Section~\ref{sec:gcrtransport}).  Thus, impulsive SEP events
reveal the early time behavior of a collection of energetic charged
particles moving in the heliospheric magnetic field.''

\subsection{Galactic cosmic rays}\label{sec:gcrtransport}
\ors[II:9.5.2] ``GCRs \indexit{cosmic ray}are cosmic rays
that pervade interstellar space and enter the heliosphere from the
outside.  The vast majority of them are swept out of the heliosphere
before ever reaching Earth's orbit.  [\ldots\ For] the purpose of a
simple illustration of modulation, consider the steady-state Parker
transport equation in one-dimensional spherical coordinates given by
\begin{equation} {\partial f \over \partial t} = {1\over
r^2}{\partial\over\partial r}\bigg( r^2\kappa{\partial f\over\partial
r}\bigg) - v{\partial f\over\partial r} + {2vp\over 3r}{\partial
f\over\partial p} = 0,
\label{eq:Giacalone_Eq1.26}
\end{equation}
\noindent [Note that this derives from
Eq.~(\ref{eq:boltzmanndiffusion}) with \tc{a} set to zero and for
$D_{ij}=0$ and uses that in spherical symmetry with an assumed
constant solar wind speed $u$ (and neglecting gravity) one has
$\partial v / \partial x_i = 2v/r$] (this simple illustration neglects
the effect of the heliosheath and termination shock).  Here we have
taken the diffusion tensor to be symmetric and $\kappa_{rr}=\kappa$.

\begin{figure}[t] \centering \includegraphics[clip=true,trim=0cm 0cm
0cm 0cm,width=11cm]{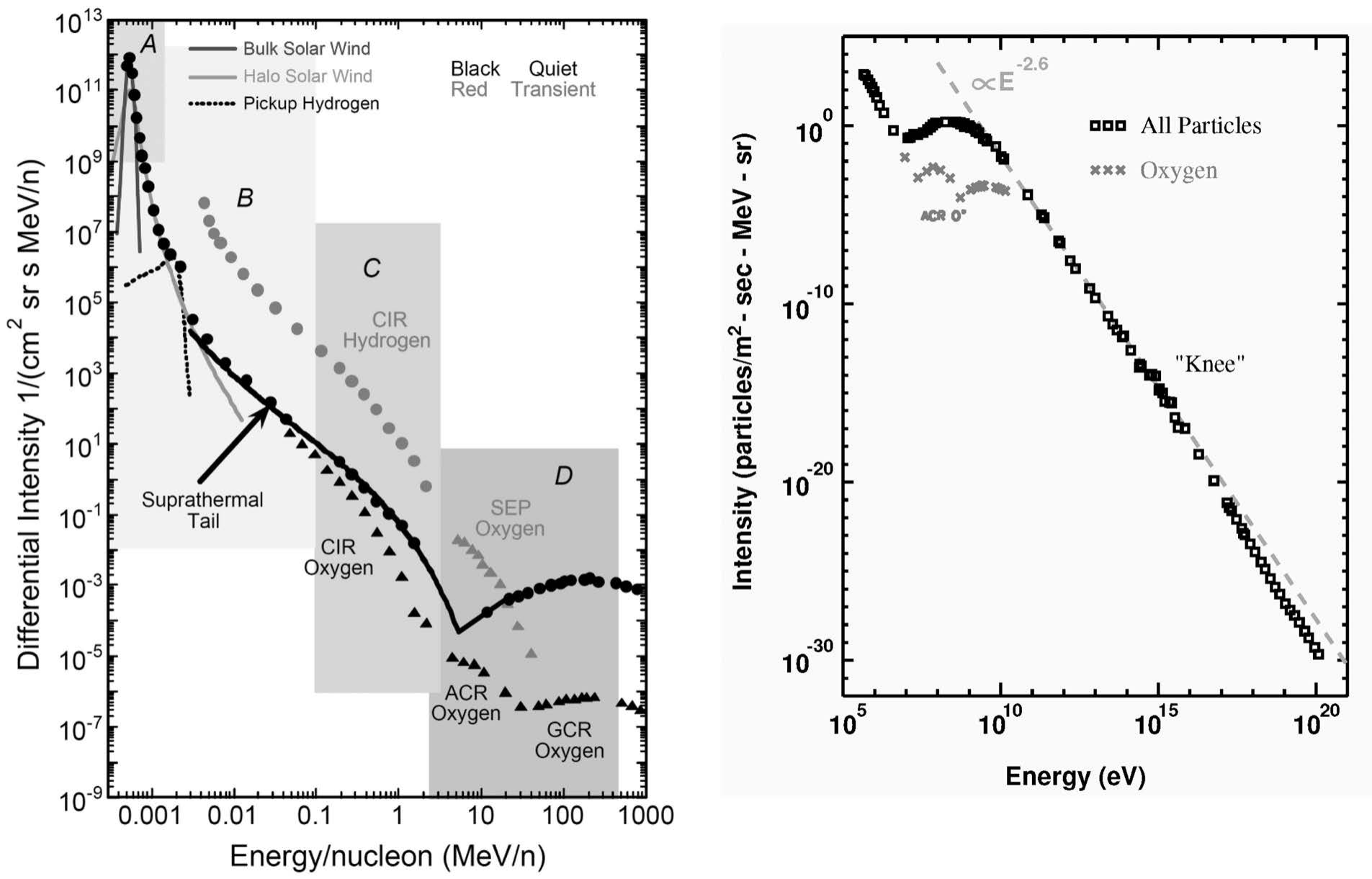}
\caption[Energy spectra of energetic particles in the heliosphere and
for GCRs.]{Energy spectra of energetic particles in the heliosphere
(left) and for cosmic rays (right). The curves illustrate the energy
spectra during quiet time and disturbed solar wind conditions. The
dots and triangles represent the supra-thermal part of the spectrum
and the particles accelerated at corotating interaction regions
(CIRs), galactic cosmic rays (GCRs), and the anomalous cosmic rays
(ACRs) together with Solar Energetic Particles (SEPs). The right
figure shows the high-energy part of the galactic cosmic ray energy
spectrum to TeV energies. Note the characteristic peak at about
10\,MeV and the $E^{-2.6}$ power-law dependence for energies above the
peak.  [Fig.~IV:12.2]}
\label{fig:Krupp_HelioIV_EnergySpectra}\label{fig:jok2}
\end{figure}\figindex{../krupp/art/Krupp_HelioIV_EnergySpectra.eps} It
is convenient to rewrite Eq.~(\ref{eq:Giacalone_Eq1.26}) in the
following form:
\begin{equation} {1\over r^2}{\partial\over\partial
r}r^2\bigg(\kappa{\partial f\over\partial r} - vf\bigg) + {2v\over
3rp^2}{\partial\over\partial p}(p^3f) = 0.
\label{eq:Giacalone_Eq1.27}
\end{equation} Generally this equation is not easy to solve, but if we
assume that the last term on the left (describing the energy change of
diffusing particles) is negligible, the resulting equation is readily
solved to yield
\begin{equation} f(r,p) = f(R,p)\exp\bigg( -\int_r^R
{v\over\kappa(r^\prime,p)}{\rm d}r^\prime\bigg).
\label{eq:Giacalone_Eq1.28}
\end{equation} Equation~(\ref{eq:Giacalone_Eq1.28}) gives an
exponential decay of particles from the source ($r=R$) inward, into
the the Solar System (where $r<R$).  Moreover, it is reasonable to
expect the diffusion coefficient to increase with momentum $p$ so that
higher-energy particles have a larger diffusion coefficient than
lower-energy particles.  Thus, higher-energy particles have a longer
exponential-decay length, or diffusive skin depth, than do lower
energy ones.  Thus, they more easily reach the inner heliosphere than
lower-energy cosmic rays.  This leads to a turnover in the spectrum
that is due to modulation.  This is in qualitative agreement with the
observed cosmic-ray spectrum at Earth as shown in
Fig.~\ref{fig:jok2}.''

\begin{figure}[t] \centering \centerline{\hspace{0.3cm}
$A>0$\hspace{.6cm} $A<0$\hspace{.6cm} $A>0$\hspace{.6cm}
$A<0$\hspace{.6cm} $A>0$ } \includegraphics[width=27pc]{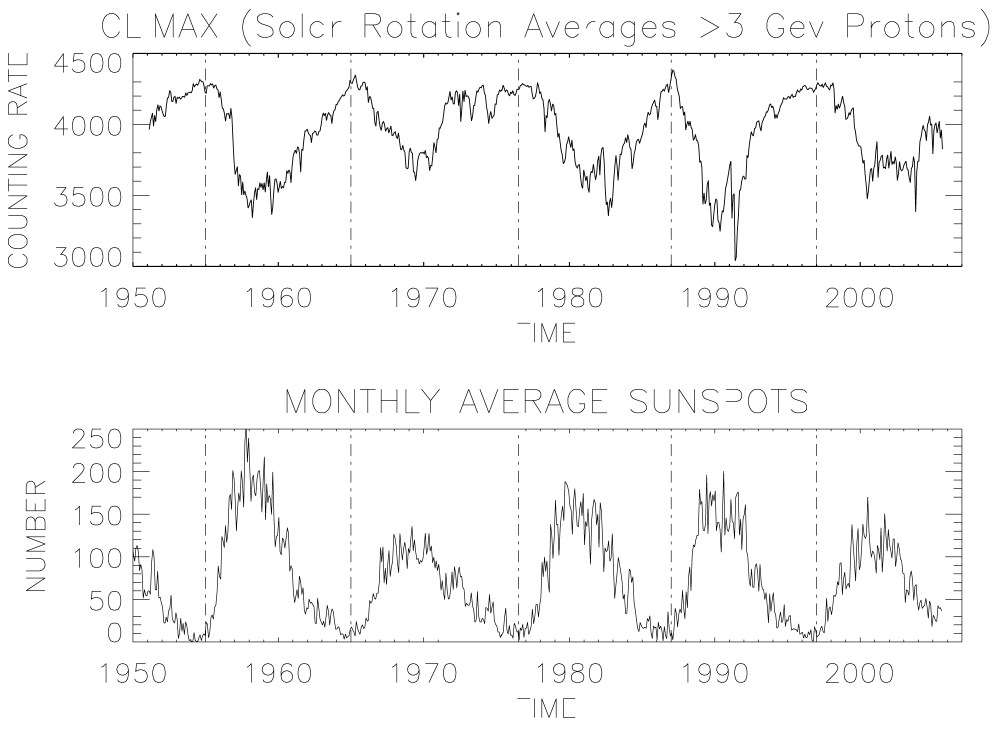}
\caption[Modulation of galactic cosmic rays during five sunspot
cycles.]{[{\rm top}) Climax neutron monitor daily count rate of
neutrons produced by the interaction of a primary cosmic ray with
Earth's atmosphere, which quantifies the] modulation of galactic
cosmic rays [near Earth orbit] during five sunspot cycles (shown in
the {\rm bottom} panel).  Note the alternation in the cosmic-ray
maxima between sharply peaked and more-rounded shapes.  [The meaning
of $A$ is defined in
Fig.~\ref{Giacalone_Fig14}. Fig.~III:9.4] \label{Giacalone_Fig13}}
\end{figure}

\label{sec:Giacalone_Sec5.4} The GCR intensity at a given orbital
distance from the Sun is not a constant but varies with the solar
cycle. \ors[II:9.5.4] ``Shown in Fig.~\ref{Giacalone_Fig13} is the
daily count of neutrons produced by the impact of cosmic rays on the
upper atmosphere, from ground-based neutron monitors.  This is an
indirect measure of the cosmic-ray flux in near-Earth orbit.  The
time-intensity profile shows a clear 11-year\indexit{cosmic ray!11-year cycle} cycle that is coincident with the sunspot-number cycle.
During periods of high solar activity, sunspot maximum, the cosmic-ray
flux is low, and during periods of low solar activity, or solar
minimum, the cosmic-ray flux is high.  In addition to this, there is
also 22-year\indexit{cosmic ray!22-year cycle} cycle present (the
alternating 'leveled' {\em vs.}\ 'rounded' cosmic-ray flux), which, as
we discuss below, is related to the drift motions of cosmic rays.

The increased modulation during periods of solar maximum is related to
a combination of effects related to the shedding of magnetic flux by
the Sun at solar maximum.  One the one hand, increased solar activity
leads to more magnetic turbulence which decreases the diffusion
coefficient in the outer heliosphere leading to more modulation.  On
the other hand, and in addition to this, the merging of more numerous
transient shocks and coronal mass ejections in the distant heliosphere
creates magnetic barriers\indexit{global merged interaction region}
(so-called global merged interaction regions, or GMIRs) which also
reduce the transport of cosmic rays into the inner heliosphere.  There
is a lower level of magnetic turbulence and fewer magnetic barriers
for cosmic rays to propagate through during solar minimum.  This is a
qualitative explanation for the 11-year cosmic-ray cycle and its
relation to the sunspot cycle.

The 22-year cosmic-ray cycle seen in Fig.~\ref{Giacalone_Fig13} is
related to the 22-year solar magnetic polarity cycle [because the]
polarity of the Sun's magnetic field is important for the cosmic-ray
drift that arises from the antisymmetric part of the diffusion tensor
in Parker's transport equation. [\ldots]

\begin{figure} \centering %\includegraphics[trim=15 300 305 540, clip,
                          %width=7cm]{figures/Fig14.ps}\includegraphics[width=11pc]{figures/fig8.eps}
%\includegraphics[viewport=0 0 295 200, clip, width=9.15cm]{figures/Fig14}\includegraphics[width=3.5cm]{figures/fig8}
%\includegraphics[width=9.15cm,bb= 15 300 297 490,clip=true]{figures/Fig14}\includegraphics[width=3.5cm]{figures/fig8}
\includegraphics[width=9.15cm]{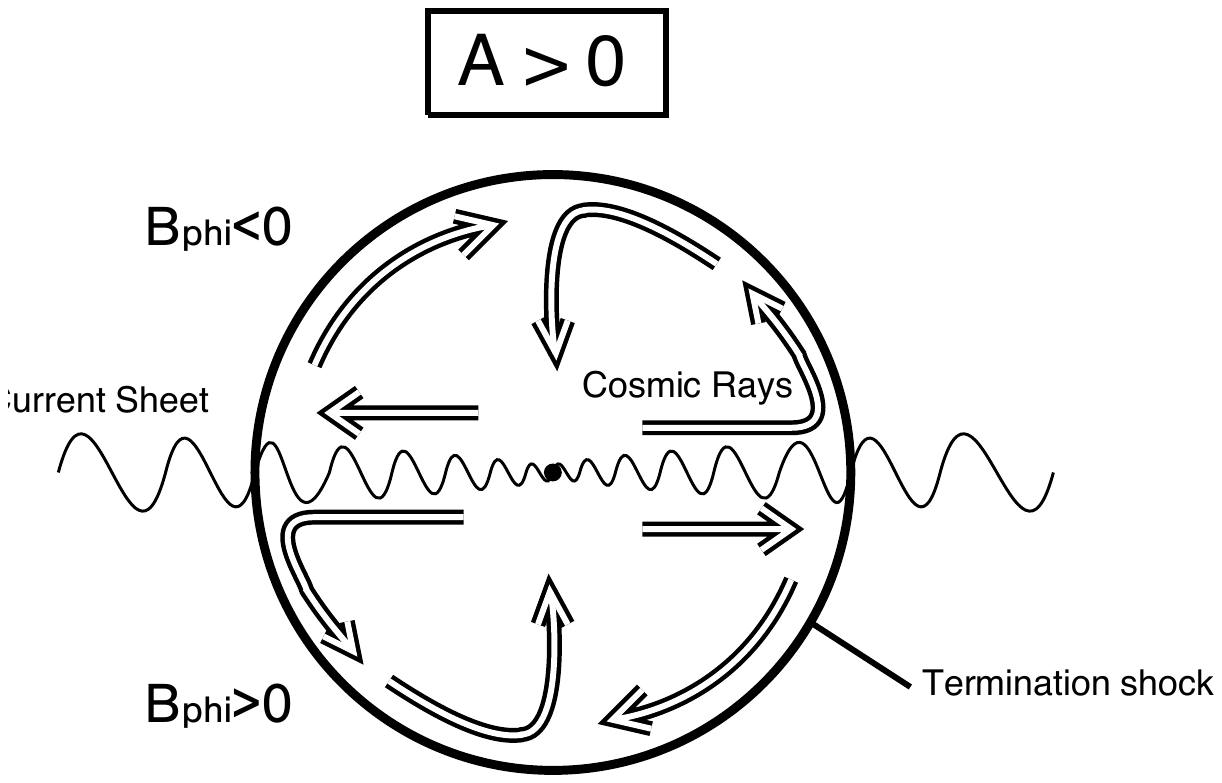}\includegraphics[width=3.5cm]{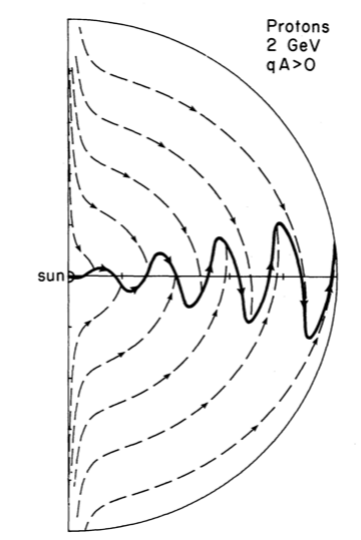}
\caption[Drift motion of cosmic rays in a Parker-spiral heliospheric
field.]{[{\em (left)} Simplified side view of the heliosphere, with
the current sheet depicted by wavy lines, to illustrated the drift]
motion of cosmic rays in the heliosphere for [one solar magnetic
polarity.  The two polarities of the heliospheric] field are separated
by the heliospheric current sheet. The value of $A>0$ is for the
period during which the solar magnetic field is outward in the north
and inward in the south. During the next sunspot cycle, $A<0$, the
heliospheric field polarities are reversed, along with the direction
of cosmic-ray advection.]  The termination of the solar wind is also
shown. [This is a cropped version of Fig.~II:9.14.]  {\em [(right)]}
Cosmic-ray drift motions in a Parker spiral magnetic field with a
current sheet.  The arrows shown correspond to the time when the
northern-hemisphere heliospheric magnetic field is outward from the
Sun ([$A>0$, as in] 1975, 1996) for positively charged particles.  The
arrows [in both panels] reverse for the alternate sign of the magnetic
field ([$A<0$, as in] 1986, 2007) and for the opposite sign of the
particle's electric charge. [Fig.~III:9.8;
\href{https://ui.adsabs.harvard.edu/abs/1981ApJ...243.1115J/abstract}{source
right panel: \cite{1981ApJ...243.1115J}}.]}
\label{Giacalone_Fig14}
\end{figure}

Including the drifts of cosmic rays has led to the widely accepted
paradigm for cosmic ray transport shown in Fig.~\ref{Giacalone_Fig14}.
Drift motions for protons during two different solar polarity cycles
are shown.  [\ldots] During the period in which the solar magnetic
field spirals outward in the north and inward in the south ($A>0$) the
GCR protons drift into the heliosphere from the polar regions of the
heliosphere and outward along the heliospheric current sheet (which
separates the two hemispheres and where the field reverses direction,
hence the term 'current sheet').  During the opposite polarity, in
which the solar field is inward in the north and outward in the south
($A<0$), galactic cosmic-ray protons drift into the heliosphere along
the current sheet.  Note that in addition to the drift along the
current sheet, there is also a gradient-B drift along the termination
shock resulting from the jump in the magnetic field strength across
the shock.

The explanation for the alternating leveled and rounded cosmic ray
intensity involves both the drift motions of the cosmic rays shown in
Fig.~\ref{Giacalone_Fig14}, and the 'waviness' of the heliospheric
current sheet due to the offset of the solar magnetic axis and its
rotation axis.  When the 'tilt' is large, the current sheet is very
warped, whereas, when it is small, the current sheet is much flatter
(imagine the current sheet [forming above the rotating Sun with a
tilted axial dipole, as in Fig.~\ref{fig:modelcs}]).  The current
sheet is generally known to be relatively flat during the center of
the solar cycle minimum.  So, during the cycle in which the cosmic
rays come into the heliosphere along the current sheet [{when $A<0$,
so opposite to the depiction in Fig.\,\ref{Giacalone_Fig14}], only
when it is very flat will the full cosmic-ray flux be reached at
Earth's orbit.  Thus, during this phase, the cosmic-ray intensity will
exhibit a rounded or 'peaked' time-intensity profile.  When the cosmic
rays come in along the poles of the heliosphere, the full intensity is
reached much sooner and remains at a high level throughout solar
minimum, and hence, during this phase, the time-intensity profile is
more level, or flat.''

\section{Particle acceleration in shocks}\label{sec:shockacceleration}
Shocks provide an effective means to \index{shock!particle acceleration}increase the kinetic energy of
individual particles. \indexit{energetic!particle!acceleration}For an ensemble of particles, this may shift
their Maxwellian velocity distribution to a higher temperature, may
distort that Maxwellian outside its core range, or lead to pronounced
high-energy tails. Some form of shock heating and shock acceleration
may play a role in processes as diverse as coronal heating (see
Ch.~\ref{ch:heating}) and \indexit{energetic!particles!shock
acceleration} the formation of solar energetic \indexit{shock!heating
versus acceleration} particles. \ors[II:8.3.2] ``Some of the processes
that heat particles at thermal energies will also elevate the energy
at the upper part of the range.  However, such enhancements are often
only by a more or less constant, relatively minor factor.  An example
of this is the adiabatic heating of ions due to the magnetic field
compression at the relatively narrow oblique or nearly-perpendicular
shocks associated with the compression of the plasma'' (for a
quantitative example for low-$\beta$ conditions as they occur, for
example, in the solar corona, see Fig.~\ref{fig:kv0}).  \ors[II:8.2]
``[T]he shocks of most interest to particle acceleration are MHD fast
mode shocks, which compress both the density and the magnetic
field. [Here, acceleration may span several orders of magnitude, and
may significantly alter the shape of the energy
distribution. However,] slow mode shocks\indexit{shock!slow mode} may
also play a role under certain circumstances.''

\ors[III:8.3.1] ``All mechanisms that contribute to the acceleration
of charged particles at shocks rely on the particle orbits in the
spatial and temporal features of electric and magnetic field
environment of the shock.  Roughly speaking, such processes are called
kinetic when\indexit{kinetic process!description} they go beyond the
fluid (MHD) properties of the shock, when they are related to the
scales associated with the charged particle motion, and when they
require some self-consistent back-reaction between the charged
particles and the plasma, {\em e.g.,} in the form of wave generation.
For the highest particle energies, gyro-radii are so large that the
size of the shock transition and even that of many local waves no
longer matter.  Conversely, for the thermal and so-called
supra-thermal particles (just above the thermal energy to several
thermal energies), the intrinsic shock scales and locally-generated
waves do matter.  As a consequence, the intrinsic shock scales and
associated mechanisms play an important role not only for the general
dissipation at the shock (the conversion to thermal energy), but also
in providing a first, background level of energetic particles from
'seed particles' in the thermal and supra-thermal energy
range. [\ldots]

\def\comp{\lambda_{\rm pi}} \def\rhop{r_{\rm gp}}
\def\betap{\beta_{\rm p}} The two most important scales in
collisionless shocks are the proton inertial length $\comp =
c/\omega_{\rm pi}$ (see Eq.~\ref{eq:ioninertial}; [$\omega_{\rm pi}$
the ion plasma frequency]) and the proton gyro-radius $\rhop=m_{\rm
p}vc/eB$, which are related via the proton beta by $\rhop / \comp =
\sqrt(\betap)$. [\ldots\ The width of the transition for many shocks
is the larger of $\comp$ and the distance $v_1 / \omega_{\rm gp}$
which the upstream flow, moving at speed $v_1$ in the normal incidence
frame (NIF, see Fig.~\ref{fig:kv1}), travels during the time
$1/\omega_{\rm gp}$ for a single gyration of a proton.] Exceptions are
the almost perpendicular shock [see footnote~\ref{note:shockterm} on
terminology)], which can be cyclically reforming and steepen to
electron scales, and quasi-parallel shocks, which are not only
reforming, but at sufficient Mach number have extended regions of
steepening upstream waves, and highly non-linear turbulence
downstream.''

\ors[II:8.3.2] ``In most shocks in the heliosphere, the thermalization
of the upstream flow is primarily achieved via the ion dynamics,
whereas the electrons mostly 'just go along for the ride,' {\em i.e.,}
they move almost adiabatically, with some subsequent scattering that
fills otherwise inaccessible regions in the downstream velocity space.
Any heating of the electrons (which can be quite small) is important
in regulating the so-called cross-shock potential, because much of the
electron phase space needs to be confined to the downstream by a
potential, to prevent escape of the highly mobile electrons and to
preserve overall charge neutrality. [\ldots]

In typical shocks of the interplanetary medium, and in planetary bow
shocks, it has been established that the reflection and gyration of
the incoming ions plays a dominant role.  At oblique shocks, part of
the incoming ion phase space is reflected, but then convected back
into the downstream.  That is, after reflection, at sufficient Mach
number, any upstream-directed parallel velocity of most thermal and
even of many supra-thermal particles is not sufficient to overcome the
general plasma drift into the shock.  Much of the converted flow
energy is initially stored in these gyrating ions, which during this
process have attained elevated perpendicular temperatures from the
magnetic field jump. Depending on parameters, it may take a while
before these protons are thermalized downstream, typically in
Alfv{\'e}n wave turbulence driven by the temperature anisotropy
$T_\perp > T_\parallel$.  Generally speaking, the closer to
perpendicular the shock, the more difficult it is for both particles
and waves to escape upstream.

In contrast, in quasi-parallel shocks reflected (and partially
gyrating) ions also play a role, but they can much more easily escape
upstream against the flow, because the magnetic field direction is
close to the shock normal.  There, they generate both
obliquely-propagating, compressional fast-mode waves, and
parallel-propagating Alfv{\'e}n waves.  These waves can grow to large,
non-linear amplitudes while convected back towards the shock, where
the beam density and growth rate are largest.  However, below
Alfv{\'e}n mach numbers of about $M_{\rm A} < 2.8$, the majority of
resonantly generated waves are no longer convected back and therefore
do not steepen as easily and do not impact the shock any longer, thus
resulting in fewer ions making it upstream to generate waves in the
first place.  [The] resulting lower level of turbulence also has a
negative impact on ion acceleration to higher energies.''

\ors[II:8.3.3] ``For most heliospheric shocks, proton acceleration is
of prime interest.  Protons can easily reach energies of tens, if not
hundreds of MeV, and as such have a large range of societal
consequences such as malfunction or destruction of equipment in space,
and posing danger to astronauts or crew and passengers of high-flying
aircraft.  Electrons, on the other hand, are rarely accelerated to
comparable fluxes at these energies, except perhaps at processes well
inside the magnetosphere that periodically lead to huge enhancements
of trapped populations (see Section~\ref{sec:radbelts}).''

\begin{figure}[t]
%\centerline{\hbox{\psfig{figure=figures/DKV_fig_2BW.eps,width=\textwidth,clip=}}}
\centerline{\hbox{\includegraphics[width=0.8\textwidth]{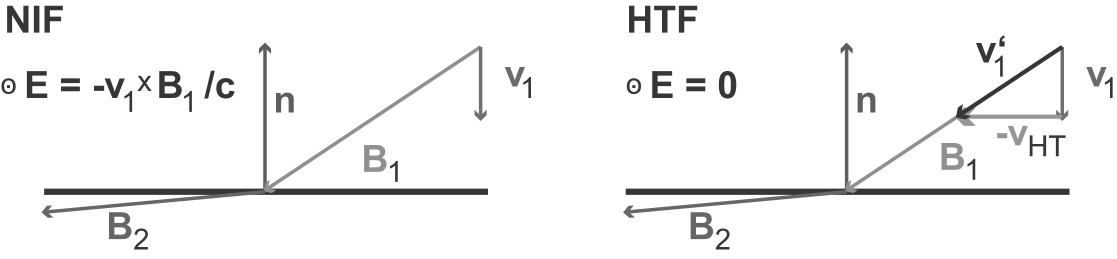}}}
\caption[Normal-incidence frame (NIF) {\em vs.}\ de Hoffman-Teller
frame (HTF).]{Comparison of the normal-incidence frame (NIF) and de
Hoffman-Teller frame (HTF) at fast-mode shocks.  The NIF is the shock
frame in which the upstream flow is aligned with the shock
normal. [\ldots] Transformation to the HTF is along the plane shock
surface until the upstream flow vector coincides with the magnetic
field. [\ldots\ Fig.~II:8.2] }\label{fig:kv1}
\end{figure} \ors[II:8.4] ``For ions, and for the energy range
typically observed in the heliosphere, it is well accepted that two
distinct acceleration mechanisms are at play:'' (a) shock-drift
acceleration and (b) diffusive shock acceleration.
 
\subsection*{(a) Shock-drift acceleration}\label{sec:kv4.2}
\ors[II:8.4.2] ``[R]eflection \indexit{shock-drift acceleration}of a portion of the incoming proton
phase space [by the shock], and subsequent convection downstream, is
the prime mechanism that eventually provides the [acceleration and
heating] at quasi-perpendicular shocks.  Even at highly oblique
shocks, a small fraction of these ions will have sufficient parallel
speed to make it upstream instead of being convected downstream, but
the flux of such ions is strongly diminishing [with increasing angle
between the shock normal and the upstream magnetic field,] making
upstream wave generation increasingly difficult.  Although the thermal
proton gyroradius is typically comparable to the shock width, and that
of supra-thermal ions clearly larger than the shock transition,
surprisingly, many ions approximately behave adiabatically in simple
shock transitions with sufficiently homogeneous upstream and
downstream fields.  A portion of the ion phase space then gains energy
through their gyromotion under consideration of the shock electric
fields.  The family of such processes is called shock-drift
acceleration (SDA).''

SDA \ors[II:8.4.1] ``is a 'kinematic' process in the sense that the
particles simply perform their usual, mostly adiabatic orbits in the
given, static or average electric and magnetic fields of the shock
transition, neglecting any scattering.  [\ldots]'' \ors[II:8.4.2]
``Consider a steady-state, one-dimensional shock.  In this case, in
the normal-incidence frame (NIF, see Fig.~\ref{fig:kv1}), there will
be an out-of-plane electric field given by the cross-product of the
upstream flow and magnetic field [\ldots] ${\bf E}_{\rm p} = -{\bf
v}_1 \times {\bf B}_1 / c$.  [The MHD Rankine-Hugoniot jump conditions
are such that the strength of this electric field is the same upstream
as downstream of the shock.  \activity{{\em Show:} Review the
Rankine-Hugoniot jump conditions (Eqs.~\ref{eq0} and
\ref{eq1}--\ref{eq5}) and show that the motional electric field ${\bf
E}_{\rm p}=-{\bf v}_1 \times {\bf B}_1 / c$ is constant across a
steady-state, one-dimensional shock.} This motional electric field is
aligned with the direction of both the curvature and gradient drifts
associated with the jump in ${\bf B}$ across the shock (see
Sect.~\ref{sec:singleparticle}, and around Eqs.~(\ref{eq:drifts})
and~(\ref{eq:Giacalone_Eq1.8})), in such a way that ions gain energy
by the gradient drift and lose energy through curvature drift.]  It
turns out that at quasi-perpendicular shocks, gradient drift wins out
for most ions, which then gain energy proportional to the distance
they drift along ${\bf E}_{\rm p}$.''

A perspective change to another inertial system provides an
alternative interpretation: in the so-called de~Hoffmann-Teller frame
(HTF, see Fig.~\ref{fig:kv1}) a translational velocity of $v_{\rm HT}
= v_1 \tan\thetabn$ is introduced so that the upstream flow and
magnetic field are aligned. As a result, there is no motional electric
field in this reference frame, and only energy conservation and
magnetic moment conservation come into play.  In the HTF
\ors[II:8.4.2] ``energy is conserved in the absence of other
processes, and the only allowed change absent scattering is between
the perpendicular and parallel velocity components.  For close to
perpendicular shocks, the field-aligned velocity component becomes
increasingly larger due to the transformation into the HTF.  [\ldots]
Because the perpendicular energy gain under magnetic moment
conservation is simply a factor based on $B_2 / B_1$, only ions with
sufficient initial perpendicular energy may exchange large fractions
of their velocity components, while slowing down significantly or
reflecting in the magnetic field gradient and in the cross-shock
potential.  Subsequent back-transformation shows that they have gained
energy proportional to the squared transformation velocity.  While
this energy gain can be huge close to $\thetabn \sim 90^\circ$, an
increasingly smaller subset of phase space has sufficient
perpendicular energy to effectively participate.''

Other mechanisms have been proposed, for electrons and ions alike,
some, like 'shock surfing' acceleration (SSA) rely on the differences
in gyro-radii and on the cross-shock potential; see Sect.~II:8.4 for
some more information. The challenge with all is that without
additional scattering mechanisms, all these processes are too limited
in the portion of phase space that is affected, and in the amount of
energy gain, to explain the large, highly-energized populations often
observed. It has been argued, however, that mechanisms like SDA and
SSA can add energy for particles already energized by another
mechanism, or that they provide the seed particles for such other
mechanism to continue the energization. Turbulence or particle-induced
waves may provide the scattering required to access more of the phase
space.

\subsection*{(b) Diffusive shock acceleration}\label{sec:kv4.3}

Diffusive \indexit{diffusive shock acceleration}shock acceleration
(also known as \indexit{first-order Fermi acceleration}first-order
Fermi acceleration) \ors[II:8.4.1] ``is of 'kinetic' nature, in the
sense that wave-particle interactions play the decisive role.  As
explained above, reflected or otherwise energized ions can easily
escape into the upstream at quasi-parallel shocks, where they
self-consistently'' generate waves. Once grown sufficiently, these
waves, and existing turbulence, diffusively scatter particles into a
population that ranges from the far upstream to the far downstream. As
the scattering centers converge owing to the compression associated
with the shock, repeated scattering results in energization until they
escape from the shock zone.

\ors[II:8.4.3] ``First-order Fermi \indexit{particle!acceleration!first-order Fermi} acceleration produces a\indexit{first-order Fermi acceleration|see{particle}} power-law distribution and intensities
that depend on the shock strength (compression ratio
$\rho_{2}/\rho_{1}$).  Power-law distributions are as ubiquitous for
SEPs as they are in cosmic plasmas, in general. [The] restricted
temporal and spatial dimensions available lead to an upper cut-off of
the spectra at high energies --~typically between 10\,MeV and 100\,MeV
for SEPs escaping interplanetary shocks. [\ldots] For particles that
are already significantly faster than the flow speed, the associated
momentum gain of a returning particle is: $\delta p/p = (v_1 - v_2) /
v (\cos \theta -\cos \theta\prime)$, where the prime denotes the new
pitch angle. [Note that the particles involved are quite energetic and
therefore have a mean free path length exceeding the shock width; thus
they sense the shock as a delta function, with the value of $(v_1 -
v_2)$ in the above expression reflecting the step associated with term
\tc{e} in Eq.~(\ref{eq:boltzmanndiffusion}).]

If one now assumes an almost isotropic distribution of particles, one
can average over all pitch angles, and the $\cos$ terms simply convert
into a constant factor.  One then proceeds to calculate the
probability of escape downstream (which is simply given by the ratio
of the downstream to upstream flux) versus the probability of an
acceleration cycle.  From the calculation it follows that the particle
distribution assumes a power law with index $q$, which depends on the
shock compression ratio: $q = 3 r / (r -1)$, where from mass
continuity in the assumed one-dimensional shock: $r = v_1 / v_2 = n_2
/ n_1$, {\em i.e.,} the compression ratio between the downstream and
upstream densities. [\ldots]

Because waves that make up efficient scattering centers should be
generated self-consistently by the energetic ions, must exist for
extended regions upstream and downstream of the shock, and should not
be convected towards or away from the shock too quickly, diffusive
shock acceleration is most efficient and easiest understood for fairly
high Mach number, almost parallel shocks.  Conversely, it is much less
understood how this process can be so efficient at the low-to-medium
Mach number, oblique shocks that make up most interplanetary shocks.
In particular, at nearly perpendicular shocks, diffusive acceleration
may require effective scattering across the magnetic field.''

There are multiple challenges to overcome in the study of diffusive
shock acceleration, including the large number of particles that need
to be tracked in numerical models, the relative roles of the various
processes and their dependence on specific geometries, the generation
of adequate turbulence to scatter particles, the complexities of the
self-generated upstream wave field with multiple possible modes, the
escape of the particles that have been energized from the upstream
wave field as well as their further propagation through the turbulence
in the solar-wind field, and the role of second-order Fermi
acceleration in which particles scatter off counter-propagating waves,
and, of course, the vast range of scales that needs to be treated.
\ors[II:8.4.4] ``It is also known that multiple shocks generate a much
more efficient acceleration environment.  Not only does the first
shock leave a much more turbulent and seed-particle rich upstream for
the following shock, but particles may scatter multiple times in both
shocks.  Of course, the upstream seed particle spectrum and background
turbulence are highly variable in the solar wind in general, and will
have an impact on achieved fluxes.''

\begin{figure}[t]
%\centerline{\hbox{\psfig{figure=figures/DKV_fig_3BW.eps,width=8cm,clip=}}}
%\centerline{\hbox{\psfig{figure=figures/DKV_fig_3.eps,width=8cm,clip=}}}
\centerline{\hbox{\includegraphics[width=9cm]{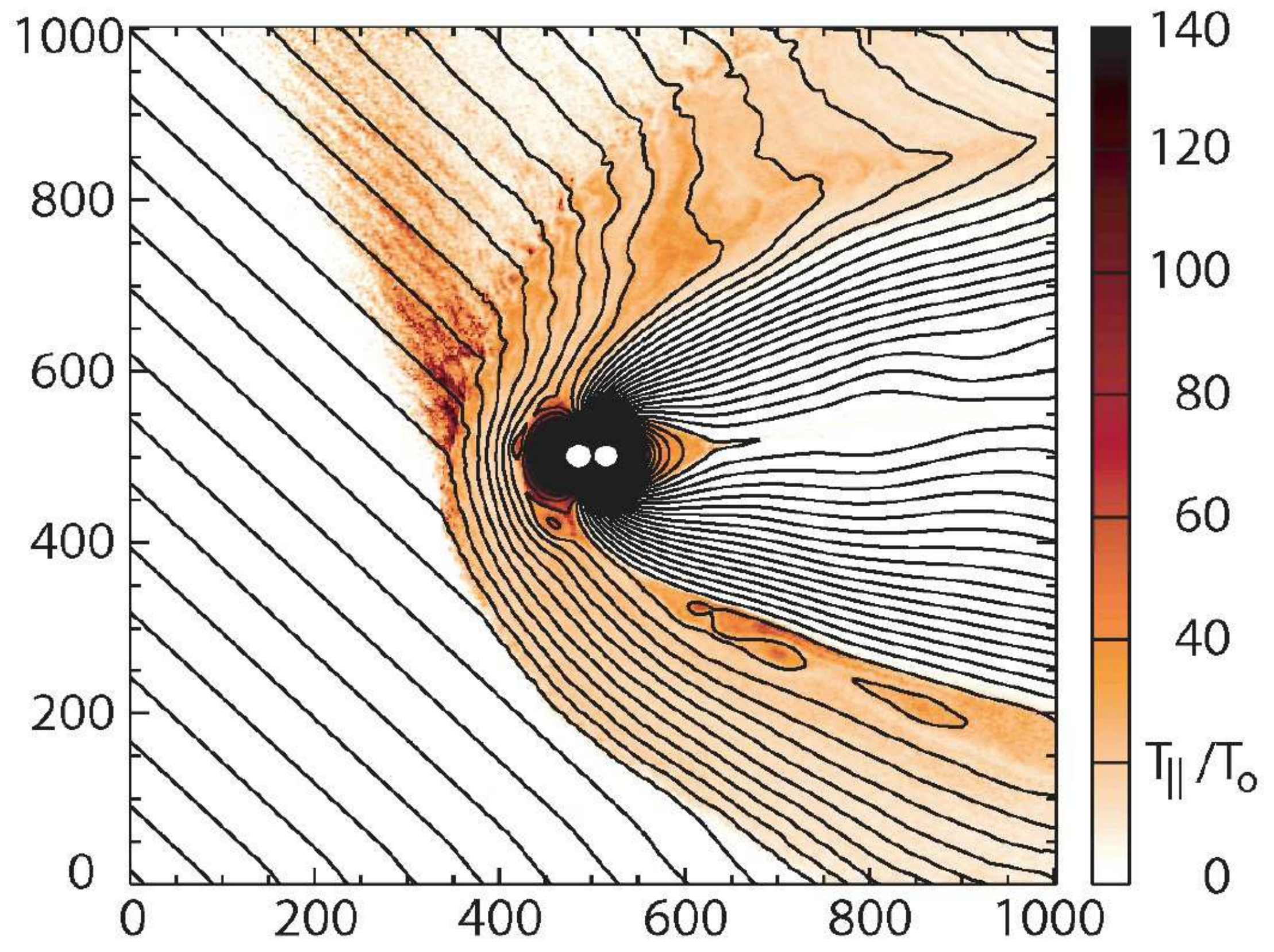}}}
\caption[2-D hybrid simulation of the solar wind -- magnetosphere
interaction.]{Example of a two-dimensional (2-D) hybrid simulation of
the solar wind -- magnetosphere interaction.  Shown are magnetic field
lines (upstream IMF angle $\theta = 45^\circ$) and, the normalized
parallel ion temperature $T_\parallel$, as a proxy of ion
acceleration.  As well-documented in many observations of the Earth's
bow shock, the ion foreshock starts close to $\thetabn = 45^\circ$
with energized and back-streaming ions, and simultaneous excitation of
waves (visible in the field line undulations).  Conversely, at this
scale, and with the number of pseudo-particles used in the simulation,
there are virtually no upstream ions at larger shock-normal
angles. [Fig.~II:8.3;
\href{https://ui.adsabs.harvard.edu/abs/2008AIPC.1039..307K/abstract}{source:
\citet{2008AIPC.1039..307K}}.] \colorfig }\label{fig:kv2}
\end{figure}

\subsection*{The Earth's bow shock}\label{sec:kv5.1}

\ors[II:8.5.1] ``[P]lanetary \indexit{Earth!bow shock}bow shocks are
of finite size, and as such, any production of energetic particles is
both localized and highly non-local: some regions ({\em i.e.,} the
quasi-parallel portion) are much more able to easily generate
energetic ions, while any ions propagating upstream, or waves excited
upstream of the oblique portion are quickly convected to a different
portion of the finite-size bow shock, or around the obstacle,
altogether.  The general scale size of the Earth's bow shock is of the
order of $20R_{\rm E}$ (Earth radii); the stand-off distance is
[typically some $15R_{\rm E}$. \ldots\ T]here is an ion foreshock that
starts somewhere below $\thetabn \sim 45^\circ$ and permeates the
quasi-parallel domain, while the faster electrons form a foreshock
boundary close to the perpendicular shock.

Figure~\ref{fig:kv2} shows a snapshot of a 2-D bow shock simulation to
further demonstrate this point.  [\ldots] The turbulence upstream and
downstream of the quasi-parallel portion is clearly visible, as is the
large enhancement of upstream-propagating, energetic protons.
Conversely, there is virtually no upstream activity at or beyond
$45^\circ$.  [This approximate description conforms with the general
state of the terrestrial environment and] also illustrates why there
is so little activity upstream of the oblique portion beyond $\thetabn
\sim 45^\circ$: any ions that manage to make it upstream of the
oblique portion, and any waves generated there, are either convected
into the quasi-parallel portion of the bow shock, or instead move past
the finite-sized obstacle altogether.''

\subsection*{Interplanetary shocks}\label{sec:kv5.2}

\ors[II:8.5] ``Both \indexit{interplanetary!shock}co-rotating
interaction regions and CME-driven shocks are capable of accelerating
charged particles; however, not surprisingly, the largest events are
associated with the fastest CMEs and can reach Alfv{\'e}n Mach numbers
of 5 to 6, and occasionally even higher.  These Mach numbers are
comparable to the Earth's bow shock; yet, energetic particle energies
and fluxes observed at the bow shock are almost dismal compared to
those at the largest CME-driven events.  Yet, while the Earth's bow
shock virtually always generates upstream energetic ions, the same
cannot be said for IP shocks.  [\ldots] Finally, the heliospheric
termination shock is also generally viewed as capable of producing
highly-energized ions.''

\begin{figure}[t] \centering \includegraphics[clip=true,trim=0cm 0cm
0cm 0cm,width=\textwidth]{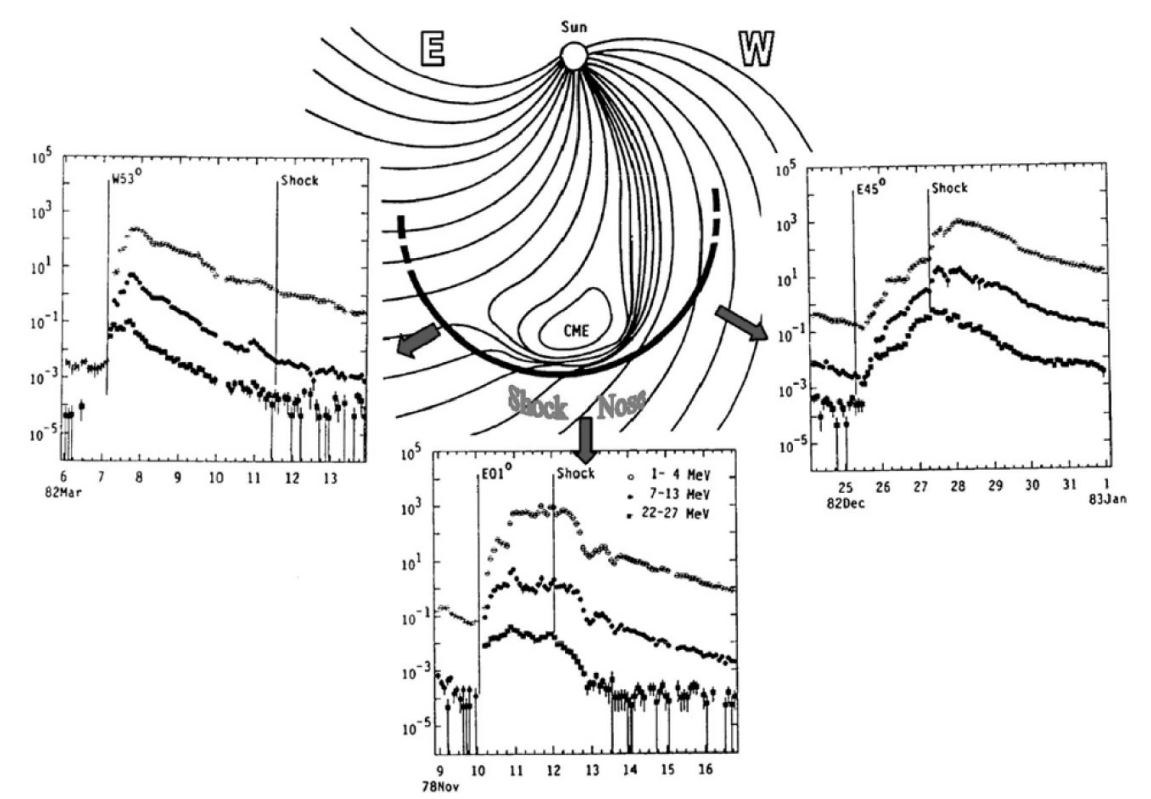}
\caption[Particle flux profiles from three different solar
longitudes.]{Intensity-time plots of particle fluxes ejected from
three different solar longitudes with respect to the nose of the shock
front. [IV:12.7;
\href{https://ui.adsabs.harvard.edu/abs/2013SSRv..175...53R/abstract}{source:
\citet{2013SSRv..175...53R}}.]}
\label{fig:Krupp_HelioIV_Reames2013-1}
\end{figure}\figindex{../krupp/art/Krupp_HelioIV_Reames2013-1.eps}
\nocite{Reames:13b} %\nocite{Reames:99a} \nocite{CaneHV:88a}
\ors[II:8.5.2] ``Interplanetary (IP)\indexit{interplanetary!shock!particle acceleration} shocks\indexit{particle!acceleration!interplanetary shocks} have a great variety of strength, and most of
them are actually not particularly active when it comes to energetic
particles.  At the other extreme are IP shocks that are associated
with strong solar energetic particle events (SEPs).  Today, it is
thought that SEPs are generated both by flare processes deep in the
solar corona, and by shocks driven by coronal mass ejections (CMEs)
[\ldots] In fact, it is estimated that almost all of the magnetic
energy released in flares goes into energetic particles, with perhaps
approximately equal share between the ions and electrons.  These
particles show up as 'prompt' events when observed at Earth: extremely
energetic ions can traverse the distance from the Sun in minutes, with
little delay compared to observed X-ray flare signatures at the Sun.

Conversely, so-called 'gradual' solar energetic particle events are
generally accepted to be associated with coronal and interplanetary
(IP) shock acceleration, driven by coronal mass ejections (CMEs).
Even in this case, the most energetic particles are produced when the
shock is in the corona, with resulting hard spectra that are observed
at Earth within tens of minutes.  However, production of energetic
ions continues to 1\,AU and beyond, and peak fluxes, with a softer
spectrum [(meaning dropping faster with increasing particle energy)],
often arrive at Earth with the shock itself --~historically called
energetic storm particle (ESP) events
[(Fig.~\ref{fig:Krupp_HelioIV_Reames2013-1})]. \activity{{\em
Consider:} Interpret the flux profiles as a function of time shown in
Fig.~\ref{fig:Krupp_HelioIV_Reames2013-1} for the three different
perspectives (with Earth in the direction of the arrows for three
different events). Argue for the differences in timing of the solar
event (first vertical line in each panel) and the passage of the shock
(second line) relative to the timing of the peak fluxes.}

[F]orward propagating interplanetary (fast mode) shocks near 1 AU
typically have an Alfv{\'e}n Mach number $M_{\rm A} < 3$, [only rarely
reaching $M_{\rm A} > 4$. D]espite their low Mach number, about one
half of [the sample of] observed shocks had identifiable (albeit
relatively low energy) upstream, energized ions and associated waves.
[The] distribution of these ions [are] fairly isotropic, whereas in
only a few cases, upstream beams where observed.  This behavior also
extends to higher energies and may be interpreted as a consequence of
the large spatio-temporal scales of IP shocks, which rarely allows one
to see the initial evolution of wave-particle interactions.  While the
large scales provide an important clue, and energetic seed particles
may play an additional role, currently no scenario self-consistently
accounts for the observed energetic ion environment of the weaker and
oblique shocks.''

\section{Relativistic particles in planetary radiation
belts}\label{sec:radbelts}
\subsection{Electron acceleration mechanisms} Earth's electron
radiation \indexit{radiation belt!acceleration}belts are located at
$L\approx 3-10$, typically peaking around $L\approx 4-5$ (where $L$ is
the radial distance in Earth radii where magnetic field lines of an
unperturbed dipole would cross the magnetic equator).  \ors[II:11.4.1]
``Many acceleration mechanisms have been proposed
to\indexit{relativistic electron acceleration mechanism} explain
electron radiation belt flux increases at Earth but their exact
contributions are still debated. Proposed acceleration mechanisms are
often separated into two categories: internal (or local) source
acceleration and external source acceleration. External source
acceleration mechanisms are so named because they move electrons from
outside geosynchronous orbit (at 6.6\,$R_{\rm E}$) to the inner
magnetosphere accelerating electrons through the transport
process. They operate over large spatial and temporal scales that
violate the particles' third adiabatic invariant. Internal source
acceleration mechanisms, on the other hand, locally accelerate
electrons in the inner magnetosphere inside of 6.6\,$R_{\rm E}$. They
operate on fast timescales and small spatial scales and violate all
three adiabatic invariants. The most prominent of the proposed
mechanisms in each category are listed below. [\ldots]''

\subsubsection{External acceleration mechanisms} \ors[II:11.4.1.1]
``The\indexit{relativistic electron acceleration mechanism! external}
manner in which external mechanisms accelerate particles can be
illustrated starting with the assumption that the first adiabatic
invariant [(Eq.~(\ref{eq:firstinvariant})] is conserved. These
mechanisms move electrons radially inward where the magnetic field is
stronger. Because $\mu_{\rm m}$ is conserved during the transport
process, the increase in field strength requires that the particles'
perpendicular energy also increase. The total energy gain is directly
related to the amount of radial transport.  The relationship between
transport and acceleration is easy to describe using the conservation
of the first adiabatic invariant but the explanation hides the complex
physics of the acceleration. Ultimately, it is an electric field that
transports and accelerates the electrons because the magnetic field
cannot change the particle energy. What separates the acceleration
mechanisms is the exact form and timescale of that electric field.
The electric field in both shock-induced acceleration and substorm
induced\indexit{shock-induced acceleration} acceleration is a
large-scale inductive electric field that sweeps through the
magnetosphere as the global magnetic field changes.

The shock-induced electric field is caused by the compression of the
magnetosphere as shocked solar wind passes Earth. [\ldots] However,
such large sudden events are rare [while] smaller more pervasive
compressions do not contribute significantly to electron radiation
belt flux increases. Thus, shock acceleration is usually only
discussed for specific events and not the very common flux increases
that occur with most geomagnetic storms.

The\indexit{substorm!electric field} substorm electric field is
produced when the stretched magnetotail is pinched off near 10 $R_{\rm
E}$ and the remaining plasma is hurled Earthward resulting in a more
dipolar magnetic field configuration. [With this mechanism, numerical
models have difficulty in transporting electrons inside of 10\,$R_{\rm
E}$; it may be that substorms] contribute to a seed population of
electrons at large radial distance but some other mechanism, such as
radial diffusion, is necessary to bring the electrons into the inner
magnetosphere. Hence, much of the acceleration debate focused on
radial diffusion.

In the case of radial diffusion, the electric field is that of
ultra-low frequency (ULF; 300\,Hz to 3\,kHz) waves that continuously
agitate the magnetosphere. [\ldots] The basic premise of the mechanism
is that electric fluctuations induce small random perturbations of the
electrons' position causing them to diffuse radially throughout the
magnetosphere. The process is similar to diffusion in a gas only in
this case the random walk motion of the particles is caused by
electric fields instead of collisions. [\ldots]

\begin{figure}[t] \centering
\includegraphics[width=8cm]{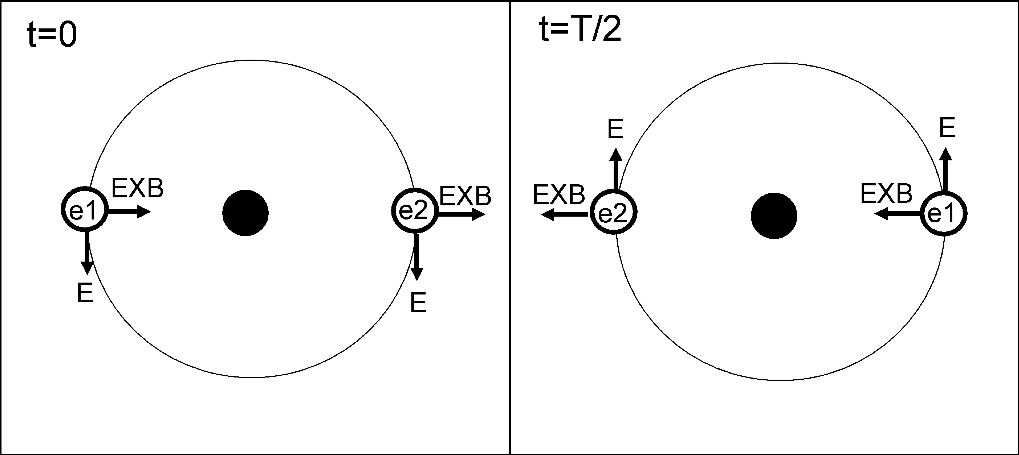}
\caption[An electron in drift resonance with a ULF
wave.]{\label{fig:green4.1}Schematic diagram of an electron in drift
resonance with a ULF wave. The left panel shows two electrons labeled
e1 and e2, the direction of the wave electric field, and the direction
of the particle's ${\bf E} \times {\bf B}$ drift at time $t=0$. The
right panel shows the same properties half a wave period and electron
drift period later. [Fig.~II:11.7]}
\end{figure} [Time] varying fields fluctuating specifically at the
same frequency of an electron drifting about Earth [cause] rapid
acceleration through a 'drift resonance'. Figure~\ref{fig:green4.1}
gives a pictorial explanation of an electron drift resonance. [\ldots]
The electron drifting around Earth labeled $e_1$ [\ldots] experiences
an azimuthal electric field that continuously moves it inward causing
the electron to gain energy. However, the electron, $e_2$, that began
at time $t=0$ on the opposite side would have seen an electric field
that pushed it radially outward in the same manner. Thus, the drift
resonance causes electrons to diffuse radially inward and outward and
decelerates as well as accelerates electrons.

[\ldots] If electrons are acted on by [a randomly varying electric
field, the net energy gain of the distribution of electrons] is
determined by the phase-space density as a function of $L$. If
electrons are uniformly distributed in $L$ then the same number of
electrons moves inward and gain energy as those that move outward and
lose energy with no net energy gain. If the slope of $f$ versus $L$ is
positive more particles move inward and gain energy than particles
move outward and lose energy and the distribution of electrons gains
energy. If the slope of $f$ versus $L$ is negative then the opposite
occurs. [\ldots\ The radial diffusion has been described by an
approximation to the Fokker-Plank equation]
\begin{equation} {\partial f(L,\mu_{\rm m},K,t) \over \partial t} =
L^2 {\partial \over \partial L} \left ( {D \over L^2} {\partial \over
\partial L} \left [ f(L,\,\mu_{\rm m},K,t) \right ]\right ).
\end{equation} Here $f(L,\mu_{\rm m},K,t)$ [(with $K$ defined in
Eq.~\ref{eq:kdef})] is the phase-space density of electrons and $D$ is
the diffusion coefficient which is calculated separately for electric
and magnetic field perturbations. [Later,] the theory was revisited
and elaborated to include higher-order resonances caused by electron
drift motion in more realistic non-dipolar fields that increase
diffusion. However, doubt about the ability of radial diffusion to
fully explain observations led to the development of new competing
ideas regarding electron acceleration including the internal source
acceleration mechanisms.''

\begin{figure}[t] \centering
\includegraphics[height=7cm]{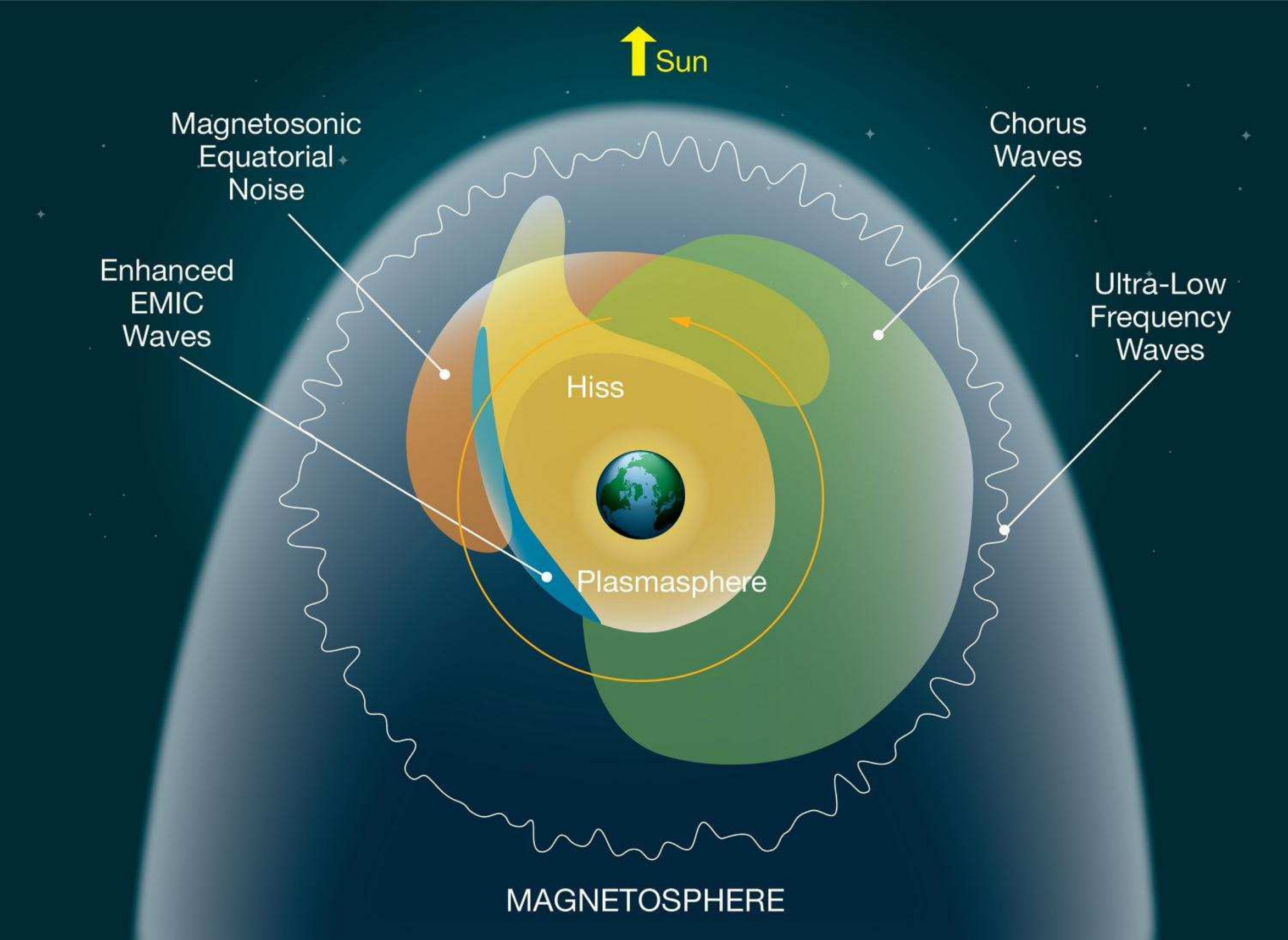}
\caption[Characteristic wave types within a
magnetosphere.]{\label{fig:magnwaves} Characteristic wave types within
a magnetosphere, here visualized for the terrestrial case, viewed from
above the arctic. Credit: NASA's Goddard Space Flight Center/Mary Pat
Hrybyk-Keith. \colorfig}
\end{figure}
\subsubsection{Internal source acceleration mechanisms}
\ors[II:11.4.1.2] ``The\indexit{relativistic electron acceleration
mechanism!internal} internal source acceleration mechanisms discussed
here accelerate electrons through interaction with the electric field
of a VLF (3\,kHz to 30\,kHz) wave.\indexit{particle-wave interaction}
The interaction is similar to the ULF wave resonance, but in this case
the resonance occurs between the wave electric field and the gyration
of the particle about the magnetic field instead of the drift about
Earth. The EMIC-Chorus wave mechanism assumes the interaction with the
wave can be described as a random walk diffusive process very similar
to radial diffusion [(Fig.~\ref{fig:magnwaves} illustrates
characteristic domains of various waves in the terrestrial
magnetosphere)]. This assumption is only valid when wave amplitudes
are small. The non-linear whistler wave acceleration mechanisms
describe how electrons interact with a monochromatic set of large
amplitude waves when diffusion is no longer valid. \activity{{\em Look
up} the properties of (whistler) chorus waves, (ELF/VLF) hiss, and
EMIC waves. Make a list or table of these properties. How do they
differ? What do they have in common?  \mylabel{act:wavecomp}}

The resonance between an electron and a VLF wave can be illustrated by
considering a VLF wave propagating at an angle $\theta$ from the
direction of the magnetic field with magnetic and electric field
perturbations perpendicular to the direction of propagation. The
electron gyrating about the magnetic field will experience a constant
electric field from the wave when the gyrofrequency of the electron
equals the Doppler-shifted frequency of the wave [\ldots] In contrast
to the ULF wave resonance, the VLF wave resonance will affect both the
electron's energy and pitch angle. [\ldots]

The {\em Chorus-EMIC wave mechanism}\indexit{wave-particle
interaction!Chorus-EMIC} proposes that electrons interact with both
whistler Chorus and EMIC waves as the electron drifts about Earth in
such a fortuitous way that the distribution is steadily pushed to
higher energy. In this model, EMIC waves at dusk interact with
electrons to produce an isotropic pitch angle distribution. The
electrons continue their drift to the dawn side of the magnetosphere
where Chorus waves are predominantly found. The diffusion curves for
Chorus waves are such that an isotropic distribution will diffuse
towards higher energy and larger pitch angles.  The energized
electrons now peaked near 90 degrees continue around to the dusk side
of the magnetosphere where the EMIC waves are found. The EMIC waves
interact with the electrons to again produce an isotropic pitch angle
distribution but with no energy loss. This isotropic distribution is
now primed to interact with the Chorus waves once again and gain
energy. Because the electrons traverse the magnetosphere in less than
10 minutes, the mechanism can effectively increase the energy over
periods of days.

The {\em non-linear whistler wave mechanisms} assume that electrons
are energized through a resonant interaction with whistler
waves. However, the previous diffusion model requires that wave
amplitudes are small in order for diffusion to be an adequate
approximation. If this is not the case, the interaction must be
described in a more detailed manner. [\ldots\ Under] the right
conditions a 100\,keV electron could be accelerated to MeV energies
within minutes. These mechanisms have yet to be compared in detail
with observations or included in any kind of global model of electron
flux. However, new measurements of whistler waves suggest that the
small amplitude assumption is very often invalid making non-linear
modeling an active area of interest.''

\subsection{Proton acceleration in the radiation belt} \ors[II:11.4.3]
``The\indexit{radiation belt!proton acceleration} structure and
temporal variability of the proton radiation belt is strikingly
different from its electron counterpart. Yet, some of the same
mechanisms are proposed to explain the acceleration of these
particles. The protons normally form only one belt [around $L=1-3$]
with fluxes that peak near $L=1.5$ and they tend to be more
stable. However, during highly geomagnetically active periods, such as
brought about by the passage of a large shock and sometimes an
accompanying solar energetic particle (SEP)event, fast and dramatic
changes occur. Often these changes mean a complete reconfiguration
where entirely new, sometimes transient proton belts are formed that
may last days to years.

Simulations of proton motion in both analytical and MHD magnetic field
models suggest that the new proton belts are formed when protons are
transported radially inward by large induced electric fields that
arise as a large shock passes the Earth. The mechanism is
almost the same as proposed for  some electron radiation belt
acceleration events at Earth, except that forming a new
proton belt requires an additional source of protons from the solar
wind. Often large shocks are accompanied by very high fluxes of
protons that are released from the Sun and further accelerated by the
shock. Normally, Earth's magnetic field acts as a protective bubble
that only allows these solar protons to enter over the polar caps
where they are absorbed into the atmosphere. However, as the shock
passes Earth, the magnetic field is distorted such that the
accompanying protons can gain access to the inner magnetosphere and
become trapped in the field. Once trapped, they are swept up by the
induced field and pushed to small radial distances and higher energies
to form a new belt.''

\subsection{Radiation belt losses at Earth}
A \ors[II:11.5.1] ``survey of electron radiation belt changes [\ldots]
found that only 53\%\ of storms cause radiation belt flux levels to
increase even though these storms signify increased energy input to
the magnetosphere. In 19\%\ of storms the flux actually decreased and
in 28\%\ the flux did not change [by more than a factor of two]. The
variable response to energy input suggests that loss and acceleration
rates are often comparable and ultimately compete to determine final
flux levels. [\ldots] The mechanisms that have been proposed to
explain the loss of relativistic electrons are: drift out the
magnetopause boundary, outward radial diffusion, and scattering into
the atmosphere. Scattering can be caused by interactions with a thin
current sheet, EMIC waves, or Chorus waves.

Loss of electrons through the magnetopause boundary occurs when the
drift paths of electrons are altered as the magnetic field changes
from a quiet time configuration to more disturbed conditions. During
quiet times, the drift motion of an electron starting in the
magnetotail is dominated by an electric potential field directed from
dawn to dusk [(see Fig.~\ref{fig:MC})] that moves electrons
Earthward. As the electrons get closer to Earth, the magnetic radial
gradient causes a westward drift. Some of these drift trajectories
will cross Earth's magnetopause and the electron will be swept away by
the solar wind. Closer to Earth, the trajectories of the electrons
will be dominated by the gradient drift. Undisturbed, electrons in
this near-Earth region will simply drift about continuously on closed
almost circular paths. [Model] results suggest that during geomagnetic
storms, most of the outer electron radiation belts are emptied into
the solar wind and replaced by an entirely new belt of accelerated
electrons. [This plausible suggestion has not been observationally. It
appears] that loss to the magnetopause was not an adequate explanation
for electron flux depletions observed during more quiet conditions
because the flux of energetic protons on similar drift paths did not
decrease.

Radial diffusion [\ldots] has also been proposed as a loss process. Radial diffusion
acts to reduce gradients by pushing particles from high phase-space
density to low phase-space density. The outermost closed drift orbit
of the radiation belts represents a very steep gradient where the
phase-space density goes to zero. If ULF waves are present, then
radial diffusion will push particles outward to the
magnetopause. [\ldots]

Losses into the atmosphere occur when some mechanism scatters
electrons to smaller pitch angles causing them to travel farther down
the field line and collide with the neutral atmosphere. The current
sheet, that forms in the magnetotail as the lobes are stretched and
forced together by solar wind dynamics, is an effective scattering
region. Scattering occurs when the magnetotail becomes stretched to
the point that an electron bouncing along a field line can no longer
make it around the kinked field without violating its first
invariant. Traversing the kink changes the particle's pitch
angle. Under certain conditions the pitch angle changes can be
described as a diffusive process. [The] significance of this loss
contribution has yet to be verified.

Chorus and EMIC waves [\ldots] may also cause rapid loss into the
atmosphere. Whether or not the waves produce net acceleration or loss
depends on the initial gradients of the electron distribution as a
function of pitch angle. [If] the appropriate distribution exists,
EMIC waves are expected to cause losses on the timescales of several
hours to a day. Whistler Chorus may cause losses on timescales of one
day, but these estimates are sensitive to parameters such as the cold
plasma density. Loss rates may increase to timescales less than a
day during storm main phase when the plasma density is expected to
vary [\ldots]''

\ors[II:11.5.2]
``The\indexit{radiation belt!proton losses at Earth} proton losses
from the radiation belts have not been analyzed in the same details as
the dramatic formation of new belts. New belts last from days to
years. Mechanisms proposed to explain the disappearance of these belts
include scattering caused by the kinked field, and interaction with
EMIC waves. [There is no firm understanding of the proton loss
mechanisms, and] it may be that more than one mechanism plays a role
in each event.''
\clearpage

\chapter{Convection, heating, conduction, and radiation}%9
\label{ch:heating}
{\narrower\narrower{
{\bf Chapter topics:}
\begin{itemize}
  \customitemize
\item Convective, radiative, and conductive energy transport 
\item Ubiquitous but structured magnetic field in the solar photosphere
\item Coronal loop atmospheres
\item Power laws for stellar atmospheric losses
\end{itemize}

\noindent{\bf Key concepts:}
\begin{itemize}
  \customitemize
\item Convective 'small-scale' dynamo action
\item Magnetohydrostatic flux tubes: bright points to dark spots
\item Optically thick/thin
\item Wilson depression
\end{itemize}

}}

\section{Convective and radiative energy transport}\label{sec:energytransport}
One topic at the foundations of heliophysics remains to be introduced
before we move on to comparative stellar and planetary astrophysics:
which processes lead to a relatively steady background heating of the
solar atmosphere in 'quiescence', {\em i.e.,}  outside of obvious
impulsiveness? These processes have their origin in the convection
that occurs below the solar surface and in the diversity of waves that
are generated by these convective motions.

 The solar convection zone persists from a depth of about 200,000\,km
all the way to the surface.  \ors[I:8.1] ``Looking at the solar
photosphere, we see the top of the convection zone in the form
of\indexit{granulation} granulation: Hot gas rising from the solar
interior as part of the energy transport process reaches a position
where the opacity is no longer sufficient to prevent the escape of
radiation. The gas radiates and cools, and in doing so loses its
buoyancy and descends.\activity{{\em Look up:} (a) What drives tropospheric convection?
  (b) Why is there no significant convection in the stratosphere (consider
  the role of ozone)? (c) Is there an equivalent of a stratosphere on
  Venus? On Mars?  (d) Is there lower atmospheric convection? Formulate
  your arguments. The Web can help. The answers are 'yes'
  (with a role for ozone only on Earth).} At the surface the gas
density is of order $10^{-7}$~g/cm$^3$ while the pressure is of order
$10^5$~dyne/cm$^2$, but decreasing exponentially with height with a
scale height of some 100\,km. (This small scale height is the reason
that the solar limb appears sharp as viewed with the naked eye.)
Granular cells have dimensions of order 1~Mm, but numerical
simulations indicate that convective length scales rapidly become
larger as one proceeds below the solar surface. \sactivity{$\circledS$ {\em Show:} The scale of
  the granulation in the photosphere of the Sun (and analogously of
  other cool stars) follows from a comparison of energy loss by
  radiation (effective once the plasma can radiate into space,
  with a time scale of order 20\,s) and supply by upflows. Work
  through this estimate: just below the photosphere, the largest
  contribution to energy being carried upward resides in latent heat
  of recombination of ionized hydrogen (with an ionization fraction of
  order 0.1); balance that with photospheric black-body radiation; use
  this to derive a minimum upward flow $v_z$ needed to 
  balance radiative losses. Then match timescales, and use that
  $v_{\rm h}\le c_{\rm s}=(kT/m_{\rm p})^{1/2}$: for overturning
  convective flows, the horizontal time scale of
  $\ell_{\rm h}/v_{\rm h}$ should equal the vertical one
  $\ell_{\rm z}/v_{\rm z}$, for a typical horizontal granular scale
  $\ell_{\rm h}$ and overturning depth
  $\ell_{\rm z}\approx H_p \approx 400$\,km somewhat below the
  photosphere. (This argument is developed in III:5.2.1)
  \mylabel{act:granscale} \solution{granscale}} These
motions, ultimately driven by the requirement that the energy
generated by nuclear fusion in the Sun's core be transported in the
most efficient manner, represent a vast reservoir of 'mechanical'
energy.

Looking closer, we see that granulation is not the only phenomenon
visible at the solar surface. The quiet and semi-quiet photosphere is
also threaded by magnetic fields that appear as \indexit{bright point}bright points, as well
as darker micro-pores and pores. These small-scale magnetic structures
are, while able to modify photospheric emission, subject to granular
flows and seem to be passively carried by the convective
motions. Bright points are organized in a honeycomb-like pattern on
the solar surface with a size scale larger than granulation, roughly
20\,Mm; this pattern defines the so-called \indexit{supergranulation}supergranular network and is suggestive of convective
cells larger than granulation extending deeper into the solar interior.

Convective flows are also known to generate the perturbations that
drive solar oscillations. Oscillations, \indexit{solar!oscillations}sound waves, with frequencies mainly in the band
centered roughly at 3~mHz or 5~minutes are omnipresent in the solar
photosphere and are collectively known as \indexit{p modes}$p$-modes ('p' for
pressure). These $p$-modes are a subject in their own right and studies
of their properties have given solar physicists a unique tool in
gathering information on solar structure --- the variation of the
speed of sound $c_{\rm s}$, the rotation rate, and other important
quantities --- at depths far below those accessible through direct
observations. [\ldots] \sactivity{$\circledS$ {\em Show:} Sound waves
  in an isothermal, hydrostatically-stratified atmosphere are
  'evanescent', {\em i.e.,} non-propagating, at frequencies below the
  acoustic cutoff frequency $\omega_{\rm a} = c_{\rm s}/2H_p$ (for
  sound speed $c_{\rm s}$ and pressure scale height $H_p$, see
  Eqs.~\ref{eq:10.13} and~\ref{eq:hp}). (a) Can you argue intuitively
  why (think of the need for a restoring force roughly within a
  wavelength)?  (b) Estimate the value of $\omega_{\rm a}$ for the
  solar atmosphere. Why are $p$-modes only observed at frequencies below
  about $\omega_{\rm a}$? (c) Now derive an approximate dispersion
  relation in simplified geometry for {\em isothermal non-dissipating}
  perturbations: Start from Eqs.~(\ref{continuity}), (\ref{momentum})
  and~(\ref{eq:hp}) for a hydrostatic 1-d ({\em i.e.}, plane-parallel
  with only vertically-propagating waves), non-magnetized plasma (mind
  the sign of $g$) and combine them retaining terms to first order for
  perturbations $\rho=\rho_0+\rho_1$ and $v=v_0+v_1$, where $\rho_0$
  and $v_0\equiv 0$ describe the background stratified atmosphere at
  rest. Then factor out the exponential growth of the amplitude with
  height by substituting $v=u \exp(z/2H_p)$ and use
  $u\propto \exp(i[kz-\omega t])$ to obtain a dispersion relation that
  has propagating waves (real values of $k$) only for frequencies
  above $\omega_{\rm a}$ (a somewhat different approach can be found
  in I:8.3 in which the approximation of isothermal perturbations is
  dropped). Lower frequency waves reflect and can form standing
  $p$-modes if they meet the criteria for global resonance, while higher
  frequency waves can propagate, but will steepen (readily into shock
  waves) as they propagate into the lower-density
  atmosphere. \mylabel{act:acousticcutoff} \solution{acousticcutoff}}

On average the photospheric gas pressure of $p_{\rm g}=10^5$~dyne/cm$^2$ is much
greater than the pressure represented by an average unsigned magnetic
field strength of 1 -- 10~Gauss ($p_B={B^2/8\pi}<10$~dyne/cm$^2$) that is
observed.  However, in the largely isothermal chromosphere, the gas
pressure falls exponentially with a scale height of some 200\,km, while
the magnetic field strength falls off much less rapidly, even as the
field expands [from the compact flux tubes that characterize it in the
photosphere] and fills all space. Thus, depending on the actual
magnetic field geometry, the magnetic pressure and energy density will
surpass the gas pressure some 1500\,km or so above the photosphere in
the mid chromosphere.  Another 1000\,km, or 5 scale heights above the
level where $\beta=1$ (see Eq.~\ref{eq:betadef}), the plasma's
ability to radiate becomes progressively worse, while the dominance of
the magnetic field becomes steadily greater. As we will explain below,
any given heat input in this region cannot be radiated away, and will
invariably raise the gas temperature to 1~MK or greater; a corona is
formed. A corona that is bound to follow the evolution of magnetic
field as the field in turn is bound to photospheric driving.''

\begin{figure}[t]
\centering
\includegraphics[height=6cm]{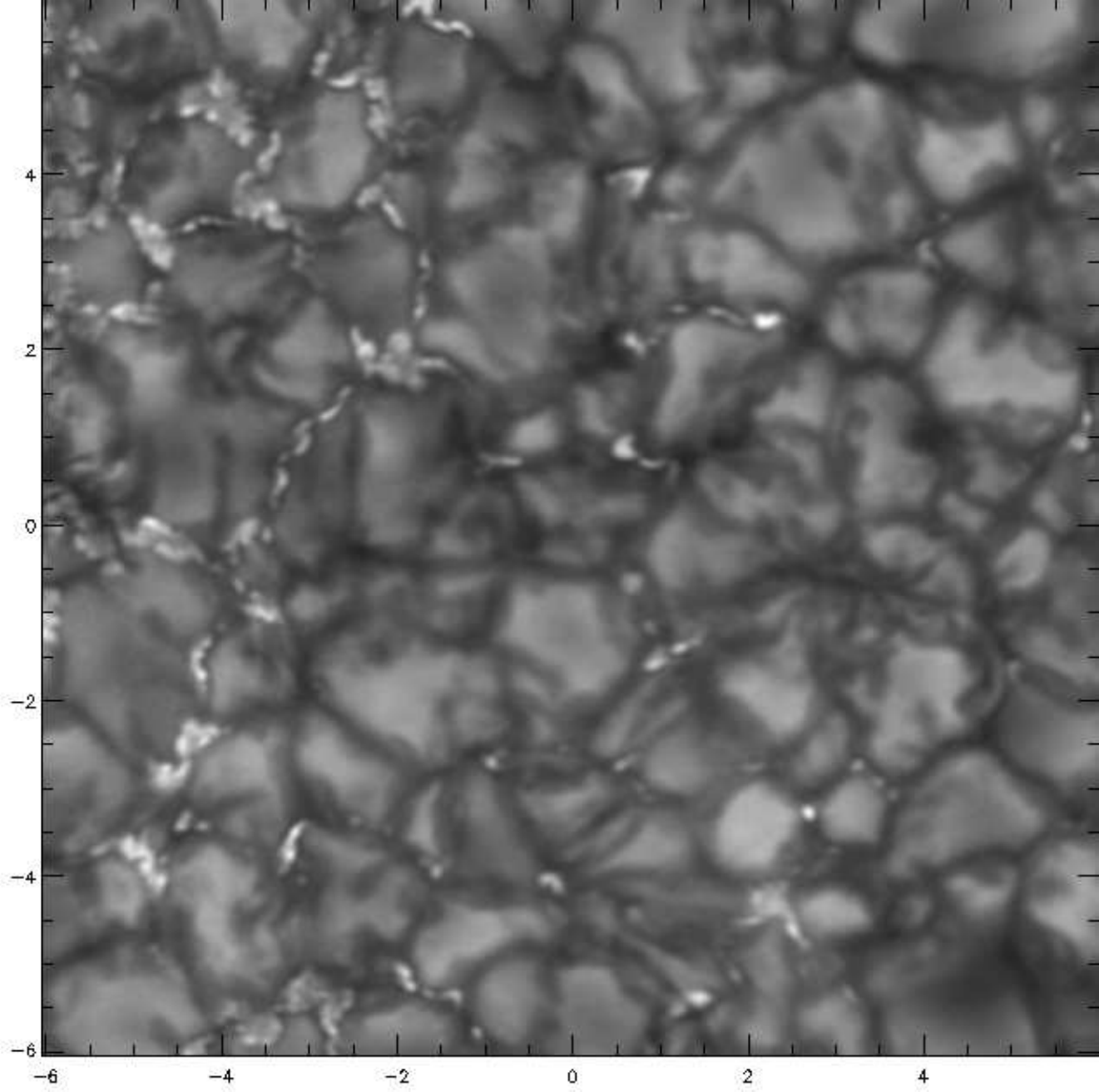}
\includegraphics[height=6cm]{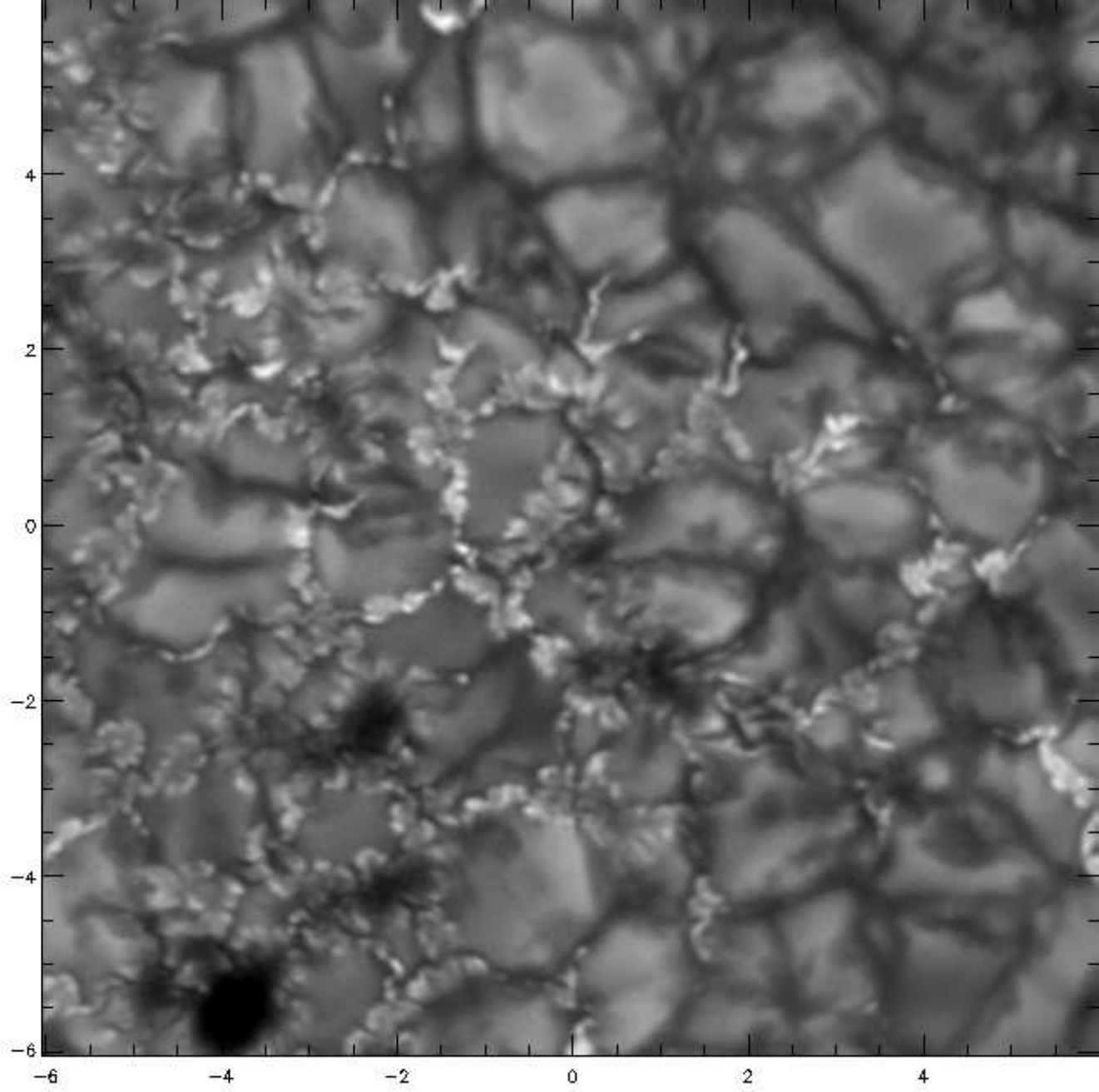}
\caption[Granulation in quiet-Sun and plage observed in the G-band.]{The
image on the {\em left} shows a typical quiet photospheric region observed
in the G-band with the Swedish 1-meter Solar Telescope. The image on
the {\em right} shows a plage region where the total amount of magnetic flux
penetrating the photosphere is larger.  The axes of both panels are
numbered in arcseconds measured on the Sun;
1~arcsec\indexit{arcsecond, equivalent length} is approximately
725\,km. The G-band near 4300\,\AA\ contains several spectral lines,
notably the lines of the CH-molecule, and is formed near the solar
surface (the height where the optical depth $\tau_{5000{\AA}}=1$); the granulation and intergranular lanes some 100\,km above
this height, bright points some 200\,km below --- as explained in the
text. Bright points are regions of enhanced magnetic field
embedded between the granular motions. Notice also that
bright points are pulled into ribbons and may fill the entire
intergranular lane.  The image on the {\em right} shows a photospheric plage
region. Notice the large number of phenomena showing complex
structure; ribbons, 'flowers', micropores, as well as isolated and
seemingly simple \indexit{bright point}bright points. The magnetic field in this image is in
places strong enough to perturb granulation dynamics and the granules
appear 'abnormal' while displaying a slower evolution than in the
quieter photosphere. [Fig.~I:8.3]}
\label{fig:bp-simple}
\end{figure}

\ors[I:8.2] ``In Fig.~\ref{fig:bp-simple} we show typical images of the
quiet to semi-quiet photosphere as well as a plage region.\regfootnote{'Plage' is
formally the bright chromospheric area over regions of enhanced magnetism
in the photosphere, but is commonly also used to 
describe the interior of active regions, {\em i.e.,}  regions of
strongly enhanced mean magnetic field in the photosphere, outside of,
but commonly in the immediate vicinity of, sunspots.} These images
are made in the so called\indexit{G-band} G-band centered around
4300\,\AA\ which is formed some 100\,km above the nominal
photosphere. [\ldots] \activity{{\em Look up:} The 'G band' is a narrow spectral
  interval centered on electronic transitions of the CH molecule,
  mixed in with spectral lines from multiple metals. For the
  interested: look up the 'Fraunhofer lines' and their
  designations. This old nomenclature from the days in which the solar
  spectrum was first studied is still used for some of these 'lines'.
%  , most frequently for the Ca~II H and K lines, the Na D lines, and the G band.
  Which elements and ions are responsible for these lines and
  bands: C, D, H, K, and G? \mylabel{act:fraunhofer}}

\begin{figure}
%\centerline{\psfig{figure=figures/tubemod.eps,height=45mm}} 
\centerline{\includegraphics[width=45mm]{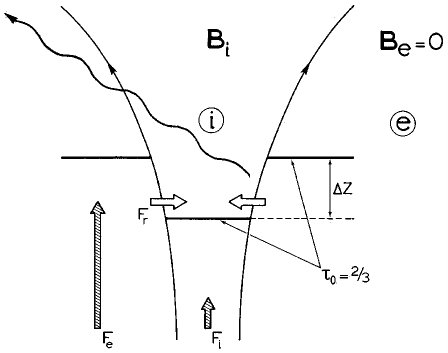}} 
\caption[Solar flux-tube model.]{Concept of the \indexit{flux!tube!concept}magnetohydrostatic flux-tube
  model.  One level of constant optical depth in the continuum,
  $\tau_0 = 2/3$, is shown, with the Wilson depression $\Delta z$.
  The hatched arrows $F_{\rm i}$ and $F_{\rm e}$ stand for the flux
  densities in the (non-radiative) energy flows inside and outside the
  flux tubes, respectively. The horizontal arrows indicate the influx
  of radiation into the transparent top part of the tube.  The
  resulting bright walls are best seen in observations toward the
  solar limb (as seen along the oblique wavy
  arrow. [\href{https://ui.adsabs.harvard.edu/abs/2000ssma.book.....S/abstract}{From
    \citet{2000ssma.book.....S}}]\label{fig:tubemod}}
\end{figure}
In [the left panel of] Fig.~\ref{fig:bp-simple} we show examples of
simple \indexit{bright point}bright points in a fairly quiet region of the photosphere, {\em
i.e.}, a region in which the magnetic field is too weak to
significantly modify granular motions. Isolated bright points are seen
to be constrained to intergranular lanes and do not seem to have any
internal structure on the scales that are visible on this resolution.
Isolated bright points appear to be passively advected towards the
periphery of \indexit{supergranulation}supergranular cells, where they gather
and form the collection of bright points that define the supergranular
network mentioned above. [\ldots]

The right panel [of Fig.~\ref{fig:bp-simple}] shows a region of stronger
average\indexit{magnetic!plage}\indexit{active
region}\indexit{pore}\indexit{micropore} field strength, a {\em plage}
region, than that found in the left-hand panel.  Flux concentrations
with larger spatial extent are embedded in (micro-)pores with
distinctly dark centers. Such small dark micropores may be the
smallest manifestation of the phenomena that produce pores and
sunspots.  The bright points in the plage region are not simple points
but are seen to have structure and appear to modify the granular flow
itself: The image shows that granules near network bright points and
in plage regions appear smaller, have lower contrast and in addition
display slower temporal behavior than granules in weak field regions.
Coalescing bright points in plage and network regions can form dark
centers and thus become micropores if their density is large enough,
or indeed brighten again if the granular flow breaks them apart into
smaller flux elements. [\ldots]''

Substantial bundles of strong-field magnetic flux form dark pores and
even larger ones form the dark sunspots. Relatively smaller bundles on
the other hand are bright compared to the surrounding photosphere,
particularly when seen somewhere between the center of the solar disk
and the edge (or 'solar limb').  Qualitatively, this can be understood
by the combination of radiative transport and MHD. Let us start with
an essentially vertical flux bundle with an intrinsic field of
$\approx 1$\,kG as commonly observed for such solar field
structures. Such a tube has settled into pressure equilibrium between
inside and outside, but as the field adds its own pressure, the gas
pressure inside a tube at a given height is lower than
outside.\sactivity{$\circledS$ {\em Show:} Stars have a range of
  surface gravities, typically increasing monotonically along the main
  sequence toward lower effective temperatures and lower mass (see the
  table associated with Fig.~\ref{fig:acthrd}), and substantially
  lower in evolved ('giant' and 'supergiant') stars than in
  main-sequence (or 'dwarf') stars. Qualitative insight is provided by
  the following exercise: using the concepts of optical depth (and the
  fact that the stellar photosphere is around unit optical depth for
  continuum emission) and hydrostatic equilibrium
  (Ch.~\ref{ch:particles}), show that the photospheric pressure would
  scale proportional with gravity in the idealized case of an
  isothermal atmosphere; focus explicitly on the optical depth
  relationship in Eq.~\ref{eq:optdepth}. In reality, radiative
  transport and convective motions modify that scaling for a real
  non-isothermal atmosphere, but the trend is in the correct
  direction. With this insight, argue for the trend of intrinsic field
  strength of photospheric magnetic concentrations with gravity: from
  $\sim 1.4$\,kG in mid F-type dwarf stars to $\sim 3.2$\,kG in late
  K-type dwarf stars, and well below $1$\,kG for cool
  giants. \mylabel{act:pressgrav} \solution{pressgrav}} Radiative
losses lead to cooling inside and outside the tube, but the field
inside lowers the ability for convection to resupply heat relative to
outside, so that the interior of the tube is cooler. Being cooler and
less dense, the level at which the optical depth reaches unity inside
the tube is lower than outside, leading to a 'depression' in the solar
'surface' (called the \indexit{Wilson depression}Wilson
\indexit{definition!Wilson depression}depression). The line of sight into the
tube from a somewhat slanted perspective looks down on the flux-tube
walls, which shows layers effectively under the surface, and thus
brighter than the surrounding photosphere (appearing as what are known
as \indexit{facula}'faculae'; compare Fig.~\ref{fig:tubemod}), [provided the tube is
relatively narrow compared to the characteristic photon mean free
path. For such a narrow tube, looking down into it shows a 'bright
point' as we view the tube's photosphere that lies below, and is
somewhat hotter, than the surrounding photosphere.] If the tube is
wide, however, such as is the case for a pore or sunspot, the sideways
heat transport cannot keep the atmosphere near the wall as hot as deep
into the wall, causing a cooler and thus darker wall around a dark
interior, which is the manifestation of the penumbra and umbra of a
sunspot. \activity{{\em Consider:} The transition from bright to dark
  magnetic structures occurs at a scale of roughly $200-300$\,km. What
  does that say about the typical photon-mean free path
  $\ell_{\rm ph}$ in the photosphere? Compare that value to the
  corresponding pressure scale height, and argue why
  $\ell_{\rm ph} \ga H_p$ just at the photosphere.} \activity{{\em
    Consider:} Explain why observed field strengths inside flux tubes
  exceed the equipartition field strength (the field strength in an
  imaginary completely evacuated tube in pressure equilibrium with the
  surrounding field-free gas) at the level of the photosphere outside
  the tube. Hint: where is the photosphere inside a
  partially-evacuated tube relative to that in the surrounding gas?
  \mylabel{act:lookintotube}} (A description of numerical models of
radiative magneto-convection that reveals the quantitative details is
given in Sect.~I:8.2.1). Flux tubes typically rise into the
photosphere with lower intrinsic field strengths, but the process of
radiative cooling and hampered internal energy transport from below
then lead to a 'convective collapse' by which a more or less isolated
flux tube forms (from small faculae to large sunspots) with final
field strengths of order $1-3$\,kG (larger at the center of larger
structures).\activity{{\em Consider:} When the total solar irradiance
  (TSI) is smoothed over time scales of, say, a week, the Sun is
  brighter at sunspot maximum than at sunspot minimum, but when
  looking at TSI curves with a resolution of a day or so, the presence
  of large sunspots leads to dips when these are near the central
  meridian. Explain this qualitatively by the mix of faculae, pores,
  and spots in and around active regions. Once you formulate a
  hypothesis, verify this by looking at, for example, Fig.~III:10.7
  and figures such as Fig.~3 in
  \href{https://ui.adsabs.harvard.edu/abs/2016SoPh..291.2951K/abstract}{the
    study} by \citet{2016SoPh..291.2951K} compared to sunspot records
  that can be found at
  \href{https://www.sidc.be/silso/monthlyssnplot}{https://www.sidc.be/silso/monthlyssnplot}. \mylabel{act:irradiance}}

\begin{figure}[t]
\begin{center}
%\centerline{\hbox{\psfig{figure=figures/Abbett_combined1.eps,width=\textwidth,clip=}}}
%\centerline{\hbox{\psfig{figure=figures/CISM_cycle_bw.eps,height=6.2cm,clip=}\psfig{figure=figures/CISM_movie16-3_bw.eps,height=6.2cm,clip=}}}
\centerline{\hbox{\includegraphics[width=\textwidth]{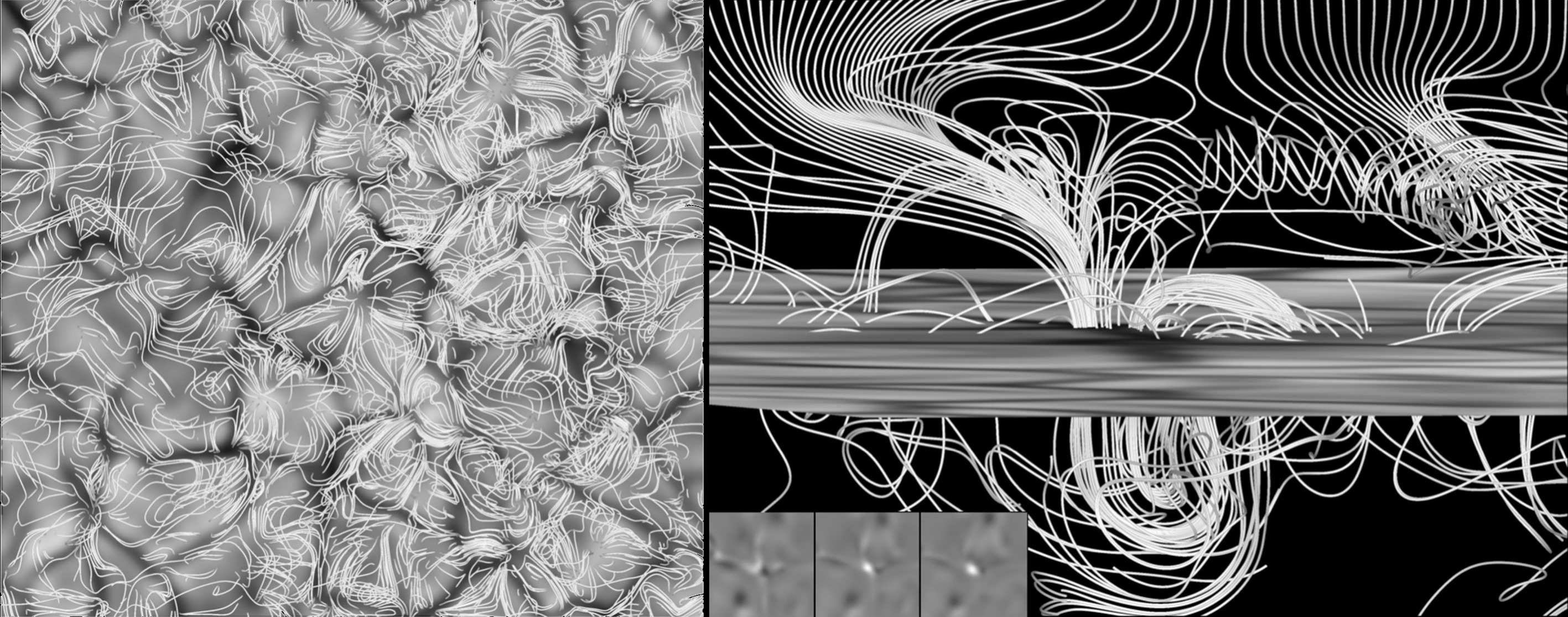}}}
\centerline{\hbox{\includegraphics[height=6.2cm]{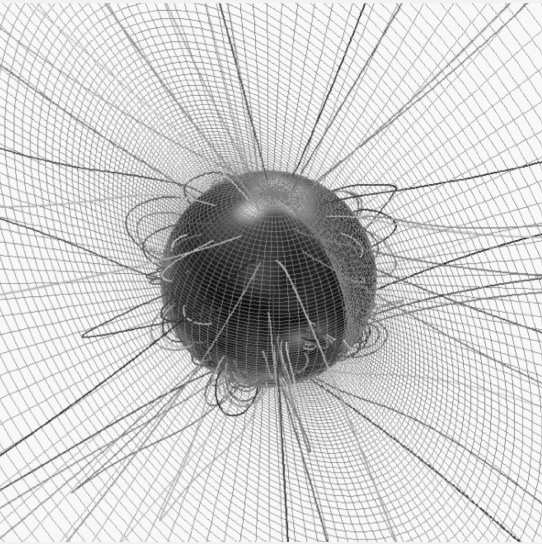}\includegraphics[height=6.2cm]{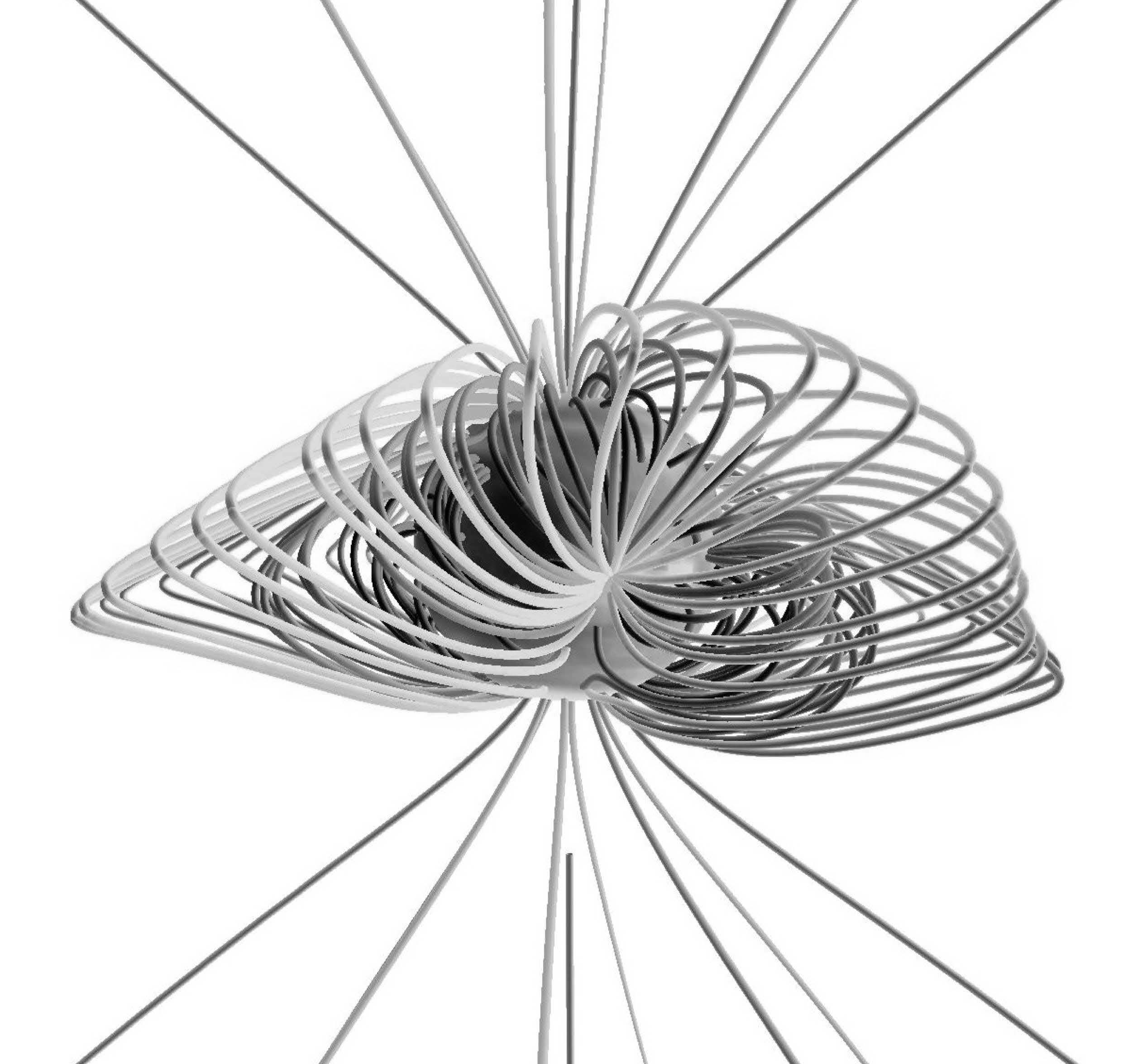}}}

\caption[The multitude of scales in the solar magnetic
field.]{Illustration of the multitude of scales in the solar magnetic
  field. The {\em top panel} shows a \indexit{solar!magnetic field}model computation of the magnetic field
  on the scale of the dominant convective motions at the solar
  surface, the 1000-km scale of the granulation (see also
  Fig.~\ref{fig:bp-simple}); the {\em left panel} is a top view of the solar
  surface with sample magnetic field lines overplotted, while the
  {\em right panel} shows a vertical cut through one of the convection cells
  to illustrate how the field in this model threads the surface
  sometimes multiple times, evolving on a time scale of a few
  minutes. The {\em lower two panels} show models of the global solar field,
  tracing field lines up to the cusps of the \indexit{streamer belt}streamers that outline
  the topologically distinct regions of closed field and the field
  that is open to the heliosphere; this global scale field evolves on
  time scales of months to a decade. [Fig.~I:1.2; sources for:
  \href{https://ui.adsabs.harvard.edu/abs/2007ApJ...665.1469A/abstract}{top
  (from \citep{2007ApJ...665.1469A}}) and \href{https://www.bu.edu/cism/}{bottom} panels.]}
\label{figure:solarfieldexamples}
\end{center}
\end{figure}\nocite{2007ApJ...665.1469A}
\section{Heating and cooling of the solar outer atmosphere}\label{sec:solarouter}
The motions of the Sun's \indexit{solar!atmosphere!heating}near-surface convection are a good fraction
of the local sound speed, and thereby they generate a lot of acoustic
power. Waves at frequencies below the acoustic \indexit{acoustic
  cutoff frequency}cutoff \indexit{cutoff frequency}frequency
$\omega_{\rm a}$ (defined in Activity~\ref{act:acousticcutoff}) are
trapped inside the Sun and can, in resonance patterns, set up one of
millions of $p$-modes. But because the solar atmosphere above the
surface has a temperature reversal into the chromosphere and corona,
some degree of tunneling occurs, while higher-frequency waves can
simply propagate upward through the atmosphere. All such waves steepen
as they move into lower-density regions. The enhanced radiative losses
during compression phases as well as dissipation through heating in
the developing shocks together cause energy conversion from wave
motion into thermal energy. Some of the heating of the solar
chromosphere is due to such acoustic phenomena. However, the most
pronounced non-radiative heating occurs at locations where magnetic
field penetrates the solar surface.

Owing to the insulating atmosphere, the magnetic field of the Earth's
dynamo couples to the terrestrial magnetosphere only through
induction. The Sun's magnetic field, in contrast, is directly coupled
from interior to atmosphere through the conduction of the plasma
throughout. Consequently, the movements of the plasma in the
convective envelope, from meridional circulation all the way down to
sub-granular motions, drive processes in the solar atmosphere from the
photosphere out to the distant heliosphere on time scales commensurate
to the length scale involved, {\em i.e.,}  from a decade down to minutes
(Fig.~\ref{figure:solarfieldexamples}).

The convective motions of the solar plasma, and the waves that these
drive, couple into the magnetic field that threads the
photosphere. The wave-like plasma motions transform into various
magnetohydrodynamic wave types, while \indexit{corona!heating
  mechanism}the convective motions 'braid' the higher field by the
random walk of the footpoints \indexit{line tying}'line-tied' to the convective
motions. In the higher atmosphere, these lead to wave interference and
mode coupling, resonances, and field discontinuities (in a cascade of
'current sheets'). Add to that the insertion of new magnetic field
emerging into the atmosphere as well as the removal from surface
layers through reconnection between opposite polarities. All of these
processes lead to high gradients in field and dynamics, and those to
dissipation into heat (in the literature generally differentiated into
low-frequency, braiding-dominated 'DC' mechanisms in contrast to
higher-frequency, wave-dominated 'AC' mechanisms). It remains an open
question as to which of the proposed heating mechanisms dominates (or
more comprehensively the question of which mechanism dominates under
what specific conditions), but observations make it clear that the
magnetic field is both the conduit and the agent involved in heating
the solar atmosphere.

These processes lead to the $\sim 20,000$\,K chromosphere and at
sufficient height to the multi-million degree corona. In order to
understand why such high temperatures arise, we need to \ors[I:8.4]
``realize that the temperature of a plasma is set not only by the heat
dissipated but also by the plasma's ability to lose
energy. \activity{{\em Consider:} The processes of electromagnetic
  radiation from a plasma involve three fundamentally distinct
  processes: bound-bound, free-bound (radiative recombination), and
  free-free (Bremsstrahlung) emission. (a) The Sun's coronal emission,
  caused by excitation collisions of partially ionized with thermal
  electrons, is dominated by the first, except for flares when the
  last is also important; why? (b) Which ions are typically strong
  contributors to the coronal X-ray and EUV emission from an active
  region at $\sim 3$\,MK? Hint: combine elemental abundances with
  ionization energies (such as given on the 
  \href{https://en.wikipedia.org/wiki/Ionization_energies_of_the_elements_(data_page)}{Wikipedia
  page} on ``ionization energies of the elements'').
%  For this rough estimate, ignore oscillator strengths for the transitions involved.
  For the solar corona under most conditions, the dominant
  radiative losses are from C, N, O (below about 0.5\,MK), and Fe
  (above about 0.5\,MK).  \mylabel{act:coronallines}}

The coronal plasma has essentially three possible ways to shed energy:
\begin{enumerate}\customitemize
\item Through \indexit{corona!radiative loss}optically thin radiation, mostly from carbon, oxygen, nitrogen,
iron, and neon (and at lower temperatures from effectively thin
hydrogen Lyman-$\alpha$), given by
\begin{equation}\label{eq:radloss}
\Lambda(T_{\rm e})=n_{\rm e}n_{\rm H}f(T_{\rm e})
\end{equation}
where $n_{\rm e}$ and $n_{\rm H}$ are the electron and total hydrogen
densities and $f_{\rm rad}(T_{\rm e})$ is a function of temperature
dependent mainly on line emission and, at higher temperatures, on
thermal bremsstrahlung.\activity{{\em Consider:} Eq.~(\ref{eq:radloss}) contains a product
  of electron and hydrogen densities, but hydrogen is fully ionized at
  coronal temperatures and thus has no spectral lines that can be
  excited through collisions with electrons. Why is it acceptable to
  express it this way? Hint: consider elemental the mix of ions. \mylabel{act:nenh}}
\item Through thermal \indexit{corona!conductive loss}conduction
  \indexit{thermal conduction!coronal loop}along the magnetic field, with a
  (Spitzer) thermal
  conduction coefficient \regfootnote{Note the steep dependence on
    temperature of the \indexit{thermal conduction}thermal conductivity $\kappa_\parallel$ in
    Eq.\,(\ref{eq:conduction}). In, say, a metal in terrestrial
    conditions $\kappa$ is generally a constant, but in a low-density
    plasma in which Coulomb collisions dominate, this temperature
    dependence is a consequence of the increase in the characteristic
    mean-free path of the electron-ion collisions
    ($\lambda_{\rm mfp} = v_{T{\rm e}} / \nu_{\rm ei} \propto T^{2}$,
    see Table~\ref{tab:dimensionlessnumbers}) and the increase in the
    characteristic electron velocity ($\propto
    T^{1/2}$). \mylabel{footspitzer}}
\begin{equation}\label{eq:conduction}
%  \kappa(T_{\rm e})= -\kappa_\parallel(T_{\rm e}) \nabla_\parallel T_{\rm e} =
%                                -\kappa_0 T_{\rm e}^{5/2}   \nabla_\parallel T_{\rm e}.
  \kappa(T_{\rm e})= \kappa_0 T_{\rm e}^{5/2} 
\end{equation}
\item The magnetically open corona can also lose energy through the
acceleration of a solar wind.  [\ldots]
\end{enumerate}

In short, when the plasma is dense, $n_{\rm e}n_{\rm H}$ is large
and variations in the heat input can be dealt with by small
changes in the plasma temperature which will remain on order
$10^4$~K or less (similar to the photospheric radiation
temperature). Conduction on the other hand is very inefficient at
these temperatures. However, the density drops exponentially with
height, with a scale height of only some hundreds of kilometers
for a $10^4$~K plasma. The efficiency of radiative losses
therefore drops very rapidly with height and {\em any} mechanical
energy input will raise the temperature of the plasma. The
temperature will continue to rise until \indexit{thermal conduction}thermal conduction can
balance the energy input. Because thermal conduction varies with a
high power of the temperature this does not happen until the
plasma has reached 1\,MK or so. Thus, we expect any and every
heating mechanism to give coronal temperatures of this order
[\ldots]''

The energy equation balances the thermal energy content of the plasma
with energy input by heating (with a volumetric rate of
$\epsilon_{\rm heat}$), loss by radiation,
and transport through thermal conduction. Looking back at
Eq.~(\ref{energy}), most terms vanish in this approximation, while
radiative losses, there not yet introduced, need to be
added. Realizing that energy conduction occurs along the magnetic
field, and with the cross section of a coronal loop \indexit{corona!loop}is inversely
proportional to the field strength ($A(s)\propto 1/B(s)$, with $s$ the
coordinate along the coronal field loop) we have:
\begin{equation}\label{eq:loop}
\rho \frac{{\rm d}e}{{\rm d}t} +{p\di\vv}= 
    \epsilon_{\rm heat} - \frac{1}{A(s)} \frac{{\rm d}}{\rm ds} \left (
    A(s) \kappa \frac{{\rm d}T}{{\rm d}s} \right ) - n_{\rm e}n_{\rm
    H}{f_{\rm rad}(T_{\rm e})}.
\end{equation}
We can use the following rough approximations: $\kappa \approx
8\times 10^{-7} T_{\rm e}^{5/2}$\,erg/cm/s/K and ${f_{\rm rad}(T_{\rm e})}\approx
1.5\times 10^{-18} T^{-2/3}$\,erg/cm$^3$/s for $T_{\rm e}$ in the
range of 0.4\,MK to 30\,MK. The terms on the left of this equation are
small in case the flows along the loop are sufficiently slow while any
short-term variability in the heating should be relatively small
compared to the internal energy content of the local plasma [for a
loop in quasi-hydrostatic equilibrium]. Using
that approximation along with Eq.~(\ref{momentum}) describes the
general appearance of any slowly evolving coronal loop (but note that
the radiative losses should not be approximated as above once looking
into the 'transition region', {\em i.e.,}  at the base of coronal loops where
the temperature transitions from coronal to chromospheric
values). Eq.~(\ref{eq:loop}) can be used to estimate, for example, the
conductive time scale $\tau_{\rm cond}$ (the ratio of thermal energy to
the rate at which conduction over a thermal gradient transports
energy) or the radiative time scale $\tau_{\rm rad}$ (the ratio of
internal energy to the time scale of radiative energy loss). The ratio
of these two shows the conditions in which conduction is more
important than radiation, or vice versa. This is most useful by
combining these time scales with a relation that emerges from the
modeling of quasi-static \indexit{corona!quasi-static}loops as a whole, made
(\href{https://ui.adsabs.harvard.edu/abs/1978ApJ...220..643R/abstract}{by
  \citet{1978ApJ...220..643R}}) in the approximation of a constant cross
section:
\begin{equation}\label{eq:rtv}
T_{\rm a,6} \approx 2.8 (n_{\rm a,10} L_9)^{1/2}\,\,\,\,;\,\,\,\, T_{\rm a,6} \approx 7.3
(\epsilon_{\rm heat} L_9^2)^{2/7},
\end{equation}
for a loop apex temperature of $T_{\rm a,6}$\,MK, half length $L_9$
(in units of $10^9$\,cm) and apex density $n_{\rm a,10}$ (in units of
$10^{10}$\,cm$^{-3}$). Using the left-hand equation yields:
\begin{equation}\label{eq:looptimescales}
\frac{\tau_{\rm cond}}{\tau_{\rm rad}} = {0.3 \over T_{\rm a,6}^{1/6}}.
\end{equation}
This shows that for coronal loops as a whole, conduction tends to be
somewhat more important than radiation in their response to energy
input fluctuations, but that both need to be involved in
modeling. \sactivity{$\circledS$ {\em Show:} (a) Use Eq.~(\ref{eq:rtv}) to estimate
  typical volumetric heating rates $\epsilon_{\rm heat}$ for a coronal
  region over 'quiet Sun' ({\em i.e.,} outside of active regions; with
  coronal field strengths of order 5\,G, loop-top temperatures of
  $\approx 1$\,MK, and loop half lengths $L\sim 4\,10^9$\,cm) and for
  an active (sunspot-bearing) region (with coronal field strengths of
  order 100\,G, loop-top temperatures of $\approx 3$\,MK, and loop
  half lengths $L\sim 15\,10^9$\,cm). (b) Compare these to the thermal
  energy densities of the gas based on densities estimated from
  Eq.~(\ref{eq:rtv}) with the above loop properties. (c) Also compute the
  ion gyro-radius, the ion sound speed, the loop crossing time, and the ratio of 
  the plasma pressure to the magnetic field pressure ({\em i.e.,}
  compute values of plasma-$\beta$
  (Eq.~\ref{eq:betadef})). (d) Are these numbers consistent with the
  concept of an isolated, field-dominated loop with an atmosphere is
  (near-)hydrostatic equilibrium? \mylabel{act:loopprop} \solution{loopprop}}

\section{Magnetic activity and atmospheric radiation}\label{sec:actrad}
\ors[III:2.2.3.2] ``The magnetic field in the solar atmosphere is associated with the
transport and dissipation of \indexit{solar!atmosphere!radiation}non-thermal energy; about one part in
$10^4$ of the Sun's luminosity is radiated from the quiet
chromosphere, and an order of magnitude less than that from the
corona.  For the most active stars [--~or, largely equivalently, for
the youngest stars (see Ch.~\ref{ch:evolvingplanetary})~--]
in\indexit{solar!activity!radiative losses} contrast, a total of about
1\%\ of the luminosity can be converted into outer-atmospheric
heating. [\ldots]

When measured for relatively large areas --~{\em i.e.,}  when averaging
over an ensemble of similar atmospheric components~-- the radiative losses
from the outer atmosphere increase with the [unsigned] magnetic flux density
at the base. A variety of heating mechanisms has been proposed for
the chromosphere, the corona, or --~for many scenarios~-- both.
Non-thermal\indexit{solar!atmospheric heating}
energy is likely deposited into the corona in the form of
electrical currents that are the result of the motion of
the field's photospheric footpoints that are moved about by
convective flows. The cascade of such
currents to smaller scales, and the details of the eventual
dissipation continue to be debated, as is the relative importance
of wave dissipation. For the chromosphere,
the situation is even less clear: waves of both predominantly
magnetic and predominantly acoustic nature have been proposed
to play a dominant role, but numerical simulations suggest that
electrical currents and reconnection phenomena contribute if not
dominate. [\ldots] 
With the high degree of structure in the magnetic field within the
chromosphere, different mechanisms may dominate in different
environments. [\ldots] 

The chromospheric and coronal emissions [scale essentially]
as power laws \indexit{atmosphere!losses!scaling}with each other and with the average magnetic
flux density of the underlying field: $F_{\rm i}\propto <|{\bf B}|>^{b_{\rm i}}$.
The power-law index $b_{\rm i}$ between
radiative and magnetic flux densities
appears to be
an essentially monotonic function of the formation temperature of the radiation
observed, increasing from about 0.5 for chromospheric emission from
$\sim 15,000$\,K plasma to
just over unity for X-ray emission from $\sim 3$\,MK plasma; these
power laws hold over a contrast in X-ray surface flux densities
from $100\times$ below the quiet Sun to $100\times$ above the active Sun,
spanning a total of nearly five orders of magnitude
(much of which will be
covered by the Sun over its lifetime [\ldots\ (see Ch.~\ref{ch:evolvingplanetary})]).

The chromospheric and coronal heating of the Sun and of stars\indexit{stellar!outer atmospheres!heating}
like the Sun are a function only of the magnetic flux density [\ldots] 
In other words, once the
magnetic field is in the stellar atmosphere, the dissipation
of that energy and the distribution of the energy over the outer-atmospheric
domains are independent of stellar properties: stars with masses
from about $0.09\,M_\odot$ (equivalent to $\approx 90$ Jupiter masses)
to a few solar masses, with radii of \hbox{$<0.5\,R_\odot$} to
\hbox{$>50\,R_\odot$}, and with coronal X-ray flux densities
ranging over a factor of $10^5$ all adhere to the same
scaling relationship within the measurement uncertainties and
the intrinsic stellar variability.''

\clearpage

%\part{{\bf Comparative eco-astrophysics}}
%\addtocontents{toc}{\vspace{0.5cm}{\centerline{\bf Comparative eco-astrophysics}}\par }

\chapter{{\bf Evolution of stars, activity, and asterospheres}}%10
\label{ch:evolvingstars}
{\narrower\narrower{
{\bf Chapter topics:}
\begin{itemize}
  \customitemize
\item Stellar evolution from formation to end of life
\item Scalings for stellar magnetic activity with age and rotation rate
\item Activity cycles
\item Stellar winds and astrospheres over time
\end{itemize}

\noindent{\bf Key concepts:}
\begin{itemize}
  \customitemize
\item Spectral type
\item T\,Tauri, main-sequence, red giant
\item White dwarf, supernova,
  neutron star
\item Asterospheric hydrogen wall
\end{itemize}

}}

\section{Evolution of stars}\label{sec:evolstruc}

This section summarizes stellar evolution in the context of
heliophysics, {\em i.e.,}  for stars like the Sun (defined below), stopping
short of the final evolutionary phases (white
dwarfs and neutron stars, as well as black holes and the supernovae on the path
to their creation). \ors[III:2.3.1] ``Here, we introduce only some
principles, terminology, and properties needed within the present
context:

In the strict definition, a star is a self-gravitating body in which
gravity is countered by gas pressure that is maintained by nuclear
fusion balancing the loss of thermal\indexit{definition!star} energy
through the stellar surface. Before a star forms, a contracting cloud
forms opaque but still nebulous Herbig-Haro objects associated
with\indexit{Herbig-Haro object} collapsing clouds, and then pre-main
sequence T\,Tauri stars\indexit{T\,Tauri star} (the subject of
Ch.~\ref{ch:formation}).  Once a balance between contraction and
internal pressure has been found, stars are on the 'main
sequence',\indexit{main sequence|seealso{definition}}
where\indexit{definition!main sequence} they spend by far the largest
fraction of their lifetime.  The term \indexit{main sequence}main sequence refers to the
well-defined clustering of stars in any one of a variety of
Hertzsprung-Russell\indexit{Hertzsprung-Russell diagram} diagrams, in
which the stellar luminosity or a logarithmic equivalent (the
[absolute] 'magnitude') is plotted against the surface temperature or
some filter ratio that measures the relative brightness in
differently-colored filters (often the $B-V$ value is used, referring
to the [absolute] {\em B\/}lue and {\em V\/}isible magnitudes,
respectively). Examples of such brightness-color diagrams (often
referred to as H--R diagrams) are shown in Figs.~\ref{fig:acthrd}
and~\ref{figure:evolmodel}(left). Stars are generally characterized by
their color, or an equivalent descriptor of their spectral properties
called 'spectral type' (see\indexit{spectral type} the top of
Fig.~\ref{fig:acthrd}). [Stars on the high-temperature, left side of
the HR diagram are called 'early' and those toward the right side
'late'; surface temperature and mass decrease monotonically along the
main sequence through the spectral type series: O, B, A, F, G, K, M.]

When stars run out of hydrogen fuel in their cores, they evolve off
the main sequence in the H-R diagram (see
Fig.~\ref{figure:evolmodel}(left)) to become giant or supergiant
stars. Their eventual fate depends on their mass: low-mass stars fade
into ever-cooling white dwarfs, heavier stars eject some of their
outer layers, while very heavy stars become \indexit{supernova}supernovae and leave
neutron stars or black holes behind. Objects that are too light to
sustain hydrogen fusion during any stage of their evolution (although
they may have phases with deuterium fusion) are called 'brown
dwarfs',\indexit{brown dwarf} which have masses of [$\la 0.07 M_{\rm
Sun}$ or ] $\la 75 M_{\rm Jupiter}$.  These cool very slowly, taking
billions of years to lose their thermal energy. Even cooler objects
merge into the realm of the (heavy) jovian planets.

Before stars reach the main sequence, they migrate through the H-R
diagram from the top right (as red giants), initially moving down (to
become red subgiants), then curving towards the main sequence
(increasing their temperature to become orange, yellow, white, or even
blue stars) with a much weaker change in their luminosity than during
their initial contraction phase [(see Figs.~\ref{fig:Ievolve}
and~\ref{figlh:fighr})]. All stars cooler than a surface temperature
of about 10,000\,K [(roughly from spectral type late-A)] have a
'convective envelope,' or\indexit{convective envelope} mantle,
immediately below their surfaces, and the coolest stars, be they young
or old, are fully convective.  All of these stars make up the ensemble
of cool stars, [all of which display some degree of \indexit{cool star}magnetic
activity.]\indexit{cool star|seealso{definition}}
Beneath\indexit{definition!cool star} the convective mantles, if any,
lie the 'radiative interiors' in which energy is transported
diffusively by photons; fusion occurs within this interior in the deep
'core' of main-sequence stars (see Fig.~\ref{fig:msstruct} for a
graphic comparison --~not to scale~-- of internal structure along the
main sequence).\activity{{\em Look up:} \label{act:starinabox} For an interactive
  visualization of stellar evolution and their tracks in an HR diagram
  see
  \href{https://starinabox.lco.global/}{https://starinabox.lco.global/}.
Use the data tables to create a diagram of main-sequence life time
vs.\ stellar mass, and formulate a power-law approximation $\tau_{\rm
  MS} \approx \epsilon (M_\ast/M_\odot)^\psi$ (save that for
comparison with Activity~\ref{act:snexposure}). Note how the heaviest
stars outlive this power law approximation, in part because they lose
a lot of mass while on the main-sequence.}

\begin{figure}[t]
\begin{center}
%\centerline{\hbox{\psfig{figure=figures/evolmodelSM.eps,width=14.cm,clip=}}}
\centerline{\hbox{\includegraphics[width=\textwidth]{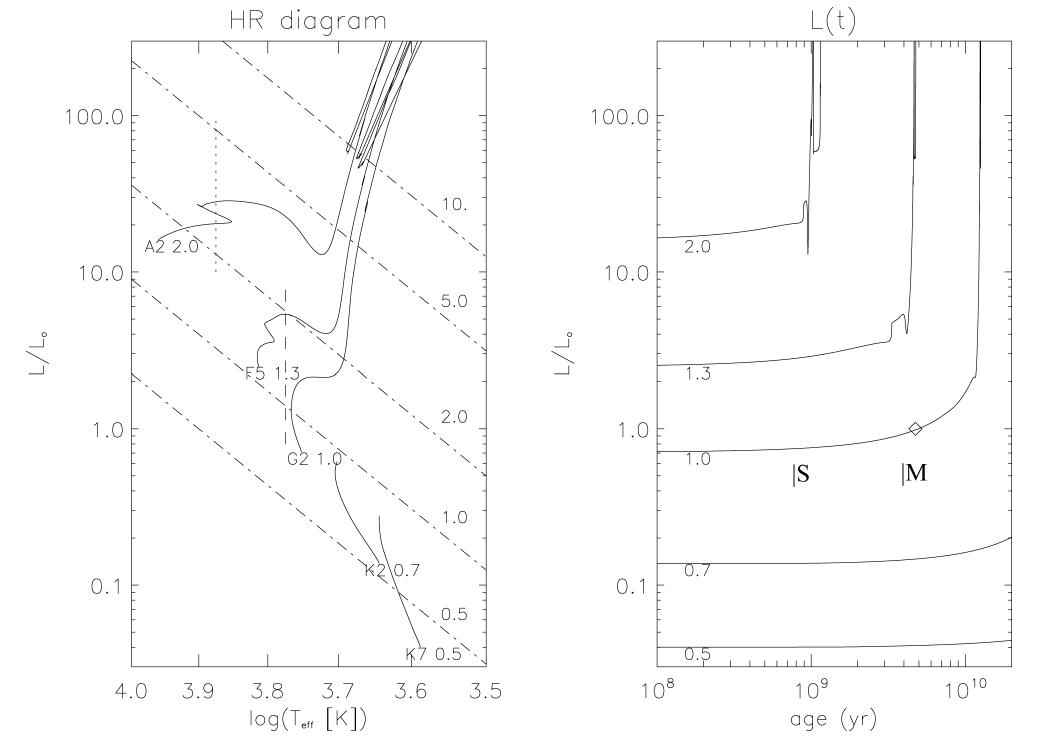}}}

\caption[Luminosity, surface
temperature, and age of evolving Sun-like stars.]{Evolutionary diagrams for luminosity, surface
temperature, and \indexit{Hertzsprung-Russell diagram}age from the mature, main-sequence phase onward. The
diagram on the {\em left\/} relates the stellar luminosity $L$ (in
present-day solar units) with the surface effective temperature
($T_{\rm eff}$; K) in
a Hertzsprung-Russel diagram (for an initial helium abundance of $Y=0.2734$ and
'metal' abundance (everything heavier than helium) of
$Z=0.0198$). Evolutionary tracks start on the \indexit{zero-age main
  sequence (ZAMS)}'zero-age main sequence'
(ZAMS), and are labeled with the spectral type on the main sequence
and the stellar mass (in solar units). The dashed line segment
indicates where dynamo action reaches its full strength; for shallower
convective envelopes (warmer stars), the activity level weakens until
it has dropped by a factor of 100 at the dotted line segment relative
to a Sun-like star at the same angular velocity. The slanted
dashed-dotted lines indicate stellar radii, with labels in solar
units. The diagram on the {\em right\/} shows the evolution of the
stellar luminosity with stellar age (yr since ZAMS). The diamond shows
the present-day Sun (see Fig.~\ref{fig:brownlee5} for details on the
Sun's red-giant phase).  The approximate ages for which the oldest
fossils of single-cell microbial life (S) and multi-cellular plants
and animals (M) have been found on Earth are indicated (see
Ch.~III:4). [Fig.~III:2.9]} \label{figure:evolmodel}
\end{center}
\end{figure}\nocite{pietrinferni+etal2004}\nocite{ref11}
The\indexit{stellar!evolution!time scales} evolutionary time scales
are a sensitive function of mass. A star with a mass of, say, three
solar masses evolves towards the \indexit{main sequence}main sequence in a few million
years[, exhibiting magnetic activity except in the final birth phase
near the main sequence.]  On the main sequence, where they stay for
'only' $\sim 0.4$\,Gyr, these stars have no magnetic activity, and
they only resume magnetic activity after they evolve off the main
sequence when they develop convective envelopes again for another 100
million years or so, until they rapidly evolve into what eventually
[becomes a white dwarf after ejecting much of the outer layers; a star
heavier than about nine solar masses ultimately] explodes as a
supernova.  A star of solar mass [($M_\odot$)] remains magnetically
active to some degree throughout the $\sim 10$\,Gyr that it spends on
the main sequence, \activity{{\em Show:} What fraction of the Sun's
  hydrogen would need to be converted to helium to keep it at
  (roughly) its current brightness throughout the time it spends on
  the main sequence? Once core hydrogen is consumed, the stellar
  internal structure changes considerably, enough to ignite fusion in
  higher layers as the star moves into its giant phase. Use
  $E=mc^2$. Start by making some simplifying assumptions. Be sure to
  state your assumptions and explain your approach. Which assumptions
  could you refine?  Could you use a different approach? \mylabel{act:hfrac}} and during
the subsequent $\sim 0.8$\,Gyr giant phase (its maximum radius may
reach $\sim 0.99$\,AU, and the maximum luminosity is likely to be
around 5,200\,$L_\odot$) until it evolves into an ever-cooling white
dwarf [after phases as giant star, ending in a series of pulses as the
star gasps for fuel, during which time an appreciable fraction of its
outer layers is ejected (compare Fig.~\ref{fig:brownlee5})].  An
$0.2\,M_\odot$ M9 brown dwarf takes $\sim 1$\,Gyr merely to contract
to the \indexit{main sequence}main sequence, changing little in effective temperature as it
descends in the H-R diagram.

\begin{figure}[t]
\centering
\includegraphics[width=7.5cm]{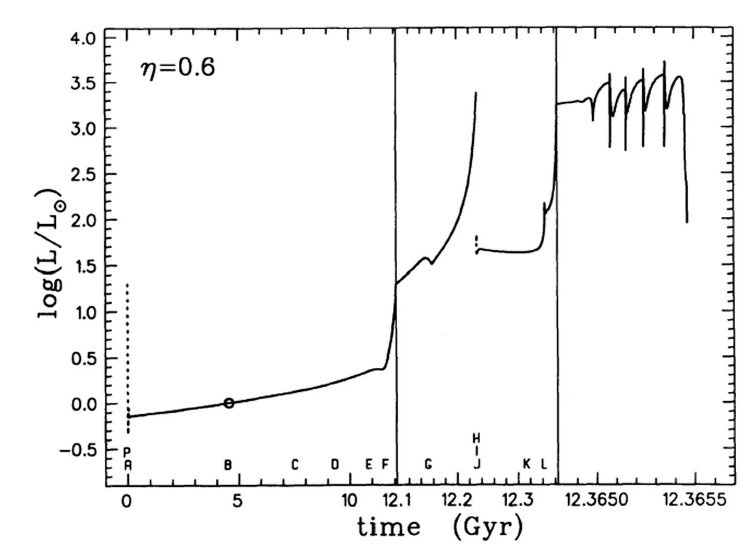}
\caption[Evolution of the luminosity of the Sun over
its full life span.]{\label{fig:brownlee5}
Evolution of the luminosity of the Sun over
its full life span.  The first 12 billion years show the gradual
brightness as hydrogen is depleted in the core of the Sun's
main-sequence lifetime as a hydrogen-fusing star ({\em cf.} 
Fig~\ref{figure:evolmodel}).  The large 
luminosity increases and pulses that follow this period include
both the red-giant and asymptotic giant branch (AGB) phases when the Sun
swells in size and loses appreciable mass to space. [Fig.~III:4.5;
\href{https://ui.adsabs.harvard.edu/abs/1993ApJ...418..457S/abstract}{source:
\citet{1993ApJ...418..457S}}].}
\end{figure}
During this evolution, the stellar luminosity and its associated color
([characterized by the 'effective temperature', and with it the
spectral irradiance]) continually change. Examples of evolutionary
changes are given in Fig.~\ref{figure:evolmodel} for stars from
$0.5 M_\odot$ to $2.0 M_\odot$.

The young Sun should have been some 25\%\ fainter at the start of the
Archean Eon (at $\sim 3.8\times 10^9$\,yr ago, when life is presumed
to have originated) than the current mature Sun according to
stellar-evolution models. This should have resulted in a much cooler
Earth, covered in ice.  Yet, geological records show that there was
liquid water on Earth even in the first billion years of its
existence.  How this could happen continues to be studied. The
greenhouse effect as a result of the high concentration of carbon
dioxide may have compensated the lower energy input from the young
Sun. Alternatively, the Sun may have been significantly more massive,
and therefore brighter, early in its life; if there has been
substantial mass\indexit{Sun!mass loss} loss in a strong wind in the
first billion years, this paradox would be resolved. [Analyses suggest
that it is possible] that a more massive young Sun (brighter, and with
somewhat tighter planetary orbits) is compatible with the present
internal structure for a Sun with a mass up to about 1.07 present
solar masses.  [One such model starts] with a mass of 1.07 solar
masses, has an initial irradiance at Earth that is 5\%\ higher than at
present (compared to 50\%\ lower for the Standard Solar Model), would
subsequently decrease to about 10\%\ lower, and then increase again to
the present value.'' \activity{{\em Show:} (a) What is the current mass
  loss rate due to the solar wind? Compare that to the mass loss due
  to fusion. Assuming these rates to be constant over the Sun's life
  time (although neither is: discuss the causes of these changes),
  what would be the mass of the early Sun (in units of the current
  solar mass). (b) How much greater would the mass loss have to be,
  averaged over time, to increase the early Sun's mass to
  1.07$M_\odot$?\mylabel{act:sunmassloss}}

\section{Stellar activity and its evolution}

\subsection{Overall activity level}\label{sec:overallactivity}

\ors[III:2.3.2] ``The \indexit{stellar!activity}defining properties of
stellar magnetic activity are the existence of \indexit{star!magnetic
  activity}variable coronal (X-ray) and chromospheric (UV-optical)
emissions. These characteristics are observed for a wide variety of
stars [(indicated in Fig.~\ref{fig:acthrd}). \ldots] In single stars
or in wide binaries, the activity level measured by emission from the
chromospheres or coronae of these stars, or by the coverage by
starspots, increases monotonically with increasing angular velocity to
rotation periods as short as a few days. Rather than using the
rotation period per se, however, studies of the rotation-activity
relationships frequently use the Rossby\indexit{Rossby number} number:
\begin{equation}\label{eq:rossby}
\Ro = {v_{\rm t} \over 2\Omega \sin(\theta) L_{\rm t}} \sim
{P_{\rm rot} \over \tau_{\rm conv}},
\end{equation}
which is defined such that it measures the relative importance of the
inertial to Coriolis forces (${\bf v} \cdot \nabla{\bf v}$ and
${\bf \Omega}\times{\bf v}$, respectively) acting on a parcel of
plasma of scale $L_{\rm t}$ moving with velocity ${\bf v}_{\rm t}$ in
a rotating system with angular velocity $\Omega$[; the Rossby number
is an important metric in the theory of astrophysical dynamos, see
around Eq.~\ref{eq:ross} and also Activity~\ref{act:rossby}].  The
central expression is a definition that includes the latitude, which
is often neglected when estimating the global effect of
rotation. When, moreover, the convective turnover time scale
$\tau_{\rm conv}=\pi L_{\rm t}/v_{\rm t}$ for characteristic length
scales and velocities of the deepest (largest and slowest) convective
motions in a stellar convective envelope is introduced, the commonly
used final expression results. When using the Rossby number [estimated
for the deepest layers of the convective envelope], the activity is
seen to increase with rotation up to a value of $\Ro \sim 0.1$ (see
Fig.~III:2.11). \regfootnote{A cautionary intermezzo:
  Sect.~\ref{sec:actrad} gives power-law scalings between radiative
  losses from chromospheres and coronae over steller surface areas
  with mean magnetic flux densities over these areas (which hold
  approximately without changes for areas up to entire
  hemispheres). The values of the power-law indices in these
  relationships depend on the formation temperatures of the
  spectral lines or bandpasses used (thus, for example, steepening towards
  higher-energy X-ray channels), while published values also depend on
  the correction for a reference level (there is a minimum or 'basal'
  level of chromospheric emission that needs to be subtracted first
  but different authors use different corrections). This dependence on
  the details of the diagnostics used are one cause behind the
  somewhat different power-law scaling between coronal and
  chromospheric radiative losses you find in the literature. There are
  other reasons why you may find other approximate
  parameterizations. For one thing, although the scaling in
  rotation-activity diagrams between a relative brightness in terms of
  luminosity or surface flux density
  ($L_{\rm i}/L_{\rm bol}\equiv F_{\rm i}/F_{\rm bol}$) versus Rossby
  number works fairly well, it does not work perfectly, and other
  authors, using other steller samples, might prefer using $F_{\rm i}$
  versus $P_{\rm rot}$. As long as the stellar sample contains stars
  of rather comparable internal properties, the choice of metric does
  not matter, but for more diverse samples, scalings with these
  properties matter --~no simple multiplicative scaling seems to lead
  to a single tight rotation-activity relationship for all cool
  stars. Other reasons for differing results from different studies
  include the fact that the relationships are not simple power laws
  and fits thus depend on the parameter range covered in stellar
  samples, and, of course, uncertainties in models for, {\em e.g.,} stellar
  ages, and intrinsic stellar variability combined with relatively
  small samples. You could review, for example,
  \href{https://ui.adsabs.harvard.edu/abs/2017MNRAS.471.1012B/abstract}{the study} by
  \citet{2017MNRAS.471.1012B} for more discussion and for
  references. \mylabel{act:powerlaws}}

For even more rapidly rotating stars, the activity reaches
a\indexit{stellar!activity!saturation}
saturation\indexit{dynamo!saturation} level, and for stars with
rotation periods of only a fraction of a day, supersaturation sets in,
with activity decreasing with increasing angular velocity. It appears
that when proceeding towards shorter rotation periods, the coronal
activity saturates first, followed by chromospheric activity, and
finally by starspot coverage. This has led to the suggestion that
different processes set in at successively shorter rotation periods:
centrifugal stripping (see Activity~\ref{act:centrifugal})
of\indexit{centrifugal stripping} the high corona, saturation of the
level to which non-thermal heating can be extracted from the
near-surface convection or deposited into the chromosphere, and
finally saturation of the dynamo process itself possibly by the
coupling of the magnetic field and the plasma flows (see
Section~\ref{sec:alphaquenching}) or because the Coriolis force
changes the large-scale circulation patterns that are involved in
efficient dynamo action.

\begin{figure}
%  \centerline{\psfig{figure=figures/Boothetal_Fig3_snap.eps,width=12cm,clip=}}
  \centerline{\includegraphics[width=12cm]{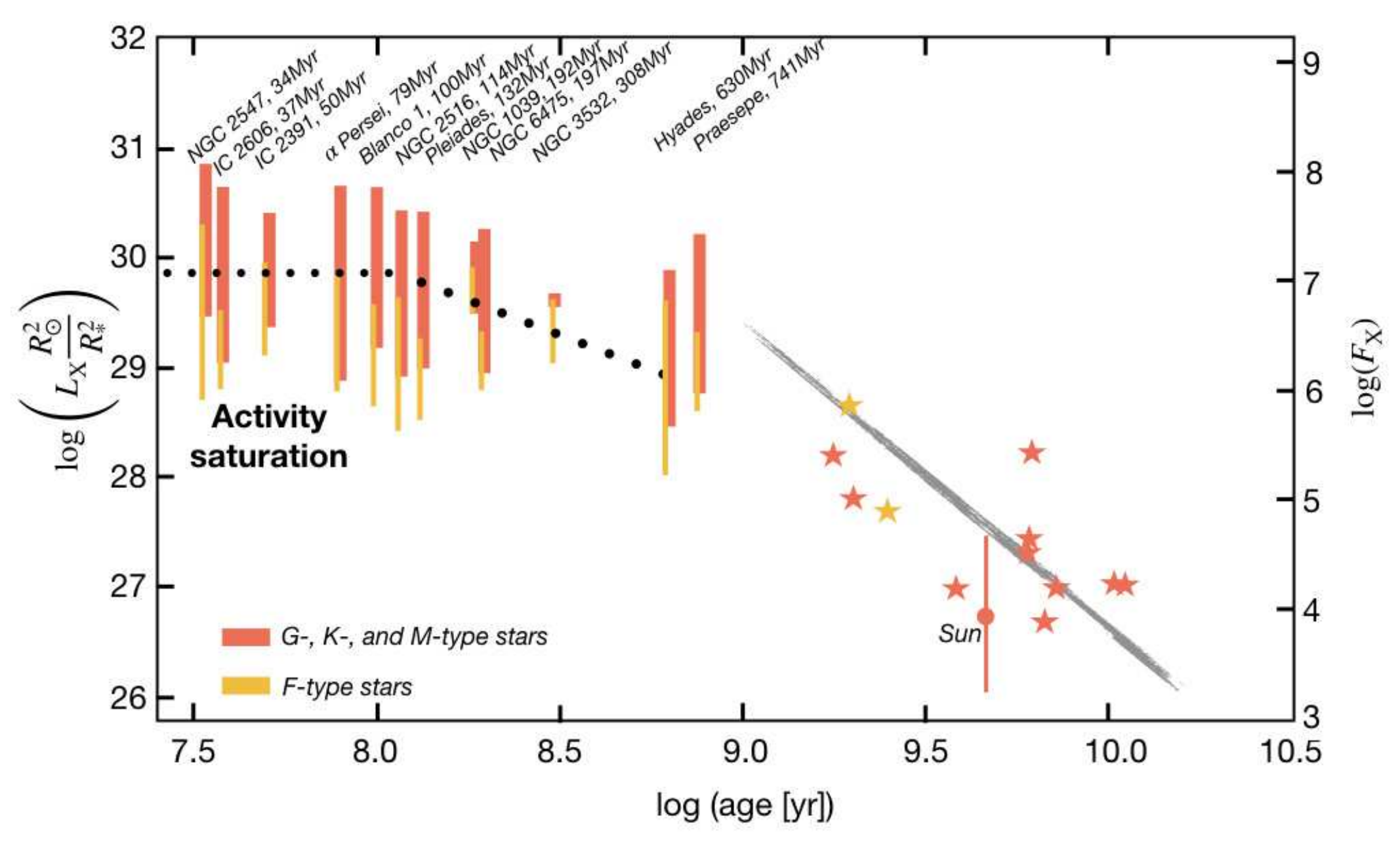}}
  \caption[Age-activity relationship for main-sequence
  stars.]{\label{figure:rotact} Relationship between stellar coronal
    X-ray brightness \indexit{atmosphere!losses!scaling}(in a passband from 6\,\AA\ to 60\,\AA, expressed as a
    luminosity scaled \indexit{rotation-activity relationship}to the equivalent solar surface area on the
    {\rm left}, and as a surface flux density on the {\rm right}) versus
    age. Bars show data for open clusters: in orange for the range of
    G-, K-, and M-type stars observed, and in yellow for the generally
    less active F-type stars. Individual stars with ages above
    1\,Gyr are shown by stars, and the Sun as a filled circle (with
    its range over the solar cycle). The relationship shown by the
    gray line segment is a fit by Booth {\em et al.}\ (2017). The dotted
    line is a relationship for stars with $(B-V) \in
    [0.56,0.79]$. [\href{https://ui.adsabs.harvard.edu/abs/2017MNRAS.471.1012B/exportcitation}{After
      \citet{2017MNRAS.471.1012B}} and references therein.] \colorfig }
\end{figure}
  
%\begin{figure}
%\centerline{\psfig{figure=figures/randich.ps,bbllx=191pt,bblly=433pt,bburx=522pt,bbury=682pt,width=6.5cm,clip=}\psfig{figure=figures/skumanichplotBW.eps,clip=,width=6.1cm}}
%%\psfig{figure=../schrijver/skumanichplotBW.ps,bbllx=191bp,bblly=225bp,bburx=531bp,bbury=495bp,clip=,width=6.1cm}}
%\caption[Activity-rotation-age relationships for main-sequence
%stars.]{\label{figure:rotact} {\em Left:\/} Relationship between coronal
%X-ray emission relative to the stellar luminosity,
%$R_{\rm X}=L_{\rm X}/L_{\rm bol}$ --~shown vertically~-- and
%Rossby number $\Ro$ 
%%(Eq.~2.1)
%(Eq.~\ref{eq:rossby})
%--~shown horizontally~--
%for F5 through M5 main-sequence
%stars in four clusters and a selection of
%field stars. Symbols: filled triangles for
%IC\,2391 ($\sim 40$\,Myr), filled squares for $\alpha$~Per ($\sim 70$\,Myr), filled
%diamonds for Pleiades ($\sim 115$\,Myr), open circles for Hyades ($\sim 600$\,Myr),
%and open squares for field stars.
%The Sun ($\odot$) is located near the low end
%of the activity measure for the sample of stars in this figure.
%[Cluster ages in the original figure are updated in this caption.] 
%{\em Right:\/}
%Activity-age
%relationship for main-sequence stars for the chromospheric Ca\,II\,H,K
%emission for field stars (diamonds and squares) and average
%properties for associations of young stars. [Fig.~III:2.11]}
%\end{figure}\nocite{baliunas+etal98}\nocite{patten+simon96}\nocite{denissenkov+etal2009}
%%\indexit{star cluster: Pleiades}
%%\indexit{star cluster: $\alpha$ Persei}
%%\indexit{star cluster: IC\,2391}
Main-sequence stars warmer than the Sun have \indexit{star!convective
  envelope}shallower convective envelopes.  Their magnetic activity is
markedly suppressed compared to cooler stars with the same rotation
period.  This has been argued to be either because of the shallowness
of their envelope or because of the short average turnover time of
convection resulting in little influence of the Coriolis force that
otherwise would introduce a preferential direction into the system.
By spectral type F2 significant magnetic activity is observed, which
rapidly increases in efficiency towards G0 as the convective envelope
becomes deeper and the time scales of deep convective motions approach
or exceed the [characteristic mean] rotation period.

Magnetic fields are observed along the \indexit{main sequence!magnetic
activity}main sequence as far down the
spectral scale as we
have been able to identify and apply Zeeman sensitive spectral lines,
{\em i.e.,}  down to at least M9.5. At that point we have already reached
the brown dwarfs, {\em i.e.,}  astrophysical objects that are too small
to have sustained hydrogen fusion in their cores.

For stars above the \indexit{star!magnetic activity}main sequence, activity is seen
both in stars that have recently formed and are still contracting
to the main sequence (pre-main-sequence stars, which include fully
convective T\,Tauri stars) and stars that have exhausted their
core hydrogen supply and are moving away from the main sequence,
once again en route to a fully-convective giant phase, now
sustained by nuclear fusion of helium and heavier elements in
either their core or in shells surrounding a burned out core.

\begin{figure}[h!]
%\centerline{\psfig{figure=figures/HK_stars.eps,width=10cm}}
\centerline{\includegraphics[width=10cm]{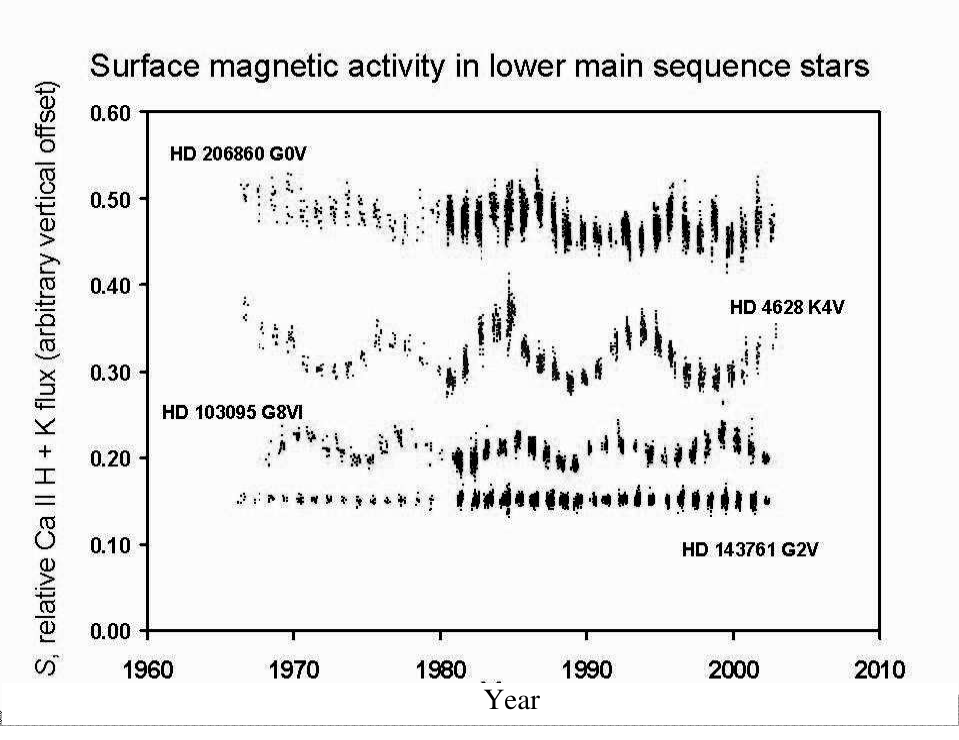}}
\caption[Examples of cycles in stellar chromospheric
activity.]{Examples of chromospheric activity cycles (as observed in
  the H and K resonance lines of \indexit{activity!cycle!stars}singly ionized Ca, or Ca~{\sc II}
  H+K).  Surface magnetic activity records of four stars on or near
  the lower main sequence from a survey begun by O.C.\ Wilson in 1966
  at Mount Wilson Observatory. [The ratio of the flux in the emission
  cores of singly-ionized calcium lines in the violet (the Fraunhofer
  H and K lines at 393.3 and 396.8 nm) and photospheric flux in nearby
  regions of the spectrum, necessarily integrated over the unresolved
  stellar disks, is used as a proxy for stellar magnetic activity.]
  The strength of the H and K fluxes increases as the coverage by and
  intensity of magnetic surface features increases; on the Sun the H
  and K fluxes vary nearly in phase with the sunspot cycle. The four
  records show the counterpart of the Sun approximately 2 billion
  years ago (upper curve, HD\,206860; $P_{\rm rot}=4.7$\,d), and then
  three Sun-like stars, which show records similar to the present-day
  Sun, HD\,4628 ($P_{\rm rot}=38$\,d), HD\,103095
  ($P_{\rm rot}=31$\,d, or $P_{\rm rot}=60$\,d) and HD\,143761
  ($P_{\rm rot}=21$\,d). Both HD\,4628 and HD\,103095 display decadal
  periodicities similar to the sunspot cycle. The star HD\,143761 may
  be in a state like the Sun's Maunder Minimum. The star HD\,103095 is
  an extremely old (approximately 10 billion years) metal deficient
  subdwarf, and is shown as an example of the persistence of decadal
  magnetic activity cycles in a star of extreme age compared to the
  Sun. The spectral types are listed next to each record's star
  name. Arbitrary vertical shifts in the average value of the H and K
  relative fluxes have been applied to show the records
  without overlap; the offsets are 0.0 (HD\,143761), 0.02
  (HD\,103095), 0.09 (HD\,4628) and 0.15
  (HD\,206860). [Fig.~III:2.12] \label{fig:HK_stars} }
\end{figure}\nocite{frick+etal2004}
During their main-sequence phases, cool stars exhibit a variety of
activity patterns. A clear activity cycle,\indexit{stellar!activity!cycle} as exhibited by the \indexit{sunspot!cycle!stars}Sun
ever\indexit{activity cycle|see{sunspot}}
since the Maunder minimum [(a period from about 1645~CE to 1715\,CE
during which sunspots were very infrequent and sometimes absent for
multiple years)], is relatively rare, even for solar analogs: only
roughly 60\%\ of solar-like stars show a clear activity cycle, and the
reasons for this and for those that set the cycle duration are still
being researched (see Fig.~\ref{fig:HK_stars}).

A few\indexit{flat-activity star} Sun-like stars in the solar
neighborhood are so-called flat-activity stars,
showing\indexit{stellar!activity!flat-activity} no clear cycle at all,
yet they rotate with a period similar to that of the Sun. Such stars
have been argued to be in a state similar to the solar\indexit{Maunder
Minimum} Maunder minimum [\ldots]''

\subsection{Flares} 

\ors[III:2.3.3] ``Solar flares define \indexit{star!flare}power
laws\indexit{flare!power-law spectra} in spectra of frequency,
$N_{\rm f}$,
versus peak brightness or overall energy, $E_{\rm f}$. The spectrum of
$N_{\rm f}(E_{\rm f})$\indexit{flare!stellar} can be approximated by a power law
$N_{\rm f}(E_{\rm f}){\rm d}E_{\rm f}\propto E_{\rm f}^{\alpha_{\rm f}} {\rm d}E_{\rm f}$ with $\alpha_{\rm f}
\approx -2$; the value of $\alpha_{\rm f}$ reported in the literature
depends on the instrument and wavelengths used, and on the sample used
(active region flares, EUV quiet-Sun brightenings, etc.), and ranges
from about $-2.4$ to $-1.5$. The flare energies studied range from
$\sim 10^{24}$\,ergs to $\sim 10^{32}$\,ergs [for the present-day Sun].

The relatively small solar flares drown into a quasi-steady background
emission if the Sun is observed as a star. It is not surprising,
therefore, that stellar flare spectra are limited to large flares that
stand out above the surface-integrated X-ray fluxes. [O]bservations of
F through M type main-sequence stars [reveal] ubiquitous power laws
with power-law indices near $\alpha_{\rm f}=-2$ (with a possible mild
steepening from cool to warm stars). Flare X-ray [energies for some
cool stars] range up to $10^{35}$\,ergs, {\em i.e.,} up to $\sim 1000$ times
brighter than the largest solar flares, with no evidence for a cutoff
energy. [For the more active stars in the population,] flare
frequencies for energies exceeding $10^{32}$\,ergs scale
proportionally to the time-averaged X-ray emission, saturating as the
X-ray activity saturates, and contribute some 10\%\ of the total X-ray
luminosity. [\ldots\ A]dopting $\alpha_{\rm f}\equiv -2$, and using a
characteristic solar X-ray luminosity [around cycle maximum] of
$L_{\rm X,\odot} = 3\,10^{27}$\,erg\,s$^{-1}$, supports a scaling for
the frequency of large flares with energy $E_{\rm f}$ exceeding a
threshold value of $E^\ast_{{\rm f},32}$ (in units of $10^{32}$\,ergs,
characteristic of a large solar flare) of
\begin{equation}\label{eq:flarefreq}
N_{\rm f}^\ast(E_{\rm f}>E^\ast_{{\rm f},32}) \approx 0.26 \left( L_{\rm X} \over
L_{\rm X,\odot} \right)^{0.95\pm 0.1} \left( 1 \over E^\ast_{{\rm f},32} \right)\,\,{\rm /day}.
\end{equation}
[Note that this relationship derived from observations of very active stars lies some two
orders of magnitude above the cycle-averaged frequency distribution observed for the
time-average present-day Sun. This mismatch remains a mystery, as
does the problem of establishing whether flares of $>10^{33}$\,ergs
can still occur on the current Sun, or whether that was only possible
in the distant past. For young, active stars we can use the above
expression to find that when] the Sun was
only 0.1\,Gyr old [\ldots] flares 
with energies\indexit{flare!frequency} exceeding $10^{35}$\,ergs would
likely have occurred once per week, and those with energies exceeding
$10^{38}$\,ergs may have occurred about once per
decade.\activity{{\em Consider:} Note that integration over the power in flares as
  parameterized in Eq.~(\ref{eq:flarefreq}) diverges when the lower
  and upper limits extend to $[0,\infty]$. Compile a list of processes that
  could be at play in introducing cutoffs to the integral on either
  side and of your reasons to include them.
  The answer remains under study: it is not clear over what
  range Eq.~(\ref{eq:flarefreq}) holds its slope, or what determines
  the energy of the 'largest flare', or how and  how  much relatively
  tiny 
  'nano-flares' contribute to coronal heating. But considering the
  possibilities should prove educational. \mylabel{act:integraldivergence}}

It appears that quiescent activity and flaring activity on stars scale
with each other, as also seen in the rise and fall of quiescent and
impulsive heating through the solar cycle. One result of this is that
more active stars have a stronger high-temperature coronal component, so that
the effective X-ray 'color temperature' or spectral hardness increases
with activity. It also appears that larger flares are associated with
higher characteristic temperatures, going from solar micro-flares to
large flares on very active cool stars [\ldots]''

\subsection{Rotation rates} 

\ors[III:2.5] ``The primary\indexit{magnetic!activity!angular momentum
loss} stellar property \indexit{stellar!rotation rate}that determines the level of magnetic activity
is the rate of rotation. The rate at which\indexit{stellar!moment of inertia} a\indexit{moment of inertia!stars} star
spins is influenced by the evolutionary changes in (1) the moment of
inertia, (2) the angular momentum loss through a stellar wind, and (3)
the angular momentum exchange in tidally interacting binaries.

\begin{figure}
\centering
%\includefigure{\psfig{figure=figures/inertia.ps,width=11truecm,bbllx=45bp,bblly=230bp,bburx=522bp,bbury=630bp,clip=}}
\includegraphics[width=13truecm,bb=45 230 522 630]{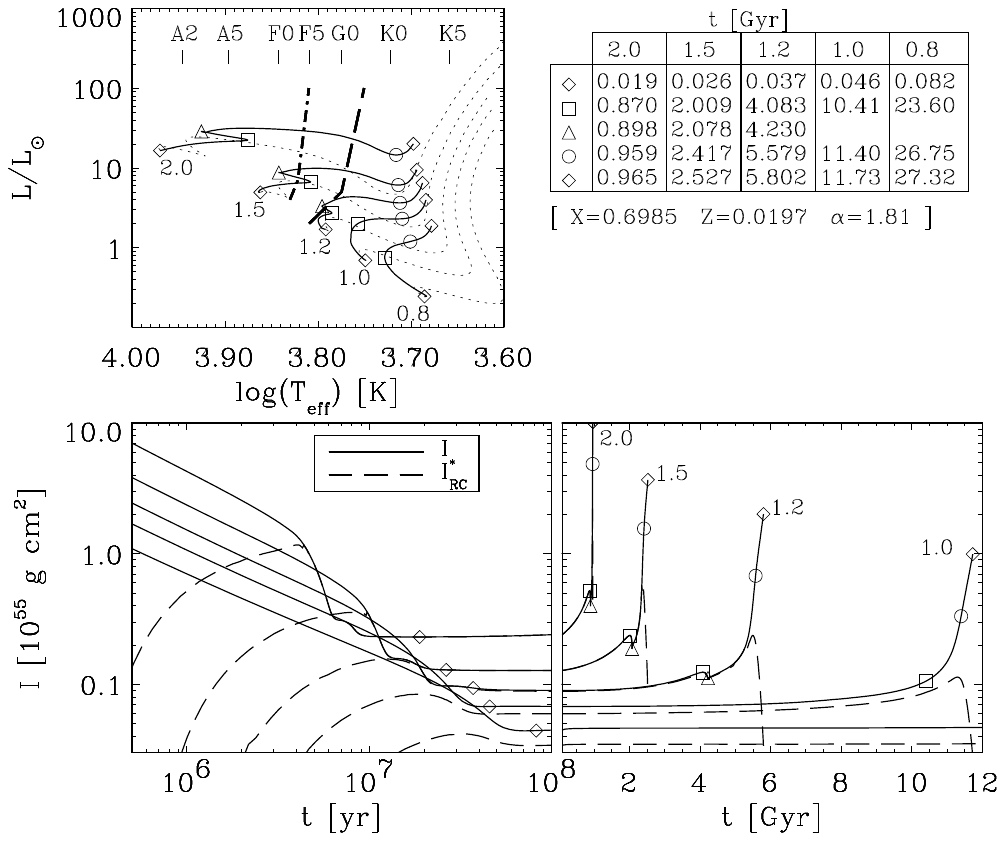}
\caption[Moments of inertia of evolving stars.]{ Evolutionary tracks
  ({\em top panel}) for \indexit{Hertzsprung-Russell diagram}late-type stars of various masses, from the pre-main
  sequence to main \indexit{star!evolution!time scale}sequence (dotted curves in the top panel), and from
  there to the base of the giant branch (solid curves in the top
  panel).  The diamonds indicate the zero-age main sequence (at the
  lower-left end of the solid curves) and the end of model
  computations. Stellar masses are given in units of the solar
  mass. The dashed-dotted curve marks the onset of envelope
  convection.  Ages at selected points along the tracks are listed in
  the table in the top right of this figure for stellar masses
  indicated in the top row. The evolutionary variations of moments of
  inertia of the entire star (solid curves) and of the radiative
  interior below any convective envelope (dashed curves) are shown in
  the {\em lower panels}. [Fig.~III:2.15]}\label{fig:Ievolve}
\end{figure}\nocite{charbonneau+etal96}
(1) The evolutionary changes in the global moment of inertia are
readily computed from stellar evolutionary models (see the example in
Fig.~\ref{fig:Ievolve}). These changes amount to several orders of
magnitude during the first tens of millions of years of a star (when
magnetic coupling with surrounding accretion disks are also important,
see Ch.~\ref{ch:formation}) and the final fraction of a Gyr, but
during the main-sequence phase, they are generally negligibly small
compared to the loss of angular momentum through the outflowing wind.

(2) The\indexit{mass loss!angular momentum loss}
outflowing\indexit{magnetic!braking} stellar wind is coupled to the
stellar magnetic field, which introduces a relatively long arm over
which the stellar wind can extract angular momentum, so that it
eventually carries far more than its own specific angular momentum
[(see Sect.~\ref{sec:braking})]. The torque on the star is
applied by the magnetic field into the stellar interior, and the rapid
convective motions cause the angular momentum to be extracted from the
entire convective envelope. How much radial and differential rotation
this sets up within the convective envelope remains under active
study, but the argument is generally made that the convective envelope
spins down as a whole. The coupling to the radiative interior
underneath it occurs somehow by coupling to a primordial field, wave
exchange, or slow flows. In rapidly evolving stars with shallow
convective envelopes this may lead to a (temporary) strong
differential rotation between envelope and interior. For the Sun,
however, helioseismic measurements have shown that interior and
envelope rotate at very nearly the same rate, with the interior
matching\indexit{differential rotation!internal} the angular velocity
of the differentially rotating envelope at a latitude of about
$30^\circ$ (see Fig.~\ref{convection1}b).

The\indexit{angular momentum!loss} angular\indexit{activity-age
  relationship} momentum loss leads to a spin down relative to the
evolution in which only the total moment of inertia $I$ is evolved as
the star ages.  During the main-sequence phase, $I$ changes little
(Fig.~\ref{fig:Ievolve}) so that most of the change of $P_{\rm rot}$
with age [is associated with magnetic braking.  \sactivity{$\circledS$
  {\em Show:} Review Fig.~\ref{fig:Ievolve} to estimate the mass of
  the least-massive post-main-sequence star (say, past the phase
  marked by a turn towards cooler surfaces marked by the squares, and
  only considering single stars) that could exist in the present-day
  Universe. Identify the symbols with the various evolutionary phases
  of a star of less then 2 solar masses, including: main sequence, red
  giant, and white dwarf. Note: this activity is a warm up for
  activity \ref{act:globclus}.\mylabel{act:hrevol}
  \solution{hrevol}} After the first Gyr when dynamo
saturation is important, the dependence of rotation rate on age $t$
for Sun-like stars can be approximated by what is often referred to as
the \indexit{Skumanich law}Skumanich law:]
\begin{equation}\label{eq:skumanich}
\Omega_{\rm rot} \propto t^{-1/2}.
\end{equation}
For the present-day Sun, the time scale of angular momentum loss is
$\sim 1$\,Gyr. [\ldots] \activity{{\em Show:} Fig.~\ref{fig:Ievolve} can
  be used to illustrate how astronomers determine the ages of 'open
  clusters' of stars (other than by the modern means of
  asteroseismology): assuming that all stars in a cluster are formed
  at about the same time, the shape of the HR diagram for stars in a
  cluster reveals the age when compared to theoretical evolutionary
  tracks as in the panel on the upper left. Try this: assume the stars
  are all close to 1\,Gyr old, as in the open cluster called NGC~2355,
  then mark the approximate positions of the stars at that age in the
  upper-left panel (the HR diagram with evolutionary tracks) of
  Fig.~\ref{fig:Ievolve} estimating also where stars of intermediate
  masses might show up, and realize how the turnoff from the main
  sequence in such a cluster HR diagram reveals the age of the cluster
  (see the discussion of Activity~\ref{act:hrevol} in
  Sect.~\ref{hrevol}.  Open clusters all have the low-mass end of
  this HR diagram in common, so even if the distance to a cluster is
  not known, the distance can be determined by shifting that low-mass
  tail to overlap with that of a cluster of known distance. Also: look
  up the definition of 'open cluster' in contrast to a 'globular
  cluster'. \mylabel{act:globclus}} 

(3) \indexit{magnetic!activity!binary stars} [\ldots] Even though the
Sun is a single star, there are interesting lessons to be learned from
close binaries. [\ldots] It is the population of tidally-interacting
binaries [\ldots] that unambiguously showed us that activity is
related causally to rotation, and only indirectly to stellar age
[\ldots\,:]'' the combination of angular momentum loss by magnetic
activity of one or both of the binary components and their tidal
interaction drains the system of angular momentum, which results in
tightening of the orbits and that, in turn, to a spin-up of the stars'
rotation rates and an increase of activity level with advancing age,
contrary to what happens in single stars (see Ch.~\ref{ch:torques}).

At present, there are no instruments capable of detecting stellar
CMEs, leaving us only --~for the time being, at least~-- with the Sun
as the guiding star, and with assumptions for scalings of CME
properties for other stars with different properties and ages.

\subsection{Stellar infancy: birth to the zero-age main sequence }

\ors[IV:2.2.1] ``Although \indexit{star!activity!birth to ZAMS}the age range in this first age category is
only a percent or so \indexit{zero-age main
  sequence (ZAMS)}of the total main-sequence lifetime of the star,
there are several important steps to the life of the star which occur
during this time. [\ldots] For this section we concentrate on ages
ranging from stellar birth to the time it takes the star to reach the
zero-age main sequence, at which point the star is in stable
hydrostatic equilibrium, and there is negligible contribution to the
stellar luminosity from any accretion-related processes.  This time
scale is a function of stellar mass, being approximately 50\,Myr for a
solar-mass star, and longer than 160\,Myr for a star of 0.5
M$_{\odot}$ or less.  For the purposes of discussion, and because
stellar ages can be uncertain by factors of two or more, we include
stars of ages up to $\sim$100\,Myr. [\ldots]

Magnetic activity in general is at a high level in these young stars
because of their rapid rotation, but the interpretation can be
confused by other processes occurring in the system which have similar
observational characteristics to magnetic reconnection processes [as
discussed in Ch.~\ref{ch:formation}. \ldots] Flares some 100-1000
times more energetic than the biggest solar flares occur roughly once
a week on these young, rapidly spinning stars. [\ldots] Over slightly
longer evolutionary time scales there is a decrease in flare rate.''

\subsection{Stellar teenage years: ZAMS - 1\,Gyr}\label{sec:teenage}

\ors[IV:2.2.2] ``At this \indexit{star!young-star activity}phase in a
star's evolution, rapid \indexit{zero-age main
  sequence (ZAMS)}rotation is still an important factor,
although it has declined since the star's youth.  According to
[Eq.~(\ref{eq:skumanich}), a 'teenage'] solar mass star would have a
rotation rate that is only a factor of 2--7 above the Sun's
present-day rotation; activity that accompanies the faster rotation
should be enhanced, but below the extremes represented by the youngest
stars.  [G- and K-type main-sequence stars spin down faster than their
M-type counterparts (see Fig.~\ref{fig:acthrd} for their properties),]
so by these ages, M dwarfs dominate the samples of active stars.  The
general decrease in activity levels compared to the extremes seen at
young ages means that capturing flaring activity on stars of this age
range (with the exception of M dwarfs) is more difficult to do
systematically, and consequently there is a heavy bias towards the
lower mass end in observations of flares on stars of this age range.
The fact that M dwarfs are the most common type of star based on mass
functions also contributes to this bias.  There are open clusters
(notably the Hyades at an age of $\sim$800\,Myr) which are nearby
enough for sensitive studies of explosive events, although they are
spatially dispersed compared to star forming regions and this makes it
difficult to capture more than one or two objects in the field of view
of typical astronomical telescopes.

The possible dependence of stellar flare rate on evolutionary age can
be explored by combining scaling relations between flare frequency and
underlying coronal emission with those relating coronal and
chromospheric emission, and others describing the decline of
chromospheric emission with time. [The empirical scaling in
Eq.~(\ref{eq:flarefreq})] between coronal flare rate and underlying
stellar X-ray luminosity [appears to hold also for stars] with ages in
this age range[. \ldots\ There are] scalings between coronal emission and
different chromospheric emission indicators for cool main-sequence
dwarfs, $L_{X} \propto L_{\rm chrom}^{y}$ where $y\sim$1.5 for C~IV
emission [from triply-ionized carbon typical of the transition
region], $y\sim2$ for Ca~II HK emission and $y\sim$3 for Mg~II h
emission[, the latter two being characteristic of singly-ionized Ca
and Mg which are strong emitters from the chromosphere. \ldots\
Chromospheric emission declines with rotation rate, which can be
transformed into a relationship with stellar age using
Eq.~(\ref{eq:skumanich}) to be \indexit{atmosphere!losses!scaling}roughly
\begin{equation}\label{eq:chromt}
L_{\rm chrom} \propto t^{-1/2}.
\end{equation}
Simplifying] these relations to
\begin{equation}
N_{\rm f}(>E_{\rm f,c}) \propto L_{X},
\end{equation}
\begin{equation}
L_{X} \propto L_{\rm chrom}^{y},
\end{equation}
where $y$ takes on different values depending on the chromospheric
emission being considered, and [with Eq.~(\ref{eq:chromt})
\regfootnote{Note that the power laws shown in this chapter relating
  stellar activity, wind, and rotation are not all consistent with
  each other. This is not all attributable to the sensitivity to the
  diagnostics used (see Activity~\ref{act:powerlaws}), which tells us
  something is missing in how these various parameters really scale
  with each other, but observations and/or theory have yet to reveal
  what it is that is missing. Part of the discrepancy is likely the use of
  different stellar samples in different studies; compare, for
  example, the slopes of the power-law fits in
  Fig.~\ref{figure:rotact}: the slope for
  $L_{\rm X}(t>800\,{\rm Myr})$ differs depending on the age range of
  the stars that is included in a study. Another reason may be a
  change in the dependence of the loss of mass and angular momentum
  somewhere around the age of 1\,Gyr --~come back to this after
  reading Sect.~\ref{sec:stellarwinds}.}], suggests that the flare rate
may decline with age anywhere from
$N_{\rm f}(>E_{\rm f,c}) \propto t^{-0.75}$ to
$N_{\rm f}(>E_{\rm f,c}) \propto t^{-1.5}$.  [The] above scaling
between flare rate and coronal luminosity cannot be used to 'correct'
the flare rate of [young, active stars] to the solar flare rate via
their coronal luminosity.  This suggests a breakdown in the validity
of a scaling relation approach [at ages, and commensurate rotation
rates, between the 'teenage years' and the Sun's present age].

Single G stars in this age range exhibit flares at least as powerful
as the largest solar flares, but occurring several times per day.
[One example is $\kappa$~Cet, a G5V star with an age of 300--400\,Myr
that exhibits 6.7 flares per day with energies of at least
10$^{32}$\,erg. The fraction of time that stars are clearly flaring in
their coronal X-ray emission tends to decrease with age, from
approximately 10\%\ around 1\,Myr to about 3\%\ approaching 1\,Gyr
(see Fig.~IV:2.8).]
%Figure~\ref{fig:xrate}
%depicts the combined flare rates of objects as a function of age, up
%to about 1\,Gyr, separated by evolutionary status at the youngest
%ages. [\ldots]'' 
%\begin{figure}[t]
%%\includegraphics[scale=2.5]{../osten/art/Stelzer2000_fig10.ps}
%\includegraphics[scale=1.5]{figures/Stelzer2000_fig10new.eps}
%\caption[X-ray flare time fraction for stars of different age.]{
%X-ray flare rate expressed as a percent of observing time for stars of
%different ages.  The sensitivity to flares is set to a uniform value
%to account for the differing distances. Diamonds indicate classical T
%Tauri stars, triangles weak-lined T Tauri stars. The horizontal line
%gives the age spread of the young stellar objects. There is a clear
%drop in flare rate by the age of the Hyades
%(800\,Myr). [Fig.~IV:2.8; \href{http://articles.adsabs.harvard.edu/pdf/2000A\%26A...356..949S}{source}.]
%\label{fig:xrate}
%}
%\end{figure}
%\nocite{2000A&A...356..949S}
%\figindex {../osten/art/Stelzer2000_fig10new.eps}

\subsection{Stellar adulthood: 1-5\,Gyr}

\ors[IV:2.2.3] ``The Solar System,
\indexit{star!activity!adulthood}and thus the Sun's, age measurement
of 4568\,Myr fits squarely within the 'stellar adulthood' phase of its
life.  Detections of flares on stars in this age range are much fewer.
The decline of flaring with age is generally assumed to follow the
trends of other activity indicators, but whether this is in fact the
case is an open question.  Evidence that magnetic activity may not
decline monotonically at Gyr ages comes from a few sources: [\ldots]
chromospheric activity in M dwarfs [may] not decline in the 1-10\,Gyr
range as fast as predicted based on extrapolating from objects with
ages $<$ 1\,Gyr. [For stars older than a few Gyr, it appears that
there is no evidence for further decay in quiescent chromospheric
activity after the major decline in activity seen] in objects at ages
of the Hyades and earlier (0.6\,Gyr), [while] for clusters of about
2\,Gyr and older (up to 4.5\,Gyr) the same activity level was seen.
[\ldots]

Because the flare rate is expected to be low on older stars, a
systematic search for flares in an older stellar population needs a
large number of stars, and involves a relatively long stare coupled
with fast cadence to detect and resolve the flaring emission from any
other variability.  The {\em Keplers} spacecraft's exquisite photometry can
be re-purposed from finding evidence of transiting extrasolar planets
around stars to looking for rare short-timescale flaring events on the
stars themselves. [Energetic flares have been found in G-type
main-sequence stars, even on apparently single solar-type stars] with
rotation periods of 21.8 and 25.3 days, near the solar value, and thus
approximately solar age.  The energetics of these flares is large,
with minimum flare energies in the range 10$^{33}$ erg, and extending
up to 10$^{36}$ erg. [\ldots]'' \activity{{\em Consider:} The minimum flare energy
  given here is instrumental, not intrinsic. Argue why the empirical
  lower limit of flares detectable by an instrument like {\em Kepler} is
  limited to of order $10^{33}$\,ergs. Note that this lower limit
  exceeds the energies observed (to date, at least) for solar flares.}

\section{Evolution of astrospheres}\label{sec:evolastrospheres}

\subsection{Effects of a variable ISM on heliospheric structure}\label{sec:pastism}

\ors[IV:3.1] ``The solar \indexit{astrosphere}wind
\indexit{interstellar medium}does not expand
indefinitely.  Eventually it runs into the interstellar medium (ISM),
the extremely low particle density environment that exists in between
the stars [({\em cf.}  Fig.~\ref{Wood_f3})].  Our Sun is moving
relative to the ISM that surrounds the planetary system, so we see a
flow of interstellar matter in the heliocentric rest frame, coming
from the direction of the constellation Ophiuchus.  The interaction
between the solar wind and the ISM flow determines the large scale
structure of our heliosphere, which basically defines the extent of
the solar wind's reach into our Galactic environment.  Other stars are
naturally surrounded by their own 'astrospheres' (alternatively
'asterospheres') defined by the strength of their stellar winds, the
nature of the ISM in their Galactic neighborhoods, and their relative
motion.''

\begin{figure}[t]
%\plotfiddle{/home/wood/marmoset/props/htids13/HTIDS13_fig1.ps}{3.2in}{0}{50}{50}{-190}{-20}
%\includegraphics[scale=0.6]{figures/wood_linsky_f6.ps}
\centering
\includegraphics[width=12cm]{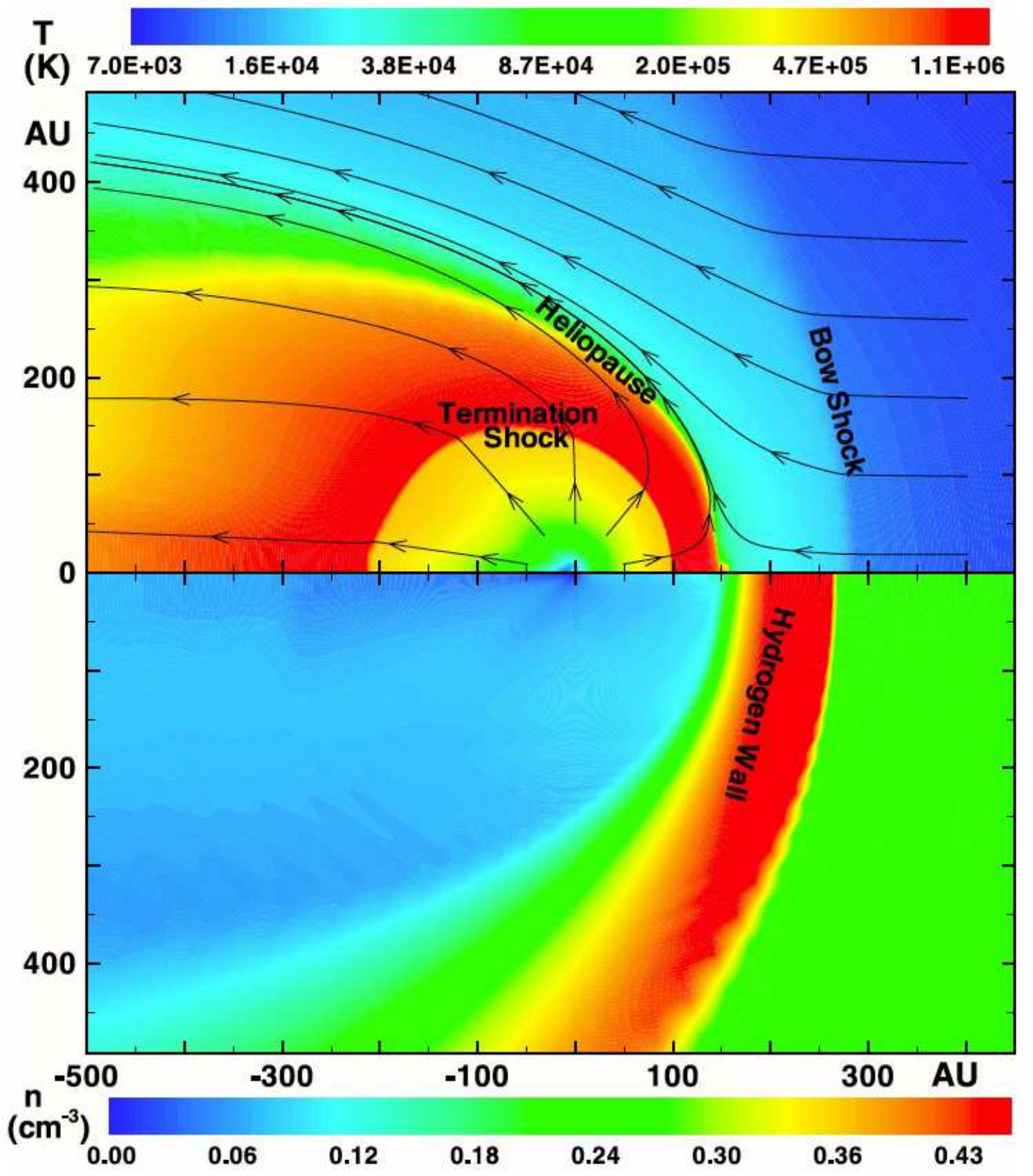}
\caption[A 2.5D axisymmetric, hydrodynamic model of the
heliosphere.]{A 2.5D axisymmetric, hydrodynamic model of the
  heliosphere and the surrounding ISM.  The {\rm upper panel} shows plasma
  temperature and ISM streamlines, and the {\rm bottom panel} shows neutral
  hydrogen density. [Note: the effective solar wind plasma temperature,
  increasing with heliocentric distance, is dominated by the energy
  density of pickup ions from several AU outward to the termination
  schock. Fig.~IV:3.6;
  \href{https://ui.adsabs.harvard.edu/abs/2004JGRA..109.7104M/abstract}{source:
  \citet{2004JGRA..109.7104M}}.] \colorfig }
\label{Wood_f3}
\end{figure}\figindex{../wood/art/wood_linsky_f6.ps}
\ors[IV:3.5] ``The Sun is now traveling through the ISM at a rate of
16--20~[parsecs (or pc, a unit of 3.26\,light years]) per million years
(Myr) compared to the average motion of nearby stars about the
Galactic Center. \activity{{\em Show:} Just to get an impression of relative
  velocities: what is the ratio of the characteristic speed of the Solar
  System relative to the local ISM to the speed of at
  which the Solar System orbits the Galactic center, 8\,kpc away, once
  every 230\,Myr? \mylabel{act:speeds}}  The ISM has
densities ranging from $10^4$~cm$^{-3}$ or higher in dense molecular
clouds down to about 0.005~cm$^{-3}$ in very low-density hot gas
regions.  Because the heliosphere will contract or expand by large
factors when the Sun enters such high- or low-density regions, it is
important to investigate when such environmentally driven changes
could have occurred and will possibly occur by considering the Sun's
historic and future path through the ISM.

     At present, the heliosphere resides inside of the \indexit{local
       interstellar cloud}partially
ionized [local interstellar cloud (LIC)], with properties likely
similar to other warm partially ionized clouds within 15\,pc of the
Sun.  The Sun likely entered this cluster of local warm clouds about
1\,Myr ago.  However, on a larger scale, the Sun actually lies in a
region called the Local Cavity, or Local Bubble, which is $\sim
200$\,pc across and is filled mostly with fully ionized, low-density
ISM. [No evidence has been found that the Sun has traveled through
significantly denser regions over the last 30\,Myr (500~pc) until about
7.5\,Myr ago (120\,pc) when the Sun was at the edge of the Local Cavity.
It appears that the Sun will] leave the LIC in less than 3000\,yr.
What will be the properties of this new environment? [\ldots] 

Our ideas concerning the properties of the gas located between the
warm local ISM clouds have undergone a radical change in the last 20
years.  The gas between the clouds, extending out to roughly 100\,pc
from the Sun in what is now called the Local Cavity was originally
assumed to be hot (roughly $10^6$\,K), fully ionized, and low density
(roughly 0.005\,cm$^{-3}$).  This conclusion was based upon the
predictions of the classical models and observations of diffuse soft
X-ray emission consistent with the properties of the hot gas.  This
picture has since been complicated by the realization that X-ray
emission from charge-exchange (CX) reactions between the solar wind
ions and inflowing interstellar neutral hydrogen can explain much of
the observed diffuse X-ray emission, except for the Galactic pole
regions. [It has instead been] proposed that the Local Cavity is an
old supernova \indexit{supernova}remnant with photo-ionized gas at a temperature of about
20,000~K. The likely photo-ionizing sources are the hot stars
$\epsilon$~CMa and $\beta$~CMa and nearby hot white dwarfs like
Sirius~B. \activity{{\em Show:} With average values for solar wind density and
  velocity (assuming a radial outflow at constant velocity and with a
  density as specified in Table~\ref{tab:wind-stats}), at what
  distance from the Sun does the total pressure of the solar wind
  equal the total pressure of the local interstellar medium (LISM) for
  estimated values of $B_{\rm LISM}\approx 3\,\mu$G,
  $T_{\rm LISM}\approx 8000$\,K,
  $n_{\rm p, LISM}\approx 0.06$\,cm$^{-3}$,
  $n_{\rm H, LISM}\approx 0.18$\,cm$^{-3}$, and a movement of 25\,km/s
  of the heliosphere relative to the LISM? These values of densities
  and temperatures assume the Sun to still be inside the Local
  Interstellar Cloud rather than in the surrounding Local Cavity or
  Local Bubble (see, {\em e.g.,} Sect.~IV:3.2).  Use
  $B_\phi \approx B_r$ near Earth. How do the contributions
  from the thermal pressure, magnetic pressure, and dynamic (or ram)
  pressure of the solar wind and the LISM compare at the heliopause
  (computed in the rest frame of the heliopause)? Look at
  Sect.~\ref{sec:impinging} for the physics involved.
  \mylabel{act:LISM}}\activity{{\em Show:} Repeat Activity~\ref{act:LISM} for the
  range of ISM densities of $0.005$ to $10^4$\,cm$^{-3}$ (as described
  in the beginning of Sect.~\ref{sec:pastism}), assuming a present-day
  spherically-symmetric, constant-velocity solar wind, and the same
  temperature and field strength for the ISM as in the LISM used in
  Activity~\ref{act:LISM}? Which pressure components dominate in the
  solar wind and ISM for the highest and lowest densities. Compare your result with
  Fig.~\ref{Linsky-f5}. \mylabel{act:varism}}

\begin{figure}[t]
%\plotfiddle{Muller2009fg1.ps}{2.7in}{-90}{40}{40}{-200}{230}
\centering
\includegraphics[scale=0.44]{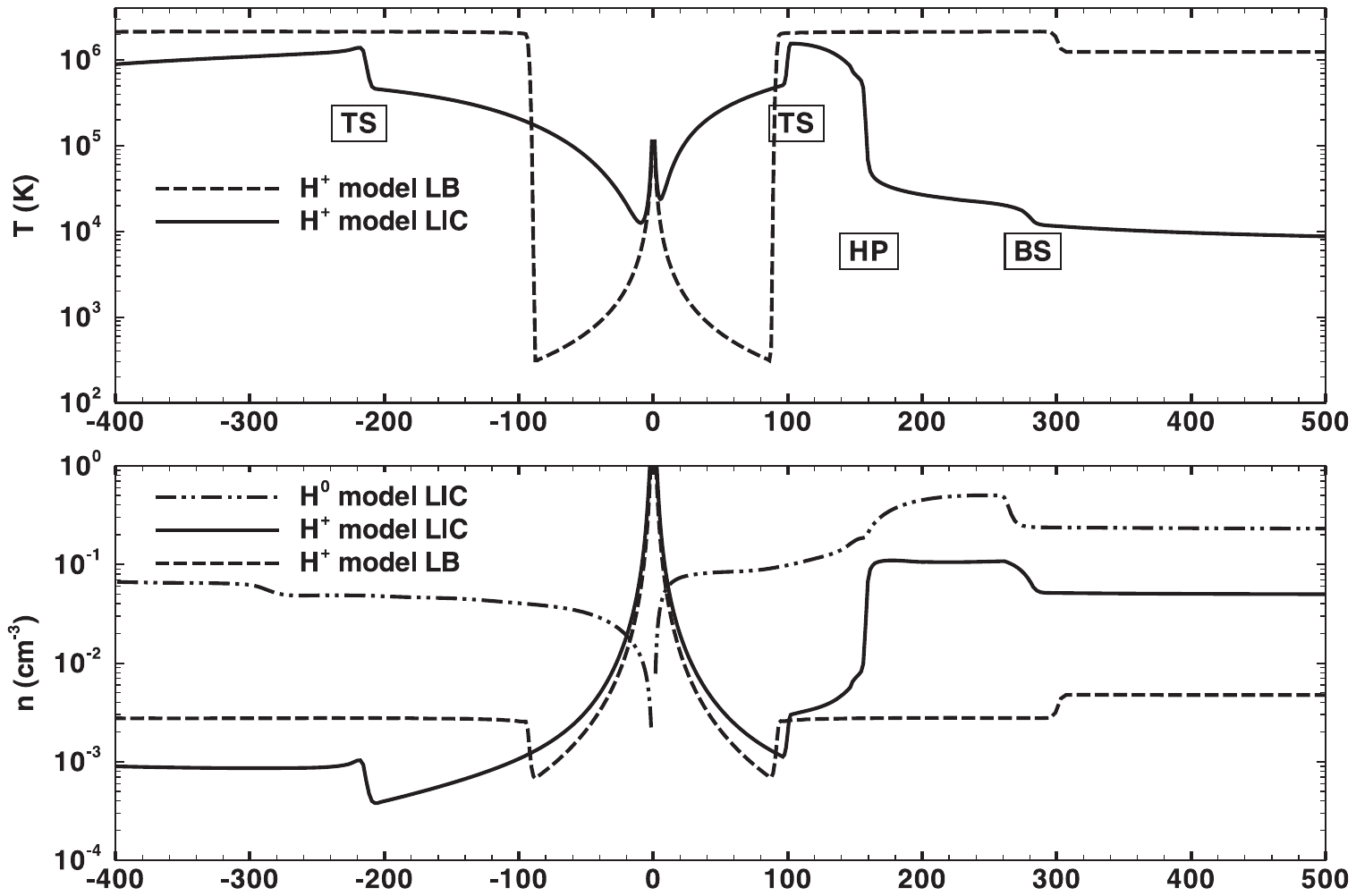}
\caption[Heliosphere-ISM models for 
different ISM parameters.]{{\em Top:}
  Plots of the temperature vs.\ distance in Sun-Earth distances
  (astronomical units, or AU) relative to the Sun (interstellar flow
  upwind direction to the right and downwind to the left) for a
  heliosphere model with the Sun located inside of the local
  interstellar cloud (LIC; solid line) or inside a $10^6$~K hot
  interstellar medium (dashed line, LB). The heliosphere in the LIC model
  has a \indexit{termination shock}termination shock (TS),
  \indexit{heliopause}heliopause (HP) and \indexit{bow shock}bow shock (BS)
  structure. {\em Bottom:} Density structures for the LIC neutral
  hydrogen (solid line), LIC protons (dot-dash line), and hot [Local
  Bubble (LB)] interstellar model protons (dashed line). Note that the
  hydrogen wall at 150--280~AU exists when the heliosphere is located
  inside partially neutral interstellar gas but not when it is inside
  fully ionized interstellar gas. [Fig.~IV:3.8;
  \href{https://ui.adsabs.harvard.edu/abs/2009SSRv..143..415M/abstract}{source:
  \citet{2009SSRv..143..415M}}.] }
\label{Linsky-f5}
\end{figure}\figindex{../wood/art/wood_linsky_f8.ps}

     How will the heliosphere change as the Sun passes through very
different regions of the interstellar medium?  [\ldots]
Figure~\ref{Linsky-f5} compares today's heliosphere properties with
the Sun located inside of the partially ionized warm LIC to a model
computed for the Sun surrounded by $10^6$~K fully ionized interstellar
plasma.  The main difference between these models is that the hydrogen
wall does not exist when the inflowing interstellar gas contains no
neutral hydrogen atoms.  The locations of the termination shock (TS),
heliopause (HS), and bow shock (BS) are determined by pressure balance
between the solar wind ram pressure and the thermal and ram pressure
of the surrounding interstellar gas.  In this comparison, the
locations of the TS, HP, and BS are about the same in the two models
because the high temperature and low density of the interstellar gas
produce a pressure that is about the same as in the LIC.

When the Sun enters a region of much higher density or speed, and
therefore higher ram pressure, the effect is to compress the
heliosphere.  For example, a model for $n_{\rm HI}=15$~cm$^{-3}$, roughly
100 times that of the LIC, has a TS at 9.8\,AU such that Uranus would
move in and out of the TS and Neptune would be surrounded by hot,
shocked plasma beyond the HP (upwind) or heliotail (downwind).  Models
of the heliosphere inside of a high-speed interstellar wind with
corresponding high ram pressure would compress the heliosphere in a
similar way.  [A potential] cloud encounter that results in a stellar
astrosphere being compressed to less than the size of the star's
habitable zone [\ldots\ has been described as \indexit{de-screening event}a
'de-screening event'. This] should happen when a star encounters an
interstellar cloud with a number density of $600(M_{\odot}/M)^{2}$
cm$^{-3}$, where M is the mass of the star.  Only the densest ISM
clouds are capable of this de-screening, with such clouds being
relatively rare.  The densest clouds are cold ($T\sim 100$\,K)
molecular clouds, with many of the refractory elements depleted onto
dust grains.  In addition to increased GCR exposure (see
Ch.~\ref{ch:evolvingexposure}), a de-screening event caused by a
molecular cloud encounter would also expose planetary atmospheres to
high fluxes of interstellar dust, with potentially dramatic
consequences [that include potential 'snowball Earth' states for
climate. 
Given what we know about mass-loss and wind properties
of stars (as discussed below), it appears] that habitable zone planets
orbiting stars significantly less massive than the Sun (with spectral
types of late K to M) are virtually never exposed to de-screening
events, but de-screening may happen occasionally for stars with the
Sun's mass or larger.  However, these calculations assumed that the
relative velocity of these encounters is only 10 km~s$^{-1}$.
Assuming a faster encounter speed would increase the estimated
frequency of de-screening events.''

\subsection{Long-term evolution of stellar winds}

\label{sec:stellarwinds}

In Sect.~\ref{sec:impinging} \indexit{stellar!wind evolution}we
described how neutrals moving toward to the heliosphere leads to a
\indexit{hydrogen wall}'hydrogen wall' outside of the heliopause
through
\indexit{charge exchange}charge-exchange
collisions in that region (see Figs.~\ref{fig:ophercomposite}
and~\ref{Wood_f3}). \activity{{\em Show:} For charge exchange only,
  and assuming (very approximately, as done
  \href{https://ui.adsabs.harvard.edu/abs/1972JGR....77.5407H/abstract}{initially
    (\citep{1972JGR....77.5407H})} decades ago) a velocity-independent
  cross section for resonant-charge exchange of solar wind protons
  with ISM neutral hydrogen of
  $\sigma_{\rm CX} \approx 2\,10^{-15}$\,cm$^2$, what fraction of
  H$^0$, looking at the population after passing through the 'hydrogen
  wall' and moving in a straight line towards the Sun, would reach
  Earth orbit for present-day slow wind conditions? In reality, other
  processes are major players: radiation pressure (for neutral
  hydrogen primarily by repeated Lyman~$\alpha$ absorption followed by
  isotropic re-emission) pushes outward on the atoms, and
  photo-ionization in the Sun's EUV and X-rays presents a significant
  loss term. It appears that Lyman $\alpha$ radiation pressure on ISM
  H$^0$ just balances solar gravity,
  \href{https://ui.adsabs.harvard.edu/abs/2013ApJ...775...86S/abstract}{see
    \citet{2013ApJ...775...86S}} \mylabel{act:cxsigma}} \ors[IV:3.4]
``The importance of this hydrogen wall is that it is actually
detectable in UV spectra from [the {\em Hubble Space Telescope} ({\em
  HST})], not only around the Sun but around other stars as well.

The effect of heliospheric and
\indexit{astrosphere!absorption}astrospheric absorption on stellar
H\,I Lyman-$\alpha$ spectra [(emitted in the stellar atmosphere by
de-excitation from the first excited to the ground state in neutral
hydrogen atoms and absorbed en route to Earth by the inverse process)]
is described by Fig.~\ref{Linsky-f4}, showing the journey of
Lyman-$\alpha$ photons from the star to the observer.  Most of the
absorption is by interstellar gas in the line of sight from the star
to the Sun, but the astrosphere and heliosphere provide additional
absorption on the [red] and [blue] sides of the interstellar
absorption, respectively.  The effect of the hydrogen wall around the
Sun is to provide additional red-shifted absorption on the right side
of [relative to] the interstellar absorption feature because the
neutral hydrogen gas in the solar hydrogen wall is slowed down and
deflected relative to the inflowing interstellar gas.  Conversely, the
absorption by the hydrogen wall gas around the star is seen as
blue-shifted relative to the interstellar flow from our perspective
outside the astrosphere, and is therefore seen on the left side of the
absorption line.  [\ldots\ By way of an example observation, the]
bottom panel of Fig.~\ref{Linsky-f4} shows the H\,I (and [equivalent
deuterium] D\,I) Lyman-$\alpha$ spectrum of the lower-brightness
component of [the star] $\alpha$~Cen~B.  Most of the intervening H\,I
and D\,I between us and the star is interstellar, but the ISM cannot
account for all of the H\,I absorption.  As mentioned above, the
red-shifted excess on the right side is heliospheric and the
blue-shifted excess on the left is astrospheric.'' \activity{{\em
    Consider:} The Lyman\,$\alpha$ line (1215.67\,\AA), as the
  brightest UV emission line of late-type stars, is an important
  diagnostic in the study the energy budget of stellar
  chromospheres. But, as discussed in the text, much of the strongest
  emission is taken away by absorption of the line core by
  circumstellar and interstellar hydrogen. What is the common property
  of late-type stars that enable us to observe the Lyman\,$\alpha$
  line mostly as emitted by these stars? Quantify a threshold value
  for that property to be useful. Once you formulate the answer, you
  can look at \citet{2022arXiv220101315Y} for a study of suitable
  target stars. \mylabel{act:intrinsiclya}}
%\begin{figure}[t]
%\plotfiddle{/home/wood/marmoset/conf/bern13/Wood_f2.ps}{2.0in}{90}{35}{35}{132}{-10}
%\caption{HST Lyman-$\alpha$ spectrum of $\alpha$~Cen~B, showing broad
%  H~I absorption at 1215.6~\AA\ and D~I absorption at 1215.25~\AA.  The
%  upper solid line is the assumed stellar emission profile and the dashed
%  line is the ISM absorption alone.  The excess absorption is due to
%  heliospheric H~I (vertical lines) and astrospheric H~I (horizontal
%  lines).  From Linsky \& Wood (1996).}
%\label{Wood_f4}
%\end{figure}
\begin{figure}[t]
%\plotfiddle{wood_linsky_f7.ps}{4.0in}{0}{50}{50}{-155}{-80}
%\includegraphics[scale=0.60,trim=140 80 0 40]{figures/wood_linsky_f7.ps}
\centering
\includegraphics[width=10cm]{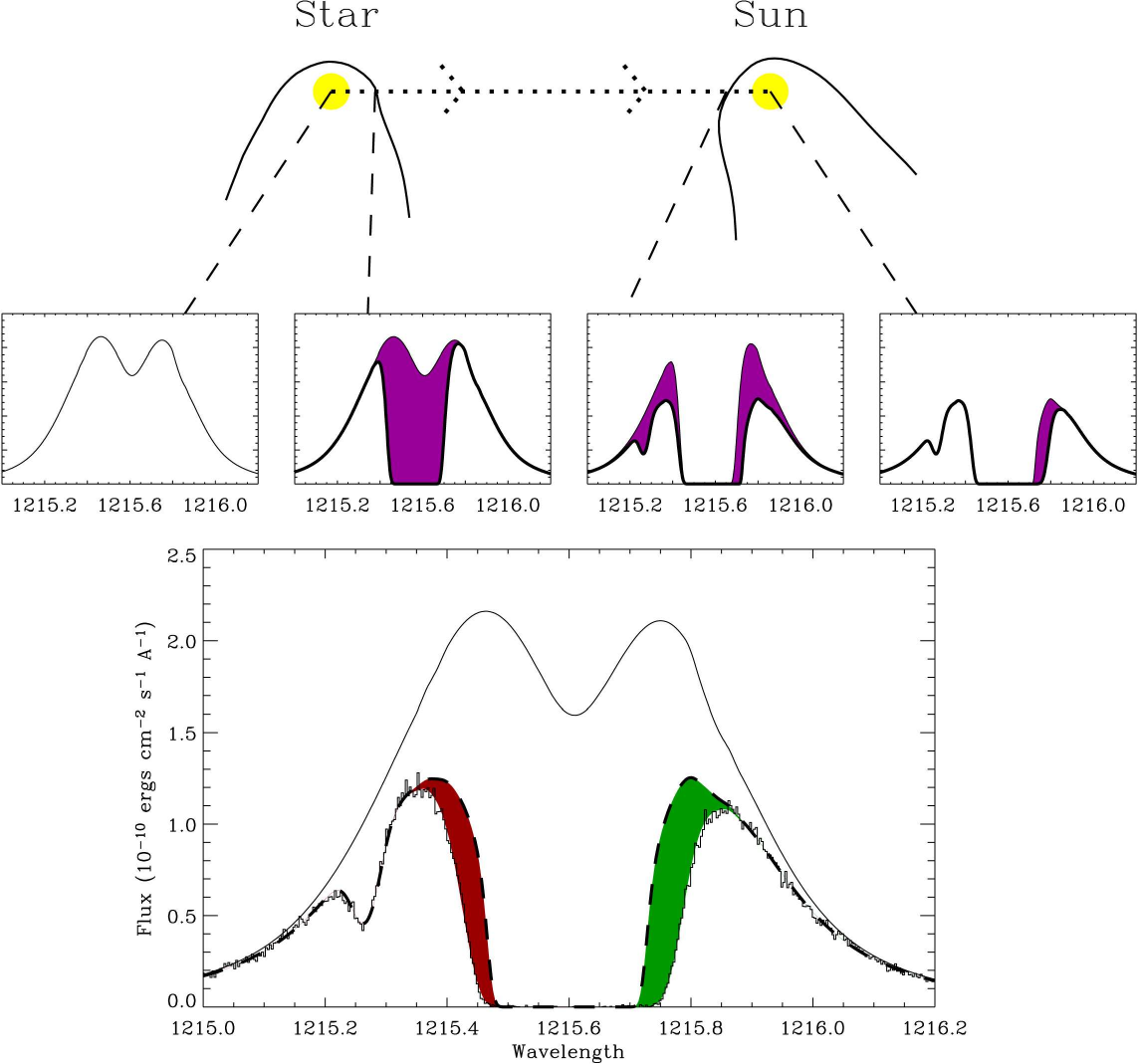}
\caption[Lyman-$\alpha$: the journey of photons; the emission line;
and $\alpha$~Cen~B]{{\em Top panel:} The journey of a Lyman-$\alpha$
  photon from a star through its astrosphere, the interstellar medium,
  and the heliosphere. {\em Middle panels from left to right:} The
  Lyman-$\alpha$ emission line emitted by the star, absorption due to
  the stellar astrosphere, additional absorption due to the
  interstellar medium, and additional absorption due to the
  heliosphere. {\em Bottom panel:} HST Lyman-$\alpha$ spectrum of
  $\alpha$~Cen~B, showing broad H\,I absorption at 1215.6~\AA\ and D\,I
  absorption at 1215.25~\AA.  The upper solid line is the assumed
  stellar emission profile and the dashed line is the ISM absorption
  alone.  The excess absorption is due to heliospheric H\,I (green
  shading, vertical lines) and astrospheric H\,I (red shading,
  horizontal lines).  [Fig.~IV:3.7;
  \href{https://ui.adsabs.harvard.edu/abs/2004LRSP....1....2W/abstract}{source:
  \citet{2004LRSP....1....2W}}.] \colorfig }
\label{Linsky-f4}
\end{figure}
\figindex{../wood/art/wood_linsky_f7.ps}

\ors[IV:3.7.2] ``Currently, the only way coronal winds can be
detected around other stars is through
\indexit{astrosphere!Lyman $\alpha$ signal}astrospheric Lyman-$\alpha$
absorption, but the number of astrospheric Lyman-$\alpha$ detections
is still very limited. [\ldots\ These measurements need to be
interpreted in terms of models that] are extrapolated from a
heliospheric model that successfully reproduces heliospheric
absorption, specifically a multi-fluid model.  These models assume the
same ISM characteristics as the heliospheric model, with the exception
of the ISM flow speed in the stellar rest frame, $v_{\rm ISM}$, which
can be computed using our knowledge of the local ISM flow vector and
each star's unique space motion vector.  [\ldots]

     The astrospheric models are computed assuming different stellar
wind densities, corresponding to different mass-loss rates, and the
Lyman-$\alpha$ absorption predicted by these models is compared with
the data to see which best matches the observed astrospheric
absorption.  [\ldots] In order to look for some correlation between
coronal activity and wind strength, Fig.~\ref{Wood_f8} shows mass-loss
rates (per unit surface area) plotted versus $F_X$ [(the ratio of
X-ray luminosity to surface area)], focusing only on the main-sequence
stars.  For the low-activity stars, mass loss increases with activity
in a manner consistent with the $\dot{M}\propto F_{X}^{1.34\pm 0.18}$
power-law relation shown in the figure.  For the $\xi$~Boo binary, in
which (like $\alpha$~Cen) the two members of the binary share the same
astrosphere, Fig.~\ref{Wood_f8} indicates how the binary's combined
wind strength of $\dot{M}=5 \dot{M}_{\odot}$ is most consistent with
the other measurements if 90\% of the wind is ascribed to $\xi$~Boo~B,
and only 10\% to $\xi$~Boo~A.
\begin{figure}[t]
%\plotfiddle{/home/wood/marmoset/conf/bern13/Wood_f5.ps}{2.6in}{0}{60}{60}{-190}{-230}
%\plotone{/home/wood/marmoset/hstarc/h72905/paper/fig4_col.ps}
\centering
\includegraphics[scale=0.60,bb=54 360 558 720]{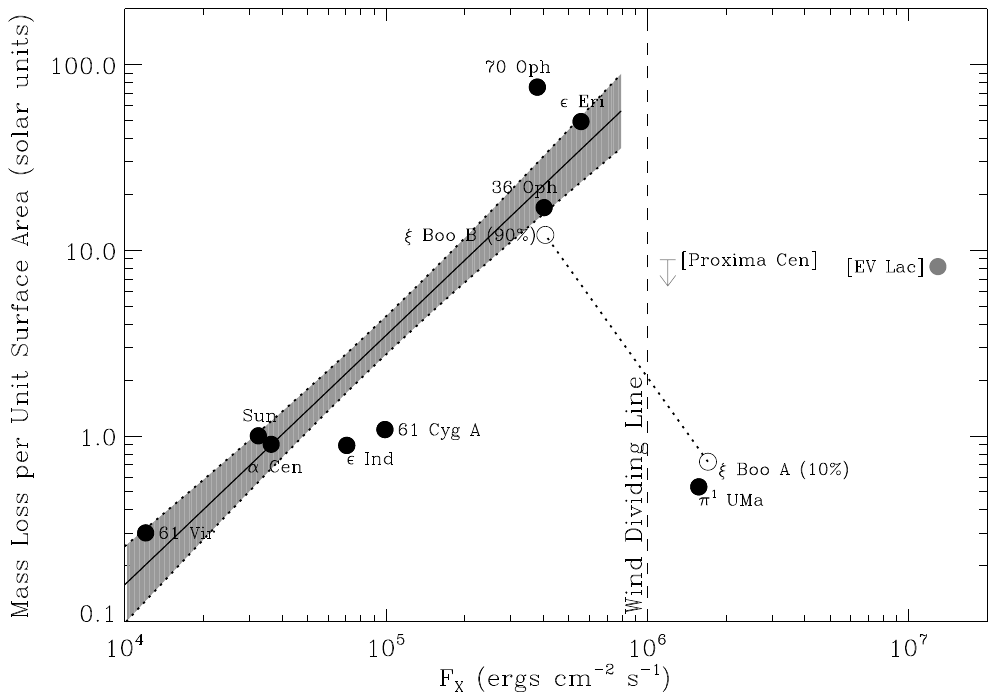}
\caption[Mass-loss rate vs.\ X-ray flux density for MS stars.]{A plot of mass-loss rate (per unit surface area)
  versus X-ray \indexit{wind!dividing line}surface flux density for all main-sequence stars with
  measured winds.  Most of these have spectral types of G (like the
  Sun) or (cooler) K, but the two with square-bracketed labels are
  (much cooler) tiny M dwarf stars.  Separate points are plotted for
  the two members of the $\xi$~Boo binary, assuming $\xi$~Boo~B
  accounts for 90\% of the binary's wind, and $\xi$~Boo~A only
  accounts for 10\%.  A power law, $\dot{M}\propto F_X^{1.34\pm
    0.18}$, is fitted to the less active stars where a wind/corona
  relation seems to exist, but this relation seems to fail for stars
  to the right of the 'Wind Dividing Line' in the figure.
  [Fig.~IV:3.12;
  \href{https://ui.adsabs.harvard.edu/abs/2014ApJ...781L..33W/abstract}{source:
  \citet{2014ApJ...781L..33W}}.]}
\label{Wood_f8}
\end{figure}\figindex{../wood/art/wood_linsky_f12.ps}

For $F_X<10^6$ erg~cm$^{-2}$~s$^{-1}$, mass loss appears to increase
with activity.  \activity{{\em Show} that the power lost in X-rays
  from the present-day solar corona (estimated from
  Fig.~\ref{figure:rotact} or \ref{Wood_f8}) is roughly twice the
  power lost in the solar wind (using the expressions in
  Sec.~\ref{sec:solwindenergy}), and that these numbers would have
  been comparable for the young Sun at \indexit{wind!dividing line}the
  'wind dividing line' if the characteristic wind speed would have
  been the same.}  However, above $F_X=10^6$ erg~cm$^{-2}$~s$^{-1}$
({\em i.e.,} for more active, and thus generally younger stars) this
relation seems to fail, a boundary identified as the 'Wind Dividing
Line' in Fig.~\ref{Wood_f8}.  Highly active stars above this limit
appear to have surprisingly weak winds.  This is suggested not only by
the two solar-like G stars above the limit, $\xi$~Boo~A and
$\pi^1$~UMa, but also by the two active M dwarfs above the limit,
which have very modest mass-loss rates.  (For Proxima~Cen we only have
an upper limit of $\dot{M}<0.2 \dot{M}_{\odot}$, while for EV~Lac
$\dot{M}=1 \dot{M}_{\odot}$.)  The apparent failure of the wind/corona
correlation to the right of the 'Wind Dividing Line' may indicate a
fundamental change in magnetic field topology at that stellar activity
level.

[\ldots] Sophisticated spectroscopic and polarimetric techniques are
also available for studying stellar [surface] magnetic fields.  One
interesting discovery is that very active stars usually have stable,
long-lived polar starspots, in contrast to the solar example where
sunspots are only observed at low latitudes.  Perhaps the polar spots
are indicative of a particularly strong dipolar magnetic field that
envelopes the entire star and inhibits stellar wind flow, thereby
explaining why very active stars have surprisingly weak winds.  Strong
toroidal fields are also often observed for active
stars. \activity{{\em Consider:} Place the coronal activity level
  corresponding to the 'wind dividing line' in the rotation-age
  diagram in Fig.~\ref{figure:rotact}. How old would the Sun have been
  when it reached the 'wind dividing line'?  What does this imply
  about the activity for the Sun? What about the solar wind?  Other
  consequences? \mylabel{act:wdl}}

Given that young stars are more active than old stars, the correlation
between mass loss and activity indicated in Fig.~\ref{Wood_f8} implies
an anti-correlation of mass loss with age.  [One parameterization of
this is given by]
$F_{X}\propto t^{-1.7\pm 0.3} [\propto \Omega^{-3.4 \pm 0.6}]$.
Combining this with the power-law relation from Fig.~\ref{Wood_f8}
yields the following relation \indexit{mass loss}between
\indexit{star!mass loss}mass-loss rate and age:
\begin{equation}\label{eq:masslossage}
\dot{M}\propto t^{-2.3\pm 0.6} [\propto \Omega^{-4.6 \pm 1.2}]
\end{equation}
[where the final expression above between brackets links to the
intrinsic 
dependence on
rotation rate via Eq.~(\ref{eq:skumanich}).]
Fig.~\ref{Wood_f9} shows what this relation suggests for the history
of the solar wind, and for the history of winds from any solar-like
star for that matter.  The truncation of the power-law relation in
Fig.~\ref{Wood_f9} near $F_X=10^6$ erg~cm$^{-2}$~s$^{-1}$ leads to
the mass-loss/age relation in Fig.~\ref{Wood_f9} being truncated as
well at about $t=0.7$~Gyr.  The plotted location of $\pi^1$\,UMa in
Fig.~\ref{Wood_f9} indicates what the solar wind may have been like
at times earlier than $t=0.7$~Gyr.
\begin{figure}[t]
%\plotfiddle{/home/wood/marmoset/conf/bern13/Wood_f6.ps}{2.6in}{0}{60}{60}{-190}{-230}
\centering
\includegraphics[scale=0.60,bb=54 360 558 720]{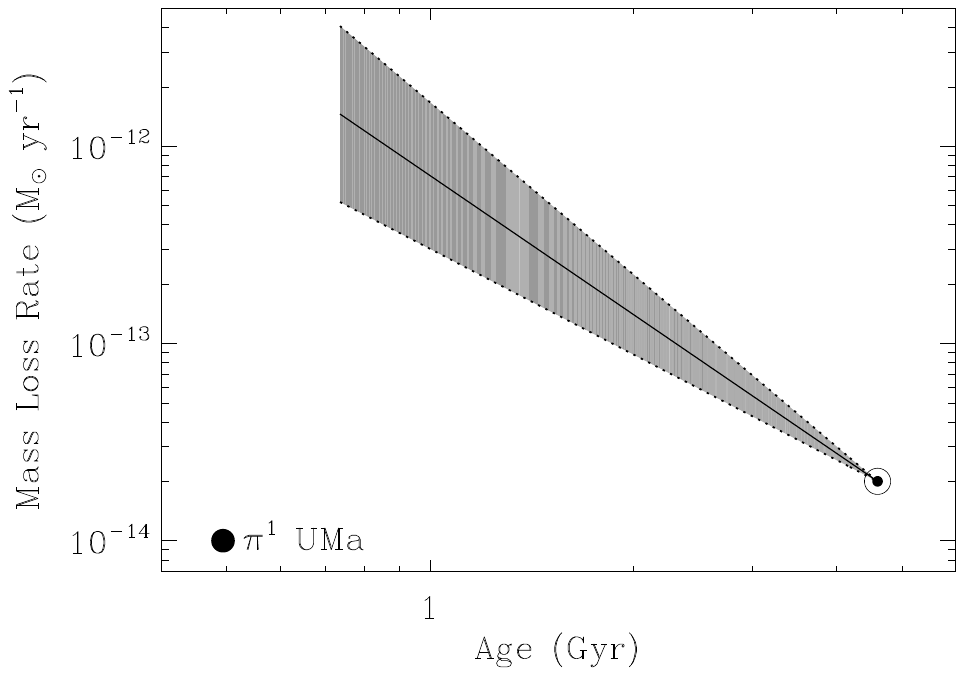}
\caption[The inferred mass-loss history of the Sun.]{The mass-loss history of the Sun inferred from the power-law
  relation in Fig.~\ref{Wood_f8}.  The truncation of the relation in
  Fig.~\ref{Wood_f8} means that the mass-loss/age relation is
  truncated as well.  The low mass-loss measurement for $\pi^1$~UMa
  suggests that the wind weakens at $t\approx 0.7$ Gyr as one goes
  back in time.  [Fig.~IV:3.13;
  \href{https://ui.adsabs.harvard.edu/abs/2005ApJ...628L.143W/abstract}{source:
  \citet{2005ApJ...628L.143W}}.]}
\label{Wood_f9}
\end{figure}\figindex{../wood/art/wood_linsky_f13.ps}

[Fig.~\ref{Wood_f9} indicates] that solar-like coronal
winds can be up to two orders of magnitude stronger than the current
solar wind at $t\approx 1$~Gyr.  This makes it more likely that the
erosive effects of stellar winds play an important role in planetary
atmosphere evolution at these later ages'' (see
Ch.~\ref{ch:evolvingplanetary}). \activity{{\em Show:} Estimate the size of the
  heliosphere and the terrestrial magnetopause distance for a young
  Sun at an age of 700\,Myr, assuming unchanged LISM conditions and
  geomagnetic properties.}

\begin{figure}[t]
%\centerline{\psfig{figure=figures/apj447726f1_hr.eps,width=\textwidth}} 
%\centerline{\psfig{figure=figures/Cohenfiles/f1a.eps,width=\textwidth}} 
\centerline{\includegraphics[width=\textwidth,bb=0 0 2376 703]{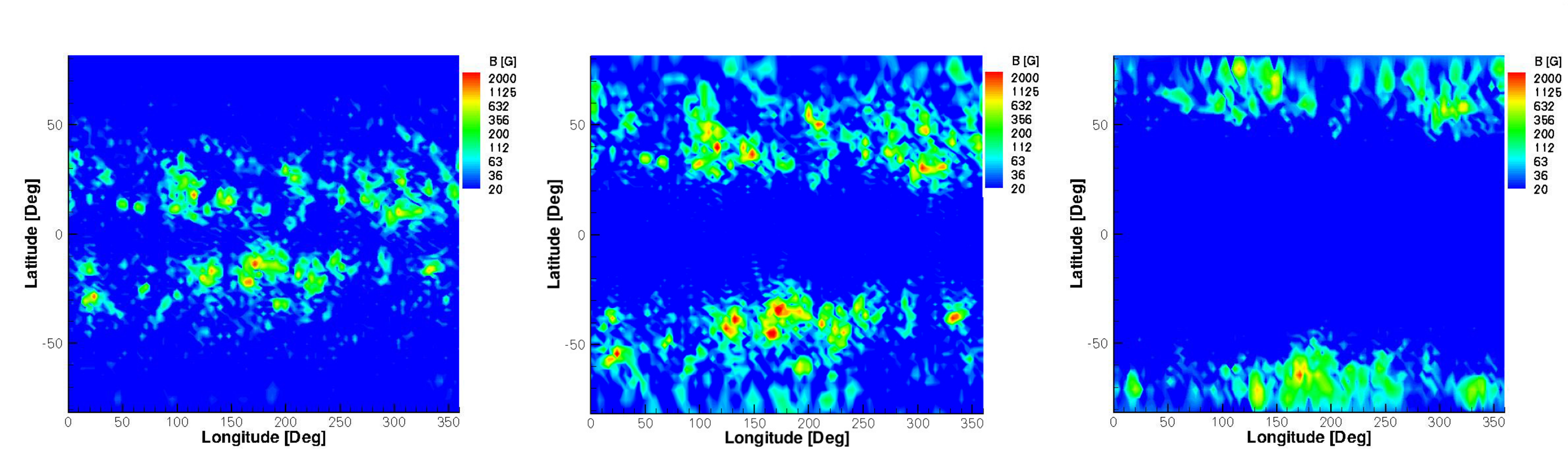}} 

%\centerline{\psfig{figure=figures/Cohenfiles/f1b.eps,width=\textwidth}} 
\centerline{\includegraphics[width=\textwidth,bb= 0 0 1782 527]{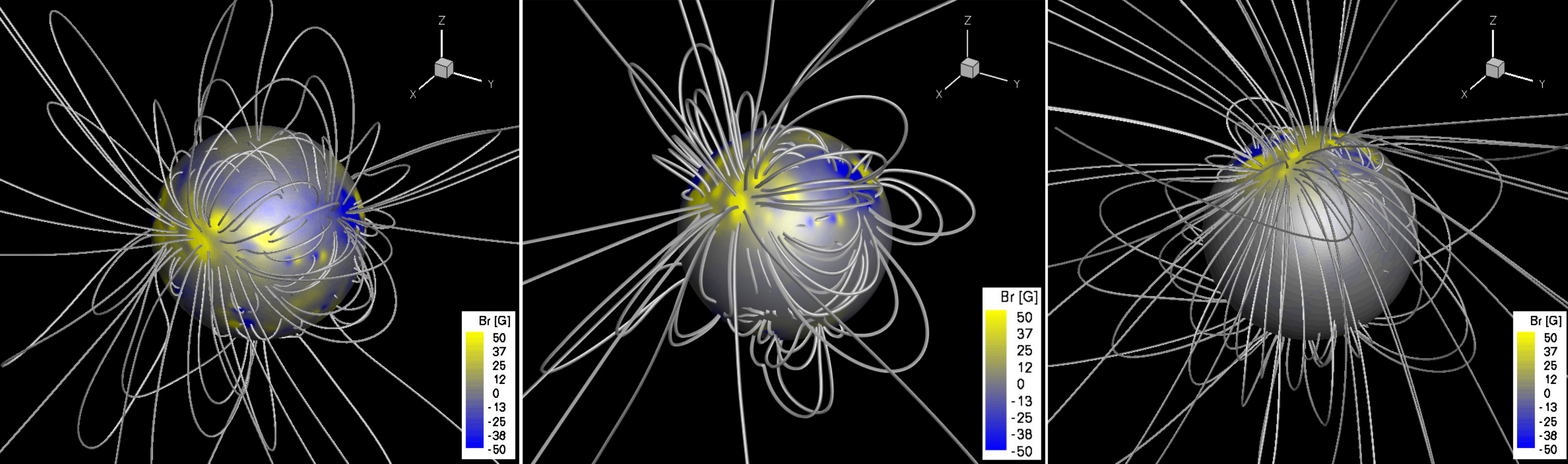}} 
\caption[Surface field maps and associated heliospheric fields.]{{\em Top row:} A map of the
  solar photospheric radial magnetic field (magnetogram) during
  Carrington Rotation 1958 (January 2000, solar maximum period) shown
  on the left. The {\rm middle and right panels} show manipulation of the
  original map, where the active regions have been shifted by 30 and
  60 degrees toward the poles, respectively. [Fig.~IV:4.4;
  \href{https://ui.adsabs.harvard.edu/abs/2012ApJ...760...85C/abstract}{source:
  \citet{2012ApJ...760...85C}}]
%\figindex{../cohen/art/ManipulatedMagnetogram_BW}
\label{fig:ManipulatedMagnetogram}
{\em Bottom row:} The three-dimensional magnetic field corresponding to
  the surface distribution of the photospheric radial magnetic field
  (shown on a sphere of $r=R_\odot$) during solar maximum ({\rm left}), and
  for manipulated photospheric filed with the active regions shifted
  by 30 ({\rm middle}) and 60 ({\rm right}) degrees toward the poles, as shown in
  the row above. [Fig.~IV:4.5] \colorfig }
\label{fig:YoungSunField}
%\figindex{../cohen/art/SpotLatitude_BWinvert.eps}
\end{figure}
%\begin{figure}[t]
%\centerline{\psfig{figure=figures/ManipulatedMagnetogram_BW,width=\textwidth}} 
%\caption[Solar surface field map, and modified versions.]{A map of the
%  solar photospheric radial magnetic field (magnetogram) during
%  Carrington Rotation 1958 (January 2000, solar maximum period) shown
%  on the left. The middle and right panels show manipulation of the
%  original map, where the active regions have been shifted by 30 and
%  60 degrees toward the poles, respectively. [Fig.~IV:4.4]}
%\figindex{../cohen/art/ManipulatedMagnetogram_BW}\label{fig:ManipulatedMagnetogram}
%%\end{figure}
%%\begin{figure}[t]
%\centerline{\psfig{figure=figures/SpotLatitude_BWinvert.eps,width=\textwidth}}
%\caption[The 3D heliospheric field corresponding for different surface
%distributions.]{The three-dimensional magnetic field corresponding to
%  the surface distribution of the photospheric radial magnetic field
%  (shown on a sphere of $r=R_\odot$) during solar maximum (left), and
%  for manipulated photospheric filed with the active regions shifted
%  by 30 (middle) and 60 (right) degrees toward the poles, as shown in
%  the figure above. [Fig.~IV:4.5]}
%\label{fig:YoungSunField}
%\figindex{../cohen/art/SpotLatitude_BWinvert.eps}\end{figure}
%\nocite{Cohen12b}
\subsection{Astrospheric field patterns in time}
\label{section:4_1}
%\label{section:4_11} %Unused in Principles...

\ors[IV:4.1.1] ``The \indexit{astrosphere!field pattern}extent and
structure of astrospheres is determined by the radially-expanding
super-Alfv\'enic stellar wind that drags the stellar magnetic field
from the stellar corona through interplanetary space, until the wind
is stopped by the Interstellar Medium (ISM). It is also determined by
the rotation of the star. As a result, each astrospheric magnetic
field (AMF) line has one end (or 'footpoint') attached to the stellar
surface, while its location at each point in the astrosphere,
$\mathbf{r}(r,\theta,\phi)$ (for co-latitude $\theta$), is given by
the following formula. It describes a spiral shape and is
\indexit{Parker spiral}known as the
'Parker Spiral' [(Sect.~\ref{sec:parker-spiral}, compare with
Eq.~\ref{eq:spiralforlarger})]:
\begin{equation}
\mathbf{B}(r)=B_0\left( \frac{r_0}{r} \right)^2 \left[
\mathbf{\uv{r}}+ \frac{(r-r_0)\Omega \sin{\theta}}{v_{\rm w}}
\mathbf{\uv{\phi}} \right].
\label{eq:4_01}
\end{equation}
Here $\Omega$ is the stellar rotation rate (angular velocity), $v_{\rm w}$ is
stellar wind speed (which is here assumed to be radial and fixed in
time and space); $r_0$ is the actual base point of the AMF, and is at a
reference distance from the stellar surface at which we assume the
stellar wind is fully developed and has achieved its asymptotic speed
and radial direction; $B_0$ is the magnetic field magnitude at that
point. We can see that the radial component of the AMF has an $r^{-2}$
dependence, while the azimuthal component has only a $r^{-1}$
dependence. As a result, throughout most of the astrospheres, the AMF is
dominated by the azimuthal field, which is a function of $\Omega$,
except for high latitudes (small $\theta$) where the AMF lines are
nearly radial.

Over time, stellar rotation periods evolve from less than one day for
very active, young stars to about 20-100 days for older, main-sequence
stars like the Sun. For very fast-rotating stars, the AMF spiral is
completely dominated by the azimuthal component: the field is highly
compressed, and its azimuthal component dominates even at relatively
small distances from the star and inside the stellar corona, which
typically extends to 10-20 stellar radii. In this case, even extended
closed magnetic loops can be bent as a result of the fast
rotation. This effect can have implications for the triggering of very
strong stellar flares, and for the mass-loss rate of the star to the
stellar wind. The right panel in Figure~\ref{fig:AMFSpiral} shows how
the compression of the AMF spiral changes for different stellar
rotation periods. The other two panels show the AMF lines close to the
star (up to 24 stellar radii). It can be clearly seen that the field
lines are nearly radial for the slow, solar-like rotation period of 25
days, while the field lines are strongly bent in the azimuthal
direction for a fast rotation period of half a
day. \activity{{\em Consider:} Section~\ref{section:4_1} mentions a solar rotation
  period of 25\,d while the caption to
  Fig.~\ref{fig:currentsheetsketch} mentions 27\,d. What is the reason
  for using these two different values in the different contexts?}

Equation~(\ref{eq:4_01}) describes how a given magnetic field line
changes with distance for a given value of $B_0$ at its base ($r_0$),
and a given asymptotic stellar wind speed $v_{\rm w}$. However, the AMF is
formed by a collection of field lines that are defined by some
spherical distribution of $B_0$ at the base of the stellar
corona. This distribution depends on the topology of the stellar
magnetic field at a given time. In addition, the value of $v$ also
varies as it empirically depends on the expansion of the magnetic flux
tubes and on the non-uniform distribution of $B_0$. [\ldots]

Over time, stellar activity appears at different latitudes, while
changing in magnitude as the behavior of surface magnetic activity is
highly tied to the rotation rate. Young active stars seem to have very
strong large-scale magnetic fields with magnitude of several
kilo-Gauss. For reference, the Sun's dipole field strength is of the
order of 5-10\,G, and while the magnetic flux density within active
regions can be high (ranging up to well over a kiloGauss in sunspots),
solar active regions are rather small in size. In addition, magnetic
activity in active stars tends to appear at high-latitude, polar
regions. This behavior is most likely related to
the role of the fast stellar rotation in the stellar dynamo and
meridional magnetic flux circulation. [\ldots] 

The appearance of stellar activity described above reflects a change
in the distribution of $B_0$. Therefore, it affects the shape of the
AMF and the astrospheric volume. It is not clear how $v_{\rm w}$
changes for young stars as we cannot directly measure stellar winds of
'cool stars', {\em i.e.,} stars with a convective envelope beneath
their surfaces such as in the case of the Sun. Some techniques to
estimate mass-loss rates from cool stars are [outlined
above]. However, these estimates do not separate the stellar wind
speed from the density, so it cannot be obtained
independently. Another cause for the lack of estimates for stellar
wind speeds of cool stars is the incomplete theory about the solar
wind acceleration. In order to demonstrate how the change in the
photospheric field affects the three-dimensional structure,
Figure~\ref{fig:YoungSunField} shows the distribution of the
photospheric magnetic field and the shape of the three-dimensional
magnetic field close to the Sun. The top-left panel is obtained using
actual data of the photospheric field during a period of high solar
activity. In the other two panels in the top row, the original data
was manipulated, so that the active regions have been shifted by 30
and 60 degrees, respectively, towards higher latitudes to
mimic the activity distribution of young active stars. It can be seen
in the bottom panels that the large-scale field topology changes
dramatically even if only the positions of the active regions are
changed.''

We can presently observe CMEs only in the heliosphere [(but see
\citep{2021NatAs...5..697V} on the
use of coronal dimmings as proxies for stellar CMEs)].  For some
discussion on CMEs in different astrospheres and their potential (but
currently speculative) role in stellar angular momentum loss and
stellar spin down, see Sect.~IV:4.2.

\clearpage

\chapter{{\bf Formation of stars and planets}}%11
%[N.B. This chapter follows stellar evolution because main-sequence needs to be known]
\label{ch:formation}
{\narrower\narrower{
{\bf Chapter topics:}
\begin{itemize}
  \customitemize
\item Formation of stars and planetary systems
\item Migration of young planets in accretion disks
\item Angular-momentum transport in accretion disks
\item Disk evaporation
\end{itemize}

\noindent{\bf Key concepts:}
\begin{itemize}
  \customitemize
\item Core-accretion model
\item Protoplanetary disk
\item Ice line
\item Kelvin-Helmholtz time scale
\end{itemize}

}}

\section{Introduction}
% http://www-ssg.sr.unh.edu/ism/LISM.html
A star\indexit{star!formation} like the Sun begins its life within a relatively dense concentration of molecular gas, called a cloud 'core', somewhere in the interstellar medium. The density in such molecular clouds is of order $10^2-10^6$\,cm$^{-3}$, to be compared with, {\em e.g.,} the density of the local interstellar medium of roughly $10^{-1}$\,cm$^{-3}$.  \ors[III:3.1] ``The mechanisms by which molecular clouds of many solar masses break up into stellar mass\indexit{definition!cloud core} pieces\indexit{cloud core|seealso{definition}} are a \indexit{cloud core}matter of debate; probably turbulence generated in the process of forming the cloud produces the denser fragments which accrete to form stars [\ldots\ G]iven the large sizes of protostellar clouds, they almost certainly contain enough angular momentum to form disks of substantial size and mass; thus, a major part of the story of star formation involves moving matter from a disk into a small, spherical protostar.\sactivity{$\circledS$ {\em Show:} (a) How many Earth masses of elements carbon or heavier are contained in a solar mass cloud of solar composition? Most of that material in the original cloud ended up inside the Sun, of course. (b) What fraction, roughly, of the original cloud would need to remain in the disk to ultimately form the planets? (c) Why are the answers to these two questions largely independent of each other (think about what mostly makes Jupiter and Saturn). \mylabel{act:emasses} \solution{emasses}}

To make a\indexit{gravitational!collapse!condition for} star of a given mass $M_\odot$ from a gas with temperature $T_{\rm c}$, gravity must overcome the pressure support; this means that [the protostellar\indexit{protostellar!cloud} cloud must have a] radius $R_{\rm c} \simgreat 2 \times 10^4$~astronomical units (AU; [see the argumentation around Eq.~\ref{eqlh:rofm}]).\activity{{\em Show:} Compare the critical protostellar cloud radius $R_{\rm c}(M_\odot) \simgreat 2 \times 10^4$~astronomical units to distances between stars in star-forming regions such as the Orion Nebula. Express that distance in light years and in parsecs, and compare those to the distances to the nearest stars for the present-day Sun. \mylabel{act:criticalcloud}}  We see pre-stellar dense concentrations of this size with properties such that they are likely to be on the verge of gravitational collapse.  As these cloud cores have sizes $\sim 10^6$ times larger than the final radius of any resulting star, it is clear that virtually all of the angular momentum of the initial cloud must be transferred somewhere else; in general, it must be to a circumstellar disk.  In this\indexit{accretion disk!angular momentum transfer} way, the formation of stars necessarily leaves behind material which can in principle form planets.'' \sactivity{$\circledS$ {\em Show:} Another way of formulating Eq.~(\ref{eqlh:rofm}) is to say that the mass of the cloud must exceed a certain value. (a) Reformulate Eq.~(\ref{eqlh:rofm}) as function of cloud temperature $T_{\rm c}$, cloud density $n_{\rm c}$, and stellar mass $M_\ast$ (Note: this is similar to what is known as the Jeans Mass, which is commonly derived from energy imbalance or by a comparison of sound and free-fall time scales in a perturbation analysis). This shows that $M_\ast \sim f M_\odot  T_{\rm c}^{3/2}/n_{\rm c}^{1/2} $. (b) Derive the value of the constant $f\approx 2$ assuming, for simplicity, that the gas consists predominantly of molecular hydrogen.
%f=8.55/2^2=2 from (3/(4\pi))^{3/2}*k^{3/2}/G^{3/2}/(\mu m_p}^2
%e.g. https://www.ast.cam.ac.uk/~pettini/STARS/Lecture11.pdf
For $n_{\rm c}$ of order 100\,cm$^{-3}$ estimate $M_\ast$ for $T_{\rm c}\approx 10$\,K, characteristic of present-day molecular clouds (realizing this is a rough order-of-magnitude estimate). Early in the life of the Universe, with only H and He in the mix, the interstellar gas lacked many of the strong emission lines of heavier elements, could therefore not cool as efficiently, leaving interstellar clouds significantly warmer, roughly of order 100\,K. (c) Use the derived expression to show that this favors the formation of much heavier stars, even when starting from a higher density of order $10^4$\,cm$^{-3}$. How this contributed to the evolution of elements heavier than H and He (known as 'metals' to astronomers) over the history of the Universe is discussed in \href{https://ui.adsabs.harvard.edu/abs/2019Sci...363..474J/abstract}{a review by \citet{2019Sci...363..474J}} \mylabel{act:jeans} \solution{jeans}} The initial phase of star formation, and the clearing of the dust-rich environment of the protoplanetary disk happens on a time scale of just a few million years, as we shall see in Sect.~\ref{lh:7}, and this means that much of the growth phase of planets, or at least the sizable planetesimals that later coalesce to form fully-grown planets, must be completed by then. 

\ors[IV:5.1] ``Confirmed and candidate \indexit{exoplanet}exoplanets number in the thousands and \indexit{exoplanet!search techniques}search techniques include Doppler measurements, transit photometry, microlensing, direct [(and since 2019 also interferometric)] imaging and astrometry.  Each detection technique has some type of observational incompleteness that imposes a biased view of the underlying population of exoplanets. In some cases, statistical corrections can be applied. For example, transiting planets can only be observed if the orbital inclination is smaller than a few degrees from an edge-on configuration. However, with the reasonable assumption of randomly oriented orbits, a geometrical correction can be applied to determine the occurrence rate for all orbital inclinations. In other cases, there is simply no information about the underlying population and it is not possible to apply a meaningful correction. For example, the number of planets with a similar mass (or radius) and a similar intensity of intercepted stellar flux as our Earth is not secure at this time because the number of confirmed detections for this type of planet [and orbit is too] small.\activity{{\em Show:} (a) Estimate the orbital Doppler swings and the fractional dimming during solar transits observed from afar of Mercury, Earth, and Jupiter. (b) Also estimate how close a Jupiter-like exoplanet (with an albedo of 0.5) should orbit for the fractional bolometric dimming during a secondary eclipse (when the planet moves behind the star) to be about 1~millimagnitude (which is the noise level for the telescope of the {\em Kepler} spacecraft for a 13th magnitude star at 1-minute exposure times). (c) Consider at what wavelength range the contrast is optimal. Use, {\em e.g.,} \href{https://nssdc.gsfc.nasa.gov/planetary/factsheet/}{the fact sheet} at https://nssdc.gsfc.nasa.gov/planetary/factsheet/. (d) Compare the Doppler signals with the thermal widths of spectral lines, and consider what to use as reference wavelengths. (e) How large is the Doppler swing added to the stellar signals owing to Earth's orbit around the Sun? \mylabel{act:exoplanetdetection}}\activity{{\em Look up} and summarize the principles of the five detection methods of exoplanets, and consider what the strengths, weaknesses, and technological challenges are for each method and how these depend on planetary mass, size, and orbital radius. Note: activities~\ref{act:doppler} and~\ref{act:exoplanetdetection} ask about Doppler signals and transit photometry. \mylabel{act:methodcomp}}
% Source: https://keplergo.arc.nasa.gov/CalibrationSN.shtml

As a result of the sample biases and observational incompleteness for each discovery technique, our view of exoplanet architectures is fuzzy at best. There are no cases beyond the Solar System where the entire parameter space for orbiting planets has been observed.  Instead, we piece together an understanding of exoplanet architectures by counting planets in the regimes where techniques are robust and then we estimate correction factors when possible. When drawing conclusions about the statistics of exoplanets, it is helpful to understand the incompleteness in this underlying patchwork of orbital parameter space.'' Sections~IV:5.2-5.6 provide brief descriptions of the methods and their limitations. 

\ors[IV:5.7.4]  ``Our view of exoplanets is still skewed by the observational sensitivities of the techniques that we use.  However, the discoveries that have been made have helped us to revise our understanding of planet formation and the formation of the Solar System.  We see that planet formation is a chaotic process and that disks are sculpted by gravitational interactions to a greater extent than we appreciated by considering our Solar System.  We now know that almost every star has planets and that planet formation is far more robust than astronomers expected.''\regfootnote{Recent reviews on the making of planets in general and on giant planet formation and migration, see Space Science Reviews (2018) volume 214, pages 38 (by Paardekooper and Johansen) and 60 (by Lammer and Blanc, referred to as $^a$ below), as well as \href{https://global.oup.com/academic/product/one-of-ten-billion-earths-9780198799894}{'One of ten billion Earths'} by \citet{2018otbe.book.....S}.} Although we do not touch on the process of forming binary star systems (or higher multiplets), the outcome of the evolution of a molecular cloud to a planetary system often involves fragmentation of the cloud into two or more stars: roughly one in every two 'stars' visible in the sky is, in fact, a double or higher-multiple star. 

It may be counterintuitive, but our knowledge of the evolution of the formative phases of stars by accretion from spinning disks of gas and dust that contracted out of huge molecular clouds has been helped greatly by the hunt for exoplanets and their story of formation. It is for that reason that this chapter begins with a very concise summary of what has been learned about (exo-)planetary systems, thereafter to go 'back in time' to the gaseous phases of the protoplanetary disks and how the gases in these formed the central stars, and how some planets ended up being ejected from the forming planetary system. \activity{{\em Look up:} Star-forming regions and disks around young stars are best observed in the near-infrared region of the spectrum. Look into what wavelengths are often used for such observations, and consider why ('Why is the sky blue?'), given that dust sizes in the interstellar medium peak around a few tenths of a micron.}

\section{(Exo-)Planets and (exo-)planetary systems}\label{sec:exo}
\ors[III:3.9] ``The ultimate\indexit{exoplanet} stage of disk evolution, in addition to accretion and photo-evaporation, \indexit{disk!evolution}involves the growth of planetary bodies.  We now know [almost 4,000 (by late 2022) exoplanetary systems], with the number continually increasing.  Of course, the first major surprise was the discovery of Jupiter-mass bodies at very small orbital radii.  This emphasized the almost certain necessity of inward {\em migration}, as it appears unlikely that disks can be sufficiently massive at $0.02-0.1$ [Sun-Earth distances (or astronomical units: AU)] to form such objects (unless disks are gravitationally unstable all the way to the central star).  The other major surprise was how eccentric [many of the orbits of close-in, large exoplanets] are.  These two features are probably related, especially if planet-planet scattering is responsible for much of the inward migration.  Before discussing migration further, it is useful to consider how the planets would form in the first place.

\begin{figure}[t]                                                                                                         
%\centerline{\psfig{figure=figures/zoom_bw.eps,width=9cm}}   
\centerline{\includegraphics[width=9cm,trim=0 2cm 0 2cm]{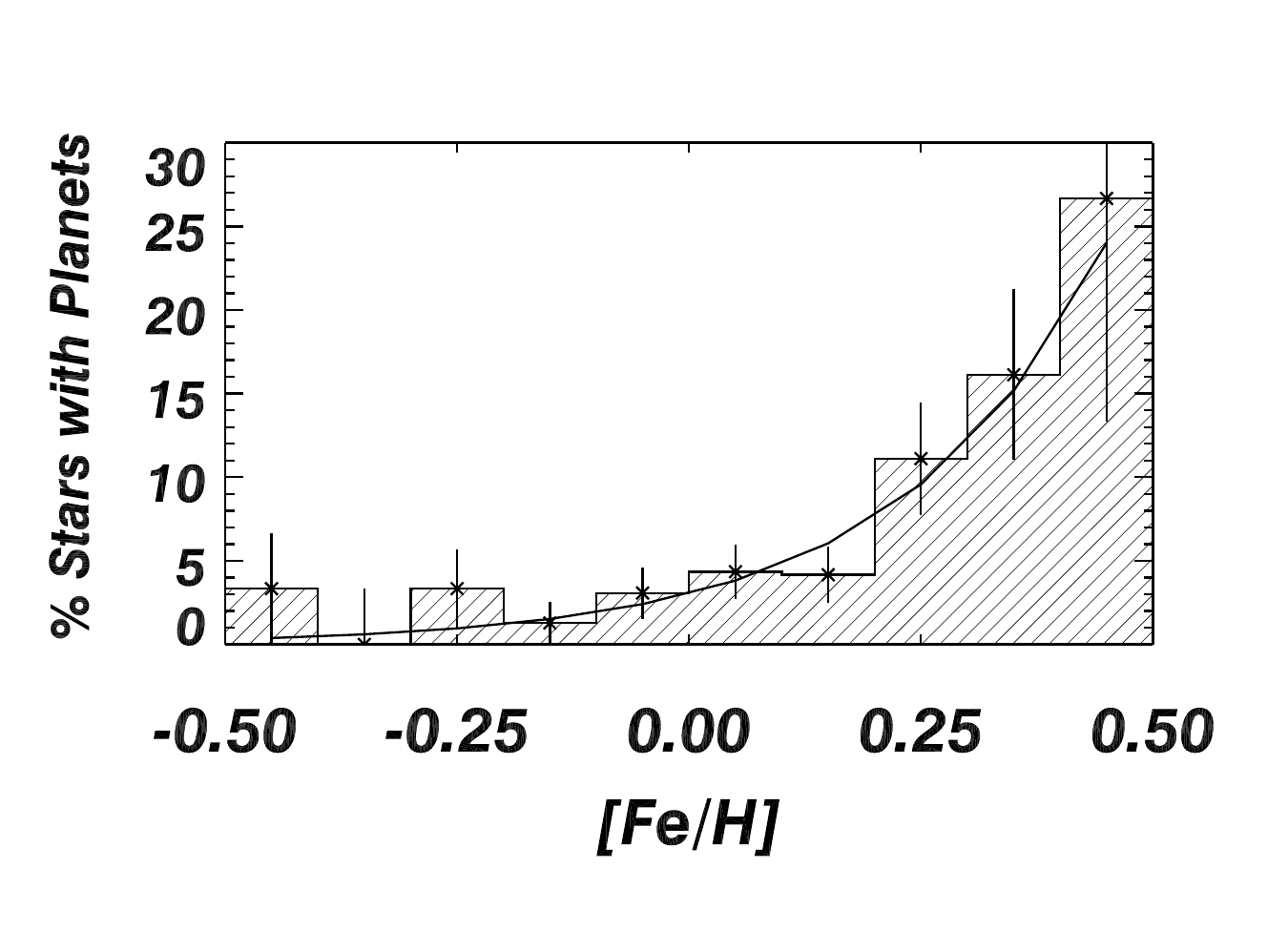}}   
\caption[Planet-metallicity correlation for gas-giant planets.]{High metallicity stars are more likely to host gas-giant planets than 
sub-solar metallicity stars.  ['Fe/H' denotes the ratio of the abundance of Fe relative to hydrogen, while the brackets mean that the logarithm has been taken of that ratio normalized to the solar value, so that the Sun would have a value of 0, and the scale reaches a factor of about three down and up on the left and right side, respectively; Fig.~IV:5.12]
\label{fig:metallicity} }                                                                                                          
\end{figure}\figindex{../fischer/art/zoom_bw.eps}
The two major scenarios\indexit{planet!formation!scenarios} of planet formation are those of core accretion [(starting with solids, and --~once heavy enough, should that indeed occur~-- also accumulating gases)] and gaseous gravitational instability.  [\ldots] One strong piece of evidence for core accretion versus gaseous gravitational fragmentation'' comes from
one \ors[IV:5.7.4] ``of the first observed statistical correlations established[:] gas-giant planets form more frequently around metal-rich stars.\regfootnote{Note that astronomers have a habit of referring to all elements heavier than helium as 'metals'.}  This planet-metallicity correlation [(see Fig.~\ref{fig:metallicity})] was used as evidence for core accretion as the formation mechanism for gas-giant exoplanets that orbit closer than a few AU around their host main-sequence stars. [\ldots] Interestingly, a similar correlation with host star metallicity has not been identified for smaller Neptune-like  [\ldots] planets. However, it remains unclear whether such correlation exists for rocky planets. [\ldots]''

\ors[III:3.9] ``In the core-accretion\indexit{planet!formation!core accretion} model for giant planet formation, solid bodies accumulate via collisions until the resulting core is sufficiently large that its gravity can pull in surrounding gas.  There is some concern that core accretion might proceed too slowly to explain the observed disk clearing on timescales as short as $1-2$\,Myr in significant numbers of stars.  There are two potential bottlenecks in the process.  One is the formation of km-sized planetesimals\indexit{planetesimal} from cm-sized objects.  Such bodies are thought to be held together lightly --~too large for effective sticking and too small for gravity to become important~-- and, as bodies of different sizes have different velocities due to gas drag, collisions between these objects might shatter them rather than build them up.  In addition, [there is growing evidence that young planets in material-rich disks are subject to rapid inward migration, which would] require fast agglomeration, especially for Earth-sized objects, though studies suggest that this inference of rapid migration may not always be correct.  Various schemes of dust concentration might help avoid shattering by reducing relative motions and increasing densities, perhaps through vortices or eddies or other turbulent structures.

Once km-sized planetesimals are made, collisions among them can lead to the building of terrestrial planets and giant-planet cores.  The remaining bottleneck[, at least significant for giant planets,] is that of accumulating gas.  The energy released by accretion of planetesimals and gravitational contraction of the envelope must be radiated by the outer envelope. If the opacity of the envelope is large, it must extend to large radii; in turn, this can limit the gas available for accretion, which must lie close enough that the tidal forces of the central star do not overcome the protoplanet's gravity.  [It appears] that, with sufficiently massive cores, giant planets can form within 1\,Myr for an opacity $\sim 2$\%\ of interstellar values, [because the opacity (dominated by dust) may be reduced due] to rainout of solid materials in the planetary envelope; as grain growth almost certainly precedes core formation, reduced dust opacity is an extremely plausible assumption. [\ldots]

There is general agreement that terrestrial planets generally (fully) form
later than the giant planets; gas drag is important in early stages
but the final growth may well occur after gas removal from the disk.
[O]nce growth to
km-sized planetesimals has occurred, gravitational effects become
important.  At first the planetesimals grow by gravitational focusing;
as they grow, eventually they excite or stir up other bodies, making
their relative velocities larger and limiting accretion.  The result
is thought to be a set of 'oligarchic' protoplanets with relatively
similar masses (at least locally). After the oligarchs have swept up
most of the available material, interactions between them dominate the
subsequent evolution, with large impacts a major feature.  This
indicates that the final state of terrestrial planet systems is
difficult to predict, as it is the result of chaotic growth.

Even after the\indexit{planetary!system!planet migration} terrestrial planets are essentially fully formed,
significant system evolution can occur, simply because multi-body
gravitating systems are generally not stable.  A particularly
interesting possibility is long-term evolution and migration due to
interactions of an outer system of gas/ice giants with the
planetesimals left in the outer disk, objects formed in regions
with such low densities that growth to large bodies was not possible.
[T]here probably has been
outward migration of at least Neptune in our outer Solar System, based
on the analysis of resonant structure in our own planetesimal system --
the Kuiper Belt.  One possible mechanism for explaining this migration
is giant planet-planetesimal interactions. Such gravitational
perturbations can result in the system becoming dynamically unstable,
resulting in ejection and scattering of many planetesimals into
high-eccentricity orbits; this has been suggested, in the so-called
'Nice' model, as\indexit{planet!formation!'Nice' model} an\indexit{Nice model!planet formation} explanation for the late heavy bombardment seen
in the impact history of the Moon (cf., [Sec.~\ref{sec:formative}) \ldots]''. It has also been proposed as an explanation of the stunted growth of Mars, and moreover of the chemical gradient in the asteroid belt: silicate-rich and carbon/water-rich populations should have been differentiated by distance to the Sun (across the 'ice line' where the temperature would have been low enough to create water-rich asteroids only further out; see footnote~\ref{note:iceline}), but actually show much overlap of the populations, albeit with a clear trend for the average chemical makeup as function of orbital radius. It is argued that this smoothed trend of what should have formed as a clear chemical segregation was introduced by gravitational interaction with migrating gas/ice giants (in what is referred to as the 'Grand Tack' model, in which the Jupiter-Satun pair first migrated inward and subsequently outward). \activity{{\em Consider:} Figure~\ref{fig:protoplanetary_disk} shows a curved 'snow line' (or 'ice line'). What is the reason behind that?}\activity{{\em Look up} the 'Grand Tack' hypothesis and describe the likely consequences for the growing Mars, for the asteroid belt, and for water distribution by scattered asteroids into the inner solar system.}

\subsection{Exoplanet formation} 

The solar \indexit{exoplanet!formation}nebula theory holds that the Sun and its attending planetary system formed out of a cloud of gas and dust (with dust making up, on average, about 1\%\ of the total mass$^a$) that contracted into a spinning \indexit{accretion disk}disk, \indexit{protoplanetary disk}with most matter migrating towards the center to form a star even as much of the angular momentum ended  up in the orbiting planets that formed out of the cool disk material before the remainder of the gases were somehow cleared out (more on that below). \ors[IV:5.7.1] ``The solar nebula theory provides a theoretical description for the formation of the Solar System.  Indeed, it has been said that this model is so elegant, that it is hard to imagine that it could be wrong.  The solar nebula theory neatly explains most observations: the planets closest to the Sun form in a hot environment and as a consequence these planets are small and comprised of refractory elements ({\em i.e.}, elements [whose solids] withstand high temperatures); the more massive gas giants form beyond the ice line (a distance where it is cold enough for dust grains to be coated with icy mantles) where the feeding ground is more voluminous; jovian planets have moons that were either captured or that form as mini-solar-systems; the planets all orbit in the same direction in the disk because they inherit the same angular momentum vector; the Solar System is littered with leftover debris such as asteroids and comets. The theory supports the idea first suggested by Kant and Laplace that the proto-Sun was surrounded by a primordial spinning disk of dust and gas.  All of the material that makes up the Sun drained through this disk. [\ldots]

The mass of the protoplanetary disk is a fraction of the stellar mass and evolves with the central star.  Our understanding of the physics and chemistry of protoplanetary disks is distilled in Fig.~\ref{fig:protoplanetary_disk}. The temperature is about 1500\,K near the inner part of the disk and along the flared outer layers.  These high temperatures are too hot for grain growth, but a few AU from the protostar, the disk mid-plane is cool enough for icy grains to stick and grow. The opacity of the disk is set by the dust, which gradually decouples from the gas and settles toward the mid-plane, increasing transparency of the disk over time.\activity{{\em Consider:} Figure~\ref{fig:protoplanetary_disk} shows a clearing near the central star. This is associated with the magnetic field of the rotating star. Consider what processes are at play there and the role of the following: accretion rate, ionization fraction, diffusion of field into the ionized gaseous disk, orbital and angular velocities, the corotation radius, winding up of magnetic field that connects the star to the disk, centrifugal force, etc. There is no easy concept for this: you can look at the literature of MHD models of T\,Tauri accretion disks to see how complex the coupling is. Store your thoughts: the star-disk interaction leading to the clearing is discussed in Sect.~\ref{lh:4}.}

\begin{figure}[t]                                                                                                         
%\centerline{\psfig{figure=figures/protoplanetary_disk.eps,width=\textwidth}}   
%\centerline{\psfig{figure=figures/protoplanetary_disk_color.eps,width=\textwidth}}   
\centerline{\includegraphics[width=\textwidth]{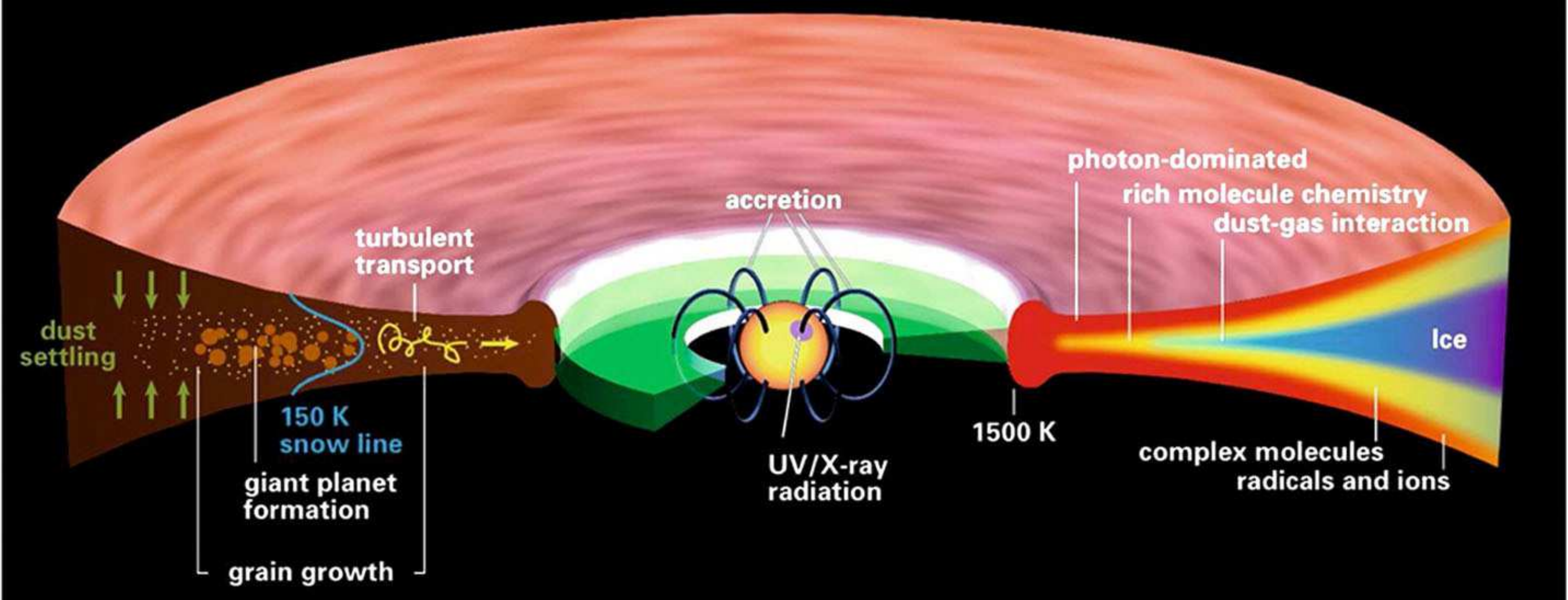}}   
\caption[Structure and processes of protoplanetary disks.]{A sketch of the structure and processes of \indexit{protoplanetary disk}protoplanetary disks [with ages in the range of about 1--5\,Myr. Fig.~IV:5.9; \href{https://ui.adsabs.harvard.edu/abs/2013ChRv..113.9016H/abstract}{source: \citet{2013ChRv..113.9016H}}. \colorfig] \label{fig:protoplanetary_disk} }  \figindex{../fischer/art/protoplanetary_disk.eps}\end{figure}        

Protoplanetary disks provide the initial conditions for planet formation. [\ldots] In the first phase of planet formation, the planet grows by runaway accretion of solid material. The second phase of growth is very slow; both solid and gas accretion are nearly time-independent and this phase sets the planet formation timescale.  Once the planet core reaches a mass of about $10 {\rm M_\oplus}$, [if indeed it succeeds in that,] the third phase of runaway gas accretion begins, growing the planet mass from 10 to a few hundred ${\rm M_\oplus}$.  [It has been] estimated that gas-giant planet formation should take roughly 10\,Myr. However, observations of protoplanetary disks in the 1990s presented a conundrum: the primordial disks appear to be nearly ubiquitous around stars that are 1\,Myr; at 2\,Myr only about half of young stars have disks and by 10\,Myr, the disks are essentially gone.  Figure~\ref{fig:disk_lifetimes} shows the fraction of protoplanetary disks found in young cluster stars.
 
One triumph that emerged from the discovery of exoplanets was a solution 
to the disagreement between theory and observations for the formation timescale of 
gas-giant planets.  The first detected gas-giant planets orbited 
close to their host stars providing evidence that exoplanets 
could undergo orbital migration. Thus, planets were not restricted to 
a planetesimal feeding ground at a fixed orbital radius; instead, the planet embryos are pushed around in 
the disk by planet-planet interactions and tidal torques. The access to a wider part of the disk suggests 
a wider feeding zone for more rapid accretion of planetesimals that would shorten the second phase of gas-giant planet formation [\ldots]''

\subsection{Exoplanet migration}

In Sect.~\ref{sec:orbitinteraction} we \indexit{exoplanet!migration}already described the possibility that planets can change \indexit{orbital!interaction}their orbits in the formation phase of a planetary system by tidal interaction with the surrounding disk. This is one scenario by which, for example, giant planets may ultimately find themselves orbiting their parent star at distances much closer than where they could readily form.  \ors[IV:5.7.2] ``Another way to push exoplanets inward is through gravitational encounters.  There are several proposed mechanisms that excite orbital eccentricity including secular migration, planet-planet scattering, and Kozai perturbation in which gravitational interactions result in coupled variations in orbital inclination and eccentricity.  High eccentricity planets with a small enough periastron passage eventually experience tidal circularization and can end up in short-period orbits.

Different migration mechanisms predict distinct observables. A particularly interesting observable is stellar obliquity, the relative angle between the stellar rotation vector and the vector normal to the planet orbital plane.  The stellar obliquity can be measured by observing the \indexit{Rossiter-McLaughlin effect}Rossiter-McLaughlin effect.  This effect is caused by a transiting object blocking some of the light from a rotating star. [In the case of a prograde, low-obliquity planet, the transiting planet first] crosses the approaching limb of the rotating star, decreasing the contribution of blue-shifted light in the spectral line and a few hours later the planet crosses the receding limb of the rotating star, decreasing the contribution of red-shifted light.  The systematic decrement of Doppler-shifted light in the composite spectral lines results in a distortion of line profile, which is (mis)interpreted as a change in the radial velocity of the star. The shape of the Rossiter-McLaughlin  curve during transit is entirely dependent on the stellar obliquity.  Consequently, the stellar obliquity is determined by modeling the anomalous radial velocity signals during a transiting event.\activity{{\em Show:} (a) Sketch and describe the observable signatures in high-resolution spectra of transiting planets for orbits of different obliquity (including effectively retrograde planets). (b) Also: estimate transit times for planets around of solar-mass star at distances such as Mercury, Earth, Jupiter, and Neptune. Use, {\em e.g.,} \href{https://nssdc.gsfc.nasa.gov/planetary/factsheet/}{this fact sheet}: https://nssdc.gsfc.nasa.gov/planetary/factsheet/.}

Disk-driven migration is expected to produce a small stellar obliquity whereas gravitational encounters that temporarily pump up the orbital eccentricity of gas-giant planets should result in a wide range of stellar obliquities including retrograde orbits. The latter has been observed for many transiting planets suggesting that high eccentricity mechanisms drive gas-giant planets inward. However, it has also been suggested that the observed stellar obliquity range may reflect a primordial stellar obliquity due to interactions between proto-planetary disk and a companion star. Interestingly, the small stellar obliquity of low-mass multi-planet systems suggests well-aligned vectors for the stellar spin and planetary orbits.  It is certainly possible that gas-giant and low-mass planets migrate by different mechanisms.

In summary, the most important revisions to the solar nebula model and our understanding of planet formation can be attributed to one source: the addition of dynamical interactions between planets and the primordial disk. These dynamical interactions speed up the accretion timescales, produce mean-motion resonances, scatter planets out of the disk into non-coplanar orbits that can be detected by the Rossiter-McLaughlin effect and even eject some planets.''

\begin{figure}[t]                                                                                                         
%\centerline{\psfig{figure=figures/mass_radius.eps,width=6cm}}  
\centerline{\includegraphics[width=5.5cm,trim=0 1cm 0 0]{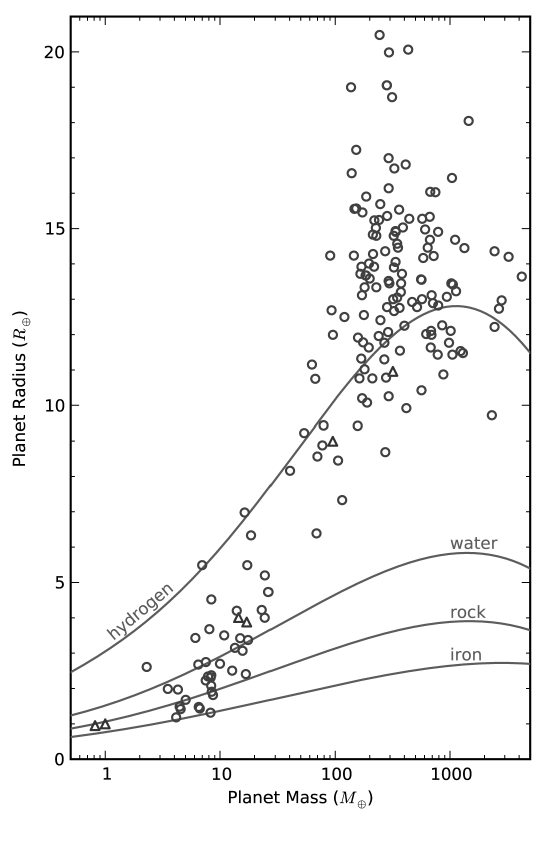}}  
\caption[Masses and radii of selected exoplanets 
and Solar-System planets.]{Masses and radii of well-characterized exoplanets (circles) 
and Solar-System planets (triangles). Curves show models for idealized planets 
consisting of pure hydrogen, water, rock (Mg$_2$SiO$_4$) or iron. [Fig.~IV:5.11; \href{https://ui.adsabs.harvard.edu/abs/2013Natur.503..381H/abstract}{source: \citet{2013Natur.503..381H}}.] \label{fig:mass_radius} }\figindex{../fischer/art/mass_radius.eps}\end{figure}        
\subsection{Exoplanet geology}

Studies suggest that there may be \ors[IV:5.7.2] ``two \indexit{exoplanet!geology}characteristic planet radii ($1.7 R_\oplus$ and $3.9 R_\oplus$) that divide planets into three populations: \indexit{exoplanet!terrestrial}terrestrial planets, \indexit{exoplanet!gas-dwarf}gas-dwarf planets and \indexit{exoplanet!gas-giant}gas-giant planets. [\ldots] Both the mass-radius relationship and the transition radius from rocky to non-rocky planets help us to better understand the formation history of small planets. Planets that form {\em in situ} in the inner part of the disk would consist primarily of rocky materials and possibly a primordial H/He atmosphere. In comparison, planets that have undergone significant migration should contain more volatile materials such as astrophysical ice ($\rm{H_2O}$, CO, and ${\rm NH_3}$). The debate of whether close-in planets form {\em in situ} or migrate should eventually gain evidence from studies of exoplanet atmospheres that add constraints on their chemical composition.''

\ors[IV:5.7.3] ``Thousands of planet candidates were discovered by the {\em Kepler} mission, allowing for precise measurements of exoplanet radii.  The combination of the radius and mass measurements (either from the Doppler technique or from transit timing variations) provides a mean density for hundreds of exoplanets and allow us to begin considering the bulk composition of unseen planets that orbit stars hundreds of light years away from us. The varying bulk composition of exoplanets results in different curves that cut through the mass-radius parameter space shown in Fig.~\ref{fig:mass_radius}.

Planets with radii smaller than 4 times that of the Earth can exhibit a remarkable diversity of compositions. [\ldots] Planets smaller than 1.5 Earth radii increase in density with increasing radius and seem to have a composition that is consistent with rock. Planets with radii between 1.5 and 4 times the radius of the Earth showed decreasing density with increasing radius, suggesting that the larger planet radius is a product of gaseous envelopes. [\ldots\ T]he significant amount of scatter in the mass-radius parameter space suggests a large diversity in planet composition at a given radius.''

With the growing number of exoplanet detections, one more thing has become abundantly clear: whereas the Solar System suggests a marked division between the four terrestrial planets (with masses of one Earth mass or less) and the four giant planets (with masses of 14.5 to 318 Earth masses), the exoplanet population overall has no such division, showing a continuum of masses from low to high$^a$.

\subsection{Exoplanets and binary star systems}

\ors[IV:5.7.4] ``Many stars \indexit{exoplanet!in binary-star systems}in the solar neighborhood are components of multiple-star systems, [and exoplanets have been found orbiting one of the two components while others have distant circumbinary orbits. \ldots] The occurrence rate of circumbinary planets is estimated to be $\sim$10\%\ assuming the orbital plane of circumbinary planets roughly align with the binary orbital plane. The occurrence rate could be much higher if the orientation of planet orbits is more isotropic.

It is expected that planet formation may be impeded in systems where the binary stars have small separations ({\em e.g.}, $\sim 10 - 200$\,AU). This is supported both by simulations and observations that find a smaller fraction of exoplanets in binary star systems.  It is not surprising that the dynamics of binary star systems stir things up and challenges planet formation. What is surprising is that the planets exist there at all.''

\begin{figure}[t]
\centering \includegraphics[width=0.45\textwidth]{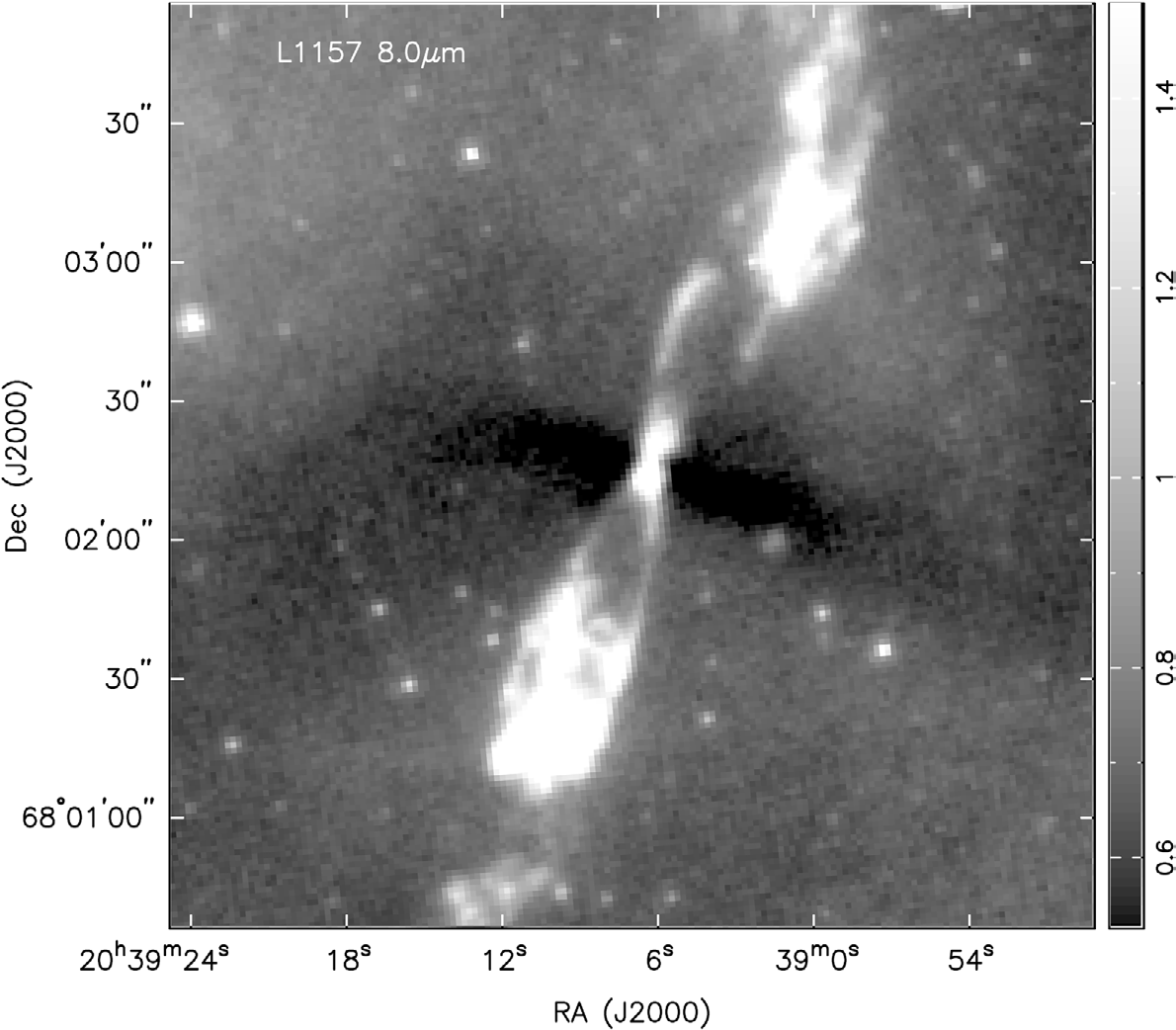}\includegraphics[width=0.53\textwidth]{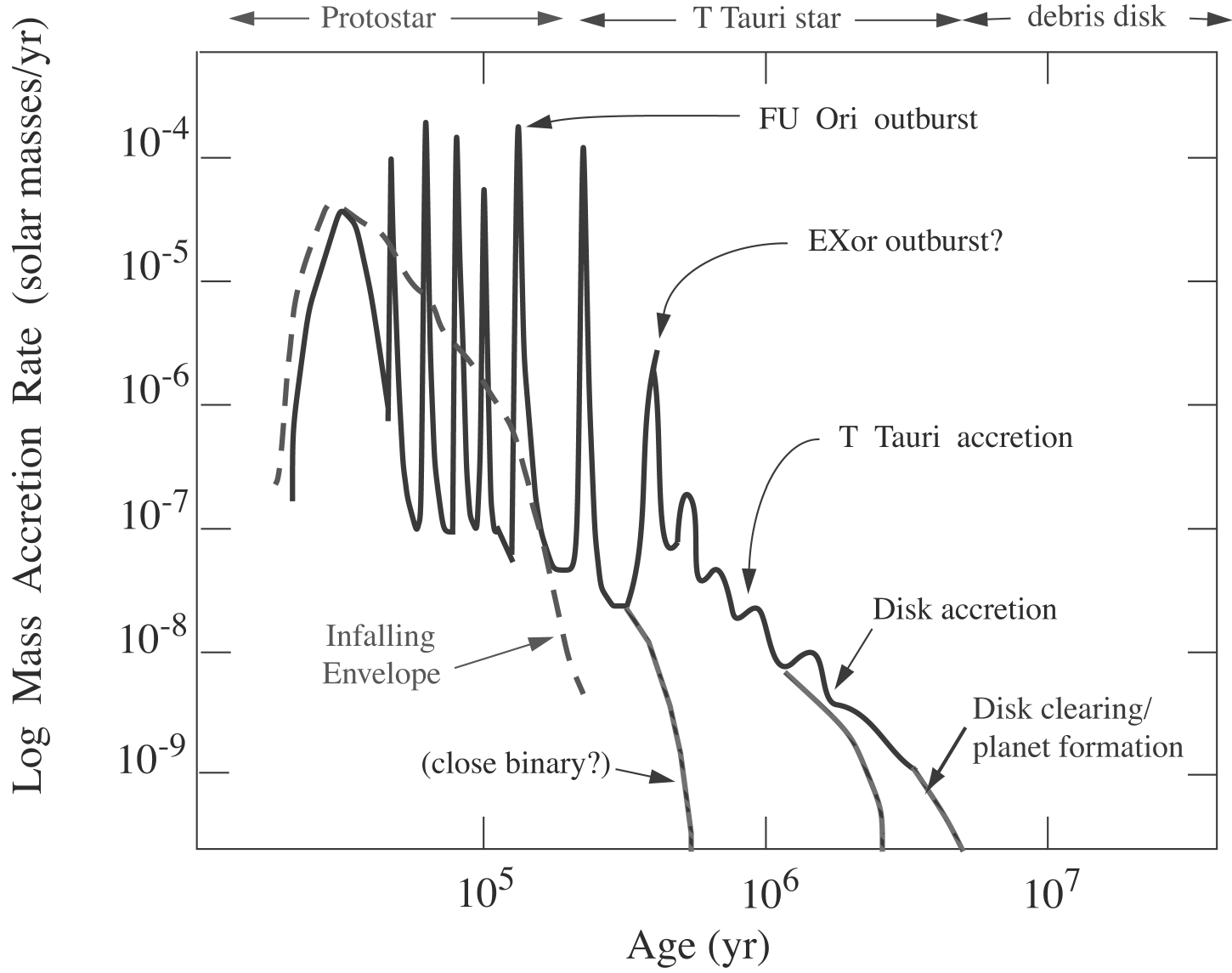}
\caption[Accreting protostar and likely accretional history of a low-mass star.]{{\em Left:} An $8\,\mu$m image of an accreting low-mass protostar.  The darker, filamentary region running east-west (horizontally in the image) represents dust extinguishing the background radiation; this indicates that the densest, most massive region of the material falling in to make disk and star is far from spherically-symmetric.  The bright regions running north-south (top to bottom) are due to protostellar continuum emission reflected from dust and molecular emission lines excited by a high-velocity, bipolar outflow thought to be driven from the innermost regions of the protostellar accretion disk.  {\em Right:} Schematic diagram of a likely accretional history of a typical low-mass star.  The dashed curve indicates the expected rate of infall of matter from the protostellar envelope ({\em e.g.,} dense region indicated in the left-hand panel).  The solid curve suggests a possible variation of accretion through the protostellar disk onto the central star, which may be steady at the earliest times but is subject to strong variations in accretion (so-called FU\,Ori outbursts).  In this picture, material piles up in the disk due to the infall rate being higher than the disk can smoothly pass on to the central star; this leads to episodic bursts of accretion which drain the excess disk mass.  Finally, after infall ceases, slower, more steady accretion occurs during the T\,Tauri phase, which may cease because either a binary companion or planets accrete the remaining mass.  This results in 'clearing' the disk, {\em i.e.,} removing most of the small dust and apparently most of the gas.  Finally, secondary production of small amounts of dust can occur during the debris disk stage, when solid bodies collide and shatter.  [Fig.~III:3.2; \href{https://ui.adsabs.harvard.edu/abs/2009apsf.book.....H/abstract}{source: \citet{2009apsf.book.....H}}.]}
\label{figlh:acc}
\end{figure}\nocite{Hartmann2009}\indexit{FU\,Orionis outburst}
\section{Formation and early evolution of stars and disks}\label{sec:star}
\subsection{Observations of star-forming processes}\label{sec:starformobs}

Before the discovery \indexit{star!formation!observations}that exoplanetary systems were about as common as stars (reached in the first decade of the 21st century) astronomers struggled to understand how angular momentum from the contracting pre-stellar cloud could be removed so that a star could form at all.  There were studies on how Alfv{\'e}n waves could carry angular momentum away, how fragmentation into multiple star systems could deal with the problem, or how winds from magnetized disks could extract angular momentum. Realizing that much of the angular momentum is left behind in the planetary system reduced the magnitude of the problem tremendously, and many of the earlier ideas about where the angular momentum would end up have been left behind or now form a lesser challenge to the formation scenario of stars. ``If we assumed that there were no planets and all the angular momentum resides in
the Sun, this \ors[III:11.2.2]  leads to an increase in the angular velocity by about a factor of
35. As a result the Sun would spin around its axis in about 18\,hours instead of
27\,days. Because the assumption of a homogeneous sphere underestimates the
effective moment of inertia, the Sun would complete a rotation within about
12\,hours which corresponds to a velocity of the photosphere in the order of
100\,km\,s$^{-1}$'' instead of the observed 2\,km\,s$^{-1}$.

Nowadays, the view is that \ors[III:3.1] ``[t]he accretion disk is basically an engine in which angular momentum is transferred outward to ever-decreasing amounts of material while the majority of the mass moves inward to the center.  \activity{{\em Show:} Estimate the total values and ratios of mass and angular momentum in the planetary system and in the Sun (use Fig.~\ref{fig:Ievolve}).} In the case of at least moderately-ionized disks, it seems increasingly certain that magnetic turbulence provides the necessary angular momentum transport for accretion.  The low ionization of protostellar disks\indexit{protostellar!disk!ionization} is likely to render this mechanism ineffective over significant radial regions; \indexit{gravitational!torque}gravitational torques can come to the rescue, moving most of the cloud mass into the central regions in any event.  However, gravitational torques alone will leave a sizable amount of mass in the disk, of order 10\%--30\% of the central star mass.  As this is much larger than estimated by many techniques, and substantially more than assumed in many models of planet formation, it may be necessary for additional angular momentum transport to occur via magnetic turbulence. On longer timescales, the remaining disk gas is probably removed by some mechanism of ejection due to stellar X-rays and extreme-ultraviolet (EUV) heating.

 \begin{figure}[t]
 \centering
 \includegraphics[width=0.4\textwidth,bb= -75 -3 689 797]{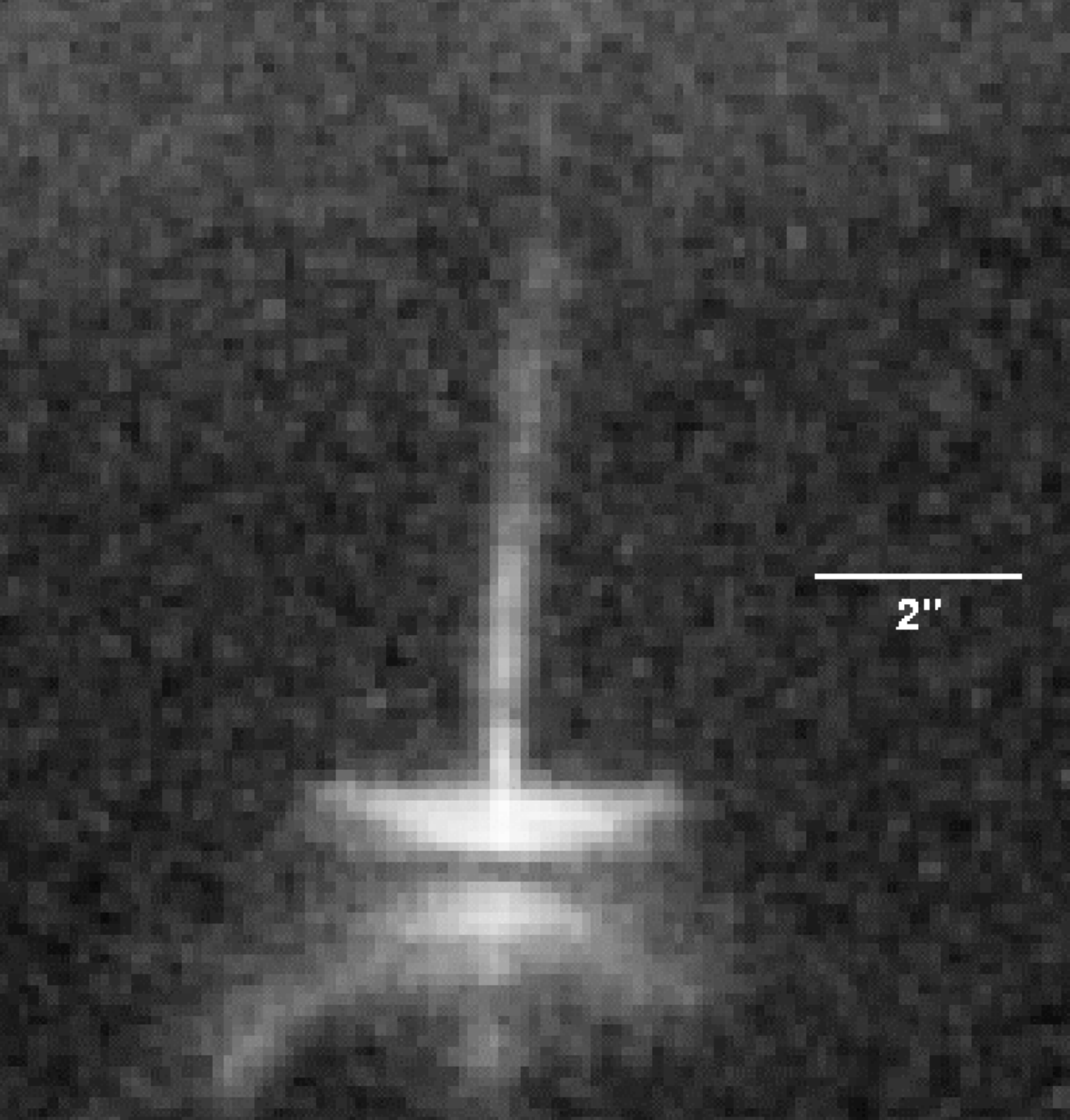} 
\caption[Optical image of the accreting young star HH\,30.]{Optical image of the accreting young star HH\,30, showing the upper surfaces of its dusty disk in scattered light (the dark lane is due to dust extinction of the central star by the disk), along with an optical, high-velocity, bipolar jet.  For scale, 2\,arcsec $=$ 280\,Astronomical Units.  [Fig.~III:3.3]}
\label{figlh:hh30}
\end{figure}\nocite{1996ApJ...473..437B}
This picture of star formation has considerable observational support.  Cold clouds of the mass and size indicated in Eq.~(\ref{eqlh:rofm}) are seen in star-forming regions, some with already growing protostars (Figure\,\ref{figlh:acc}, left).  We also observe extended circumstellar disks around many young stars (Figure\,\ref{figlh:hh30}).  The masses of these disks are at least $\sim$~1\%\ that of the central star; with radii of hundreds of AU, they clearly must contain most of the system angular momentum.\activity{{\em Show:} Iron, oxygen, and silicon make up three quarters of the Earth's mass. Iron is some 30\%\ of the total. In the interstellar medium, iron makes up about 1 part in 1,000 of total mass. How many Earth-equivalents of iron does a circumstellar disk with a mass of 1\%\ of the Sun contain?}

The implication of this picture is that most of the mass of a star must pass through
its disk; that is, stars are most directly formed by disk {\em accretion}.
As shown schematically in the right side
of Figure~\ref{figlh:acc}, disk accretion may not be steady if it
cannot keep up with the infall to the disk; instead, early stellar evolution may
be punctuated by outbursts of very rapid accretion followed by extended periods of
slow mass addition.  There is observational evidence for such accretional outbursts in the\indexit{FU\,Orionis outburst}
FU Orionis objects; their properties suggest that disks
are likely to be quite massive, at least in early stages.'' \activity{{\em Consider:} List similarities and differences with Solar-System magnetic instabilities as discussed in Ch.~\ref{ch:mhd} when reading about things like FU-Orionis outbursts and 'ballooning out' of magnetic field in ejections of mass from corona and disk, likely driven by necessarily failing attempts of the forces at play to impose corotation.\mylabel{act:instabilities}}

Magnetic fields are good candidates for the transport of angular momentum in a disk, as the differentially rotating disk (trying to have matter orbit the forming star in Keplerian orbits) would stretch embedded field, causing a back-reaction that works to reduce the differential rotation (see Sect.~\ref{sec:magrot}). That can work, provided that the disk material has a sufficient degree of ionization so that the field and the gases can effectively couple. Another process that can contribute to the transport of angular momentum is gravitational coupling; see Sect.~\ref{sec:gi} for a description of this process. 

\ors[III:3.3] ``The\indexit{disk wind} other major potential mechanism of disk angular momentum transport is that of winds.  It is now thought that most of the angular momentum of disks results in expansion of the outer disk rather than simply being lost in a wind; however, because [Sun-like,] low-mass stars become slowly-rotating early in their existence (Sect.~\ref{sec:lh5}), it is quite possible that winds from the innermost disk regions play a central role in regulating the rotation of protostars.

Young stars with disks often eject powerful, collimated, bipolar winds or jets.  These outflows are clearly the result of disk accretion.  We can say this confidently because a) young stars without disks do not show this phenomenon, and b) mass ejection rates, as best we can determine, clearly scale with the accretion rate.  Indeed, in the case of the most powerful low-mass outflows --~those of the FU~Ori objects~-- accretion is the only energy source large enough to account for the necessary driving.

The high degree of collimation seen in many jets ({\em e.g.,} Figure\,\ref{figlh:hh30})
favors magnetic fields, as well-developed theory shows that rotating fields
can provide the necessary collimation.
Moreover, the observed outflows or jets are relatively cold; that is,
the sound speeds of the gas are well below escape velocity, making thermal acceleration
unimportant; and thus magnetic acceleration is not only attractive but probably
necessary.  What is not clear is whether {\em outer} disk regions exhibit
outflows, at least at a sufficiently significant level to affect disk evolution. [\ldots]

Using the basic theory\indexit{disk wind!magneto-centrifugal acceleration} of magneto-centrifugal acceleration, spatially-resolved kinematics --~expansion, rotation ~-- of jets can be used to infer the origin of the outflow, below currently resolvable scales.  Observations of jets using the {\em Hubble Space Telescope} have suggested that the source region for the observed optical jets is $\sim 0.2$ to 2\,AU.  These estimates must be regarded as uncertain, as it is very difficult to detect the jet rotation; the analysis must assume no asymmetries in the flow, which may be questionable, given the probable presence of complex internal shocks needed to heat the radiating jet gas.

While outflows clearly emerge from the inner disk, there is little evidence for significant mass loss from outer disks, which could take away significant amounts of angular momentum.  In addition, there are difficulties with assuming that the disk wind dominates angular momentum transport even in the inner disk.  Removing all the angular momentum by the wind involves removing all the accretion energy in the wind as well, leaving no remaining energy to radiate; but this is problematic, because some rapidly-accreting pre-main sequence disks are self-luminous.  It seems more plausible that other mechanisms --~the gravitational and magneto-rotational instabilities~-- dominate the angular momentum transport of disks, with the winds being a byproduct of accretion.  However, the slow rotation of low-mass\indexit{protostar!rotation} protostars may require a powerful wind from the innermost regions to remove the final amount of angular momentum (Sect.~\ref{sec:lh5}).''

\begin{figure}[t]
\centering
\includegraphics[width=0.45\textwidth]{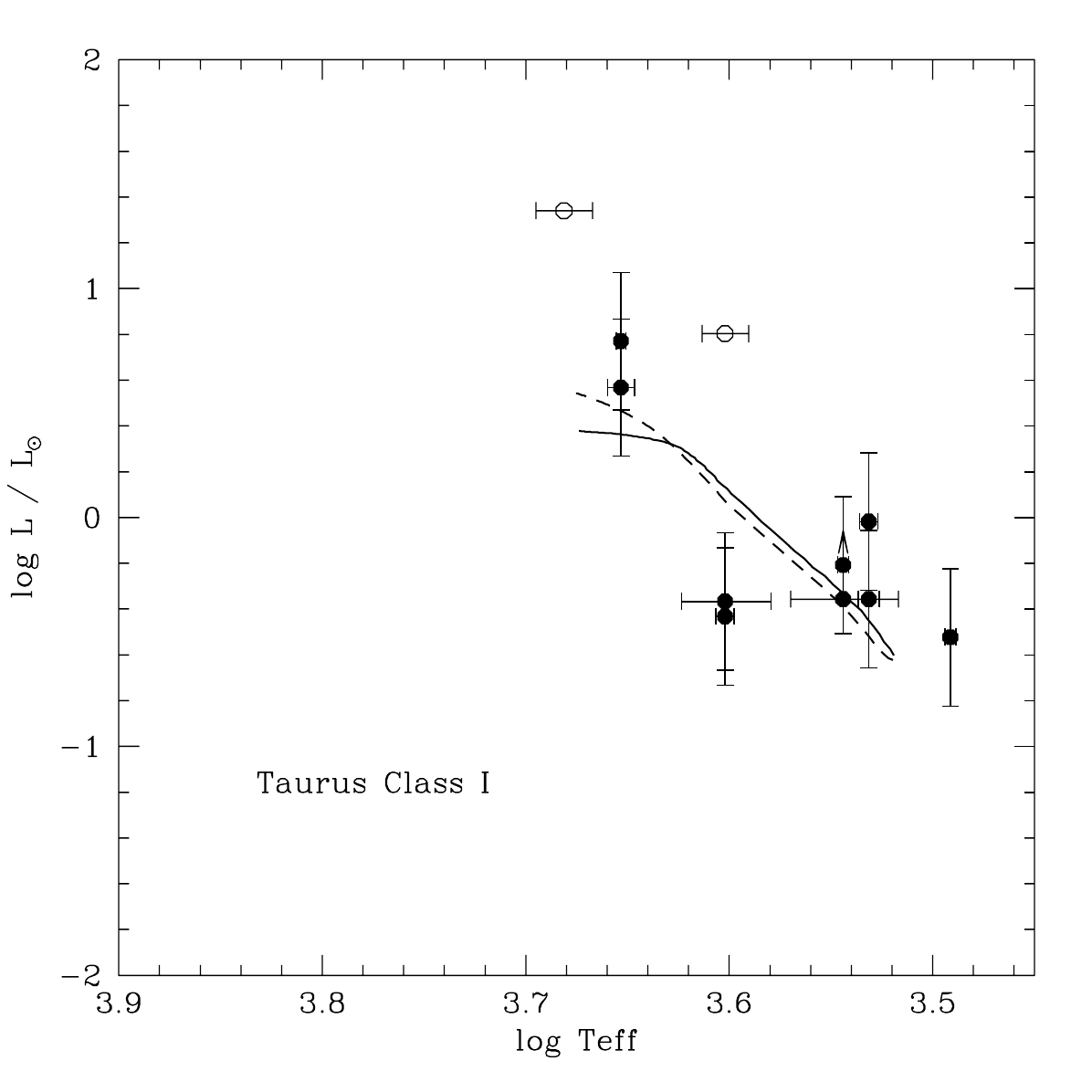}
\includegraphics[width=0.45\textwidth]{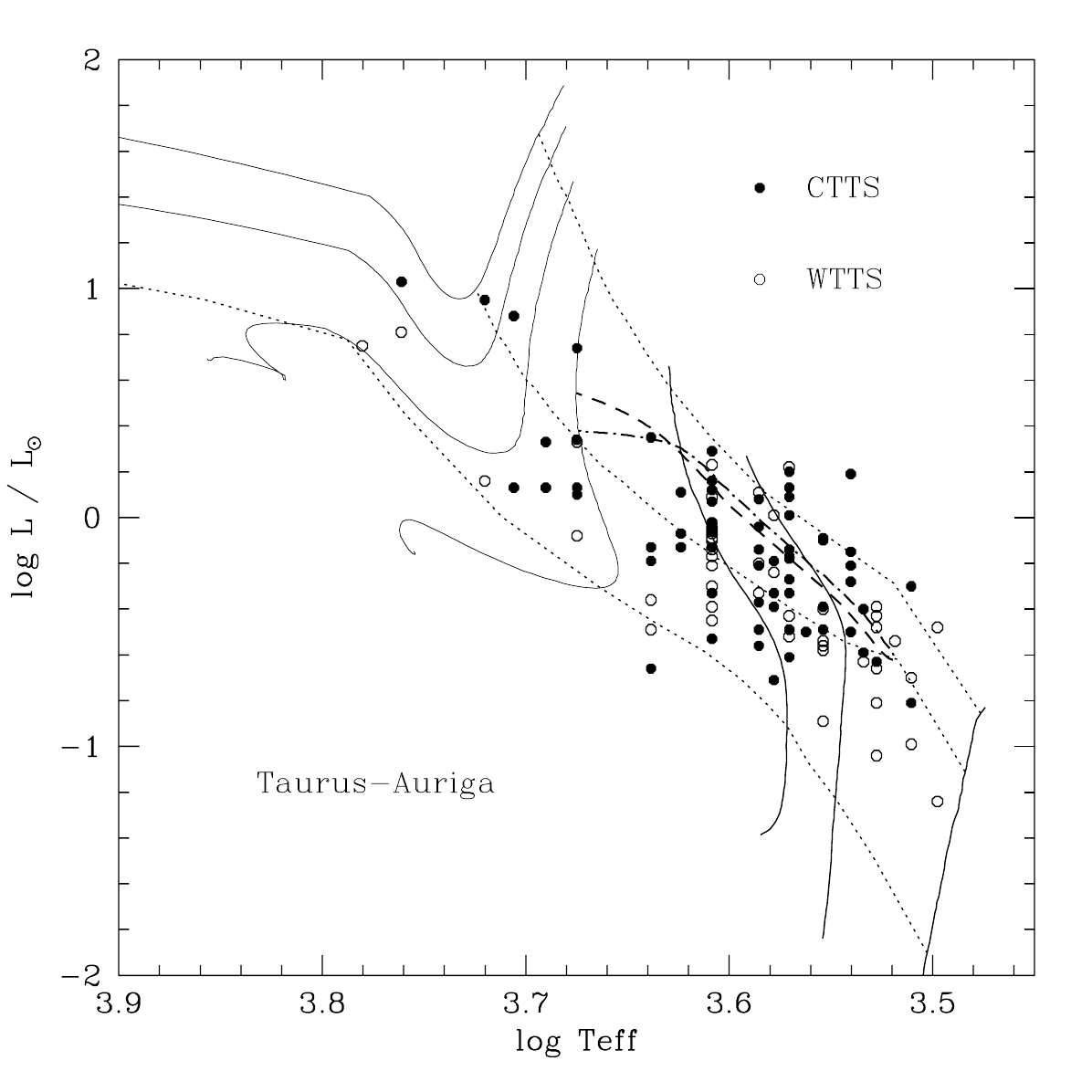}
\caption[HR diagram:  Taurus protostars  and
young (T\,Tauri) stars.]{Hertzsprung-Russell diagram positions of Taurus protostars ({\em left}) and
young ([pre-main sequence] T\,Tauri) stars ({\em right}).  These plots of two observed quantities --~the stellar luminosity $L$ (in solar units) and the effective temperature
$T_{\rm eff}$~-- can be used directly to infer the stellar radius $R$ via the equation
$L = 4\pi R^2 \sigma T_{eff}^4$, and indirectly the stellar mass via evolutionary
tracks.
{\em Left:} Solid and dashed curves correspond to theoretical estimates of initial
protostellar radii ('birth lines')
as a function of effective temperature (which corresponds
roughly to mass).  The open circles denote objects in which most of the
luminosity derives from accretion, not stellar photospheric radiation.
The agreement between theory and observation is reasonably satisfactory given the
uncertainties, showing that low-mass protostars do indeed begin their existence
with radii only a few times larger than that of the Sun's.
{\em Right:} Standard stellar evolutionary tracks
compared with observed HR diagram positions of T\,Tauri stars in the Taurus--Auriga
star-forming region.  The dashed lines show approximate isochrones for
1\,Myr and 10\,Myr, assuming
contraction from very large radii, along with the birth lines of
the left-hand panel.  Ages of young solar-type stars are thus determined by
the amount they have descended in the HR diagram from the birth line, due to
gravitational contraction.
[Fig.~III:3.8; \href{https://ui.adsabs.harvard.edu/abs/2009apsf.book.....H/abstract}{source: \citet{2009apsf.book.....H}}.]}
\label{figlh:fighr}
\end{figure}\nocite{Hartmann2009}\indexit{star!formation!birth line}

\subsection{Properties of young stars}\label{lh:4}

\ors[III:3.4] ``Solar-type \indexit{star!young star properties}stars begin their lives with only modestly-larger radii than [in the state into which they settle as 'mature' stars (referred to as the \indexit{main sequence}'main sequence' phase; [{\em e.g.,} Fig.~\ref{fig:acthrd}]).  This is a consequence of (a) the need to have a significant gas opacity to trap thermal energy, and thus produce enough pressure to halt collapse, and (b) the fact that most of the energy of accretion is radiated outward rather than being trapped.  Item (b) is ensured in general by the very high opacity of the protostar compared with the infalling material, and in particular by the angular momentum of the protostellar core, which makes much (most) of the material land first on the disk rather than onto the central star.

In the absence\indexit{Kelvin-Helmholtz contraction time scale} of\indexit{contraction time scale!Kelvin-Helmholtz} energy input, [so prior to the initiation of nuclear fusion, the star-to-be] contracts on the
Kelvin--Helmholtz time scale
\begin{equation}
\tau_{\rm KH} ~=~ {3 \over 7} { G M_*^2 \over R_* L_*}\, \label{eqlh:tkh}
\end{equation}
where $R_*$ is the protostellar radius and $L_*$ its luminosity.  This is basically the ratio of the internal energy divided by the rate at which energy is being lost, with the numerical coefficient set in this case by the assumption that the star is completely convective.\activity{{\em Show:} The internal energy of the star in Eq.~(\ref{eqlh:tkh}) is derived from the so-called 'virial theorem' which states that the total gravitational energy $E_{\rm grav}$ is related to the total thermal energy $E_{\rm thermal}$ as $E_{\rm grav}=-2 E_{\rm thermal}$ if $\gamma = 5/3$ as for a monoatomic ideal gas. Derive this from Eq.~(\ref{momentum}) assuming a field-free stationary state for a spherically symmetric ball of gas: ${\rm d}p/{\rm d}r = -GM(r)\rho/r^2$. (a) One way to do so is to multiply both sides by $4\pi r^3$, integrate (in part 'by parts') from center to surface (where $p(R)$ essentially vanishes, and realizing that the internal energy per unit volume of the gas is given by $u=p/(\gamma-1)$ for an adiabatic exponent $\gamma$. The result is equivalent to the virial theorem. Eq.~(\ref{eqlh:tkh}) can be used for the present-day Sun to show that continued gravitational contraction cannot support the solar energy budget over the age estimated for the Earth based on radio-nuclide dating (note a factor of two difference between thermal and gravitational time scales). (b) What is the present-day value of $\tau_{\rm KH}$ in Eq.~(\ref{eqlh:tkh}) for the Sun?}  More detailed calculations indicate that during protostellar accretion, the protostellar luminosity and radius have roughly those values which would yield a Kelvin-Helmholtz contraction time of the same order as the timescale for infall.  In the case of the protostellar cloud described above, this timescale is $\sim R /c_{\rm s}$, or a few times $10^5$\,yr.
%Virial theorem: http://www-astro.physics.ox.ac.uk/~podsi/stars_summary.pdf

For low-mass\indexit{protostar!deuterium fusion} protostars, fusion of\indexit{deuterium fusion} deuterium can play an important role in stopping protostellar contraction at early times.  Deuterium fusion occurs at a significant rate when the central temperature reaches $\sim 10^6$\,K; this results when $R_*/M_* \sim 5 \rsun/\msun$ for a completely convective star.  However, as D has a very low abundance, its fusion represents a significant energy source for only a modest time at low masses and very short times for higher-mass, higher-luminosity objects.  The result is that stars of masses $\simless 0.5 \msun$ may be detected initially near the D main sequence in the \indexit{Hertzsprung-Russell diagram}Hertzsprung-Russell (HR) diagram (see Sect.~\ref{sec:evolstruc} and Figs.~\ref{fig:acthrd} and~\ref{figlh:fighr}), but the youngest\indexit{star!formation!birth line} higher-mass objects will be found below this 'birth line').  After D is exhausted, the solar-type star will then undergo Kelvin-Helmholtz contraction until it reaches the main sequence, as shown in Figure\,\ref{figlh:fighr}. \activity{{\em Show:} Draw lines of equal radius (as multiples of the solar value) in Fig.~\ref{figlh:fighr}, using $\log(T_{\rm eff,\odot})=3.762$.}

Stellar ages\indexit{star!age estimate} for very young stars
are estimated from Kelvin-Helmholtz contraction timescales.  The
accuracy of these estimates depends mainly on uncertainties in two quantities:
the stellar mass and the 'starting' radius for KH contraction
(left-hand panel of Figure\,\ref{figlh:fighr}).
Masses are mostly estimated from theoretical evolutionary tracks, though progress
is being made in calibrating these from binary orbits and disk rotation; currently
there are significant uncertainties for the lowest-mass stars.  For higher masses,
calibrations are better but the starting radius or birth-line position is uncertain,
as it depends upon the precise thermal content of accreted matter
rather than on the occurrence of D fusion (see Figure~\ref{figlh:fighr}).
For solar mass stars, the upshot is that ages are uncertain by a factor of two
or more for Kelvin-Helmholtz estimates at $\sim 1$\,Myr, and perhaps 30\%\ at 10\,Myr.

Stellar magnetic\indexit{star!magnetic field} fields and activity are important for understanding the angular momentum 'problems' [\ldots] In brief, large areas of the photospheres of very young stars are covered with strong magnetic fields, with $B \sim 2$~kG and covering (or filling) factors of tens of percent.  Polar dark spots seem to be typical, though there are significant spots at other latitudes, and the spot areas/fields are not axisymmetric --~explaining why there is often substantial rotational modulation of the optical/near-IR stellar photospheric emission. [\ldots] The variability of the rotationally-modulated starspot-produced light curves --~on timescales of days, weeks, months, years~-- indicates that the fields are not fossil in origin but are produced by some sort of stellar dynamo.  [\ldots] The large-scale (dipolar) magnetic field strengths of these stars are important in understanding the interface between the accretion disk and the stellar photosphere. [\ldots\ W]hile Zeeman {\em broadening} clearly demonstrates the existence of 2\,kG photospheric fields over substantial areas of the star, the low measurements or upper limits of {\em polarization} suggest that there must be substantial [polarity] reversals to cancel out the net polarization; this would seem to indicate that the fields are of higher order than dipole, and thus that the large-scale (dipolar) component may be relatively weak, [although it appears that there are] non-negligible large scale fields nonetheless. \seeactivity{act:zeeman}
%\{$\mathcal{A}$:\ref{act:zeeman}\}

An important\indexit{stellar!magnetosphere} consequence\indexit{magnetosphere!stellar} of the large magnetic fields of pre-main sequence stars is that the stellar magnetospheric pressure and torques truncate the disk accretion disks well above the stellar photosphere [(as sketched in Figs.~\ref{fig:protoplanetary_disk} and~\ref{figlh:magnetoscheme})].  Magnetospheres are certainly present, given the strong fields found empirically.  Moreover, it is clear from observations that [young, still fully convective, pre-main sequence] T\,Tauri stars accrete through their magnetospheres.  The high H$\alpha$ emission and the strongly Doppler-broadened H$\alpha$ emission line profiles of accreting T\,Tauri stars are convincingly explained by some type of quasi-radial infall; this implies that the rapid rotation and slow radial drift of accreting material in the disk must be disrupted, most plausibly by the stellar magnetosphere (Figure\,\ref{figlh:magnetoscheme}).  The magnitude of the observed velocity line widths can be explained only if the stellar magnetic field is strong enough to truncate the disk at least a few stellar radii above the photosphere, allowing the essentially freely-infalling gas to develop a large gravitationally-produced velocity.

\begin{figure}[t]
\centering
\includegraphics[width=0.7\textwidth]{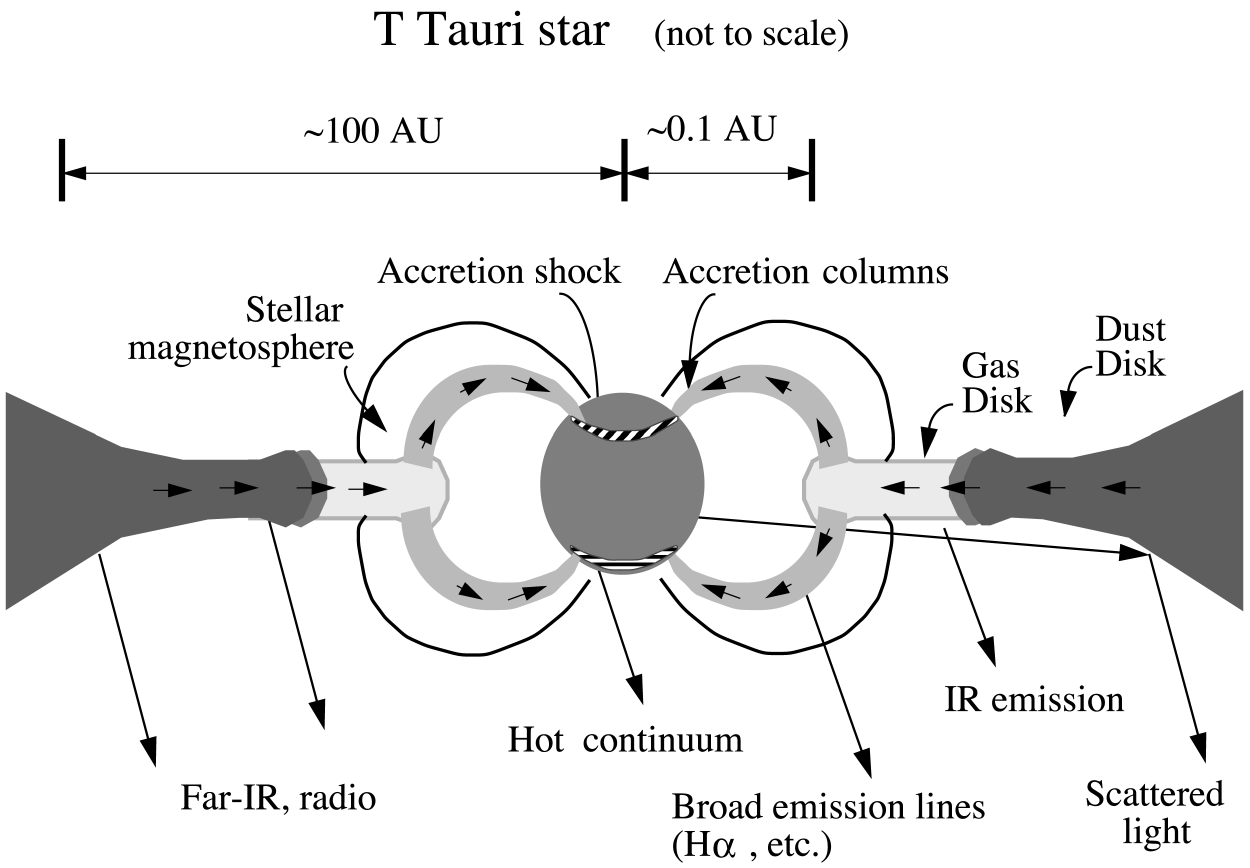}
\caption[Magnetosphere-disk interaction in low-mass, pre-main
sequence stars.]{Schematic representation of magnetosphere-disk interaction in low-mass, pre-main
sequence (T\,Tauri) stars, with diagnostics of specific regions labeled.
[Compare this to the discussion of the 'open magnetosphere' of planets in Sect.~\ref{open} and the effects of corotation and its failure in Sect.~\ref{corot}. Fig.~III:3.10; \href{https://ui.adsabs.harvard.edu/abs/2009apsf.book.....H/abstract}{source: \citet{2009apsf.book.....H}}.]}
\label{figlh:magnetoscheme}
\end{figure}
In addition to broad emission lines, accreting T\,Tauri stars exhibit significant amounts of excess\indexit{T\,Tauri star!excess continuum} continuum emission at wavelengths running from the far-ultraviolet through the optical region.  This ultraviolet-optical continuum emission is most plausibly explained as radiation produced in the accretion shock at the base of the magnetosphere, where the material in near-freefall comes to rest at the stellar photosphere.  As described in the previous paragraph, it appears that the disk must be magnetospherically truncated at a few stellar radii above the photosphere; this implies that most of the energy generated by accretion will be radiated in this accretion shock.  Estimates of mass accretion rates $\dot{M}$ for \indexit{T\,Tauri star!accretion rate}T\,Tauri stars are thus generally based on setting the UV-optical emission excess luminosity $L_{\rm acc} \sim G M_* \dot{M}/R_*$.''

\subsection{The rotation rate of very young stars}\label{sec:lh5}

\ors[III:3.5] ``One\indexit{star!formation!angular-momentum problem} of the most striking problems of angular momentum transport is that very slowly-rotating low-mass stars are produced by accretion from rapidly-rotating disks.  In general, T\,Tauri stars of masses $\simless 1 \msun$ rotate at rates from a few tens of percent to less than ten percent of their breakup values. The problem of producing slowly rotating stars somewhat older is made much more difficult by the apparent requirement of spinning down the star at the same time it is accreting high-angular-momentum material.  \activity{{\em Show:} (a) Derive the expression for the breakup rotation rate of stars as function of mass and radius. (b) What is the value for the Sun? Ignore distortion from spherical symmetry for this estimate.}  Of course, if magnetic stellar winds were intrinsically powerful enough to spin down stars rapidly, there is no problem; but spindown does not seem to be extremely rapid in non-accreting stars, at least not on timescales needed to explain the slow rotation in stars of ages $\sim 1$\,Myr.

One possible option is that the magnetospheric coupling between
the star and its disk transfers the angular momentum outwards at the necessary rate.
However, there are difficulties with applying this model.
In the first place {\em accretion}, which is observed
in essentially all T\,Tauri stars with detectable inner disks, basically requires
magnetic field lines tied to disk material inside of corotation; this spins
down the gas so that it can accrete, spinning up the star.
Spindown of the accreting star requires magnetic fields
connected to the disk outside of corotation; thus, to explain T\,Tauri stars
one would like one set of stellar magnetic field lines to be connected inside
of corotation, and another outside of corotation, and somehow balance the
angular momentum addition due to the accretion with coupling to the outer disk.
Numerical simulations indicate that a quasi-steady state [with both types of connections] may be possible with a large enough turbulent diffusivity, but whether such diffusivities are realistic is unknown. [However, some] estimates of inner gas-disk
radii are significantly inside of corotation, raising the question as to whether
there is a strong enough large-scale
magnetic field to effectively couple to the outer disk for spindown.

T\,Tauri\indexit{magnetosphere!stellar, T\,Tauri} magnetospheres\indexit{stellar!magnetosphere!T\,Tauri} are probably best thought of
as a series of individual magnetic loops, not all of which are filled with
accreting gas; this makes it
easier to explain the very small covering factors of the hot (shocked) continuum regions
on the stellar photosphere of order $\simless$~1\%.
As at least some of the loops (if not most) must connect to the disk interior to
corotation, it is almost certainly the case that magnetic field lines must tend to
become twisted.  Such twists rapidly lead to a 'ballooning out' of closed field lines, with
eventual opening up of field lines and possible ejection of mass, with reconnection following. [\ldots]'' Such processes would appear to make the angular momentum transfer from star to disk even less efficient, although processes related to winds and waves complicate the modeling and our understanding of how things work (as discussed in Ch.~III:3.5). It is also possible that heating of part of gas coming into the stellar magnetosphere enables a hot stellar wind from within the star's magnetosphere, which could lead to efficient magnetic braking. Clearly, for now, the loss of angular momentum from the material accreting onto the protostar remains an area of study. 

\subsection{Protoplanetary disks and gravity}\label{sec:gi}

\ors[III:3.6] ``The\indexit{instability!magneto-rotational} mechanisms of angular momentum transport determine the \indexit{disk!protoplanetary}mass distribution
within the protoplanetary disk.  It is important to understand whether
gravitational \indexit{instability!gravitational}instabilities dominate this transport, in which case accretion
onto the central star is likely to decay away with time, leaving a relatively
massive disk behind; or whether another mechanism not tied to gravity can
reduce disk mass distributions leading to the epoch of planet formation.

The one non-gravitational mechanism of angular momentum transport that we currently understand (at some level) is the \indexit{instability!magneto-rotational}magneto-rotational instability (MRI; Sect.~\ref{sec:magrot}).  \indexit{MRI|see{instability}}It is possible that the upper layers of the otherwise cold disk can be non-thermally ionized by stellar X-rays [(as suggested in Fig.~\ref{fig:protoplanetary_disk})] and cosmic rays [entering from outside the system], to the extent that a significant amount of mass and angular momentum transport can occur.  If large amounts of the disk can be activated magnetically in this way, then the disk can behave essentially as a standard viscous disk, with most of the mass at large radii.  However, X-ray and cosmic ray ionization are insufficient if small dust grains, which can absorb ions and electrons very efficiently, are not heavily depleted. [\ldots\ Whereas {\em Spitzer} {\em IRS} ({\em Infra-Red Spectrograph}) spectra suggest levels of depletion of $10^{-2}$ to $10^{-3}$ from interstellar medium values of small dust, it appears that] depletions of order $10^{-4}$ are needed for the MRI to operate robustly in upper disk layers.

As discussed earlier, it is plausible if not likely that protostellar disks are initially gravitationally
unstable, given\indexit{gravitational!instability}
the need to accrete most of the mass of the central star through the disk
and likely limited MRI transport in cold disks.
If the MRI is inefficient, the disk could settle into a state of
marginal gravitational \indexit{Toomre parameter}instability, with\indexit{definition!Toomre parameter} the Toomre\indexit{Toomre parameter|seealso{definition}} parameter
\begin{equation}
Q = {c_{\rm s} \Omega_{\rm e} \over \pi G \Sigma} \sim 1.4\,
\label{eqlh:toomre}
\end{equation}
where $\Omega$ is the Keplerian (presumed to be the epicyclic) angular frequency and $\Sigma$ is the disk surface mass density [(the epicyclic frequency is the frequency at which a radially displaced parcel oscillates within the disk)].  The $Q$ parameter basically results from satisfying two conditions: one, that gravity can overcome resisting gas pressure forces; and two, that gravity is stronger than the effects of angular momentum in opposing collapse.  Larger values of $Q$ mean that the disk is gravitationally-stable, while smaller values of $Q$ indicate strong instability.  In many instances disks tend to self-regulate; strong instabilities tend to produce heating via shocks which raise $c_{\rm s}$ and thus increase $Q$, until the sound speed rises sufficiently that the instabilities heating the gas begin to decay.

\begin{figure}[t]
\centering \includegraphics[width=0.65\textwidth]{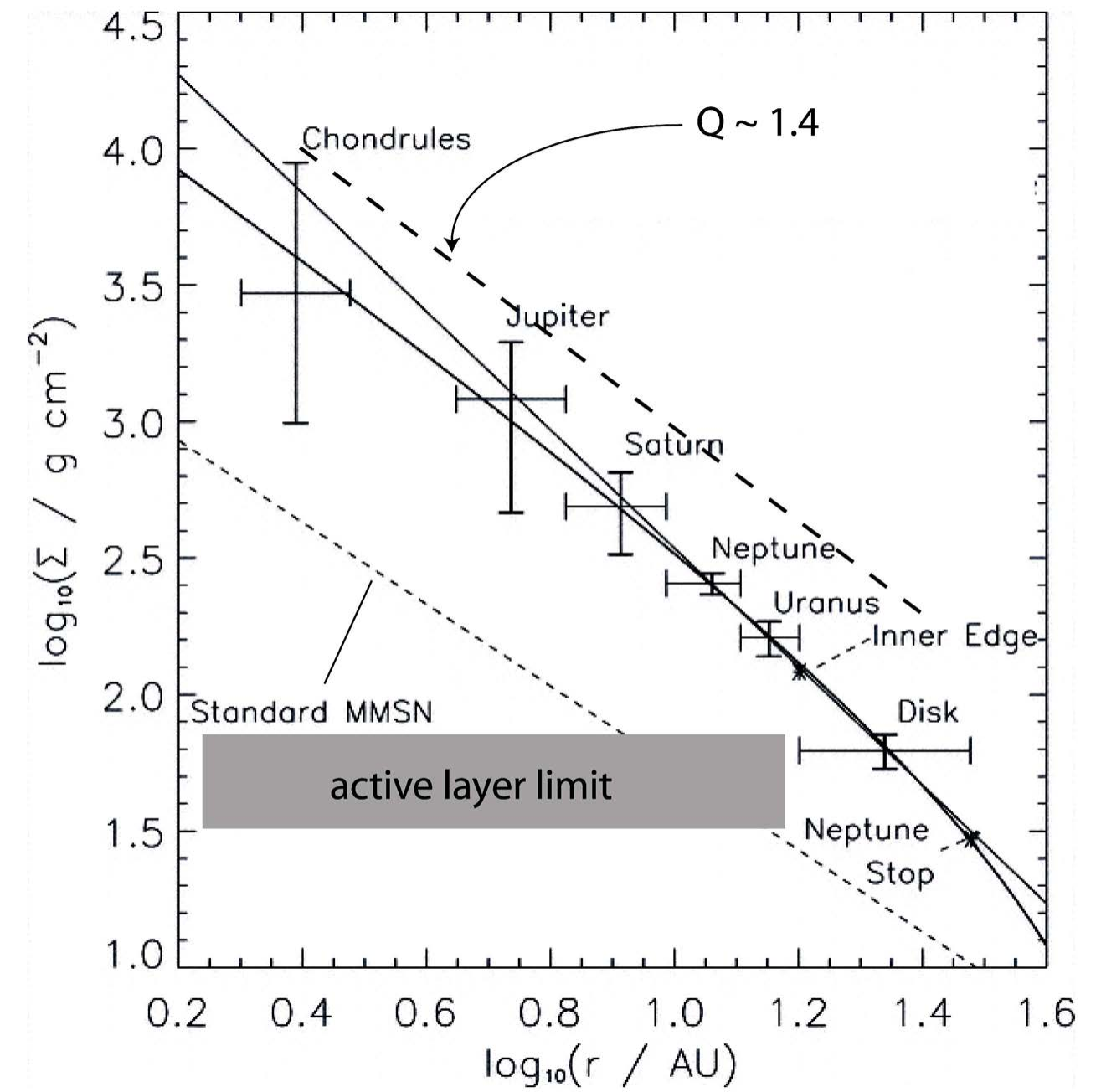} \caption[Estimates of the minimum mass solar nebula.]{Two estimates of the minimum mass solar nebula.  The lower, light-dashed curve indicates the usual estimate, derived from the current position of the giant planets and accounting for the missing light elements; the solid curves show a higher estimate based on the initial positions of the giant planets assumed in a model which has substantial outward migration of the giant planets.  Limits on the expected MRI-active surface density due to non-thermal ionization and on the surface density expected for a marginally gravitationally-unstable disk (the dashed line showing the condition for the critical value of the 'Toomre $Q$' parameter --~see
%Eq.~(11.2) 
Eq.~(\ref{eqlh:toomre})~-- are also
shown. [Fig.~III:3.12; \href{https://ui.adsabs.harvard.edu/abs/2007ApJ...671..878D/abstract}{source: \citet{2007ApJ...671..878D}}.]}
\label{figlh:desch}
\end{figure}\nocite{2007ApJ...671..878D}
Even if the MRI is reasonably well activated by non-thermal ionization,
it may easily be insufficient over the $1-10$\,AU region to transport all the mass
viscously; this could result in the general
picture in which a 'magnetically dead'
zone of the disk is sandwiched radially by MRI-active regions at small and large radii.

To develop this further, consider estimates of the mass distribution of the solar nebula.  [Figure\,\ref{figlh:desch} compares two different estimates of the \indexit{minimum mass solar nebula} so-called 'minimum mass solar nebula' (MMSN), one using the so-called 'Nice' model (falling not far below the gravitational instability result with Eq.~\ref{eqlh:toomre}), which\indexit{planet!formation!'Nice' model} posits substantial inward migration of Jupiter and Saturn and outward migration\indexit{Nice model!planet formation} of the Uranus and Neptune from their original positions, the other an older version based on the current positions of the giant planets. Both estimates lie] above the maximum $\Sigma \sim 100\, {\rm g\, cm^{-2}}$ estimated for non-thermal ionization by cosmic rays in the most optimistic scenario.  While either version of the MMSN must be considered uncertain, the possibility that the solar nebula had a '[magnetically] dead zone' must clearly be considered.  [\ldots] 

The consequence\indexit{gravitational!instability} of a disk structure with a 'dead zone', as described in the previous paragraph, may be highly time-variable accretion during the protostellar phase.  [The gravitational instability, GI] can be relatively efficient in transferring mass inward at large disk radii but tends to become inefficient at small radii; conversely, the MRI becomes increasingly important at small radii, especially at high mass accretion rates.  If matter moving inward under GI dissipates enough energy locally in the inner disk, it can 'turn on' the MRI thermally, resulting in an onrush of mass onto the central star.  This picture has been invoked to explain the FU Orionis outbursts [during which] of order $10^{-2} \msun$ gets dumped onto the central low-mass star over timescales $\sim 10^2$\,yr.  It is difficult to explain the FU Ori outbursts without having a large amount of disk mass at a few AU, well above that of the standard MMSN.

The possibility of gravitational instability [makes one reconsider] the possibility of forming giant planets directly
through gravitational fragmentation [rather than by the core-accretion scenario described near the top of Sect.~\ref{sec:exo}].
This suggestion runs into difficulty, however, because a low $Q$ is not enough;
the disk must be able to cool on something like an orbital period $P_{\rm orb}$ to continue fragmenting;
otherwise perturbations shear
out and transport angular momentum instead.
This poses a problem for protostellar disks because they are so cold, and thus do
not cool rapidly.
The cooling timescale $t_{\rm c}$ for an optically-thick disk [\ldots]
%\begin{equation}
%t_{\rm c} ~\sim ~ {\Sigma c_{\rm s}^2 \over \gamma - 1 } \, {4 \tau_{\rm R} \over 3 %\sigma T_{\rm d}^4}\,
%\end{equation}
%where $\tau_{\rm R} = \Sigma k_{\rm R}/2$ is the vertical Rosseland mean optical depth; this
is basically
the energy content divided by the blackbody radiation loss.
Numerically, for temperatures below 170\,K, one finds
\begin{equation}
{t_{\rm c} \over P_{\rm orb}} ~\sim~ 10^4 \left({M\over\msun}\right)^{3/2} R_{10}^{-9/4}\,
\end{equation}
(with $R_{10}$ is a characteristic scale in units of 10\,AU), which poses an obvious difficulty for fragmentation in that the cooling time far exceeds the Keplerian period.  (Things change on distance scales $\sim 100$~AU or larger, because the disk typically becomes optically thin, and thus cools much more rapidly than indicated by the above equation.)

Even if fragmentation could occur after infall ceases, one would still expect it to
be more important early on, when the disk is more massive.  It is not obvious
how initial gravitational instability would explain the observed clearing of
disks over millions of years.''

\begin{figure}[t]                                                                                                         
%\centerline{\psfig{figure=figures/disk_lifetimes.eps,width=9cm}}   
\centerline{\includegraphics[width=9cm]{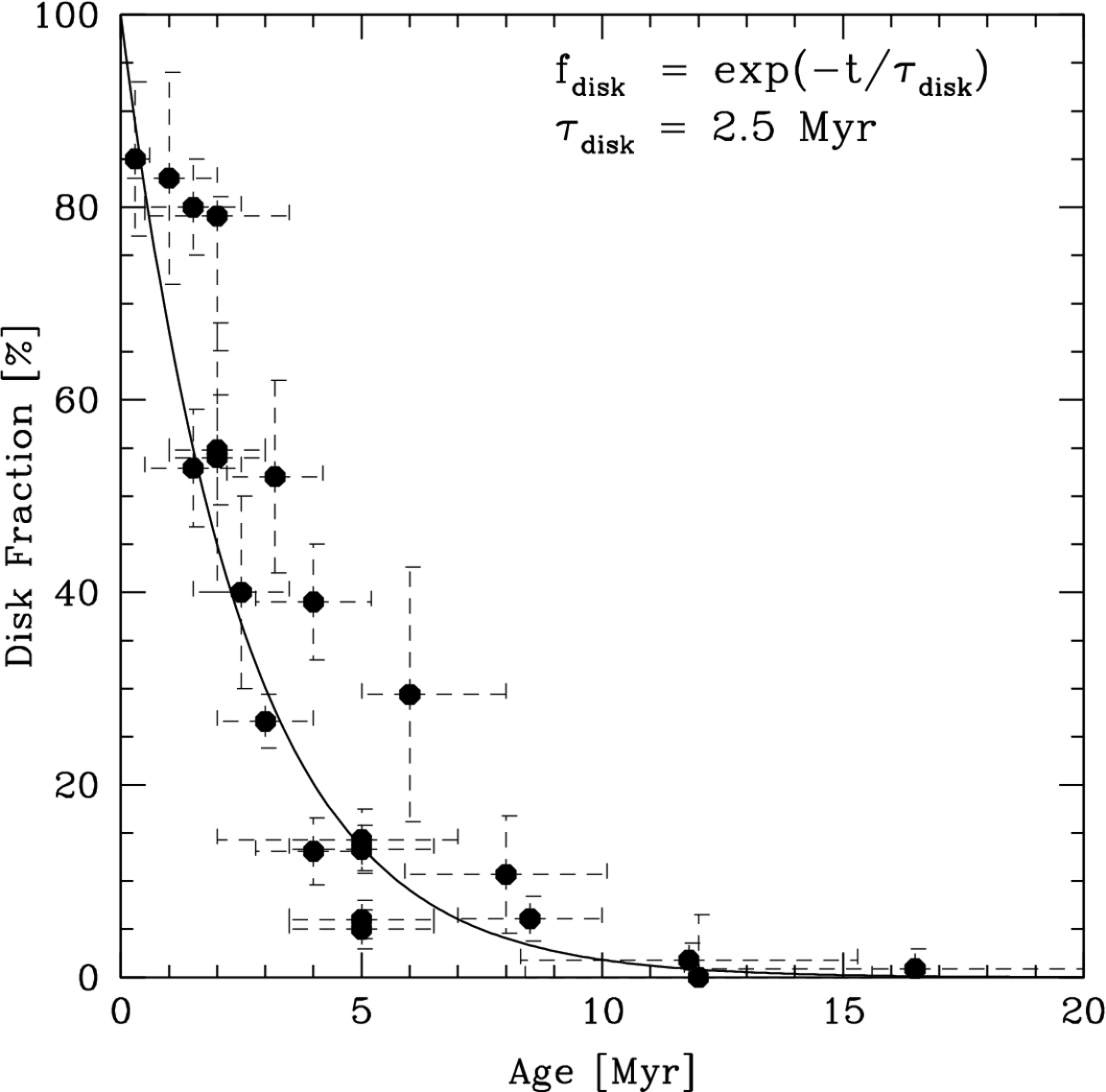}}   
\caption[Primordial disk fractions of stars in young clusters.]{Primordial disk fractions of stars in young clusters. 
These observations show that the dust disks only last for a few million years.  [Fig.~IV:5.10; \href{https://ui.adsabs.harvard.edu/abs/2009AIPC.1158....3M/abstract}{source: \citet{2009AIPC.1158....3M}}. Note that since this figure was made the ages for very young stars systematically increased by $\sim 50$-100\%\ for many of these young samples (see, {\em e.g.,} work \href{https://ui.adsabs.harvard.edu/abs/2013MNRAS.434..806B/abstract}{by \citet{2013MNRAS.434..806B}}, \href{https://ui.adsabs.harvard.edu/abs/2015MNRAS.454..593B/abstract}{by \citet{2015MNRAS.454..593B}}; \href{https://ui.adsabs.harvard.edu/abs/2012ApJ...746..154P/abstract}{by \citet{2012ApJ...746..154P}}, \href{https://ui.adsabs.harvard.edu/abs/2016MNRAS.461..794P/abstract}{and by \citet{2016MNRAS.461..794P}}.), so that the e-folding timescale for primordial disk loss is something like 4-5\,Myr rather than the $\sim$2\,Myr shown in the figure (Eric Mamajek, private communication).]                                                                                                                                              
\label{fig:disk_lifetimes} \label{figlh:disk_evol} } \figindex{../fischer/art/disk_lifetimes.eps}\end{figure}

%\begin{figure}[t]
%\centering
%\includegraphics[angle=0,width=0.6\textwidth]{figures/jesus.eps}
%\caption[Fraction of stars with near-infrared disk emission as a function
%of the age of the stellar group.]{Fraction of stars with near-infrared disk emission as a function of the age of the stellar group.  Open circles represent the disk frequency for stars in the T\,Tauri mass range, derived using observations out to 2-3 $\mu$m; solid symbols represent the disk frequency as measured to $8 \mu$m or beyond.  [Fig.~III:3.13]} \label{figlh:disk_evol} \end{figure}\nocite{2007ApJ...671.1784H} \nocite{Hartmann2009}

\subsection{Dust-disk evolution}\label{lh:7}

\ors[III:3.7] ``In the\indexit{planet!formation!dust-disk evolution} core\indexit{dust!disk} accretion model \indexit{disk!dust disk evolution}for the formation of giant planets, and in all models
of terrestrial planet formation, dust\indexit{dust} grains grow from sub-micron sizes to thousands
of km.  A starting point for thinking about how planets grow from disks is then
considering observations of the evolution of disk dust, detected through its
emission.

Figure~\ref{figlh:disk_evol} shows the estimated fractions of young stars in
various groups with large dust-disk excesses as a function
of age.  [\ldots] The overall result is that optically-thick dust
disks (with the opacity probably dominated by particles of $\mu$m size or a bit
less) disappear on timescales of a few Myr.  While less is known about the
presence of gas in the inner 10--20\,AU, clearing of small dust particles
seems generally accompanied by removal or disappearance of gas as well.
It is important to emphasize that there is no single timescale for disk clearing.
Some (inner) disks disappear immediately, perhaps because of disk disruption by
a binary companion; others take a few Myr; a small percentage last for 10\,Myr.

The disk can 'disappear' in one of three ways;\indexit{dust!disk} mass can be accreted,
ejected, or condensed into large bodies.  It is difficult to accrete all
the mass of the disk, as some must be left behind to take up the angular
momentum; the outer disk is likely to expand over time and evolve on continually
slower timescales.  Evaporation of the disk may be
important, though it is thought to take place over longer timescales than
this (Sect.~\ref{lh:8}).  Perhaps the strongest evidence for coagulation into larger
bodies is the detection, either through spectral energy distribution fitting
and/or imaging, of systems with substantive outer disks but inner disk holes
or gaps.  This is consistent with the idea that
settling, grain growth, coagulation, and formation of large solid bodies
occurs fastest in the inner disk, where the surface densities are largest.

Dust\indexit{dust!grain evolution} grains in the disk generally are thought to evolve to larger sizes, with a decreasing population of small grains with increasing age.  During this overall growth, dust is expected to settle vertically and drift radially.  \activity{{\em Consider:} The key mechanism by which dust is expected to settle into the center of an accretion disk is hydrodynamic drag. Explain how this works. Consider orbital inclination and effects of gas pressure, gravity, and stratification.} This evolution of dust in size and position in the disk can reduce and ultimately eliminate infrared excess emission, consistent with the observed disappearance of dusty disk emission over millions of years (Figure~\ref{figlh:disk_evol}).  In principle, dust growth can be extremely rapid: [the disk interior to about 10\,AU may become] optically thin on timescales of 0.1 to 1\,Myr, as dust particles settle and coalesce into larger bodies.  The evolution of the mm fluxes is slower because of longer timescales of accumulation in the outer disk, with substantial reductions in mm-wave emission on timescales of 10\,Myr.  One might expect that turbulence would lengthen settling and growth timescales, but [it may actually stimulate] growth due to turbulent mixing.  [\ldots]

As particles grow in size, many effects converge to make evolution uncertain.  For example, the difference in velocities between objects within an order of magnitude of meter size can result in their complete shattering or disruption.  Turbulent eddies or whirlpools might help collect these objects at low velocities so that they can accrete, or alternatively disperse them more widely.

Can core accretion proceed fast enough to explain the observed disk clearing on timescales as short as [several] Myr?  One problem is the formation of km-sized planetesimals\indexit{planetesimal} from cm-sized objects.  Such bodies are thought to be held together lightly --~too large for effective sticking and too small for gravity to become important~-- and, as bodies of differing sizes have differing velocities due to gas drag, collisions between these objects might shatter them rather than build them up.  Another problem is that the so-called Type\,I inward migration due to torques between the disk and the body is very rapid, making it important to grow quickly at $\sim$~1 Earth mass to avoid falling into the central star on a timescale $<1$~Myr.  These estimates have usually been made in the 'minimum' MMSN (Figure~\ref{figlh:desch}); the timescale for inward migration is inversely proportional to the surface density, so gravitationally-unstable disks may pose even bigger problems in this regard.

Once km-sized planetesimals are made, collisions among them can lead to the building of terrestrial planets and giant planet cores.  The remaining bottleneck is that of accumulating gas which depends upon the opacity; larger opacities make it difficult for the growing planet to lose energy and allow additional material to be accreted.  A reduction in opacity due to grain growth and depletion would help considerably in this regard.''

\subsection{Disk evaporation}\label{lh:8}

\ors[III:3.8] ``As the planets\indexit{planet!formation!disk evaporation} are overabundant in heavy \indexit{disk!evaporation}elements relative to the Sun, it is
clear that most of the original gas in the solar nebula has been lost.  Of course
some of it accreted into the Sun, but it is unlikely that all of this material
was removed in this way.  For some time it was thought that a powerful solar-type
wind was responsible for gas removal from the nebula.  However, we now realize that
the strong mass loss we see is not a solar wind but a disk wind; more importantly,
the wind material is ejected perpendicular to, not into, the disk (Figure\,\ref{figlh:hh30}).

The high-energy radiation emitted by T\,Tauri stars provides a mechanism by
which the gas of the disk can be evaporated rather than accreted.  In this
case, rather than generating stellar mass loss from the star via a coronal wind,
one can generate disk mass loss from a much lower temperature wind because
the material is ejected from much farther out in the gravitational potential
field, where the escape velocity is very much smaller than at the stellar
surface.  Using the usual Parker wind formula ({\em e.g.,} Eq.~\ref{eqlh:critpoint}),
and assuming photoionization and thus heating to
a typical temperature of $\sim 10^4$K, the sonic point
occurs for
\begin{equation}
R_{\rm s} ~\sim ~{G M_* \over 2 c_{\rm s}^2} ~\sim~ 3.6 {M_* \over \msun} ~ {\rm AU} \,,
\end{equation}
where the mean molecular weight is 0.67, appropriate for a gas
of cosmic abundance with ionized hydrogen and neutral helium.
Thus, ionizing photons have the potential for removing disk gas at radii of
a few to ten AU from the central star.

To see the essential physics of the problem with a minimum of geometrical
complication, assume that a volume of $4 \pi R^3$ must be ionized, where
$R$ is a characteristic radius of escape.  This estimate is justified
because the gas must maintain its ionization over the disk to a distance
comparable to its escape radius to flow out of the gravitational potential well.
The balance between photoionization and recombination leads to
\begin{equation}\label{eq:ionbal}
\Phi_{\rm i} ~=~ 4 \pi R^3 n_{\rm e} n_{\rm p} \alpha_{\rm B}\,,
\end{equation}
where
$\Phi_{\rm i}$ is the flux of ionizing photons from the central source,
$n_e$ and $n_p$ are the electron and proton densities, respectively,
and $\alpha_B$ is the Case B recombination rate for hydrogen. \regfootnote{'Case B' recombination considers only recombinations in which the recombined electron transitions to the ground state via intermediate transitions; a direct transition to the ground state would emit a photon that could be absorbed and lead to ionization in which case no net recombination would have occurred.}
Assuming complete ionization of hydrogen, the mass loss rate is
\begin{equation}
\mdot ~\sim~ 10^{-9} \Phi_{\rm i,41}^{1/2} R_{10}^{1/2} \msunyr\,,
\end{equation}
where $\Phi_{\rm i,41}$ is the Lyman continuum photon flux in units of
$10^{41} {\rm s^{-1}}$ and $R_{10}$ is a characteristic scale of the flow
in units of 10\,AU.  This estimate illustrates the potential\indexit{disk!evolution!photo-evaporation} of photo-evaporation
to remove disk gas over evolutionarily interesting timescales.
Much more sophisticated treatments of the outflow have been considered,
but this illustrates
the basic result.  [\ldots]

Unfortunately, the true ionizing fluxes of young stars are not really known because
interstellar absorption prevents direct detection [\ldots, but there are observational results that suggest] evaporation of disks due to stellar magnetic activity occurs on
timescales of order 10\,Myr or more. Whether photo-evaporation plays
a major role in the strong disk evolution from 1--10\,Myr remains
unclear.

Disks close to a hot luminous star can be photo-evaporated rapidly
due not only to EUV (Lyman continuum) radiation but also by far-UV ($\sim 1000$\,\AA)
radiation, which can heat the gas to temperatures $\sim 1000$\,K as electrons are driven
off grains.  The FUV radiation thus can drive a wind off the outer disk, and may
be more important in many systems if most of the disk mass resides at large distances.
[\ldots]  Although the solar nebula appears to have been
'polluted' by ejection from a \indexit{supernova}supernova, it is not clear that it was close
enough to the massive star such that FUV radiation was important in evaporating
Solar System gas.'' \activity{{Look up:} For further study/reading: Most stars are born in groups of substantial numbers (often in what are called 'open clusters'). In such clusters, stars of a range of masses are formed (statistically yielding the 'initial mass function'). The heaviest among these evolve fastest, and if heavy enough can end their lives in a 'supernova'. The open cluster is eventually pulled apart by the 'galactic tides', which limits the exposure of planetary systems to nearby supernovae and to gravitational perturbation of the orbits of the planets. Look up the terms between quotation marks. The occurrence of a nearby supernova appears consistent with several properties of the solar system, including one of several possible means for the early melting of small bodies (as reflected in what are known as 'chondrules'). Look at \href{https://ui.adsabs.harvard.edu/abs/2018A&A...616A..85P/abstract}{the study} by \citet{2018A&A...616A..85P} for more on this.}

\clearpage

\chapter{{\bf Irradiance, atmospheres, and habitability}}
\label{ch:evolvingplanetary}%12
{\narrower\narrower{
{\bf Chapter topics:}
\begin{itemize}
  \customitemize
\item Terrestrial atmospheric composition and ocean coverage over time
\item Comparison of the atmospheres of the terrestrial planets
\item Solar irradiance, orbital changes, and the equilibrium temperature of planets
\item The effects of stellar winds and geological activity on
  planetary atmospheres
\end{itemize}

\noindent{\bf Key concepts:}
\begin{itemize}
  \customitemize
\item Habitable zone
\item Albedo
\item Radiative forcing
\item Atmospheric loss
\end{itemize}

}}

\section{Evolving planetary habitability}
Historical \indexit{habitability!evolving}records are too short for us to see first-hand accounts of Earth in a significantly different climatic state than the present one. There are, of course, reports on the relatively recent moderate (but nonetheless impactful) excursions from the mean climatic state, such as the \indexit{climate!Earth}Medieval Warm \indexit{Earth!climate}Period, the Little Ice Age, and the modern-day onset of global warming, but there have been much larger changes over the life of the planet. Substantial modifications of climate in the past have been attributed to the formative processes of Earth and asteroid impacts, to the evolving spectral output of the Sun and the stripping effects of the solar wind, to orbital changes in response to the gravitational pull by the giant planets, to the torque applied by the Moon, to geological and geochemical activity over eons (including the geodynamo), and --~last but by no means least~-- to the emergence and evolution of life. This chapter provides brief introductions to each of these drivers of the terrestrial atmosphere and its climate system. This provides insight into the diversity of conditions on Earth over time, while also setting the stage for appreciating the challenge of establishing the 'habitability' of planets elsewhere in the universe.

\subsection{Earth's formative phase}\label{sec:formative}
\ors[III:4.5] ``Earth's \indexit{Earth!formative phase}formation, like that that of the other solid planets, occurred by accretion of solid materials.  The processes began with particles of dust, but collision and sticking processes rapidly led to the formation of larger and larger bodies. An important aspect of the growth of rocky planets is the amount of a planet's mass that is accreted in the form of large chunks. The accretional growth process yields a number of Moon to Mars-sized 'embryos' in a given radial region of the nebula.  The final assembly of a rocky planet involves both the accretion of numerous large embryos as well as gravitational ejection of some of them to other locales [(including, as we now know through gravitational microlensing, out of a planetary system altogether and into interstellar space)].

\begin{figure}[t]
\centering
\includegraphics[width=\textwidth]{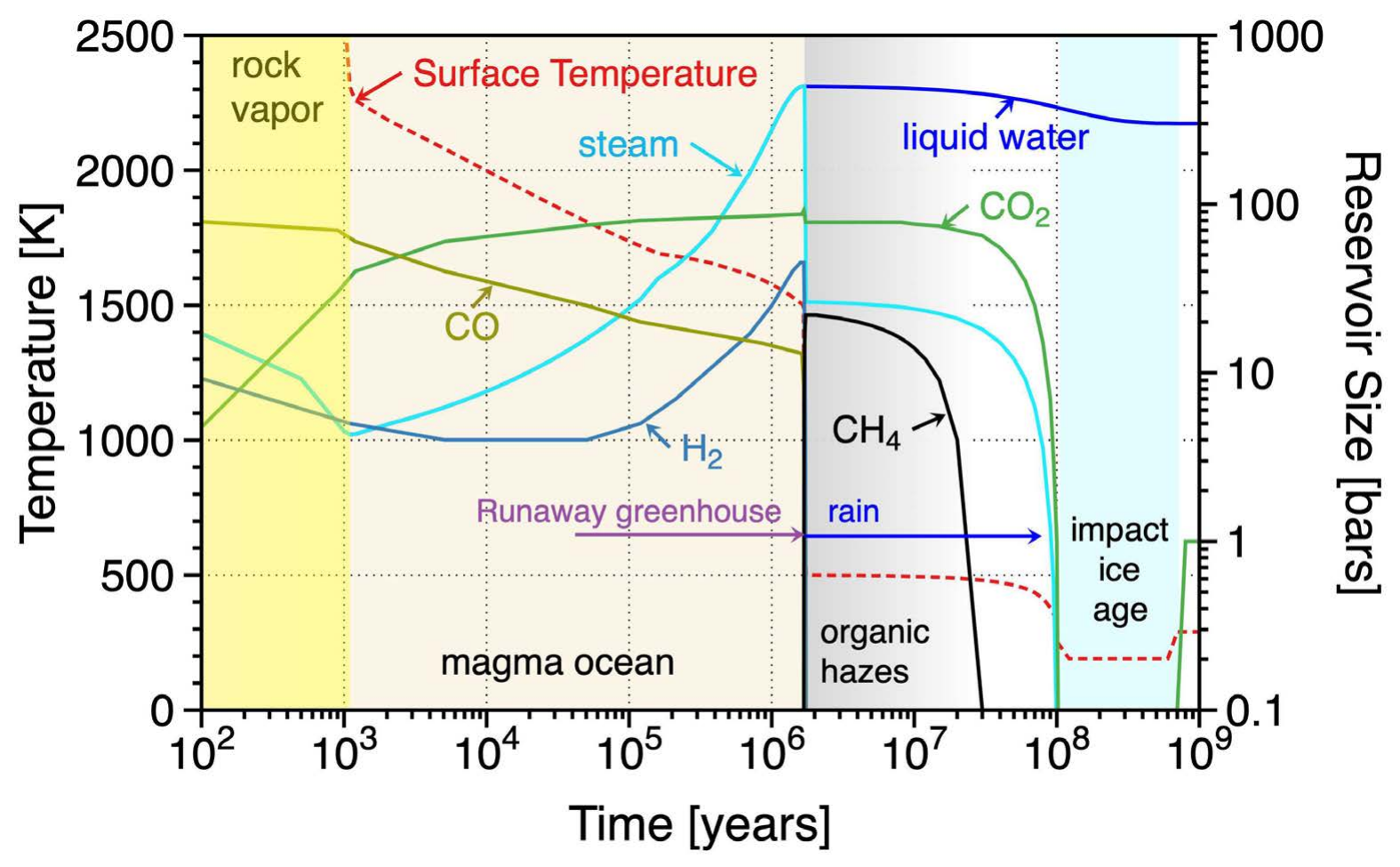}  
  \caption[Earth's $T$ and H$_2$O and CO$_2$ after Moon-forming collision.]{\label{fig:brownlee3}
The Earth's surface temperature  and above-surface reservoirs of water
and carbon dioxide after the Moon-forming collision.
% The water dip after $\sim 10^6$\,yr occurs because of storage of water in Earth's short-lived magma ocean.
The surface temperature drops below 1000\,K after a few million years when Earth's steam atmosphere condenses, and it drops below 500\,K to habitable conditions after about 100 million years when most of the atmospheric CO$_2$ is incorporated into the mantle. [This is an updated version, courtesy of Kevin Zahnle (July 2019), of the originally used Fig.~III:4.3; the latter was from this \href{https://ui.adsabs.harvard.edu/abs/2007SSRv..129...35Z/abstract}{source: \citet{2007SSRv..129...35Z}}.] \colorfig }
\end{figure}
This\indexit{Earth!pre-biotic} formation mode that includes impacts of very large bodies is indicated both by the numerical simulation of accretion processes and by evidence that our Moon formed as the result of the impact of a Mars-sized body with the growing Earth [\ldots] \activity{{\em Show:} For an impression of order-of-magnitude numbers, estimate the energy involved in a collision between an Earth-mass body and an Mars-mass body at an impact velocity of, say, 14\,km/s. Ignoring the energy going into the formation of the Moon in such a process, but rather assuming all mass and energy remain within the newly formed body, estimate the average temperature increase if all energy were distributed throughout half of the volume of the mantle, and that that material has a specific heat of approximately $10^7$\,erg/g/K.}
Following lunar formation, Earth's post-impact atmosphere of vaporized silicates may have condensed in $\sim$1000\,yr (Fig.~\ref{fig:brownlee3}).  The heat of the impact would have melted and partly vaporized Earth's mantle, but the resulting silicate magma ocean may have solidified in only a few million years.  Once the magma ocean crystallized, cooling conditions would have allowed the large amount of water vapor injected into the atmosphere to condense and thus reduce the extreme greenhouse warming of the early Earth and allow surface temperatures to drop below 1000\,K.  \activity{{\em Show:} Make an order of magnitude estimate of the cooling time of Earth's atmosphere after impact of a Mars-mass body: assume an impact velocity of 14\,km/s, that all kinetic energy remains within the near-surface layers and atmosphere; an optically  thick atmosphere of vaporized silicate; and a characteristic temperature of the radiating vapor of, say, 2000\,K. } Even with most of the water condensed, the atmosphere would still retain $\sim$100\,bars of CO$_2$ whose greenhouse warming would keep the Earth's surface temperature at $\sim$500\,K, even though the early Sun was $\sim 30$\%\ fainter that its present brightness.  The final lowering of the Earth's surface temperature to habitable conditions requires transfer of most of the atmospheric CO$_2$ to the mantle and crust, a process that can happen over a timescale of $10-100$\,Myr [\ldots\ by the process of weathering (more on that below).] 

[\ldots] Earth's oldest known rocks, whose properties could provide information about the early Earth, are just less than 3.9 billion years.  This is a curious age: the Earth's oldest surviving rocks formed just after a rock-destroying time period known as the Late Heavy Bombardment or LHB.\indexit{late heavy bombardment} [\ldots] The origin of the LHB has long remained a mystery. Solar system formation models as well as the observed crater record suggests that the LHB was not just the tail end of the planetary accretion process.  The presence of heavily cratered regions on other bodies, including Mars, suggest that the LHB may have been a Solar-System wide process. [\ldots] The 'Nice\indexit{Nice model!hypothesis} Hypothesis' [(named after the city in France at whose university this hypothesis was first formulated)] suggests that a dramatic rearrangement of the outer planets gravitationally perturbed a large number of cometary bodies into orbits that penetrated the inner Solar System and cratered the surfaces of all Solar-System bodies [(see Sect.~\ref{sec:exo}) \ldots]''

%Titan's lakes and seas: https://www.jpl.nasa.gov/news/news.php?feature=7378
\subsection{The habitable zone}
\ors[IV:4.3] ``One of the most important requirements for life as we know it is water. The\indexit{habitable zone|seealso{definition}} ability\indexit{definition!habitable zone} to retain surface water is the general basis of the concept of the Habitable Zone (HZ). As\indexit{habitable zone} most commonly used, the habitable zone is an estimate of the range of distances from a star where an Earth-like planet can maintain surface water for extended periods of time.\activity{{\em Consider:} Although the definition of 'habitability' commonly involves the requirement of liquid surface water, some definitions are more relaxed. Perhaps other surface liquids can serve as agents in support of life (such as ethane and methane lakes and seas on that cover 1.6 million square kilometers, or 2\%\  of the surface, of Saturn's moon Titan) or perhaps subsurface water (as encapsulated seas or even globe-spanning layers) can support life. With that in mind, explore the moons of the giant planets that are thought to meet at least the condition of large reservoirs of some liquid somewhere, in particular: Europa, Callisto, Ganymede, and Io at Jupiter, Enceladus and Titan at Saturn, and Triton at Neptune. Which three power sources are thought to be most important in maintaining liquid states on giant-planet moons?}  While a number of factors, including greenhouse gases, tilt of spin axis, planet composition, surface gravity and cloud properties can be important for habitability, the primary factor considered for the habitable zone is the most fundamental, just the distance from the star (see below around Eq.~\ref{eq:2.3-7}).  For the present-day Sun, the habitable zone is generally considered to be the range from just inside Earth's orbit to a region near or just beyond Mars' orbit.  The inner boundary is where surface water is lost to space by either a runaway greenhouse effect or the 'moist greenhouse' effect.  In a full runaway, the surface temperature can exceed the critical point of water (374\,$^\circ$C), {\em i.e.,}  the temperature where liquid water and steam have the same density and are not distinguishable from each other.  Due to the extreme greenhouse warming caused by an ocean mass of water vapor, the surface temperatures on an Earth-like planet can reach the melting points of rocks. In comparison, the moist greenhouse is gentle and occurs when the partial pressure of water vapor at high altitudes becomes sufficiently elevated so that a substantial flux of water can be transported into the stratosphere and beyond. At high altitudes, H$_2$O is decomposed by UV photolysis and the liberated hydrogen ultimately escapes to space.

The outer edge of the habitable zone occurs when surface water freezes.  A
commonly quoted limit is 1.37\,AU based on the onset of formation of CO$_2$
ice clouds.  A more extended limit of 1.67 AU is
based on the maximum greenhouse warming that could occur in a
cloud-free CO$_2$-H$_2$O atmosphere. The highest estimate and perhaps an
upper limit is 2.4\,AU based on a combination of cloud altitudes and
particle sizes that could optimize radiative warming by CO$_2$ clouds [\ldots]

For planets,\indexit{stellar!evolution!brightening with age} the conventional habitable zone moves outward with time
as their central stars brighten.  Typical stars brighten by a factor of $\sim$2.5
during their main-sequence lifetimes, the periods of their lives
when they are stable stars fusing hydrogen to form helium.  Main
sequence stars of all mass brighten by a similar fraction as the ratio
of He/H in their cores increases with time.  At present, the Sun is
nearing half its main-sequence lifetime and it is brightening at a
rate of about 10\%\ per billion years, and 
is currently about 30\%\ more luminous than it was 4 billion
years ago.  More massive and less massive
stars brighten at higher and lower rates proportionate to their total
main-sequence lifetimes (cf., Fig.~\ref{figure:evolmodel}) [\ldots]

\begin{figure}[t] 
\centering
\includegraphics[width=8cm]{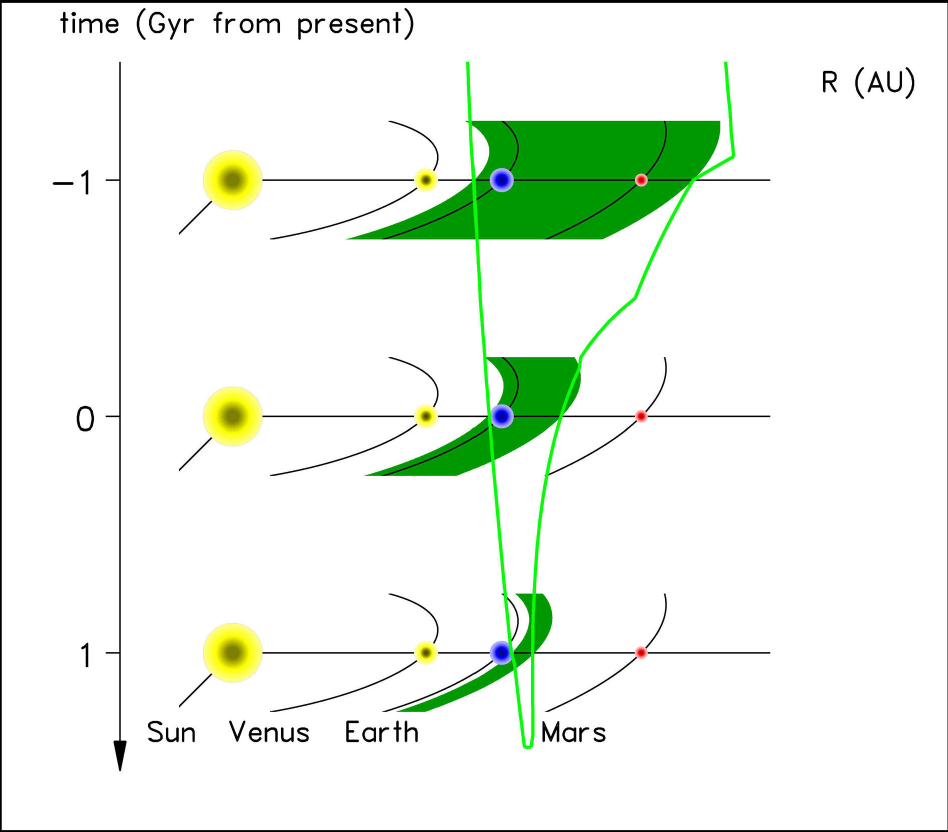}
\caption[The photosynthetic habitable zone (pHZ)
over time.]{\label{fig:brownlee1} The photosynthetic habitable zone (pHZ)
over time, from 1\,Gyr in the past to 1\,Gyr in the future.  The inner
edge of the pHZ moves outwards as the Sun becomes brighter with age and
the outer edge moves inwards as surface warming leads to decline
of CO$_2$ in the atmosphere to the point where photosynthesis is not
possible. [Fig.~III:4.1; modified from \href{https://ui.adsabs.harvard.edu/abs/2001NW.....88..416F/abstract}{source: \citet{2001NW.....88..416F}}; also  \href{http://www.pik-potsdam.de/~bloh/extrasolar/milkyway.html}{here on the web}.] \colorfig }  
\end{figure} 
The habitable zone concept becomes more complex when the ability\indexit{habitable zone!photosynthetic} to have photosynthesis is considered. \indexit{photosynthetic habitable zone} A more \indexit{pHZ|see{photosynthetic habitable zone}} restrictive consideration of surface habitability by organisms similar to plants and animals is the photosynthetic Habitable Zone or pHZ.  Photosynthesis requires atmospheric levels of CO$_2$ above some critical limit, approximately 10\,ppm for known plants.  The pHZ of a given star (see Fig.~\ref{fig:brownlee1} for the case of the Sun) narrows over time as the star gets brighter. The inner edge moves outwards and the outer edge moves inwards. For a planet with land and surface water, weathering processes remove CO$_2$ from the atmosphere. The process involves sequestering CO$_2$ in carbonates and this becomes increasingly more effective as [brightening] stars produce warmer planetary surfaces. This process can cause the pHZ to shrink to zero. Estimates indicate that the Sun's pHZ will shrink to zero width when the Sun reaches an age of 6.5 billion years. The Earth, even now close to the inner edge of the habitable zone, will be left behind the moving pHZ, and lose most of its surface water, long before this time [\ldots]'' \activity{{\em Background, for the curious:} Photosynthesis depends on the chemicals involved and as such is sensitive to the spectral energy distribution of the star. You could search the literature on developments in this area, but for stars substantially different from our Sun that work remains hypothetical. \href{https://phys.org/news/2019-01-habitable-planets-red-dwarf-stars.html}{Here} is a possible entry point. (a) Look up where the main absorption bands of chlorophyll and $\beta$-carotene lie relative to the solar spectrum at sea level. (b) How does the solar spectrum change under water for, for example, flora in the oceans?}

\subsection{Oxygen, methane, and carbon dioxide over time}
\ors[III:4.7] ``Prior to 2.4 \indexit{carbon dioxide!evolution}billion years ago, the Earth's atmosphere was
essentially devoid of free \indexit{oxygen!evolution}oxygen.  Although it was being produced by
photosynthetic organisms \indexit{meethane!evolution}such as cyanobacteria as well as the
photolysis of water vapor, it was efficiently removed from the
atmosphere as it oxidized compounds on the surface and in the
atmosphere.  Before this time, the atmosphere was dominated by N$_2$ but
it contained appreciable amounts of CO$_2$, water vapor and probably
moderate amounts of CH$_4$, possibly up to the percent level. There is
abundant evidence for low oxygen abundance on the early Earth
including the oxidation state of various minerals, including iron
oxide [\ldots]

The crossover, {\em i.e.,} the appearance of oxygen and simultaneous loss of methane, occurred 2.4 billion years ago.  At this time the Earth entered a severe ice age, also called a 'Snowball Earth' episode, during which the planet surface cooled to the point where ice formed at equatorial latitudes. It seems likely that this unusual cooling event was related to the rapid loss of significant greenhouse warming previously associated with the presence of methane.'' \activity{{Consider:} Sect.~\ref{sec:pastism} describes the possibility of the Solar System moving through dense, cold interstellar clouds, which could greatly enhance the dust environment of Earth. Review the \href{https://ui.adsabs.harvard.edu/abs/2005GeoRL..32.3705P/abstract}{study} by \citet{2005GeoRL..32.3705P} for the potential effects on terrestrial climate, including periods of strong glaciation and potentially the triggering of a 'Snowball Earth' state.}

\ors[III:4.9] ``Photosynthesis is the primary means by which life on Earth derives energy from the Sun. The complex chemical processes involved with photosynthesis depend on the availability CO$_2$ in the atmosphere and CO$_2$ can be considered an essential 'food of life' on our planet.\indexit{carbon dioxide!inevitable decline of} CO$_2$ on the present Earth is controlled by biogeochemical processes but in the future, as the Sun becomes brighter, the atmospheric CO$_2$ abundance will decline below the minimum ($\sim$10\,ppm) amount needed to support plants.  The end of CO$_2$ will mark the end of plants and animals that depend on direct contact with Earth's natural atmosphere [\ldots]

We currently have major concerns with the CO$_2$ increase from burning fossil fuels and its global warming effects. However, this is a short-term problem.  Ultimately, all of the atmospheric CO$_2$ will become locked up in carbonates and removed from the atmosphere. Even now, most of the CO$_2$ that has ever been in the atmosphere is already in carbonates. CO$_2$ is the dominant gas in the atmospheres of Venus and Mars and it must have been a major gas in the Earth's atmosphere before it declined due to carbonate formation.  If Earth's total carbonate content were decomposed, it would yield over 20 atmospheres of CO$_2$, over $4\times 10^4$ times the present CO$_2$ content of the atmosphere. As the Sun gets brighter, as all stars do as the hydrogen content of their cores is consumed, the Earth's surface temperature will increase and the CO$_2$ will decline as more and more is sequestered into carbonates.  The removal process is related to weathering of rocks, a process whose rate increases with increasing temperature.  The presence of silicates, water and atmospheric CO$_2$, leads to the formation of carbonates. Presently, this process is dominated by biological processes such as the formation of shells, corals, and microscopic organisms such as foraminifera. \activity{{\em Consider} the evolving CO$_2$ content of the atmosphere of a lifeless terrestrial planet. Which of the following parameters would influence the atmospheric CO$_2$ content over time:
  (1) atmospheric mass and composition,
  (2) chemical composition of seas and oceans,
   (3) continent sizes and placement,
   (4) fractional coverage by liquids in seas and oceans,
   (5) motion through, and density of, local interstellar medium,
   (6) orbital obliquity,
   (7) orbital period (length of the planetary 'year'),
   (8) planetary mass,
   (9) planetary radius,
   (10) planetary spin obliquity,
   (11) planetary spin rate (length of the planetary 'day'),
   (12) planets elsewhere in the planetary system,
   (13) plate tectonics,
   (14) properties of moons,
   (15) spectral type of the central star,
   (16) stellar spin rate.
  Formulate your arguments for each. You may want to read on in Ch.~\ref{ch:evolvingplanetary} and return here later to complete the activity. \mylabel{act:co2}}

When atmospheric CO$_2$ is sufficiently depleted, Earth will have lost an
important factor that has promoted the long-term stability of its
surface temperature.  Over Earth's history, the abundance of carbon
dioxide and its greenhouse warming effects have varied in ways that
have counteracted changes in atmospheric temperature.  When Earth
cools over long time periods, the CO$_2$ abundance can rise and promote
greenhouse warming.  When Earth warms, the CO$_2$ abundance can decline
and promote cooling. This effect is called the carbonate-silicate
cycle and it is a case of negative feedback where change is resisted
leading to stability. [\ldots] Carbon is removed by weathering but is involved
in a cycle because it is ultimately reintroduced back into the
atmosphere. Carbonate deposits in the ocean floor are subducted
beneath continents on $\sim$100\,Myr time scale where they are thermally
decomposed and release CO$_2$ back into the atmosphere via volcanism.
The CO$_2$ sink depends on weathering and carbonate deposition and the
CO$_2$ source depends on subduction, an ongoing process associated with
plate tectonics.''

\subsection{Water over time}
It appears that much of the water \indexit{Earth!water over time}on the terrestrial planets may have been \indexit{water!evolution}transported to the inner parts of the Solar System frozen within asteroids that were scattered from further out during phases of orbital changes of the giant planets, and then to Earth in collisions. Venus and Mars have lost their oceans a long time ago, as discussed in Sect.~\ref{sec:atmlong}. Earth, too, will eventually lose the bulk of its water 
\ors[III:4.10] ``when a critical threshold brightness is reached [in the Sun's evolution].  Ocean loss\indexit{oceans!inevitable loss} is a drastic change for a planet, and for Earth it will mean a change to a seemingly 'unearth-like' state, a planet more like Mars than the blue planet of its past. [\ldots] Even without oceans, Earth will probably always have regional ponds or lakes fed by water derived from the mantle.  The mantle is a reservoir that may contain several ocean-masses of water.

The most likely fate of Earth's oceans is loss by \indexit{moist greenhouse effect}the 'moist greenhouse' effect, a process that occurs at present but at a very low rate.  In this process, water is transported through the troposphere and stratosphere to heights where its hydrogen can be liberated by photolysis with solar UV photons [(Sect.~\ref{sec:atmlong})].  Near the exosphere the liberated hydrogen escapes to space, and forms Earth's Ly\,$\alpha$ geocorona.\activity{{\em Look up} what constitutes the geocorona.}  This process currently occurs at a rate of only a meter of ocean in a billion years due to the very low abundance of water vapor in the stratosphere. As the Sun warms, the partial pressure of water in the upper atmosphere rises and the timescale for water loss shortens. Modeling of this process indicates that the moist greenhouse effect will begin severely depleting the Earth's oceans in about a billion years or less.  If surface water is not largely depleted by the rather gentle moist greenhouse process in roughly 3 billion years, a much more severe process will take over when the Sun is about 35\%\ brighter that it is at present (Fig.~\ref{fig:brownlee5}, also Fig.~\ref{figure:evolmodel}). In a runaway, increasing temperatures introduce more greenhouse gas thereby providing positive feedback. This full runaway greenhouse advances to the critical point of water where density of water vapor equals the density of liquid water.  In a runaway, the enormous amount of water vapor in the atmosphere produces greenhouse warming sufficient to melt surface rocks.  Either the moist-greenhouse or the runaway-greenhouse process will result in the Earth's loss of its oceans to space and our planet will spend over half of its total life as an ocean-free planet, at least initially covered with salt and very oxidized rocks. [\ldots]

The loss of\indexit{plate tectonics!likely end of} oceans\indexit{oceans!and plate tectonics} is likely to also lead to the end of plate tectonics.  Hydrated minerals have lower melting points and in several ways the presence of water promotes the sinking of oceanic crust to subduct beneath continents.  Without oceans it is expected that plate movement will stop and Earth --~like all other planets in the Solar System~-- will cease to have subduction and the drift of continents.  Without subduction, the Earth's major mechanism for cycling CO$_2$ back into the atmosphere will be lost.'' When plate tectonics stops, this may also have major consequences for the planetary dynamo. \activity{{Consider:} What role does plate tectonics likely play in dynamos in terrestrial planets? Reminder: Sect.~\ref{intro-planets}.}

\section{Atmospheres and climates of Venus, Earth, and Mars}
In this volume, we \indexit{Mars!atmosphere and climate}focus on the terrestrial planets Venus, Earth, and \indexit{Venus!atmosphere and climate}Mars \ors[IV:7.1] ``because they are \indexit{Earth!atmosphere and climate}thought \indexit{climate!Venus, Earth, Mars}to have been habitable at their surfaces at some point during Solar System history. They formed under similar conditions, with early atmospheres that were more similar than they are today. The present day climates of Venus and Mars provide a useful contrast to that of Earth, and exploration of the root causes for differences in the present climates of all three planets allows us to better understand the processes that control climate on terrestrial exoplanets. Their current climates are summarized in Table\,\ref{tab:brain1} [\ldots] Despite their large differences in mass, the atmospheres of Venus and Mars have similar bulk compositions, with carbon dioxide (CO$_2$) comprising $\sim$95\%\ by volume, followed by molecular nitrogen ($\sim$3\%) and argon ($\sim$1\%). Earth's atmosphere, by contrast, is composed mainly of nitrogen and oxygen, followed by argon. \activity{{\em Consider:} The Earth's argon is predominantly Argon-40, whereas that in the universe at large, as in the Sun, is Argon-36. What is the source of Argon-40 in Earth's atmosphere?} Earth's atmospheric composition likely mirrored that of Venus and Mars early on, but much of Earth's atmospheric CO$_2$ now resides in carbonates on the ocean floors, leaving nitrogen as the most common constituent. Earth's abundant atmospheric oxygen is believed to have been contributed by photosynthetic bacteria.

The surface temperatures of the three planets also differ widely, in part due to the distance of each planet from the Sun and in part due to the quantity of greenhouse gases in each atmosphere. Earth is the only of the three planets with a surface temperature (and pressure) appropriate for liquid water to be stable for long periods of time, thanks to $\sim$30\,K of greenhouse warming. The Cytherean atmosphere is too hot for water to exist as liquid at the surface, while the Martian atmosphere has too low a surface pressure (liquid water would sublime, except at the lowest elevations). The atmosphere of Venus is very dry, indicating that any surface water driven into the atmosphere by the high temperatures no longer resides there. The atmospheric water content at Mars is an order of magnitude larger than at Venus and, given the low atmospheric pressure, is often nearly saturated. Despite the near 100\%\ Martian relative humidity, Earth still has roughly 50 times more water molecules (per number of particles of atmosphere) than Mars. The composition, temperature, and water content lead to different forms of precipitation on the three planets. Earth has a variety of forms of water precipitation, while Mars has carbon dioxide and water frost. Venus has no precipitation at the surface due to its high temperatures; any precipitation that forms higher in the atmosphere would turn to vapor before reaching the ground.

Circulation patterns on the three planets also differ. Earth possesses three circulation cells in each hemisphere, leading to prevailing winds organized by latitude. The circulation results, in a simplified sense, from an equator-to-pole temperature gradient that causes warm air to rise at the equator and fall at the poles. Earth's rotation provides a Coriolis \indexit{Coriolis effect!influence on climate}influence that breaks the circulation cells into three regions, keeping the warmest air relatively confined at low latitudes. Venus, by contrast, rotates very slowly. Thus, heat is transferred efficiently from the equator to polar regions, leading to uniform surface temperatures as a function of latitude and local time. Mars rotates at nearly the same rate as Earth but has only one circulation cell per hemisphere, though there are some arguments to suggest that while there is a net circulation, air tends to move in localized regional cells. Air at the surface of Mars moves sufficiently quickly to drive dust devil activity, while the surface of Venus is very still. At higher altitudes on Venus, however, the atmosphere super-rotates on timescales of days.

While Earth's seasonal variations, caused by a 23.5$^\circ$ tilt relative to its orbital plane, will be well known to the reader, seasonal variations on Venus and Mars are substantially different. Venus has nearly no seasonal variation due to a very small ($\sim$3$^\circ$) axis tilt. Mars has a tilt of 25$^\circ$, similar to that of Earth, but the planet's greater orbital eccentricity (a 21\%\ difference between the perihelion and aphelion distances compared to 1.4\%\ and 3.3\%\ for Venus and Earth, respectively) leads to shorter and more intense summers in the southern hemisphere compared to the north. Strong heating during southern summer drives enhanced dust devil activity, which can couple across circulation cell boundaries and grow into planet-encompassing dust storms that last several weeks.''

%\begin{figure}[t]
%\includegraphics[width=12cm]{figures/Figure01.eps}
%\caption[Evidence for climate change on the terrestrial planets.]{Evidence for climate change on the terrestrial planets. a) Determinations of D/H in the Venus atmosphere relative to terrestrial atmospheric D/H; b) Earth's atmospheric carbon dioxide and methane concentrations as a function of time, as determined from ice cores; c) A dendritic river valley network in the Warrego Valles region of Mars. [Note that the present-day CO$_2$ concentration on Earth exceeds 400\,ppm. Fig.~IV:7.1; source for panel %\href{https://www.sciencedirect.com/science/article/pii/S0019103511002983}{a},
% \href{https://royalsocietypublishing.org/doi/pdf/10.1098/rsta.2012.0294}{b}.] %\label{fig:brain1}}
%\end{figure}\figindex{figures/Figure01.eps}
\ors[IV:7.2] ``[A]bundant evidence points to changes in the climate of all three terrestrial planets on a variety of timescales. Here, we focus on evidence for climate change over tens of thousands of years or longer.  [\ldots] The most compelling evidence for climate change on Venus comes from measurements of the isotopes deuterium and hydrogen in the atmosphere today. Deuterium is far scarcer than hydrogen in the atmospheres of all planets. However, the ratio of deuterium to hydrogen (D/H) in the Venus atmosphere --~about 2 deuterium atoms for every 100 hydrogen atoms~-- is more than 100 times the same ratio calculated for Earth and most other Solar System objects. There is little reason to expect that Venus formed with a D/H ratio significantly different from that of Earth, so we infer that the D/H ratio on Venus increased after the planet formed. Specifically, it is thought that hydrogen atoms (possibly from a primordial ocean \regfootnote{Alternatives to a primordial ocean for Venus include a more recent reservoir, or perhaps one that continues to be replenished, from volcanic outgassing --~possibly clustered in major events~-- or cometary supplies --~see work cited in
\href{https://ui.adsabs.harvard.edu/abs/2018SSRv..214...10M/abstract}{this review by \citet{2018SSRv..214...10M}}}) preferentially escaped the planet's gravity compared to deuterium and were lost to space [\ldots] --~water was dissociated in the atmosphere and the hydrogen removed to space. [\ldots]

[E]vidence for climate change on Earth is abundant and comes in many different forms. [\ldots] The terrestrial climate record [derived from a diversity of sources (including growth rates of tree rings and corals, isotope ratios, gases trapped in air pockets in ice, geochemistry, fossils and sediments)] suggests that Earth's climate varies on many timescales, with departures in temperature of as much as 10--15$^\circ$C over Earth's history. There are many inferred cold (glaciation) and warm periods that have been tied with changes in atmospheric conditions and diversity of life. Similarly, there are a few major changes in atmospheric composition, the most notable of which is the oxygenation of the terrestrial atmosphere more than two billion years ago, likely caused by the rise of oxygen-producing bacteria and the subsequent depletion of sinks for oxygen at Earth's surface. Analysis of the size and depth of fossilized raindrop imprints in sedimentary rock even suggests that Earth's surface pressure has varied by as much as a factor of two over 2.7 billion years. Taken together, the evidence provides a caution against interpreting the present day climates of other terrestrial planets too finely, and assuming only monotonic changes in planetary climates over billion year timescales. At the same time, one of the most notable aspects of the terrestrial record is the fact that water has existed as liquid at the surface for most of the planet's history, suggesting that despite short term deviations Earth's climate has been relatively stable over its history, in likely contrast to Venus and Mars.

Mars also provides several lines of evidence suggesting past climate that differs from today. [\ldots] These include dry dendritic (branching) river valley networks, river delta deposits, possible regions of sedimentary rock, smoothed and rounded rocks imaged by Mars rovers, and possible ancient ocean shorelines. These features all suggest an ancient Mars where liquid water was abundant and active in shaping the surface of the planet. Further, highly eroded crater rims and a paucity of small craters relative to what might be expected from the abundance of large craters suggest that the ancient atmosphere was much more efficient at eroding surface features ({\em i.e.}, thicker) than today --~perhaps as thick as 0.5-3 bars, or even more. [\ldots\ A] number of Martian atmospheric isotope ratios (D/H, $^{38}$Ar/$^{36}$Ar, $^{13}$C/$^{12}$C, $^{15}$N/$^{14}$N, $^{18}$O/$^{16}$O) point to the stripping of atmospheric particles to space over billions of years, similar to the inference drawn from D/H measurements at Venus. [\ldots\ T]he isotope ratios suggest that 50-90\%\ of the total atmospheric content has been removed to space from stripping processes alone.''

\section{Irradiance, orbits, spin, and climate}
\subsection{Atmospheric effects and albedo}\label{sec:atmalb}
In this section we \indexit{albedo!climate}first look into equilibrium temperatures in the absence of a planetary atmosphere, and then proceed to see how an atmosphere modifies such an equilibrium. We will focus on planets orbiting our Sun, but the same arguments hold for exoplanets orbiting other stars, of course.  \ors[III:11.2.3] ``The fraction of the solar luminosity $L_\odot$ that is absorbed by a planet is given by the ratio of the planet's cross section $\pi R_{\rm p}^2$ to the area $4\pi d_{\rm p}^2$ of a sphere containing the planet at distance $d_{\rm p}$ from the Sun, corrected for the albedo $a$ ([the fraction of total incoming power that is reflected]):
\begin{equation}
    {\cal P}_{\rm p|a} = L_\odot \, \frac{\pi\,R_{\rm p}^2}{4\,\pi\,d_{\rm p}^2} \,(1-a) \equiv {\cal P}_\odot (1-a) 
    \label{eq:2.3-5}
\end{equation}
[where ${\cal P}_\odot$ is defined as the total energy per unit time intercepted by the planetary disk. \ldots]'' \ors[III:11.3.4] ``The\indexit{albedo|seealso{definition}} albedo\indexit{definition!albedo} is \indexit{albedo}defined as the ratio of diffusely reflected to incident electromagnetic radiation and, therefore, lies in the interval $0-1$. It is difficult to determine the total albedo of a planet because it is highly variable, ranging from less than 0.1 for water and forests to more than 0.8 for fresh snow. On Earth, the largest contribution comes from the clouds which cover about 50\,\% of its surface.  For the Earth an average albedo of 0.3 is usually assumed [\ldots]''

\begin{figure} %\centerline{\psfig{figure=figures/Fig3_Sun-Earth-mapping_grey.eps,width=11cm,clip=}}
%\centerline{\psfig{figure=figures/lean_1.1996472.figures.f2.eps,width=11cm,clip=}}
%\centerline{\psfig{figure=figures/energyflow_f12.3.eps,width=11cm,clip=}}
\centerline{\includegraphics[width=11cm]{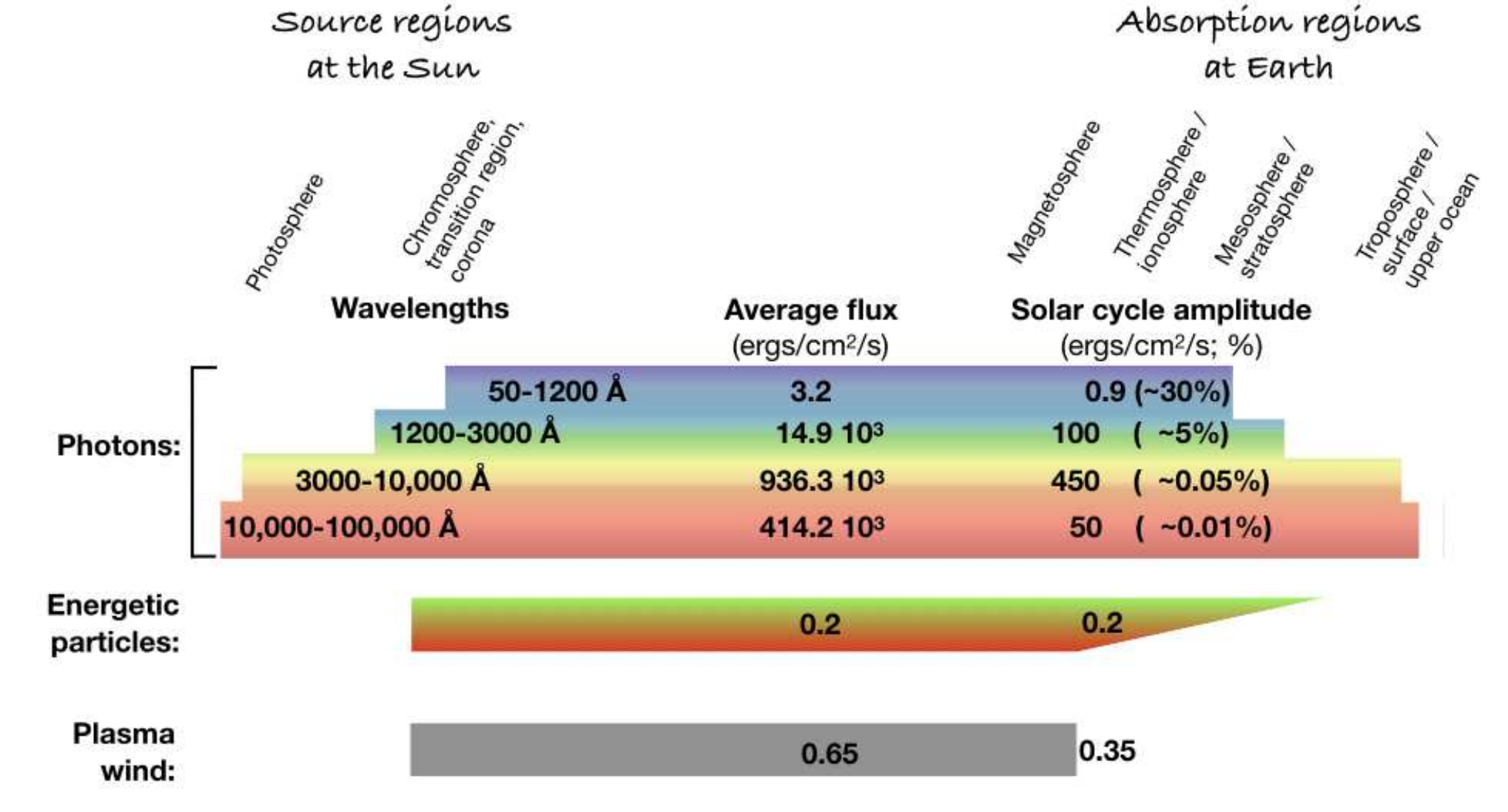}}
\caption[Sun-Earth energy flow: photons, energetic particles, 
solar wind.]{\label{lean:f3}
The flow of energy from the Sun to the Earth is compared for photons
in four different wavelength bands, energetic particles, and the
plasma wind. The numbers are approximate energies with their variations
during an 11-yr solar activity cycle, in erg/cm$^2$/s. Visible radiation
connects the surfaces of the Sun and Earth while ultraviolet radiation
connects their atmospheres. Particles and plasma connect the outer
solar atmosphere primarily with Earth's magnetosphere and high-latitude
upper atmosphere. [After Fig.~III:10.3] \colorfig } %\href{https://physicstoday.scitation.org/doi/10.1063/1.1996472}{source}
\end{figure}
\ors[III:11.2.3] ``If we assume as a first approximation that a planet is an atmosphere-free black body and that the climate machine distributes the incoming solar radiation uniformly [({\em i.e.,}  that effects of very low or very high spin rates can be ignored)], the emitted power is given by the law\indexit{planetary!radiation balance} of\indexit{radiation balance for planets} Stefan-Boltzmann:
\begin{equation}
{\cal P}_{\mathrm{emi}} = 4\,\pi\,R_{\rm p}^2\,\sigma\,T_{\rm e|0}^4.
\label{eq:2.3-6}
\end{equation}
Under steady state conditions absorption and emission are equal and the
temperature $T_{\rm e|0}$ [(with index $0$ to indicate absence of an atmosphere as we have here)] can be calculated:
\begin{equation}
 T_{\rm e|0} = \left( {\frac{L_\odot (1-a)}{16\,\pi\,\sigma\,d_{\rm p}^2}} \right)^{1/4}.
    \label{eq:2.3-7}
\end{equation}
Note that the temperature of a planet does not depend on its size [\ldots] \activity{{\em Show:} At what distance would an Earth-equivalent exoplanet need to orbit an $0.6\,M_\odot$ M0\,V star to reach the same \indexit{equilibrium temperature}global 'equilibrium temperature', all other things being equal? You may disregard effects associated with the difference in the stellar spectral energy distribution on the exoplanet, but you should not ignore the bolometric correction in estimating the total stellar irradiance. How long would a year last on such a planet compared to Earth's? Use Fig.~\ref{fig:acthrd}. Note: such close-in planets are subject to very strong tidal forces that will synchronize spin and orbital periods, causing these exoplanets to lose their day-night cycles. That, in turn, invalidates your estimate --~why?}\activity{{\em Show:} {Beyond the furthest planet:} The New Horizons spacecraft flew by Kuiper Belt Object 2014\,MU$_{69}$ on 2019/01/01, the most distant body visited by a spacecraft to date, at an orbital distance of $\sim 44$\,AU. Estimate the surface temperature of 2014\,MU$_{69}$, which has an albedo of $\sim 0.1$. Compare your estimate to the observed temperature in \href{https://ui.adsabs.harvard.edu/abs/2019Sci...364.9771S/abstract}{the paper} by \citet{2019Sci...364.9771S}.}

\begin{table}[t]
\caption[Temperatures of the planets for
different albedo and stellar luminosity.]{Comparison of the calculated temperatures of the planets for different combinations of planetary albedo $a$ and stellar luminosity $L$ in the absence of atmospheres, compared with the observed temperatures. [Table~IV:11.3]}
\label{tab:2.3-2}
\centering
\begin{center}\begin{tabular}{lcccc|c}
\hline
& \multicolumn{4}{c|}{Effective temperature absent an atmosphere ($^\circ$C)} & Observed \\
Planet &    Distance    & $a = 0.5$     & $a = 0.3$     & $a = 0.1$  & temperature \\
&   (AU)    & $L = 0.8$ &  $L = L_\odot$    &  $L = 1.3L_\odot$ &  ($^\circ$C)\\\hline
Mercury & 0.38& 77& 130&    175&    180 to  420 \\
Venus&  0.72&   -10&    30& 66& 460 \\
Earth&  1&  -50&    -18&    11& 15 \\
Mars&   1.52&   -95&    -65&    -40 & -87 to  5 \\
Jupiter&    5.2&    -175&   -160&   -150&   -130 \\
Saturn& 9.54&   -200&   -190&   -180&   -180 \\
Uranus& 19.18&  -220&   -215&   -210&   -210 \\
Neptune &30.06& -230&   -225&   -220&   -210\\
\hline
\end{tabular}\end{center}
\end{table}

In Table~\ref{tab:2.3-2} the calculated equilibrium temperatures for the eight planets in the absence of atmospheres are compared to the
measured ones. [\ldots] Overall there is a reasonable agreement between the estimated and the observed temperatures. The largest discrepancy is observed for Venus. The reason is that Venus has a very dense atmosphere which consists for 96\%\ of CO$_2$ [(see Table~\ref{tab:brain1})] with clouds of SO$_2$ generating the strongest greenhouse effect in the Solar System. In the case of Earth, the difference between calculated (using the present values $a=0.3$ and solar luminosity) and measured mean global temperature is 33\,$^\circ$C. This difference is also due to the natural greenhouse effect. It is important to note that the Earth needs the natural greenhouse effect to be habitable, but not necessarily an additional anthropogenic increase. The range of observed temperatures on Mars is very large because Mars has only a very thin atmosphere (0.3\,millibar compared to 1\,bar of Earth) and no liquid water to transport and distribute energy. Jupiter is considerably warmer than calculated (-110\,$^\circ$C instead of -160\,$^\circ$C). Most likely, this difference is due to gravitational contraction which provides an additional power at least as large as the solar insolation.''

\begin{figure}[t]
%\centerline{\psfig{figure=figures/brf2.eps,width=11cm}}
\centerline{\includegraphics{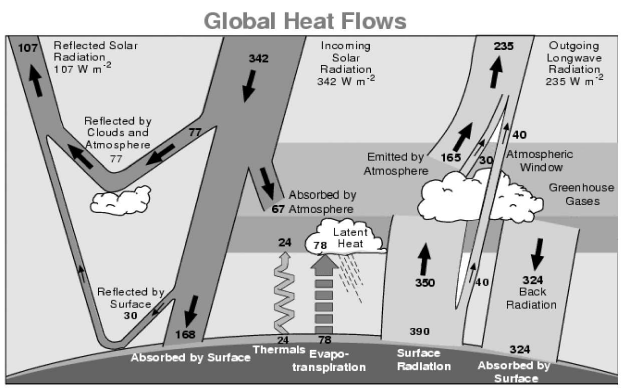}}
\caption[Exchanges of solar and terrestrial energy in the Earth's
atmosphere.]{Exchanges of solar (shortwave) and terrestrial (longwave) energy in the Earth's atmosphere. The flow of energy is expressed in\,Wm$^{-2}$ (kiloerg/cm$^2$/s), averaged over the entire Earth surface ({\em i.e.,}  over the day-night cycle). [Fig.~III:16.2; \href{https://ui.adsabs.harvard.edu/abs/1997BAMS...78..197K/abstract}{source: \citet{1997BAMS...78..197K}}.]}\label{fig:br2}
\end{figure}
The value of the planetary albedo is determined by the properties of the planetary surface and, if present, the planetary atmosphere. For the Earth, not surprisingly, the impact of the atmosphere in setting the overall albedo has been studied in great detail.
\ors[III:16.2] ``Perhaps, the best way to represent the exchanges of radiative energy in the atmosphere is\indexit{climate!energy balance} to refer to Figure~\ref{fig:br2}. This figure shows that [\ldots\ a large fraction of the infrared radiation from the planetary surface] is absorbed by greenhouse gases in the atmosphere. These gases, whose temperature is lower than the surface temperature, re-emit radiation both towards space and towards the Earth's surface [\ldots]''

\begin{figure}[t]
%\centerline{\psfig{figure=figures/brf3_symbols.eps,width=8.5cm}}
\centerline{\includegraphics[width=9.5cm]{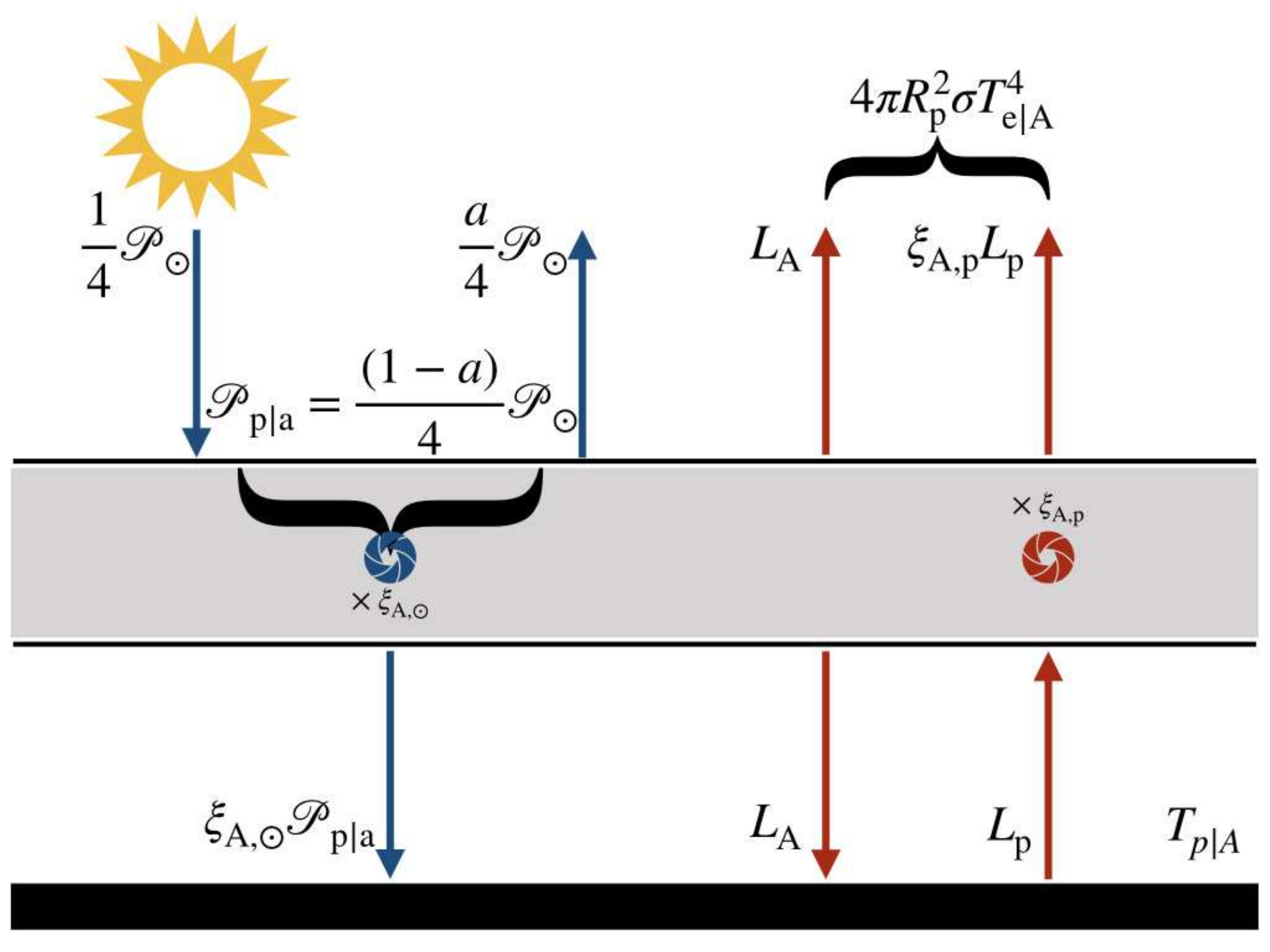}}
\caption[Simplified model of radiative exchange in the Earth's atmosphere.]{Simplified model of radiative exchange in the [atmosphere of a planet with a partially transparent atmosphere that balances incoming solar power ${\cal P}$ and outgoing planetary luminosity $L$]. The optical transmission in the atmosphere is represented by $\xi_{\rm A,\odot}$ in the shortwave ([solar, shown in blue]) spectral region and by $\xi_{\rm A,p}$ in the longwave ([planetary, shown in red]) part of the spectrum. [This figure is a modified version of Fig.~III:16.3 for consistency with symbols used in this text.] \colorfig }\label{fig:br3}
\end{figure}
For an illustrative first-order approximation of the sensitivity of the ground-level climate to the greenhouse effect of an atmosphere, we can look at a highly simplified version of the energy flow in which we disregard the mechanical and evapo-transpiration energies and assume that the atmosphere radiates equally in the upward and downward directions (which is a significant oversimplification as you can infer from Fig.~\ref{fig:br2}), as sketched in Fig.~\ref{fig:br3}.
\ors[IV:7.3] ``[The e]ffective temperature $T_{\rm e|A}$ can be related to surface temperature of a planet with an atmosphere under a few assumptions. Here, [one may approximate the planetary surface temperature in the presence of an atmosphere by]
\begin{equation}\label{eq:brain2}
T_{\rm p|A}=(1+\tau_{\rm A})^{1/4} T_{\rm e|A},
\end{equation}
where $T_{\rm p|A}$ is the [ground-level or] surface temperature and $\tau_{\rm A}$ is the optical depth of the atmosphere.''
Although proper radiative transfer is essential to quantify $\tau_{\rm A}$ and thereby how atmospheric properties combine to set the ultimate planetary temperature, let us here look at a very \ors[III:16.2] ``simple model of the radiative transfer processes described above [but allowing for a simple wavelength-dependent effect in radiative transport]; we represent the atmosphere by a single\indexit{solar!irradiance!climate driving} layer of radiatively active gases whose optical transmission is noted $\xi_{{\rm A}\odot}$ and $\xi_{\rm A,p}$ for shortwave (incoming solar) and longwave ([emitted planetary]) radiation, respectively. The radiative shortwave solar [power] and the longwave surface [luminosity] are noted by symbols ${\cal P}_{\rm p|A}$ and $L_{\rm p}$, respectively. $L_{\rm A}$ represents the radiative [power] emitted by the atmospheric layer and $T_{\rm p|A}$, an indicator of the [planet's] climate, represents the surface (ground) temperature. From Figure~\ref{fig:br3}, we derive the energy balance at the top of the atmosphere [and at the planets's surface, respectively:]
\begin{equation}
{\cal P}_{\rm p|a} = (1 - a) {\cal P}_\odot/4 = \xi_{\rm A,p} L_{\rm p} + L_{\rm A}
%\end{equation}
%and at the Earth's surface,
%\begin{equation}
{\,\,\, ; \,\,\,}
\xi_{A,\odot}  {\cal P}_{\rm p|a} = L_{\rm p} - L_{\rm A}.
\end{equation}
We deduce that the surface temperature [in the presence of an atmosphere] is given by
\begin{equation}
T_{\rm p|A} = T_{\rm e|0} \left( {{1 + \xi_{A,\odot}} \over {1 + \xi_{\rm A,p}}} \right )^{1/4},
\end{equation}
where the planetary\indexit{radiative!equilibrium temperature} equilibrium temperature $T_{\rm e|0}$ [is given by Eq.~(\ref{eq:2.3-7}). For Earth, this]
is equal to 255\,K ($-18^\circ$C) for ${\cal P}_\odot / 4 = 342$\,Wm$^{-2}$ and for an albedo $a = 0.31$. Assuming that the atmosphere is approximately transparent to solar radiation, so that the shortwave transmission $\xi_{{\rm A},\odot}$ is close to 1.0, and adopting a longwave transmission $\xi_{\rm A,p}$ of 0.2, the surface temperature [in the presence of its atmosphere] becomes
\begin{equation}
T_{\rm p|A} = T_{\rm e|0} \left({2.0\over 1.2}\right)^{1/4} = 289\,{\rm K} {\, \rm [or \,} 16^\circ{\, \rm C]}.
\end{equation}

The value calculated by this simple model, tuned by approximating choices for $\xi_{\rm A,\odot}$ and $\xi_{\rm A,p}$ [(and suggesting an effective atmospheric optical depth in Eq.~\ref{eq:brain2} of $\tau_{\rm A}\approx 0.67$)], is in agreement with the observed temperature $T_{\rm p|A, obs}$ (288\,K). More refined models account in greater detail for wavelength-dependent radiative transfer, vertical and horizontal heat transport in the atmosphere, energy and water exchanges at the Earth's surface. Absorption coefficients for different molecules in different spectral regions are measured in the laboratory. [\ldots] The simple conceptual model presented here can, however, be used to estimate to a first approximation the change in the surface temperature that would result, for example from a relative change in the solar input ${\cal P}_\odot$ of 0.1\%. We derive easily that, for constant $\xi_{\rm A,\odot}$ and $\xi_{\rm A,p}$,
\begin{equation}
{\Delta T_{\rm p|A}\over T_{\rm p|A}} = {\Delta {\cal P}_\odot\over 4{\cal P}_\odot}.
\end{equation}
For $T_{\rm p|A} = 288$\,K, we obtain a surface temperature change $\Delta T_{\rm p|A}$ of 0.07\,K for a solar-cycle TSI variation of 1500\,erg/cm$^{-2}$/s. The amplitude of the solar variation is therefore a factor of 10 smaller than the surface temperature trend observed since the beginning of the industrial era. However, over a period of a decade or so, the solar signal should be significant compared to human-driven temperature trends, and should therefore be taken into consideration in the analysis of temperature records.  [Studies] have shown that, even if the global temperature variation associated with solar forcing is small, changes in temperature patterns become significant at the regional scale.

\begin{figure}[h!]%[ph!]
\centering
\includegraphics[width=\textwidth]{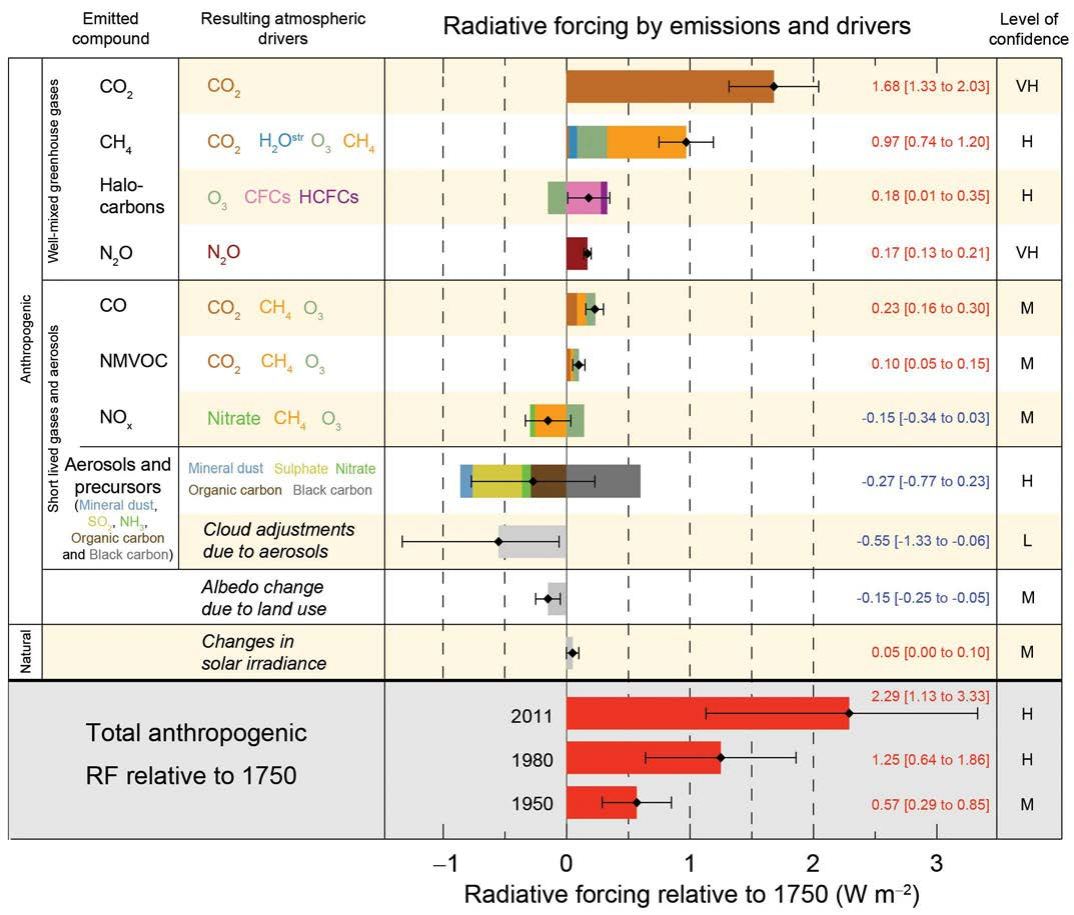}
\caption[Radiative forcing (RF) bar chart for Earth's climate.]{\label{fig:IPCC2013_18.7}
Radiative forcing estimates in 2011 relative to 1750 and aggregated uncertainties for the main drivers of climate change. Values are global average \indexit{radiative!forcing|see{climate}}radiative \indexit{climate!radiative forcing}forcing (RF), partitioned according to the emitted compounds or processes that result in a combination of drivers. The best estimates of the net RF are shown as black diamonds with corresponding uncertainty intervals; the numerical values are provided on the right, together with the confidence level in the net forcing (VH -– very high, H -– high, M -– medium, L -– low, VL -– very low). Albedo forcing due to black carbon on snow and ice is included in the black carbon aerosol bar. Small forcings due to contrails (0.05\,W\,m$^{-2}$, including contrail induced cirrus), and HFCs, PFCs and SF$_6$ (total 0.03\,W\,m$^{-2}$) are not shown. Concentration-based RFs for gases can be obtained by summing the like-colored bars. Volcanic forcing is not included. Total anthropogenic radiative forcing is provided for three different years relative to 1750. [ \href{https://www.ipcc.ch/site/assets/uploads/2018/02/WG1AR5_SPM_FINAL.pdf}{From \citet{ipcc2013}.]} \colorfig } 
\end{figure}
A more accurate treatment requires that the transmission functions and the atmospheric emissivity change with the chemical composition of the atmosphere in response to Sun-induced climatic changes, that dynamical feedbacks be taken into account and that the influence of the\indexit{climate!role of oceans} ocean be considered. [\ldots]'' Many details go into establishing the 'radiative forcing' of an atmosphere, see, for example, the publications of the Intergovernmental Panel on Climate Change (IPCC) from which Figure~\ref{fig:IPCC2013_18.7} was taken to illustrate the often counteracting effects of atmospheric constituents on radiative forcing. \activity{{\em Show} that the simple model in Fig.~\ref{fig:br3} yields an estimate consistent with Earth's global temperature rise of about one degree (observed between 1850 and 2010) based on the increase in anthropogenic radiative forcing as shown in  Figure~\ref{fig:IPCC2013_18.7} within the uncertainty
indicated in that figure.}

Equation~(\ref{eq:2.3-7}) illustrates \ors[IV:7.3] ``four \indexit{climate!factors for change}main ways in which planetary climate can be altered. First, the amount of radiation from the star ($L_\odot$) can change. The solar constant at Earth varies by only $\sim$0.1\%\ over the course of a solar cycle. [Evolutionary stellar-structure models] suggest that the Sun is $\sim$30\%\ brighter today than it was when the terrestrial planet[s formed (Fig.~\ref{figure:evolmodel}) \ldots] 

Second, changes in the albedo ($a$) of a planet will change the amount of incident energy absorbed by the surface (and atmosphere). Variation in cloud cover, the extent of polar ices, vegetation, or wind blown dust, for example, can all change the albedos of the terrestrial planets, and will have an influence on the atmospheric energy budget. Venus has an albedo of $\sim$0.9, while the albedos of Earth ($\sim$0.3) and Mars ($\sim$0.25) are considerably lower. [\ldots] \activity{{\em Show:} Compare the values of ${\cal P}_{\mathrm{abs}}$ from Eq.~(\ref{eq:2.3-5}) for Venus and Earth. Explain qualitatively why Venus' surface temperature exceeds Earth's; Section~\ref{sec:atmalb} provides the answer.}

Third, characteristics of a planet's orbit and rotation influence its energy budget. The amount of solar radiation encountering a planet varies with average orbital distance ($d_{\rm p}$), with the result that Venus encounters roughly double the energy that Earth does, while Mars encounters $\sim$45\%. Ellipticity of the orbit [(see Eq.~\ref{eq:3.3-1}) causes incident energy to vary] between 36\%\ and 52\%\ over a Martian year due to Mars' relatively high orbital ellipticity. This explains why the southern summer at Mars (near perihelion) is more extreme than the northern summer. [Realizing that incident solar energy is not uniformly distributed by the thin atmospheres of the terrestrial planets, it will be clear that t]ilt also influences the amount of sunlight that reaches each part of a planet's surface, making some portions of the planet cold and other portions warm. This effect influences where ices form at the surface, removing some gases from the atmosphere and changing albedo in some locations. Chaotic changes in the eccentricity, obliquity, and spin precession of Mars and Earth over periods of tens to hundreds of thousands of years are thought to contribute to climate variations (Sect.~\ref{sec:orbitalchanges}), though the range of variation in both orbital properties (especially tilt) and climate is estimated to be larger at Mars due to the lack of a large Moon.

Fourth, the amount of radiation-absorbing atmosphere ({\em i.e.}, greenhouse gases) influences surface temperatures. [\ldots]  The thick CO$_2$ atmosphere of Venus provides more than 500\,K of greenhouse warming compared to the theoretical surface temperature in the absence of an atmosphere. Earth's atmosphere provides approximately 30\,K of greenhouse warming. This warming, while much smaller than at Venus, is crucial to keeping our average surface temperature above the freezing point of water, making life and many aspects of our climate possible. The atmosphere of Mars, while dominated by CO$_2$, is too thin to provide substantial greenhouse warming today. The temperature is warmed only $\sim$5\,K due to greenhouse gases [\ldots].''

\subsection{Orbital changes}\label{sec:orbitalchanges}
The physical basis of orbital \indexit{climate!orbital change}changes and of tidal effects on planetary rotation were discussed in Ch.~\ref{ch:torques}. Here, we look in some detail at the orbital effects on climate. 
\ors[III:11.3.2] ``[T]he distance $d_{\rm p}$ is a prime parameter for the temperature of a planet. [S]olar power decreases with the square of the distance or in other words that the relative change of the temperature is 1/2 of the relative\indexit{planetary!insolation!distance, eccentricity} change of the distance:
\begin{equation}
\frac{\Delta T_{\rm e|0}}{T_{\rm e|0}} = -\frac{1}{2}\,\frac{\Delta d_{\rm p}}{d_{\rm p}}.
\label{eq:3.2-1}
\end{equation}
[\ldots] Because all the planets have elliptical orbits the distance is continuously changing. The eccentricity ranges from 0.0068 for Venus to 0.2056 for Mercury. The eccentricity of the Earth's orbit is 0.017. That means the\indexit{planetary!insolation!annual variation at Earth} distance between Earth and Sun is 1.017\,AU at the aphelion compared to 0.983\,AU at the perihelion. This difference results in a change of insolation by about $10^5$\,erg/cm$^{2}$/s'' and would result in $\Delta T_{\rm e|0} \approx 5$\,K throughout the year, but that is strongly dampened by the thermal inertia of Earth's oceans and land masses.

%\subsection{Orbital or Milankovic forcing}
\label{chapter:3.3}

\begin{figure}[t]
\centering
\includegraphics[width=10cm]{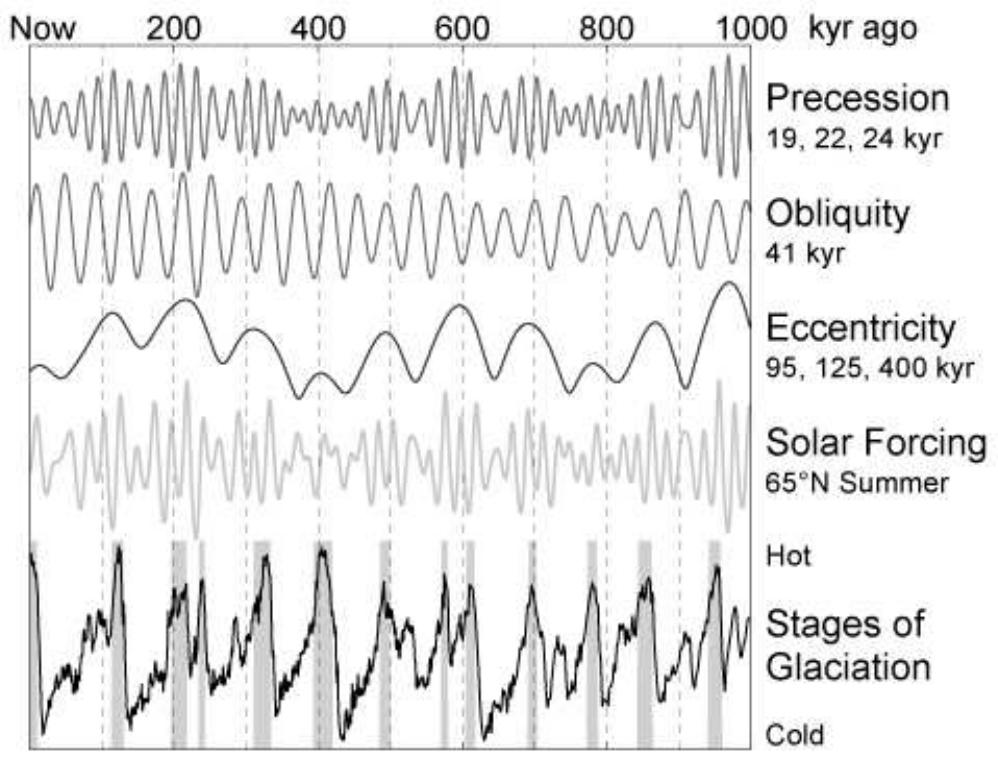}
\caption[Earth's orbital parameters for the past million years. ]{Earth's orbital parameters for the past million years. The first three panels
show the three orbital parameters influenced by the other planets (mainly Jupiter and Saturn) and the moon (precession). The fourth panel exhibits the calculated solar forcing at 65$^\circ$\,N. The lowermost panel shows sea level changes derived from stable isotope measurements on benthic foraminifera indicating glacial (cold) and interglacial (warm, grey bands) periods. [Fig.~III:11.6]}
\label{fig:3.3-3}
\end{figure}
But not only does ellipticity of orbits lead to seasonal changes, the orbits actually evolve over time. \ors[III:11.3.3] ``[A]ll the bodies in the Solar System are gravitationally coupled. This was known already since Newton's time. [\ldots\ It] was Milutin\indexit{Milankovic climate forcing} Milankovic
\indexit{climate!change!Milankovic forcing} who, for the first time, worked out the mathematical details of these disturbances [\ldots] There are three orbital parameters of the Earth which are affected by the other planets, the Sun, and the Moon.

{\bf (1) Orbital eccentricity:}
[\ldots\ Integration over a full year] \indexit{orbital!eccentricity}reveals the following relationship between the relative change in the annual amount of solar radiation $S$ received by Earth and the relative change in the\indexit{insolation!variation with orbital eccentricity} eccentricity $e$:
\begin{equation}
\frac{\mathrm{d}S}{S} = \frac{e^2}{(1-e^2)^{3/2}}\,\frac{\mathrm{d}e}{e}.
\label{eq:3.3-1}
\end{equation}
The largest change in $e$ (0.06) which the Earth experienced over the past million years (Fig.~\ref{fig:3.3-3}) therefore leads to a very small change of 0.36\,\% in the annual mean insolation which corresponds to a mean global forcing of less than $10^3$\,erg/cm$^{2}$/s [(compare Fig.~\ref{fig:IPCC2013_18.7} for present-day forcings)].  The changes in the eccentricity occur on time scales of 100,000 and 400,000 years. It is interesting to note that it is exactly this small change in the eccentricity which seems responsible for the 100,000-year cycle in the sequence of glacial and interglacial periods during the past 1,000,000 years (Fig.~\ref{fig:3.3-3}).  This is a\indexit{climate!feedback mechanisms} nice example that climate is a non-linear system and that even a small forcing can cause a large effect if feedback mechanisms are involved. Such a feedback mechanism could be that although a larger eccentricity does not change the mean annual insolation much, but with it the seasonality changes: colder summers on the northern hemisphere may result in a reduced melting of the winter snow enlarging the ice sheets and the albedo which further reduces the effective insolation.

{\bf (2) Obliquity:} The tilt \indexit{obliquity}angle of the Earth's
spin axis relative to\indexit{climate!change!obliquity} the ecliptic plane varies between 22.1$^\circ$ and 24.5$^\circ$ with a periodicity of about 41,000 years.  Contrary to the eccentricity changes the obliquity does not change the total amount of received solar radiation but only its latitudinal distribution. The larger the obliquity the stronger is the seasonality. A smaller obliquity reduces both the mean insolation and the summer insolation at high latitudes, thereby providing favorable conditions for ice ages.

{\bf (3) Precession:}
[\ldots] Because the \indexit{precession}Earth is spinning, its shape deviates slightly from a sphere leading to an equatorial bulge.  Tidal forces act on the bulge and force the [rotation] axis to precess. The periods of precession range from 19,000 to 24,000 years.

%\vskip 0.5cm

The calculated values of the three orbital parameters are plotted in Fig.~\ref{fig:3.3-3}, together with the corresponding summer insolation at 65$^\circ$\,N, a\indexit{precession!of equinoxes and climate} latitude\indexit{climate!equinox precession} which is considered as critical for the formation of ice sheets as a result of cold summers. The bottom panel shows a compilation of $\delta^{18}$O records from deep-sea sediments. Benthic foraminifera live in the deep sea and form CaCO$_3$ shells. After death, the\indexit{climate!change!ice core data} shells are buried in the sediment layer by layer for millions of years. Measuring the $^{18}$O/$^{16}$O isotope ratio with a mass spectrometer relative to a standard, expressed as $\delta^{18}$O, reflects the sea level.  Water evaporating from the sea preferentially contains the lighter molecules H$_2\,^{16}$O. If the evaporated water stays on the continents forming glacial ice sheets the ocean becomes depleted in $^{16}$O. Warm interglacial periods are indicated by grey bands. They\indexit{ice ages} normally\indexit{interglacials} last 10,000 to 20,000 years and occur with a typical periodicity of 100,000 years when the eccentricity is large.''

\section{Planetary atmospheres, geological activity, and stellar winds}
\subsection{On time scales beyond millions of years}\label{sec:atmlong}

\ors[IV:7.4] ``[Planetary s]urface temperature and climate are strongly affected by the amount of greenhouse gases in an atmosphere, which can be viewed as a combination of the total number of particles in an atmosphere (surface pressure) and its composition. Several mechanisms are capable of changing atmospheric abundance and composition (Fig.\,\ref{fig:brain3}) [\ldots]

\begin{figure}[t]
\centering
\includegraphics[width=10cm]{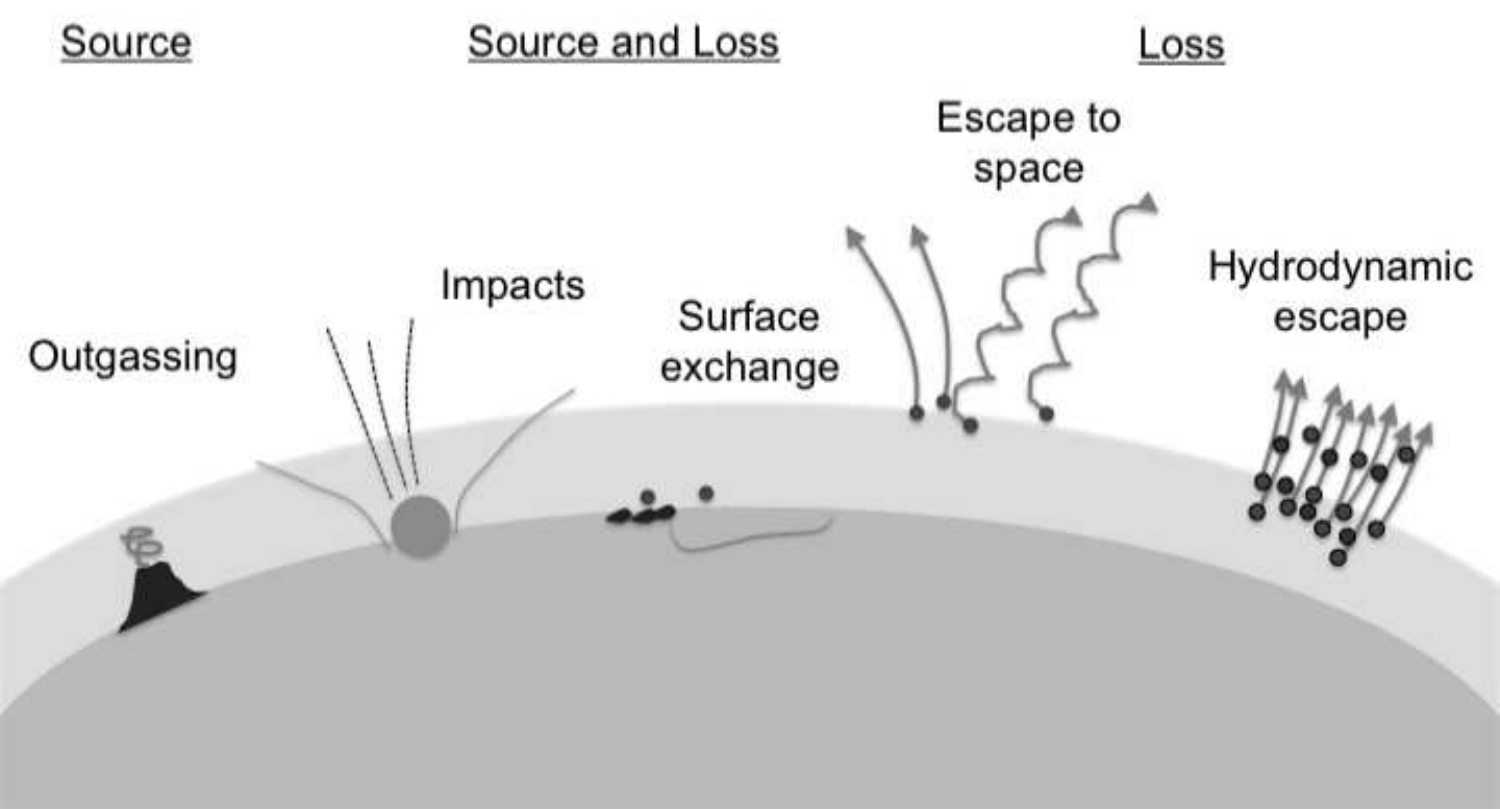}
\caption[Source and loss mechanisms for planetary atmospheres.]{Source and loss mechanisms for planetary atmospheres. [Fig.~IV:7.3] \label{fig:brain3}}
\end{figure}\figindex{../brain/art/Figure03.eps}

Volcanic outgassing \indexit{climate!volcanic outgassing}from planetary interiors is thought to be the primary source for the terrestrial planet atmospheres we observe today. Water vapor is the most common gas released in terrestrial eruptions, followed by CO$_2$. Other commonly released gases include sulfur dioxide, nitrogen, argon, methane, and hydrogen. Outgassing should be a declining source of atmospheric particles over Solar System history, as the interior heat required to generate volcanic activity declines. [Earth is evidently volcanically active today, Venus likely is (although without signatures of active plate tectonics), while there is no direct evidence for ongoing activity for Mars.] 

Atoms and molecules can be exchanged between a planet's surface layers and its atmosphere via a variety of processes and over many timescales. For example, changes in temperature can increase condensation rates to the surface, forming surface liquids or ices (evident on Earth and Mars). Chemical reactions (weathering) can also remove particles from the atmosphere, and is typically most effective in warm or wet environments (evident on Venus, Earth, and Mars). Adsorption removes atmospheric particles that stick to surface materials. Most or all of these processes can be considered to be reversible. Release of particles back to the atmosphere can involve changes in temperature, chemical reactions (including reactions with sunlight), and geologic events that allow subsurface reservoirs access to the atmosphere.

%Atmosphere: https://www.esrl.noaa.gov/gmd/ccgg/faq_cat-3.html
All planetary atmospheres are \indexit{climate!asteroid impact}subject to impact from asteroids, comets, dust, and even atoms and molecules. Impactors of all sizes can deliver volatile species to an atmosphere ({\em e.g.}, impact delivery is responsible for at least part of Earth's water inventory as well as meteoritic layers observed in terrestrial planet ionospheres). Impacts can also remove atmospheric particles via collisions, and sufficiently large impactors can additionally accelerate atmospheric particles via impact vapor plumes and lofted surface material [\ldots] Monte Carlo simulations suggest impacts have resulted in a net gain of atmospheric gases for Earth and Mars over Solar System history, and a net loss for Venus.\activity{{\em Show:} To get an idea of scales: estimate the size of a comet that would double the CO$_2$ content of Earth's atmosphere. How does that compare to, e.g. comet 1P, the target of the {\em Giotto} mission, and 67C, the target of the {\em Rosetta} mission?} 
%Answer: at 3e18g of CO2 in the atmosphere and 10% of comet by weight: ~30km if at average density of 1g/cm^3

Hydrodynamic escape \indexit{atmosphere!hydrodynamic escape}occurs when a light species escapes (thermally) in \indexit{hydrodynamic escape|see{atmosphere}}sufficient abundance that it becomes equivalent to a net upward wind, and drags heavier species with it through collisions. This process is usually enabled by high solar EUV flux or another form of heating. It should have been significant for all of the terrestrial planets during the first few hundred million years after formation, stripping away most of their primordial atmospheres. [\ldots] 

The removal of atmospheric particles to space from the upper layers of the atmosphere is commonly referred to as “escape to space”. This term typically excludes impacts by asteroids, meteoroids, and comets, and hydrodynamic escape is also often listed as a distinct process. Here, “escape to space” encompasses a set of approximately six processes, all of which provide escape energy to atmospheric particles. The energy is ultimately provided (sometimes directly, and sometimes indirectly) through interaction with the parent star and stellar wind. [\ldots] It is currently thought that atmospheric escape has played an important role in the evolution of the climates of both Venus and Mars by altering atmospheric pressure and trace gas abundance.''

\ors[IV:7.5] ``All particles escaping from a planetary atmosphere share three characteristics. The first is that they have sufficient energy to escape the gravity of the planet[, which means that their velocity should exceed the escape speed (Eq.~\ref{eq:brain3} with $r$ set to the radial distance from which the escape occurs, typically the exobase, discussed below). The values listed in Table~\ref{tab:brain1} show that Mars has a much lower escape speed than Earth or Venus.]

A second characteristic of an escaping particle is that it is unlikely to collide with other particles after acquiring sufficient escape energy. In planetary atmospheres, the region above which collisions are unlikely is \indexit{definition!exobase}termed the exobase, and is loosely defined as the location where the mean free path of a particle [(Eq.~\ref{eq:mfp})] is equal to an atmospheric scale height [(Eq.~\ref{eq:hp}) \ldots] 

Finally, any escaping particles must not be confined to the planet by planetary magnetic fields. This requires either that an escaping particle be neutral, that the planet lack a magnetic field, or that any magnetic fields are weak enough that energized charged particles are able to easily traverse magnetic field lines. Venus lacks a measurable global magnetic field like that of Earth. Mars also lacks a global magnetic field but possesses localized regions of strongly magnetized crust that may locally trap energized atmospheric ions.

Due to the highly collisional nature of planetary lower atmospheres, escape is generally limited to three regions of the upper atmosphere: the thermosphere, the exosphere, and the ionosphere. The altitude and composition of these regions are summarized for each planet in Table\,\ref{tab:brain3} [\ldots]''

\begin{figure}[t]
\centering
\includegraphics[width=10cm]{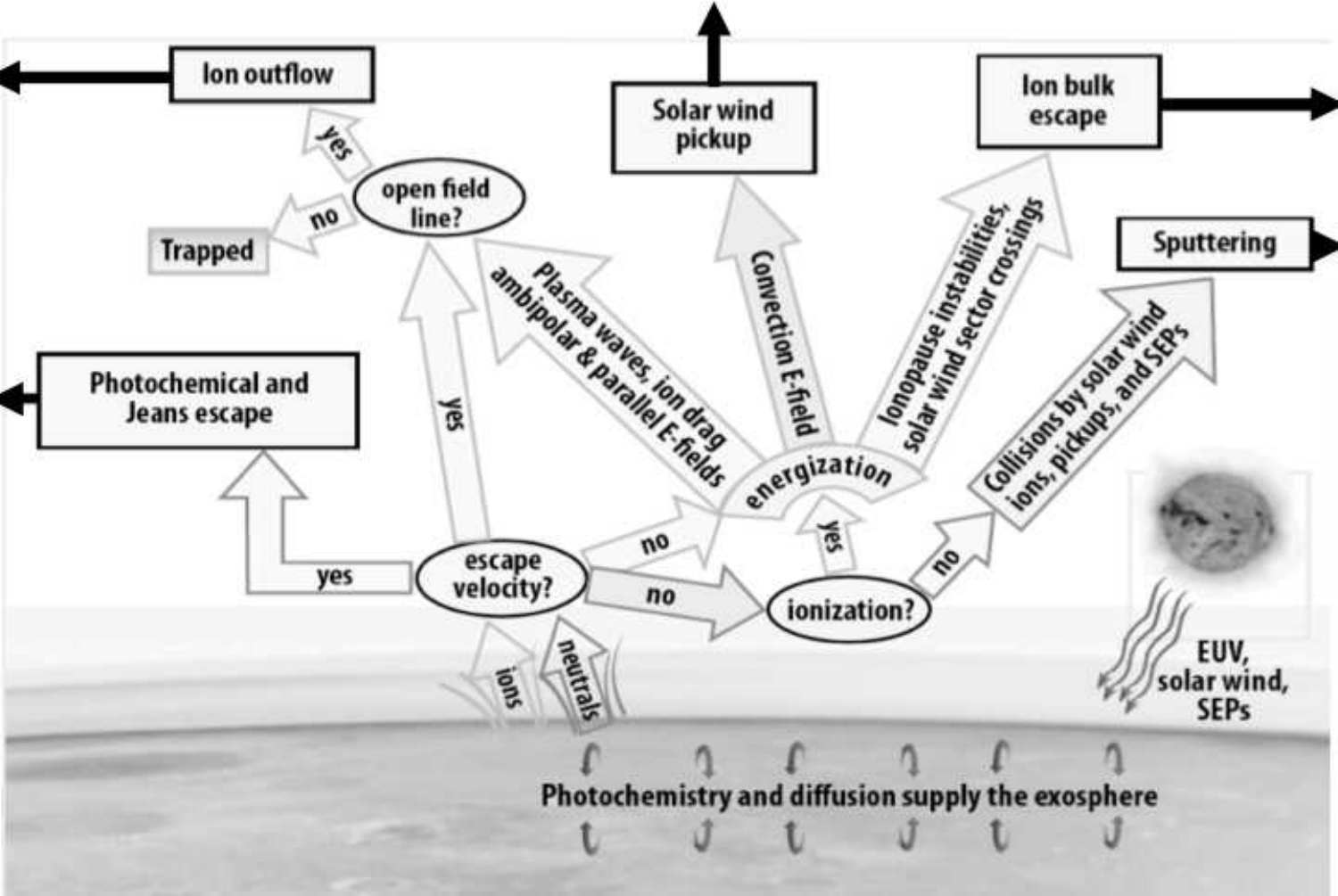}
\caption[Flowchart for energization and escape of atmospheric particles.]{Flowchart showing pathways to energization and escape of particles from a planetary atmosphere. [\ldots\ Fig.~IV:7.4] \label{fig:brain4}}
\end{figure}\figindex{../brain/art/Figure04.eps}

\ors[IV:7.6] ``A number of \indexit{atmosphere!escape!mechanisms}mechanisms are capable of giving atmospheric particles sufficient energy to escape from a planet [(see Fig.\ref{fig:brain4}.] Neutral particles can escape an atmosphere in one of three ways: {\em (1) Jeans escape, (2) photochemical escape,} and {\em (3) atmospheric sputtering}. 
[Ion loss processes can be grouped into three additional] categories: {\em (4) ion outflow, (5) ion pickup,} and {\em (6) bulk plasma escape}.]

{\em (1) Jeans (or thermal) escape} occurs because some fraction of neutral particles near the exobase will have sufficient energy to escape simply because the particles have a thermal distribution. Neutral temperatures near the exobase of all three planets are sufficiently low ($\sim$250--1000\,K) that only species with small mass (H, D, and He) can escape via this mechanism in significant quantity. The process should be more efficient for Mars (due to its low gravity) and for Earth (due to its higher exobase temperature) than for Venus.

{\em (2) Photochemical escape} refers to the escape of fast neutral particles energized by sunlight-driven chemical reactions. These reactions typically involve dissociative recombination of an ionized molecule with a nearby electron, resulting in two fast neutral atoms. Photochemical escape fluxes depend upon ionospheric molecular densities near the exobase, as well as electron density and temperature. Photochemistry is thought to be the dominant loss process for neutral species more massive than hydrogen and helium at Mars. Fast atoms produced photochemically at Venus and Earth are typically not energetic enough to escape the larger gravity.

{\em (3) Atmospheric sputtering} occurs when atmospheric particles near the exobase receive sufficient energy from collisions to escape. Collisions occur when energetic incident particles (often ionospheric particles accelerated by electric fields near the planet) encounter the exobase. There are no unambiguous observations that sputtering is actively occurring at any of the terrestrial planets, [but] it may have been important earlier in Solar System history, especially for unmagnetized planets. [\ldots]

{\em (4) Ion outflow} refers to the acceleration of low energy particles out of the ionosphere via plasma heating and outward directed charge separation (ambipolar) electric fields. In this case the ion acceleration can occur below the exobase, where collisions maintain a more fluid-like behavior. Ion outflow is the only significant ion loss process for the terrestrial atmosphere, and encompasses a number of processes referred to in the terrestrial literature, including wave heating, polar wind, and auroral outflow. [Analogues of these processes should be active for Venus and Mars.]

{\em (5) Ion pickup} refers to the situation where a neutral particle is ionized (via photons, electron impact, or charge exchange) and accelerated away from the planet by a motional electric field (${\bf E} = - \frac{1}{c}{\bf v}  \times {\bf B}$). Ion pickup occurs primarily for ionized exospheric neutrals (though some ionized thermospheric neutrals near the exobase region may escape via pickup as well). The motional electric field is usually supplied by the solar wind, so that the process is most relevant for compact magnetospheres unshielded by strong planetary magnetic fields (Venus and Mars) [\ldots]

{\em (6) Bulk plasma escape} refers to any process which removes spatially localized regions of the ionosphere {\em en masse}. Bulk escape is relevant for unmagnetized planets, where the external plasma flow can create magnetic and/or velocity shear with the ionosphere. A popular example involves the Kelvin-Helmholtz instability, which may form at the ionopause of Venus or Mars and steepen into waves which eventually detach from the ionosphere. Other bulk escape processes are possible as well, such as transport via plasmoid-style flux ropes that may remove ionospheric plasma from Martian crustal magnetic field regions. \activity{{\em Consider:} (a) What is the basis of the Kelvin-Helmholtz instability? This instability also occurs between the terrestrial magnetosphere and magnetopause flow because the magnetic tension is not strong enough to stabilize the developing waves. (b) Why is this geospace phenomenon not listed as a process for 'bulk outflow'?}

[Based on models and observations,] the \indexit{atmosphere!escape!rate}present-day global escape rate for Venus is estimated to be $10^{24}$-$10^{26}$\,s$^{-1}$. The escape rate for Earth is $10^{25}$-$10^{27}$\,s$^{-1}$, and for Mars is $10^{24}$-$10^{26}$\,s$^{-1}$. [Normalized per unit area, these rates] are on the order of $10^6$-$10^9$\,cm$^{-2}$\,s$^{-1}$. [These escape rates are a very small fraction of the column densities in the present-day atmospheres that range from $10^{23}$-$10^{27}$\,cm$^{-2}$, but they may be substantial when accumulated over $\sim$4 billion years ($\sim 10^{17}$\,s)]. For this latter point the two orders of magnitude uncertainty in escape rates are crucial; they are the difference between heliophysical drivers being the main loss mechanism for planetary atmospheres and merely an afterthought in determining present-day [atmospheres. \ldots]

Finally, it is important to keep in mind that escape to space not only influences atmospheric abundance but also atmospheric composition, which can be important in planetary evolution. One example is the aridity of the Cytherean atmosphere. The loss of atmospheric water is attributed to dissociation of the water in the atmosphere by sunlight, and the subsequent escape to space of oxygen. Water is only a trace gas in planetary atmospheres, but is an important greenhouse gas and is extremely important for habitability. So even if escape to space does not appreciably change atmospheric thickness, it may contribute in important ways to climate. Interestingly, the escape rates listed above, when converted to precipitable microns of water, amount to global layers of water only centimeters thick. More than this is assumed to have been lost from Venus, suggesting either that escape rates have changed over time (and are low today) or that other processes (such as impacts) have been important for removing water.''

\begin{table}[t]
\caption[Solar wind and interplanetary magnetic field at the terrestrial planets.]{Properties of the solar wind and interplanetary magnetic field (IMF) at the terrestrial planets. [Table~IV:7.4] \label{tab:brain4}}
\begin{center}\begin{tabular}{lccc}
\hline
&	Venus	&Earth	&Mars\\
\hline
IMF strength	&$\sim$0.10--0.12 mG	&$\sim$0.06 mG	&$\sim$0.03 mG\\
Solar wind speed&	$\sim$400 km/s&	$\sim$400 km/s&	$\sim$400 km/s\\
Solar wind density&	10--15 cm$^{-3}$&	$\sim$6 cm$^{-3}$&	$\sim$1-3 cm$^{-3}$\\
Alfv{\'e}n speed&	$\sim$70 km/s&	$\sim$55 km/s&	$\sim$45 km/s\\
Mach number&	5--7&	6--8&	8--10\\
H$^+$ gyroradius&	$\sim$1500 km&	$\sim$2500 km&	$\sim$5000 km\\
H$^+$ gyroradius / $R_{\rm p}$	&0.5	&0.4	&$\sim$3\\
\hline
\end{tabular}\end{center}
\end{table}

\ors[IV:7.7] ``Observations, simulations, and common sense all tell us that atmospheric escape rates are not constant, and are influenced by a number of heliophysical drivers that vary on both short and long timescales. [\ldots] 
The three main drivers are photons, charged particles, and electromagnetic fields. Photons deposit energy in atmospheres when they are absorbed by atmospheric particles. Extreme ultraviolet (EUV) and soft X-ray photons (generated in the solar corona and chromosphere, and not to be confused with solar luminosity) provide the dominant energy source in upper atmospheric regions. Charged particles in the solar wind also supply energy to planetary upper atmospheres and plasma environments. Table\,\ref{tab:brain4} summarizes some of the relevant quantities of the solar wind at each terrestrial planet. While density and velocity can each vary independently, studies of solar wind influences on atmospheric escape (especially the induced magnetospheres of Venus and Mars) typically use solar wind pressure ($\rho v^2$) as the organizing quantity. Finally, the solar wind carries a magnetic field, which creates a convection electric field (${\bf E}_{\rm sw}$) in the frame of the planet that depends upon solar wind velocity and interplanetary magnetic field (IMF) strength and orientation ([see around Eq.\,\ref{ohm-general}]). Magnetic and electric fields organize charged particle motion, and electric fields accelerate charged particles; both effects influence the ability of charged particles to escape a planet's atmosphere.

%\begin{figure}[t]
%\includegraphics[width=10cm]{figures/Figure05.eps}
%\caption[Evolution of solar drivers of atmospheric escape.]{Evolution of solar drivers of atmospheric escape. a) solar EUV photon flux, relative to today; b) solar mass loss rate ({\em i.e.}, solar wind flux); c) interplanetary magnetic field (curve labeled $B_{\rm S}$).  [Fig.~IV:7.5] \label{fig:brain5}}
%\end{figure}\figindex{../brain/art/Figure05.eps}
The external drivers of atmospheric escape vary on four main timescales. \indexit{atmosphere!escape rate!variability}Billion-year timescales are associated with the age of the Sun, and both theoretical calculations and observations of Sun-type stars suggest that all three drivers should have declined in intensity with age (see Figs.\,\ref{figure:rotact}, \ref{Wood_f9}, and \ref{fig:estvarragne}). EUV flux varies by factors of several over a solar cycle (from solar minimum to solar maximum), and solar wind pressure varies by factors of $2-10$. The IMF, in particular, is a function of the solar rotation period, and all three drivers also vary on more rapid timescales of minutes to hours.

Variability in the heliophysical drivers should influence atmospheric escape rates. In general, an increase in solar EUV fluxes ({\em e.g.}, a transition from solar minimum to solar maximum) is expected to result in an increase in loss rates of neutral particles. Energy from solar photons heats the upper atmospheric neutrals, so that Jeans escape rates should increase with solar EUV. This is likely to be true at Mars, but not at Earth where hydrogen escape from the exobase is limited not by the available energy, but by the supply (via diffusion) of particles from lower altitudes. Jeans escape should be negligible at Venus today, but may have been significant in the past if either exobase temperatures or solar EUV fluxes were much higher. Energy from solar photons is also used to drive the chemical reactions necessary for photochemical escape, so that contemporary Martian photochemical escape should vary with EUV flux. Neutral escape rates should be largely insensitive to changes in both the solar wind and the IMF, except for sputtering rates from Venus and Mars, which are thought to be dominated by re-impacting atmospheric pickup ions and will therefore increase as the pickup ion population increases in response to changes in solar EUV.

Ion escape rates should also vary with the three drivers. An increase in solar wind pressure will cause a corresponding decrease in the size of the magnetospheric cavity at all terrestrial planets, effectively lowering the pressure balance altitude between the solar wind and planetary obstacle to the flow. For Mars, with an extended neutral corona, an increase in solar wind pressure exposes significant additional high-altitude neutrals to ionization and stripping by the solar wind (via electron impact and charge exchange). The IMF, by contrast, chiefly organizes the trajectories of escaping particles at Venus and Mars; large-gyroradius pickup ions are preferentially accelerated away from the planet in regions where ${\bf E}_{\rm sw}$ points away from the planet. At Earth, the orientation of the IMF affects the location and extent of cusp regions, from which outflowing ions escape. EUV fluxes have a more indirect effect. In total, one might expect the ion escape rate to increase at solar maximum due to the additional energy input from EUV. At unmagnetized Venus and Mars, however, the increased ionospheric content deflects the solar wind around the planet at higher altitudes and can prevent the interplanetary magnetic field from entering the ionosphere. The escape of heavy ion species (which are concentrated at lower ionospheric altitudes) via pickup and bulk escape may therefore remain roughly constant, or even decrease during solar maximum periods, even as lighter ion species escape more efficiently.'' \activity{{\em Consider:} Make a table summarizing which atmospheric loss processes work on each of the terrestrial planets. Which two processes are most effective for the present-day Earth based on the description in Sect.~\ref{sec:atmlong}?}

\ors[IV:7.8] ``A number of characteristics of a terrestrial planet itself influence the properties and energetics of upper atmospheric reservoirs for escape, including transient events such as dust storms ({\em e.g.}, for Mars), or longer-lived phenomena such as gravity waves that couple the lower and upper atmospheres. In the context of heliophysics, the nature of a planet's intrinsic magnetic field is of the greatest relevance. [\ldots\ Present-day Earth has an intrinsic magnetosphere] that deflects the solar wind at large distances from the planet ($\sim$10\,$R_\oplus$). There is an induced magnetosphere at Venus that deflects the solar wind at much closer distances ($\sim$1.3$R_{\venus}$), and a similarly-sized (with respect to the planet) induced magnetosphere at Mars punctuated by 'mini-magnetospheres' tied to specific regions of the crust and that rotate with the planet. [\ldots]

When considering the total atmospheric loss from a planet, it has often been assumed that the presence of a \indexit{atmosphere!escape rate!magnetic field}magnetic field results in lower escape rates. [However, we mentioned] that the measured atmospheric escape rates for Venus, Earth, and Mars are comparable within the current uncertainties. It has recently been proposed that magnetic fields, rather than shielding a planetary atmosphere from stripping by the solar wind, actually collect solar wind energy and transfer it to the ionosphere along field lines. Global magnetic field lines converge near the cusps, so that the energy is more spatially concentrated than for unmagnetized planets. The escape rate for a given planet may be comparable when it is magnetized, or even greater because planetary magnetic fields extend much further than the planet's atmosphere, giving it a larger energy-collecting cross-section in the solar wind. One key difference with magnetized planets is that the concentrated energy in cusp regions is likely to lead to more efficient removal of heavy species.

There are a few caveats: [the estimated planetary atmospheric escape rates are quite uncertain, not all solar wind energy collected by a planet need go into removing atmospheric particles, and accelerated ions in Earth's cusps may not escape the planet. Clearly multiple issues need to be understood] before we can determine whether magnetic fields protect an atmosphere from being lost.''

\subsection{On time scales of up to several millennia}
The Sun's variability has affected Earth's climate and atmospheric composition on astronomical time scales, but a multitude of studies looking for both causes and effects on shorter time scales suggest that the \ors[III:12.1] ``conclusion at the time of this writing with respect to the importance of low-frequency solar variability in the most recent decades, and perhaps up to centuries, might be 'Perhaps, but probably small'. The main reasons why uncertainties persist regarding this issue include these:
\begin{enumerate}
\customitemize
\item The $\sim$150-year instrumental record is too short to draw definitive statistical conclusions about the connection of any relation existing on the multi-decadal time scale. \item Forcing \indexit{climate!forcing}from anthropogenic greenhouse gases represent a significant overprint on trends since about 1850\,CE.  Because to first order the trends in proxies for solar activity indices and in greenhouse gas concentrations are similar, there is a statistical degeneracy which leads to ambiguous, and thus potentially misleading, conclusions unless great care is taken. \item A similar problem of statistical degeneracy applies to the Little Ice Age interval of cool conditions during the last millennium (main phase about $1450-1850$\,CE), when mountain glaciers advanced in many regions and planetary temperatures were about 0.5$^\circ$C lower.  During the Little Ice\indexit{little ice age!and Maunder Minimum} Age,\indexit{Maunder Minimum!and little ice age} solar activity, as inferred from changes in radiogenic isotopes such as $^{14}$C and $^{10}$Be, appears to have varied similar to pulses in volcanism and slightly lower carbon dioxide levels. Ignoring this similarity in patterns of variability in internal and external (in planetary terms) climate drivers can lead to erroneous conclusions. \activity{{\em Background:} Human impacts on climate appear not to be limited 
to the Industrial Revolution! Have a look at a
\href{https://ui.adsabs.harvard.edu/abs/2019QSRv..207...13K/abstract}{study by \citet{2019QSRv..207...13K}}: they argue that the large population reduction in the Americas following the arrival of European conquerors and settlers, and the resulting reforestation of abandoned agricultural lands, was a significant part of the change in atmospheric CO$_2$ in the late 16th Century and in the 17th Century.} 
\end{enumerate}
[\ldots] `` There are, however, fingerprints of solar variability that locally stand out. For example, \ors[III:16.7] ``[v]ariations in solar radiation over the 11-yr cycle as well as over the 27-d solar rotation period have substantial effects in the upper atmosphere where energetic photons penetrate and directly initiate photochemical effects. In the stratosphere and the troposphere, above which shortwave radiation is absorbed, the direct impact of solar variability becomes less pronounced. Solar signal in ozone and temperature, however, can be derived from observations above approximately 25\,km altitude. Below this height, the situation becomes more complex because other dynamical signals such as those produced by climatic modes of variability ({\em e.g.,} El Ni{\~n}o) interfere with possible variations resulting from solar variability.

Several mechanisms have been proposed to explain a plausible relation between solar variability and the observed 11-yr dynamical variability in the lower atmosphere. One of them is associated with disturbances produced in the upper atmosphere and resulting from ozone variations generated by changes in shortwave solar radiation. A second mechanism is linked to the ocean-surface response to 11-yr changes in the total solar irradiance. Observed weather patterns correlated with solar \indexit{climate!forcing}forcing could result from both downward-propagating disturbances produced in the stratosphere and upward-propagating perturbations generated at the surface of the ocean. To capture the amplifying mechanisms producing a dynamical response of the troposphere to solar variability, atmospheric models must therefore account for photochemical processes in the upper atmosphere and, at the same time, must be coupled to an ocean module. Despite many remaining uncertainties, much progress has been made in the last years to better understand how solar variability could potentially affect the climate system, particularly on decadal timescales.'' \activity{{\em Consider:} Compile a list of all the processes involved in setting a planetary climate system that reflects at least all those mentioned in Chs.~\ref{ch:formation} and~\ref{ch:evolvingplanetary}. You can assimilate relevant processes from Activity~\ref{act:co2} here as a start. \mylabel{act:prochab}}

\clearpage

\chapter{{\bf Upper atmospheres and magnetospheres}}
\label{ch:transport}%13
{\narrower\narrower{
{\bf Chapter topics:}
\begin{itemize}
  \customitemize
\item Spectral irradiance and the ionospheric chemistry of terrestrial planets
\item Solar X-ray and UV emission and solar wind over time
\item The ionosphere subject to evolving solar activity
\item Impacts of the evolving geomagnetic field on the ionosphere
\end{itemize}

\noindent{\bf Key concepts:}
\begin{itemize}
  \customitemize
\item Photo-ionization and -dissociation
\item Collisional recombination
\item Gravitational stratification 
\item Gravitational differentiation
\end{itemize}

}}

\section{Upper-atmospheric chemistry and insolation}
\ors[IV:9] ``As one \indexit{atmosphere!upper atmosphere}}moves up in
altitude in a planetary atmosphere, several important changes in
composition and structure are apparent.  Most notably, as a
consequence of hydrostatic equilibrium, the gas density decreases,
{\em i.e.,} the air becomes 'thinner'. [\ldots\ W]ith decreasing
density, the frequency of collisions between atmospheric molecules
decreases to the point where bulk motions such as turbulence are no
longer able to mix the atmosphere.  Instead, molecular diffusion
becomes the more rapid process and this also leads to a composition
change whereby the lighter constituents, typically atomic species such
as atomic oxygen, diffuse upward more rapidly than their heavier
counterparts such as O$_2$, N$_2$ or CO$_2$. The region where the
atmosphere is well mixed is known as the
\indexit{definition!homosphere}homosphere; the region where diffusive
separation dominates is known as the
\indexit{definition!heterosphere}heterosphere. [\ldots]''
\ors[IV:9.1.1] Because ''molecular \indexit{atmosphere!molecular
  diffusion}diffusion coefficients ($D$) vary inversely as [the square
root of the] molecular mass, the molecular diffusion velocities are
greater for the lighter constituents and smaller for heavier
constituents. Furthermore, they vary inversely as the total density
({\em i.e.,} diffusion of a gas is more rapid if collisions are less
frequent), thus $D$ increases with altitude.''\sactivity{$\circledS$
  {\em Show:} The approximate scaling of the molecular diffusion
  coefficient $D$ (measuring the mean square displacement per unit
  time) with molecular mass $m$ and particle density $n$ follows from
  energy equilibrium of the constituent particles. Formulate $D$ as
  function of the collisional cross section $\sigma$ and of
  temperature and density in the case of self-diffusion, {\em i.e.,}
  for molecules diffusing among themselves. For a mixture of
  components, mutual diffusion needs to be
  considered. \mylabel{act:diffcoeff} \solution{diffcoeff}}

High in planetary atmospheres is also the region where the
temperatures rise (Fig.~\ref{fig:temps}) as a result of inefficient
cooling while absorbing solar UV to X-rays.  This absorption also acts
to break chemical bonds and to liberate electrons from their orbits,
thus creating the ionospheres. Earth's upper neutral atmosphere is
dominated by N$_2$ up to about 200\,km (see Fig.~\ref{fig:fr2}), and
the overlapping ionosphere is dominated by NO$^+$ and O$_2^+$ (see
Fig.~\ref{sojka:fig3.1} ).  Up to roughly 150\,km and 200\,km,
respectively, Venus' and Mars' neutral atmospheres are dominated by
CO$_2$ while O$_2^+$ dominates in the corresponding ionic components
in the lower layers of their ionospheres.  In the next 150\,km above
that, atomic oxygen is the dominant species in all three neutral
atmospheres, while O$^+$ dominates for Earth, O$_2^+$ yields dominance
to O$^+$ after the first 50\,km or so for Venus, and O$_2^+$ dominates
for Mars [(see Tables~\ref{tab:brain1} and~\ref{tab:brain3})]. High in
this domain, ions are lifted higher than simple estimates of pressure
scale heights might suggest because of the
\indexit{ambipolar!effect}ambipolar effects associated with the free
electrons.
%\activity{Charge neutrality is key in the
%  stratification of an ionized gas: what are the approximate ratios of
%  density scale heights for the constituent populations in a low-density,
%  isothermal atmosphere of mixed molecular N$_2$, atomic N, and
%  singly-ionized N?  Why?} 
The compositional differences of the ionospheres are a consequence of
the different pathways for photo-dissociation\indexit{photo-chemistry}
of molecules by solar radiation, which in turn feed a number of chemical
reactions in the atmosphere.

An example of photo-dissociation is
provided by the photo-dissociation of molecular oxygen (O$_2$)
\begin{equation}\label{eq:br1}
{\rm O}_2 + h\nu \rightarrow  {\rm O} + {\rm O},
\end{equation}
which leads to the formation of two oxygen atoms. These atoms may
react with molecular \indexit{atmosphere!ozone}oxygen to
produce\indexit{ozone!photo-chemistry} ozone molecules (O$_3$)
\begin{equation}\label{eq:br2}
{\rm O} + {\rm O}_2 + {\rm M} \rightarrow  {\rm O}_3 + {\rm M}.
\end{equation}
Here, ${\rm M}$ represents a 'third body' ({\em e.g.,} N$_2$, O$_2$, Ar), which
removes the thermal energy released by this exothermic reaction.''
\ors[III:16.4] ``This photo-chemical process constitutes the only
significant ozone production mechanism above 20\,km altitude'' in
Earth's atmosphere. 
\ors[III:16.3] ``In this example, the rate of ozone production [per unit volume] is
directly proportional to the rate at which oxygen molecules are
photo-dissociated:
\begin{equation}\label{eq:br3}
\Pi({\rm O}_3) = 2 J_{\rm O2} [{\rm O}_2],
\end{equation}
where $J_{\rm O2}$ represents the photo-dissociation coefficient of
O$_2$ and $[O_2]$ the number density of this molecule. The
photo-dissociation frequency depends on the [local intensity of the
solar radiation after having traversed the higher absorbing layers
($I(\lambda,z,\chi)$ for wavelength $\lambda$, height $z$, and slant
or zenith angle $\chi$)] and on the ability of the molecule to
absorb solar photons at particular wavelengths. This last parameter is
generally expressed as a wavelength-dependent absorption cross-section
$\sigma_{\rm X}(\lambda)$, which can also vary with temperature. In
more general terms, the photo-dissociation frequency of a molecule
${\rm X}$ is expressed as an integral over all wavelengths that
contribute to the decomposition of the molecule. The upper bound of
this integral corresponds to the minimum energy required to break the
molecular bond. The probability that the absorption of a photon leads
to the dissociation of molecule ${\rm X}$ is expressed by the quantum
efficiency $\eta_{\rm X}$, which also varies with wavelength and
in some cases with temperature. Thus,
\begin{equation}\label{eq:br4}
J_{\rm X} = \int \sigma_{\rm X}(\lambda,T(z)) I(\lambda,z,\chi) \eta_{\rm X}(\lambda,T(z)){\rm d}\lambda.
\end{equation}
\activity{{\em Show:} Work through the units of Eqs.~(\ref{eq:br3})
  and~(\ref{eq:br4}) to show that $\eta_{\rm X}$ is an efficiency
  per unit energy per unit wavelength.} The solar actinic flux $I$
must be calculated by a radiative transfer model that accounts for (1)
absorption processes, (2)
multiple scattering by air molecules and atmospheric particles, (3)
cloud radiative transfer and (4) surface reflection. When considering
upper and middle atmosphere processes, the most important contribution
to photo-dissociation is the direct solar flux, so that the value of the
actinic flux can be approximated by considering only absorption
processes. In the lower atmosphere, multiple scattering and
specifically cloud effects cannot be ignored. [\ldots] The depth of
penetration of solar radiation varies substantially with wavelength
(Figure~\ref{fig:br4}). [\ldots]

The relative amplitude of the \indexit{atmosphere!variable F/E/UV
  flux}changes in the solar flux over the 11-yr solar cycle or the
27-d mean synodic solar rotation period decreases with increasing
wavelengths (see Figures~\ref{lean:f1} and~\ref{lean:f3}) and, as a
result, the influence of solar variability is considerably more
pronounced in the upper atmosphere [(where EUV and FUV are absorbed)]
than in the lower layers [(where longer wavelength radiation is
absorbed)]. Strong solar signals associated with the solar cycle are
visible in the thermospheric temperature and air density, with
impacts, for example, on satellite drag [(and, of course, on
ionospheric densities)]. Substantial changes have also been reported
in the concentration of nitric oxide (NO); these changes, however, are
also related to the modulation of energetic particle precipitation
associated with geomagnetic activity. Solar-related changes in the
temperature, water vapor and polar mesospheric clouds have also been
reported in the mesosphere. In the stratosphere, solar-driven changes
in temperature and ozone concentrations have been observed. The
influence of solar variability in the troposphere is [touched upon in
Ch.~\ref{ch:evolvingplanetary}]. A major forcing function for many of
these changes is the variation of
\indexit{photo-dissociation}photo-dissociation rates. Together with
the solar-induced changes in atmospheric heating resulting from the
absorption of solar radiation by ozone and molecular oxygen,
atmospheric models designed to simulate the response of the atmosphere
account for the changes in the photo-dissociation coefficients of the
different chemical compounds.''

%\section{Ozone chemistry in the stratosphere}\label{sec:br4}

Because ozone is an efficient absorber of solar UV radiation in the
stratosphere it has received much attention. Having been generated by
reaction (\ref{eq:br2}), ozone is in principle
\ors[III:16.4] ``photo-dissociated
\begin{equation}\label{eq:br5}
{\rm O}_3 + h\nu \rightarrow  {\rm O} + {\rm O}_2,
\end{equation}
but, in most cases, this reaction does not constitute a net loss for
stratospheric ozone because the oxygen atoms that result from this
photo-decomposition usually recombine with molecular oxygen
(reaction~\ref{eq:br2}) to reproduce ozone. The net loss of ozone
results from the reaction between oxygen atoms and ozone molecules
that produce two oxygen molecules
\begin{equation}\label{eq:br6}
{\rm O} + {\rm O}_3 \rightarrow  2 {\rm O}_2.
\end{equation}
The simple scheme presented here provides a first-order description of
the ozone chemistry in the stratosphere and mesosphere. Photo-chemical
models that account only for [the above] reactions tend to
substantially overestimate the concentration of ozone in the middle
atmosphere, as shown by numerous atmospheric observations. The
discrepancy can be eliminated by considering several additional
reactions that catalyze ({\em i.e.,}  accelerate) the net loss mechanism
represented by reaction (\ref{eq:br6}). [T]he presence of the hydrogen
atoms and hydroxyl radicals, produced in the upper atmosphere from the
photo-dissociation of water vapor (H$_2$O), could generate an efficient
catalytic cycle such as
\begin{eqnarray}\label{eq:br7}
{\rm H} + {\rm O}_3 & \rightarrow &  {\rm OH} + {\rm O}_2, \label{eq:br7a}\\
{\rm OH} + {\rm O} & \rightarrow & {\rm H} + {\rm O}_2. \label{eq:br7b}
\end{eqnarray}
[T]he most effective ozone destruction in the stratosphere results
from a catalytic cycle involving nitrogen oxides
\begin{eqnarray}
{\rm NO} + {\rm O}_3 & \rightarrow &  {\rm NO}_2 + {\rm O}_2, \label{eq:br8a} \\
{\rm NO}_2 + {\rm O} & \rightarrow &  {\rm NO} + {\rm O}_2. \label{eq:br8b}
\end{eqnarray}
NO is produced in the stratosphere by the oxidation of nitrous oxide
(N$_2$O), a long-lived compound released from soils by bacterial
activity. It can also be produced in the upper layers of the
atmosphere by the dissociation and ionization of molecular nitrogen
(N$_2$) by energetic particles.

Additional destruction mechanisms must be considered including
catalytic processes involving halogen compounds\indexit{ozone!role of
bromine, chlorine} including chlorine (Cl) and bromine (Br). For example
\begin{eqnarray}
{\rm Cl} + {\rm O}_3 & \rightarrow &  {\rm ClO} + {\rm O}_2, \label{eq:br9a} \\
{\rm ClO} + {\rm O} & \rightarrow &  {\rm Cl} + {\rm O}_2. \label{eq:br9b}
\end{eqnarray}
Before the 1960s, the contribution of this cycle was relatively
small. However, its importance has grown in the last decades as the
atmospheric abundance of Cl has increased steadily due to the
production of industrially manufactured chlorofluorocarbons
(CFCs). The atmospheric lifetime of CFCs varies typically from 50 to
100 years, so that anthropogenic chlorine will remain for several
decades in the stratosphere.

In the cold polar regions, and
specifically in Antarctica, ozone can be efficiently destroyed in a
layer between 12\,km and 25\,km where polar stratospheric clouds are
formed. The solid or liquid tiny particles inside these thin and often
invisible clouds that are present during winter provide surfaces for
heterogeneous chemical reactions to operate. Chemical chlorine
reservoirs such as HCl and ClONO$_2$, which are very slow to react in
the gas phase, are rapidly converted on the surface of these cloud
particles to form less stable molecules such as Cl$_2$ or HOCl. Large
quantities of reactive chlorine atoms (Cl) are liberated via
photo-dissociation as soon as the Sun returns in early spring. This chlorine
activation leads to rapid ozone destruction with the formation of the
springtime Antarctic ozone\indexit{ozone!hole} hole in September and
October. These mechanisms are less efficient in the Arctic, where the
winter temperature is usually $10^\circ-15^\circ$C higher than at the
opposite pole, and the presence of polar stratospheric clouds is
therefore less frequent.

A full description of the ozone behavior requires that large-scale
transport processes are taken into consideration, specifically in the
lower stratosphere, where the photo-chemical lifetime of this molecule
becomes much longer than the time constant associated with
transport. Below approximately 25\,km altitude, ozone can be regarded
as a quasi-inert tracer that is more sensitive to advection and mixing
processes than to photo-chemical transformations. This highlights why
the global ozone distribution in the atmosphere is strongly affected
by the meridional circulation, and specifically why the ozone column
abundance reaches a maximum value at high latitudes at the end of the
winter. The poleward meridional circulation transports ozone towards
the Arctic where it accumulates from December to April before it is
slowly destroyed by photo-chemical processes after the Sun returns in
early spring. [\ldots] The same dynamical process occurs in the
southern hemisphere with a lag of 6 months. However, ozone does not
easily penetrate poleward of 60$^\circ$S due to the existence of a
strong dynamical barrier provided by the intense southern polar
vortex. The ozone maximum is therefore located in a latitude band
located at about 60$^\circ$S. Large-scale planetary waves that
characterize the northern hemisphere winter dynamics do not allow the
northern hemisphere polar region to be isolated from lower latitudes
as is the case in the less dynamically disturbed Southern
hemisphere stratosphere. The ozone maximum in the Northern hemisphere
is thus located near the Pole.''

\section{Maintaining ionospheres}
\subsection{Ionization}\label{sec:ss2}

\ors[III:13.2] ``The\indexit{ionosphere!photo-ionization} ionosphere is
created by ionizing radiation, including extreme ultraviolet (EUV)
and\indexit{photo-ionization} X-ray photons from the Sun, and [--~in
magnetized planets~--] corpuscular radiation that is mostly energetic
electrons, [which for Earth occurs mostly] at high magnetic latitude as \indexit{aurora}auroral
'precipitation.'  The solar photon output at these wavelengths, from
$\sim$10\,\AA\ to the H\,I~Lyman-$\alpha$ line \regfootnote{Ions are
  denoted by their electrical charge, such as doubly-ionized C:
  C$^{2+}$. The line spectrum of such an ion is identified by a roman
  numeral that is one higher than the ionization charge, so the
  spectrum of C$^{2+}$ is written in shorthand as C~III; the numeral I
  is reserved for the spectrum of the neutral species, {\em e.g.,} C\,I for
  neutral carbon. Some spectral sequences have a proper name
  associated with them: for example, the H~I Lyman sequence is a
  series of spectral lines absorbed or emitted when excited electrons
  transition from or to the ground state, respectively.} at 1216\,\AA,
varies by factors ranging from $\sim$2 to $>100$ over the 11-yr solar
activity cycle ({\em cf.}, Fig.~\ref{lean:f1}), and is additionally variable
on shorter time scales, including especially the 27-d solar rotation
period.  This causes dramatic variations in the temperature and
density of the thermosphere and ionosphere.  Changes in the solar wind
and in the interplanetary magnetic field also affect the
thermosphere/ionosphere through geomagnetic perturbations that result
in transfer of energy from the magnetosphere, both in the form of
auroral \indexit{aurora}particle ionization and in the form of heat from the resulting
currents imposed in the polar regions.  An additional form of energy
transfer is the generation of energetic electrons released in the
ionization process.  These electrons, referred to as photo-electrons
in the case of photo-ionization and secondary electrons in the case of
particle-impact ionization, have enough energy to excite, dissociate,
and further ionize the neutral atmosphere as well as heat the ambient
plasma.  Solar ionization and its byproducts provide most of the
ionization and heating of the thermosphere, and account for most of
its 11-yr cyclic and 27-d rotational variation, but geomagnetic
activity accounts for much of the shorter term variation on time
scales from hours to days.

The details\indexit{solar!irradiance!penetration depth} of
ionospheric formation can be explained
through\indexit{ionosphere!photo-ionization} examination of the
photo-ionization and photo-absorption cross sections of thermospheric
constituents.  The ionization continua of N$_2$, O, and O$_2$ all peak
in the vicinity of 600\,\AA\ at tens of Megabarns
(1\,Mb\,$=10^{-18}$\,cm$^2$).  This causes their energy to be
deposited largely in the $F_1$\indexit{ionosphere!$F_1$ region} region
(compare [Figs.~\ref{fig:br4} and~\ref{fig:ss1}, and see
Eq.~\ref{eq:chapmanprofile} and the text preceding it).] Short-ward of
600\,\AA, cross sections decrease and the radiation penetrates to lower
altitude.  At Earth, the intense\indexit{ionosphere!$E$ region} solar He\,II
emission at 304\,\AA\ deposits most of its energy near 150\,km and
$10-100$\,\AA\ soft X-rays can penetrate to 100\,km.  Most of the $E$
region is produced by longer wavelength radiation,
particularly the C III line at 977\,\AA\ and the H I Lyman-$\beta$ line
at 1027\,\AA.  These do not have enough energy to ionize N$_2$ and O
but penetrate through gaps in the N$_2$ absorption spectrum to ionize
O$_2$ to O$^+_2$.  Longward of 1030\,\AA, only the important minor
species NO has a low enough ionization potential to be ionized by
solar radiation.  H~{\sc I} Lyman-$\alpha$ happens to fall at a low
point in the O$_2$ absorption spectrum and so penetrates below 90\,km,
where ionization of NO to NO$^+$ and subsequent products create the
$D$\indexit{ionosphere!$D$ region} region.  Thus, while the Chapman
production function [(Eq.~\ref{eq:chapmanprofile})] is approximately
correct for any species at each wavelength, ionized regions are
created by the superposition of many such functions
[\ldots]''\activity{{\em Show:} Indicate the various wavelengths mentioned in
  Sect.~\ref{sec:ss2} on Figure\,\ref{fig:br4}, add the equivalent wavelengths
  for the first ionization energies of H, C, N, O, N$_2$, O$_2$ and
  CO$_2$, and indicate height ranges of the three main layers of the
  Earth atmosphere as shown in Figure\,\ref{fig:fr1}. \mylabel{act:ionoenergies}}

\subsection{Recombination}\label{sec:ss3}

\ors[III:13.3] ``Positive ions have\indexit{ionosphere!recombination}
generally fast collision rates with electrons, so one would suppose
that ionospheric production would be balanced by recombination and
that the ions would be short-lived after sunset.  However, atomic ions
colliding with electrons have the problem common to all two-body
reactions that a single atom is unlikely to result, because there is
nothing to carry away surplus kinetic energy.
Photon\indexit{radiative!recombination} emission following collision
of an atomic ion with an electron can stabilize\indexit{dissociative
  recombination} the resulting atom; this radiative recombination is
quite slow, with\indexit{charge exchange} rate coefficients of the
order of $10^{-12}$\,cm$^3$\,s$^{-1}$.  Although\indexit{atom-ion
  interchange} radiative recombination occurs and is important in the
highest reaches of the ionosphere, it is insufficient as a loss
mechanism for ions and electrons given their observed $F$ region
densities.  Because the solar ionization frequency is
$\sim 10^{-6}$\,s$^{-1}$ at 1\,AU, ion densities would be several
orders of magnitude larger than observed if radiative recombination
were the only loss mechanism.  [What commonly happens is that atomic
ions yield their charge to molecular ions in order to undergo rapid
dissociative recombination, while in addition there is loss through
diffusive transport.]  {\em Dissociative recombination}, schematically
$XY^+ + e^- \rightarrow X + Y$, has rate coefficients of the order of
$10^{-7}$\,cm$^3$\,s$^{-1}$ and is the fundamental loss mechanism for
ions in dense planetary ionospheres.  In order for an atomic ion to become a
molecular ion, {\em atom-ion interchange}, schematically
$X^+ + YZ \rightarrow XY + Z^+$, or {\em charge exchange},
schematically $X^+ + YZ \rightarrow X + YZ^+$, must occur.  Charge
exchange reactions are typically fast if energetically possible, but
atom-ion interchange rates depend on the nature of the reacting
molecule, because a bond must be broken. \activity{{\em Consider} the
  similarities and differences between the charge-exchange reactions
  described here and two- and three-body gravitational interactions,
  specifically what is needed for the capture of interplanetary
  spacecraft into closed orbits, or the capture of planetary bodies as
  moons of planets. For the latter, look up the concepts proposed for
  the capture of Triton, the largest moon of Neptune, orbiting that
  planet in a retrograde orbit (which implies it has to involve a
  capture well after the formation of the planet).}
	
In regions of the atmosphere where\indexit{ionosphere!molecules and
  ionization balance} molecules dominate, recombination chemistry is
simplified because it is essentially a balance between ionization and
dissociative recombination.  [A] common approximation is the use of an
effective recombination rate coefficient $\alpha_{\rm eff}$, the ion
density-weighted average of the ion recombination rates.  In
photo-chemical equilibrium, the production rate [per unit volume]
$\Pi({\rm e}^-) = \alpha_{\rm eff} [M^+] [e^-]$, where $M^+$ is the
sum of the ions, and where square brackets denote number densities.
Assuming charge neutrality, this yields
\begin{equation}\label{eq:ionizationscaling}
[e^-]= (\Pi({\rm e}^-)/\alpha_{\rm eff})^{1/2}.
\end{equation}  
\activity{{\em Consider:} Note the equivalence between
  Eq.~(\ref{eq:ionizationscaling}) and Eq.~(\ref{eq:ionbal}) for a
  volumetric ionization rate of
  $\Pi({\rm e}^-)\propto \Phi_{\rm i}/(4\pi R^3)$. This means that
  $\alpha_{\rm eff}$ is, in effect, for a 'case B' recombination, {\em
    i.e.,} excluding the possibility that emitted photons in
  recombination are absorbed to lead to another ionization
  event. Consider what could happen to avoid that. Also see a parallel
  with the formulation of what can be viewed as the inverse in
  Eq.~(\ref{eq:loop}): for a stationary, isothermal case, the
  'incoming' volumetric heating $\epsilon_{\rm heat}$ balances the
  outgoing radiation $n_{\rm e}n_{\rm H}{f_{\rm rad}}$ in which the
  product of ion and electron densities is a measure for the number of
  collisions leading to excitation, to compare with the ionizing
  radiation in the ionosphere which balances the recombination in
  which the product of ion and electron densities is a measure for the
  number of collisions leading to recombination.}  Applying the
Chapman\indexit{Chapman!layer} production function for solar radiation
[(Eq.~\ref{eq:chapmanprofile})] to obtain $\Pi$ results in a Chapman
'layer', considering as above the caveats associated with use of that
term.  Thus,\indexit{ionosphere!ionization 'chemistry'} in molecular
ionospheres, electron density varies approximately as the square root
of the ionization rate profile.  Eq.~(\ref{eq:ionizationscaling}) is a
particularly useful form for \indexit{aurora}auroral ionization, where electrons (and
sometimes protons or heavier ions) penetrate to $\sim$100\,km or
deeper into Earth's atmosphere. \activity{{\em
    Show:} (a) Use
  Figs.~\ref{lean:f1} and \ref{fig:br4} to argue why the solar-cycle
  variability in the ionospheric electron density seen in
  Fig.~\ref{fig:ss1} is much higher at high altitudes than near the
  base of the terrestrial ionosphere. (b) Also: argue, roughly, why the contrast in
  electron densities between day and night in Fig.~\ref{fig:ss1} is
  larger at lower altitudes. \mylabel{act:ionvar}}\activity{{\em Background:} One might think that
  collisions between particles that can 'bond' and thereby be taken
  out of a population under study, such as electrons and
  positively-charged ions that combine into a neutral particle, might
  have a good analogy in how flux concentrations in the solar
  photosphere behave: the concentrations perform a random walk and in
  collisions opposite magnetic polarities 'cancel', {\em i.e.,}
  disappear from the population of magnetic charges. Yet the scaling
  behavior between the strength of the source (the total of emerging
  bipoles per unit time) and sinks (the total of canceling flux per
  unit time) is different: the square root dependence reflected in
  Eq.~(\ref{eq:ionizationscaling}) does not show up, but instead a
  near-linear dependence appears (as shown
  \href{https://ui.adsabs.harvard.edu/abs/2001ApJ...547..475S/abstract}{in
   a study
    by \citet{2001ApJ...547..475S}}). Consider the reasons: when the
  Sun's activity increases, flux concentrations grow larger by
  collision thereby countering the increase in collision frequency
  expected; larger concentrations are less mobile within the evolving
  convective motions; fragmentation and coagulation are seeking a
  balance; while in general the large-scale meridional flow aids in
  separating polarities (a process that is countered in an ionosphere
  by the tendency towards charge neutrality).}

Although the $F_2$\indexit{ionosphere!$F_2$ region} region has some of
the morphological appearance of this type of layer, it is at the wrong
altitude, and in the atom-dominated region.  It is not a Chapman layer
at all, but a result of diffusive processes.  $O^+$ has an
increasingly long lifetime as altitude increases and the molecular
fraction of the thermosphere decreases.  Above 200\,km it becomes
subject to diffusion, but is still chemically controlled up to the
peak of the $F_2$ layer near 300\,km.  Above this altitude,
ambipolar\indexit{ambipolar!diffusion} diffusion takes over, where
'ambipolar' refers to the effect of electrical attraction between the
ions and nearly massless electrons, resulting in a scale height for
O$^+$ about twice that of O.  \activity{(a) {\em Show} for the simplified case
  of a fully-ionized stationary gas that the scale height for ions is
  twice that for the corresponding atoms in a neutral atmosphere by
  combining the momentum equations for ions and electrons (see
  Eq.~\ref{eq:emomentum} and Activity~\ref{act:derivmomentum}) in
  comparison to that equation for a neutral species. (b) And remind
  yourself how this is consistently incorporated in the MHD
  equations. \mylabel{act:pressscalestation}} The $F_2$ region varies in response to thermospheric
winds and electric fields, so the mid-latitude and equatorial
ionosphere can be greatly influenced by auroral processes at high
latitudes through their effect on thermospheric dynamics.

\begin{figure}[t]
%\centerline{\hspace{3cm} Earth \hspace{6cm} Venus and Mars \hspace{2cm}}
%\centerline{\psfig{figure=figures/ssf3.eps,width=6.0cm}{\hspace{1.8cm}}\psfig{figure=figures/ssf8.eps,width=5.2cm}}
\centerline{\hspace{3cm} Earth \hspace{6cm} Venus and Mars \hspace{2cm}}
\centerline{\includegraphics[width=6.0cm]{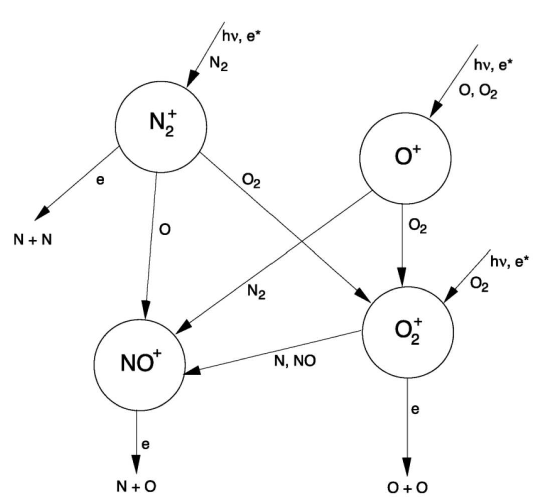}{\hspace{1.8cm}}\includegraphics[width=5.2cm]{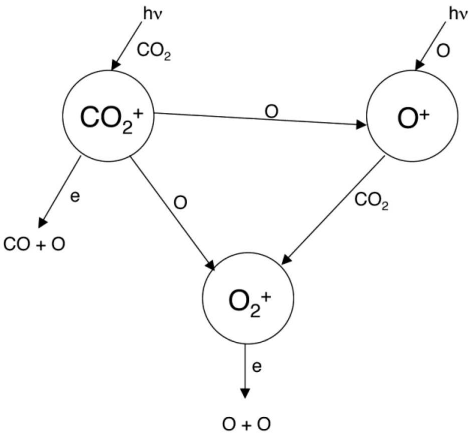}}
\caption[Ionospheric chemistry for Earth, Venus, Mars.]{Simplified 
diagram of ionospheric chemistry in the upper atmosphere of Earth
({\em left)} and of Venus and Mars ({\em right}). [The relative ionization
potential of these species is roughly indicated by the position in
this diagram: higher position indicates higher ionization potential. Fig.~III:13.4]}\label{fig:ss3}\label{fig:ss8}
\end{figure}

Figure~\ref{fig:ss3}(left) is provided as a guide to understanding the
ion-neutral chemical processes described above.  It is a greatly
simplified schematic, but contains the essential species and reactions
necessary to describe ionospheric photo-chemistry from $100 - 600$\,km.
Ionization occurs primarily on the three major species N$_2$, O, and
O$_2$ by photon, photo-electron, and auroral particle impact.  $N_2^+$
quickly loses its charge through dissociative recombination, atom-ion
interchange with O to make NO$^+$, and, at lower altitude, charge
exchange with O$_2$ to make O$^+_2$.  Thus it is always low in density
and negligible in the absence of production.  O$^+$ loses its charge
by molecule-ion interchange with N$_2$ and O$_2$.  Reaction with N$_2$ is
slow, $\sim 10^{-12}$\,cm$^3$\,s$^{-1}$, because of the high strength of
the triple $N\equiv N$ bond.  Reaction with O$_2$ is faster, $\sim
10^{-11}$\,cm$^3$\,s$^{-1}$, but there is far less O$_2$ available.
This is why O$^+$ is long-lived in the Earth's ionosphere, and why the
$F_2$ region exists.  O$^+_2$ loses its charge through dissociative
recombination or through reaction with the odd-nitrogen species NO and
N, which control the balance between O$^+_2$ and NO$^+$ in the $E$
region.  NO$^+$, daughter of all the above and the 'terminal ion',
is subject only to dissociative
recombination. Figure~\ref{fig:ss3}(left) thus describes a mechanism for
dissociating molecular gases.  Ionization goes in the top [involving
high energies], and
dissociation comes out the bottom, because dissociative recombination
is the only significant way out [of the cascade that is accessible
given the energies involved].

Figure~\ref{fig:ss3}(left) [neglects the effects of hydrogen at high
altitudes, where that] reacts by charge exchange with O$^+$ to make
H$^+$.  N$^+$ is also a significant minor ion, created by
photo-dissociative ionization of N$_2$, that is neglected here.
Doubly-ionized species are also ignored in this simplification.
Although ground state O$^+$(4s) does not have enough energy to make
N$_2^+$, metastable O$^+$(2d) and O$^+$(2p) are created by photon and
electron impact ionization, and these can charge exchange with N$_2$
to form N$_2^+$.  It is possible that vibrational excitation of
thermospheric N$_2$ can also accelerate the reaction of O$^+$ with
N$_2$.

In Earth's $E$ region,\indexit{ionosphere!$E$ region} there is a complex
interplay between O$^+_2$ and NO$^+$, due to the involvement of
odd-nitrogen species with the ion chemistry, because O$^+_2$ is
converted to NO$^+$ by reaction with NO and N.  NO in particular is
highly variable with solar activity, geomagnetic activity, and
location, so this is a complicated problem.  Older empirical and
theoretical models which assumed that O$^+_2$ is the dominant $E$
region ion, due to its production by solar H I Lyman-$\beta$
radiation, have been superseded by evidence that NO$^+$ is generally
observed to be the dominant $E$ region ion, and considerable recent
observational and modeling advances in understanding the high levels
of NO and its importance to radiative cooling as well as ion
chemistry have occurred.

In the $D$\indexit{ionosphere!$D$ region chemistry} region, ion
chemistry is entirely different due to the higher neutral density
which allows three-body attachment, particularly $2O_2 + e^-
\rightarrow O_2 + O_2^-$.  This sets in motion a complicated
negative-ion chain involving carbon, nitrogen, and hydrogen compounds,
including water, that finally results in mutual neutralization of
negative and positive ions, schematically $M^+ + M^- \rightarrow M +
M$.  NO$^+$ created by H I Lyman-$\alpha$ ionization also initiates an
involved positive-ion sequence, again involving hydration processes. [\ldots]''

\subsection{Venus and Mars}\label{sec:ss4}

\ors[III:13.4] ``The terrestrial\indexit{ionosphere!Venus}
planets\indexit{ionosphere!Mars} Venus, Earth, and Mars, are so named
because of their fundamental\indexit{Venus!ionosphere}
similarity,\indexit{Mars!ionosphere} and are presumed to have had
common elemental origins.  However, their subsequent evolution differed,
due to their differing distance from the Sun, the smaller size of
Mars, and the lack of rotation of Venus ([see
Ch.~\ref{ch:evolvingplanetary}]).  Thus, their atmospheres are
entirely different, and so are their upper atmospheres and
ionospheres.  Early exploration of Venus and Mars found that instead
of persistent, high-altitude, $F_2$-type ionospheres, these planets
had less dense, lower-altitude ionospheres ([Fig.~\ref{fig:fr2}]) that
more resembled Chapman 'layers', that were greatly attenuated at
night, and consisted mostly of O$^+_2$ and other molecular ions.  The
presence of O$^+_2$ seems especially perplexing, because Earth is the
planet we generally associate with the unusual and quite reactive
oxygen molecule. At higher altitude, O$^+$ becomes an important
species in the ionospheres of Venus and Mars, as on Earth, but at
significantly lower density and without the same degree of persistence
throughout the night.  CO$_2^+$ is a minor ion on both planets.  There
is a basic similarity in their ionospheres, despite the vastly
different density of their lower atmospheres.  [\ldots]

The reason that the ionospheres of Venus and Mars are different from
that of Earth is that the molecular compositions of their atmospheres
are different, and therefore the compositions of their thermospheres
are different ([Fig.~\ref{fig:fr2}]).
Table~[\ref{tab:brain1}] gives a simple overview of the abundance
of\indexit{Venus!atmospheric composition} the\indexit{Mars!atmospheric
composition} primary\indexit{atmosphere!composition!terrestrial
planets} atmospheric gases in the three terrestrial planets.

Aside from the large differences in surface pressure, the atmospheres
of Venus and Mars are similar in composition, and N$_2$ is an important
species on all three planets.  N$_2$ requires more energy to dissociate
than the oxygen compounds, however, so at thermospheric altitudes,
atomic oxygen becomes important on all three planets.  The Venus and
Mars thermospheres are distinguished by high levels of CO$_2$ (and
also CO) due to the underlying atmospheric composition, as shown in
[Fig.~\ref{fig:fr2}].

%\begin{figure}[t]
%\centerline{\psfig{figure=figures/ssf6.eps,width=9cm,height=6cm}}
%\caption[Typical composition of the Venus thermosphere.]{Typical
%composition of the Venus thermosphere. [Fig.~III:13.7]}\label{fig:ss6}
%%\end{figure}
%%\begin{figure}[t]
%\centerline{\psfig{figure=figures/ssf7.eps,width=9cm,height=6cm}}
%\caption[Typical composition of the Mars thermosphere.]{Typical
%composition of the Mars thermosphere. [Fig.~III:13.8]}\label{fig:ss7}
%\end{figure}

On Earth, O$^+$ is a long-lived species in the high ionosphere because
the $O^+ + N_2$ reaction is so slow and there are few other molecules
to react with to make a short-lived molecular ion.  On Venus and Mars,
the reaction $O^+ + CO_2$ is quite fast, $\sim
10^{-9}$\,cm$^3$\,s$^{-1}$, because CO$_2$ is much less strongly bound
than N$_2$.  (The triple bond in $N\equiv N$ is [among the strongest
known chemical bonds in nature, along with the triple bond in CO].)
The reaction of $O^+ + CO_2$ produces O$^+_2$, which is also produced
by the reaction $CO_2^+ + O$.  Ionization of the major thermospheric
gases on Venus and Mars, O and CO$_2$, is thus quickly converted into
O$^+_2$, which dissociatively recombines, resulting in the observed
ionospheric morphology, lacking a significant $F_2$ region.  A
simplified schematic of these processes is shown in
Figure~\ref{fig:ss8}(right).  Thus, curiously, although
life-supporting Earth is the planet associated with O$_2$, Venus and
Mars are the planets with O$^+_2$ ionospheres.

The $F_2$ ionosphere is unique\indexit{ionosphere!$F_2$ region unique
  to Earth} to Earth among the known planets.  This is due to its
peculiar atmosphere, lacking in CO$_2$, dominated by N$_2$, and
carrying its oxygen in unusual and reactive states.  Venus and Mars
have nitrogen as well, but carbon and oxygen dominate their upper
atmospheres, so it has little effect.  Earth has a significant carbon
budget, and once had much higher levels of CO$_2$ in its atmosphere,
but most of its carbon is currently locked up in the crust in the form
of carbonate rocks.  Thus, the $F_2$ ionosphere may be a recent event
in the history of Earth, an artifact of geology and biology.''
\activity{{\em Consider:} Review Figs.~\ref{fig:fr2}, \ref{fig:ss8},
  and \ref{sojka:fig3.1} and think through the dominant reactions
  described in Sects.~\ref{sec:ss3} and~\ref{sec:ss4}.  (a) Convert the
  ionization energies to equivalent wavelengths and then use
  Fig.~\ref{fig:br4} to estimate approximate depths of maximum
  penetration. (b) Considering photo-ionization and the reaction pathways
  in Fig.~\ref{fig:ss8} argue why the O$_2^+$ height profile in
  Fig.~\ref{sojka:fig3.1} might have a double peak.  Ionization energies:
  O 13.6\,eV, % 912A, 130km
  O$_2$ 12.1\,eV, % 1025A, 110km
  N$_2$ 15.6\,eV, % 795A, 170km
  CO$_2$ 13.8\,eV. % 898A 130km
  \mylabel{act:ionreactions} }

\section{Setting geospace climate}\label{sec:sojkaintro}
%\label{sojka:sec1}\label{sojka:sec2}
\subsection{Geospace climate response to solar photon irradiation}\label{sojka:sec3}

\ors[III:14.3] ``The \indexit{climate!irradiation}solar spectrum provides a [fairly] stable
irradiance of $\sim$1360\,W/m$^{2}$ ($\sim1.36\,10^6$\,erg/s/cm$^{2}$)
to the Earth's \indexit{Earth!geospace climate}upper atmosphere.  Geospace [(the region of space near
Earth down to, and including, the ionosphere-thermosphere))] responds to 
[roughly 2--6\,ppm of that] this fraction of the
solar spectrum lies between 30 and 3600\,\AA\ which extends from the
X-ray through the ultraviolet part of the spectrum.  The photons in
this spectral range may ionize atoms and molecules or may deposit
their energy directly into the thermal reservoir of the upper
atmosphere.  These\indexit{geospace climate!dependence on photon
  irradiation} processes are responsible for the 'climate' of the
geospace-atmosphere interface whose regions are labeled the ionosphere
and the thermosphere (IT).  The IT in this sense is only weakly
dependent on the Earth's magnetic field or the solar wind.  In this
'climate' scenario, the role of the terrestrial dipole field can be
viewed as defining the boundary for the plasmasphere [(the inner
magnetosphere filled with low-energy, cool plasma)] and then with the
solar wind the magnetosphere.  In the case of the Earth's sister
planets, Venus and Mars, the absence of a significant intrinsic
magnetic field confirms that the IT development has been based on
these three photo-chemical processes.

\begin{figure}[t]
\hbox{\includegraphics[height=8.0cm]{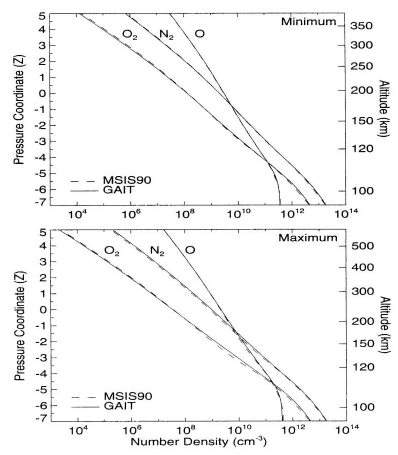}\includegraphics[height=8.5cm]{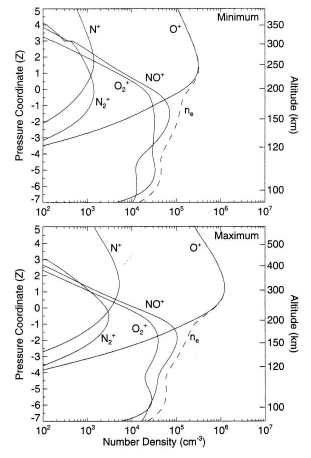}}
\caption[Model of the Earth's ionospheric density profiles.]{{\em
    Left:} Global mean number density profiles under the same
  conditions for the three major neutral species in Earth's upper
  atmosphere (N$_2$, O$_2$, and O) calculated using the GAIT model
  (solid lines) and the MSIS-90 empirical model (dashed lines).  {\em
    Right:} Global mean number density profiles for five ion species
  in Earth's ionosphere (O$^+$, NO$^+$, O$^+_2$, N$^+$, and N$_2^+$)
  and the total electron density ($n_e$) calculated using the
  [1-dimensional Global Averaged Ionosphere and Thermosphere (GAIT)]
  model. Solar minimum ({\em top}) and solar maximum ({\em bottom};
  assuming quiet geomagnetic conditions with $Ap = 4$).  The
  discontinuity observed in the NO$^+$ and N$_2^+$ profiles at $Z = 3$
  corresponds to where the photo-electron calculation stops.
  [Fig.~III:14.1;
  \href{https://ui.adsabs.harvard.edu/abs/2005JGRA..110.8305S/abstract}{source:
  \citet{2005JGRA..110.8305S}}.]}\label{sojka:fig3.1} %\label{sojka:fig3.2}
\end{figure}

\begin{figure}[t]
\centerline{\includegraphics[height=8.5cm]{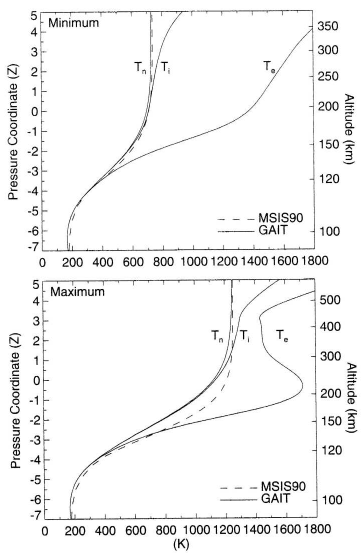}}

\caption[Model for the mean Earth-iono-/thermospheric
temp.\ profiles.]{\label{sojka:fig3.3} Global mean temperature
  profiles of Earth's ionosphere/thermosphere calculated using the
  GAIT model [(solid lines).  The three profiles correspond to neutral
  ($T_{\rm n}$), ion ($T_{\rm i}$), and electron ($T_e$) gases.  The
  dashed lines show $T_{\rm n}$ for the MSIS-90 empirical model.]
  Solar minimum ({\em top}) and solar maximum ({\em bottom}) assuming
  quiet geomagnetic conditions ($Ap = 4$).  [Fig.~III:14.2;
  \href{https://ui.adsabs.harvard.edu/abs/2005JGRA..110.8305S/exportcitation}{source:
    \citet{2005JGRA..110.8305S}}.]}
\end{figure}

Now let us explore \indexit{climate!ionosphere!response to solar
  variation}the question how geospace climate responds to extremes of
the solar photon radiation. Figures~\ref{sojka:fig3.1}
and~\ref{sojka:fig3.3} provide a comparison of the [1-D Global
\indexit{ionosphere!solar variability}Averaged Ionosphere and
Thermosphere (GAIT)] solar cycle climate of the thermosphere's neutral
densities, of the ionosphere's plasma densities, and of the neutral
and plasma temperatures respectively.  Each panel is shown as a
function of pressure level defined by
%\begin{equation}
$Z = \log{p_0/p}$,
%\end{equation}
where $p_0$ is the reference pressure of $0.5\,10^{-3}$\,dyn/cm$^2$,
%50\,$\mu$Pa.
The corresponding altitude scale is also provided. For reference to
observations, the dashed lines where present in
Figure~\ref{sojka:fig3.1} and~\ref{sojka:fig3.3} correspond to
profiles obtained from the MSIS-90 empirical model of the
thermosphere.

A simple interpretation of this solar minimum to solar maximum climate
change in the IT is that the effective IT energy deposition has almost
quadrupled; hence, the neutral atmosphere, which at these heights is
in hydrostatic equilibrium, leads to a hotter thermosphere; compare
$T_{\rm n}$ in the two panels of Figure~\ref{sojka:fig3.3}.  In turn,
the hotter thermosphere has redistributed neutrals now with relatively
higher densities at higher altitudes; compare neutral densities in the
two left-hand panels of Figure~\ref{sojka:fig3.1} using the right side
altitude scale.  [\ldots] A secondary but also important additional
effect is that the composition is also being modified because of the
different neutral masses.  For the ionosphere, the consequences can
readily be seen by comparing the two right-hand panels of
Figure~\ref{sojka:fig3.1}.  [\ldots] A comparison of the $E$ and $F$
layer peak density provides another useful scaling law, rule-of-thumb,
in that the $F$-layer density scales linearly with the appropriate
photon wavelength energy flux while that of the $E$ layer is more like
a square root dependence on energy flux [(compare with
Eq.~\ref{eq:ionizationscaling})].

That the different ionospheric layers respond differently to the solar
spectrum creates the problem of deciding what the most suitable solar
spectral representation is.  In fact, even over the limited solar
cycle energy flux range of a factor of about 4, the spectrum itself is
variable and the $E$ and $F$ layers respond to different parts of the
spectrum. \activity{{\em Consider:} Trace which part of the solar
  spectrum provides the predominant power to the $E$ and $F$ layers of
  the terrestrial ionosphere and overlapping thermospheric regions,
  and note that the power going into the $F$ layer exhibits a larger
  variation over the solar cycle than that going into the $E$
  region. See Sect.~\ref{sec:ss2} and Figs.~\ref{lean:f1},
  \ref{fig:br4} and~\ref{fig:ss1}.} The thermosphere is a somewhat
better integrator as seen [from a] study in which four distinctly
different representations of the solar spectrum were used as drivers
for the GAIT model.  As each spectral model was run over the solar
cycle range of 2--8\,erg/s/cm$^{2}$ [going into the
ionospheric/thermospheric height range,] the GAIT exospheric
temperature was determined.  The results are that the GAIT-model
thermosphere responded linearly to each spectral model, and the same
linear dependence is found for each.  Note that the exospheric neutral
temperature refers to the asymptotic, altitude-independent,
temperature found at higher altitudes, see Figure~\ref{sojka:fig3.3}
for specific solar minimum and maximum examples. The exosphere refers
to the ionospheric plasma whose composition is light ions of hydrogen
and helium that is located in altitude above the $F$
layer. 

\begin{figure}[t]
\centerline{\includegraphics[width=\textwidth]{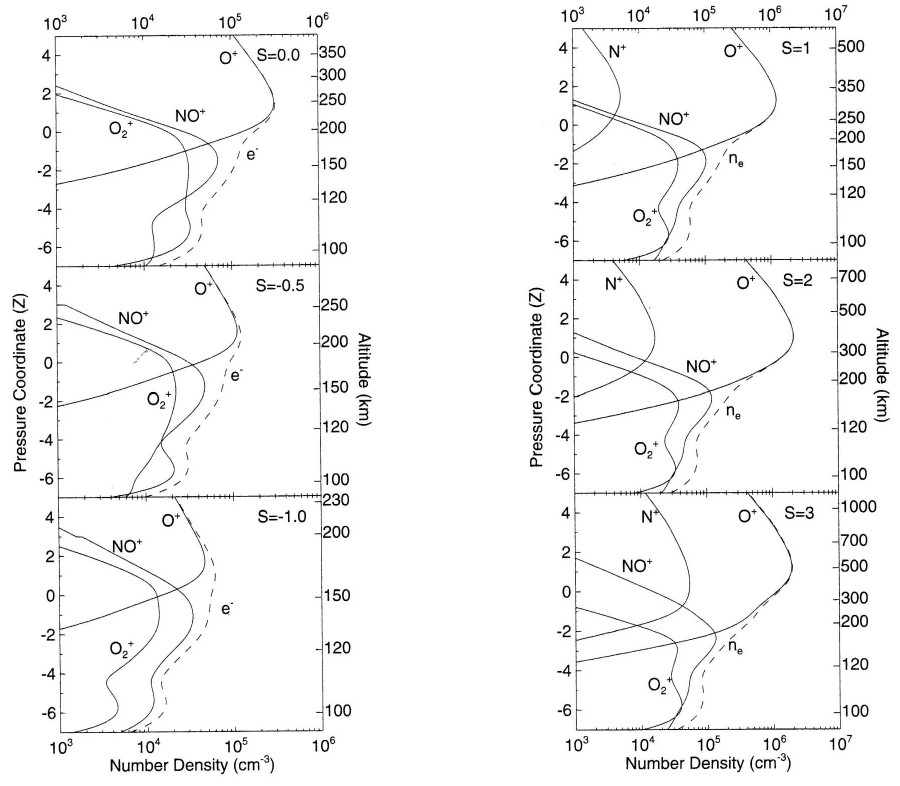}}

\caption[Global mean ion and electron profiles for Earth {\em vs.}\
solar activity.]{ Global mean concentration of the ion and electron
  (dashed line) gases in Earth's upper atmosphere, calculated using
  the GAIT model for six different levels of the solar activity
  increasing from $S_{\rm EUV} = -1.0$ to $S_{\rm EUV} = 3$, clockwise
  from the bottom left.  The profiles are plotted as a function of the
  pressure coordinate, $Z = \log{p_0/p}$, with the corresponding
  altitudes provided on the right-hand axis.  [Fig.~III:14.4;
  \href{https://ui.adsabs.harvard.edu/abs/2005JGRA..110.8305S/exportcitation}{source:
    \citet{2005JGRA..110.8305S}}.] \label{sojka:fig3.5}}
\end{figure}
%unused: \label{sojka:sec3.1}
[By combining various irradiance models computed for solar activity
levels throughout the sunspot cycle] it is possible to extrapolate how
the thermospheric exospheric temperature would trend for lower and
higher levels of the solar EUV flux. [\ldots\ The] procedure assumed
that a linear dependence existed in the relevant EUV energy flux
between solar minimum and solar maximum.  An index $S_{\rm EUV}$ is
defined to be 0 at solar minimum (energy flux of 3\,erg/s/cm$^{2}$)
and $S_{\rm EUV} = 1$ at solar maximum (energy flux of
7\,erg/s/cm$^{2}$).  Then \indexit{Maunder Minimum}Maunder-Minimum
type conditions correspond to $S_{\rm EUV}<0$ and grand maximum values
correspond to $S_{\rm EUV}>1$.  Note that the specific response to the
solar cycle of each wavelength is different, hence $S_{\rm EUV}$ is
applied to each wavelength separately to generate extreme solar
spectra.''  \ors[III:14.3.1] ``[T]he Maunder Minimum $S_{\rm EUV}$
value would be between $S_{\rm EUV} = -0.5$ and $-1.0$.
Figure~\ref{sojka:fig3.5} shows the GAIT ionospheric plasma
composition for solar minimum ($S_{\rm EUV} = 0$),
$S_{\rm EUV} = -0.5$, and $S_{\rm EUV} = -1.0$.  The earlier trends
concerning the $E$ and $F$ layer are continued as the $S_{\rm EUV}$
value [is lowered.  \ldots] The most significant ionospheric
modification during the Maunder-Minimum period is that the molecular
ion NO$^+$ peak below 200\,km becomes significant; compare the
top-left panel for $S_{\rm EUV} = 0$ and the bottom-left panel
$S_{\rm EUV} = -1$ in Figure~\ref{sojka:fig3.5}.  This additional
structure in the electron density profile is referred to as the $F_1$
layer.  Indeed, in Figure~\ref{sojka:fig3.5} the $S_{\rm EUV} = -1.0$
case almost has this $F_1$ electron density equal to that of the
higher altitude $F_2$ peak.  [\ldots] These Maunder-Minimum scenarios
provide significant problems for modern-day technology. For example:
(1) using the ionosphere to propagate radio waves over the horizon is
restricted to much lower frequencies because the maximum ionospheric
density has decreased; and (2) because the $F_1$ layer is located
significantly lower than the $F_2$ layer propagation, paths for radio
waves are also modified significantly. [On the other hand, (3)] with
less ionospheric density in the path of GPS radio waves, the adverse
role of the ionosphere in geolocation analysis is reduced.''

% \subsection{Grand maximum geospace climate}\label{sojka:sec3.2}

Now for a much more active Sun: \ors[III:14.3.2] ``[a value of
$S_{\rm EUV} = 3$ can be used to characterize] the upper range of
enhanced solar EUV flux to simulate grand-maximum type conditions.
The grand maximum existed between 1100 and 1250~CE.  An $S_{\rm EUV}$
value of 3 corresponds to doubling the solar maximum solar energy flux
from 7\,erg/s/cm$^{2}$ to just over 14\,erg/s/cm$^{2}$.  The
right-hand column in Figure~\ref{sojka:fig3.5} shows the GAIT-model
ionospheric plasma distributions at solar maximum ($S_{\rm EUV} = 1$),
$S_{\rm EUV} = 2$, and $S_{\rm EUV} = 3$ from top panel to bottom
panel.  In all cases, the $F_2$ layer is the dominant layer with O$^+$
the dominant ion.\indexit{geospace climate!during grand maximum} As
predicted from the normal solar cycle trend, this layer will rise, in
this case from 300\,km ($S_{\rm EUV} = 1$) to about 500\,km
($S_{\rm EUV} = 3$).  The $F_2$ peak density does not increase
linearly with $S_{\rm EUV}$!  Between $S_{\rm EUV} = 2$ and
$S_{\rm EUV} = 3$ the $F_2$ peak density has remained at
$2\times 10^6$\,cm$^{-3}$. \activity{{\em Show:}
  Upward traveling
  radio waves with frequencies below the plasma frequency are
  evanescent within the ionosphere and are reflected downward, thus
  enabling 'over the horizon' or 'skywave' communication. Look up the
  plasma frequency (Eq.~\ref{eq:plasmafrequency}), typical ionospheric
  electron densities within the ionosphere, and resulting values for
  the radio frequencies useful for such communication. See, {\em
    e.g.,} Fig.~\ref{fig:tsb1} for an overview of the EM spectrum with
  an indication of various radio bands (including what differentiates
  propagation of AM and FM bands).  Similarly, radio waves from
  outside Earth cannot penetrate the ionosphere at frequencies below
  that corresponding to the layer of the largest electron
  density. Estimate the lowest frequency of, say, solar radio emission
  that can be observed by a ground-based observatory. Mind the factor
  of $2\pi$. \mylabel{act:ionradio}}

This maximum in the $F_2$ peak density is by far the most significant
change in the geospace climate in response to solar photon radiation.
The processes responsible for this effect are: (i) the production of
neutral O and, hence, its concentration has non-linearly decreased at
altitudes at which the $F_2$ peak is created now as the thermosphere
heats up as $S_{\rm EUV}$ increasing from 2 to 3; (ii) the O$^+$
production rate does increase linearly as $S_{\rm EUV}$ increases from
2 to 3; and (iii) the competition between these two processes leads to
a maximum peak $F_2$ density at $S_{\rm EUV} = 2$, and then as
$S_{\rm EUV}$ increases further even a slight decrease in the peak
density.  The consequences for modern-day technologies under enhanced
solar maximum, grand maximum conditions are: (1) the
changing\indexit{space weather!impact of grand maximum} altitude of
the $F_2$ layer leads to modified radio wave propagation paths; (2)
that the peak $F_2$ density saturates only slightly above solar
maximum values implies that the 'radio' reflection characteristics of
the ionosphere are consistent with today's 'radio climate'; (3) the
impact on trans-ionospheric radio applications such as GPS geolocation
is somewhat adverse since the total electron content (the electron
density integrated over a line of sight, {\em i.e.,} a column
density)\indexit{total electron content (TEC)} continues
\indexit{total electron content (TEC)|seealso{definition}} to
increase even though the $F_2$ peak density becomes constant; and (4)
because the ionosphere is significantly more dense, the absolute
magnitude of plasma density irregularities would increase which would
lead to greater scintillation problems with radio propagation.''

% \subsection{Other geospace climate impacts}\label{sojka:sec3.3}

\ors[III:14.3.3] ``In modeling the ionosphere and thermosphere as the solar EUV
energy flux is changed, there are at least two impacts of
significance for the outer reaches of geospace.  First, assuming
that the magnetosphere is somewhat similar to the state that we
are familiar with, then the IT contributes plasma to the
magnetosphere/plasmasphere and second, the IT electrical
conductivity is a component of the magnetosphere-ionosphere (M-I)
electrical coupling.  Under the Maunder-Minimum type conditions,
ionospheric outflow of plasma into the magnetosphere/plasmasphere
will decrease because the ionospheric topside is colder and less
dense.  Under extreme conditions such as $S_{\rm EUV} = -1$, the
composition may also begin to change from atomic to molecular.  In
contrast, under $S_{\rm EUV} = 2$ and upward, during grand-maximum
conditions with hotter topside, the outflow would increase and be
very much O$^+$ dominated.  Note that in these GAIT-type modeling
studies the light ion, H$^+$, has not been included, and
therefore the remarks pertain to O$^+$ and heavier molecular
ions.  In contrast, the ionospheric conductivity changes are
smaller because the major contribution comes from the $E$ layer
whose composition remains molecular.  However, the decreasing
dayside conductivity during Maunder-Minimum conditions would raise issues
about how this impacts the M-I electric circuit response, {\em i.e.,} 
would this modify present-day concepts of voltage versus current
generator descriptions of the M-I system?  Under the grand maximum
with enhanced conductivities and also the assumption of increased
solar wind energy, would M-I coupling be characterized by
significantly enhanced currents and electric field? Both scenarios
would probably impact the morphology of \indexit{aurora}auroral displays!  This
may lead to the most significant human experience of the geospace
climate.''

\subsection{Geospace climate at earlier terrestrial ages}\label{sojka:sec4}

Now let us look at the \indexit{climate!young Earth}far larger range
of solar activity as that \indexit{Earth!geospace climate!younger Earth}evolved over the 4.6\,Gyr history of Venus,
Earth and Mars. \ors[III:14.4] ``In earlier times, the solar EUV was
more intense and, consequently, the thermosphere was much hotter,
leading to the dominance of significantly different processes.  [In
particular, we look at] an early period when atomic hydrogen was in a
blow-off phase as well as periods when high escape\indexit{geospace
  climate!variation over geological time scales} rates for the fastest
particles in the energy distribution ('Jeans escape')\indexit{Jeans
  escape process} of\indexit{atmosphere!escape!Jeans loss} heavier
species like H$_2$, He, C, N, O existed [(compare
Ch.~\ref{ch:evolvingplanetary}). S]tudies also show that IR-radiating
molecules like CO$_2$, NO, OH, etc., control the exospheric
temperature that, in turn, controls the Jeans escape rates for the
neutral constituents.  Hence, the results depend not only on a
knowledge of solar EUV but also on the contribution of molecules such
as CO$_2$ and H$_2$O in the earlier terrestrial atmosphere [that can
be addressed, for example, with] a diffusive-gravitational equilibrium
and thermal balance model to study heating of the earlier
thermosphere.

\begin{figure}[t]
\raisebox{2.7cm}{
\begin{minipage}[t]{6.2cm}
\begin{tabular}{llrl}
\multicolumn{4}{c}{Sun-like stars of different ages}\\
\hline
Name & Spectral & $P_{\rm rot}$ & Age \\
     & type     & (days)        & (Gyr)\\
\hline
EK Dra &G1.5 {\sc V} &2.7 &0.1$^*$\\
$\pi^1$ UMa &G1.5 {\sc V} &4.9 &0.3\\
$\chi^1$ Ori &G1 {\sc V} &5.2 &0.3\\
$\kappa^1$ Cet &G5 {\sc V} &9.2 &0.7\\
$\beta$ Com &G0 {\sc V} &12. &1.6\\
Sun &G2 {\sc V} &25.4 &4.6\\
$\beta$ Hyi &G2 {\sc IV} &$\sim 28$ &6.7\\
\hline
\end{tabular}
\noindent $^*$ [Another study reports] an age
of $0.03-0.05$\,Gyr.
\end{minipage}}\ \begin{minipage}[t]{6.5cm}
\includegraphics[width=6.5cm]{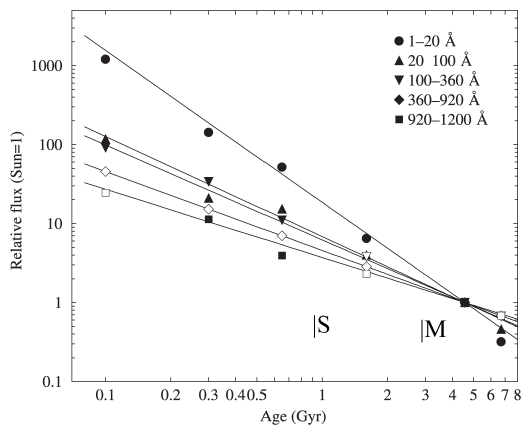}
\end{minipage}
\caption[Spectral radiance versus age of solar-type stars.]{Spectral
  radiance \indexit{atmosphere!losses!scaling}versus age of solar-type stars (identified
in the table on the left, with spectral type, rotation period, and
estimated age), in solar units.
Measurements are shown by filled symbols; missing data (open symbols)
are derived from power-law fits (solid lines)
for passbands from 1 to 1200\AA. The approximate ages for which the oldest
fossils of
single-cell microbial life (S) and multi-cellular plants and animals (M)
have been found on Earth are indicated.
[Fig.~III:2.14;
\href{https://ui.adsabs.harvard.edu/abs/2005ApJ...622..680R/abstract}{source:
\citet{2005ApJ...622..680R}}.]
\label{fig:estvarragne}}
\end{figure}\nocite{ribas+etal2005}\nocite{jarvinen+etal2007}
In an initial simulation, this model was used to evaluate the
terrestrial exospheric temperature over the past 4.6\,Gyr.  Significant
assumptions were made that the present-day composition as well as that
of the lower atmosphere up to 90\,km were the same then as they are
today.  The increased solar flux values at earlier ages [were
estimated as] summarized in [Fig.~\ref{fig:estvarragne}].

\begin{figure}[t]
%\centerline{\psfig{figure=figures/sojka4.1.eps,height=5cm,clip=}}
\centerline{\includegraphics[height=5cm]{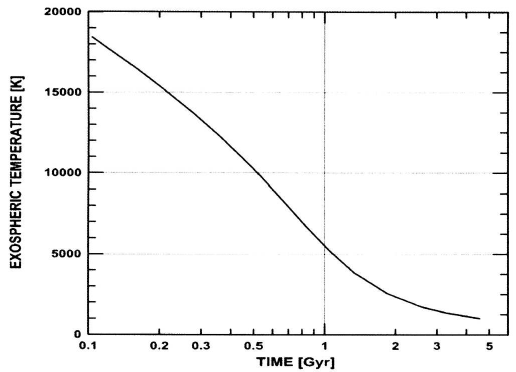}}

\caption[Evolution of the exospheric
temperature.]{\label{sojka:fig4.1}  Evolution of the exospheric
  temperature, assuming Earth's present atmospheric composition, over
  the planet's history as a function of the solar XUV flux for a
  strongly limited hydrogen blow-off rate. [Fig.~III:14.5;
  \href{https://ui.adsabs.harvard.edu/abs/2007SSRv..129..207K/abstract}{source:
  \citet{2007SSRv..129..207K}}.]}
\end{figure}
\begin{figure}[t]
\raisebox{5.7cm}{
\begin{minipage}[t]{4.8cm}
\label{tab:sojka1} {\begin{flushleft} Historical values of the solar EUV fluxes
relative to the present-day value. [Table~III:14.1] \end{flushleft}}
\begin{tabular}{|l|l|}
\hline
Time &Solar flux \\
& multiplier \\
\hline
3.5 Gyr ago  & factor $\sim $6 \\
3.8 Gyr ago  & factor $\sim$10 \\
4.33 Gyr ago & factor $\sim50$ \\
4.5 Gyr ago  & factor $\sim$100 \\
\hline
\end{tabular}
\end{minipage}}\ \hskip 3mm \begin{minipage}[t]{8cm}
\includegraphics[width=8.cm]{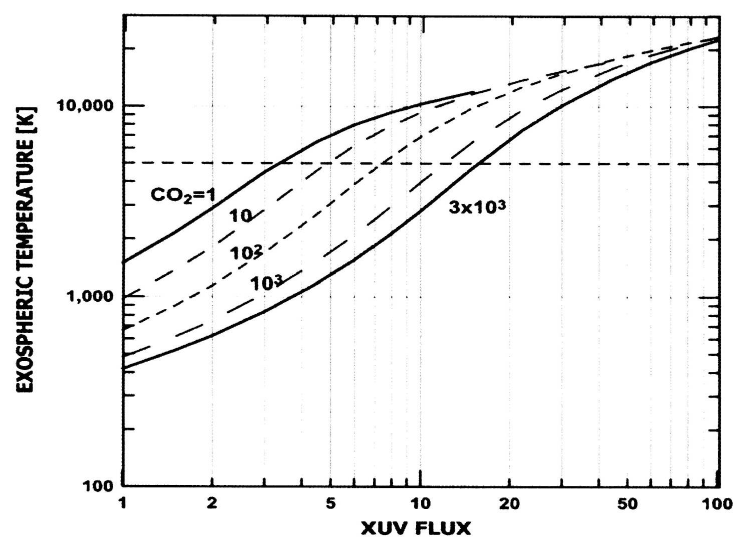}
\end{minipage}
\caption[Earth's exospheric temp.\ with CO$_2$ and
solar XUV flux.]{\label{sojka:fig4.2} Earth's exospheric temperatures
  for different levels of CO$_2$ abundance in units of PAL (Present
  Atmospheric Level: 1 PAL for CO$_2 = 330$\,ppm) in the
  thermosphere as a function of solar XUV flux.  The numbers by the
  curves correspond to CO$_2$ volume mixing ratios expressed in PAL.
  The horizontal dashed line shows the blow-off temperature of atomic
  hydrogen.  [The Table lists estimated XUV flux levels relative to
  the present day. Fig.~III:14.6;
  \href{https://ui.adsabs.harvard.edu/abs/2007SSRv..129..207K/exportcitation}{source:
  \citet{2007SSRv..129..207K}}.]}
\end{figure}

Figure~\ref{sojka:fig4.1} shows [the simulated] history of the Earth's
exospheric temperature.  Assuming that the blow-off temperature for
atomic hydrogen is about 5000\,K, the Earth's first Gyr would exhibit
a markedly\indexit{atmosphere!blow-off} different upper atmosphere
where even the atomic atoms and molecular hydrogen would be
approaching their thermal escape speeds.  [T]his simple model becomes
a rough estimate when in the life of the Sun (smoothing over time
scales long compared to solar cycles) the exospheric temperatures exceed 10,000\,K.

The major assumption that would be questioned for these earlier Earth
ages would be the density of the IR radiating molecules such as
CO$_2$.  In significantly earlier times, these would be expected to be
larger.  If this is the case, then their role in 'cooling' the
thermosphere would increase.  Defining the CO$_2$ mixing ratio
relative to present atmospheric level (PAL)\indexit{exosphere!cooling
by CO$_2$} as 1, [\ldots] Figure~\ref{sojka:fig4.2} shows how, indeed,
significant\indexit{thermosphere!temperature profile over geological
time scales} increases in CO$_2$ will cool the upper atmosphere.  In
this figure, the 'XUV flux' is the scaling ratio of earlier age solar
EUV compared to today.  The current situation is shown at unit XUV
flux.  This shows that increasing CO$_2$ by a factor of 10 (10\,PAL)
leads to a drop of almost 600\,K in the exospheric temperature from
1600\,K to 1000\,K. [\ldots\ Further work has shown that] by solar fluxes
that are about 5 times the present average EUV energy flux of
5.1\,erg/s/cm$^{2}$, the composition of the upper thermosphere will be
dramatically different from today as the Jeans escape mechanism becomes
effective for hydrogen as well as other atomic species.  [Model
studies (see III:12.3.3 for some details and references)] indicate
that at earlier ages of the Earth's upper atmosphere-ionosphere, the
response to increased solar EUV flux was a heating of this part of
geospace\indexit{ionosphere-thermosphere!conditions for young Earth}
with the following impacts on the geospace climate: (1) for
exospheric temperatures of\indexit{geospace climate!impact of
sustained EUV increase} 5000\,K and above, the upper atmospheric
composition would be dramatically different due to Jeans escape
fluxes of hydrogen and other atomic species; (2) the altitude of the
$F_2$ layer would increase to heights above 2000\,km; (3) the $F_2$
layer peak density would become constant; (4) the total electron
content of the ionosphere would increase linearly with increasing
solar EUV flux.

The geospace climate would change from an ionospheric standpoint when
the solar energy flux slightly exceeds levels of the present-day solar
maximum.  From the thermospheric point of view, the geospace climate
would change when the solar flux EUV reaches about 20\,erg/s/cm$^{2}$
(three times the present-day solar maximum values).  [\ldots]''
\activity{{\em Show:} Using information from this chapter and
  Ch.~\ref{ch:evolvingstars}, (a) estimate when the solar EUV flux dropped
  to a level that the thermospheric climate became comparable to the
  present-day state; and (b) summarize the ionospheric changes over
  geological time scales as far as the models discussed here are
  concerned. \mylabel{act:geoact}}

\subsection{Geospace climate and Earth's magnetic field}\label{sojka:sec5}

\ors[III:14.5] ``[T]he outer \indexit{climate!ionosphere!magnetic
  field}boundary of \indexit{Earth!geospace climate!magnetic field}geospace is defined to be the magnetosphere, and
specifically, the magnetopause.  It is created by the solar wind that
interacts with an intrinsic property of the Earth, the magnetic field.
Consequently, in this section questions concerning how long-term
trends of the solar wind and Earth's magnetic field are considered in
discussing the long-term geospace climate.  Of specific interest are
the conditions under which the geospace would be dramatically
changed. [\ldots]''

%\subsubsection{Geospace climate response to dipole flips}\label{sojka:sec5.1}

\ors[III:14.5.1] ``Perhaps the geospace response to flips in the Earth's
dominant dipolar field is the most frequently discussed geospace 'what
if' scenario.  Geological evidence obtained in the last century has
clearly proven that the Earth's magnetic field, especially its
dominant dipole component, has reversed many times during geological
times.  The\indexit{geomagnetic!field!reversals} most recent reversal
occurred 0.78\,Myr ago.  Prior to this reversal the 6 most recent
occurred at 0.99, 1.07, 1.19, 1.2, 1.77, and 1.95\,Myr ago.
[Reversals] occur at quite irregular intervals with the shortest time
between reversals being at the Cobb Mountain reversal pair separated
by only about 10,000 years. [\ldots\ R]eversals occur when the dipole
field strength is relatively weak.

[\ldots] The specific 'N-S' or 'S-N' dipole orientation itself
would not introduce significant geospace climate changes.  Perhaps,
the most obvious would be that the solar wind northward versus
southward reconnection morphology would be reversed.  What is
significantly more important would be the magnitude of the Earth's
field and the orientation of its dipole component.''

%\subsubsection{Geospace climate dependence on dipole strength}\label{sojka:sec5.2}

\ors[III:14.5.2] ``Over\indexit{geomagnetic!field!dipole strength
variation} the past 100 years, the Earth's dipole moment has decreased
by about 5\% from
%$8.3 \times 10^{22}$ to $7.8 \times 10^{22}$\,Am$^2$,
$8.3 \times 10^{25}$ to $7.8 \times 10^{25}$\,erg/G,
while three-thousand years ago, it was at almost
%$12 \times 10^{22}$\,Am$^2$,
$12 \times 10^{25}$\,erg/G,
at its highest value during the Holocene era.
From [Ch.~\ref{ch:flows} it is clear that] it is the balance between
the Earth's magnetic field and the solar wind pressure that determines
the outer boundaries of the magnetosphere/geospace.  Hence, a larger
(or smaller) dipole moment with otherwise the same solar wind conditions
would increase (or reduce) the size of geospace.  In\indexit{geospace
climate!dependence on dipole strength} turn, this would reduce
(or increase) the size of the polar cap, and auroral regions would move
poleward (or equatorward).  However, a 5\%\ change in the [virtual axial
dipole moment (VADM)] would probably not have a dominant impact on
geospace because the solar wind pressure varies by more than this over
its normal solar cycle.  Considering earlier times when the VADM did
decrease to values as low as, if not lower than,
%$2 \times 10^{22}$\,Am$^2$,
%$2 \times 10^{25}$\,erg/G,
[a quarter of the present-day value,]
the geospace climate may well have been dramatically
different, especially during solar maximum type conditions.  The
magnetosphere would have been severely reduced, and in volume regions
such as the plasmasphere it would have almost been reduced to
ionospheric altitudes and in the 'open' polar regions would extend to
mid-latitudes.  The effectiveness of plasma sheet energization
processes would also have been changed, causing impacts on ring
currents, electrojets, as well as the visible \indexit{aurora}aurora.  Perhaps, the
energy transfer to geospace would simply decrease as the
magnetosphere's cross section to the solar wind decreases, and
consequently, all internal energy processes would be similarly scaled
down.

The extreme scenario of the dipole reversal is the idea that the VADM
for a time period is extremely small, approximately zero.  If the
higher-order multipole terms are also negligible, then the Earth's
atmosphere is unprotected.  But this is the Venus and Mars type
scenario and extensive analysis has been done on these planetary
atmospheres.  At present, the scientific
techniques that provide information on the reversals are unable to be
specific on this question, but a near-zero magnetic field appears to last no
longer than a few thousand years, if that.''

\begin{figure}[t]
%\centerline{\psfig{figure=figures/Tilt0.eps,width=6.cm,clip=}\psfig{figure=figures/Tilt30.eps,width=6.cm,clip=}}
\centerline{\includegraphics[width=6.cm]{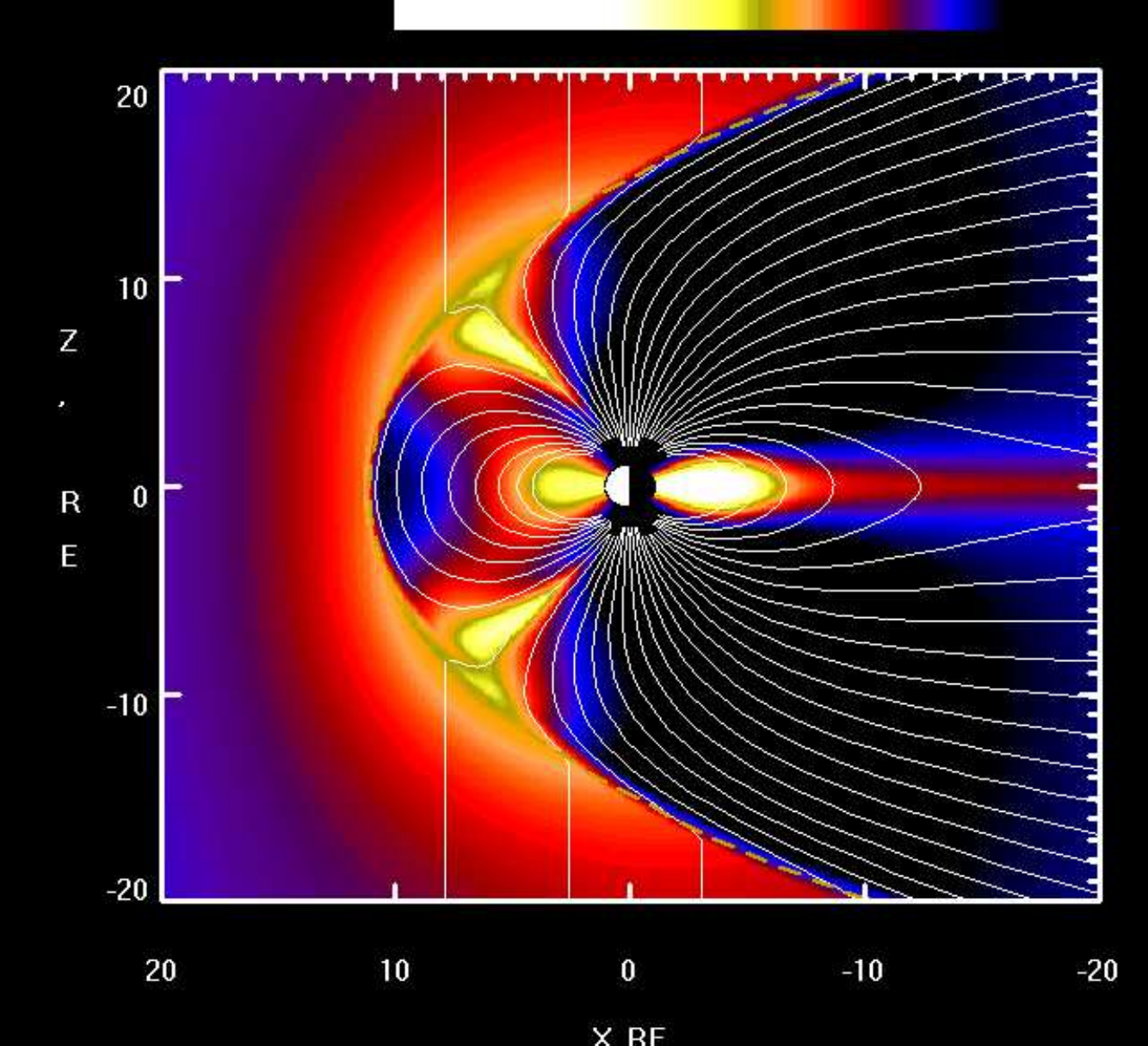}\includegraphics[width=6.cm]{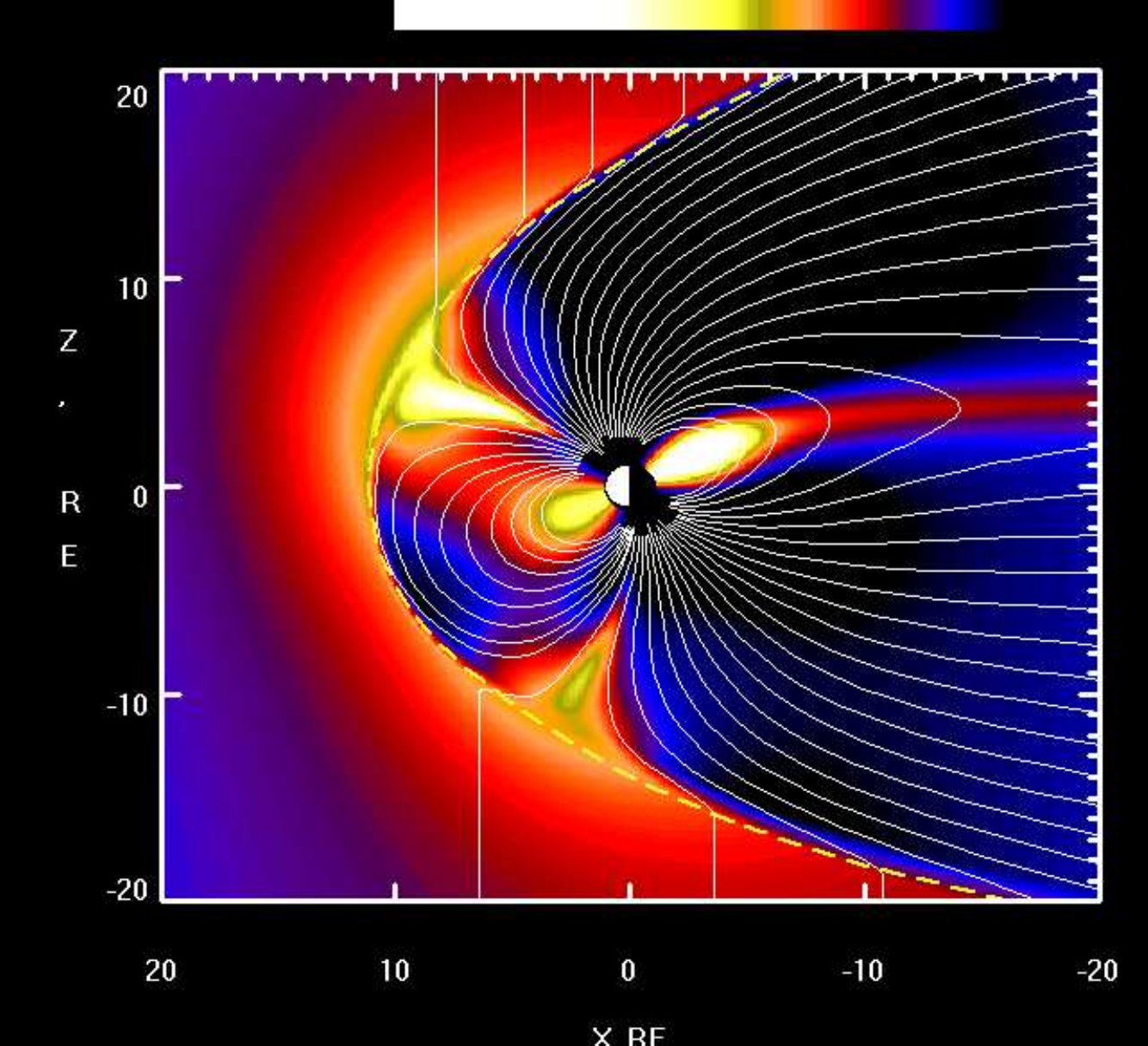}}

\caption[Field lines for 
untilted and tilted planetary dipoles in the solar
wind.]{\label{sojka:fig5.4}   Field lines plotted in the noon-midnight
  meridian plane for an untilted planetary dipole embedded in the
  solar wind ({\em left}) and a dipole tilted in the noon-midnight
  meridian by 30$^\circ$ ({\em right}). The colors show the difference
  in the magnitudes of the field in the $x-z$ plane of a model field
  including the effects of a solar wind compared to that of a dipole
  field; positive differences are shown in blue to black, negative
  differences in red through yellow to
  white. [\href{https://ccmc.gsfc.nasa.gov/modelweb/magnetos/data-based/whatnew4.html}{source
  NASA's CCMC}; see also Fig.~III:14.13] \colorfig }
\end{figure}
%\begin{figure}[t]
%\centerline{\psfig{figure=figures/sojka5.4top.eps,width=6.cm,clip=}\psfig{figure=figures/sojka5.4bottom.eps,width=6.cm,clip=}}
%
%\caption[Field lines for 
%untilted and tilted planetary dipoles in the solar
%wind.]{\label{sojka:fig5.4}   Field lines plotted in the noon-midnight
%  meridian plane for an untilted, closed planetary dipole embedded in the solar wind (left) and the dipole tilted in the noon-midnight meridian at 35$^\circ$ (right). [Fig.~III:14.13]}
%\end{figure}
%\subsubsection{Geospace climate and the orientation of the Earth's dipole}\label{sojka:sec5.3}

\ors[III:14.5.3] ``The scenarios for the geospace climate dependence on
tilt angle between the Earth's rotational\indexit{geomagnetic!dipole!geospace effect of tilt} axis\indexit{geospace climate!effect
of dipole tilt} and dipole axis provide vivid geometries of geospace
regions such as the plasmasphere, plasma sheet, cusps, auroral zones,
and open/closed field line regions.  For extreme tilt angles, a
significant question would be how rapidly these region can evolve and
replenish themselves.  [\ldots] Figure~\ref{sojka:fig5.4} provides a
pair of noon-midnight cross sections through the Earth's magnetosphere
for a 0$^\circ$ and 30$^\circ$ tilt.  In the left panel, all the
conventional magnetosphere regions can be identified and their
evolution over a day would be shown at each time, as seen in this
panel for a constant solar wind.  Our present-day tilt scenario is
somewhat different; in the northern hemisphere it is approximately
10$^\circ$ while in the southern hemisphere it is almost 15$^\circ$.

However, even with this tilt, the fundamental magnetospheric regions
found in Figure~\ref{sojka:fig5.4}(left) are present all day with
relatively small wobbles in the geocentric-solar-ecliptic coordinate
system (GSE; $x$, Earth-Sun line; $z$, ecliptic north pole) of this
figure.  Both cusps are dayside and wobble in latitude.  The
plasmaspheric equatorial plane is that of the 'average' dipole and
would wobble in 24 hours about the GSE-$x$ axis.  Even today, the
concept of the plasmasphere's 'average' dipole orientation is not
fully explored since it is well known that the Earth's equatorial
fields are not well represented by a pure dipole component.

Over time scales of decades and more, the tilt angle as well as its
geographic longitude wander.  Indeed, this has been identified as a
major factor in complicating the historic \indexit{aurora}auroral observation data
base.  For example, when an aurora was observed at lower mid-latitudes
as described in Sect.~\ref{sec:sojkaintro}, was this due to an
especially strong or geo-effective solar storm (CME) or did the Earth's
dipole tilt have a particularly large value at that time, making this
terrestrial location a much higher geomagnetic latitude?

The right-hand panel in Figure~\ref{sojka:fig5.4} shows the
magnetospheric geometry for a specific 'UT' during northern-hemisphere
summer solstice when the tilt can reach some 30$^\circ$.  At other times of
the day, as the Earth rotates, this geometry changes significantly.
Six hours earlier or later, the $x-z$ GSE cross section might look
similar to the symmetric geometry in the left panel.  However, the
cusps would be displaced in the $y$ GSE direction and the plasma sheet
would have a large tilt in $y-z$ GSE cross section.  As the tilt angle
increases beyond 35$^\circ$, would the normal diurnal
independences of the magnetospheric morphologies remain?  For example,
would auroral zones still be referred to as a north and a south
auroral oval?  In the extreme case of a tilt approaching 90$^\circ$,
does the plasma sheet in the $x-z$ GSE cross section have two plasma
sheets at certain UTs?  Under these conditions with the same VADM and
solar wind, dramatically different geospace climate would be observed
in the form of auroral sightings as well as terrestrial magnetic field
records of the electrojets and ring currents.''

\begin{figure}[t]
%\centerline{\psfig{figure=figures/sojka5.5.eps,height=5.7cm,clip=}}
\centerline{\includegraphics[height=5.7cm]{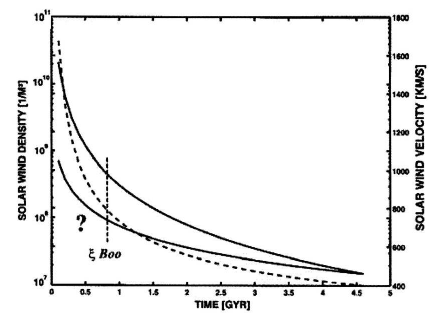}}
\caption[Evolution of the min. \&\ max. stellar wind densities
at 1\,AU.]{\label{sojka:fig5.5} Evolution of the observation-based
  minimum and maximum stellar wind densities (in units of
  m$^{-3}$) scaled to 1\,AU (left scale; solid lines) obtained from
  several nearby solar-like stars.  On the right scale is the solar
  wind speed for the stellar wind evolution (dashed line).  [The star
  $\xi$\,Boo sits on the 'wind dividing line'; mass loss appears
  strongly reduced for stars younger than that, contrary to the simple
  extrapolation shown here, see Fig.~\ref{Wood_f8}. Fig.~III:14.14;
  \href{https://ui.adsabs.harvard.edu/abs/2007SSRv..129..245L/abstract}{source:
  \citet{2007SSRv..129..245L}}.]}
\end{figure}
\subsection{Geospace climate dependence on the solar wind}\label{sojka:sec5.4}

What are the effects of the long-term \indexit{climate!solar wind}evolution of the solar wind on
the geospace climate? \ors[III:14.4] ``Most
\indexit{magnetosphere!solar wind}of today's knowledge of the
early Sun's history, \indexit{Earth!geospace climate!solar wind}normally referred to times that the Sun reached
its zero-age \indexit{main sequence!stellar wind}main sequence (ZAMS) has been obtained from studies of
Sun-like stars, {\em i.e.,}  main-sequence G and\indexit{solar!wind!variation
on geological time scale} K stars.  [The] time dependences for the
solar wind velocity ($v_{\rm sw}$) and density ($n_{\rm sw}$) at 1\,AU
[have been coarsely approximated by]:
\begin{equation}\label{eq:windintime}
v_{\rm sw}= v_\ast \left [ 1+ {t \over \tau_{\rm sw}}\right ]^{-0.4} \,\,\,\,;\,\,\,\,
n_{\rm sw}= n_\ast \left [ 1+ {t \over \tau_{\rm sw}}\right ]^{-1.5},
\end{equation}
where $v_\ast = 3200$\,km/s, $n_\ast = 2.4 \times 10^{4}$\,cm$^{-3}$
and $\tau_{\rm sw} = 2.56 \times 10^7$\,yr.  [\ldots]'' \sactivity{$\circledS$ {\em
    Show:} (a) Assuming a
  similar geomagnetic field, use the expressions in
  Eq.~(\ref{eq:windintime}) to derive an estimate of the magnetopause
  distance over time. (b) Show that for a young Sun this comes down to
  $\sim 1.25 R_\oplus$ (with Eq.~\ref{eq:PB}). \mylabel{act:mpdt} \solution{mpdt}}
% using these relationships with the assumption that the geomagnetic
% field remains unchanged, yields a subsolar magnetopause distance
% (with Eq.~\ref{eq:PB}) that is as small as $\approx 1/8$ times the
% present-day value for a young Sun, or $\sim 1.25 R_{\rm E}$.

Figure~\ref{sojka:fig5.5} shows the large spread in the range of the
$n_{\rm sw}$ dependence since the Sun reached ZAMS about 4.6\,Gyr ago.
Note that the [above approximations and the curves] shown in
Figure~\ref{sojka:fig5.5} have as a reference a present-day
$n_{\rm sw}$ of 20\,cm$^{-3}$ ($2 \times 10^7$\,m$^{-3}$) and a
$v_{\rm sw} = 400$\,km/s.  Today's solar cycle and solar storms have
periods when the density can almost be a factor of 10 higher and the
velocity reaches 1000\,km/s.  These enhanced conditions are associated
with storms and superstorms in geospace that can persist for days
while the solar wind remains perturbed.  If the Earth's intrinsic
magnetic field were then as it is today, would geospace at 2 to 3\,Gyr
ago be in a continuous superstorm state?  Figure~\ref{sojka:fig5.5}
shows that at these times the solar wind's pressure would permanently
be at, or exceed, superstorm solar wind conditions.  Would the auroral
phenomena be permanent displays and exist to very low latitudes, or
would perhaps M-I coupling require an unsustainable flow of
ionospheric plasma into the magnetosphere?  The past geospace climate
over the Holocene, {\em i.e.,} the human time period, was not significantly affected
by long-term changes in the solar wind while at very early ages it
could well have been a very illuminating dynamic M-I coupling
environment.''

\clearpage

\chapter{{\bf Cosmic rays and magnetic fields over time}}
\label{ch:evolvingexposure}%14
{\narrower\narrower{
{\bf Chapter topics:}
\begin{itemize}
  \customitemize
\item Evolution of terrestrial energetic-particle exposure
\item Radionuclides as proxies of long-term magnetic variability
\item Transport and deposition of radionuclides
\item Exposure to supernovae
\end{itemize}

\noindent{\bf Key concepts:}
\begin{itemize}
  \customitemize
\item Cosmogenic radionuclides
\item Tree rings, ice sheets, and rocks as records
\item Cut-off rigidity
\item Force-field approximation
\end{itemize}

}}

\section{Introduction}
 Energetic particles can affect electronics components  
and presents a health hazard for astronauts, particularly
when outside the magnetosphere, such as en route to the Moon or to
Mars (as described in Chs.~II:13 and~II:14).  The energetic
particles discussed in this chapter originate at the Sun, in the solar
wind, and in the Galaxy beyond the heliosphere. Observation of their
variability tells us about their sources and about conditions en route
to Earth.

Collisions of the most energetic
among these particles with bodies in the Solar System lead to the
formation of radionuclides that subsequently decay with half-lives of
various durations. If such radionuclides are stored in suitably
'stratified' natural archives --~such as in long-lived snow deposits
or growth rings of trees~-- their concentrations measured through such
archives shine a light on intensities and variability in times from
before instrumental records. Even unstratified 'archives', such as
lunar and meteoric rocks, serve as dosimeters in which depth profiles
of cosmogenic radionuclides contain information on particle energies
and fluxes, and the different decay time scales some information on
the exposure history.

In the present chapter, we focus on changes in exposure over time
scales up to billions of years, and on what these changes tell us
about solar activity, galactic cosmic rays, the state of the
heliosphere, and the terrestrial magnetic field.

\begin{figure}[t]
\centering
\includegraphics[width=7cm]{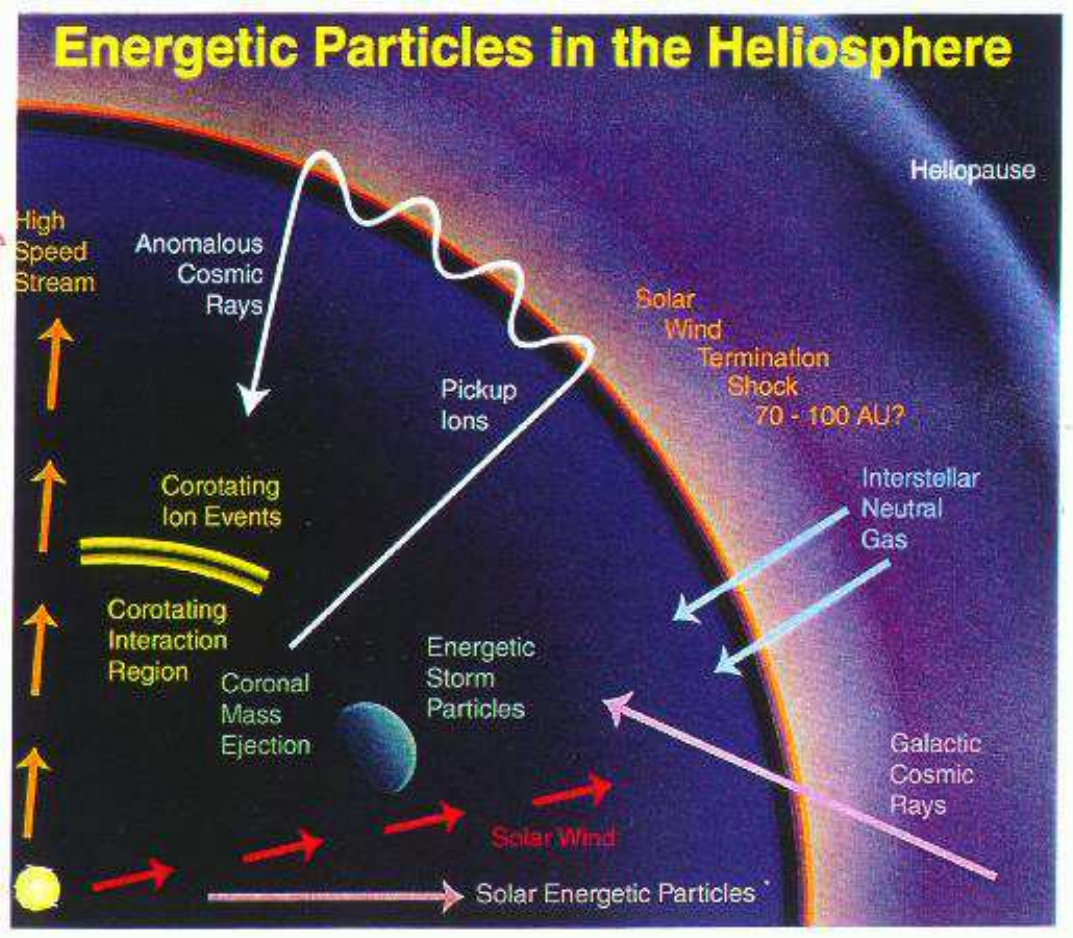}
\caption[Sources of energetic particles in the heliosphere.]{Sources
  of energetic \indexit{energetic!particles!sources}particles in the heliosphere. [IV:12.1;
  \href{http://www.srl.caltech.edu/ACE/CRIS_SIS/scitech.html}{source
    CalTech}.] \colorfig }
\label{fig:Krupp_HelioIV_Overview}
\end{figure}\figindex{../krupp/art/Krupp_HelioIV_Overview.eps}
\ors[IV:12] ``Energetic particles in the energy range between $1$\,eV to
$10^{20}$\,eV can be found everywhere in our Solar System as sketched
in Fig.~\ref{fig:Krupp_HelioIV_Overview}.  Their sources can be either
outside our Solar System from galactic [and extra-galactic
interstellar space or inside our Solar System, including the Sun,
interplanetary space, and planetary magnetospheres]. The types of
energetic particles range from electrons to charged atoms and
molecules to neutral atoms and molecules as well as dust particles.
Fig.~\ref{fig:Krupp_HelioIV_EnergySpectra} shows the particle
intensity versus energy of various types of energetic
particles (left) and for cosmic rays (right).''

\section{Long-term energetic-particle exposure of Earth}
The exposure of \indexit{energetic!particle!exposure}Earth to energetic particles over many millions of
years can be derived from the study of various cosmogenic
radionuclides found in rocks (including those brought back from the
Moon by the Apollo astronauts) and ice deposits (in Greenland and
Antarctica). On time scales up to thousands of years we find signals
in the biosphere (primarily in tree rings).

\subsection{Generation of cosmogenic radionuclides}
\ors[III:11.4.2] \label{sec:beerradio} ``Cosmogenic\indexit{cosmogenic radionuclides!formation}radionuclides are produced by nuclear interactions of
the galactic cosmic rays (GCR) with atoms (N, O, Ar) in the
atmosphere; the contribution of solar cosmic rays is negligible
because of their low energies.'' \ors[II13.2.1] ``Galactic cosmic
radiation originates outside our\indexit{ionizing radiation!galactic
  cosmic rays} Solar System but generally within our Milky Way galaxy
and is treated as [isotropic. This radiation consists of atoms
that] have been ionized and accelerated to very high energies,
probably by [shock fronts of] \indexit{supernova}supernova remnants. The GCR population
consists of about 87{\%} protons and 12{\%} $\alpha$ particles, with
the remaining 1-2{\%} heavier nuclei with charges ranging from 3
(lithium) to about 28 (nickel). Ions heavier than nickel are also
present, but they are rare. Electrons and positrons constitute about
1{\%} of the overall GCR.''

\ors[III:11.4.2]  ``To reach the atmosphere, the GCR have to propagate
through the heliosphere which forms a bubble with a radius of about
100\,AU around the Sun that is filled with solar plasma carrying
magnetic field [as discussed in Ch.~\ref{ch:conversion}. It is
appropriate] to use the transport equation [Eq.~(\ref
{eq:Giacalone_Eq1.17})] to parameterize the intensity of the GCR,
however the so-called force-field approximation has proven to be a
[reasonable simplification] near Earth.  This approximation describes the
modulation effect of the Sun on the energy spectrum of the GCR in
terms of a parameter $\Phi$ called the solar modulation function'' and
comes about as follows:

\ors[III:9.5] ``If one assumes spherical symmetry, and then considers
only high energy cosmic rays, for which the dimensionless modulation
quantity $ r v/ \kappa $, which measures the strength of the
modulation, is small, a very simple analytic solution can be obtained.
The form of this solution corresponds exactly to that obtained for
charged particles influenced by [an electric] field\indexit{cosmic ray!force field solution} with a potential given as a
function of heliocentric radius $r$ by
\begin{equation}\label{eq:gcrpot}
\Phi(r) \propto \int_r^D \left[ v_{\rm sw} / \kappa \right] {\rm d}r
\end{equation}
[with the solar wind velocity $v_{\rm sw}$ assumed constant and
$\kappa$ an equivalent radial diffusion coefficient, also assumed to
be a simple constant, for the full
expression in Eq.~(\ref{eq:boltzmanndiffusion}).] \activity{{\em Advanced/Background:}
  If you are interested in how Eq.~(\ref{eq:boltzmanndiffusion}) can be
  approximated by something like Eq.~(\ref{eq:gcrpot}) you can find
  the origin of this transformation in a 
  \href{https://ui.adsabs.harvard.edu/abs/1968ApJ...154.1011G/abstract}{study
    by \citet{1968ApJ...154.1011G}}.}
Note that this is not a real electrostatic potential because it
affects positively and negatively charged particles in the same
way. [Moreover, observations show a strong
dependence of the cosmic-ray intensity on heliographic latitude which
cannot develop  in the force-field approximation.]  
Attempts to fit the data yield values of $\Phi \approx $ 300\,MeV near
1\,AU.  Because of the use of an effective potential energy, this
approximation is called the 'force-field' solution.  [\ldots]''
\ors[III:11.4.2] ``The solar
modulation\indexit{cosmic ray!solar modulation function} function
$\Phi$ basically corresponds to the average energy lost by a cosmic
ray proton on its way to the Earth.

\begin{figure}[t]
\centering
\includegraphics[width=8cm]{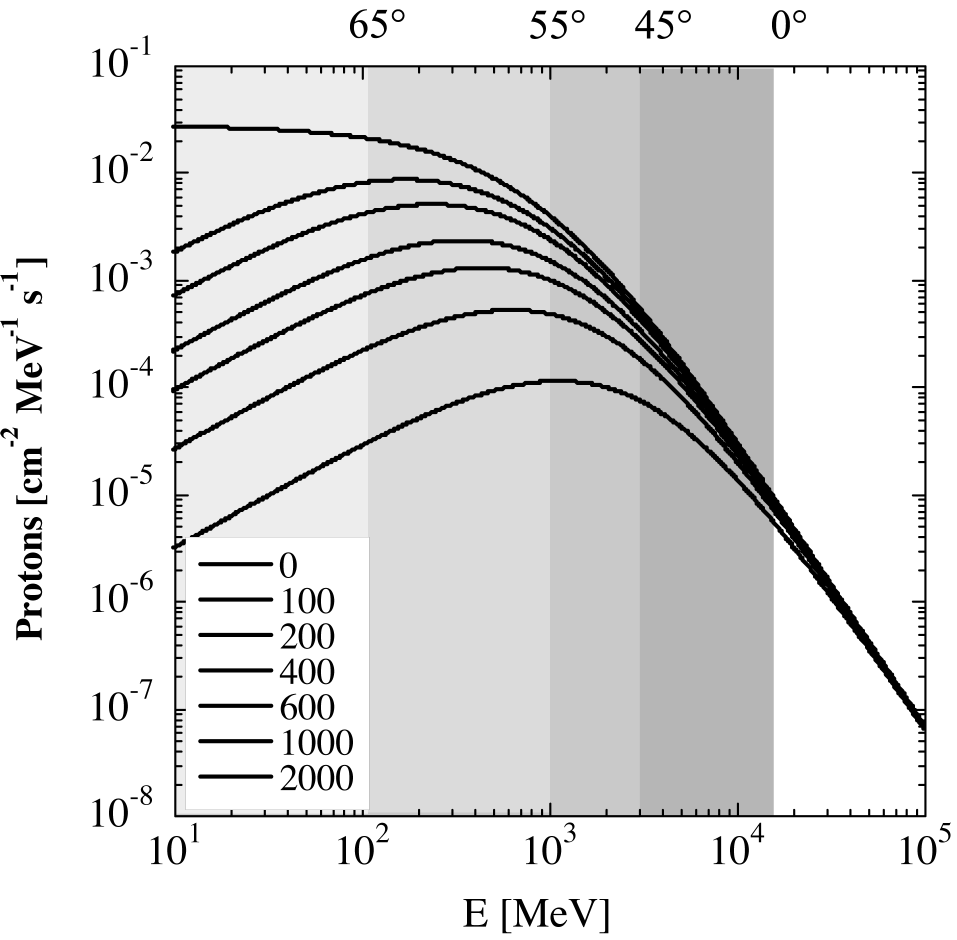}
\caption[Differential GCR
proton fluxes for different levels of solar activity.]{Differential
galactic cosmic ray
proton fluxes for different levels of solar activity
ranging from a value of the solar modulation
function $\Phi = 0$\,MeV 
(Eq.~\ref{eq:gcrpot}), 
%(Eq.~\ref{eq:gcrpot}), 
corresponding to the local interstellar spectrum arriving
at Earth without any solar influence, to $\Phi = 2000$\,MeV which corresponds to a
very active Sun.  There are similar curves for cosmic ray alpha particles and
heavier nuclides.  The vertical bands illustrate the effect of the geomagnetic
field which cuts off all protons approaching vertically with an energy below
about 100\,MeV for a geomagnetic latitude of 65$^\circ$; below 1\,GeV for 55$^\circ$, and
below 3\,GeV for 45$^\circ$.  
At 0$^\circ$ the cut-off energy is 13.9\,GeV for the present geomagnetic
field. [Fig.~III:11.11]
\label{fig:4.2-1}}
\end{figure}

Figure~\ref{fig:4.2-1} shows the differential energy spectrum of the
GCR proton flux for different levels of solar activity. A value of
$\Phi = 0$\,MeV corresponds to the local interstellar spectrum outside
the heliosphere. [The spectrum shown here is an estimate (dependent on
the model approximation and on the 
properties of the interstellar medium and the solar wind) made before
Voyagers 1 and 2 had reached the interstellar medium, which appears to
have happened in late 2018.] Figure~\ref{fig:4.2-1} shows
that the shielding effects of the open solar magnetic field and the
advecting solar wind are most pronounced at the low energy end of the
spectrum. As a consequence, GCR particles above about 20\,GeV are
hardly affected by the varying heliospheric magnetic field.

Before reaching [the terrestrial atmosphere], the cosmic ray particles
have to overcome a second barrier, the geomagnetic field. This field
prevents particles with too low rigidity (momentum per unit charge)
from reaching the top of the \indexit{cosmic ray!rigidity}atmosphere.
In a first\indexit{cosmic ray!rigidity|seealso{definition}}
approximation,\indexit{definition!cosmic ray
  rigidity} the\indexit{cosmic ray!cut-off rigidity} geomagnetic
field is considered as a dipole and in this case the cut-off rigidity
depends only on the angle of incidence and the geomagnetic
latitude. At low latitudes the cut-off rigidity for vertical incidence
is presently $\sim$14.9\,GV. This means that a cosmic ray proton needs
a kinetic energy of at least 14\,GeV (14.9-m$_{\rm p}$\,c$^2$) to reach the
top of the atmosphere (see shaded bands in Fig.~\ref{fig:4.2-1})
[\ldots]

\begin{table}[t]
\caption[Production reactions and rates
for cosmogenic radionuclides.]{Properties for some cosmogenic radionuclides including
nuclear production \indexit{cosmogenic radionuclides}reactions and globally-averaged production rates for the present
geomagnetic field and a solar modulation of $\Phi=550$\,MeV. (EC:
electron capture). 
All nuclear reactions  are induced by high-energy secondary particles
generated by the primary  cosmic-ray particles (so-called spallation
reactions). The only exception is $^{14}$C which is almost totally produced
by thermal neutrons interacting with nitrogen. [Table~III:11.4]
\label{tab:beergcrgen}}
\begin{center}\begin{tabular}{rccccc}
  \hline
  Isotope & half life & decay & target & nuclear reaction & production \\
          & (yr)       &       &        &   & rate (cm$^{-2}$s$^{-1}$) \\
  \hline
  $^{14}$C & 5730 & $\beta^-$ & N,O & $^{14}$N(n,p)$^{14}$C & 2.02 \\
 & & & & $^{16}$O(p,3p)$^{14}$C & \\
 & & & & $^{16}$O(n,2p1n)$^{14}$C& \\
  \hline
  $^{10}$Be & $1.5\times 10^6$ & $\beta^-$ & N,O & $^{14}$N(n,3p2n)$^{10}$Be & 0.018 \\
    &   &  &  & $^{14}$N(p,4p1n)$^{10}$Be &  \\
    &   &  &  & $^{18}$O(n,4p3n)$^{10}$Be &  \\
    &   &  &  & $^{18}$O(p,5p2n)$^{10}$Be &  \\
  \hline
  $^{36}$Cl & $0.30\times 10^6$ & $\beta^-$, EC & Ar &  $^{40}$Ar(n,1p4n)$^{36}$Cl  & 0.0019\\
    &   &  &  & $^{40}$Ar(p,2p3n)$^{36}$Cl &  \\
    &   &  &  & $^{36}$Ar(n,p)$^{36}$Cl &  \\
  \hline
\end{tabular}\end{center}
\end{table}
If a primary cosmic ray particle makes its way through the heliosphere
and the geomagnetic field and enters the atmosphere it will interact
quickly with an atomic nucleus of oxygen, nitrogen, or argon. Because the
energies of incoming particles are generally very high, only part of
their kinetic energy is transferred to the first atom they hit. They
continue their travel and hit a few more atoms until their energy is
dissipated. Each collision results in the generation of secondary
particles covering the full spectrum of hadrons and leptons, which
either decay or interact with other atoms of the atmosphere. In this
way a cascade of secondary particles develops which can be simulated
using Monte Carlo techniques. Table~\ref{tab:beergcrgen} shows the
different production reactions for the radionuclides $^{14}$C,
$^{10}$Be, and $^{36}$Cl, and the resulting mean global production
rates for the present geomagnetic field intensity and a solar
modulation function equal to $\Phi=550$\,MeV.\activity{{\em Show:} To appreciate
  how little radionuclide material there is to work with, compute the
  global annual production in kg for $^{14}$C and $^{10}$Be. That
  production rate puts roughly one $^{14}$C atom per $10^{12}$ atoms
  of $^{12}$C in living tissue through uptake of atmospheric CO$_2$ by
  plants and their subsequent consumption by animals.}

The simulations show that the majority of the secondaries are neutrons
followed by protons. Both, in turn, collide with atmospheric atoms
initiating spallation reactions that generate the cosmogenic nuclides
that are archived for us in ice ($^{10}$Be, $^{36}$Cl) or tree rings
($^{14}$C).  In addition, the cosmic-ray produced neutrons have been
monitored continuously since 1951 by so-called neutron
monitors. [These measurements show that whenever] the magnetic
activity is high (at high sunspot count) the shielding is strong and
the neutron flux is low. [\ldots] Many studies have shown that the
11-yr and longer-term variations are faithfully reproduced in the
cosmogenic data, and they and the neutron monitor data have been
inter-calibrated to yield a continuous cosmic-ray record for the past
10,000 years.

\begin{figure}[t]
\centering
\includegraphics[width=8cm]{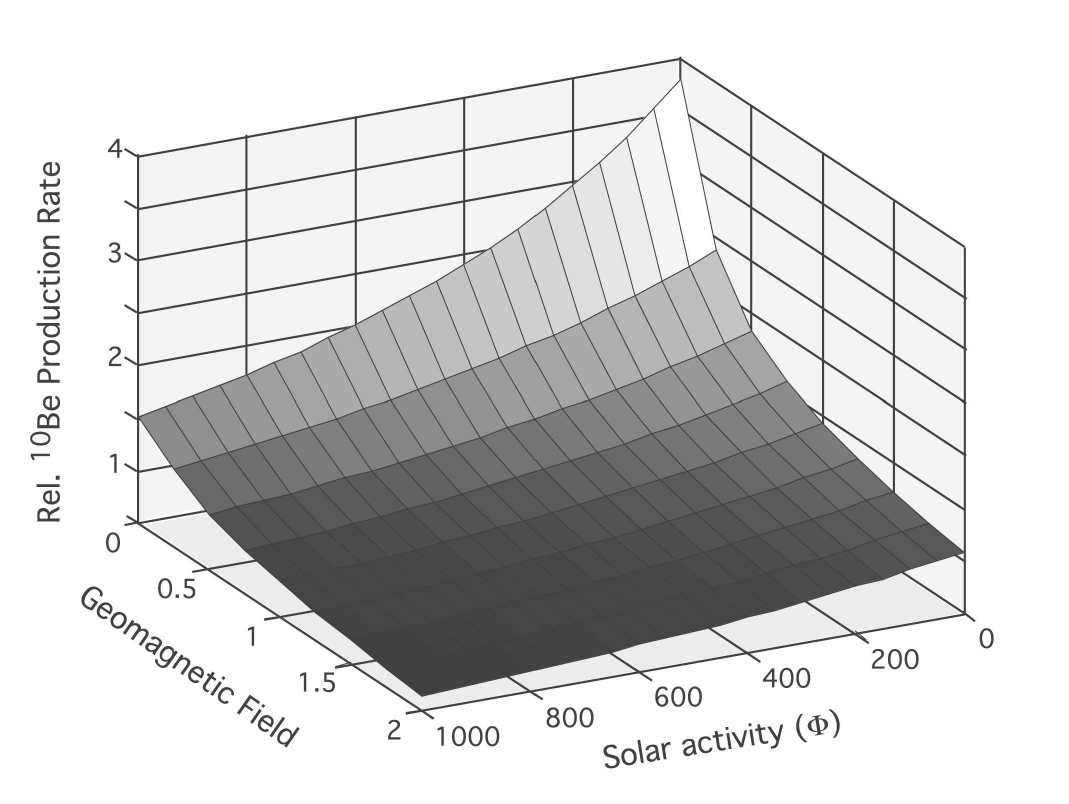}
\caption[$^{10}$Be production rate {\em vs.}\ geomagnetic
field strength and solar activity.]{\label{fig:4.2-2} Dependence of the $^{10}$Be
  production rate on the geomagnetic field intensity (in units
  relative to today's field) and the solar activity (expressed by
  the solar modulation function $\Phi$, %Eq.~9.8).
  Eq.~\ref{eq:gcrpot}). 
  The production rate is normalized to the present strength of the
  geomagnetic field and solar activity corresponding to a solar
  modulation function of 550\,MeV (matching the long-term
  average). [Fig.~III:11.12] }
\end{figure}
%%% Neue TEXT
As an example, the combined effect of solar activity and geomagnetic
field on the relative production rate of cosmogenic nuclides is shown
for $^{10}$Be in Fig.~\ref{fig:4.2-2}. The relative dipole component
of the geomagnetic field $\mu_{\rm r}$ varies between 0 and 2, 1 being
the present field. For $\mu_{\rm r}=1$ and $\Phi=$550\,MeV (long-term
average), the $^{10}$Be production rate is normalized to 1. It should
be noted that the dependence of the production rate on $\mu_{\rm r}$
and $\Phi$ is nonlinear.'' \sactivity{$\circledS$ {\em Show:} The
  $^{10}$Be production rate for Mars would be about 2.5 times the
  terrestrial rate if it had a terrestrial atmosphere. Show why based
  on data in this text. \mylabel{act:beprod} \solution{beprod}}

\subsection{Transport and deposition of cosmogenic radionuclides}
\ors[III:11.4.2] ``The fate indeit{cosmogenic radionuclides!transport and
  deposition}of a cosmogenic nuclide after its production
in the atmosphere depends strongly on its geochemical
properties. Within a short time, $^{10}$Be becomes attached to
aerosols and follows their pathways. $^{14}$C on the other hand
oxidizes to $^{14}$CO$_2$ and is exchanged between atmosphere,
biosphere and ocean. After a mean residence time of 1 to 2 years,
$^{10}$Be is removed from the atmosphere mainly by wet
precipitation. The flux $F$ of cosmogenic nuclides from the atmosphere
into, for example, a polar ice sheet is proportional to the atmospheric production
rate $\Pi$:
%\begin{equation}
$F = \psi\,\Pi$.
%\label{eq:4.2.1-1}
%\end{equation}
Locally and temporally, $\psi$ can vary due to changes in the
atmospheric transport and deposition processes. The
degree\indexit{cosmogenic radionuclides!transport and deposition} of
variability depends very much on how well the atmosphere is mixed. In
the case of $^{14}$C, the large atmospheric $^{14}$CO$_2$ reservoir
leads to an atmospheric residence time of 6 to 7 years and therefore
to a complete mixing. In the case of the aerosol-bound nuclides the
residence times are shorter, roughly $1-2$ years; mixing in the
troposphere is not complete. After deposition, some of the
nuclides become incorporated into natural archives such as ice sheets,
glaciers, sediments, and tree rings.

For our purpose, a useful archive stores the complete flux of nuclides
from the atmosphere in a stratigraphically undisturbed way and records
the time accurately. Excellent archives in this respect are ice sheets
which directly collect the atmospheric precipitation containing
$^{10}$Be. Typically, they cover the last several $10^5$ years with a
time resolution per sample ranging from 1\,year at the top to decades
or centuries near the bottom. However, due to the flow characteristics
of ice, dating is difficult, especially in the deeper part of ice
cores.

Tree rings represent\indexit{cosmogenic radionuclides!ice cores and
  tree rings} an ideal archive for the atmospheric $^{14}$C/$^{12}$C
ratio. So far, by chronologically matching trees of different ages,
the atmospheric $^{14}$C/$^{12}$C ratio has been reconstructed back to
approximately [14\,kyr BP (before present, relative to
1950\,CE)]. Potentially, the full range covered with today's measuring
techniques (40 to 50\,kyr) will be traceable in tree rings in the
future.''\activity{{\em Consider:} Over the past century the concentration of $^{14}$C
  in the biosphere has dropped because of
  fossil-fuel burning. (a) Why? (b) Argue why this
  leads to an ambiguity in $^{14}$C dating if no other information on
  the age of an object is known.}

\section{Radionuclides as proxies of magnetic variability}\label{sec:archives}
\ors[III:11.4.2.1]
``What can be learned by measuring \indexit{cosmogenic
  radionuclides!proxies of magnetic activity}cosmogenic nuclides in ice?
[\ldots] In an archive, changes in the concentration can result from
changes either in the production rate $\Pi$ or in the Earth-system
processes $\psi$ (transport and deposition). Changes due to
radioactive decay can be corrected for, if a reliable time scale is
available. Changes in the production rate can be caused by
heliomagnetic and geomagnetic modulation of the cosmic-ray
flux. Episodic solar proton events can cause short but intense cosmic
radiation, but do not contribute much to the total production rate due
to the relatively low proton energies. Changes in the system on the other hand
are related to the atmospheric transport and mixing processes as well
as to the local precipitation rate.

The question arises how the different causes of concentration changes
can be separated. A straightforward answer to this question is to
combine several nuclide records from different sites. Comparing
$^{10}$Be with $^{14}$C permits separating production from system
effects. Changes in the production rate due to helio- and geomagnetic
modulation of the cosmic-ray flux are reflected both in $^{10}$Be and
in $^{14}$C in a very similar way. Changes within the Earth system,
however, are expected to affect $^{10}$Be and $^{14}$C in a completely
different way because the geochemical behavior of these nuclides is
fundamentally different. [\ldots\ Of course, this argument is useful
only over] the range of the radiocarbon dating (last 50\,kyr) and
requires a high precision [record of $\Delta^{14}$C that] is not yet
available for the period $13-50$\,kyr BP. The next step is to separate
heliomagnetic and geomagnetic signals. In principle, these two signals
could be separated by looking at two radionuclide records, one from
the equator and one from the regions of the magnetic poles. Without
latitudinal atmospheric mixing, the record from the magnetic pole
would only reflect solar modulation because geomagnetic shielding
disappears at high latitudes, whereas the signal in the equatorial
record would be dominated by geomagnetic modulation. However, as a
result of atmospheric mixing, this is not the case.

Solar modulation effects have been found in cores from Greenland and
Antarctica. The same is true for geomagnetic modulation effects like
for the Laschamp\indexit{Laschamp event} event at about 40\,kyr BP,
when the magnetic dipole field was close to zero. This event is
present in the high latitude ice-cores from the Arctic and from
Antarctica (GRIP, Vostok, Byrd, Dome C, Taylor Dome). Radionuclide
records from low-latitude ice cores are still rare [and have a
smaller potential due to dating and other problems.]

\begin{figure}[t]
\centering
\includegraphics[width=9cm]{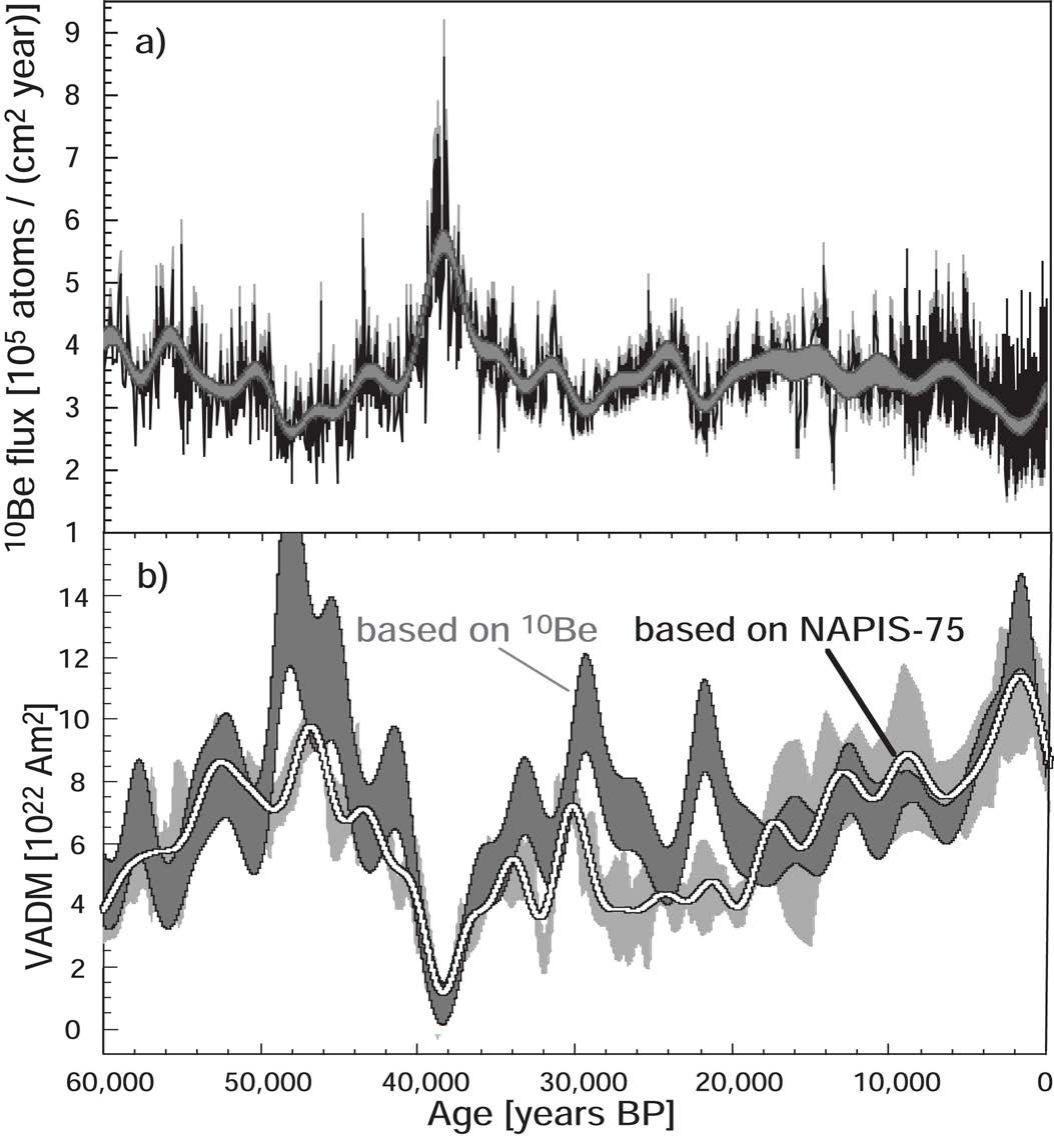}
\caption[$^{10}$Be data and geomagnetic dipole field for the past 60,000 years.]{
Comparison of {\em (a)}
$^{10}$Be data with {\em (b)} the geomagnetic dipole field for the
past 60,000 years.  Panel {\em (a)}
shows a compilation of $^{10}$Be data from the GRIP and GISP ice cores
in Greenland.  Panel {\em (b)} compares the dipole field derived from
$^{10}$Be (panel {\em a}) to that from remanence data (NAPIS-75) measured in
ocean sediment cores. [Fig.~III:11.14]
}
\label{fig:4.2-4}
\end{figure}
Another approach is to assume that solar modulation effects generally
occur on shorter time scales than geomagnetically induced production
changes. Applying low-pass filters with cut-off frequencies in the
range of 1/2000 and 1/3000\,yr$^{-1}$ on cosmogenic nuclide fluxes
provides production signals in good agreement with paleomagnetic
intensity records based on remanence measurements.\activity{{\em Look up}
  'paleomagnetic dating' in relation to the 'remanence measurements'
  mentioned in Sect.~\ref{sec:archives}.}

The task of separating the different causes of variability observed in
radionuclide records is complicated by the fact that some of the
causes are coupled. For example, changes in solar activity affect
atmospheric processes and possibly also induce, to a smaller extent, climatic
changes. Therefore, additional information from other
measured parameters should be included to obtain a complete and
consistent picture of what happened during the period of
investigation. In the following, we discuss how the intensity of
the geomagnetic dipole field and the solar variability can be derived
from cosmogenic nuclides.''

\subsection{Geomagnetic field} 
Fig.~\ref{fig:4.2-4}a shows \ors[III:11.4.2.2] ``a compilation of
$^{10}$Be data from the GRIP and the GISP ice cores drilled in central
Greenland [\ldots] covering the past 60,000 years. To correct for
the lower precipitation rate during glacial times ($10-60$\,kyr BP) the
$^{10}$Be flux has been calculated and smoothed (gray band). The
plot\indexit{geomagnetic field!long-term variations} shows a
significant peak at about 40\,kyr BP. To check whether the smoothed
curve does reflect the geomagnetic dipole field as expected from
Fig.~\ref{fig:4.2-4} the corresponding changes in the dipole field
intensity have been calculated based on its relationship with the
$^{10}$Be production shown in Fig.~\ref{fig:4.2-2}. The result is
compared in Fig.~\ref{fig:4.2-4}b with the completely independent
reconstruction NAPIS-75 which was derived from remanence measurements
in Atlantic sediment cores. Overall the agreement is good and confirms
that the $^{10}$Be peak at 40\,kyr BP corresponds to the Laschamp event
when the dipole field intensity was almost zero but did not reverse.''

\begin{figure}[t]
\centering
\includegraphics[width=9cm]{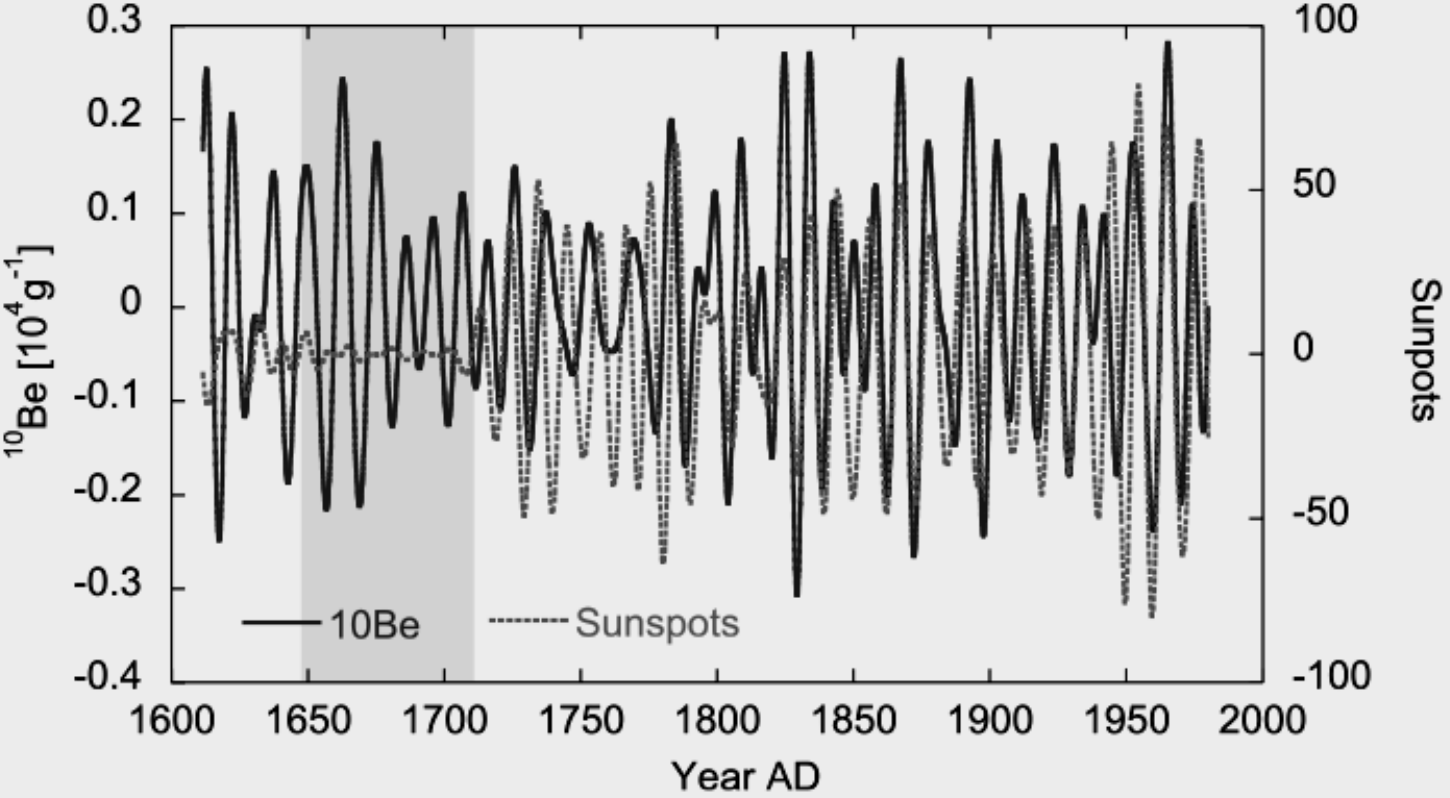}
\caption[Bandpass-filtered $^{10}$Be concentration and sunspot
number.]{ Comparison of the $^{10}$Be concentration measured in the
  Dye 3 ice core from Greenland with the sunspot number after applying
  a band-pass filter ($8-16$ years). Note that during the Maunder
  Minimum $1645-1715$ (shaded area) when almost no sunspots were
  observed $^{10}$Be shows a clear 11-yr sunspot
  cycle. [Fig.~III:11.15;
  \href{https://ui.adsabs.harvard.edu/abs/1994svsp.coll..291B/abstract}{source:
  \citet{1994svsp.coll..291B}}.] }
\label{fig:4.2-5}
\end{figure}
\subsection{Solar variability}
 \ors[III:11.4.3] ``We return now to the
discussion of solar variability and discuss to what extent cosmogenic
radionuclides can expand our knowledge about long-term solar
variability. In a first step, we compare annual $^{10}$Be data with the
sunspot record which represents the longest observational data of
solar variability. A resolution of one year is about the limit because
it corresponds to the mean travel time for a $^{10}$Be atom produced
in the atmosphere to reach the\indexit{solar!activity!long-term
  variations} Earth surface where it is stored in, for example, an ice
sheet. Fig.~\ref{fig:4.2-5} shows a comparison of the $^{10}$Be
concentration from Dye 3, Greenland, with the sunspot number. Both
records have been band-pass filtered ($8-16$\,yr). While during the
Maunder Minimum (shaded area between 1645 and 1715) hardly any
sunspots were observed, the solar dynamo clearly continued to
produce open magnetic field modulating the cosmic rays and the
$^{10}$Be production.

The overall good agreement between $^{10}$Be and sunspot numbers gives us
confidence to extend the time interval over the Holocene, {\em i.e.,}  about
the last 10,000 years. During this period the climate was relatively
stable compared to glacial times and therefore we can assume that
transport and deposition effects did not disturb the production signal
in the $^{10}$Be record.  This assumption is confirmed by global
circulation model (GCM) runs which show that the transport effects
were relatively stable during the climatic conditions prevailing
during the Holocene. So, indeed, to a first approximation they can be
neglected. This is not the case for the geomagnetic field which
exhibits significant long-term changes.

\begin{figure}[t]
\centering
\includegraphics[width=10cm]{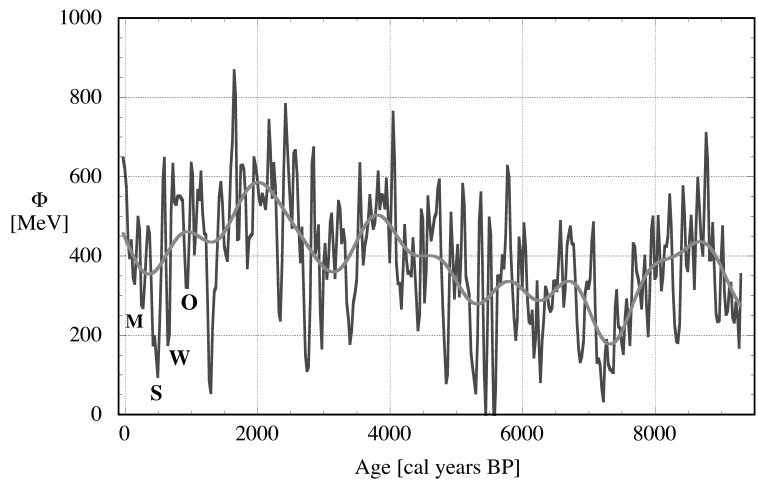}
\caption[Solar modulation function $\Phi$ from the present to
9350\,BP]{ Solar modulation function $\Phi$ from the present (0 BP
  corresponds to 1950) back to 9350\,BP. The
  black curve shows data that have been low-pass filtered with a
  cut-off of 150\,years; the smooth grey curve with 1000\,years. The
  most recent solar minima are indicated: M: Maunder; S: Sp{\"o}rer;
  W: Wolf, and O: Oort. [Fig.~III:11.16] }
\label{fig:4.2-6}
\end{figure}
Using Monte Carlo simulations, the
effect of the geomagnetic dipole field has been removed and we are
left with the solar modulation function $\Phi$ [\ldots]
The data of Fig.~\ref{fig:4.2-6} have been low-pass
filtered with a 150\,yr cutoff.  The most striking features of the
$\Phi$ record are the many distinct minima which correspond to grand
solar minima such as the Maunder (M), Sp{\"o}rer (S), Wolf (W), and
Oort (O) minima. The fact that $\Phi$ never reaches zero means that there is
always some residual open magnetic flux; in other words the solar
dynamo seems to weaken from time to time, but, as a close inspection
of the unfiltered data shows, it never stops. The two exceptions in
Fig.~\ref{fig:4.2-6} are due to uncertainties in the data.

The maxima are less pronounced. It is interesting to note that the
present level of solar activity is comparatively high, although there
were earlier periods with similar or possibly even higher activity
around 2000, 4000, and 9000\,BP. There is also a clear long-term trend
indicated by the thick line that is low-pass filtered with a cut-off
of 1000 years.''

For periods covering $10^5$\,yr to over $10^6$\,yr the obvious radionuclide
archives are ocean sediments that go back many millions of years, such
as $^{10}$Be and $^{26}$Al. The price one pays for the long records is
the reduced temporal resolution owing to the very small sedimentation rates
and additional processes related to the transport of the
radionuclide into the sediment. As is the case for tree ring records,
these radionuclides are sequentially stored as they are taken from the
atmosphere, in contrast to radionuclides measured in rocks, which we
discuss next, that are continually produced and that therefore provide
only an integral measure of the production rate. The only time
information available comes from the different half-lives. It is
therefore important to measure as many distinct radionuclides as
feasible.

\subsection{Very-long time scale variability in cosmic-ray exposure}
\ors[III:9.2.4.2] ``The record of the galactic cosmic-ray
\indexit{cosmic ray!variability}flux on a million-year time
scale can be inferred from induced nuclear reactions in
extraterrestrial matter of known exposure geometry, such as
lunar rocks or meteorites.  Nuclear reactions produce a variety of
radioactive and stable nuclei that can be measured and related to
the incident cosmic ray flux.  The
radionuclides $^{81}$Kr (2.1 x 10$^2$\,yr half-life),
$^{36}$Cl (3.0 x 10$^5$\,yr), $^{26}$Al (7.2 x 10$^5$\,yr),
$^{10}$Be (1.6 x 10$^6$\,yr) and $^{53}$Mn (3.7 x 10$^6$\,yr)
represent a good set of monitors for cosmic ray flux variations on this
time scale.  Among the chondritic meteorites which were studied extensively,
the production rates of the above radionuclides can vary because of
differences in size and shielding conditions.  These, when
analyzed, reflect a constant ($\pm 10-15$\%)
galactic flux over the 10$^5$ - 10$^7$\,yr time scale,
which matches the average present-day flux.''

\ors[III:9.2.4.3] ``There are few radioisotopes with appropriate half-lives that can
be used for [times scale of $10^7$--$10^9$\,yr] and only
$^{129}$I (1.6 x 10$^7$\,yr) and $^{40}$K (1.3 x 10$^9$\,yr)
have been studied so far.

Chondritic meteorites cannot be used to study variations in the cosmic
ray flux on longer time scales, because their exposure ages (time
between being formed and striking the Earth) are typically less than a
few tens of million years.  Fortunately, there are numerous recovered
iron meteorites which were exposed in space as small bodies for up to
two billion years since being formed, and which are well suited for
this purpose.  The measurement of all three isotopes of potassium
permits the detection of the cosmic-ray-produced component which is
superimposed to potassium initially present in the meteorite.  For the
period of $0.2 - 1.0$\,Gyr ago, essentially constant $^{38}$Ar
production rates are observed, and agreement between ages determined
from $^{38}$Ar and from $^{40}$K and $^{41}$K.'' 
\href{https://ui.adsabs.harvard.edu/abs/2013SSRv..176..351W/abstract}{It
  appears} (\citep{2013SSRv..176..351W}) that not much has changed in terms of long-term variability
in solar activity or long-term trends in GCR fluxes coming into the
heliosphere over the past billion years, to within a factor of
$\sim 1.5$, based on a variety of radionuclides in meteoritic samples
and terrestrial sediments combined \ldots\ but realize that
variability below time scales of hundreds of thousands to millions of
years cannot be detected within these records. \activity{{\em
    Consider/Show:} How are the
  decrease of stellar rotation speed, magnetic activity, and mass-loss
  rate on long time scales compatible with the 'essentially constant'
  GCR exposure over the past $\approx 1$\,Gyr? The answer has to do
  with the fact that the Sun is already an aged star, and can be
  traced to its relatively weak magnetic braking over the past 1\,Gyr,
  and thus relatively little decrease in coronal activity and
  mass-loss rate. The limited impact on GCRs at Earth orbit over time
  also suggests that the heliospheric variability (leading to
  diffusive GCR scattering) has not changed too much. Estimate the
  changes over time using Eqs.~(\ref{eq:skumanich})
  and~(\ref{eq:masslossage}), and Fig.~\ref{figure:rotact}.}

\section{Exposure to supernovae}
There is evidence of at \indexit{supernova!exposure}least one nearby
supernova in the very early formation phases of the Solar System when
the terrestrial planets had yet to fully take shape. Then, the Sun was
still embedded within its birth cluster, and it appears that one of
its heavy siblings exploded prior to the cluster falling
apart.\sactivity{$\circledS$ {\em Look up/Show:} Heavy stars evolve
  much faster than low-mass stars, and can, if heavy enough, explode
  in a supernova even as lower-mass stars and their planetary system
  within the same molecular cloud are still in their formative
  phases. Stars spend most of their lives (typically some 85--90\%\ of
  its full life time prior to the supernova or white-dwarf
  phases)fusing hydrogen on the 'main sequence' which runs diagonally
  through a Hertzsprung-Russell (luminosity-temperature) diagram ({\em
    e.g.}, Figs.~\ref{fig:acthrd} and \ref{figure:evolmodel}). The
  heaviest stars are of order 100\,$M_\odot$ at which mass their outer
  layers are almost blown off by radiation pressure. The least massive
  stars are about $0.08\,M_\odot$ below which sustained hydrogen
  fusion is impossible. Stars heavier than about $8\,M_\odot$ end
  their lives in supernovae (close binaries with mass transfer muddy
  that simple threshold).  (a) Estimate the life times $\tau_\ast$ of
  stars on the main sequence at 100, 8, 0.5 and 0.08 solar masses
  based on a simple scaling assuming that luminosity $L_\ast$ is
  maintained by using a fixed fraction of stellar mass $M_\ast$ so
  that $\tau_\ast\propto M_\ast/L_\ast$ and that along the main
  sequence --~very roughly~-- $L_\ast \propto M_\ast^{7/2}$ while
  $\tau_\odot\approx 10^{10}$\,yr.  (b) Assuming there were only
  single stars in the Sun's birth cluster, what is the highest age
  $\tau_{\odot{\rm SN}}$ of the Sun when a supernova could have
  happened within that cluster?  (c) For stars heavier than about
  0.5\,$M_\odot$ the initial mass function (IMF, the number of stars
  formed as a function of their mass) can be approximated by
  $\xi(M){\rm d}M\approx\xi_0M^{-7/3}{\rm d}M$. For there to have been
  at least one supernova within $\tau_{\odot{\rm SN}}$\,yr, what is
  the minimum number of stars with masses between, say, 0.5\,$M_\odot$
  and 2\,$M_\odot$ (Sun-like stars) that should have been in the Sun's
  birth cluster. That number lies comfortably below the content of
  many young clusters, such as the Orion Nebula, Hyades, and Pleiades
  clusters with roughly 100 to 1,000 member stars. But there is
  radionuclide evidence that the young Sun was in fact exposed to a
  supernova when it was a mere 1.8\,Myr old.  (d) Repeat the above for
  the lowest-mass star to go supernova within that time, and estimate
  the corresponding minimum number of stars with masses between 0.5
  and 2 solar masses (which is almost an order of magnitude larger
  than the number under (3) above, but compatible with the estimated
  solar birth cluster mass of 500 to 3000\,$M_\odot$).  These are, of
  course, rather rough approximations and stellar structure and
  evolution models can do a much better job, but it gives an
  impression of what it takes to make such estimates and yields
  approximations that are good for an initial exploration; if you
  want, you can do somewhat better with piece-wise approximations that
  you can find on the web.  For more on this, and on where the Sun's
  siblings have gone, see
  \citet{2009ApJ...696L..13P}. \mylabel{act:snexposure}
  \solution{snexposure}
  % https://www.ucolick.org/~bolte/AY4_04/class4_04bwd.pdf
  } There are some stars relatively nearby that
are candidates to go supernova in the distant future, but none so
close that the explosion would directly affect the solar
system. Indirect effects, however, are possible: in the case of a
blast wave from a nearby supernova, for present-day conditions of the
solar wind, pressure \ors[IV:3.5] ``balance between the \indexit{supernova}supernova shock
wave and the solar wind produces extreme heliosphere models that have
the same physical structures as the models with the heliopause at 1.4
times the distance of the termination shock in the upwind direction
but with both located very close to the Sun.  [A] supernova located at
$\approx 9$~pc from the Sun would create a heliopause that penetrates
to within 1~AU, subjecting the Earth to an infusion of supernova
debris including iron and other heavy atoms.  \activity{{\em Show:} Estimate how
  much stronger the dynamic pressure of the incoming supernova wave
  front needs to be than the present-day IMF, assuming comparable
  solar-wind properties, to push the heliospheric boundary to within
  1\,AU. See Activity~\ref{act:LISM}.} The discovery of the
radioisotope $^{60}$Fe with a half-life of 1.5\,Myr in a deep-sea
ferromanganese crust and dated to $2.8\pm0.4$\,Myr ago indicates that a
nearby supernova explosion likely occurred [around that time (and
perhaps
\href{https://ui.adsabs.harvard.edu/abs/2016Natur.532...69W/abstract}{another
(\citep{2016Natur.532...69W}})
some $6-9$\,Myr ago).]
 
The effect on the Earth of a nearby supernova and the effect on more
distant planets from supernovae at distances up to 30~pc will include
an increase in the amount of neutral hydrogen atoms, dust, supernova
metals, and Galactic cosmic rays reaching the planet's atmosphere.
The latter would influence the planet's magnetosphere and change the
planet's atmospheric chemistry, including the important molecule
ozone.''

\clearpage
	    
\chapter{{\bf Applied heliophysics, {\em mutatis mutandis}}}
\label{ch:beyond}
%\vspace{-3.8cm}

\begin{figure}[t]
\includegraphics[width=\textwidth]{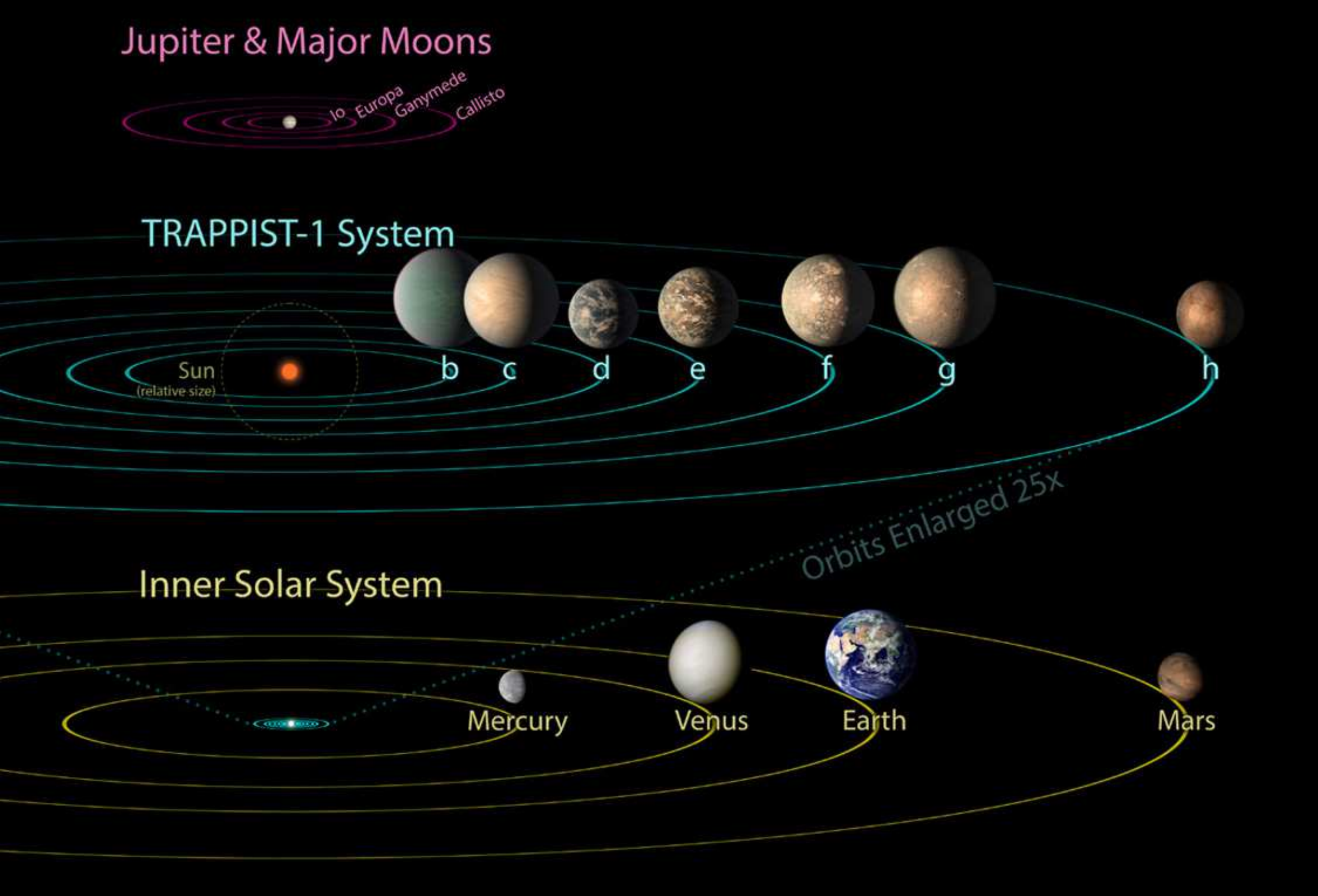}
\caption[Systems to scale: Jupiter, TRAPPIST-1, and the inner
Solar System.]{\label{fig:trappistcomp} Comparison of the orbits of
the major moons of Jupiter, of the known TRAPPIST-1 planets, and the
inner Solar System. The planets and moons are shown on the same scale,
but vastly enlarged relative to the orbital scales. Courtesy:
NASA/JPL-Caltech. \colorfig }
\end{figure}
Our advancing understanding of the processes that are part of the network of Sun-planet connections in heliophysics is being applied and tested in innovative studies of star-exoplanet couplings. A small sampling from the already extensive and rapidly growing literature illustrates the fertile and diverse fields that heliophysics finds in astrophysics at large:
\begin{itemize}\customitemize
\item For ultracool stars near the boundary of the stellar and substellar regimes, where coronal heating fails as the photosphere decouples from the magnetic field because of weak ionization, an analogy has been identified with Jupiter's magnetosphere in which failure of co-rotation \indexit{magnetosphere!ultra cool stars}introduces stresses in the field that ultimately lead to an auroral radio signature that can have a counterpart in ultracool stars (e.g., \href{https://ui.adsabs.harvard.edu/abs/2009ApJ...699L.148S/abstract}{a paper by \citet{2009ApJ...699L.148S}}, and \href{https://ui.adsabs.harvard.edu/abs/2017ApJ...846...75P/abstract}{work by \citet{2017ApJ...846...75P}}).

\item For cool stars in general, signatures of stellar winds are evasive other than indirectly through magnetic braking over many millions of years, but where such a wind collides with the interstellar medium a neutral-hydrogen wall forms whose optical depth for, e.g., Lyman $\alpha$ radiation \indexit{astrosphere!Lyman $\alpha$ signal}provides a field-independent measurement of mass-loss rates (\href{https://ui.adsabs.harvard.edu/abs/2005ApJ...628L.143W/abstract}{see \citet{2005ApJ...628L.143W}}).

\item Another signature of stellar winds may be found in observations of the bow shock around \indexit{bow shock!stellar wind}the magnetosphere/exosphere of a transiting exoplanet during pre-transit phases (\href{https://ui.adsabs.harvard.edu/abs/2015ApJ...810...13C/abstract}{see \citet{2015ApJ...810...13C}}).

\item The stellar winds and their coupling with exoplanetary atmospheres can be modeled using codes developed for our Solar System, providing insight into star-planet couplings \indexit{atmosphere!stellar influence}and exoplanetary magnetospheric activity for systems well outside the parameter domain of our own home in the local cosmos (such as for TRAPPIST-1 --~see Fig.~\ref{fig:trappistcomp}~-- \href{https://ui.adsabs.harvard.edu/abs/2017ApJ...843L..33G/abstract}{see \citet{2017ApJ...843L..33G}}).

\item The exposure of exoplanets to the stellar equivalent of solar energetic particles is being estimated by\indexit{energetic!particles!other stars} applying astrospheric models that include turbulence by which energetic particles are scattered, which enables, for example, quantitative estimates of particle radiation for exoplanets in stellar habitable zones (\href{https://ui.adsabs.harvard.edu/abs/2019ApJ...874...21F/abstract}{see \citet{2019ApJ...874...21F}}).

\item The galactic cosmic \indexit{cosmic ray!exoplanets}ray exposure for exoplanets can be quantified based on modified heliospheric models, such as for the TRAPPIST-1 planets that orbit deep inside a strong-field astrosphere  
(\href{https://ui.adsabs.harvard.edu/abs/2018Ge&Ae..58.1108S/abstract}{see \citet{2018Ge&Ae..58.1108S}}).

\item The effects of stellar radiation and of stellar and galactic cosmic rays on the chemistry of exoplanetary atmospheres is being analyzed guided by, and calibrated against, solar-terrestrial spectral models and terrestrial tropospheric and ionospheric models  (\href{https://ui.adsabs.harvard.edu/abs/2018ApJ...863....6S/abstract}{see \citet{2018ApJ...863....6S}}).

\item Evolution of the chemical makeup of planetary atmospheres subject to differences in insolation, rotation, \indexit{atmosphere!exoplanets}atmospheric and oceanic circulations, and chemical weathering provide insight of the impacts of each of these on planetary habitability (\href{https://ui.adsabs.harvard.edu/abs/2019ApJ...875...79J/abstract}{see \citet{2019ApJ...875...79J}}) and help guide target selection for the search for biosignatures. The same is true on evolutionary timescales for the role of plate tectonics and volcanism  (\href{https://ui.adsabs.harvard.edu/abs/2018AsBio..18..873F/abstract}{see \citet{2018AsBio..18..873F}}).
\end{itemize}

It should be no surprise that stellar and exoplanetary-system sciences conversely continue to provide crucial information for heliophysics:
\begin{itemize}\customitemize
\item The very formation of stars and their planetary systems is being observed, models being refined, and \indexit{star!formation}results \indexit{planetary!system!formation}compared to empirical evidence from within the Solar System (e.g., \href{https://ui.adsabs.harvard.edu/abs/2018SSRv..214...60L/abstract}{see \citet{2018SSRv..214...60L}}, and other papers in that volume).

\item Over the past few decades, \indexit{magnetic!activity!evolution}observations of stars of various spectral types and evolutionary phases have revealed the fundamental ingredients for stellar dynamos --~rotation and convection~--and quantitative information on how these set stellar atmospheric activity (e.g., \href{https://ui.adsabs.harvard.edu/abs/2000ssma.book.....S/abstract}{see \citet{2000ssma.book.....S}}) and how the Sun's activity and wind have changed over its lifetime (\href{https://ui.adsabs.harvard.edu/abs/2007LRSP....4....3G/abstract}{see \citet{2007LRSP....4....3G}}, and \href{https://ui.adsabs.harvard.edu/abs/2019MNRAS.483..873O/abstract}{see \citet{2019MNRAS.483..873O}}), and from that how the Earth's magnetopause distance would have changed over time (\href{https://ui.adsabs.harvard.edu/abs/2018ApJ...856...53P/abstract}{see \citet{2018ApJ...856...53P}}).

\item The multitude of exoplanetary systems continues to clarify the formative processes of planetary systems in general, and of the Solar System in particular, with strong evidence for migration \indexit{planetary!system!orbit migration}of planetary orbits, the impact of that on the formation of Mars, and the possible cause of the Late Heavy Bombardment and the transport of water to the terrestrial planets by gravitational scattering (e.g., \href{https://ui.adsabs.harvard.edu/abs/2014Icar..239...74O/abstract}{see \citet{2014Icar..239...74O}}).
\end{itemize}

The exchange of concepts, knowledge, and models between the domains of heliophysics, stellar astrophysics, and exoplanetary sciences is only in its beginning phases: expanding observational and computational capabilities will propel these fields forward, catalyzed by a joint approach. One such area that requires a joint approach is already developing under the name \indexit{transit light source effect}of 'transit light source effect' (\href{https://ui.adsabs.harvard.edu/abs/2018ApJ...853..122R/abstract}{see \citet{2018ApJ...853..122R}}, \href{https://ui.adsabs.harvard.edu/abs/2019AJ....157...96R/abstract}{and \citet{2019AJ....157...96R}}, \href{https://ui.adsabs.harvard.edu/abs/2019arXiv190306152R/abstract}{and \citet{2019BAAS...51c.328R}}): exoplanetary transit spectroscopy (e.g., \href{https://ui.adsabs.harvard.edu/abs/2019PASP..131a3001D/abstract}{see \citet{2019PASP..131a3001D}}) to study the chemical makeup and dynamics of an exoplanetary atmosphere is unavoidably linked with the analysis of the non-magnetic atmosphere (\href{https://ui.adsabs.harvard.edu/abs/2017A%26A...605A..90D/abstract}{see \citet{2017A&A...605A..90D}},\href{https://ui.adsabs.harvard.edu/abs/2017A&A...605A..91D/abstract}{and \citet{2017A&A...605A..91D}}) and of the magnetic structures on the stellar atmosphere (e.g., \href{https://ui.adsabs.harvard.edu/abs/2018MNRAS.480.5314P/abstract}{see \citet{2018MNRAS.480.5314P}}; \href{https://ui.adsabs.harvard.edu/abs/2018AJ....156..178Z/abstract}{and \citet{2018AJ....156..178Z}}). Thus as we learn about exoplanetary atmospheres we will at the same time learn about starspots and stellar active regions at a resolution that has so far been unobtainable. That will provide information on how stellar dynamos structure their magnetic field from the scale of starspots upward that may prove critical to the development and validation of a predictive model for solar and stellar magnetic activity. That, in turn, will diminish the uncertainties on properties of stellar winds and their impacts over time on planetary atmospheres.

\section{Activities to take you beyond the present-day solar system}
The purpose this book is to
give you fundamental insights into the couplings between the Sun, its
planets, and the interstellar medium, with particular focus on Venus,
Earth, and Mars.  The text could only give you an introduction to the
multitude of aspects of the field of heliophysics and the links to
other disciplines, although many of the related fields --~such as
nuclear physics, geophysics, biochemistry, meteorology, and radiative
transfer~-- were only mentioned or implied in passing. With this book,
you have the basic tools in hand to explore the Solar System over
time, from its formation to its ultimate demise (although it stops
short of the very end when the Sun transforms into a white dwarf,
while the Earth may end up either evaporated and blown into
interstellar space or pulled apart and spiraling into the white
dwarf's atmosphere, or a combination of these). This book also gives
you the tools to look outward and into the future: you have the basic
concepts available now to explore the multitude of new worlds that are
being discovered, analyzed, and inspected for potential signs of life
(such as the \indexit{exoplanet!TRAPPIST\,1 b--h}TRAPPIST-1 system sketched in
Fig.~\ref{fig:trappistcomp}). It is an exciting time for all of
that. Why not start exploring with the final set of activities?

% \vspace{1 cm}
% http://www.spectralcalc.com/blackbody/CalculatingBlackbodyRadianceV2.pdf
\activity{{\em Background/Advanced/Group:} {\bf 'What if' scenarios:}
  If you would like to think well 'outside the box' of things
  explicitly discussed in this book and in the Heliophysics volumes
  then consider this: what are things like when settings change?  You
  could think of exoplanets with different host stars, orbits, and
  atmospheres, but there will be limited guidance by what we actually
  know from the literature. (1) For an example not too far from home,
  you could consider
  \href{https://en.wikipedia.org/wiki/Titan_(moon)}{Titan}, the only
  moon (natural satellite) in the solar system with a substantial
  atmosphere that is mostly N$_2$ (some 97\%) and CH$_4$ (much of the
  remainder). {\em Activity 1}: Find Titan's equivalent values for the
  quantities listed in Table~\ref{tab:brain1}; use, example,
  \href{https://en.wikipedia.org/wiki/Titan_(moon)}{Wikipedia} and its
  references for a start. {\em Further reading on Titan:} You can find
  publications on the \hbox{(photo-)}chemistry of its atmosphere
  leading to an ionosphere rich in HCNH$^+$ and C$_2$H$_5^+4$. The
  chemical network in the high atmosphere leads to heavy organic
  molecules and aerosols that are deposited onto Titan's frozen
  surface and into its hydrocarbon lakes. Titan orbits within Saturn's
  magnetosphere, generally shielded from the direct impacts of the
  solar wind. However, the solar wind causes Saturn's magnetosphere to
  be highly asymmetric, and thus the environment through which Titan
  orbits is highly
  \href{https://eos.org/research-spotlights/how-saturn-alters-the-ionosphere-of-titan}{dependent
    on its orbital phase}. Cosmic rays and energetic particles from
  Saturn's magnetosphere penetrate deep into Titan's atmosphere
  causing ionization and influencing chemical pathways. Titan has no
  intrinsic magnetic field ({\em i.e.,} no functioning magnetic
  dynamo) but an induced magnetosphere that changes as the moon orbits
  the rotating giant planet Saturn. There may be subsurface areas of
  liquid water, a water-ammonia mixture, or different mixtures in
  different locations and at different depths. Life might exist under
  these circumstances, and the traditional definition of
  'habitability' as involving liquid surface water may need rethinking
  as we learn more. (2) For something far from home, consider the
  compact 7(?)-planet system of TRAPPIST-1 (see
  Fig.~\ref{fig:trappistcomp}) on which much is being written: execute
  an \href{https://ui.adsabs.harvard.edu}{ADS search} (at
  https://ui.adsabs.harvard.edu) for refereed
  papers with 'TRAPPIST-1' in the title. {\em Activity 2}: For at
  least one of the TRAPPIST-1 exoplanets, complete the first six lines
  of Table~\ref{tab:brain1} --~the rest remains unknown.  {\em
    Activity 3}: Consider how those environmental
  conditions would affect the overall space
  environment. \mylabel{act:titan}}

\activity{{\em Show:} {\bf
    Observing exoplanetary atmospheres:} (1) Approximate the contrast
  $C_{\ast,{\rm p}}(\lambda)$ between exoplanetary atmospheric
  radiation and stellar surface radiation, assuming both star and
  planet radiate as black bodies (using Planck's law $B(\lambda,T)$;
  ignoring center-to-limb effects), as function of wavelength, of the
  temperatures of star ($T_\ast$) and planet ($T_{\rm p}$), and of the
  effective radii of star ($R_\ast(\lambda)$) and planet
  ($R_{\rm p}(\lambda)$).  Show that
  $C_{\ast,{\rm p}}(\lambda=10\,\mu{\rm m}) \approx
  (20*R_\ast(\lambda)/R_{\rm p(\lambda)})^2$ for $T_\ast=T_\odot$ and
  $T_{\rm p}=T_\oplus$. (2) What is
  $C_{\odot,\oplus}(\lambda=10\,\mu{\rm m})$? This value shows how
  hard it is to separate stellar and planetary signals (it is easier
  for closer-in, warmer, planets and for larger planets, such as hot
  Jupiters).  (3) Why does the IR domain of the wavelength spectrum
  provide optimal access to the exoplanetary spectrum using a
  secondary eclipse (when the planet moves behind the star)? (4) At
  what wavelength does $B(\lambda,T){\rm d}\lambda$ peak for a planet
  at $T_{\rm p}=T_\oplus$ (use Wien's displacement law)? (5) For
  transit spectroscopy, in contrast, optical wavelengths are most
  suitable for G- and K-type stars; why?  See
  \href{https://ui.adsabs.harvard.edu/abs/2019PASP..131a3001D/abstract}{the
    tutorial} by \citet{2019PASP..131a3001D} on exoplanet transit
  spectroscopy for answers and for more on this
  topic. \mylabel{act:transitspectroscopy}}

\activity{{\bf
    Exoplanetary atmospheric spectroscopy:} How does
  wavelength-dep\-en\-dent transparency of an exoplanetary atmosphere
  lead to wavelength-dependent transit depths ${\cal T}(\lambda)$, and
  thereby yield spectral signatures of atmospheric chemicals?
  Basically, the apparent radius of exoplanet-plus-atmosphere depends
  on wavelength because the atmospheric opacity does. But the transit
  depth also depends on whether there are features on the stellar disk
  within the transit path. This provides information on, {\em e.g.,}
  starspot properties. (a) Sketch how these two signals combine into the
  observed transit depth signal ${\cal T}(\lambda,t)$ over a
  transit. (b) Consider how one might go about disentangling these two
  signals.  See the reference in
  Activity~\ref{act:transitspectroscopy} for more information.  }

\activity{{\em Consider:} {\bf Comparative heliophysics:} Look up the
  properties of the stars $\alpha$\,CMa\,A, the Sun, and
  TRAPPIST-1. Consider how the following properties differ for a
  planet (unconfirmed to exist in the case of $\alpha$\,CMa\,A)
  orbiting each of these stars within the continuously habitable zone:
  (a) the size and color of the star, (b) the maximum possible age of
  the planet, (c) the duration of the orbital year and constraints on
  the length of the planetary day, (d) possible constraints on the
  planetary dynamo (subject to what we know about these at present),
  (e) the Alfv{\'e}n Mach number of the stellar wind, (f) the
  magnetopause distance (assuming comparable planetary dynamos), (g)
  constraints on loss of planetary water, (h) the potential of
  measuring interstellar neutral hydrogen from an orbit near that
  planet, (i) the spectrum of the stellar and galactic cosmic rays
  (assuming the same spectrum external to the planetary system).  }

\activity{{\em Consider:} {\bf Energetic particles in
    TRAPPIST-1:}
  \href{https://en.wikipedia.org/wiki/TRAPPIST-1}{TRAPPIST-1} is a
  very different world from our Solar System. The central star
  –-~itself only first observed in 1999~-– is merely 1/8th the size of
  the Sun, only slightly larger than Jupiter. Its brightness is almost
  2,000 times less than that of the Sun. The star is orbited by seven
  known exoplanets (first published on in 2016), much like Earth in
  size and mass, but all very close to their star. At least three of
  these seven planets are estimated to orbit within the liquid-water
  habitable zone. You can start reading up on TRAPPIST-1 using an ADS
  search, but for this Activity review
  \href{https://ui.adsabs.harvard.edu/abs/2017ApJ...843L..33G/abstract}{the
    study} by \citet{2017ApJ...843L..33G} on the astrosphere, and
  \href{https://ui.adsabs.harvard.edu/abs/2019ApJ...874...21F/abstract}{the
    study} by \citet{2019ApJ...874...21F} of the possibly very intense
  radiation environment of the planets. The role of heliophysics in
  this is evident throughout these studies: (1) identify the processes
  you have read about in this book that are elements of these
  studies. (2) Use what you learned in this text to explain why the
  wind is mostly sub-sonic around the seven planets. (3) With dynamic
  pressures 3 to 6 orders of magnitude higher than for Earth, what
  does that do for the planetary magnetopause distances? (4) Although
  this system is diminutive, its astrosphere is potentially huge:
  estimate the distance to the astropause assuming the system is
  subject to ISM conditions similar to those for the Solar
  System. \mylabel{act:trappist1}}

\sactivity{$\circledS$ {Consider/Advanced/Group:} {\bf Arriving at Earth's climate
    from scratch: } In Activity~\ref{act:prochab} you compiled a list
  of all the processes involved in setting a planetary climate system
  that reflected at least all those mentioned in
  Chs.~\ref{ch:formation} and~\ref{ch:evolvingplanetary}. Now,
  complement that list with the additional topics discussed in
  Ch.~\ref{ch:evolvingexposure}. Do not forget to add relevant
  thoughts from your notes for Activity~\ref{act:titan}! Then review
  that list and flag those processes that are beneficial to life as we
  know it on Earth and those that are detrimental to it. The duality
  of many, perhaps most, of the entries on your list should make you
  think about how our Earth, as it is in its present state, is a
  consequence of a remarkable interplay of often simultaneously
  beneficial and detrimental processes, including, perhaps, a series
  of fortuitous developments. Consider the extraordinary challenge of
  thinking about 'habitability' of any of the other thousands of
  exoplanets found to date, including what the phrase 'habitability'
  itself adds to that challenge given how little we know about life
  itself. Better yet, write an essay on this to share with fellow
  students, with teachers, and perhaps a much larger
  readership. \mylabel{act:thefinalquestion} \solution{thefinalquestion}}

%reset final sactivity mark to activity mark for enotes summary:
\renewcommand{\makeenmark}{\hbox{{\tiny \,\{A\theenmark\}\,}}}

\clearpage

\addtocontents{toc}{\vspace{0.5cm}{\centerline{\bf Activities}}\par }
\chapter{\bf Activities}
\label{ch:activities}
\addtoendnotes{\unexpanded{\enotedivision{}{}}}
\enotesize
\setcounter{notelabelcount}{0} % this is necessary for hyperref with endnotes
\theendnotes

\setcounter{secnumdepth}{2} % Reset after enotes
\chapter{\bf Solutions and supplemental text for selected activities}
\label{ch:solutions}
\section{Activity\,\ref{act:windquestions}: Solar wind travel time}\label{windquestions}%Number in V1.3: 13
\indexit{solar!wind!travel time} (a,b) The solar wind speed is
$v_{\rm sw}=(4.3 {\, \rm to \,} 9)\,10^7$\,cm/s, and the Sun-Earth
distance is $d_\oplus=1.5\,10^{13}$\,cm, so the time for the solar
wind to reach Earth is
$\Delta t_\oplus = d_\oplus /v_{\rm sw} = 1.9 {\, \rm to \,}
4.0$\,days (during which the Sun has rotated
$25^\circ {\, \rm to \,} 53^\circ$).

(c) $d_{\rm Neptune}=30$\,AU, so
$\Delta t_{\rm Neptune}=30 t_\oplus = 57 {\, \rm to \,} 120$\,days (or
$\approx 2 {\, \rm to \,} 5$ solar rotations).

(d) The angle of the incoming wind at Earth relative to the Sun-Earth line
is given by $arctan(v_{\rm orb}/v_{\rm sw})$ for an orbital velocity
of $v_{\rm orb}$: $1.9^\circ {\, \rm to \,} 4.0^\circ$.

Note: Because the Sun has rotated $25^\circ {\, \rm to \,} 53^\circ$
in the 2 to 4\,d that it takes for the solar wind to reach Earth, the
field lines that connect the Sun to the Earth typically start from
around that longitude west of the Sun's central meridian (using the
common geocentric reference for directions on the Sun as seen in the
sky).  This is important to understand the typical source region on
the Sun for solar energetic particles reaching Earth: as they follow
the Parker spiral (Sect.~\ref{sec:parker-spiral}), they preponderantly
originate from the western side of the Sun; this is discussed in
Ch.~\ref{ch:conversion}.

\section{Activity\,\ref{act:windenergy}: Parker solar wind; basics}\label{windbasics}%Number in V1.3: 15

\indexit{solar!wind!energy source}Start with Eq.~(\ref{eq:solwind}):
\begin{equation}
\frac{1}{v}\frac{{\rm d}v}{{\rm d}r}\left\{ v^2-\frac{2kT}{m_{\rm p}}\right\} =
\left\{\frac{4kT}{m_{\rm p}r}-\frac{GM_\odot}{r^2}\right\}. \ntag
\label{eq:i} \end{equation}
With $e_{\rm kin}={1\over 2}\rho v^2$, $e_{\rm grav}=-GM_\odot\rho/r$, and
$p_{\rm gas}= 2 \rho kT/m_{\rm p}$ for a fully-ionized
hydrogen-dominated gas, this can be rephrased as a balance of specific
energies and work by isothermal expansion:
\begin{equation}
  {{\rm d}\over {\rm d}r} \left\{ {e_{\rm kin}\over\rho} + {e_{\rm grav}\over\rho} \right\} =
  {p_{\rm gas}\over\rho} \left\{  {1\over v} {{\rm d}v \over {\rm d}r} + {2\over r} \right\} =
  -{p_{\rm gas}\over \rho} \left\{{1\over \rho} {{\rm d}\rho \over {\rm d}r}\right\}, \ntag \label{eq:windenergetics}
\end{equation}
where the rightmost term follows from the central expression using
the time-independent continuity equation Eq.~(\ref{continuity}) absent
sources and sinks in a spherical geometry:
\begin{equation}
{\rho {\bf \nabla}\cdot {\bf v}} = -({\bf v}\cdot{\bf \nabla})\rho \,\,\rightarrow\,\,
{1\over r^2} {{\rm d}\over {\rm d} r} r^2v = -v {{\rm d}\over {\rm d} r}\rho. \ntag
\end{equation}

In words: the changes in kinetic and potential energy are compensated
by the work done by the change in volume of the isothermal gas
(expressed in the central expression of Eq.~(\ref{eq:windenergetics})
in terms of radial expansion by the velocity gradient and lateral
expansion because of the geometry). And that work/energy is supplied
by what keeps the plasma (near-)isothermal: electron heat conduction
from the coronal heat source.

Problem: for an isothermal wind, the terminal velocity is essentially
unbounded, and thus the required energy to power it is also formally
unbounded. That issue goes away if we would allow the temperature to
decrease with distance so that actual heat conduction can do its job
properly (and satisfy the energy equation) at the expense of somewhat
more complicated math. If you did that, then the energy available to
power the solar wind would be set by $\kappa(T) {\rm d}T/{\rm d}r$ at
the coronal base (with $\kappa(T)$ the Spitzer thermal conductivity as
in Eq.~\ref{eq:conduction}).

\subsection*{Analogy to the Laval nozzle}
% http://www.physics.usyd.edu.au/~cairns/teaching/2010/lecture8_2010.pdf

Some \indexit{solar!wind!de Laval nozzle} people draw a comparison between the solar wind subject to gravity and a {\em de Laval} nozzle absent gravity. For a stationary, one-dimensional, isothermal flow through a geometry with cross section $A(r)$, and without magnetic field, gravity, viscosity, or sources or sinks for particles, the continuity and momentum equations (Eqs.~\ref{continuity} and~\ref{momentum}) simplify to:

\begin{equation}
  {{\rm d} \over {\rm d}r} (\rho v A) =0 \,\,\,;\,\,\,\,%\ntag
%\end{equation}
%and 
%\begin{equation}
  \rho v {{\rm d} v\over {\rm d}r}  = - {2kT \over m_{\rm p}}  {{\rm d} \rho \over {\rm d}r}, \ntag
\end{equation}
which can be combined to read
\begin{equation}
\frac{1}{v}\frac{{\rm d}v}{{\rm d}r}\left\{ v^2-\frac{2kT}{m_{\rm p}}\right\} =
{2kT \over m_{\rm p}} {1 \over A}  {{\rm d}A \over {\rm d}r}. \ntag
\end{equation}
Comparison with Eq.~(\ref{eq:i}) suggests:
\begin{equation}
  {2kT \over m_{\rm p}} {1 \over A}  {{\rm d}A \over {\rm d}r} = \left\{\frac{4kT}{m_{\rm p}r}-\frac{GM_\odot}{r^2}\right\}. \ntag
\end{equation}
This has a solution
\begin{equation}
  A(r) \propto r^2 \exp\left( -{e_{\rm grav}\over e_{\rm gas} } \right),\ntag
\end{equation}
with a minimum where $-2e_{\rm grav} = e_{\rm gas}$ below
which gravitational potential energy acts like a constricting nozzle
and beyond which the 'nozzle' eventually 'expands' as $r^2$.

{Insights}: (1) The solar wind is enabled by the high coronal
temperature (electron conduction and high conductive flux - see
Eq.~(\ref{eq:conduction}) and Note~\ref{footspitzer}: high mean-free
path in Coulomb collisions), which supplies energy to beyond the point
where the thermal energy equals the (unsigned) gravitational potential
energy (= kinetic energy of the escape velocity); (2) there is a
hidden energy transport in the assumption made in the model under
discussion that the gas is isothermal, which violates the equation for
internal energy (Eq.~\ref{energy}) because you cannot have conductive
energy transport without a thermal gradient; (3) a different
formulation of the same equation(s) provides different insights.

\section{Activity\,\ref{act:mfps}: Mean free paths and MHD}\label{mfps}%Number in V1.3: NEW
(a) The \indexit{mean free path!heliosphere}'mean' or 'characteristic'
mean-free path length (in cm) for heliospheric electron-electron
\indexit{heliosphere!mean free path}interactions (because we are
talking about electron thermal conduction) as derived from quantities
in Table~\ref{tab:dimensionlessnumbers}:
\begin{equation}
  {\overline{\lambda}}_{\rm mfp,e} \approx {v_{T{\rm e}} \over \nu_{\rm ee}} \sim 1.1\,10^4 {T^2_{\rm e} \over n_{\rm e}}.  \ntag
\end{equation}
Near Earth, with Table~\ref {tab:atmos-param}, for slow and fast wind:
\begin{equation*}
\overline{\lambda}_{\rm mfp,e} \approx 2\,10^{13} - 4\,10^{13}\,{\rm
  cm\, or\,} 1.3 - 2.6\,{\rm AU}.
\end{equation*}
Note: the relevant velocity here is the random velocity $v_{T{\rm e}}$ of the electrons superposed on the much lower bulk flow $v_{\rm sw}$. What is the characteristic ratio $v_{T{\rm e}}/v_{\rm sw}$ at 1\,MK?

(b) Closer to the Sun, $\overline{\lambda}_{\rm mfp,e}$ is much
smaller (because of the higher density) and Maxwellian distributions
are a fair approximation. The electron mean-free path equals the
density scale height roughly around the critical point in the
isothermal Parker solution, which at $10^6$\,K lies at
$\approx 6R_\odot$ (which defines the base of the largely
collisionless exosphere). Why does MHD still provide a fair
approximation for the solar wind beyond that point? We can look at one
of Gene
%Also: https://ui.adsabs.harvard.edu/abs/2021JGRA..12629666C/abstract
Parker's \indexit{MHD!collisionless conditions}lessons: ``\ldots\ it
is widely believed that the large-scale bulk motion within a body of
collisionless gas or plasma is not described by the Newtonian
equations of hydrodynamics \ldots\
%In the absence of interparticle collisions the pressure may be
%anisotropic, of course, represented by a tensor $p_{ij}$ rather than
%a single scalar $p$.
But whether interparticle collisions happen or not,
the bulk flow conserves particles, momentum, and energy, and when
those three conservation conditions are written down, they provide the
equations of hydrodynamics, with the familiar gradient of [pressure],
%$p_{ij}$,
compressibility, etc.  Most textbooks derive these hydrodynamic
equations by computing the zero, first, and second velocity moments of
the collisionless Boltzmann equation, but the simple idea of flux
conservation of particles, momentum, and energy can be used directly
(\citep{2007cemf.book.....P}).''
%It is important to understand that the pressure $p_{i}$j
%represents the momentum flux in the $i$th direction transported by the
%thermal motions in the $j$th direction and has nothing to do with collisions.''

Moreover, in the collisionless domain of the heliosphere, you need to
realize that there are collective electromagnetic interactions and
magnetic perturbations (gradients, including waves: turbulence) that
scatter particles and help maintain fair validity of the concepts
of temperature and pressure as in thermodynamics.  See
Sect.~\ref{sec:mhdintro} and Table~\ref{fig:mhdvalidity} for more on
the assumptions behind MHD and its general applicability to bulk
properties even in collisionless settings.

There is also the following argument that has been explored:

\subsection*{Push or pull? Push and pull?}
An (extreme) alternative to the (equally extreme) hydrodynamic (or
fluid) approximation is the {\em collisionless approximation}. In the
latter, an ambipolar electrostatic potential builds up between the low-mass
electron population and the high-mass ion population. One way to think
about the resulting wind is this: The fastest electrons (which, by the
way, are least affected by collisions) can overcome the gravitational
potential barrier, \indexit{ambipolar!effect}but they cannot flow out
in bulk without taking the ions (and lower-energy electrons) lest they
increase the electric potential as they would leave a charged Sun
behind. So any sustained bulk electron flux must be balanced by an ion
flux (which has to deal with its much higher gravitational potential
that is only partially countered by the electric potential), which
happens when the electrostatic field that builds up sufficiently
counters the gravity on the ions to pull them along.

A very rough approximation and the assumption of Maxwellian tails
(maintained by the hot electron reservoir in the corona below the
'exosphere') shows that the electric potential energy at the exobase
would be roughly double the enthalpy of a Maxwellian electron-proton
plasma: $\sim 5kT$ vs.\ $\sim 5kT/2$ (see
\href{https://ui.adsabs.harvard.edu/abs/1999EJPh...20..167M/abstract}{\citet{1999EJPh...20..167M}}
who provides a very readable introduction of the collisionless
'exospheric' model of the solar wind)

You can expect that the resulting solar wind (mass flux and speed) is
determined by the high-energy tail of the velocity distribution
\ldots\ which is not likely to be that of a true Maxwellian precisely
because the fastest particles have an even lower cross section for
Coulomb collisions and thus interact even less: plasmas are notorious
for having non-thermal high-energy tails.

The reality of the solar wind is neither fully hydrodynamic nor fully
collisionless, that of course also includes a magnetic field and its
perturbations. But, as it turns out, the net behavior of the fluid and
exospheric approximations is much the same because it is governed by
unavoidable conservation laws:

\vskip \baselineskip
N.B. A historical note:
\href{https://ui.adsabs.harvard.edu/abs/1960ApJ...132..175P/abstract}{\citet{1960ApJ...132..175P}} took the wind density at Earth to be 100\,cm$^{-3}$
(interestingly much lower than in his \href{https://ui.adsabs.harvard.edu/abs/1958ApJ...128..664P}{\citeyear{1958ApJ...128..664P}} paper where he says that
``Biermann infers densities at the orbit of earth ranging from 500
hydrogen atoms/cm$^3$ on magnetically quiet days [to much higher
during storms]''). In his 1960 paper, he simply noted that ``the
mean-free path for interparticle collisions is small compared with the
dimensions of the flow'' and used standard hydrodynamic equations,
assuming collective behavior and Maxwellian statistics.
Also, at the
time, he could assume ``that the extension of solar gas into
interplanetary space comes from the entire corona. Hence the
observations altogether suggest that the whole corona flows
hydrodynamically outward into space''.

Interestingly, \href{https://ui.adsabs.harvard.edu/abs/2014RAA....14....1P/abstract}{\citet{2014RAA....14....1P}} says: ``Hardly anyone
believed the trans-sonic expansion of the solar corona. So I had the
field to myself for about four years, elaborating the analytic theory
of the expanding corona, producing two hydrodynamic models of the
heliosphere depending on the existence or absence of an interstellar
wind.''

%\vfill And also: ``The calculation [by Lemaire and Sherer (1971)]
%reminds us once again that the large-scale bulk motions of
%collision-free and collision-dominated plasmas obey the same general
%hydrodynamic equations.  Each approach provides its own insights into
%the dynamics of the solar wind phenomenon.'' (See, e.g., Echim {\em et
% al.}, Surv Geophys (2011) 32:1–70 for a comparison of hydrodynamic
%(or fluid) and kinetic (or exospheric) descriptions of the solar wind)

\section{Activity\,\ref{act:energeticpartpen}: Penetration depth of energetic
  particles}\label{energeticpartpen}%Number in V1.3: 21
Assuming an \indexit{energetic!particles!penetration depth}isothermal
atmosphere dominated by particles of mass $m_{\rm a}$ at temperature
$T_{\rm a}$, the density $n(h)$ as a function of height $h$ above a
reference level with density $n_o$ and the column density $N(h)$ from
infinity to altitude $h$ are given by
\begin{equation}
  n(h)= n_0 e^{-{h \over H_{p}}} \,\, ; \,\, N(h)=\int^\infty_h n(x)
  {\rm d}x= n_o H_{p} e^{-h \over H_{p}} ,\ntag
\end{equation}
for a density scale height $H_{p}$ given by Eq.~(\ref{eq:hp}). The
altitude at which a vertically-moving incoming energetic particle has
penetrated a column mass of $\mu_{\rm col}$ is thus given by:
\begin{equation}
h =   H_{\rm p} {\rm ln}( m_a n_0 H_{p} / \mu_{\rm col}). \ntag
\end{equation}
For energetic particles that penetrate to a column depth of
$\mu_{\rm col}=10$\,g/cm$^2$:
\vskip 0.5\baselineskip
\begin{center}\vbox{
  \begin{tabular}{llrrrcrr}
      \hline
      &Species&$m_a$&$n_0$&$T_a$&$g$&$H_{\rm p}$&$a_p$\\
      &&(g)&(cm$^{-3}$)&(K)&(cm/s$^2$)&(km)&(km)\\
      \hline
  Earth   & N$_2$  &$4.7\,10^{-23}$&$3\,10^{19}$&288&980&8.6&47\\
Mars   & CO$_2$&$7.3\,10^{-23}$&$3\,10^{17}$&210&370&11.&17\\
Sun     & H          &$1.7 \,10^{-24}$&$10^{17}$&5780&$2.7\,10^4$&175.&-90\\
      \hline
    \end{tabular}
  }\end{center}
{\footnotesize{You can combine typical densities and scale heights
  for the solar corona and chromosphere to show that these domains
  cannot efficiently stop particles of these energies.}}
\vskip 0.5\baselineskip

Note that for Earth and Mars, these penetration altitudes are well
above where the isothermal approximation is valid, while the negative
value for the Sun shows that such particles can just penetrate the
photosphere (in the process leading to $\gamma$-ray emission and
nuclear reactions resulting in, among others, neutron-capture line
emission and positron-annihilation line emission; see {\em e.g.},
\citep{2006ApJ...650.1184S} for a multi-mission study of a bright
solar flare). For Al, with a mass density of 2.7\,g/cm$^3$, the
penetration depth required to reach $\mu_{\rm col}=10$\,g/cm$^2$ is 3.7\,cm.

\section{Activity\,\ref{act:hallped}: Collisions and conductivities}\label{hallped}%Number in V1.3: 22

(a) High \indexit{conductivity!collisions}in the atmosphere, collision
frequencies are low, and thus magnetization high, while deeper in the
atmosphere the magnetization is small. As the Pedersen conductivity
$\sigma_{\rm P}$ is, as stated in the text, dominated by the ion term,
a first-order Taylor expansion yields, for singly-ionized ions high
and low altitudes, respectively:
\begin{eqnarray}
  {\rm high\,collision\,frequency:} & & {\rm low\,collision\,frequency:} \nonumber \\
  \sigma_{\rm P}\approx {n_{\rm e}ec \over B} {1\over M_{\rm i}} &\,\,\,&
   \sigma_{\rm P}\approx {n_{\rm e}ec \over B} {M_{\rm i}} \nonumber \\
 \sigma_{\rm H}\approx {n_{\rm e}ec \over B} \left ({-{1\over M^2_{\rm
  e}} +{1\over M^2_{\rm i}}} \right ) &\,\,\,&
   \sigma_{\rm H}\approx {n_{\rm e}ec \over B} \left ({{M^2_{\rm e}}
                                               -{M^2_{\rm i}}} \right
                                               ) \nonumber
\end{eqnarray}

For ${\bf E}$ as specified in the Activity:
\begin{equation}
  {\bf E} \equiv 
  \left ( \begin{array}{c} 0 \\ E \\ 0 \end{array} \right ) \ntag
\end{equation}
so that,:
\begin{eqnarray}
  {\rm high\,(\, to\, first\, order):} & \null & {\rm low:} \nonumber        \\
  {\bf j} \approx \left ( 
  \begin{array}{c}
    0\\ {1 \over M_{\rm i}} \\ 0
\end{array}
  \right ) \parallel {\bf E} &\,\,\,&
   {\bf j} \approx \left ( 
\begin{array}{c}
     +M_{\rm e}^2- M_{\rm i}^2 \\ M_{\rm i} \\ 0
\end{array} \right ) \nonumber                   
\end{eqnarray}
where the righthand result shows ${\bf j}$ rotated in the direction of
$-{\bf E} \times {\bf B}$ assuming that $M_{\rm i} > M_{\rm e}$.

(b) When $\sigma_{\rm H}=\sigma_{\rm P}$, the current direction is
horizontal at
45$^\circ$ from the electric field direction in the geometry
specified in the Activity.

\section{Activity\,\ref{act:unitconv}: cgs to SI}\label{unitconv}%Number in V1.3: 28
The momentum equation in the two unit systems read:
\begin{eqnarray}
  \text{cgs:}&  \rh \frac{\dd\vv}{\dt} +\rh (\vv\cdot\grad)\vv &=
                       +{\rh\vec{g}}-{\grad p} +{\frac{1}{4\pi}(\curl\B)\times\B} \nonumber\\
  \text{SI:}&  \rh \frac{\dd\vv}{\dt} +\rh (\vv\cdot\grad)\vv &=
                       +{\rh\vec{g}}-{\grad p} +{\frac{1}{\mu_0}(\curl\B)\times\B} \nonumber
\end{eqnarray}

The induction equation is the same in cgs as in SI units, as are the
other equations in Table~\ref{fig:mhdset} because they either have
$\B$ in all terms so that the conversion factors cancel, or they have
no term with $\B$ and no conversion is needed.

The
  \href{https://www.nrl.navy.mil/ppd/content/nrl-plasma-formulary}{online}
  NRL Plasma Formulary not only has a conversion table from cgs to SI
  and {\em vice versa}, it is full of other information on plasmas.

\section{Activity\,\ref{act:estimatebeta}: Plasma $\beta$ in the Parker solar wind}\label{estimatebeta}%Number in V1.3: NEW

(a) For a stationary state and ignoring viscosity and plasma sources/sinks (modified Eq.~\ref{momentum}):
\begin{equation}
\rho ({\bf v}\cdot\nabla) {\bf v} =+{\rho{\bf g}}-{{\bf \nabla} p} 
+{\frac{1}{4\pi}({\bf \nabla}\times{\bf B})\times {\bf B}} + \ldots \ntag
\end{equation}
which for orders of magnitude reads like:
\begin{equation}\label{eq:oommom}
{\cal{O}}({1\over 2}\rho v^2/\ell) = \ldots - {\cal{O}}(p/\ell) + {\cal{O}}(B^2/8\pi\ell) + \ldots \ntag
\end{equation}

Here, the plasma-$\beta_{\rm gas}$ naturally arises from a comparison of the final two
terms on the right ({\em cf.},\ Eq.~\ref{eq:betadef}):
\begin{equation}
\beta_{\rm gas}= {p/\ell \over  B^2/8\pi\ell } = 3.5\,10^{-15} {nT \over B^2}, \ntag
\end{equation}
(where $n$ is the total number density of particles, {\em i.e.,} ions plus
electrons).

(b) Approximating the solar wind as a fully-ionized hydrogen plasma,
$\beta_{\rm gas}\approx 2.5 - 0.6$ near Earth and (with $n\approx
(4600-1400)$\,cm$^{-3}$ and $B\approx 0.03$\,G), $\beta_{\rm gas} \approx 0.04 -
0.01$ at $10 R_\odot$. That might suggest that the magnetic field is
not important near Earth  orbit but not quite ignorable close to the
Sun. However, see (d) below.

(c) Alternatively, a plasma $\beta_{\rm ram}$ can be based on the dynamic
(or ram) pressure of the flow, i.e. in a comparison of the term on the
left with the final term on the right in Eq.~(\ref{eq:oommom}):
\begin{equation}
\beta_{\rm ram}= 1.1\,10^{-23} {nv^2 \over B^2}, \ntag
\end{equation}
Using the same velocities (ignoring the wind's acceleration with solar
distance), this yields
$\beta_{\rm ram} \approx 6. - 20.$ near Earth and $\beta_{\rm ram} \approx 0.1$ at
$10 R_\odot$, leading to the same conclusion
as above. However:

(d) In Parker's initial approximation (\href{https://ui.adsabs.harvard.edu/abs/1958ApJ...128..664P}{\citet{1958ApJ...128..664P}}), the flow
is strictly radial, and he ignores rotation. Thus
${\bf \nabla}\times{\bf B}\equiv 0$, so there is no effect of the
magnetic field: the approximation is internally consistent.  In
reality, the field is not ignorable, hence models such as
that discussed in Section~\ref{sec:parker-spiral} (in which a wind can act as a magnetic brake).

The plasma $\beta_{\rm ram}$ based on the ram pressure is
commonly used to assess the balance between, for example, the solar
wind and a planetary magnetic field (see Sect.~\ref{mp}).

{N.B}: dimensionless numbers provide insight into the relative
importance of terms, but note: you have to assess them {\em at a given
  scale} (although in this particular case, the scale length cancels
out when comparing terms but that is not generally so; see, for
example, Activity~\ref{act:meanfieldequation}), so be careful with
them in, {\em e.g.} turbulent spectra, such as in solar convection where a
term may seem ignorable at one scale but is important nonetheless
through scale couplings, and also where small scales enable
large-scale evolution, as in reconnection.

\section{Activity\,\ref{act:buoy}: Thin flux tube and hydrostatic equilibrium}\label{buoy}%Number in V1.3: 38
(a) The MHD momentum \indexit{flux!tube!hydrostatic equilibrium}equation Eq.~(\ref{momentum}) for a static situation reads:
\begin{equation}
+{\rh\vec{g}}-{\grad p} + {\frac{1}{4\pi}(\curl\B)\times\B}= 0\ntag
\end{equation}
Take the dot product of that equation with $\hat{\ell}$, a unit vector
anywhere along the field $\B$, and realize that the Lorentz force is
strictly perpendicular to $\B$, to find;
\begin{equation}
{\rh\vec{g}\cdot \hat{\ell}}={\grad p} \cdot \hat{\ell}, \ntag
\end{equation}
which describes the hydrostatic equilibrium inside the tube.

(b) In a direction $\hat{n}$ normal to gravity we have
\begin{equation}
  -{\grad p}\cdot\hat{n} -
  \frac{1}{8\pi} (\grad B^2\cdot\hat{n}) +\frac{1}{4\pi}((\B \cdot \grad)\B \cdot\hat{n})= 0. \ntag
\end{equation}
  Because we assume a thin flux tube small compared to variations in
  the surrounding magnetic field, we can compare the conditions inside
  the magnetized tube to the unmagnetized outside (with a thin current
  sheet forming the transition). The final term in the above equation
  is approximately the same inside and outside the tube and thus
  cancels in the following when comparing interior $i$ and exterior
  $e$ of the  tube:
\begin{equation}
  {\grad p_{\rm i}} +
  \frac{1}{8\pi}\grad B^2\approx {\grad p}_{\rm e}, \ntag
\end{equation}
which shows that the total pressure within the tube follows the
stratification of the atmosphere around it, with the internal field
strength (and thus flux tube diameter) adjusting by expansion of the
tube as set by the gas pressures inside and outside the tube.

\section{Activity\,\ref{act:meanfieldequation}: Mean-field induction
  equation}\label{meanfieldequation}%Number in V1.3: 51

(a) Substituting and rearranging as specified in the Activity yields:
\begin{eqnarray}
  \frac{\dd\avr{\B}}{\dt}&=&
  \curl(\avr{\vv}\times \avr{\B}) - \curl((\beta+\eta) \curl\avr{\B})
                             + \curl(\alpha\avr{\B}) \nonumber \\
  {\cal{O}}\left({B\over \tau}\right) &=& {\cal{O}}\left({vB \over \hat{\ell}}\right) -
                               {\cal{O}}\left ({(\beta+\eta) B \over
                               \hat{\ell^2}}\right) + {\cal{O}}\left ({\alpha B
                               \over \hat{\ell}}\right) \nonumber
\end{eqnarray}

(b) Using the provided numbers gives: $\tau_{\rm corr}\sim 1$\,d,
$\alpha\sim -0.01$\,cm/s. The time scales for the three terms on the
righthand side (for a characteristic length scale
$\hat{\ell}=R_\odot$) are: $\hat{\ell}/\avr{v}\approx 4$\,yr for the
advection term, $\hat{\ell}^2/\beta \approx 50$\,yr for the diffusive
term (as $\beta >> \eta$),
and $\hat{\ell}/|\alpha|\approx 2$\,yr for the growth/decay term.

\section{Activity\,\ref{act:bl}: Babcock-Leighton surface flux dispersal}\label{bl}%Number in V1.3: 53
(a) Start with the induction \indexit{flux!dispersal!Babcock-Leighton}equation Eq.~(\ref{induction}) assuming that the diffusion
coefficient is (1) a constant and (2) describing the dispersal by
random convective motions, so here represented by $\beta$ from
Eq.~(\ref{eq:SOCA}):
\begin{equation}
  \frac{\dd\B}{\dt}=\curl(\vv\times\B)-\beta \curl\,({\curl\B)}. \ntag \label{sol:induction}
\end{equation}
Using vector identities and the divergence-free nature of the magnetic field we can use
\begin{equation}
  \curl(\vv\times\B)=-\B(\grad\cdot \vv)+(\B\cdot \grad)\vv-(\vv\cdot\grad)\B, \ntag
\end{equation}
and
\begin{equation}
  \curl\,({\curl\B})=-\nabla^2B. \ntag 
\end{equation}
Substituting these final two into Eq.~(\ref{sol:induction}) and regrouping yields:
\begin{equation}
  \frac{\dd\B}{\dt} +(\vv\cdot\grad)\B=-\B(\grad\cdot \vv)+\beta \nabla^2B+(\B\cdot \grad)\vv. \ntag \label{solution:bldiff}
\end{equation}
Reading from the very left side, the first three terms are analogous
to the continuity equation for the scalar density in
Eq.~(\ref{continuity}), and the next term reflects the diffusive
nature of the (super-)granular motions. For a purely radial field
moving in strictly horizontal flows, the final term disappears, and
then Eq.~(\ref{solution:bldiff}) becomes a diffusion
equation for the scalar quantity $B$ reflecting the purely radial
field.

(b) The characteristic time scale for flux to diffuse over the solar
surface is $R_\odot^2/\beta \approx 60$\,yr. That is significantly
longer than the solar cycle, which puzzled the early researchers of
this mechanism because, at the time, the slow meridional advection had
not yet been discovered. But its time scale of
$R_\odot/v_\theta\approx 20$\,yr for a flow that peaks around
10\,m/s. Together, the large-scale advection and the diffusive
dispersal can result in a 22-year cycle period
(\href{https://link.springer.com/article/10.12942/lrsp-2005-5}{see}
\citep{2005LRSP....2....5S} for a historical review of the
flux-transport mechanism).

\begin{figure}[ph!]
  \centerline{\includegraphics[width=\textwidth]{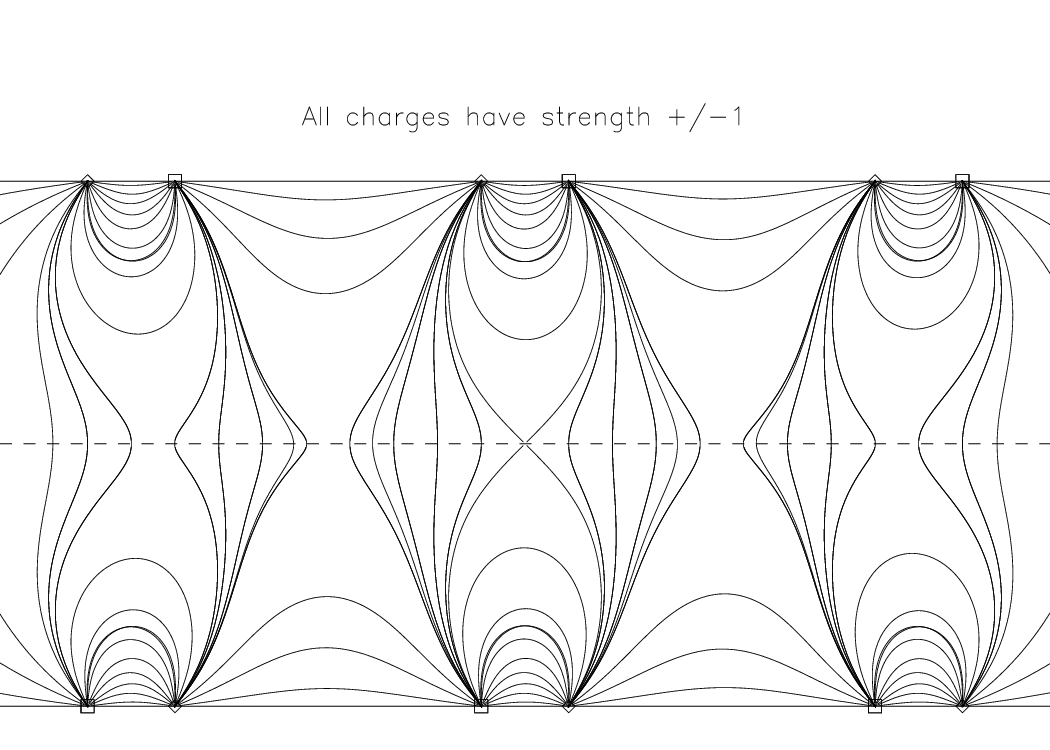}}
\centerline{\includegraphics[width=\textwidth]{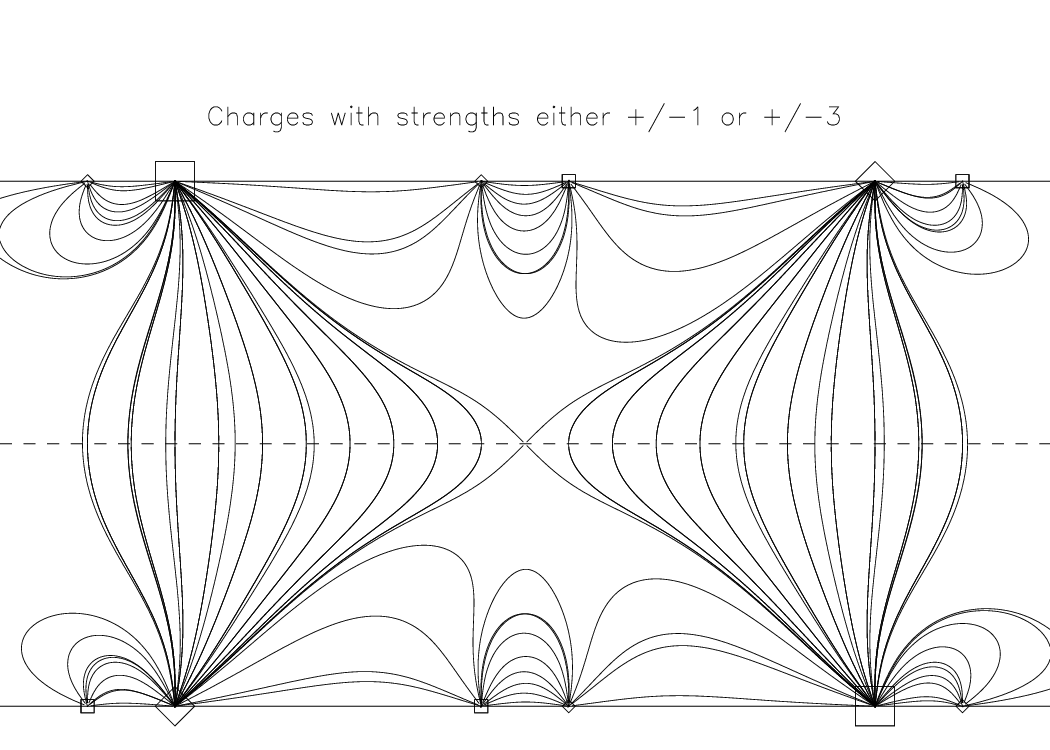}}
\caption[Concept of mirror charges in a potential field ]{Concept of
  mirror charges in a potential field of mixed-polarity monopoles
  leading to nulls and 'helmets.' The size of the charge symbols
  reflects their strength: equal in the top panel and with a few
  larger ones in the bottom panel.  The dashed line, centered between
  the charges and their mirror charges, is where the field is strictly
  vertical everywhere or null as in the $X$ points. In the analogy
  with the spherical PFSS model for the coronal base of the
  heliospheric field (in which, by definition for a potential field,
  there are no currents), the portion below the dashed line is
  equivalent to the coronal domain while above it the field is assumed
  to be fully radial (while spiraling outward further away for a
  rotating star). \label{fig:cartesianpfss}}
\end{figure}
\section{Activity\,\ref{act:sourcesurface}: Solar surface to heliosphere: PFSS}\label{pfssplots}%Number in V1.3: 63
\indexit{PFSS model!surface to heliosphere} {\bf On part (a)}: What
does \indexit{potential field!source surface model|seealso{PFSS}}the streamer \indexit{streamer belt}belt look like in
source-surface models for the Sun?

A figure like that which you were asked to draw is shown in
Fig.~\ref{fig:cartesianpfss}. The top panel (with equal charges apart
from their sign or polarity) gives the impression that each bipolar
area (such as active regions on the Sun) is associated with a helmet
touching on the plane of symmetry but that is not true in general; the
bottom panel has some larger monopoles on either side (like the polar
caps on the Sun around solar cycle minimum) which results in an
overarching field because of the large dipole moment involved (larger
charge and longer base). That field 'caps' some of the potential
'helmets' underneath.

Additional question: from {\em Wikipedia (2022/08/01)}: ``Helmet streamers are bright
  loop-like structures which develop over active regions on the Sun.''
  Is the Wikipedia entry correct?  
  NO! ``\ldots\ the most prominent white-light features (i.e., the bright
streamer stalks) occur at those longitudes where the latitude of the
plasma sheet reaches a local maximum or minimum
${\rm d}\lambda/{\rm d}\phi\approx 0$; when these "stationary points"
are located close to the sky plane, the sheet is viewed edge-on, and
the number of scatterers in the line of sight is greatest.'' (From \href{https://ui.adsabs.harvard.edu/abs/2000JGR...10525133W/abstract}{\citet{2000JGR...10525133W}}).

Look at Fig.~\ref{fig:modelcs} for a 3D representation of model PFSS extrapolations and compare that to what you have learned from your drawing and Fig.~\ref{fig:modelcs}.

{\bf On part (b)}: Simplifications: (1) only one pair of mirror
charges needs to be considered: if the field is radial on some sphere
for that pair, than any other similar pair added will conserve that
property; (2) we can look at this in the $(x,y)$ plane and can rotate
the charges to lie on the $x$ axis; (3) we can think of this as a
problem in electrostatics (which allows monopoles but the magnetic Sun
does not). And we can simplify it to a 2d analysis here because of
points (1) and (2) above.

The field of a point charge of charge $q$ at ${\bf r}_i$ is given by:
\begin{equation}
 {\bf E}_i = q_i {{\bf r}-{\bf r}_i\over |{\bf r}-{\bf r}_i|^3}.
 \ntag
\end{equation}
Let charges 1 and 2 be located at $(1,0)$ and $(\alpha,0)$, respectively.

The requirement that the field becomes radial at a distance
$r_{\rm SS}$ from the origin can be expressed as the requirement that
the vector field anywhere on that circle is normal to the
tangent to that circle.
For a vector and its unit-length normal
\begin{equation}
  {\bf r} =\left(\begin{array}{r}  r \cos\theta \\ r \sin\theta \end{array}\right);
  \hat{\bf r}_\perp =\left(\begin{array}{r} \sin\theta \\ - \cos\theta \end{array}\right)\ntag
\end{equation}
this translates to the requirement that
\begin{equation}
  \left ( {\bf E}_1({\bf r}) + {\bf E}_2({\bf r})\right )\cdot \hat{\bf r}_\perp = 0. \ntag
\end{equation}
Use a unit radius for the Sun and a TBD radius $\alpha$ for the mirror
surface and put the charges at $(1,0)$ and $(\alpha,0)$. Then, working
through the math, you end up with:
\begin{equation}
  (\alpha^2+2\alpha r \cos\theta +r^2)^{3/2} q_1 +
   (r^2        +2r \cos\theta +1)^{3/2} \alpha q_2 =0,  \ntag
%  (r^2+\alpha^2+2\alpha r \cos\theta)^{3/2} q_1 +
%   (r^2+1+2r \cos\theta)^{3/2} \alpha q_2 =0,  \ntag
\end{equation}
which holds true at any $\theta$ for $r_{\rm SS}=\sqrt{\alpha}$ and
$q_2=-\sqrt{\alpha} q_1$. So, for a source surface radius (where the
field becomes radial -- sorry for the confusing term here!) at
$r_{\rm SS}=2.5R_\odot$, the mirror charges need to be placed at
$6.25R_\odot$ and be of strength $-2.5$ times the surface charges.

N.B. An alternative (equivalent) approach is to quantify the fact that
a surface that is normal to a field anywhere on it is an
equipotential surface.

{Insights}: (1) the model naturally explains the formation of streamer cusps (see
Fig.~\ref{fig:modelcs}); (2) below the source surface, the expansion
of the magnetic field can be sub-radial or super-radial
(which is an important factor in determining wind speed). But: the
PFSS model does not describe field dynamics, and by
definition there are no currents within the PFSS field.

\begin{figure}[th!]
  \centerline{\includegraphics[width=\textwidth]{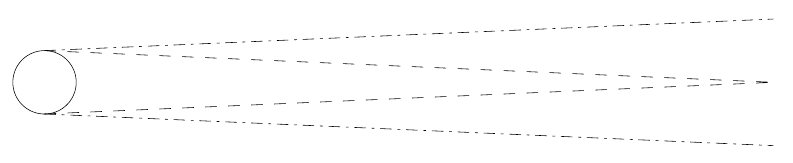}}
\caption[Sketch of the lunar solar-wind wake]{Sketch of a solar-wind
  wake behind the non-conducting Moon. The refilling of the wake is
  outlined by the dashed lines, the expansion of the rarefaction front
  by the dashed-dotted lines. \label{fig:wake}}
\end{figure}
\section{Activity\,\ref{act:noncondsphere}: Solar wind behind a
  non-conducting body}\label{noncondsphere}%Number in V1.3: 66

The wake behind \indexit{solar!wind!non-conducting body}the Moon (see
Fig.~\ref{fig:wake}) is filled at the slow mode speed
${\rm min}(v_{\rm A},c_{\rm s})$ while \indexit{solar!wind!wake}the
rarefaction front moving outward from the object moves at a speed
dependent on the direction relative to the magnetic field:
perturbations along the field move at ${\rm max}(v_{\rm A},c_{\rm s})$
while those perpendicular to the field move at
$(v_{\rm A}^2+c_{\rm s}^2)^{1/2}$ (see Sect.~\ref{sec:mhdwaves}). For
typical slow solar wind conditions around the Moon taken from
Table~\ref{tab:wind-stats} (using the ion temperature), using
expressions from Table~\ref{tab:dimensionlessnumbers}, we have
$v_{\rm A}=30$\,km/s and $c_{\rm s}=18$\,km/s (ignoring $\gamma$
here), giving us parallel and perpendicular fast-mode and slow-mode
velocities of $v_{\rm fm,\parallel}=$30\,km/s,
$v_{\rm fm,\perp}=35$\,km/s and $v_{\rm sm}=18$\,km/s, respectively.

With these numbers, the wake behind the Moon should disappear at a
distance of $v_{\rm sw}/v_{\rm sm}\approx 20$ lunar radii (with an
opening angle at the top of
$2*{\rm atan}(v_{\rm sm}/v_{\rm sw}) \sim 4.8$\,degrees (dashed lines in Fig.~\ref{fig:wake}), while the
rarefaction front should move away from the edge of the Moon
at angles of ${\rm atan}(v_{\rm fm,\parallel}/v_{\rm sw}) \sim 4.0$\,
degrees and ${\rm atan}(v_{\rm fm,\perp}/v_{\rm sw}) \sim 4.7$\,
degrees, respectively (dashed-dotted lines in Fig.~\ref{fig:wake}).

See, {\em e.g.}, \citet{2014JGRA..119.5220Z} for observations and MHD simulations
of the lunar wake.

\section{Activity\,\ref{act:stagnation}: Temperature at the wind stagnation point}\label{stagnation}%Number in V1.3: 69

If we approximate the solar wind as a fully-ionized hydrogen gas and
ignore the relatively small amount of kinetic energy in the electrons,
then from energy conservation it follows that once the solar wind has
come to a standstill at the stagnation point just outside the
terrestrial magnetopause, and after sufficient collisional
interactions in the dense setting there, its temperature is
approximately given by the energy comparison per hydrogen ion:
\begin{equation}\label{eq:spt1}
 T_{\rm sp}={1 \over 4} {m_{\rm H}\over k} v_{\rm sw}^2+T_{\rm sw}. \ntag
\end{equation}
With $T_{\rm sw}\approx 2.4\,10^5$\,K for the characteristic ion
temperature of the fast solar wind (Table~\ref{tab:wind-stats}) at
$v_{\rm sw}=800$\,km/s, we find $T_{\rm sp}\sim 20$\,MK.

This approximation ignores the fact that the electron and ion
temperatures are not the same in the solar wind and also that the
parallel and perpendicular temperatures are the same. We can make the
point, however, that this should not matter much in estimating $T_{\rm
  sp}$ because it is clear from the magnitude of the terms on the
righthand side of Eq~(\ref{eq:spt1}) that the wind's thermal energy is
small  compared to its bulk kinetic energy. What matters more is that
we are not dealing with a pure hydrogen plasma: helium adds extra mass
and extra electrons. We could make the approximation for $T_{\rm
  sp}$ a little better by taking that into account.

With the fractional hydrogen abundance $X=0.73$, helium abundance
$Y=0.26$, and ignoring heavier elements, again assuming a
fully-ionized plasma, and ignoring the thermal energy in the wind,
Eq~(\ref{eq:spt1}) transforms to:
\begin{equation}\label{eq:spt2}
 T_{\rm sp}={1 \over 2} {m_{\rm H}\over k} v_{\rm sw}^2 \left ( {1+Y/X
     \over 2+3Y/X} \right ), \ntag
\end{equation}
from which, again for the fast solar wind as above, $T_{\rm sp}\sim
15$\,MK.

This looks perhaps unexpected: $T_{\rm sp}$ is substantially larger
than the temperature of the corona from which the energy is drawn, but
there is a conversion step that occurs in the solar wind --~thermal
energy to work by expansion and additional conducted thermal energy to
thermal energy~-- that makes this possible.

\section{Activity\,\ref{act:cfradius}: Magnetopause distances}\label{cfradius}%Number in V1.3: 70

(a) The magnetopause \indexit{magnetopause!distance}distance is
determined (see Sect.~\ref{mp}) primarily by the requirement that the
total pressure (plasma plus magnetic) must have the same value on both
sides of the discontinuity. At the Chapman-Ferraro distance of
Eq.~(\ref{eq:PB}), the linear momentum flux density (or dynamic
pressure) in the undisturbed solar wind,
\mbox{$\rho_{\rm sw}{v_{\rm sw}}^2$}, at the sub-solar region equals
the interior magnetic pressure of the dipole field,
\mbox{$B(r) = (1/8\pi)(\mu_{\rm p}/{\mathrm r}^3)^2$} with
$\mu_{\rm p}=B_{\rm p}R_{\rm p}^3$ the magnetic dipole moment of the
planet with equatorial field $B_{\rm p}$ and radius $R_{\rm p}$, with
an extra factor $\xi \simeq 2$ to roughly correct for the added field
from magnetopause currents:
\begin{equation}
  R_{\rm CF} \approx R_{\rm p} \left( \frac{ B^2_{\rm p}}{2 \pi \rho_{\rm sw} {v_{\rm sw}^2}} \right)^{1/6} = 1.7\,10^{10} {R_{\rm p} \over R_\oplus}
  \left( \frac{ B^2_{\rm p}}{ n_{\rm sw}}\right)^{1/6}, \ntag
\end{equation}
where the solar wind speed has been set to 400\,km/s in the righthand expression.

Note the weak dependence on $n_{\rm sw}$: $1000^{1/6}\approx 3$.

With data from Tables~~\ref{tab:fran2} and \ref{tab:fran3}:
\vskip 0.5\baselineskip
\begin{center}\begin{tabular}{lclcclrcl}
  \hline
  Planet & $B_{\rm p}$  (G) & $n_{\rm sw}$ (cm$^{-3})$& $R_{\rm p}/R_\oplus$ & $R_{\rm CF}/R_{\rm p}$ & $R_{\rm CF}$ (cm)\\
 \hline
  Earth&0.3&7.&1&9&$6.3\,10^9$\\
  Jupiter&4.3&0.2&11.2&43&$3.1\,10^{11}$\\
  Saturn&0.21&0.07&9.4&19&$1.1\,10^{11}$\\
  Uranus&0.23&0.02&4.0&24&$6.1\,10^{10}$\\
  Neptune&0.14&0.006&3.9&24&$6.2\,10^{10}$\\
  \hline
\end{tabular}\end{center}
\vskip 0.5\baselineskip

N.B This does it not work well for Jupiter (and also Saturn): $R_{\rm
  CF, Jup}$ is some 43 planetary radii but the observed magnetosphere is $50-100$ planetary radii, and it also varies substantially with changing solar wind parameters. For the size difference, see Fig.~\ref {fig:MP} and its discussion. On the size variability, Fran Bagenal in [HI:13.2.2] writes: ``This greater compressibility of the jovian magnetosphere is due to a significant contribution of the plasma pressure in the equatorial plasma sheet as well as a substantial system of azimuthal currents that weaken the radial gradient of the magnetic field compared to a dipole.''

(b) For orbital radii $d_{\rm orb}$ around their planet, we have
$r\equiv R_{\rm CF}/d_{\rm orb} \approx 0.16$ for Earth's Moon,
$r=4.6$ for Enceladus, and  $r=0.9$ for Titan. Consequently, Earth's
Moon is always outside the magnetosphere except when it crosses Earth's
geotail, Enceladus is always well inside the magnetosphere, and Titan
can be in or out depending on the solar wind conditions.

\section{Activity\,\ref{act:loopemission}: Coronal loop cooling time scales}\label{loopemission}%Number in V1.3: 85

(a) The EUV and X-ray \indexit{coronal!loop!cooling time
  scale}emissions from the solar corona are dominated by de-excitation
from electron-ion collisions (line emission) with a weak contribution
from free-free emission, both of which scale with $n_{\rm e}n_{\rm i}$
(for a given spectral line, you have to bear in mind that $n_{\rm i}$
refers to the density of the corresponding ion so that the ionization
balance comes into the line strength; here, however, we look at the
total energy loss from the plasma for which
$n_{\rm r}\appropto n_{\rm e}$).

(b) The radiative cooling time scales as the ratio of the thermal energy content of the plasma to the rate of energy loss, so:
\begin{equation}
\tau_{\rm rad} \propto \frac{nT}{n^2 P(T)} \propto \frac{T^{5/3}}{n}
\propto {L \over T^{1/3}},
\ntag
\end{equation}
where the final expression results by introducing the scaling law
Eq.~(\ref{eq:rtv}), which you will encounter in Ch.~\ref{ch:heating},
to eliminate $n$.

The conductive time scale for a loop with a length scale $L$ and with
constant cross section scales as the ratio on the thermal energy
content of the plasma to the order-of-magnitude estimate of the rate
of heat conduction (based on the central term on the righthand side of
Eq.~\ref{eq:loop}):
\begin{equation}
\tau_{\rm cond} \propto \frac{nT}{\frac{T^{7/2}}{L^2}} = \frac{n
  L^2}{T^{5/2}} \propto {L \over T^{1/2}},  \ntag
\end{equation}
again using Eq.~(\ref{eq:rtv}) for the final expression.

For plasma at 1\,MK and 5\,MK, respectively,
$\tau_{\rm rad}(1\,{\rm MK})/\tau_{\rm rad}(5\,{\rm MK}) \approx 1.7$
and
$\tau_{\rm cond}(1\,{\rm MK})/\tau_{\rm cond}(5\,{\rm MK}) \approx
2.2$. The latter shows a somewhat larger contrast because of the steep
dependence of electron conduction on the thermal gradient.

\section{Activity\,\ref{act:verbrake}: Magnetic braking of the
  Sun}\label{verbrake}%Number in V1.3: 92

The typical slow \indexit{magnetic!braking}solar wind carries a
kinetic energy flux density from the Sun given by
\begin{equation}
  \ntag
  F_{\rm kin,sw}={1\over 2} n_{\rm sw,\oplus} m_{\rm H} v_{\rm sw,\oplus}^3
  {d_\oplus^2 \over R_\odot^2} = 3\,10^4 {\rm \,erg/cm^2/s} \approx {1\over 3} F_{\rm corona},
\end{equation}
with $\oplus$ denoting values at Earth.

Approximating the field as radial (so that it scales as $1/r^2$, as
does the solar wind density), and the solar wind speed as constant
within the distance $d_\oplus$ to the Earth, the Alfv{\'e}n radius
is found by equating the solar wind speed to the Alfv{\'e}n speed,
yielding for the slow solar wind, using values for wind density and
speed at Earth:
\begin{equation}
  \ntag
  r_{\rm A} \approx {B_\oplus d_\oplus \over (4\pi n_\oplus
  m_{\rm  H})^{1/2} v_{\rm sw} } \approx 20 R_\odot.
\end{equation}

The time scale for angular momentum loss for the present-day Sun
through the solar wind (using numbers for the slow wind in the
quantitative estimates) is, with Eq.~(\ref{eq:4_02}), given by:
\begin{equation}
  \ntag
  \tau_{\rm J}= {J \over |\dot{J}|} = {3 I \over 8 \pi r_{\rm A}^2
    n_\oplus m_{\rm H} v_{\rm sw}^3} =
  {24\pi I v_{\rm sw} \over \Phi_{\rm open}^2} \approx 3\,10^{17}\,{\rm s} \approx
  10^{10}\,{\rm yr},
\end{equation}
where the final expression with symbols shows the dependence on solar
wind speed and the level of solar activity expressed as function of
the open flux $\Phi_{\rm open}$ in the heliosphere.

\section{Activity\,\ref{act:breakup}: Tidal
  breakup}\label{breakup}%Number in V1.3: 94

(a) In order for an \indexit{tide!breakup}object to be pulled apart by tidal forces, the
gradient of the gravitational pull of the central object (with mass
$M$ at a distance $d$) across the object (with mass $m$ and radius
$R$) has to exceed that object's own gravitational pull, {\em i.e.},
\begin{equation}
  \ntag
  {GM \over (d-R)^2} - {GM \over d^2} > {Gm \over R^2}.
\end{equation}
With first-order Taylor expansion for $d\gg R$ and resorting of terms:
\begin{equation}
  \ntag
  d_{\rm breakup} = \left ({2M\over m} \right )^{1/3} R.
\end{equation}
For the Earth, with $m=6\,10^{27}$\,g, $R=6.4\,10^{8}$\,cm, and
$M=M_\odot=2\,10^{33}$\,g: $d_{\rm breakup}= 6.\,10^{10}$\,cm or
0.0035\,AU.

(b) For an
object like comet 67P: $d_{\rm breakup}= 0.015$\,AU or $3R_\odot$.

Note: With this simplest of descriptions, for a 1.5\,km diameter comet with
a density of some $0.4$\,g/cm$^3$, the breakup distance for an
approach to Jupiter is estimated as $d_{\rm breakup}= 1.3\,10^{10}$\,cm or 1.8
Jupiter radii. For a detailed study of the 1992 breakup of comet
Shoemaker-Levy 9 on its approach of Jupiter to 1.33 planetary radii,
see \citet{2012ApJ...759...93M}.

%\section{Activity\,\ref{act:exbdrift}: ${\bf E} \times {\bf B}$ drift}\label{exbdrift}%Number in V1.3: 100

\section{Activity\,\ref{act:driftvelocities}: Drift velocity}\label{driftvelocities}%Number in V1.3: 101

(a) Given the specifications in the Activity, the magnitude of the
gradient drift can be expressed as
\begin{equation}
  \ntag
  |v_{\rm G,}(r)| = {c W_\perp \over qB}{3\over r} = {3 c W_\perp \over
    qB_{\rm e} r_{\rm p}} \left ( {r\over r_{\rm p}}\right )^2.
\end{equation}

For Earth, $|v_{\rm G,\oplus}(2R_\oplus)|\approx 3.$\,km/s and
$|v_{\rm G,\oplus}(10R_\oplus)|\approx 80$\,km/s (with an orbital
period of $P_{\rm G,\oplus}(10R_\oplus)=5$\,ks. For Mercury,
$|v_{\rm G,Merc}(2R_{\rm Merc})|\approx 8.$\,km/s and
$|v_{\rm G,Merc}(10R_{\rm Merc})|\approx 200.$\,km/s (with an orbital
period of $P_{\rm G,Merc}(10R_{\rm Merc})\approx 0.7$\,ks.

For Jupiter,
$|v_{\rm G,Jup}(2R_{\rm Merc})|\approx 0.02$\,km/s and
$|v_{\rm G,Jup}(10R_{\rm Merc})|\approx 0.5$\,km/s (with an orbital
period of $P_{\rm G,Jup}(10R_{\rm Jup})\approx 100$\,d.

(b) Relativistic
energies for ions and electrons are in the range of 100\,MeV and
100\,keV, so the non-relativistic approximation is warranted at 50\,keV.

(c) Except
for the direction of the drift, the results are the same for electrons
because there is no mass dependence in $v_{\rm G,}(r)$.

\section{Activity\,\ref{act:sepsun}: SEP paths and
  scattering}\label{sepsun}%Number in V1.3: 108

\paragraph{(1a,b)} The \indexit{energetic!particles!diffusion}earliest-arriving solar energetic particles have
undergone little if any scattering. The particles' energy yields a
velocity, and with the known Sun-Earth distance, that might be seen to
suffice to derive an estimated time for the explosive event. BUT:
particles follow the Parker spiral, the tightness of which depends on
the history of the solar wind speed. So instead of working with a
single timing, we can use the difference $\Delta t$ of two arrival
times $t_{1,2}$ and energies $E_{1,2}$ to determine both event time
$t_0$ and path length $\ell$ (which is the same for these particles),
from which:
\begin{equation}
  \ntag
  \ell = \Delta t / ( v_2^{-1} -  {v_1^{-1}})\,\,;\,\,t_0=t_1-(\ell/v_1).
\end{equation}

Looking at the first arrivals at 0.3\,Mev/n
($v_1\approx 7500$\,km/s for protons) in Fig.~\ref{Giacalone_Fig11} we
see $t_1\approx$15\,UT, while at 0.04~MeV (with
$v_2\approx 2800$\,km/s for protons) we estimate $t_2\approx$01\,UT on the
next day. With $\Delta t \approx 3.6\,10^4$\,s, $\ell\approx 160\,10^6$\,km,
as expected somewhat longer than the distance from the solar surface
to the Earth: $149.6-0.7$\,km. From this: $t_0$ should be about 6\,h
prior to $t_1$ so 1999/01/09 9\,UT.

\paragraph{(2)} Differentiating Eq.~(\ref{eq:Giacalone_Eq1.25}) for $t$
yields:
\begin{equation}
  \ntag
  {\partial f(x,t) \over \partial t} = {1 \over 2t} f(x,t) \left (
    {x^2 \over 2\kappa t} - 1 \right ),
\end{equation}
which has an extremum at $\kappa =x^2/(2t)$. The estimated peak time
in particle density at 0.04\,MeV/n is 1999/01/10 4\,UT, some 19\,h after
the estimated event time, so that $\kappa \sim 1.9\,10^{21}$\,cm$^2$/s.

\section{Activity\,\ref{act:granscale}: Setting the scale of granulation}\label{granscale}%Number in V1.3: 114

{\em Based on the text in [III:5.2.1]:}
\indexit{granulation!scale}Consider a single convection cell, or
granule, with an upflow region surrounded by narrow downflow
lanes. The vertical velocity at its center $v_z$ must be large enough
to balance the radiative losses at the surface.  The convective
enthalpy flux just under the photosphere is dominated by the latent
heat flux associated with hydrogen ionization so to get an idea for
what $v_z$ is needed we can set
\begin{equation}%\label{photosphere}
\rho N_A v_z y \chi_H = \sigma T^4, \ntag
\end{equation}
where $y$ is the ionization fraction, $N_A$ is Avogadro's number,
$\chi_H$ is the ionization energy for hydrogen, $\sigma$ is the
Stefan-Boltzmann constant, and $\rho$ and $T$ are the density and
temperature.  For photospheric conditions ($y \sim 0.1$,
$T \sim 5800$\,K, $\rho \sim 2\times 10^{-7}$\,g/cm$^3$), this gives a
minimum vertical velocity $v_z$ of about 2 km\,s$^{-1}$.

With this we can estimate the maximum horizontal size $L$
of granules from mass conservation.  If a granule simply overturns
without altering the local density appreciably, then the continuity equation
implies
\begin{equation}\label{eq:vzbal}
  L \sim D v_h / v_z, \ntag
  \end{equation}
where $D$ is the vertical scale and $v_h$
is the horizontal velocity (assuming a cylindrical geometry with upflow in
the center and downflow around the periphery adds a factor of 1/2).
The pressure-driven horizontal flows are unlikely to exceed the sound
speed $c_s \sim$ 10 km\,s$^{-1}$,
so with $v_z$ from Eq.(\ref{eq:vzbal}) and $D \sim H_p \sim $400\,km, we find that
the horizontal scale of granules should not exceed about 1-2\,Mm.

\section{Activity\,\ref{act:acousticcutoff}: Acoustic
  cutoff}\label{acousticcutoff}%Number in V1.3: 115

(a) In an isothermal, \indexit{cutoff frequency}gravitationally
stratified atmosphere, a wave at the acoustic cutoff frequency
propagates over a distance of $c_{\rm s}/\omega_{\rm a}=2H_p$ within
one period, which means that it is moving almost the entire mass in
the atmosphere above. Consequently, there is no substantial force to
counter the expansion of the wave, so no restoring pressure force,
only the gravitational force of the lifted atmosphere above, so the
wave cannot propagate (but is evanescent) and is reflected. In the
Sun, with $H_p(z)$ a function of $z$ as the temperature drops outward
(up to a near-photospheric temperature minimum below the chromosphere)
waves with frequencies below $\omega_{\rm a}$ for around (roughly) the
temperature minimum are reflected and can form resonant standing
($p$-mode) waves within the Sun, while waves at higher frequencies can
propagate out of the solar interior.

(b) For the solar photosphere and an ideal mono-atomic gas with $\gamma=5/3$:
\begin{equation}
  \ntag
  \omega_{\rm a,\odot} = {c_{\rm s} \over 2H_p} = {\gamma g_\odot \over
    2 c_{\rm s}} = 0.02\,{\rm s}^{-1},
\end{equation}
with an acoustic cutoff period of
$P_{\rm a} =2\pi/\omega_{\rm a}\approx 5$\,min. Oscillations at lower
frequencies are trapped, while higher-frequency ones can reach into
the chromosphere. For comparison, the acoustic cutoff period for the
actual non-isothermal solar photosphere is about 200\,s
(\citep{2011ApJ...743...99J}).

\newcommand{\ddvh}[2]{{{\partial#1}\over{\partial#2}}}
\newcommand{\ddtvh}[1]{\ddvh{#1}{t}}
\newcommand{\ddzvh}[1]{\ddvh{#1}{z}}
\newcommand{\dddzvh}[1]{\ddvh{^2#1}{z^2}}
(c) Now, let us look at the idealized situation from the Activity.
Using that the background is stationary ($v_0=0$) and hydrostatically
stratified:
\begin{equation}
  \ntag
  \rho_0 g = - \ddvh{p_0}{z} \,\,\rightarrow \,\, \rho_0(z)=\rho_{\rm
    b} e^{-{z\over H_p}}
\end{equation}
the linearized continuity and momentum equations are:
\begin{equation}
  \ddtvh{\rho_1}=-\rho_0\ddzvh{v_1}-v_1\ddzvh{\rho_0}
  =-\rho_0\ddzvh{v_1}+{\rho_0 v_1\over H_p}, \ntag
\end{equation}
\begin{equation}
  \rho_0\ddtvh{v_1}=-g\rho_1-c^2_s \ddzvh{\rho_1},  \ntag
\end{equation}
because with $\gamma=1$ (for isothermal perturbations) 
$\rho_1=c_{\rm s}^2 p_1$.

Taking the time derivative of the linearized momentum equation and
substituting the linearized continuity equation leads to:
\begin{equation}
 {\partial^2 v_1\over \partial t^2}-c_s^2{\partial^2 v_1\over\partial
  z^2}= -g {\partial v_1 \over \partial z}.  \ntag
\end{equation}

Because in this non-dissipative scenario the kinetic energy in the
wave is a conserved quantity, we can anticipate a solution that scales
like
\begin{equation}
  \ntag
v_1 = u e^{{z/2H_p}} 
\end{equation}
so that $\rho_0 \langle v_1^2 \rangle$ is independent of $z$.
Combining these last two equations leads to:
\begin{equation}
 {\partial^2 u\over \partial t^2}-c_s^2{\partial^2 u\over\partial
  z^2}= -{c_{\rm s}^2 \over 4H_p^2} u.  \ntag
\end{equation}
With $u=u_0 \exp(-i(kz-\omega t))$ this  yields the dispersion relationship
\begin{equation}
  \ntag
  c_{\rm s}^2k^2=\omega^2-\omega_{\rm a}^2,
\end{equation}
which has propagating solutions only for $\omega>\omega_{\rm a}$.

\section{Activity\,\ref{act:pressgrav}: Optical depth and field strength}\label{pressgrav}%Number in V1.3: 118
Eq.~(\ref{eq:optdepth}) \indexit{flux!tube!field strength and optical
  depth}discussed the optical depth, $\tau$, in terms of incoming
radiation, but it is equally valid of course for outgoing radiation
and it can help establish the column depth of a stellar
photosphere. Looking at it for light propagating vertically through an
isothermal atmosphere, we thus have:
\begin{equation}
\tau=\sigma_{\rm a} n(h) {H_{p} }, \ntag
\end{equation}
where the product of the number density $n(h)$ at height $h$ with the
scale height $H_{p}=kT/mg$ at that level represents the integrated
content of an isothermal stratified column of gas above that point. If we focus on the lower main sequence, then (approximating data in \citet{2018MNRAS.479.5491E} in their
Fig. 6 (top) below 1.2 solar masses):
\begin{equation}
{R_\ast \over R_\odot} \approx {M_\ast \over M_\odot}. \ntag
\end{equation}
Consequently, at $\tau\approx 1$ and with $p\propto nT$ we have
\begin{equation}
  p \propto {1 \over M_\ast}. \ntag
\end{equation}
For a maximally evacuated flux tube the external gas pressure balances the internal magnetic pressure so that $p \approx B^2/8\pi$ and therefore
\begin{equation}
  B \propto {1 \over M_\ast^{1/2}}. \ntag
\end{equation}
For stellar masses of 1.7\,$M_\odot$ at F0 and 0.6\,$M_\odot$ at
late-K, and with a field strength of some 2\,kG for the Sun, we can
estimate $1.5$\,kG for F0\,{\sc V} stars and $2.6$\,kG for late-K MS
stars. That is not quite the contrast supported by observations (from
about 1.4\,kG to 3.2\,kG), but then opacities depend on density and
temperature (increasing with both for near-photospheric conditions),
and atmospheres are neither isothermal nor plane parallel.

\section{Activity\,\ref{act:loopprop}: Properties of coronal loops}\label{loopprop}%Number in V1.3: 118
(a,b,c) Quiet-Sun, \indexit{coronal!loop!properties}mixed-polarity
area: $B=5$\,G, $T=$1\,MK, $L=4\,10^9$\,cm: \\$\epsilon_{\rm
  heat}\approx 6\,10^{-5}$\,erg/cm$^2$/s, $n_{\rm e}\approx
3\,10^8$\,cm$^{-2}$, $e\approx 0.1$\,erg/cm$^3$,
\\$r_{\rm gi}=200$\,cm, $c_{\rm s}=30$\,km/s, $\beta \approx 0.1$, and
loop crossing time $L/c_{\rm s}\approx 1300$\,s.
  
Active region: $B=100$\,G, $T=$3\,MK, $L=15\,10^9$\,cm:
\\$\epsilon_{\rm heat}\approx 2\,10^{-4}$\,erg/cm$^2$/s, $n_{\rm
  e}\approx 8\,10^8$\,cm$^{-2}$, $e\approx 1.$\,erg/cm$^3$,
\\$r_{\rm gi}=16$\,cm, $c_{\rm s}=160$\,km/s, $\beta \approx 0.002$,
and loop crossing time $L/c_{\rm s}\approx 500$\,s.

(d) The values of the ion gyroradii are very small compared to the loop
length, the plasma $\beta$ substantially below unity, and the loop
crossing times of pressure perturbations shorter (although marginally
for the quiet Sun) than the cooling and heating time scales of the loops (which
should equal, on average, $e/\epsilon_{\rm heat}$ for a
quasi-equilibrium state). So the concept of an isolated
quasi-stationary loop is warranted for the corona, provided that
heating occurs fairly steadily (or in sufficiently small and frequent events).

\section{Activity\,\ref{act:hrevol}: Least-massive post-main-sequence
  star}\label{hrevol}%Number in V1.3: 128

For a star of $1M_\odot$ it takes 10.4\,Gyr to turn substantially off the main
sequence (marked by the squares in Fig.~\ref{figure:evolmodel}). For
the next less-massive star in the table in that figure at $0.8M_\odot$
it takes already 24\,Gyr, so given the age of the Universe of some
13.8\,Gyr, the least-massive post-main-sequence
star cannot be substantially lighter than some $0.95M_\odot$.

When star formation started after the Big Bang is an active field of
study. The analysis by \citet{2022Natur.603..599X} suggests that in
our own Galaxy it started some 13\,Gyr ago, only 800\,Myr after the
Big Bang. These authors searched for stars in the subgiant phase, just
beyond the main-sequence turnoff: these are good markers to study
stellar ages because that phase is relatively brief thus narrowing the
uncertainty interval. But because they are short-lived, a large
sampling of stars with known distances is required to build up a
sufficiently-large sample. Fortunately, the Gaia astrometry mission
combined with the LAMOST spectroscopy survey provided such a view of
the Galaxy, which was then combined with model isochrones. 
Spectroscopy showed that that first phase
of Galactic thick-disk and halo star formation continued for some 5~to
6\,Gyr during which the interstellar medium (ISM) out of which these
stars formed was enriched with heavier elements (by well over an order
of magnitude for Fe, for example).

\section{Activity\,\ref{act:emasses}: Earth masses in the Sun-forming cloud}\label{emasses}%Number in V1.3: 140

(a) One source of solar-system elemental abundances is
\citet{2019arXiv191200844L} who lists these in her Table~8. A
mass-weighted sum of the abundances for all elements truncating at
no.\,92) compared to that for elements heavier than B yields the
requested mass fraction for elements $N$ and heavier (for elements $i$
with abundance $A_i$ and atomic mass $m_i$):
\begin{equation}
  \ntag
  f_N = {\sum_6^{92} A_i m_i\over \sum_1^{92} A_i m_i}
  \approx 0.007.
\end{equation}
With these abundances, the number of Earth masses of elements heavier
than C contained in a solar-mass cloud is given by:
\begin{equation}
  \ntag
  n_\odot = { f_N *M_\odot \over M_\oplus} \sim 2000.
\end{equation}

(b) The fraction $f_{\rm pl}$ of the original cloud that would need to
remain in the disk to ultimately form the planets is given by the sum
of the masses of all planets relative to that of the Sun:
$f_{\rm pl}=0.0013$.

(c) These two numbers, $f_N$ and $f_{\rm pl}$, are
largely independent of each other because the giant planets are
predominantly made up of H and He.

\section{Activity\,\ref{act:jeans}: The Jeans Mass}\label{jeans}%Number in V1.3: 142
(a) \indexit{Jeans mass}With
\begin{equation}
  \ntag
M_\ast = {4\over 3} \pi R_{\rm c}^3 \mu m_{\rm p} n_{\rm c},  
\end{equation}
the radius can be eliminated from Eq.~(\ref{eqlh:rofm}) so that the
limiting mass can be expressed as
\begin{equation}
  \ntag
  M_\ast \approx \left ({3 \over 4\pi}\right )^{1/2} \left ( k \over G \right )^{3/2} 
  \left( {1\over \mu m_{\rm p}}\right )^2 {T_{\rm c}^{3/2} \over n_{\rm c}^{1/2}} =
    8.2 M_\odot {T_{\rm c}^{3/2} \over \mu^2 n_{\rm c}^{1/2}} .
\end{equation}

(b) For molecular hydrogen, with $\mu=2$, that means that $f\approx
2$.

(c) At 10\,K and with $n=10^2$\,cm$^{-3}$ we have
$M_\ast \approx 7 M_\odot$. At 100\,K and with $n=10^4$ we have
$M_\ast \approx 20 M_\odot$.

\section{Activity\,\ref{act:diffcoeff}: Molecular diffusion
  coefficients}\label{diffcoeff}%Number in V1.3: 176

A \indexit{diffusion!molecular!diffusion coefficient}diffusion
coefficient, measuring the rms displacement per unit time, scales as
the product of the mean free path length $\hat{\ell}$ and the
characteristic velocity $\hat{v}$:
\begin{equation}
  D \sim {\hat{\ell}^2 \over \hat{\tau}} = \hat{\ell} \hat{v}, \ntag
\end{equation}
with $\hat{v}$ the characteristic time between collisions. With
\begin{equation}
\hat{\ell} =  {1 \over \sigma n} \,\,\,{\rm and} \,\,\,  kT = {1\over 2} m v_{\rm th}^2,  \ntag
\end{equation}
we have
\begin{equation}
D \sim \left ( {2kT\over \mu} \right )^{1/2}   {1 \over \sigma n} \propto
{\sqrt{T}\over \sqrt{m}n}\ntag
\end{equation}

\section{Activity\,\ref{act:mpdt}: Earth's magnetopause distance over time}\label{mpdt}%Number in V1.3: 188

(a) Combining the \indexit{magnetopause!evolution}various approximation leads to
\begin{equation}
{R_{\rm mp}(t) \over R_{\rm p}} \approx {(\xi B_{\rm p})^{1/3}\over (8\pi
  m_{\rm H} n_\ast v_\ast^2 )^{1/6}} \left (1+{t\over\tau_{\rm sw}}
\right )^{0.38}. \ntag
\end{equation}

(b) With $\xi=2$ and $B_{\rm p}=0.31$\,G as for the present-day Earth:
\begin{eqnarray}
  t < \tau_{\rm sw} & \,\, &   t >> \tau_{\rm sw} \nonumber \\
                             {R_{\rm mp}(t) \over R_\oplus} \approx 1.25 &\,\,&
                             {R_{\rm mp}(t) \over R_\oplus} \approx 5
                                                                           t_{\rm Gyr}^{0.38}.       \nonumber                                        
\end{eqnarray}

\section{Activity\,\ref{act:beprod}: $^{10}$Be production on a hypothetical Mars}\label{beprod}%Number in V1.3:191

Read Fig.~\ref{fig:4.2-2} down to a zero geomagnetic field and compare
the result with that for the present-day Earth.

\section{Activity\,\ref{act:snexposure}: Exposure to supernovae}\label{snexposure}%Number in V1.3:195
{(a,b)} Combination of \indexit{supernova!exposure}the scalings provided in the Activity
yields an approximate main-sequence life time $\tau_\ast$ for stars 
of mass $M_\ast$
\begin{equation}
\tau_\ast(M_\ast/M_\odot) \sim 10^{10} (M_\ast/M_\odot)^{-5/2} \,{\rm yr}, \ntag
\end{equation}
(which is only some 10\%\ below the full life time)  which gives
\begin{equation}
  \tau_\ast(100) \sim 10^{5} \,{\rm yr},\,
  \tau_\ast(8)=\tau_{\odot{\rm SN}} \sim 6\,10^{7} \,{\rm yr},\,
  \tau_\ast(0.5) \sim 6\,10^{10} \,{\rm yr},\,
  \tau_\ast(0.08) \sim 6\,10^{12} \,{\rm yr}.
  \nonumber
  \end{equation}
  Note that the lowest-mass stars can in principle live longer than
  the universe with its age of
  $1.4\,10^{10}$\,yr. \indexit{Sun!formation in star cluster}Beware:
  stars heavier than about 20\,$M_\odot$ live appreciably longer than
  estimated from the above scaling law (compare
  Activity~\ref{act:starinabox}), but that has no impact on the next
  sub-Activity:

{(c)} Integrating the initial mass function $\xi(M)$ from
$8M_\odot$ to $100M_\odot$ shows that $\xi_0\gtrsim 21$ in order for
there to be at least one supernova within $\tau_{\odot{\rm
    SN}}$. Combining that with the integral of $\xi(M)$ from
$0.5M_\odot$ to $2M_\odot$ shows that there should be
$N(0.5-2)\gtrsim 40$ stars in the neighborhood for the Sun (and its
solar system) to be exposed to a supernova within its birth cluster.

{(d)} For a star to go supernova within 1.8\,Myr, its mass
should exceed about $32M_\odot$. Doing as above then yields
$\xi_0\gtrsim 133$ and $N(0.5-2)\gtrsim 250$ stars.
  
\section{Activity\,\ref{act:thefinalquestion}: Reaching Earth's climate
    from scratch}\label{thefinalquestion}%Number in V1.3:200
\def\mp#1{\begin{minipage}[t]{0.35\linewidth} \raggedright {#1}\par \end{minipage}}
\def\mpn#1{\begin{minipage}[t]{0.22\linewidth} \raggedright {#1}\par \end{minipage}}
Table~\ref{tab:habprop} is a non-exhaustive list of properties that influence the
habitability of planets. 

\begin{table}[hp!]
  \centering
  \caption[Properties that influence the
habitability of planets]{Some of the properties that influence the
habitability of planets\label{tab:habprop}}

{\footnotesize
\begin{tabular}{lcc}
\hline
{\em Property }& {\em Beneficial} & {\em Detrimental}\\ \hline

  \mpn{Dense stellar neighborhood} & \mp{Supernova at suitable
                                    distance can trigger star
                                     formation in a molecular cloud} &
                                                                  \mp{Supernovae. Tidal
                                                                  disruption of planetary
                                                                  orbits or system} \\ \hline
  \mpn{Metallicity of initial cloud} & \mp{Higher metallicity
                                       correlates with higher
                                       likelihood of planetary system} & \mp{\null} \\ \hline
  \mpn{Multiple star system} & \mp{\null} & \mp{Orbital instability} \\ \hline
\mpn{Stellar mass} & \mp{Lower mass: longer life time} & \mp{Very low
                                                      mass: dim stars
                                                      have
                                                      habitable zones
                                                      so close that
                                                      spin-orbit
                                                      synchronization
                                                      occurs, and
                                                         stellar
                                                         activity
                                                         decays much
                                                         more slowly. Low-energy
                                                      photons may hamper
                                                      photosynthesis. High
                                                      mass: short
                                                      life time} \\ \hline
  \mpn{Evolution of stellar brightness} & \mp{\null} & \mp{Migration of
                                                 habitable zone} \\ \hline
  \mpn{Stellar magnetic activity: photons, wind, CMEs} & \mp{Energetic radiation may
                                                        benefit the
                                                        early evolution of 
                                                        life. Wind
                                                        shields from galactic
                                                        cosmic rays. Sustained
                                                        drop in activity
                                                        by magnetic braking}
                     & \mp{Irradiation of life with energetic photons
                       and particles. Stripping of planetary atmosphere} \\ \hline
  \mpn{Star-planet distance} & \mp{Habitable if at the correct
                              distance} & \mp{Tidal spin-orbit
                                          synchronization
                                          when too close in} \\ \hline
\mpn{Giant neighbor planets in the system} & \mp{Shield against comet/asteroid
                                              impacts well after
                                             system formation} & \mp{Orbital migration
                                                         could lead to
                                                         planetary
                                                         destruction. Effective
                                                         shielding
                                                         from impacts
                                                         may supply
                                                         too little
                                                         water to
                                                         terrestrial planets} \\ \hline
  \mpn{Asteroid belt} & \mp{May be important for creation of
                        settings suitable for life$^1$} &
                                                                      \mp{Frequent
                                                                      asteroid
                                                                      impacts
                                                                      can
                                                                      destroy
                                                                      life} \\ \hline
  \mpn{Orbital eccentricity} & \mp{Circular orbits provide stable irradiance} & \mp{Strong seasonal effects for
                                      eccentric orbit} \\ \hline
  \mpn{Planetary mass} & \mp{Intermediate mass retains
                         chemically diverse atmosphere} & \mp{Low
                                                          mass: loss of
                                                       atmosphere. High
                                                          mass could mean
                                                       predominant H/He atmosphere} \\ \hline
   \mpn{Satellite companion} & \mp{Stability of spin axis} & \mp{\null} \\ \hline
%\end{tabular}
%
%\begin{tabular}{lcc}
%  \hline
%{\em Property }& {\em Beneficial} & {\em Detrimental}\\ \hline
  \mpn{For a satellite of a giant planet} & \mp{Tidal effects may keep
                                           water liquid and dynamo
                                           functioning for a long
                                           time, even far from the
                                            star, and for 'rogue planet'} & \mp{Potential for
                                                   strong radiation environment} \\ \hline
  \mpn{Planetary dynamo} & \mp{Protects against energetic particles} & \mp{\null} \\ \hline
  \mpn{Planetary water content} & \mp{Water essential to known life} &
                                                                      \mp{Absence
                                                                      of
                                                                      near-surface
                                                                      water
                                                                      may
                                                                      cause
                                                                      shutdown
                                                                      of
                                                                      planetary
                                                                      dynamo,
                                                                      with
                                                                      consequences
                                                                      for 
                                                                      atmosphere} \\ \hline
\mpn{Plate tectonics} & \mp{Important for dynamo action and atmospheric
                       composition ({\em e.g.,} CO$_2$ weathering can
                       be balanced by volcanic activity)} & \mp{Too
                                                           much
                                                           activity
                                                           can lead to
                                                           an inhospitable climate} \\ \hline

%\mpn{\null} & \mp{\null} & \mp{\null} \\ \hline
%\mpn{\ldots} & \mp{\ldots} & \mp{\ldots} \\ \hline
\end{tabular}
$^1$ See, {\em e.g.}, \citet{2020AsBio..20.1121O}, \citet{2022arXiv220902860C}}
\end{table}

\clearpage

\addtocontents{toc}{\vspace{0.5cm}{\centerline{\bf Appendices}}\par }
\chapter*{Version history}
\fancyhead[RE]{\leftmark}
\markboth{Version history}{Version history}
\addcontentsline{toc}{chapter}{Version history}
\label{ch:versionhistory}
\begin{itemize}
\item[1.1] 2019/10/24: Original version.

\item[1.2] 2019/10/30: Deleted cover figure and compressed
  large-volume figures to fit into ar$\chi$iv's 15MB limit. Added
  explicit references and associated bibliography, corrected a few
  URLs to figure sources, and made the Astrophysics Data System (ADS)
  the primary path to the literature in URLs throughout. Added version
  history. Added explanation of Heliophysics volume numbers to Preface.
\item[1.3] 2022/01/15: {\bf Note: the numbering of some activities has
    changed because several new ones have been added in this version
    while others have been omitted or moved from the chapter they were
    originally in! For most Activities, this leaves their numbers
    unaffected or changes their number by 1 or 2 up or down.}

  Modified caption to Fig.~\ref{fig:conditions}.  Final paragraph of
  Sect.~6.1.2: removed '[a Dst index of]'. Added reference to list of
  space weather texts by \refcite{knippcade2018} in the
  Preface. Modified caption to Table~\ref{tab:brain3}.  Added
  geomagnetic (sub-)storm to the glossary in
  Tables.~\ref{fig:glossary} (split up from the original single page,
  continuing in Table~\ref{fig:glossaryb}), and differentiated
  magnetic and the (traditional) chromospheric plage in the definition
  of active region. Clarified captions to
  Fig.~\ref{fig:ophercomposite} and Fig.~\ref{fig:kv0}. Clarified text
  following Eq.~(\ref{eq:fieldenergy}). Corrected
  several minor typographical errors and descriptions of omitted or
  modified figures. Modified the table in Fig.~\ref{fig:acthrd} to
  contain stellar masses. Modified the text immediately following
  Activity~\ref{act:imagecycle}. Modified
  Activity~\ref{act:stormstrength}. Included Parker reference in the
  text leading up to Eq.~(\ref{eq:brain3}).
  Corrected sign for $\beta$ term in Eq.~(\ref{eq:emf2}). Corrected
  'interplanetary' to read 'interstellar' in the second line of Sect.~\ref{sec:impinging}.

  Added a classification for all activities: ``Look up,''
  ``Consider,'' ``Show,''  ``Background,'' or ``Advanced/Group.''

  Added 9 new Activities: \ref{act:yourdefinition},
  \ref{act:inductcomp}, \ref{act:georossby}, \ref{act:oblatedipole},
  \ref{act:dstindex}, \ref{act:sunmassloss}, \ref{act:intrinsiclya},
  \ref{act:ionoenergies}, \ref{act:ionvar}.

  Added or modified tasks within 78 Activities: \ref{act:defplanet},
  \ref{act:starsystems}, \ref{act:maxwellunits}, \ref{act:imagecycle},
  \ref{act:windquestions}, \ref{act:windenergy}, \ref{act:parkerplot},
  \ref{act:bondenergy}, \ref{act:opticaldepth},
  \ref{act:energeticpartpen}, \ref{act:spotvsar}, \ref{act:rgi},
  \ref{act:hmimovie}, \ref{act:fieldlinenote}, \ref{act:unitconv}, 
  \ref{act:addlinestofig}, \ref{act:speedcomp},
  \ref{act:convenergytransport}, \ref{act:zeeman},
  \ref{act:magpatterns}, \ref{act:convcells}, \ref{act:diffrot},
  \ref{act:cutoff}, \ref{act:fieldwrap}, \ref{act:meanfieldequation},
  \ref{act:rossby}, \ref{act:joy}, \ref{act:buoy2},
  \ref{act:sourcesurface}, \ref{act:noncondsphere},
  \ref{act:russellmcpheron}, \ref{act:corotation}, %\ref{act:magcycle},
  \ref{act:stardungey}, \ref{eq:nonpotform}, \ref{act:auroracolors},
  \ref{act:cmecme}, \ref{act:swphenomena}, \ref{act:bcs},
  \ref{act:evaporation}, \ref{act:cmevolume}, \ref{act:ttauri},
  \ref{act:radwindpressure}, \ref{act:breakup}, \ref{act:eclipses},
  \ref{act:binarieclasses}, \ref{act:bulge}, \ref{act:eqorigin},
  \ref{act:sepprop}, \ref{act:wavecomp}, \ref{act:fraunhofer},
  \ref{act:pressgrav}, \ref{act:lookintotube}, \ref{act:irradiance},
  \ref{act:coronallines}, \ref{act:nenh}, \ref{act:loopprop},
  \ref{act:hfrac}, \ref{act:sunmassloss}, \ref{act:integraldivergence}, \ref{act:hrevol},
  \ref{act:globclus}, \ref{act:speeds}, \ref{act:LISM},
  \ref{act:varism}, \ref{act:cxsigma}, 
  \ref{act:wdl}, \ref{act:emasses}, \ref{act:criticalcloud}, \ref{act:methodcomp},
  \ref{act:diffcoeff}, \ref{act:pressscalestation},
  \ref{act:ionreactions}, \ref{act:ionradio}, \ref{act:geoact},
  \ref{act:snexposure}, \ref{act:titan}, \ref{act:trappist1}.

  Omitted 7 Activities from
  V1.2 in V1.3: Nos.~30, 36, 94, 99, 110, 116, 125 (made into a footnote), 129,
  130, and 134.

  Moved within the chapter: \ref{act:unitconv}.  Moved to another
  chapter: \ref{act:titan}.  

\item[2.0] 2022/09/21: {\bf Note: the numbering of activities has
    changed once more because several new ones have been added in this
    version while others have been omitted or moved from the chapter
    they were originally in!}

  Activities are no longer shown as footnotes
  while their compilation in Ch.~\ref{ch:activities} is now shown at
  regular font size. New are solutions and supplemental text to
  selected activities; these are marked with
  $\circledS$\marginpar{$\circledS$} in the Activity identifier
  in-line and in margin. New also are itemized key topics and concepts
  at the start of each chapter, and appendices with useful numbers and
  vector identities. A subject index was added. References to the
  original text were moved into the margin.

  Added new Activities: \ref{act:mfps}, \ref{act:estimatebeta},
  and\ref{act:starinabox}, and removed Activities~28, 73, 105, and 134
  (as numbered in V1.3).  Corrected minor typographical errors. Added
  the official name to the provisional designation of 2014\,MU$_{69}$
  in Ch.~\ref{ch:introduction}, increased the number of exoplanets
  'now' known to over 5,000, and modified how the JWT is
  described. Added 'circulating' to items (3) and (4) in
  Activity~\ref{act:maxwellunits}. Corrected expression for magnetic
  energy density in Eq.~(\ref{eq:helwind}). Clarified that ion
  densities are given in Tables~\ref{tab:wind-stats}
  and~\ref{tab:fran2}. Added note $^d$ in
  Table~\ref{tab:dimensionlessnumbers} for clarity. Corrected Eq.~(\ref{eq:conduction}).

  Rephrased tasks (1) and (2) in
  Activity~\ref{act:sourcesurface}. Added option to rewrite
  Eq.~(\ref{eq:mom}) to Activity~\ref{act:windenergy}. Corrected
  expression for specific internal energy in the caption of
  Table~\ref{fig:mhdset}. Corrected mirror charge strength in
  Activity~\ref{act:sourcesurface}, clarified that distances are
  expressed in solar units, and separated the activity more clearly
  into two parts. Added a quantitative question to
  Activity~\ref{act:cfradius}.  And modified Activity~\ref{act:buoy},
  Activity~\ref{act:loopemission} (to point ahead to the RTV scaling
  law), Activity~\ref{act:acousticcutoff} (to apply to isothermal
  perturbations), and Activity~\ref{act:loopprop} (to clarify what the
  numbers imply, and for lowered field strengths).

  Added Figs.~\ref{fig:suncomp} and \ref{fig:ionolayers}.

\item[2.1] 2024/03/26: Made Activity labels active links to the
  respective Activity and added page numbers to the Activity.
  Added navigation hint for Mac OS Preview in
  Preface. Corrected $R_\odot$ in the list of constants and a few
  minor typographical errors.
  
\end{itemize}

\listoffigures
\addcontentsline{toc}{chapter}{\listfigurename}
\listoftables
\addcontentsline{toc}{chapter}{\listtablename}

\bibliography{master}

\begin{thebibliography}{}

\bibitem[\protect\citeauthoryear{{Abbett}}{2007}]{2007ApJ...665.1469A}
{Abbett}, W.~P.: 2007,
\newblock \apj 665(2), 1469,
\newblock doi:10.1086/519788

\bibitem[\protect\citeauthoryear{{Asai} {\em
  et~al.\/}}{2004}]{2004ApJ...611..557A}
{Asai}, A., {Yokoyama}, T., {Shimojo}, M., {Masuda}, S., {Kurokawa}, H., {\&}
  {Shibata}, K.: 2004,
\newblock \apj 611(1), 557,
\newblock doi:10.1086/422159

\bibitem[\protect\citeauthoryear{{Beer} {\em
  et~al.\/}}{1994}]{1994svsp.coll..291B}
{Beer}, J., {Baumgartner}, S.~T., {Dittrich-Hannen}, B., {Hauenstein}, J.,
  {Kubik}, P., {Lukasczyk}, C., {Mende}, W., {Stellmacher}, B., {\&} {Suter},
  M.: 1994,
\newblock in J.~M. {Pap}, C. {Frohlich}, H.~S. {Hudson}, and S.~K. {Solanki}
  (Eds.), {\em Invited Papers from IAU Colloquium 143: The Sun as a Variable
  Star: Solar and Stellar Irradiance Variations\/},  291

\bibitem[\protect\citeauthoryear{{Behar} {\em
  et~al.\/}}{2017}]{2017MNRAS.469S.396B}
{Behar}, E., {Nilsson}, H., {Alho}, M., {Goetz}, C., {\&} {Tsurutani}, B.:
  2017,
\newblock \mnras 469, S396,
\newblock doi:10.1093/mnras/stx1871

\bibitem[\protect\citeauthoryear{{Bell} {\em
  et~al.\/}}{2015}]{2015MNRAS.454..593B}
{Bell}, C.~P.~M., {Mamajek}, E.~E., {\&} {Naylor}, T.: 2015,
\newblock \mnras 454(1), 593,
\newblock doi:10.1093/mnras/stv1981

\bibitem[\protect\citeauthoryear{{Bell} {\em
  et~al.\/}}{2013}]{2013MNRAS.434..806B}
{Bell}, C.~P.~M., {Naylor}, T., {Mayne}, N.~J., {Jeffries}, R.~D., {\&}
  {Littlefair}, S.~P.: 2013,
\newblock \mnras 434(1), 806,
\newblock doi:10.1093/mnras/stt1075

\bibitem[\protect\citeauthoryear{Benz}{2002}]{2002ASSL..279.....B}
Benz, A: 2002,
\newblock {\em {Plasma Astrophysics, second edition}\/},
\newblock Kluwer, Springer,
\newblock Astrophysics and Space Science Library, Vol.\ 279

\bibitem[\protect\citeauthoryear{{Booth} {\em
  et~al.\/}}{2017}]{2017MNRAS.471.1012B}
{Booth}, R.~S., {Poppenhaeger}, K., {Watson}, C.~A., {Silva Aguirre}, V., {\&}
  {Wolk}, S.~J.: 2017,
\newblock \mnras 471(1), 1012,
\newblock doi:10.1093/mnras/stx1630

\bibitem[\protect\citeauthoryear{{Bougher} and
  {Roble}}{1991}]{1991JGR....9611045B}
{Bougher}, S.~W. {\&} {Roble}, R.~G.: 1991,
\newblock \jgr 96(A7), 11045,
\newblock doi:10.1029/91JA01162

\bibitem[\protect\citeauthoryear{{Brun} {\em
  et~al.\/}}{2004}]{2004ApJ...614.1073B}
{Brun}, A.~S., {Miesch}, M.~S., {\&} {Toomre}, J.: 2004,
\newblock \apj 614(2), 1073,
\newblock doi:10.1086/423835

\bibitem[\protect\citeauthoryear{{Burkepile} {\em
  et~al.\/}}{2004}]{2004JGRA..109.3103B}
{Burkepile}, J.~T., {Hundhausen}, A.~J., {Stanger}, A.~L., {St. Cyr}, O.~C.,
  {\&} {Seiden}, J.~A.: 2004,
\newblock Journal of Geophysical Research (Space Physics) 109(A3), A03103,
\newblock doi:10.1029/2003JA010149

\bibitem[\protect\citeauthoryear{{Carron}}{2007}]{2007ipep.book.....C}
{Carron}, N.~J.: 2007,
\newblock {\em {An Introduction to the Passage of Energetic Particles through
  Matter}\/}

\bibitem[\protect\citeauthoryear{{Cauley} {\em
  et~al.\/}}{2015}]{2015ApJ...810...13C}
{Cauley}, P.~W., {Redfield}, S., {Jensen}, A.~G., {Barman}, T., {Endl}, M.,
  {\&} {Cochran}, W.~D.: 2015,
\newblock \apj 810(1), 13,
\newblock doi:10.1088/0004-637X/810/1/13

\bibitem[\protect\citeauthoryear{{Charbonneau}}{2010}]{2010LRSP....7....3C}
{Charbonneau}, P.: 2010,
\newblock Living Reviews in Solar Physics 7(1), 3,
\newblock doi:10.12942/lrsp-2010-3

\bibitem[\protect\citeauthoryear{{Charbonneau}}{2014}]{2014ARA&A..52..251C}
{Charbonneau}, P.: 2014,
\newblock \araa 52, 251,
\newblock doi:10.1146/annurev-astro-081913-040012

\bibitem[\protect\citeauthoryear{{Cheung} {\em
  et~al.\/}}{2019}]{2019NatAs...3..160C}
{Cheung}, M.~C.~M., {Rempel}, M., {Chintzoglou}, G., {Chen}, F., {Testa}, P.,
  {Mart{\'\i}nez-Sykora}, J., {Sainz Dalda}, A., {DeRosa}, M.~L.,
  {Malanushenko}, A., {Hansteen}, V., {De Pontieu}, B., {Carlsson}, M.,
  {Gudiksen}, B., {\&} {McIntosh}, S.~W.: 2019,
\newblock Nature Astronomy 3, 160,
\newblock doi:10.1038/s41550-018-0629-3

\bibitem[\protect\citeauthoryear{{Childs} {\em
  et~al.\/}}{2022}]{2022arXiv220902860C}
{Childs}, Anna~C., {Martin}, Rebecca~G., {\&} {Livio}, Mario: 2022,
\newblock arXiv e-prints  arXiv:2209.02860

\bibitem[\protect\citeauthoryear{{Cohen} and
  {Drake}}{2014}]{2014ApJ...783...55C}
{Cohen}, O. {\&} {Drake}, J.~J.: 2014,
\newblock \apj 783(1), 55,
\newblock doi:10.1088/0004-637X/783/1/55

\bibitem[\protect\citeauthoryear{{Cohen} {\em
  et~al.\/}}{2012}]{2012ApJ...760...85C}
{Cohen}, O., {Drake}, J.~J., {\&} {K{\'o}ta}, J.: 2012,
\newblock \apj 760(1), 85,
\newblock doi:10.1088/0004-637X/760/1/85

\bibitem[\protect\citeauthoryear{{Crooker} {\em
  et~al.\/}}{1999}]{1999SSRv...89..179C}
{Crooker}, N.~U., {Gosling}, J.~T., {Bothmer}, V., {Forsyth}, R.~J., {Gazis},
  P.~R., {Hewish}, A., {Horbury}, T.~S., {Intriligator}, D.~S., {Jokipii},
  J.~R., {K{\'o}ta}, J., {Lazarus}, A.~J., {Lee}, M.~A., {Lucek}, E., {Marsch},
  E., {Posner}, A., {Richardson}, I.~G., {Roelof}, E.~C., {Schmidt}, J.~M.,
  {Siscoe}, G.~L., {Tsurutani}, B.~T., {\&} {Wimmer-Schweingruber}, R.~F.:
  1999,
\newblock \ssr 89, 179,
\newblock doi:10.1023/A:1005253526438

\bibitem[\protect\citeauthoryear{{Deming} and
  {Louie}}{2019}]{2019PASP..131a3001D}
{Deming}, D. {\&} {Louie}, D.and~{Sheets}, H.: 2019,
\newblock \pasp 131(995), 013001,
\newblock doi:10.1088/1538-3873/aae5c5

\bibitem[\protect\citeauthoryear{{Desch}}{2007}]{2007ApJ...671..878D}
{Desch}, S.~J.: 2007,
\newblock \apj 671(1), 878,
\newblock doi:10.1086/522825

\bibitem[\protect\citeauthoryear{{Dravins} {\em
  et~al.\/}}{2017a}]{2017A&A...605A..90D}
{Dravins}, D., {Ludwig}, H.-G., {Dahl{\'e}n}, E., {\&} {Pazira}, H.: 2017a,
\newblock \aap 605, A90,
\newblock doi:10.1051/0004-6361/201730900

\bibitem[\protect\citeauthoryear{{Dravins} {\em
  et~al.\/}}{2017b}]{2017A&A...605A..91D}
{Dravins}, D., {Ludwig}, H.-G., {Dahl{\'e}n}, E., {\&} {Pazira}, H.: 2017b,
\newblock \aap 605, A91,
\newblock doi:10.1051/0004-6361/201730901

\bibitem[\protect\citeauthoryear{{Drury}}{1983}]{1983RPPh...46..973D}
{Drury}, L.~Oc.: 1983,
\newblock Reports on Progress in Physics 46(8), 973,
\newblock doi:10.1088/0034-4885/46/8/002

\bibitem[\protect\citeauthoryear{{Durran} and
  {Arakawa}}{2007}]{2007CRMec.335..655D}
{Durran}, D.~R. {\&} {Arakawa}, A.: 2007,
\newblock Comptes Rendus Mecanique 335(9), 655,
\newblock doi:10.1016/j.crme.2007.08.010

\bibitem[\protect\citeauthoryear{{Eker} {\em
  et~al.\/}}{2018}]{2018MNRAS.479.5491E}
{Eker}, Z., {Bak{\i}{\c{s}}}, V., {Bilir}, S., {Soydugan}, F., {Steer}, I.,
  {Soydugan}, E., {Bak{\i}{\c{s}}}, H., {Ali{\c{c}}avu{\c{s}}}, F., {Aslan},
  G., {\&} {Alpsoy}, M.: 2018,
\newblock \mnras 479(4), 5491,
\newblock doi:10.1093/mnras/sty1834

\bibitem[\protect\citeauthoryear{{Foley} and
  {Smye}}{2018}]{2018AsBio..18..873F}
{Foley}, B.~J. {\&} {Smye}, A.~J.: 2018,
\newblock Astrobiology 18(7), 873,
\newblock doi:10.1089/ast.2017.1695

\bibitem[\protect\citeauthoryear{{Forbes} and
  {Priest}}{1995}]{1995ApJ...446..377F}
{Forbes}, T.~G. {\&} {Priest}, E.~R.: 1995,
\newblock \apj 446, 377,
\newblock doi:10.1086/175797

\bibitem[\protect\citeauthoryear{{Franck} {\em
  et~al.\/}}{2001}]{2001NW.....88..416F}
{Franck}, S., {Block}, A., {Bloh}, W., {Bounama}, C., {Garrido}, I., {\&}
  {Schellnhuber}, H.~J.: 2001,
\newblock Naturwissenschaften 88(10), 416,
\newblock doi:10.1007/s001140100257

\bibitem[\protect\citeauthoryear{{Fraschetti} {\em
  et~al.\/}}{2019}]{2019ApJ...874...21F}
{Fraschetti}, F., {Drake}, J.~J., {Alvarado-G{\'o}mez}, J.~D., {Moschou},
  S.~P., {Garraffo}, C., {\&} {Cohen}, O.: 2019,
\newblock \apj 874(1), 21,
\newblock doi:10.3847/1538-4357/ab05e4

\bibitem[\protect\citeauthoryear{{Garraffo} {\em
  et~al.\/}}{2017}]{2017ApJ...843L..33G}
{Garraffo}, C., {Drake}, J.~y~J., {Cohen}, O., {Alvarado-G{\'o}mez}, J.~D.,
  {\&} {Moschou}, S.~P.: 2017,
\newblock \apjl 843(2), L33,
\newblock doi:10.3847/2041-8213/aa79ed

\bibitem[\protect\citeauthoryear{{Gleeson} and
  {Axford}}{1968}]{1968ApJ...154.1011G}
{Gleeson}, L.~J. {\&} {Axford}, W.~I.: 1968,
\newblock \apj 154, 1011,
\newblock doi:10.1086/149822

\bibitem[\protect\citeauthoryear{{G{\"u}del}}{2007}]{2007LRSP....4....3G}
{G{\"u}del}, M.: 2007,
\newblock Living Reviews in Solar Physics 4(1), 3,
\newblock doi:10.12942/lrsp-2007-3

\bibitem[\protect\citeauthoryear{{Hartmann}}{2009}]{2009apsf.book.....H}
{Hartmann}, L.: 2009,
\newblock {\em {Accretion Processes in Star Formation: Second Edition}\/},
\newblock Cambridge University Press, Cambridge, UK

\bibitem[\protect\citeauthoryear{{Hathaway} {\em
  et~al.\/}}{2000}]{2000SoPh..193..299H}
{Hathaway}, D.~H., {Beck}, J.~G., {Bogart}, R.~S., {Bachmann}, K.~T., {Khatri},
  G., {Petitto}, J.~M., {Han}, S., {\&} {Raymond}, J.: 2000,
\newblock \solphys 193, 299,
\newblock doi:10.1023/A:1005200809766

\bibitem[\protect\citeauthoryear{{Henning} and
  {Semenov}}{2013}]{2013ChRv..113.9016H}
{Henning}, T. {\&} {Semenov}, D.: 2013,
\newblock Chemical Reviews 113(12), 9016,
\newblock doi:10.1021/cr400128p

\bibitem[\protect\citeauthoryear{{Holzer}}{1972}]{1972JGR....77.5407H}
{Holzer}, T.~E.: 1972,
\newblock \jgr 77(28), 5407,
\newblock doi:10.1029/JA077i028p05407

\bibitem[\protect\citeauthoryear{{Howard} {\em
  et~al.\/}}{2013}]{2013Natur.503..381H}
{Howard}, A.~W., {Sanchis-Ojeda}, R., {Marcy}, G.~W., {Johnson}, J.~A., {Winn},
  J.~N., {Isaacson}, H., {Fischer}, D.~A., {Fulton}, B.~J., {Sinukoff}, E.,
  {\&} {Fortney}, J.~J.: 2013,
\newblock \nat 503(7476), 381,
\newblock doi:10.1038/nature12767

\bibitem[\protect\citeauthoryear{IPCC}{2013}]{ipcc2013}
IPCC: 2013,
\newblock in T.F. Stocker, D. Qin, G.-K. Plattner, M. Tignor, S.K. Allen, J.
  Boschung, A. Nauels, Y. Xia, V. Bex, and P.M. Midgley (Eds.), {\em Summary
  for Policymakers\/}, Cambridge University Press, Cambridge, United Kingdom
  and New York, NY, USA

\bibitem[\protect\citeauthoryear{{Jansen} {\em
  et~al.\/}}{2019}]{2019ApJ...875...79J}
{Jansen}, T., {Scharf}, C., {Way}, M., {\&} {Del Genio}, A.: 2019,
\newblock \apj 875(2), 79,
\newblock doi:10.3847/1538-4357/ab113d

\bibitem[\protect\citeauthoryear{{Jia} {\em
  et~al.\/}}{2008}]{2008JGRA..113.6212J}
{Jia}, X., {Walker}, R.~J., {Kivelson}, M.~G., {Khurana}, K.~K., {\&} {Linker},
  J.~A.: 2008,
\newblock Journal of Geophysical Research (Space Physics) 113(A6), A06212,
\newblock doi:10.1029/2007JA012748

\bibitem[\protect\citeauthoryear{{Jim{\'e}nez} {\em
  et~al.\/}}{2011}]{2011ApJ...743...99J}
{Jim{\'e}nez}, A., {Garc{\'\i}a}, R.~A., {\&} {Pall{\'e}}, P.~L.: 2011,
\newblock \apj 743(2), 99,
\newblock doi:10.1088/0004-637X/743/2/99

\bibitem[\protect\citeauthoryear{{Johnson}}{2019}]{2019Sci...363..474J}
{Johnson}, J.~A.: 2019,
\newblock Science 363(6426), 474,
\newblock doi:10.1126/science.aau9540

\bibitem[\protect\citeauthoryear{{Jokipii} and
  {Thomas}}{1981}]{1981ApJ...243.1115J}
{Jokipii}, J.~R. {\&} {Thomas}, B.: 1981,
\newblock \apj 243, 1115,
\newblock doi:10.1086/158675

\bibitem[\protect\citeauthoryear{{Kallenrode}}{2003}]{2003JPhG...29..965K}
{Kallenrode}, M.-B.: 2003,
\newblock Journal of Physics G Nuclear Physics 29(5), 965

\bibitem[\protect\citeauthoryear{{Karak} and
  {Miesch}}{2017}]{2017ApJ...847...69K}
{Karak}, B.~B. {\&} {Miesch}, M.: 2017,
\newblock \apj 847(1), 69,
\newblock doi:10.3847/1538-4357/aa8636

\bibitem[\protect\citeauthoryear{{Kiehl} and
  {Trenberth}}{1997}]{1997BAMS...78..197K}
{Kiehl}, J.~T. {\&} {Trenberth}, K.~E.: 1997,
\newblock Bulletin of the American Meteorological Society 78(2), 197,
\newblock doi:10.1175/1520-0477(1997)078<0197:EAGMEB>2.0.CO;2

\bibitem[\protect\citeauthoryear{Knipp and Cade}{2020}]{knippcade2018}
Knipp, D.~J. {\&} Cade, W.~T.: 2020,
\newblock doi:10.5281/zenodo.3843629

\bibitem[\protect\citeauthoryear{{Koch} {\em
  et~al.\/}}{2019}]{2019QSRv..207...13K}
{Koch}, A., {Brierley}, C., {Maslin}, M.~M., {\&} {Lewis}, S.~L.: 2019,
\newblock Quaternary Science Reviews 207, 13,
\newblock doi:10.1016/j.quascirev.2018.12.004

\bibitem[\protect\citeauthoryear{{Krauss-Varban} {\em
  et~al.\/}}{2008}]{2008AIPC.1039..307K}
{Krauss-Varban}, D., {Li}, Y., {\&} {Luhmann}, J.~G.: 2008,
\newblock in Gang {Li}, Qiang {Hu}, Olga {Verkhoglyadova}, Gary~P. {Zank},
  R.~P. {Lin}, and J. {Luhmann} (Eds.), {\em American Institute of Physics
  Conference Series\/}, Vol. 1039, p.~307

\bibitem[\protect\citeauthoryear{{Kulikov} {\em
  et~al.\/}}{2007}]{2007SSRv..129..207K}
{Kulikov}, Y.~N., {Lammer}, H., {Lichtenegger}, H.~I.~M., {Penz}, T., {Breuer},
  D., {Spohn}, T., {Lundin}, R., {\&} {Biernat}, H.~K.: 2007,
\newblock \ssr 129(1-3), 207,
\newblock doi:10.1007/s11214-007-9192-4

\bibitem[\protect\citeauthoryear{{Lammer} and
  {Blanc}}{2018}]{2018SSRv..214...60L}
{Lammer}, H. {\&} {Blanc}, M.: 2018,
\newblock \ssr 214(2), 60,
\newblock doi:10.1007/s11214-017-0433-x

\bibitem[\protect\citeauthoryear{{Lemerle} and
  {Charbonneau}}{2017}]{2017ApJ...834..133L}
{Lemerle}, A. {\&} {Charbonneau}, P.: 2017,
\newblock \apj 834(2), 133,
\newblock doi:10.3847/1538-4357/834/2/133

\bibitem[\protect\citeauthoryear{{Linsky}}{1985}]{1985SoPh..100..333L}
{Linsky}, J.~L.: 1985,
\newblock \solphys 100, 333,
\newblock doi:10.1007/BF00158435

\bibitem[\protect\citeauthoryear{{Lodders}}{2019}]{2019arXiv191200844L}
{Lodders}, Katharina: 2019,
\newblock arXiv e-prints  arXiv:1912.00844

\bibitem[\protect\citeauthoryear{{Lundin} {\em
  et~al.\/}}{2007}]{2007SSRv..129..245L}
{Lundin}, R., {Lammer}, H., {\&} {Ribas}, I.: 2007,
\newblock \ssr 129(1-3), 245,
\newblock doi:10.1007/s11214-007-9176-4

\bibitem[\protect\citeauthoryear{{MacNeice} {\em
  et~al.\/}}{2004}]{2004ApJ...614.1028M}
{MacNeice}, P., {Antiochos}, S.~K., {Phillips}, A., {Spicer}, D.~S., {DeVore},
  C.~R., {\&} {Olson}, K.: 2004,
\newblock \apj 614(2), 1028,
\newblock doi:10.1086/423887

\bibitem[\protect\citeauthoryear{{Mamajek}}{2009}]{2009AIPC.1158....3M}
{Mamajek}, E.~E.: 2009,
\newblock in Tomonori {Usuda}, Motohide {Tamura}, and Miki {Ishii} (Eds.), {\em
  American Institute of Physics Conference Series\/}, Vol. 1158 of {\em
  American Institute of Physics Conference Series\/}, p.~3

\bibitem[\protect\citeauthoryear{{Marcq} {\em
  et~al.\/}}{2018}]{2018SSRv..214...10M}
{Marcq}, E., {Mills}, F.~P., {Parkinson}, C.~D., {\&} {Vandaele}, A.~C.: 2018,
\newblock \ssr 214(1), 10,
\newblock doi:10.1007/s11214-017-0438-5

\bibitem[\protect\citeauthoryear{{Mazur} {\em
  et~al.\/}}{2000}]{2000ApJ...532L..79M}
{Mazur}, J.~E., {Mason}, G.~M., {Dwyer}, J.~R., {Giacalone}, J., {Jokipii},
  J.~R., {\&} {Stone}, E.~C.: 2000,
\newblock \apjl 532(1), L79,
\newblock doi:10.1086/312561

\bibitem[\protect\citeauthoryear{{McComas} {\em
  et~al.\/}}{2013}]{2013ApJ...771...77M}
{McComas}, D.~J., {Dayeh}, M.~A., {Funsten}, H.~O., {Livadiotis}, G., {\&}
  {Schwadron}, N.~A.: 2013,
\newblock \apj 771(2), 77,
\newblock doi:10.1088/0004-637X/771/2/77

\bibitem[\protect\citeauthoryear{{Meyer-Vernet}}{1999}]{1999EJPh...20..167M}
{Meyer-Vernet}, N.: 1999,
\newblock European Journal of Physics 20(3), 167,
\newblock doi:10.1088/0143-0807/20/3/006

\bibitem[\protect\citeauthoryear{{Moraal}}{1976}]{1976SSRv...19..845M}
{Moraal}, H.: 1976,
\newblock \ssr 19(6), 845,
\newblock doi:10.1007/BF00173707

\bibitem[\protect\citeauthoryear{{Movshovitz} {\em
  et~al.\/}}{2012}]{2012ApJ...759...93M}
{Movshovitz}, Naor, {Asphaug}, Erik, {\&} {Korycansky}, Donald: 2012,
\newblock \apj 759(2), 93,
\newblock doi:10.1088/0004-637X/759/2/93

\bibitem[\protect\citeauthoryear{{M{\"u}ller} {\em
  et~al.\/}}{2009}]{2009SSRv..143..415M}
{M{\"u}ller}, H.-R., {Frisch}, P.~C., {Fields}, B.~D., {\&} {Zank}, G.~P.:
  2009,
\newblock \ssr 143(1-4), 415,
\newblock doi:10.1007/s11214-008-9448-7

\bibitem[\protect\citeauthoryear{{M{\"u}ller} and
  {Zank}}{2004}]{2004JGRA..109.7104M}
{M{\"u}ller}, H.~R. {\&} {Zank}, G.~P.: 2004,
\newblock Journal of Geophysical Research (Space Physics) 109(A7), A07104,
\newblock doi:10.1029/2003JA010269

\bibitem[\protect\citeauthoryear{{Ngwira} {\em
  et~al.\/}}{2014}]{2014JGRA..119.4456N}
{Ngwira}, C.~M., {Pulkkinen}, A., {Kuznetsova}, M.~M., {\&} {Glocer}, A.: 2014,
\newblock Journal of Geophysical Research (Space Physics) 119(6), 4456,
\newblock doi:10.1002/2013JA019661

\bibitem[\protect\citeauthoryear{{{\'O}~Fionnag{\'a}in} {\em
  et~al.\/}}{2019}]{2019MNRAS.483..873O}
{{\'O}~Fionnag{\'a}in}, D., {Vidotto}, A.~A., {Petit}, P., {Folsom}, C.~P.,
  {Jeffers}, S.~V., {Marsden}, S.~C., {Morin}, J., {do Nascimento}, J.~D., {\&}
  {BCool Collaboration}: 2019,
\newblock \mnras 483(1), 873,
\newblock doi:10.1093/mnras/sty3132

\bibitem[\protect\citeauthoryear{{O'Brien} {\em
  et~al.\/}}{2014}]{2014Icar..239...74O}
{O'Brien}, D.~P., {Walsh}, K.~J., {Morbidelli}, A., {Raymond}, S.~N., {\&}
  {Mandell}, A.~M.: 2014,
\newblock Icarus 239, 74,
\newblock doi:10.1016/j.icarus.2014.05.009

\bibitem[\protect\citeauthoryear{{Osinski} {\em
  et~al.\/}}{2020}]{2020AsBio..20.1121O}
{Osinski}, G.~R., {Cockell}, C.~S., {Pontefract}, A., {\&} {Sapers}, H.~M.:
  2020,
\newblock Astrobiology 20(9), 1121,
\newblock doi:10.1089/ast.2019.2203

\bibitem[\protect\citeauthoryear{{Parker}}{1958}]{1958ApJ...128..664P}
{Parker}, E.~N.: 1958,
\newblock \apj 128, 664,
\newblock doi:10.1086/146579

\bibitem[\protect\citeauthoryear{{Parker}}{1960}]{1960ApJ...132..175P}
{Parker}, E.~N.: 1960,
\newblock \apj 132, 175,
\newblock doi:10.1086/146910

\bibitem[\protect\citeauthoryear{{Parker}}{2007}]{2007cemf.book.....P}
{Parker}, Eugene~N.: 2007,
\newblock {\em {Conversations on Electric and Magnetic Fields in the Cosmos}\/}

\bibitem[\protect\citeauthoryear{{Parker}}{2014}]{2014RAA....14....1P}
{Parker}, E.~N.: 2014,
\newblock Research in Astronomy and Astrophysics 14(1), 1,
\newblock doi:10.1088/1674-4527/14/1/001

\bibitem[\protect\citeauthoryear{{Pavlov} {\em
  et~al.\/}}{2005}]{2005GeoRL..32.3705P}
{Pavlov}, A.~A., {Toon}, O.~B., {Pavlov}, A.~K., {Bally}, J., {\&} {Pollard},
  D.: 2005,
\newblock \grl 32(3), L03705,
\newblock doi:10.1029/2004GL021890

\bibitem[\protect\citeauthoryear{{Pecaut} and
  {Mamajek}}{2016}]{2016MNRAS.461..794P}
{Pecaut}, M.~J. {\&} {Mamajek}, E.~E.: 2016,
\newblock \mnras 461(1), 794,
\newblock doi:10.1093/mnras/stw1300

\bibitem[\protect\citeauthoryear{{Pecaut} {\em
  et~al.\/}}{2012}]{2012ApJ...746..154P}
{Pecaut}, M.~J., {Mamajek}, E.~E., {\&} {Bubar}, E.~J.: 2012,
\newblock \apj 746(2), 154,
\newblock doi:10.1088/0004-637X/746/2/154

\bibitem[\protect\citeauthoryear{{Pineda} and
  {Hallinan}}{2017}]{2017ApJ...846...75P}
{Pineda}, J.~S. {\&} {Hallinan}, G.and~{Kao}, M.~M.: 2017,
\newblock \apj 846(1), 75,
\newblock doi:10.3847/1538-4357/aa8596

\bibitem[\protect\citeauthoryear{{Pinhas} {\em
  et~al.\/}}{2018}]{2018MNRAS.480.5314P}
{Pinhas}, A., {Rackham}, B.~V., {Madhusudhan}, N., {\&} {Apai}, D.: 2018,
\newblock \mnras 480(4), 5314,
\newblock doi:10.1093/mnras/sty2209

\bibitem[\protect\citeauthoryear{{Pognan} {\em
  et~al.\/}}{2018}]{2018ApJ...856...53P}
{Pognan}, Q., {Garraffo}, C., {Cohen}, O., {\&} {Drake}, J.~J.: 2018,
\newblock \apj 856(1), 53,
\newblock doi:10.3847/1538-4357/aaaebb

\bibitem[\protect\citeauthoryear{{Portegies Zwart} {\em
  et~al.\/}}{2018}]{2018A&A...616A..85P}
{Portegies Zwart}, S., {Pelupessy}, I., {van Elteren}, A., {Wijnen}, T.~P.~G.,
  {\&} {Lugaro}, M.: 2018,
\newblock \aap 616, A85,
\newblock doi:10.1051/0004-6361/201732060

\bibitem[\protect\citeauthoryear{{Portegies Zwart}}{2009}]{2009ApJ...696L..13P}
{Portegies Zwart}, S.~F.: 2009,
\newblock \apjl 696(1), L13,
\newblock doi:10.1088/0004-637X/696/1/L13

\bibitem[\protect\citeauthoryear{{Quenby}}{1984}]{1984SSRv...37..201Q}
{Quenby}, J.~J.: 1984,
\newblock \ssr 37(3-4), 201,
\newblock doi:10.1007/BF00226364

\bibitem[\protect\citeauthoryear{{Rackham} {\em
  et~al.\/}}{2019a}]{2019BAAS...51c.328R}
{Rackham}, B., {Pinhas}, A., {Apai}, D., {Haywood}, R., {Cegla}, H.,
  {Espinoza}, N., {Teske}, J., {Gully-Santiago}, M., {Rau}, G., {Morris},
  B.~M., {Angerhausen}, D., {Barclay}, T., {Carone}, L., {Cauley}, P.~W., {de
  Wit}, J., {Domagal-Goldman}, S., {Dong}, C., {Dragomir}, D., {Giampapa},
  M.~S., {Hasegawa}, Y., {Hinkel}, N.~R., {Hu}, R., {Jord{\'a}n}, An.,
  {Kitiashvili}, I., {Kreidberg}, L., {Lisse}, C., {Llama}, J.,
  {L{\'o}pez-Morales}, M., {Mennesson}, B., {Molaverdikhani}, K., {Osip},
  D.~J., {\&} {Quintana}, E.~V.: 2019a,
\newblock Bull. Am. Astron. Soc. 51(3), 328

\bibitem[\protect\citeauthoryear{{Rackham} {\em
  et~al.\/}}{2018}]{2018ApJ...853..122R}
{Rackham}, B.~V., {Apai}, D., {\&} {Giampapa}, M.~S.: 2018,
\newblock \apj 853(2), 122,
\newblock doi:10.3847/1538-4357/aaa08c

\bibitem[\protect\citeauthoryear{{Rackham} {\em
  et~al.\/}}{2019b}]{2019AJ....157...96R}
{Rackham}, B.~V., {Apai}, D., {\&} {Giampapa}, M.~S.: 2019b,
\newblock \aj 157(3), 96,
\newblock doi:10.3847/1538-3881/aaf892

\bibitem[\protect\citeauthoryear{{Reames}}{2013}]{2013SSRv..175...53R}
{Reames}, D.~V.: 2013,
\newblock \ssr 175(1-4), 53,
\newblock doi:10.1007/s11214-013-9958-9

\bibitem[\protect\citeauthoryear{{Ribas} {\em
  et~al.\/}}{2005}]{2005ApJ...622..680R}
{Ribas}, I., {Guinan}, E.~F., {G{\"u}del}, M., {\&} {Audard}, M.: 2005,
\newblock \apj 622(1), 680,
\newblock doi:10.1086/427977

\bibitem[\protect\citeauthoryear{{Rosner} {\em
  et~al.\/}}{1978}]{1978ApJ...220..643R}
{Rosner}, R., {Tucker}, W.~H., {\&} {Vaiana}, G.~S.: 1978,
\newblock \apj 220, 643,
\newblock doi:10.1086/155949

\bibitem[\protect\citeauthoryear{{Sackmann} {\em
  et~al.\/}}{1993}]{1993ApJ...418..457S}
{Sackmann}, I.-J., {Boothroyd}, A.~I., {\&} {Kraemer}, K.~E.: 1993,
\newblock \apj 418, 457,
\newblock doi:10.1086/173407

\bibitem[\protect\citeauthoryear{{Schatten} {\em
  et~al.\/}}{1969}]{1969SoPh....6..442S}
{Schatten}, K.~H., {Wilcox}, J.~M., {\&} {Ness}, N.~F.: 1969,
\newblock \solphys 6(3), 442,
\newblock doi:10.1007/BF00146478

\bibitem[\protect\citeauthoryear{{Scheucher} {\em
  et~al.\/}}{2018}]{2018ApJ...863....6S}
{Scheucher}, M., {Grenfell}, J.~L., {Wunderlich}, F., {Godolt}, M., {Schreier},
  F., {\&} {Rauer}, H.: 2018,
\newblock \apj 863(1), 6,
\newblock doi:10.3847/1538-4357/aacf03

\bibitem[\protect\citeauthoryear{{Schrijver}}{2001}]{2001ApJ...547..475S}
{Schrijver}, C.~J.: 2001,
\newblock \apj 547(1), 475,
\newblock doi:10.1086/318333

\bibitem[\protect\citeauthoryear{{Schrijver}}{2009}]{2009ApJ...699L.148S}
{Schrijver}, C.~J.: 2009,
\newblock \apjl 699(2), L148,
\newblock doi:10.1088/0004-637X/699/2/L148

\bibitem[\protect\citeauthoryear{{Schrijver} {\em
  et~al.\/}}{2016}]{2016hasa.book.....S}
{Schrijver}, C.~J., {Bagenal}, F., {\&} {Sojka}, J.~J. (Eds.): 2016,
\newblock {\em {Heliophysics: Active Stars, their Astrospheres, and Impacts on
  Planetary Environments}\/},
\newblock Cambridge University Press, Cambridge, UK,
\newblock (Volume IV)

\bibitem[\protect\citeauthoryear{{Schrijver} {\em
  et~al.\/}}{2006}]{2006ApJ...650.1184S}
{Schrijver}, C.~J., {Hudson}, H.~S., {Murphy}, R.~J., {Share}, G.~H., {\&}
  {Tarbell}, T.~D.: 2006,
\newblock \apj 650(2), 1184,
\newblock doi:10.1086/506583

\bibitem[\protect\citeauthoryear{{Schrijver} {\em
  et~al.\/}}{2015}]{2015AdSpR..55.2745S}
{Schrijver}, C.~J., {Kauristie}, K, {Aylward}, A.~D., {Denardini}, C.~M.,
  {Gibson}, S.~E., {Glover}, A., {Gopalswamy}, N., {Grande}, M., {Hapgood}, M.,
  {Heynderickx}, D., {Jakowski}, N., {Kalegaev}, V.~V., {Lapenta}, G.,
  {Linker}, J.~A., {Liu}, S., {Mandrini}, C.~H., {Mann}, I.~R., {Nagatsuma},
  T., {Nandy}, D., {Obara}, T., {O'Brien}, P.~T., {Onsager}, T., {Opgenoorth},
  H.~J., {Terkildsen}, M., {Valladares}, C.~E., {\&} {Vilmer}, N.: 2015,
\newblock Advances in Space Research 55(12), 2745,
\newblock doi:10.1016/j.asr.2015.03.023

\bibitem[\protect\citeauthoryear{{Schrijver} and
  {Siscoe}}{2011}]{2011hppl.book.....S}
{Schrijver}, C.~J. {\&} {Siscoe}, G.~L. (Eds.): 2011,
\newblock {\em {Heliophysics: Plasma Physics of the Local Cosmos}\/},
\newblock Cambridge University Press, Cambridge, UK,
\newblock (Volume I)

\bibitem[\protect\citeauthoryear{{Schrijver} and
  {Siscoe}}{2012a}]{2012hesa.book.....S}
{Schrijver}, C.~J. {\&} {Siscoe}, G.~L. (Eds.): 2012a,
\newblock {\em {Heliophysics: Evolving Solar Activity and the Climates of Space
  and Earth}\/},
\newblock Cambridge University Press, Cambridge, UK,
\newblock (Volume III)

\bibitem[\protect\citeauthoryear{{Schrijver} and
  {Siscoe}}{2012b}]{2012hssr.book.....S}
{Schrijver}, C.~J. {\&} {Siscoe}, G.~L. (Eds.): 2012b,
\newblock {\em {Heliophysics: Space Storms and Radiation: Causes and
  Effects}\/},
\newblock Cambridge University Press, Cambridge, UK,
\newblock (Volume II)

\bibitem[\protect\citeauthoryear{{Schrijver} and {Siscoe}}{2015}]{heliophyicsv}
{Schrijver}, C.~J. {\&} {Siscoe}, G.~L. (Eds.): 2015,
\newblock {\em {Heliophysics: Space Weather and Society}\/},
\newblock published online at NASA's Heliophysics Summer School site,
\newblock
  https://cpaess.ucar.edu/sites/default/files/heliophysics/documents/HSS5.pdf
  (Volume V)

\bibitem[\protect\citeauthoryear{{Schrijver} and
  {Zwaan}}{2000}]{2000ssma.book.....S}
{Schrijver}, C.~J. {\&} {Zwaan}, C.: 2000,
\newblock {\em {Solar and Stellar Magnetic Activity}\/},
\newblock Cambridge University Press, Cambridge, UK

\bibitem[\protect\citeauthoryear{{Schrijver}}{2018}]{2018otbe.book.....S}
{Schrijver}, K.: 2018,
\newblock {\em {One of ten billion Earths: How we Learn about our Planet's Past
  and Future from Distant Exoplanets}\/},
\newblock Oxford University Press, Oxford, UK

\bibitem[\protect\citeauthoryear{{Schr{\"o}der} and {Connon
  Smith}}{2008}]{2008MNRAS.386..155S}
{Schr{\"o}der}, K.~P. {\&} {Connon Smith}, Robert: 2008,
\newblock \mnras 386(1), 155,
\newblock doi:10.1111/j.1365-2966.2008.13022.x

\bibitem[\protect\citeauthoryear{{Schwadron} {\em
  et~al.\/}}{2013}]{2013ApJ...775...86S}
{Schwadron}, N.~A., {Moebius}, E., {Kucharek}, H., {Lee}, M.~A., {French}, J.,
  {Saul}, L., {Wurz}, P., {Bzowski}, M., {Fuselier}, S.~A., {Livadiotis}, G.,
  {McComas}, D.~J., {Frisch}, P., {Gruntman}, M., {\&} {Mueller}, H.~R.: 2013,
\newblock \apj 775(2), 86,
\newblock doi:10.1088/0004-637X/775/2/86

\bibitem[\protect\citeauthoryear{{Sheeley}}{2005}]{2005LRSP....2....5S}
{Sheeley}, Neil~R., Jr.: 2005,
\newblock Living Reviews in Solar Physics 2(1), 5,
\newblock doi:10.12942/lrsp-2005-5

\bibitem[\protect\citeauthoryear{{Smithtro} and
  {Sojka}}{2005}]{2005JGRA..110.8305S}
{Smithtro}, C.~G. {\&} {Sojka}, J.~J.: 2005,
\newblock Journal of Geophysical Research (Space Physics) 110(A8), A08305,
\newblock doi:10.1029/2004JA010781

\bibitem[\protect\citeauthoryear{{Spruit}}{2013}]{2013arXiv1301.5572S}
{Spruit}, H.~C.: 2013,
\newblock arXiv e-prints  arXiv:1301.5572

\bibitem[\protect\citeauthoryear{{Stern} {\em
  et~al.\/}}{2019}]{2019Sci...364.9771S}
{Stern}, S.~A., {Weaver}, H.~A., {Spencer}, J.~R., {\&} {\em et al.}, {\null}:
  2019,
\newblock Science 364(6441), aaw9771,
\newblock doi:10.1126/science.aaw9771

\bibitem[\protect\citeauthoryear{{Struminsky} {\em
  et~al.\/}}{2018}]{2018Ge&Ae..58.1108S}
{Struminsky}, A.~B., {Sadovski}, A.~M., {\&} {Zharikova}, M.~S.: 2018,
\newblock Geomagnetism and Aeronomy 58(8), 1108,
\newblock doi:10.1134/S0016793218080169

\bibitem[\protect\citeauthoryear{{Thompson} {\em
  et~al.\/}}{2003}]{2003ARA&A..41..599T}
{Thompson}, M.~J., {Christensen-Dalsgaard}, J., {Miesch}, M.~S., {\&} {Toomre},
  J.: 2003,
\newblock \araa 41, 599,
\newblock doi:10.1146/annurev.astro.41.011802.094848

\bibitem[\protect\citeauthoryear{{Titov} and
  {D{\'e}moulin}}{1999}]{1999A&A...351..707T}
{Titov}, V.~S. {\&} {D{\'e}moulin}, P.: 1999,
\newblock \aap 351, 707

\bibitem[\protect\citeauthoryear{{T{\"o}r{\"o}k} {\em
  et~al.\/}}{2004}]{2004A&A...413L..27T}
{T{\"o}r{\"o}k}, T., {Kliem}, B., {\&} {Titov}, V.~S.: 2004,
\newblock \aap 413, L27,
\newblock doi:10.1051/0004-6361:20031691

\bibitem[\protect\citeauthoryear{{Tsurutani} {\em
  et~al.\/}}{2006}]{2006JGRA..111.7S01T}
{Tsurutani}, B.~T., {Gonzalez}, W.~D., {Gonzalez}, A.~L.~C., {Guarnieri},
  F.~L., {Gopalswamy}, N., {Grande}, M., {Kamide}, Y., {Kasahara}, Y., {Lu},
  G., {Mann}, I., {McPherron}, R., {Soraas}, F., {\&} {Vasyliunas}, V.: 2006,
\newblock Journal of Geophysical Research (Space Physics) 111(A7), A07S01,
\newblock doi:10.1029/2005JA011273

\bibitem[\protect\citeauthoryear{{Vasyliunas}}{1976}]{1976mgpa.proc...99V}
{Vasyliunas}, V.~M.: 1976,
\newblock in {\em Magnetospheric Particles and Fields\/}, p.~99

\bibitem[\protect\citeauthoryear{{Veronig} {\em
  et~al.\/}}{2021}]{2021NatAs...5..697V}
{Veronig}, Astrid~M., {Odert}, Petra, {Leitzinger}, Martin, {Dissauer}, Karin,
  {Fleck}, Nikolaus~C., {\&} {Hudson}, Hugh~S.: 2021,
\newblock Nature Astronomy 5, 697,
\newblock doi:10.1038/s41550-021-01345-9

\bibitem[\protect\citeauthoryear{{Wallner} {\em
  et~al.\/}}{2016}]{2016Natur.532...69W}
{Wallner}, A., {Feige}, J., {Kinoshita}, N., {Paul}, M., {Fifield}, L.~K.,
  {Golser}, R., {Honda}, M., {Linnemann}, U., {Matsuzaki}, H., {Merchel}, S.,
  {Rugel}, G., {Tims}, S.~G., {Steier}, P., {Yamagata}, T., {\&} {Winkler},
  S.~R.: 2016,
\newblock \nat 532(7597), 69,
\newblock doi:10.1038/nature17196

\bibitem[\protect\citeauthoryear{{Wang} {\em
  et~al.\/}}{2000}]{2000JGR...10525133W}
{Wang}, Y.-M., {Sheeley}, N.~R., {Socker}, D.~G., {Howard}, R.~A., {\&} {Rich},
  N.~B.: 2000,
\newblock \jgr 105(A11), 25133,
\newblock doi:10.1029/2000JA000149

\bibitem[\protect\citeauthoryear{{Wieler} {\em
  et~al.\/}}{2013}]{2013SSRv..176..351W}
{Wieler}, R., {Beer}, J., {\&} {Leya}, I.: 2013,
\newblock \ssr 176(1-4), 351,
\newblock doi:10.1007/s11214-011-9769-9

\bibitem[\protect\citeauthoryear{{Wood}}{2004}]{2004LRSP....1....2W}
{Wood}, B.~E.: 2004,
\newblock Living Reviews in Solar Physics 1, 2,
\newblock doi:10.12942/lrsp-2004-2

\bibitem[\protect\citeauthoryear{{Wood} {\em
  et~al.\/}}{2014}]{2014ApJ...781L..33W}
{Wood}, B.~E., {M{\"u}ller}, H.-R., {Redfield}, S., {\&} {Edelman}, E.: 2014,
\newblock \apjl 781(2), L33,
\newblock doi:10.1088/2041-8205/781/2/L33

\bibitem[\protect\citeauthoryear{{Wood} {\em
  et~al.\/}}{2005}]{2005ApJ...628L.143W}
{Wood}, B.~E., {M{\"u}ller}, H.~R., {Zank}, G.~P., {Linsky}, J.~L., {\&}
  {Redfield}, S.: 2005,
\newblock \apjl 628(2), L143,
\newblock doi:10.1086/432716

\bibitem[\protect\citeauthoryear{{Xiang} and {Rix}}{2022}]{2022Natur.603..599X}
{Xiang}, Maosheng {\&} {Rix}, Hans-Walter: 2022,
\newblock \nat 603(7902), 599,
\newblock doi:10.1038/s41586-022-04496-5

\bibitem[\protect\citeauthoryear{{Youngblood} {\em
  et~al.\/}}{2022}]{2022arXiv220101315Y}
{Youngblood}, A., {Pineda}, J.~S., {Ayres}, T., {France}, K., {Linsky}, J.~L.,
  {Wood}, B.~E., {Redfield}, S., {\&} {Schlieder}, J.~E.: 2022,
\newblock arXiv e-prints  arXiv:2201.01315

\bibitem[\protect\citeauthoryear{{Zahnle} {\em
  et~al.\/}}{2007}]{2007SSRv..129...35Z}
{Zahnle}, K., {Arndt}, N., {Cockell}, Ch., {Halliday}, A., {Nisbet}, E.,
  {Selsis}, F., {\&} {Sleep}, N.~H.: 2007,
\newblock \ssr 129(1-3), 35,
\newblock doi:10.1007/s11214-007-9225-z

\bibitem[\protect\citeauthoryear{{Zhang} {\em
  et~al.\/}}{2014}]{2014JGRA..119.5220Z}
{Zhang}, H., {Khurana}, K.~K., {Kivelson}, M.~G., {Angelopoulos}, V., {Wan},
  W.~X., {Liu}, L.~B., {Zong}, Q.~G., {Pu}, Z.~Y., {Shi}, Q.~Q., {\&} {Liu},
  W.~L.: 2014,
\newblock Journal of Geophysical Research (Space Physics) 119(7), 5220,
\newblock doi:10.1002/2014JA020111

\bibitem[\protect\citeauthoryear{{Zhang} {\em
  et~al.\/}}{2018}]{2018AJ....156..178Z}
{Zhang}, Z., {Zhou}, Y., {Rackham}, B.~V., {\&} {Apai}, D.: 2018,
\newblock \aj 156(4), 178,
\newblock doi:10.3847/1538-3881/aade4f

\end{thebibliography}
\addcontentsline{toc}{chapter}{Bibliography}

\markboth{Index}{Index}
{\footnotesize
  \printindex
}

\setcounter{chapter}{18}
\chapter*{Physical constants, plasma quantities, and vector identities}
\section*{Physical constants and other numbers}
\setcounter{table}{0}
\addcontentsline{toc}{chapter}{Physical constants, plasma quantities, and vector
identities}
\label{ch:constants}
  \begin{table}[h!]
    \caption[Selected physical constants.]{Selected physical constants (cgs).
    \label{tab:constants}}
    \begin{center}\begin{tabular}{llll}
      \hline \hline
      Quantity & Symbol & Value & Units \\
      \hline
      1\,eV equivalent wavelength & -  & $1.2\,10^{-4}$ & cm \\
      1\,eV equivalent energy & -  & $1.6\,10^{-12}$ & erg \\
      1\,eV equivalent temperature & -  & $1.2\,10^{4}$ & K \\
      arcsecond on the Sun & - & $\approx 725$ & km \\
      Astronomical Unit & $AU$ & $1.5\,10^{13}$ & cm \\
      Avogadro's number & $N_{\rm A}$ & $6.0\,10^{23}$ & mol$^{-1}$ \\
      Boltzmann constant & $k$ & $1.4\,10^{-16}$ & erg/K\\
      Earth mass & $M_\oplus$ & $6.0\,10^{27}$ & g \\
      Earth radius & $R_\oplus$ & $6.4\,10^8$ & cm \\
      Elementary charge   & $e$ & $4.8\,10^{-10}$ & esu \\
      Electron mass & $m_{\rm e}$ & $9.1\,10^{-28}$ & g \\
      Gravitational constant & $G$ & $6.7\,10^{-8}$ & dyne cm$^2$/g$^2$ \\
      Planck constant & $h$ & $6.6\,10^{-27}$ & erg s \\
      Proton mass & $m_{\rm p}$ & $1.7\,10^{-24}$ & g \\
      Solar Luminosity & $L_\odot$ & $3.8\, 10^{33}$ & erg/s \\
      Solar mass & $M_\odot$ & $2.0\, 10^{33}$ & g \\
      Solar radius & $R_\odot$ & $7.0\,10^{10}$ & cm \\
      Speed of light (vacuum) & $c$ & $3.0\,10^{10}$ & cm/s \\
      Stephan-Boltzmann constant & $\sigma$ & $5.7\,10^{-5}$ & erg/cm$^2$/s/K$^4$ \\
    \hline
    \end{tabular}\end{center}
    \end{table}

\markboth{Quantities and vector identities}{Quantities and vector identities}
\vfill\clearpage
\section*{Plasma quantities}
\label{ch:plquan}

% See NRL formulary and https://en.wikipedia.org/wiki/Plasma_parameters
%(CGS and eV) and
%http://plasma.physics.swarthmore.edu/brownpapers/basics.pdf (cgs and K).
\begin{table}[bh!]
  \indexit{dimensionless numbers}\indexit{plasma!quantities}
  \caption[Selected plasma quantities;
  summary.]{\label{tab:dimensionlessnumbers} Plasma
    quantities, mostly for thermal motions in fully-ionized plasmas.$^a$}
\begin{center}
\begin{tabular}{lll}
\hline\hline
Name/Symbol       & Value$^b$  & Description     \\
\hline
\multicolumn{3}{c}{{Frequencies and rates}}\\
$f_{\rm ge}$ (Hz)&$2.8\,10^6\,B$&electron gyrofrequency\\
$f_{\rm gi}$ (Hz)&$1.5\,10^3\,B\,\frac{Z}{\mu}$&ion gyrofrequency\\
&&\\
\hline\multicolumn{3}{c}{{{Thermal collision frequencies for fully ionized plasmas:}}}\\
$\nu_{\rm ee}$ (s$^{-1}$)&$3.6 \frac{n_{\rm e}}{T_{\rm e}^{3/2}} \ln{(\Lambda)}$ &electron-electron collision rate\\
$\nu_{\rm ii}$ (s$^{-1}$)&$0.06 \frac{n_{\rm i}}{T_{\rm
    i}^{3/2}} \frac{Z^4}{\mu^{1/2}} \ln{(\Lambda)}$ &ion-ion collision rate\\
$\nu_{\rm ei}$ (s$^{-1}$)&$\approx 0.5 \nu_{\rm ee}$ &electron-ion collision rate\\
&&\\
\hline\multicolumn{3}{c}{{Thermal length scales}}\\
$r_{\rm ge}$ (cm)&$0.022 \frac{T_{\rm e}^{1/2}}{B}$& electron \indexit{gyroradius}gyroradius\\
$r_{\rm gi}$ (cm)&$0.95 \frac{T_{\rm i}^{1/2}}{B} \frac{\mu^{1/2}}{Z}$& ion gyroradius\\
$\lambda_{\rm D}$ (cm)&$6.9 \frac{T^{1/2}}{n^{1/2}}$&Debye length\\
&&\\
\hline\multicolumn{3}{c}{{Velocities}}\\
$v_{\rm Te}$ (km/s)&$3.9\, T_{\rm e}^{1/2}$&electron thermal velocity\\
$v_{\rm Ti}$ (km/s)&$0.091\, T_{\rm i}^{1/2}\frac{1}{\mu^{1/2}}$&ion thermal velocity\\
$c_{\rm s}$ (km/s)&$0.091\, T_{\rm e}^{1/2}\frac{\gamma^{1/2}
  Z^{1/2}}{\mu^{1/2}}$&ion sound velocity\\
$v_{\rm A}$ (km/s) &$2.2\,10^{6} B \frac{1}{\mu^{1/2} n_{\rm
                     i}^{1/2}}$&Alfv{\'e}n velocity\\
&&\\
\hline\multicolumn{3}{c}{{Dimensionless numbers}}\\
electron Hall coeff.&$\frac{f_{\rm e}}{\nu_{\rm
    ee}}\approx 8.\, 10^{5} \frac{B T_{\rm
    e}^{3/2}}{n_{\rm e}}$&electron gyro- to collision freq. $^c$\\
ion Hall coefficient&$\frac{f_{\rm i}}{\nu_{\rm ii}}\approx 2.2\,10^{4} \frac{B T_{\rm
    i}^{3/2}}{n_{\rm i} Z^3 \mu^{1/2}}$&ion gyro- to collision frequency$^c$\\
plasma $\beta$&$3.5\,10^{-15} \frac{n_{\rm t} T}{B^2}$&thermal to magnetic energy$^d$\\
&&\\
\hline \hline
\end{tabular}
\end{center}
\vspace{-0.5\baselineskip}
{\em \small $^a$ See \url{https://www.nrl.navy.mil/ppd/content/nrl-plasma-formulary}; 
$^b$ in cgs-Gaussian units. Symbols: 
$B$ magnetic field strength (G); $T$ temperature in K, $n$
density in cm$^{-3}$; 
$\gamma$ the adiabatic index;
$\ln{(\Lambda)}$ the Coulomb logarithm (typically
10 to 20, {\em cf.}\ Table~\ref{tab:5.1});
$\mu$ ion mass in units of the proton mass;
$n_{e,i}$ electron or ion density;
$Z$ ion charge state;
$^c$ for $\ln{(\Lambda)}\approx 10$;
$^d$ where $n_{\rm t}$ the total particle density (ions plus electrons).}
\end{table}

\vfill\clearpage
\section*{Vector identities}
\label{ch:vidents}

\newcommand{\beq}{\begin{equation}}
\newcommand{\nb}{\bs{\nabla}}
\newcommand{\bs}{\boldsymbol}

\begin{table}[ph!]
  \caption{Vector identities. \label{tab:vecident}}
    \begin{center}\begin{tabular}{c}
      \hline \hline
$ 
{\bf A\cdot}({\bf B\times C})={\bf C\cdot}({\bf A\times B})={\bf B\cdot}({\bf C\times A})
$\\

$
{\bf A\times}({\bf B\times C})=({\bf A\cdot C}){\bf B}-({\bf A\cdot B}){\bf C}
$ \\

$ \bs{\nb}\times(\bs{\nb}\bf\times A)={\bs{\nb}}({\bs{\nb}\cdot A})-\nb^2{\bf A}$\\

$ 
{\bs{\nb}\cdot}(f{\bf A})=f{\bs{\nb}\bf\cdot A}+{\bf A\cdot\bs{\nb}}f
$\\

$
{\bs{\nb}\times}(f{\bf A})=\bs{\nb}f{\bf\times A}+f{\bs{\nb}\bf\times A}
$\\

$
{\bs{\nb}\cdot}({\bf A\times B})={\bf B\cdot}({\bs{\nb}\bf\times A})-{\bf A\cdot}({\bs{\nb}\bf\times B})
$\\

$ 
{\bs{\nb}\times}({\bf A\times B})={\bf A}({\bs{\nb}\bf\cdot B})-{\bf B}({\bs{\nb}\bf\cdot A})+({\bf B\cdot\bs{\nb}}){\bf A}-({\bf A\cdot\bs{\nb}}){\bf B}
$\\

$ \bs{\nb}\times\bs{\nb}f=0 $\\

$ {\bs{\nb}\cdot}({\bs{\nb}\bf\times A})=0$\\
    \hline
\end{tabular}\end{center}

\end{table}

\end{document}